\RequirePackage{lhchiggs}
\documentclass[letterpaper]{cernyrep}
\usepackage{color}
\usepackage{subfigure}
\usepackage{lscape}
\usepackage{axodraw}
\begin{document}

  

\thispagestyle{empty}
{
\setlength{\unitlength}{1mm}
\begin{picture}(0.001,0.001)
\put(135,8){CERN--2011--002}
\put(135,3){17 February 2011}
\put(0,-40){\large\bf ORGANISATION EUROP\'EENNE POUR LA RECHERCHE NUCL\'EAIRE}
\put(0,-50){\huge\bf CERN}
\put(23.35,-50){\large\bf EUROPEAN ORGANIZATION FOR NUCLEAR RESEARCH}

\put(18,-100){\LARGE\bfseries
            Handbook of LHC Higgs cross sections:}
\put(38,-110){\LARGE\bfseries
             1. Inclusive observables}
\put(9,-145){\Large\bfseries
             Report of the LHC Higgs Cross Section Working Group}
\put(117,-185){\Large Editors:}
\put(135,-185){\Large S.~Dittmaier}
\put(135,-191){\Large C.~Mariotti}
\put(135,-197){\Large G.~Passarino}
\put(135,-203){\Large R.~Tanaka}

\put(70,-250){\makebox(0,0){GENEVA}}
\put(70,-255){\makebox(0,0){2011}}
\end{picture}
}


\pagenumbering{roman}




\newpage

\leftline{\bf Conveners}


\noindent \emph{Gluon-Fusion process:}
         M.~Grazzini,\,  	 
         F.~Petriello,\,
         J.~Qian,\,  	 
         F.~St\"ockli 	    	 
\vspace{0.1cm}

\noindent \emph{Vector-Boson-Fusion process:} 	
         A.~Denner,\, 	
         S.~Farrington,\, 	
         C.~Hackstein,\,
         C.~Oleari,\,
         D.~Rebuzzi 
\vspace{0.1cm}

\noindent{$\PW\PH/\PZ\PH$ \emph{production mode}:} 	
         S.~Dittmaier,\, 	
         R.~Harlander,\,
         C.~Matteuzzi,\, 	
         J.~Olsen,\, 	
         G.~Piacquadio
\vspace{0.1cm}

\noindent{$\PQt\PQt\PH$ \emph{process}:} 	
         C.~Neu,\, 	  	
         C.~Potter,\, 	
         L.~Reina,\, 	
         M.~Spira
\vspace{0.1cm}

\noindent\emph{MSSM neutral Higgs:} 	
         M.~Spira,\, 	
         M.~Vazquez Acosta,\, 	  	
         M.~Warsinsky,\, 	
         G.~Weiglein 
\vspace{0.1cm}

\noindent\emph{MSSM charged Higgs:} 	
         M.~Flechl,\, 	
         M.~Kr\"amer,\, 	
         S.~Lehti,\, 	  	
         T.~Plehn
\vspace{0.1cm}

\noindent\emph{PDF:} 	
         S.~Forte,\, 	
         J.~Huston,\, 	
         K.~Mazumdar,\, 	  	
         R.~Thorne
\vspace{0.1cm}

\noindent\emph{Branching ratios:} 	
         A.~Denner,\, 	
         S.~Heinemeyer,\,
         I.~Puljak,\, 	  	
         D.~Rebuzzi
\vspace{0.1cm}

\noindent\emph{NLO MC:}
         M.~Felcini,\, 	  	
         F.~Maltoni,\, 	
         P.~Nason,\,
         J.~Yu	
\vspace{0.1cm}

\noindent\emph{Higgs Pseudo-Observables:} 	
         M.~D\"uhrssen,\, 	
         M.~Felcini,\, 	  	
         S.~Heinemeyer,\, 	
         G.~Passarino 


\vfill
\begin{flushleft}
\begin{tabular}{@{}l@{~}l}
  ISBN & 978--92--9083--358-1 \\
  ISSN & 0007--8328\\ 
\end{tabular}\\[1mm]
Copyright \copyright{} CERN, 2011\\[1mm]
\raisebox{-1mm}{\includegraphics[height=12pt]{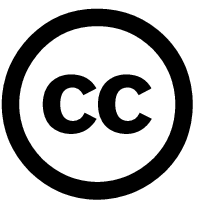}}
 Creative Commons Attribution 3.0\\[1mm]
Knowledge transfer is an integral part of CERN's mission.\\[1mm]
CERN publishes this report Open Access under the Creative Commons
Attribution 3.0 license (\texttt{http://creativecommons.org/licenses/by/3.0/})
in order to permit its wide dissemination and use.\\[3mm]
This Report should be cited as:\\[1mm]
LHC Higgs Cross Section Working Group,
S.~Dittmaier, C.~Mariotti, G.~Passarino, R.~Tanaka (Eds.), \\
\emph{Handbook of LHC Higgs Cross Sections: 1. Inclusive Observables}, \\
CERN-2011-002 (CERN, Geneva, 2011).\\[3mm]
\end{flushleft}

\newpage
\vspace*{10cm}
\begin{center} 
 \bf {Abstract}
\end{center}
\vspace{0.5cm}
This Report summarizes the results of the first 10 months' activities of 
the LHC Higgs Cross Section Working Group.
The main goal of the working group was to present the state of the art of
Higgs Physics at the LHC, integrating all new results that have appeared in the
last few years.
The Report is more than a mere collection of the proceedings of
the general meetings. The subgroups have been working in different 
directions. An attempt has been made to present the first Report from these
subgroups in a complete and homogeneous form. The subgroups' contributions
correspondingly comprise the main parts of the Report.
A significant amount of work has been performed in providing higher-order
corrections to the Higgs-boson cross sections and pinning down the
theoretical uncertainty of the Standard Model predictions.
This Report comprises explicit numerical results on total cross sections,
leaving the issues of event selection cuts and differential distributions to future publications.
The subjects for further study are identified.

\newpage
\vspace*{10cm}
\begin{center}
We, the authors, would like to dedicate this Report to the memory of \\
Nicola Cabibbo and Georges Charpak.
\end{center}

\newpage

\begin{flushleft}

S.~Dittmaier$^{1}$,\,
C.~Mariotti$^{2}$,\,
G.~Passarino$^{2,3}$\,and\,
R.~Tanaka$^{4}$\,(eds.);\\
J.~Baglio$^{5}$,\,
P.~Bolzoni$^{6}$,\,
R.~Boughezal$^{7}$,\, 
O.~Brein$^{1}$,\,
C.~Collins-Tooth$^{8}$,\,
S.~Dawson$^{9}$,\,
S.~Dean$^{10}$,\,
A.~Denner$^{11}$,\,
S.~Farrington$^{12}$,\, 
M.~Felcini$^{13}$,\,
M.~Flechl$^{1}$,\,
D.~de Florian$^{14}$,\,
S.~Forte$^{15}$,\,
M.~Grazzini$^{16}$,\, 
C.~Hackstein$^{17}$,\,
T.~Hahn$^{18}$,\, 
R.~Harlander$^{19}$,\,
T.~Hartonen$^{20}$,\,
S.~Heinemeyer$^{13}$,\,
J.~Huston$^{21}$,\,
A.~Kalinowski$^{22}$,\, 
M.~Kr\"amer$^{23}$,\,
F.~Krauss$^{24}$,\,
J.S.~Lee$^{25}$,\, 
S.~Lehti$^{20}$,\,
F.~Maltoni$^{26}$,\,
K.~Mazumdar$^{27}$,\,
S.-O.~Moch$^{28}$,\,
A.~M\"uck$^{23}$,\,
M.~M\"uhlleitner$^{17}$,\,
P.~Nason$^{29}$,\,
C.~Neu$^{30}$,\,
C.~Oleari$^{29}$,\,
J.~Olsen$^{31}$,\, 
S.~Palmer$^{30}$,\,
F.~Petriello$^{7,32}$,\,
G.~Piacquadio$^{33}$,\,
A.~Pilaftsis$^{34}$,\, 
C.T.~Potter$^{35}$,\,
I.~Puljak$^{36}$,\,
J.~Qian$^{37}$,\, 
D.~Rebuzzi$^{38}$,\,
L.~Reina$^{39}$,\,
H.~Rzehak$^{1,17}$,\,
M.~Schumacher$^{1}$,\,
P.~Slavich$^{40}$,\, 
M.~Spira$^{41}$,\,
F.~St\"ockli$^{33}$,\,
R.S.~Thorne$^{10}$,\,
M.~Vazquez Acosta$^{42}$\,
T.~Vickey$^{12,43}$,\,  
A.~Vicini$^{15}$,\,
D.~Wackeroth$^{44}$,\,
M.~Warsinsky$^{1}$,\,
M.~Weber$^{18}$,\,
G.~Weiglein$^{45}$,\, 
C.~Weydert$^{46}$,\, 
J.~Yu$^{47}$,\,
M.~Zaro$^{26}$,\,
and
T.~Zirke$^{19}$.
\end{flushleft}

\begin{itemize}

\item[$^{1}$] 
  Physikalisches Institut, Albert-Ludwigs-Universit\"at Freiburg,
  D-79104 Freiburg, Germany

\item[$^{2}$] 
 INFN, Sezione di Torino, Via P. Giuria 1, 10125 Torino, Italy

\item[$^{3}$] 
  Dipartimento di Fisica Teorica, Universit\`a di Torino,  Via P.
  Giuria 1, 10125 Torino, Italy

\item[$^{4}$] 
  Laboratoire de l'Acc\'el\'erateur Lin\'eaire, CNRS/IN2P3, 
  F-91898 Orsay CEDEX, France


\item[$^{5}$] 
 Laboratoire de Physique Th\'eorique, Universite Paris XI et CNRS, 
 F-91405 Orsay, France 

\item[$^{6}$] 
  II. Institut f\"ur Theoretische Physik, Universit\"at Hamburg, \\
  Luruper Chaussee 149, D-22761 Hamburg, Germany

\item[$^{7}$] 
  High Energy Physics Division, Argonne National Laboratory, \\
  Argonne, IL 60439, USA 

\item[$^{8}$] 
  Department of Physics and Astronomy, University of Glasgow, \\
  Glasgow G12 8QQ, UK

\item[$^{9}$] 
 Department of Physics, Brookhaven National Laboratory, \\
 Upton, NY 11973, USA 

\item[$^{10}$] 
  Department of Physics and Astronomy, University College London, \\
  Gower Street, London WC1E 6BT, UK 

\item[$^{11}$] 
  Institut f\"ur Theoretische Physik und Astrophysik, Universit\"at W\"urzburg, \\ 
  Am Hubland, D-97074 W\"urzburg, Germany

\item[$^{12}$] 
  Department of Physics, University of Oxford, Denys Wilkinson Building, \\
  Keble Road, Oxford OX1 3RH, UK

\item[$^{13}$] 
  Instituto de F\'isica de Cantabria (IFCA), CSIC-Universidad de
  Cantabria, \\
  Santander, Spain

\item[$^{14}$] 
  Departamento de F\'isica, Facultad de Ciencias Exactas y Naturales\\ 
  Universidad de Buenos Aires, 
  Pabellon I, Ciudad Universitaria (1428) \\
  Capital Federal, Argentina 

\item[$^{15}$] 
  Dipartimento di Fisica, Universit\`a degli Studi di Milano and
  INFN,\\ Sezione di Milano, Via Celoria 16, I-20133 Milan, Italy

\item[$^{16}$] 
  INFN, Sezione di Firenze and Dipartimento di Fisica e Astronomia, Universit\`a di Firenze,\\
  I-50019 Sesto Fiorentino, Florence, Italy 

\item[$^{17}$] 
  Institut f\"ur Theoretische Physik und Institut f\"ur Experimentelle
  Teilchenphysik, \\ 
  Karlsruhe Institut of Technology, D-76131 Karlsruhe, Germany

\item[$^{18}$] 
  Max-Planck-Institut f\"ur Physik, Werner-Heisenberg-Institut,\\
  F\"ohringer Ring 6, D-80805 M\"unchen, Germany

\item[$^{19}$] 
  Bergische Universit\"at Wuppertal, D-42097 Wuppertal, Germany  

\item[$^{20}$] 
  Helsinki Institute of Physics, P.O. Box 64, FIN-00014 University of Helsinki, Finland

\item[$^{21}$] 
  Department of Physics and Astronomy, Michigan State University, \\
  East Lansing, MI 48824, USA

\item[$^{22}$] 
  Faculty of Physics, University of Warsaw, Hoza 69, 00-681 Warsaw, Poland 

\item[$^{23}$] 
 Institut f\"ur Theoretische Teilchenphysik und Kosmologie,
 RWTH Aachen University, \\
 D-52056 Aachen, Germany 

\item[$^{24}$] 
  Institute for Particle Physics Phenomenology, Department of Physics, \\
  University of Durham, Durham DH1 3LE, UK

\item[$^{25}$] 
  National Center for Theoretical Sciences, \\ 
  101, Section 2 Kuang Fu Road Hsinchu, Taiwan 300, Republic of China

\item[$^{26}$] 
  Centre for Cosmology, Particle Physics and Phenomenology (CP3), \\ 
  Universit\'e Catholique de Louvain, B-1348 Louvain-la-Neuve, Belgium

\item[$^{27}$] 
  Tata Institute of Fundamental Research, Homi Bhabha Road, Mumbai 400 005, India 

\item[$^{28}$] 
  DESY, Zeuthen, Platanenallee 6, D-15738 Zeuthen, Germany

\item[$^{29}$] 
  Universit\`a di Milano-Bicocca and INFN, Sezione di Milano-Bicocca,\\
  Piazza della Scienza 3, 20126 Milan, Italy

\item[$^{30}$] 
  University of Virginia, Charlottesville, VA 22906, USA 

\item[$^{31}$] 
  Department of Physics, Princeton University, Princeton, NJ 08542, USA

\item[$^{32}$] 
  Department of Physics \& Astronomy, Northwestern University, Evanston, IL 60208, USA

\item[$^{33}$] 
  CERN, CH-1211 Geneva 23, Switzerland 

\item[$^{34}$] 
  School of Physics and Astronomy, University of Manchester, Manchester M13 9PL, UK

\item[$^{35}$] 
  Department of Physics, University of Oregon, Eugene, OR 97403-1274, USA 

\item[$^{36}$] 
  University of Split, FESB, R. Boskovica bb, 21 000 Split, Croatia

\item[$^{37}$] 
  Department of Physics, University of Michigan, Ann Arbor, MI 48109, USA 

\item[$^{38}$] 
  Universit\`a di Pavia and INFN, Sezione di Pavia, Via A. Bassi, 6, 27100 Pavia, Italy

\item[$^{39}$] 
  Physics Department, Florida State University, Tallahassee, FL 32306-4350, USA

\item[$^{40}$] 
  Laboratoire de Physique Th\'eorique et des Hautes Energies, \\
  4 Place Jussieu, F-75252 Paris CEDEX 05, France

\item[$^{41}$] 
  Paul Scherrer Institut, CH--5232 Villigen PSI, Switzerland

\item[$^{42}$] 
  Physics Dept., Blackett Laboratory, Imperial College London, \\
  Prince Consort Rd, London SW7 2BW, UK

\item[$^{43}$] 
  School of Physics, University of the Witwatersrand, Private Bag 3, \\
  Wits 2050, Johannesburg, South Africa

\item[$^{44}$] 
  Department of Physics, SUNY at Buffalo, Buffalo, NY 14260-1500, USA

\item[$^{45}$] 
  DESY, Notkestrasse 85, D-22607 Hamburg, Germany

\item[$^{46}$] 
  Laboratory for Subatomic Physics and Cosmology, Universit\'e Joseph
  Fourier, \\ 
  (Grenoble 1),  F-38026 Grenoble CEDEX, France

\item[$^{47}$] 
   Department of Physics, Univ. of Texas at Arlington, 
   SH108, University of Texas,\\
    Arlington, TX 76019, USA 

\end{itemize}

\newpage
\begin{center}
 {\bf Prologue}
\end{center}
\vspace{0.5cm}
The implementation of spontaneous symmetry breaking in the framework
of gauge theories in the 1960s triggered the breakthrough in the
construction of the standard electroweak theory, as it still persists
today. The idea of driving the spontaneous breakdown of a gauge
symmetry by a self-interacting scalar field, which thereby lends mass
to gauge bosons, is known as the {\it Higgs mechanism} and goes back
to the early work of 
\Brefs{Higgs:1964ia,Higgs:1964pj,Higgs:1966ev,Englert:1964et,Guralnik:1964eu}. 
The postulate of a
new scalar neutral boson, known as the {\it Higgs particle}, comes as
a phenomenological imprint of this mechanism. Since the birth of this
idea, the Higgs boson has successfully escaped detection in spite
of tremendous search activities at the high-energy colliders
LEP and Tevatron, leaving open the crucial question whether the
Higgs mechanism is just a theoretical idea or a `true model'
for electroweak symmetry breaking. The experiments at the Large Hadron
Collider (LHC) will answer this question, either positively upon detecting 
the Higgs boson, or negatively by ruling out the existence of a particle with
properties attributed to the Higgs boson within the Standard Model.
In this sense the outcome of the Higgs search at the LHC will
either carve our present understanding of electroweak interactions
in stone or will be the beginning of a theoretical revolution.


\newpage
\mbox{}

\newpage
\tableofcontents

\newpage
\pagenumbering{arabic}
\setcounter{footnote}{0}

\section{Introduction\footnote{S.~Dittmaier, C.~Mariotti, G.~Passarino
    and R.~Tanaka}}

After the start of $\Pp\Pp$ collisions at the LHC the natural question is: Why 
precision Higgs physics now? The LHC successfully started at the end of 2009 colliding two proton 
beams at centre-of-mass energies of $\sqrt{s}= 0.9$\UTeV\  
and $2.36$\UTeV. 
In 2010 the energy has been raised up to $7$\UTeV. 

By the end of the $7$\UTeV\ run in 2011 and (likely at $8$\UTeV) in 2012  
each experiment aims to collect an integrated luminosity of a few inverse femtobarns. 
Then a long shutdown will allow the implementation of necessary modifications to 
the machine, to restart again at the design energy of $14$\UTeV. 
By the end of the life of the LHC, each experiment will have collected
$3000$\ifb on tape. 
The luminosity that the experiments expect to collect with the $7$\UTeV\ run 
will allow us to probe a wide range of the Higgs-boson mass. Projections of 
ATLAS and CMS 
when combining only 
the three main channels ($\PH \rightarrow \PGg\PGg, \PH \rightarrow \PZ\PZ,\PH 
\rightarrow \PW\PW$), indicate that in case of no observed excess, the 
Standard Model (SM) Higgs boson can be excluded in the range between
$140$\UGeV\ and $200$\UGeV.
A $5\sigma$ significance can be reached for a Higgs-boson mass range 
between $160$\UGeV\ and $170$\UGeV. 
The experiments (ATLAS, CMS, and LHCb) are now analysing more channels 
in order to increase their potential for exclusion at lower and higher 
masses.
For these reasons an update of the discussion of the proper definition of 
the Higgs-boson mass and width has become necessary.
Indeed, in this scenario, it is of utmost importance to access the best 
theory predictions for the Higgs cross sections and branching ratios,
using definitions of the Higgs-boson properties that are objective 
functions of the experimental data while respecting first principles of 
quantum field theory.
In all parts we have tried to give a widely homogeneous summary for the
precision observables. Comparisons among the various groups of authors
are documented reflecting the status of our theoretical knowledge.
This may be understood as providing a common opinion about the present
situation in the calculation of Higgs cross sections and their theoretical 
and parametric errors.

The experiments have a coherent plan for using the input suggestions
of the theoretical community to facilitate the combination of the 
individual results. Looking for precision tests of theoretical models 
at the level of their quantum structure requires the highest standards on the 
theoretical side as well.
Therefore, this Report is the result of a workshop started as an appeal
by experimentalists. Its progress over the subsequent months to its 
final form was possible only because of a close contact between the 
experimental and theory communities. 

The major sections of this Report are devoted to discussing the
computation of cross sections and branching ratios for the SM 
Higgs and for the Minimal Supersymmetric Standard Model (MSSM) Higgs bosons,
including the still-remaining theoretical uncertainties. 
The idea of presenting updated calculations on Higgs physics was triggered by 
experimentalists and is substantiated as far as possible in this Report.
The working group was organized in $10$ subgroups. The first four 
address different Higgs production modes: gluon--gluon fusion, 
vector-boson fusion, Higgs-strahlung, and associated production with 
top-quark pairs. Two more groups are focusing on MSSM neutral and MSSM charged 
Higgs production. One group is dedicated to the prediction of the 
branching ratios (BR) of Higgs bosons in the
SM and MSSM. Another group studies predictions from
different Monte Carlo (MC) codes at next-to-leading order (NLO) and  
their matching to parton-shower MCs. The definition of 
Higgs pseudo-observables is also a relevant part of this analysis, in 
order to correctly match the experimental observables and the 
theoretical definitions of physical quantities.
Finally, a group is devoted to parton density functions (PDFs), 
in particular to the issue of new theoretical input related to PDFs, 
in order to pin down the theoretical uncertainty on cross sections.

To  discover or exclude certain Higgs-boson mass regions 
different inputs are needed:
\begin{itemize}
\item SM cross sections and BR in order to produce predictions;
\item theoretical uncertainties on these quantities. These 
uncertainties enter also the determination of systematic errors of the 
mean value.
\end{itemize}

Furthermore, common and correlated theoretical inputs 
(cross sections, PDFs, SM and MSSM parameters, etc.) require the
highest standards on the theoretical side.
The goal has been to give precise common inputs to the experiments to 
facilitate the combination of multiple Higgs search channels.

The structure of this Report centres on a description of cross sections 
computed at next-to-next-to-leading order (NNLO) or NLO, for each of the production 
modes. Comparisons among the various groups of authors for the central value
and the range of uncertainty are documented and reflect the 
status of our theoretical knowledge. Note that all the central values 
have been computed using the same SM parameters input, as presented in table 
\refT{tab:SMinput} of the Appendix.
An update of the previous discussions of theoretical uncertainties has 
become necessary for several reasons:
\begin{itemize}
\item The PDF uncertainty has been computed following the PDF4LHC prescription 
as described in \Sref{pdfsection} of this Report.
\item The $\alphas$ uncertainty has been added in quadrature to the 
PDF variation.
\item The renormalization and factorization QCD scales have been varied 
following the criterion of pinning down, as much as possible, the 
theoretical uncertainty. It often remains the largest of the uncertainties.
\end{itemize}
A final major point is that, for this Report, all cross sections have
been computed within an inclusive setup, not taking into account the
experimental cuts and the acceptance of the apparatus. A dedicated study 
of these effects (cuts on the cross sections and on $K$-factors) will be 
presented in a future publication.

The final part of this Report is devoted to describing a new direction of 
work: what the experiments observe in the final state is not always 
directly connected to a well defined  theoretical quantity. 
We have to take into account the acceptance of the detector, the
definition of {\em signal}, the interference {\em signal--background}, 
and all sorts of approximations built into the Monte Carlo codes.
As an example at LEP, the line shape of the $\PZ$ for the final state with 
two electrons has to be extracted from the cross section of the process 
($\Pep\Pem\rightarrow \Pep\Pem$), after having subtracted the contribution 
of the photon and the interference between the photon and the $\PZ$. 
A corrected definition of the Higgs-boson mass and width is needed. Both are
connected to the corresponding complex pole in the $p^2$ plane
of the propagator with momentum transfer $p$. We claim that the correct 
definition of mass of an unstable particle has to be used in Monte 
Carlo generators.

Different Monte Carlo generators exist at LO and NLO. It was important 
to compare their predictions and to stress the corresponding differences, 
also taking into account the different algorithms used for 
parton shower. 
Note that NLO matrix-element generators matched with a parton shower 
are the tools for the future.
Beyond the goals of this Report remains the agreement between NLO MC 
predictions and NNLO calculations within the acceptance of the detectors.
The next step in the activities of this working group will be
the computation of cross sections that include acceptance cuts and differential
distributions for all final 
states that will be considered in the Higgs search at the LHC.
Preferably this should be carried out with the same set of (benchmark) cuts for
ATLAS and CMS. The goal is to understand how the $K$-factors from (N)LO 
to (N)NLO will change after introduction of cuts and to compare the NNLO 
differential distributions with the ones from Monte Carlo generators at 
NLO.
There is a final comment concerning the SM background: we plan to estimate
theoretical predictions for the most important backgrounds in the 
signal regions. This means that a {\em background control region} 
has to be defined, and there the experiments will measure a given 
source of background directly from data. 
The {\em control region} can be in the bulk of the background production 
phase space, but can also be in the tail of the distributions. Thus it 
is important to define the precision with which the SM background will be 
measured and the theoretical precision available for that particular 
region. Then the background uncertainty should be extrapolated back to 
the {\em signal region}, using available theoretical predictions and their 
uncertainty.
It will be important to compute  the interference between signal and 
background and try to access this at NLO.
The (N)LO Monte Carlos will be used to simulate this background and 
determine how the $K$-factor is changing with the chosen kinematic cuts.

The present documentation is the result of a workshop that started in
January 2010 as a new joint effort for Higgs cross sections between ATLAS, CMS,
and the theory community.

In this Report the Higgs-boson cross section calculations are presented 
at the energy of the first $\Pp\Pp$ run, $7$\UTeV, as well as at the nominal 
one ($14$\UTeV). Updated tables at the future energy will be made
available at the twiki page: 
https://twiki.cern.ch/twiki/bin/view/LHCPhysics/CrossSections .



\newpage

\newcommand{\ccaption}[2]{
\caption[#1]{{{#2}}}
}

\providecommand{\lsim}
{\;\raisebox{-.3em}{$\stackrel{\displaystyle <}{\sim}$}\;}
\providecommand{\gsim}
{\;\raisebox{-.3em}{$\stackrel{\displaystyle >}{\sim}$}\;}

\section{Gluon-Fusion process\footnote{M. Grazzini, F. Petriello,
J. Qian, F. Stoeckli (eds.); J. Baglio, R. Boughezal and D. de Florian.}}
\label{ggFsection}

\subsection{Higgs-boson production in gluon--gluon fusion}

Gluon fusion through a heavy-quark loop~\cite{Georgi:1977gs} (see \Fref{fig:triangle}) is the main production mechanism of the Standard Model 
Higgs boson at hadron colliders.  When combined with the decay channels $\PH\to\PGg\PGg$, $\PH\to \PW\PW$, and $\PH\to \PZ\PZ$, this production
mechanism is one of the most important for Higgs-boson searches and studies over the entire mass
range, $100\UGeV \lsim \MH \lsim 1\UTeV$, to be investigated at the LHC.

\begin{figure}[hbt]
\begin{center}
\SetScale{0.8}
\begin{picture}(120,80)(0,0)
\Gluon(0,20)(50,20){-3}{5}
\Gluon(0,80)(50,80){3}{5}
\ArrowLine(50,20)(50,80)
\ArrowLine(50,80)(100,50)
\ArrowLine(100,50)(50,20)
\DashLine(100,50)(150,50){5}
\Vertex(50,20){2}
\Vertex(50,80){2}
\Vertex(100,50){2}
\put(124,37){$\PH$}
\put(17,37){$\Pt,\Pb$}
\put(-12,14){$\Pg$}
\put(-12,62){$\Pg$}
\end{picture}  \\
\caption{\label{fig:triangle} Feynman diagram
contributing to $\Pg\Pg\to\PH$ at lowest order.}
\end{center}
\end{figure}
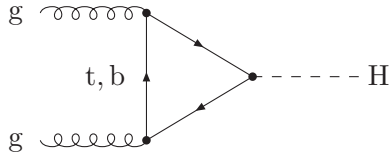

The dynamics of the gluon-fusion mechanism is controlled by strong interactions. Detailed studies
of the effect of QCD radiative corrections are thus necessary to obtain accurate theoretical predictions.
In QCD perturbation theory, the leading order (LO) contribution \cite{Georgi:1977gs}
to the gluon-fusion cross section is proportional to $\alphas^2$, where $\alphas$ is the QCD coupling constant.  The 
main contribution arises from the top quark, due to its large Yukawa coupling to the Higgs boson.  The QCD radiative corrections to this 
process at next-to-leading order (NLO) have been known for some time, both in the large-$\Mt$ limit~\cite{Dawson:1991zj,Djouadi:1991tka} and maintaining the full top- and bottom-quark mass dependence~\cite{Graudenz:1992pv,Spira:1995rr}. They increase the 
LO cross section by about $80{-}100\%$ at the LHC.  The exact calculation is very well approximated by the
large-$\Mt$ limit. When the exact Born cross section with the full dependence on the mass of the top quark is used
to normalize the result, the difference between the exact and the approximated NLO
cross sections is only a few percent.  The next-to-next-to-leading order (NNLO) corrections have been computed only in this 
limit~\cite{Harlander:2000mg,Catani:2001ic,Harlander:2001is,Harlander:2002wh,Anastasiou:2002yz,Ravindran:2003um,Blumlein:2005im},
leading to an additional increase of the cross section of about 
$25\%$.
The NNLO calculation has been consistently improved by resumming the soft-gluon contributions
up to NNLL \cite{Catani:2003zt}. The result leads to an additional
increase of the cross section of about $7{-}9\%$ ($6{-}7\%$) at
$\sqrt{s}=7$ $(14)\UTeV$.
The NNLL result is nicely confirmed by the evaluation of the leading soft contributions at N$^3$LO \cite{Moch:2005ky,Laenen:2005uz,Idilbi:2005ni,Ravindran:2005vv,Ravindran:2006cg}.

Recent years have seen further progress in the computation of radiative corrections and in the assessment
of their uncertainties. The accuracy of the large-$\Mt$ approximation at NNLO has been studied in
\Brefs{Marzani:2008az,Harlander:2009bw,Harlander:2009mq,Harlander:2009my,Pak:2009bx,Pak:2009dg}. These papers have definitely shown that if the Higgs 
boson is relatively light 
($\MH\lsim 300\UGeV$), the large-$\Mt$ approximation works extremely well,
to better than $1\%$. As discussed below, these results allow us to formulate accurate theoretical predictions where the top and bottom loops 
are treated exactly up to NLO, and the higher-order corrections to the top contribution
are treated in the large-$\Mt$ approximation~\cite{Anastasiou:2008tj}.

Considerable work has also been done in the evaluation of electroweak (EW) corrections.
Two-loop EW effects are now known \cite{Djouadi:1994ge,Aglietti:2004nj,Degrassi:2004mx,Actis:2008ts,Actis:2008ug}.  
They increase the cross section by a factor that strongly depends on the Higgs-boson mass, changing from 
$+5\%$ for $\MH=120$\UGeV\ to about $-2\%$ for $\MH=300$\UGeV\ \cite{Actis:2008ug}.
The main uncertainty in the EW analysis comes from the fact that it is not obvious how to combine them with the large QCD corrections.
In the {\it partial factorization} scheme of \Bref{Actis:2008ug}
the EW correction applies only to the LO result. In the {\em complete factorization} scheme, the EW correction instead multiplies the full 
QCD-corrected cross section.
Since QCD corrections are sizeable, this choice has a non-negligible
effect on the actual impact of EW corrections in the computation.
The computation of the dominant mixed QCD--EW effects due to light quarks~\cite{Anastasiou:2008tj},
performed using an effective-Lagrangian approach,
supports the complete factorization hypothesis, suggesting that
EW corrections become a multiplicative factor times the full QCD expansion.  This result should be interpreted carefully 
since the effective theory is strictly valid only when $\MH\muchless\MW$.  However, as discussed later, it is expected to be a good approximation to 
the exact result 
for Higgs-boson masses below several hundred \UGeV\ for the same reasons that the large-$\Mt$ limit furnishes a good approximation to the 
exact top-mass dependent calculation up to nearly $\MH = 1$\UTeV.  Very recently, EW effects for
Higgs production at finite transverse momentum \cite{Keung:2009bs,Brein:2010xj} 
have also been studied. Their effect is at the $1\%$ level or smaller.

In the following we present the results of three updated computations, based 
on the work presented in \Brefs{Anastasiou:2008tj,deFlorian:2009hc}
(see \Sref{se:ggfXS1}) and \Brefs{Baglio:2010um,Baglio:2010ae}
(see \Sref{se:ggfXS2}).%
\footnote{The central values of these cross-section predictions are in 
good mutual agreement, but the error assessment -- in particular
of theoretical errors that go beyond mere scale uncertainties --
is still under debate, leading to this splitting
of cross-section predictions into parts I and II.
It is worth noting that both calculations (ABPS and dFG) include the 
exact NLO mass dependence already. Also, the $\PQb$ mass parametric 
error should be accounted for by scale variations.
The combined numbers in the Summary, \Sref{se:summary}, are based
on the two predictions (ABPS and dFG) of the next section;
the inclusion of the BD analysis described in \Sref{se:ggfXS2},
with a common combination of all uncertainties, is in progress.}
These calculations use MSTW2008 NNLO parton distribution functions (PDFs)~\cite{Martin:2009iq}.

\subsection{Cross-section predictions I}
\label{se:ggfXS1}

The following predictions are based on calculations by
Anastasiou/Boughezal/Petriello/Stoeckli and by de Florian/Grazzini. 

The calculation by Anastasiou, Boughezal, Petriello and Stoeckli (ABPS) \cite{Anastasiou:2008tj}
starts from the exact NLO cross section with full dependence on the top- and bottom-quark masses
and includes the NNLO top-quark contribution
in the large-$\Mt$ limit. The result includes EW contributions~\cite{Aglietti:2004nj,Degrassi:2004mx,Actis:2008ug,Actis:2008ts} according 
to~\Brefs{Actis:2008ug,Actis:2008ts}, evaluated in the complete factorization scheme.  Mixed QCD--EW contributions \cite{Anastasiou:2008tj} are 
also accounted for, together with some effects from EW corrections at finite transverse momentum~\cite{Keung:2009bs}. The effect of soft-gluon 
resummation is mimicked
by choosing
the central value of the renormalization and factorization scales
as $\mu_R=\mu_F=\MH/2$.
The latter choice is also motivated by an improved convergence of the fixed-order QCD perturbative expansion.


The calculation by de Florian and Grazzini (dFG)
is a slightly improved version on the calculation presented in \Bref{deFlorian:2009hc}.
The starting point is the exact NLO cross section with full dependence on
the top- and bottom-quark masses, computed with the program {\sc HIGLU} \cite{Graudenz:1992pv,Spira:1995rr}, on top of which the NLL resummation of soft-gluon contributions is included.
Then, the top-quark contribution is considered and the
NNLL+NNLO corrections \cite{Catani:2003zt} are consistently added in the large-$\Mt$ limit. The result is finally corrected for EW 
contributions~\cite{Aglietti:2004nj,Degrassi:2004mx,Actis:2008ug,Actis:2008ts} according to~\Brefs{Actis:2008ug,Actis:2008ts} in 
the complete factorization scheme. The central value of factorization and renormalization scales is chosen
to be $\mu_F=\mu_R=\MH$.  The results of this calculation are available through an online calculator \cite{calculator}.

\begin{figure}[htb]
\vspace{5pt}
\begin{center}
\begin{tabular}{cc}
\includegraphics[width=.46\linewidth]{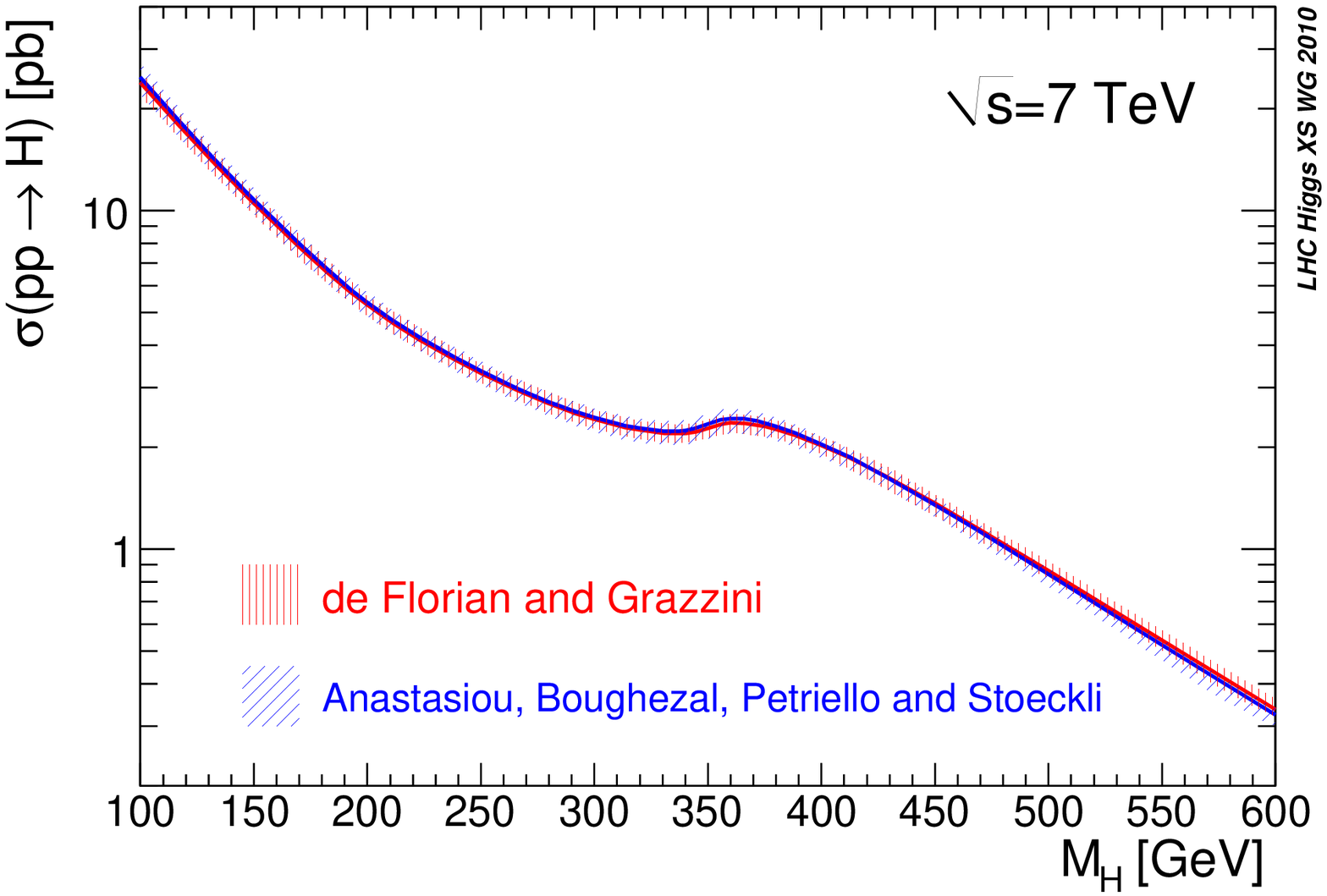} &
\includegraphics[width=.46\linewidth]{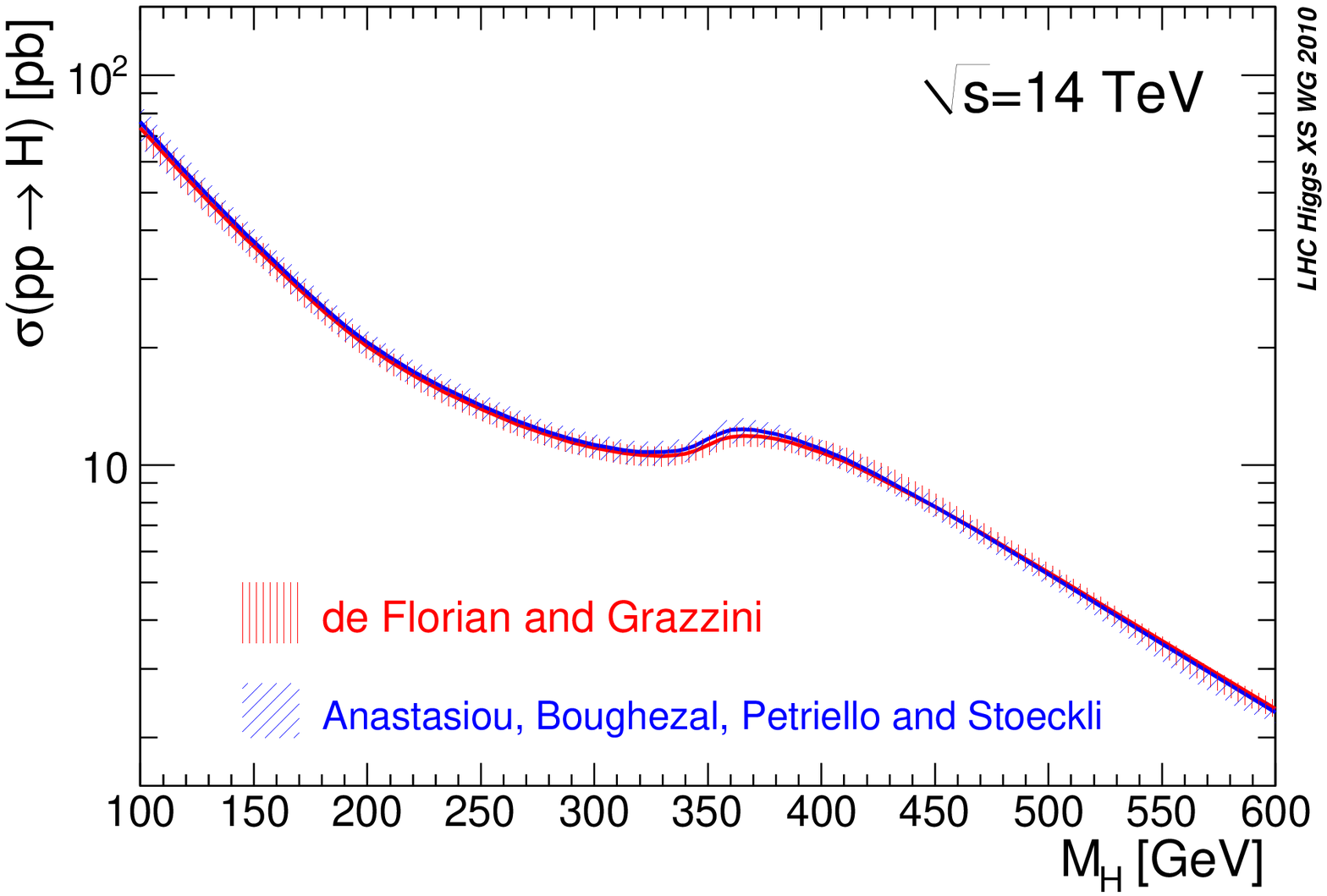}\\
\end{tabular}
\end{center}
\caption{Comparison of ABPS \cite{Anastasiou:2008tj} and dFG \cite{deFlorian:2009hc} results, including scale uncertainty bands.}
\label{fig:comp}
\end{figure}

The results of the dFG and ABPS calculations are reported in \Trefs{tab:dFG7},\ref{tab:dFG14} and \ref{tab:ABPS7},\ref{tab:ABPS14}, respectively.
For each Higgs-boson mass the corresponding cross section is reported.
We also quote three uncertainties: Scale uncertainty, PDF+$\alphas$ uncertainty,
and the latter uncertainty according to the PDF4LHC recipe, computed as discussed below.
In \Fref{fig:comp} we present a comparison of ABPS and dFG results, including scale uncertainties. We see that the results are perfectly 
consistent and show a very good agreement over a wide range of Higgs-boson masses.
At $\sqrt{s}=7$\UTeV\ the difference between ABPS and dFG central values ranges from $+3.5\%$ for $\MH=100$\UGeV\
to $-6\%$ for $\MH=1$\UTeV. In the range $\MH=115{-}300$\UGeV\ the difference ranges from $+3\%$ to $+1\%$.
At $\sqrt{s}=14$\UTeV\ the difference between ABPS and dFG central values ranges from $+3.7\%$ for $\MH=100$\UGeV\
to $-3\%$ for $\MH=1$\UTeV. In the range $\MH=115{-}300$\UGeV\ the difference ranges from $+3\%$ to $+2\%$.


\begin{table}
   \begin{center}
   \ccaption{}{\label{tab:dFG7}{Results on $\Pp\Pp(\Pg\Pg)\to \PH+X$ cross sections with $\sqrt{s}=7$\UTeV\ based on dFG calculation, using MSTW2008 NNLO PDFs.}}
   \small
   \begin{tabular}{ccccc}
   \hline
   $\MH[\UGeVZ]$ & $\sigma[\UpbZ]$ & Scale [\%] & PDF+$\alphas$ [\%] & \small{PDF4LHC} [\%]\\
   \hline
$  90  $&$ 29.48 $& ${+ 8.2}   \;{- 8.7} $ & $ {+ 4.0}  \;{- 3.1} $ & ${+ 7.8}   \;{- 6.7} $ \\
$  95  $&$ 26.48 $& ${+ 8.0}   \;{- 8.6} $ & $ {+ 4.0}  \;{- 3.0} $ & ${+ 7.8}   \;{- 6.7} $ \\
$ 100  $&$ 23.97 $& ${+ 7.8}   \;{- 8.4} $ & $ {+ 4.0}  \;{- 3.0} $ & ${+ 7.7}   \;{- 6.8} $ \\
$ 105  $&$ 21.74 $& ${+ 7.7}   \;{- 8.3} $ & $ {+ 4.0}  \;{- 3.0} $ & ${+ 7.7}   \;{- 6.9} $ \\
$ 110  $&$ 19.81 $& $ {+ 7.5}  \;{- 8.1} $ & $ {+ 4.0}  \;{- 3.0} $ & $ {+ 7.7}  \;{- 6.9} $ \\
$ 115  $&$ 18.12 $& $ {+ 7.4}  \;{- 8.0} $ & $ {+ 4.0}  \;{- 3.0} $ & $ {+ 7.7}  \;{- 7.0} $ \\
$ 120  $&$ 16.63 $& $ {+ 7.2}  \;{- 7.9} $ & $ {+ 4.0}  \;{- 3.0} $ & $ {+ 7.6}  \;{- 7.0} $ \\
$ 125  $&$ 15.31 $& $ {+ 7.1}  \;{- 7.8} $ & $ {+ 4.0}  \;{- 3.1} $ & $ {+ 7.6}  \;{- 7.1} $ \\
$ 130  $&$ 14.13 $& $ {+ 7.0}  \;{- 7.7} $ & $ {+ 4.0}  \;{- 3.1} $ & $ {+ 7.6}  \;{- 7.2} $ \\
$ 135  $&$ 13.08 $& $ {+ 6.9}  \;{- 7.6} $ & $ {+ 3.9}  \;{- 3.1} $ & $ {+ 7.6}  \;{- 7.3} $ \\
$ 140  $&$ 12.14 $& $ {+ 6.8}  \;{- 7.5} $ & $ {+ 3.9}  \;{- 3.1} $ & $ {+ 7.6}  \;{- 7.3} $ \\
$ 145  $&$ 11.29 $& $ {+ 6.7}  \;{- 7.5} $ & $ {+ 3.9}  \;{- 3.1} $ & $ {+ 7.6}  \;{- 7.4} $ \\
$ 150  $&$ 10.52 $& $ {+ 6.6}  \;{- 7.4} $ & $ {+ 3.9}  \;{- 3.1} $ & $ {+ 7.6}  \;{- 7.5} $ \\
$ 155  $&$  9.80 $& $ {+ 6.5}  \;{- 7.3} $ & $ {+ 3.9}  \;{- 3.1} $ & $ {+ 7.5}  \;{- 7.5} $ \\
$ 160  $&$  9.08 $& $ {+ 6.4}  \;{- 7.2} $ & $ {+ 3.9}  \;{- 3.1} $ & $ {+ 7.5}  \;{- 7.6} $ \\
$ 165  $&$  8.35 $& $ {+ 6.4}  \;{- 7.2} $ & $ {+ 3.9}  \;{- 3.2} $ & $ {+ 7.5}  \;{- 7.7} $ \\
$ 170  $&$  7.76 $& $ {+ 6.3}  \;{- 7.1} $ & $ {+ 3.9}  \;{- 3.2} $ & $ {+ 7.5}  \;{- 7.8} $ \\
$ 175  $&$  7.24 $& $ {+ 6.2}  \;{- 7.0} $ & $ {+ 3.9}  \;{- 3.2} $ & $ {+ 7.5}  \;{- 7.8} $ \\
$ 180  $&$  6.76 $& $ {+ 6.2}  \;{- 7.0} $ & $ {+ 3.9}  \;{- 3.2} $ & $ {+ 7.5}  \;{- 7.8} $ \\
$ 185  $&$  6.32 $& $ {+ 6.1}  \;{- 6.9} $ & $ {+ 3.9}  \;{- 3.2} $ & $ {+ 7.5}  \;{- 7.8} $ \\
$ 190  $&$  5.92 $& $ {+ 6.1}  \;{- 6.9} $ & $ {+ 3.9}  \;{- 3.3} $ & $ {+ 7.5}  \;{- 7.8} $ \\
$ 195  $&$  5.57 $& $ {+ 6.1}  \;{- 6.8} $ & $ {+ 4.0}  \;{- 3.3} $ & $ {+ 7.5}  \;{- 7.8} $ \\
$ 200  $&$  5.27 $& $ {+ 6.0}  \;{- 6.8} $ & $ {+ 4.0}  \;{- 3.3} $ & $ {+ 7.6}  \;{- 7.8} $ \\
$ 210  $&$  4.74 $& $ {+ 6.0}  \;{- 6.7} $ & $ {+ 4.0}  \;{- 3.4} $ & $ {+ 7.5}  \;{- 7.9} $ \\
$ 220  $&$  4.29 $& $ {+ 6.5}  \;{- 6.6} $ & $ {+ 4.0}  \;{- 3.4} $ & $ {+ 7.6}  \;{- 7.9} $ \\
$ 230  $&$  3.92 $& $ {+ 5.9}  \;{- 6.5} $ & $ {+ 4.0}  \;{- 3.4} $ & $ {+ 7.7}  \;{- 8.0} $ \\
$ 240  $&$  3.59 $& $ {+ 5.9}  \;{- 6.4} $ & $ {+ 4.0}  \;{- 3.5} $ & $ {+ 7.7}  \;{- 8.0} $ \\
$ 250  $&$  3.32 $& $ {+ 5.8}  \;{- 6.3} $ & $ {+ 4.1}  \;{- 3.5} $ & $ {+ 7.8}  \;{- 8.1} $ \\
$ 260  $&$  3.08 $& $ {+ 5.8}  \;{- 6.3} $ & $ {+ 4.1}  \;{- 3.6} $ & $ {+ 7.8}  \;{- 8.1} $ \\
$ 270  $&$  2.87 $& $ {+ 5.8}  \;{- 6.2} $ & $ {+ 4.1}  \;{- 3.6} $ & $ {+ 7.9}  \;{- 8.1} $ \\
$ 280  $&$  2.70 $& $ {+ 5.8}  \;{- 6.1} $ & $ {+ 4.2}  \;{- 3.7} $ & $ {+ 7.9}  \;{- 8.2} $ \\
$ 290  $&$  2.55 $& $ {+ 5.8}  \;{- 6.1} $ & $ {+ 4.2}  \;{- 3.7} $ & $ {+ 8.0}  \;{- 8.3} $ \\
$ 300  $&$  2.42 $& $ {+ 5.8}  \;{- 6.0} $ & $ {+ 4.2}  \;{- 3.8} $ & $ {+ 8.0}  \;{- 8.3} $ \\
$ 320  $&$  2.25 $& $ {+ 5.8}  \;{- 6.0} $ & $ {+ 4.3}  \;{- 3.9} $ & $ {+ 8.2}  \;{- 8.4} $ \\
$ 340  $&$  2.20 $& $ {+ 5.8}  \;{- 5.9} $ & $ {+ 4.4}  \;{- 4.0} $ & $ {+ 8.3}  \;{- 8.4} $ \\
$ 360  $&$  2.36 $& $ {+ 5.8}  \;{- 5.9} $ & $ {+ 4.5}  \;{- 4.1} $ & $ {+ 8.4}  \;{- 8.5} $ \\
$ 380  $&$  2.26 $& $ {+ 5.9}  \;{- 5.6} $ & $ {+ 4.5}  \;{- 4.2} $ & $ {+ 8.4}  \;{- 8.6} $ \\
$ 400  $&$  2.03 $& $ {+ 5.9}  \;{- 5.4} $ & $ {+ 4.7}  \;{- 4.3} $ & $ {+ 8.8}  \;{- 8.6} $ \\
$ 450  $&$  1.37 $& $ {+ 5.9}  \;{- 5.3} $ & $ {+ 5.0}  \;{- 4.5} $ & $ {+ 9.2}  \;{- 8.7} $ \\
$ 500  $&$  0.865 $& $ {+ 6.0}  \;{- 5.2} $ & $ {+ 5.4}  \;{- 4.8} $ & $ {+ 9.5}  \;{- 8.9} $ \\
$ 550  $&$  0.538 $& $ {+ 6.0}  \;{- 5.2} $ & $ {+ 5.8}  \;{- 5.0} $ & $ {+ 9.7}  \;{- 9.0} $ \\
$ 600  $&$  0.336 $& $ {+ 6.1}  \;{- 5.2} $ & $ {+ 6.2}  \;{- 5.3} $ & $ {+10.1}  \;{- 9.4} $ \\
$ 650  $&$  0.212 $& $ {+ 6.2}  \;{- 5.2} $ & $ {+ 6.5}  \;{- 5.5} $ & $ {+10.4}  \;{- 9.7} $ \\
$ 700  $&$  0.136 $& $ {+ 6.3}  \;{- 5.3} $ & $ {+ 6.9}  \;{- 5.8} $ & $ {+10.7}  \;{- 9.9} $ \\
$ 750  $&$  0.0889 $& $ {+ 6.4}  \;{- 5.4} $ & $ {+ 7.2}  \;{- 6.1} $ & $ {+10.9}  \;{-10.1} $ \\
$ 800  $&$  0.0588 $& $ {+ 6.5}  \;{- 5.4} $ & $ {+ 7.6}  \;{- 6.3} $ & $ {+11.2}  \;{-10.4} $ \\
$ 850  $&$  0.0394 $& $ {+ 6.5}  \;{- 5.5} $ & $ {+ 8.0}  \;{- 6.6} $ & $ {+11.8}  \;{-11.0} $ \\
$ 900  $&$  0.0267 $& $ {+ 6.7}  \;{- 5.6} $ & $ {+ 8.3}  \;{- 6.9} $ & $ {+12.6}  \;{-11.8} $ \\
$ 950  $&$  0.0183 $& $ {+ 6.8}  \;{- 5.7} $ & $ {+ 8.8}  \;{- 7.2} $ & $ {+13.5}  \;{-12.7} $ \\
$1000  $&$  0.0127 $& $ {+ 7.0}  \;{- 5.7} $ & $ {+ 9.1}  \;{- 7.5} $ & $ {+14.2}  \;{-13.5} $ \\

\hline
\end{tabular}
\end{center}
\end{table}


\begin{table}

   \begin{center}
   \ccaption{}{\label{tab:ABPS7}{Results on $\Pp\Pp(\Pg\Pg)\to \PH+X$ cross sections with $\sqrt{s}=7$\UTeV\ based on ABPS calculation, using MSTW2008 NNLO PDFs.}}
   \small
   \begin{tabular}{ccccc}
   \hline
   $\MH[\UGeVZ]$ & $\sigma[\UpbZ]$ & Scale [\%] & PDF+$\alphas$ [\%] & \small{PDF4LHC} [\%]\\
   \hline
 $ 90 $&$ 30.70 $& $ {+10.2}  \;{-11.9} $ & $ {+ 4.2}  \;{- 3.1} $ & $ {+ 8.0}  \;{- 6.9} $ \\
 $ 95 $&$ 27.54 $& $ {+ 9.9}  \;{-10.8} $ & $ {+ 4.1}  \;{- 3.1} $ & $ {+ 8.0}  \;{- 6.9} $ \\
 $100 $&$ 24.81 $& $ {+ 9.7}  \;{-10.5} $ & $ {+ 4.1}  \;{- 3.1} $ & $ {+ 7.9}  \;{- 7.0} $ \\
 $105 $&$ 22.47 $& $ {+ 9.4}  \;{-10.3} $ & $ {+ 4.1}  \;{- 3.1} $ & $ {+ 7.9}  \;{- 7.0} $ \\
 $110 $&$ 20.44 $& $ {+ 9.2}  \;{-10.1} $ & $ {+ 4.1}  \;{- 3.1} $ & $ {+ 7.9}  \;{- 7.1} $ \\
 $115 $&$ 18.67 $& $ {+ 8.9}  \;{-10.0} $ & $ {+ 4.1}  \;{- 3.1} $ & $ {+ 7.9}  \;{- 7.2} $ \\
 $120 $&$ 17.12 $& $ {+ 8.7}  \;{- 9.8} $ & $ {+ 4.1}  \;{- 3.1} $ & $ {+ 7.8}  \;{- 7.2} $ \\
 $125 $&$ 15.74 $& $ {+ 8.6}  \;{- 9.7} $ & $ {+ 4.0}  \;{- 3.1} $ & $ {+ 7.8}  \;{- 7.3} $ \\
 $130 $&$ 14.52 $& $ {+ 8.3}  \;{- 9.6} $ & $ {+ 4.0}  \;{- 3.1} $ & $ {+ 7.8}  \;{- 7.4} $ \\
 $135 $&$ 13.43 $& $ {+ 8.2}  \;{- 9.4} $ & $ {+ 4.0}  \;{- 3.1} $ & $ {+ 7.7}  \;{- 7.4} $ \\
 $140 $&$ 12.45 $& $ {+ 8.1}  \;{- 9.3} $ & $ {+ 4.0}  \;{- 3.1} $ & $ {+ 7.8}  \;{- 7.5} $ \\
 $145 $&$ 11.58 $& $ {+ 8.0}  \;{- 9.3} $ & $ {+ 4.0}  \;{- 3.2} $ & $ {+ 7.8}  \;{- 7.5} $ \\
 $150 $&$ 10.79 $& $ {+ 7.9}  \;{- 9.3} $ & $ {+ 4.0}  \;{- 3.2} $ & $ {+ 7.8}  \;{- 7.6} $ \\
 $155 $&$ 10.08 $& $ {+ 7.7}  \;{- 9.2} $ & $ {+ 4.0}  \;{- 3.2} $ & $ {+ 7.7}  \;{- 7.7} $ \\
 $160 $&$  9.36 $& $ {+ 7.6}  \;{- 9.2} $ & $ {+ 4.0}  \;{- 3.2} $ & $ {+ 7.7}  \;{- 7.7} $ \\
 $165 $&$  8.54 $& $ {+ 7.5}  \;{- 9.2} $ & $ {+ 4.0}  \;{- 3.2} $ & $ {+ 7.7}  \;{- 7.8} $ \\
 $170 $&$  7.92 $& $ {+ 7.5}  \;{- 9.2} $ & $ {+ 4.0}  \;{- 3.2} $ & $ {+ 7.7}  \;{- 7.9} $ \\
 $175 $&$  7.40 $& $ {+ 7.4}  \;{- 9.2} $ & $ {+ 4.0}  \;{- 3.3} $ & $ {+ 7.7}  \;{- 7.9} $ \\
 $180 $&$  6.93 $& $ {+ 7.3}  \;{- 9.1} $ & $ {+ 4.0}  \;{- 3.3} $ & $ {+ 7.7}  \;{- 7.9} $ \\
 $185 $&$  6.44 $& $ {+ 7.2}  \;{- 9.1} $ & $ {+ 4.0}  \;{- 3.3} $ & $ {+ 7.7}  \;{- 8.0} $ \\
 $190 $&$  6.03 $& $ {+ 7.2}  \;{- 9.1} $ & $ {+ 4.0}  \;{- 3.3} $ & $ {+ 7.7}  \;{- 8.0} $ \\
 $195 $&$  5.67 $& $ {+ 7.2}  \;{- 9.1} $ & $ {+ 4.0}  \;{- 3.4} $ & $ {+ 7.7}  \;{- 8.0} $ \\
 $200 $&$  5.36 $& $ {+ 7.1}  \;{- 9.1} $ & $ {+ 4.1}  \;{- 3.4} $ & $ {+ 7.8}  \;{- 8.0} $ \\
 $210 $&$  4.82 $& $ {+ 7.0}  \;{- 9.1} $ & $ {+ 4.0}  \;{- 3.4} $ & $ {+ 7.7}  \;{- 8.0} $ \\
 $220 $&$  4.37 $& $ {+ 7.0}  \;{- 9.0} $ & $ {+ 4.1}  \;{- 3.5} $ & $ {+ 7.8}  \;{- 8.1} $ \\
 $230 $&$  3.98 $& $ {+ 6.8}  \;{- 9.0} $ & $ {+ 4.1}  \;{- 3.5} $ & $ {+ 7.8}  \;{- 8.1} $ \\
 $240 $&$  3.65 $& $ {+ 6.8}  \;{- 9.0} $ & $ {+ 4.1}  \;{- 3.5} $ & $ {+ 7.9}  \;{- 8.2} $ \\
 $250 $&$  3.37 $& $ {+ 6.7}  \;{- 9.0} $ & $ {+ 4.2}  \;{- 3.6} $ & $ {+ 7.9}  \;{- 8.2} $ \\
 $260 $&$  3.12 $& $ {+ 6.6}  \;{- 9.0} $ & $ {+ 4.2}  \;{- 3.6} $ & $ {+ 8.0}  \;{- 8.3} $ \\
 $270 $&$  2.91 $& $ {+ 6.5}  \;{- 9.0} $ & $ {+ 4.2}  \;{- 3.7} $ & $ {+ 8.0}  \;{- 8.3} $ \\
 $280 $&$  2.73 $& $ {+ 6.6}  \;{- 9.0} $ & $ {+ 4.2}  \;{- 3.7} $ & $ {+ 8.1}  \;{- 8.3} $ \\
 $290 $&$  2.58 $& $ {+ 6.6}  \;{- 8.9} $ & $ {+ 4.3}  \;{- 3.8} $ & $ {+ 8.1}  \;{- 8.4} $ \\
 $300 $&$  2.45 $& $ {+ 6.5}  \;{- 8.9} $ & $ {+ 4.3}  \;{- 3.8} $ & $ {+ 8.2}  \;{- 8.4} $ \\
 $320 $&$  2.28 $& $ {+ 6.5}  \;{- 9.0} $ & $ {+ 4.4}  \;{- 3.9} $ & $ {+ 8.3}  \;{- 8.5} $ \\
 $340 $&$  2.25 $& $ {+ 6.7}  \;{- 9.2} $ & $ {+ 4.5}  \;{- 4.0} $ & $ {+ 8.4}  \;{- 8.6} $ \\
 $360 $&$  2.44 $& $ {+ 6.8}  \;{- 9.2} $ & $ {+ 4.5}  \;{- 4.1} $ & $ {+ 8.5}  \;{- 8.6} $ \\
 $380 $&$  2.31 $& $ {+ 6.1}  \;{- 8.9} $ & $ {+ 4.6}  \;{- 4.2} $ & $ {+ 8.7}  \;{- 8.7} $ \\
 $400 $&$  2.05 $& $ {+ 5.7}  \;{- 8.6} $ & $ {+ 4.8}  \;{- 4.3} $ & $ {+ 8.9}  \;{- 8.7} $ \\
 $450 $&$  1.35 $& $ {+ 4.8}  \;{- 8.2} $ & $ {+ 5.2}  \;{- 4.6} $ & $ {+ 9.5}  \;{- 8.9} $ \\
 $500 $&$  0.844 $& $ {+ 4.2}  \;{- 7.9} $ & $ {+ 5.5}  \;{- 4.8} $ & $ {+ 9.7}  \;{- 9.0} $ \\
 $550 $&$  0.522 $& $ {+ 3.8}  \;{- 7.7} $ & $ {+ 6.0}  \;{- 5.1} $ & $ {+10.0}  \;{- 9.2} $ \\
 $600 $&$  0.325 $& $ {+ 3.5}  \;{- 7.5} $ & $ {+ 6.4}  \;{- 5.4} $ & $ {+10.5}  \;{- 9.6} $ \\
 $650 $&$  0.205 $& $ {+ 3.3}  \;{- 7.4} $ & $ {+ 6.8}  \;{- 5.6} $ & $ {+10.8}  \;{- 9.9} $ \\
 $700 $&$  0.131 $& $ {+ 3.2}  \;{- 7.3} $ & $ {+ 7.1}  \;{- 5.9} $ & $ {+11.1}  \;{-10.2} $ \\
 $750 $&$  0.0850 $& $ {+ 3.1}  \;{- 7.2} $ & $ {+ 7.5}  \;{- 6.2} $ & $ {+11.3}  \;{-10.4} $ \\
 $800 $&$  0.0560 $& $ {+ 3.0}  \;{- 7.2} $ & $ {+ 7.9}  \;{- 6.5} $ & $ {+11.6}  \;{-10.8} $ \\
 $850 $&$  0.0374 $& $ {+ 2.9}  \;{- 7.1} $ & $ {+ 8.3}  \;{- 6.8} $ & $ {+12.3}  \;{-11.4} $ \\
 $900 $&$  0.0253 $& $ {+ 2.8}  \;{- 7.1} $ & $ {+ 8.7}  \;{- 7.2} $ & $ {+13.1}  \;{-12.2} $ \\
 $950 $&$  0.0173 $& $ {+ 2.8}  \;{- 7.1} $ & $ {+ 9.1}  \;{- 7.5} $ & $ {+14.0}  \;{-13.1} $ \\
$1000 $&$  0.0119 $& $ {+ 2.7}  \;{- 7.1} $ & $ {+ 9.5}  \;{- 7.8} $ & $ {+14.9}  \;{-14.0} $ \\

\hline
\end{tabular}
\end{center}
\end{table}


\begin{table}

   \begin{center}
   \ccaption{}{\label{tab:dFG14}{Results on $\Pp\Pp(\Pg\Pg)\to \PH+X$ cross sections with $\sqrt{s}=14$\UTeV\ based on dFG calculation, using MSTW2008 NNLO PDFs.}}
   \small
   \begin{tabular}{ccccc}
   \hline
   $\MH[\UGeVZ]$ & $\sigma[\UpbZ]$ & Scale [\%] & PDF+$\alphas$ [\%] & \small{PDF4LHC} [\%]\\
   \hline
 $ 90 $&$ 87.68 $& $ {+ 8.7}  \;{- 9.0} $ & $ {+ 4.0}  \;{- 3.0} $ & $ {+ 7.3}  \;{- 6.0} $ \\
 $ 95 $&$ 79.95 $& $ {+ 8.5}  \;{- 8.8} $ & $ {+ 3.9}  \;{- 3.0} $ & $ {+ 7.3}  \;{- 6.0} $ \\
 $100 $&$ 73.38 $& $ {+ 8.3}  \;{- 8.6} $ & $ {+ 3.9}  \;{- 3.0} $ & $ {+ 7.2}  \;{- 6.0} $ \\
 $105 $&$ 67.47 $& $ {+ 8.1}  \;{- 8.5} $ & $ {+ 3.9}  \;{- 3.0} $ & $ {+ 7.2}  \;{- 6.0} $ \\
 $110 $&$ 62.28 $& $ {+ 7.9}  \;{- 8.3} $ & $ {+ 3.9}  \;{- 2.9} $ & $ {+ 7.2}  \;{- 6.0} $ \\
 $115 $&$ 57.69 $& $ {+ 7.8}  \;{- 8.2} $ & $ {+ 3.8}  \;{- 2.9} $ & $ {+ 7.2}  \;{- 6.0} $ \\
 $120 $&$ 53.62 $& $ {+ 7.6}  \;{- 8.1} $ & $ {+ 3.8}  \;{- 2.9} $ & $ {+ 7.2}  \;{- 6.0} $ \\
 $125 $&$ 49.97 $& $ {+ 7.5}  \;{- 8.0} $ & $ {+ 3.8}  \;{- 2.9} $ & $ {+ 7.2}  \;{- 6.0} $ \\
 $130 $&$ 46.69 $& $ {+ 7.3}  \;{- 7.9} $ & $ {+ 3.8}  \;{- 2.9} $ & $ {+ 7.2}  \;{- 6.0} $ \\
 $135 $&$ 43.74 $& $ {+ 7.2}  \;{- 7.8} $ & $ {+ 3.7}  \;{- 2.8} $ & $ {+ 7.1}  \;{- 6.0} $ \\
 $140 $&$ 41.05 $& $ {+ 7.1}  \;{- 7.7} $ & $ {+ 3.7}  \;{- 2.8} $ & $ {+ 7.1}  \;{- 6.0} $ \\
 $145 $&$ 38.61 $& $ {+ 7.0}  \;{- 7.6} $ & $ {+ 3.7}  \;{- 2.8} $ & $ {+ 7.1}  \;{- 6.1} $ \\
 $150 $&$ 36.38 $& $ {+ 6.9}  \;{- 7.5} $ & $ {+ 3.7}  \;{- 2.8} $ & $ {+ 7.1}  \;{- 6.1} $ \\
 $155 $&$ 34.26 $& $ {+ 6.8}  \;{- 7.5} $ & $ {+ 3.7}  \;{- 2.8} $ & $ {+ 7.1}  \;{- 6.1} $ \\
 $160 $&$ 32.08 $& $ {+ 6.7}  \;{- 7.4} $ & $ {+ 3.7}  \;{- 2.8} $ & $ {+ 7.1}  \;{- 6.1} $ \\
 $165 $&$ 29.84 $& $ {+ 6.7}  \;{- 7.4} $ & $ {+ 3.6}  \;{- 2.8} $ & $ {+ 7.0}  \;{- 6.1} $ \\
 $170 $&$ 28.01 $& $ {+ 6.6}  \;{- 7.2} $ & $ {+ 3.6}  \;{- 2.8} $ & $ {+ 7.0}  \;{- 6.2} $ \\
 $175 $&$ 26.41 $& $ {+ 6.5}  \;{- 7.2} $ & $ {+ 3.6}  \;{- 2.8} $ & $ {+ 7.0}  \;{- 6.2} $ \\
 $180 $&$ 24.92 $& $ {+ 6.4}  \;{- 7.1} $ & $ {+ 3.6}  \;{- 2.8} $ & $ {+ 7.0}  \;{- 6.2} $ \\
 $185 $&$ 23.53 $& $ {+ 6.4}  \;{- 7.1} $ & $ {+ 3.6}  \;{- 2.8} $ & $ {+ 7.0}  \;{- 6.3} $ \\
 $190 $&$ 22.26 $& $ {+ 6.3}  \;{- 7.0} $ & $ {+ 3.6}  \;{- 2.8} $ & $ {+ 7.0}  \;{- 6.3} $ \\
 $195 $&$ 21.15 $& $ {+ 6.2}  \;{- 7.0} $ & $ {+ 3.6}  \;{- 2.7} $ & $ {+ 7.0}  \;{- 6.3} $ \\
 $200 $&$ 20.18 $& $ {+ 6.2}  \;{- 6.9} $ & $ {+ 3.6}  \;{- 2.7} $ & $ {+ 7.0}  \;{- 6.3} $ \\
 $210 $&$ 18.50 $& $ {+ 6.1}  \;{- 6.8} $ & $ {+ 3.6}  \;{- 2.7} $ & $ {+ 6.9}  \;{- 6.4} $ \\
 $220 $&$ 17.08 $& $ {+ 6.0}  \;{- 6.7} $ & $ {+ 3.6}  \;{- 2.8} $ & $ {+ 6.9}  \;{- 6.4} $ \\ 
 $230 $&$ 15.86 $& $ {+ 5.9}  \;{- 6.6} $ & $ {+ 3.6}  \;{- 2.8} $ & $ {+ 6.9}  \;{- 6.5} $ \\
 $240 $&$ 14.82 $& $ {+ 5.8}  \;{- 6.5} $ & $ {+ 3.5}  \;{- 2.8} $ & $ {+ 6.9}  \;{- 6.6} $ \\
 $250 $&$ 13.92 $& $ {+ 5.8}  \;{- 6.4} $ & $ {+ 3.5}  \;{- 2.8} $ & $ {+ 6.9}  \;{- 6.7} $ \\
 $260 $&$ 13.15 $& $ {+ 5.7}  \;{- 6.4} $ & $ {+ 3.5}  \;{- 2.8} $ & $ {+ 6.9}  \;{- 6.8} $ \\
 $270 $&$ 12.48 $& $ {+ 5.7}  \;{- 6.3} $ & $ {+ 3.5}  \;{- 2.8} $ & $ {+ 6.9}  \;{- 6.8} $ \\
 $280 $&$ 11.91 $& $ {+ 5.7}  \;{- 6.2} $ & $ {+ 3.5}  \;{- 2.8} $ & $ {+ 6.8}  \;{- 6.9} $ \\
 $290 $&$ 11.44 $& $ {+ 5.7}  \;{- 6.2} $ & $ {+ 3.5}  \;{- 2.8} $ & $ {+ 6.8}  \;{- 6.9} $ \\
 $300 $&$ 11.07 $& $ {+ 5.6}  \;{- 6.1} $ & $ {+ 3.5}  \;{- 2.9} $ & $ {+ 6.8}  \;{- 7.0} $ \\
 $320 $&$ 10.60 $& $ {+ 5.6}  \;{- 6.0} $ & $ {+ 3.5}  \;{- 2.9} $ & $ {+ 6.8}  \;{- 6.9} $ \\
 $340 $&$ 10.69 $& $ {+ 5.6}  \;{- 6.0} $ & $ {+ 3.5}  \;{- 2.9} $ & $ {+ 6.8}  \;{- 7.0} $ \\
 $360 $&$ 11.81 $& $ {+ 5.6}  \;{- 5.9} $ & $ {+ 3.5}  \;{- 3.0} $ & $ {+ 6.8}  \;{- 7.0} $ \\
 $380 $&$ 11.66 $& $ {+ 5.6}  \;{- 5.7} $ & $ {+ 3.6}  \;{- 3.0} $ & $ {+ 6.8}  \;{- 7.1} $ \\
 $400 $&$ 10.76 $& $ {+ 7.3}  \;{- 5.5} $ & $ {+ 3.6}  \;{- 3.0} $ & $ {+ 6.9}  \;{- 7.1} $ \\
 $450 $&$  7.80 $& $ {+ 5.5}  \;{- 5.1} $ & $ {+ 3.6}  \;{- 3.2} $ & $ {+ 6.9}  \;{- 7.2} $ \\
 $500 $&$  5.31 $& $ {+ 5.5}  \;{- 5.0} $ & $ {+ 3.7}  \;{- 3.3} $ & $ {+ 7.0}  \;{- 7.2} $ \\
 $550 $&$  3.54 $& $ {+ 5.4}  \;{- 4.9} $ & $ {+ 3.8}  \;{- 3.4} $ & $ {+ 7.3}  \;{- 7.5} $ \\
 $600 $&$  2.37 $& $ {+ 5.4}  \;{- 4.8} $ & $ {+ 3.9}  \;{- 3.5} $ & $ {+ 7.3}  \;{- 7.4} $ \\
 $650 $&$  1.60 $& $ {+ 5.3}  \;{- 4.7} $ & $ {+ 4.0}  \;{- 3.6} $ & $ {+ 7.5}  \;{- 7.5} $ \\
 $700 $&$  1.10 $& $ {+ 5.3}  \;{- 4.7} $ & $ {+ 4.1}  \;{- 3.8} $ & $ {+ 7.7}  \;{- 7.5} $ \\
 $750 $&$  0.765 $& $ {+ 5.4}  \;{- 4.7} $ & $ {+ 4.3}  \;{- 3.9} $ & $ {+ 8.0}  \;{- 7.6} $ \\
 $800 $&$  0.539 $& $ {+ 5.3}  \;{- 4.6} $ & $ {+ 4.5}  \;{- 4.0} $ & $ {+ 8.2}  \;{- 7.7} $ \\
 $850 $&$  0.385 $& $ {+ 5.3}  \;{- 4.6} $ & $ {+ 4.7}  \;{- 4.1} $ & $ {+ 8.4}  \;{- 7.8} $ \\
 $900 $&$  0.279 $& $ {+ 5.3}  \;{- 4.6} $ & $ {+ 4.9}  \;{- 4.2} $ & $ {+ 8.6}  \;{- 8.0} $ \\
 $950 $&$  0.204 $& $ {+ 5.4}  \;{- 4.7} $ & $ {+ 5.1}  \;{- 4.4} $ & $ {+ 8.8}  \;{- 8.1} $ \\
$1000 $&$  0.151 $& $ {+ 5.4}  \;{- 4.6} $ & $ {+ 5.3}  \;{- 4.5} $ & $ {+ 8.9}  \;{- 8.2} $ \\

\hline
\end{tabular}
\end{center}
\end{table}


\begin{table}

   \begin{center}
   \ccaption{}{\label{tab:ABPS14}{Results on $\Pp\Pp(\Pg\Pg)\to \PH+X$ cross sections with $\sqrt{s}=14$\UTeV\ based on ABPS calculation, using MSTW2008 NNLO PDFs.}}
   \small
   \begin{tabular}{ccccc}
   \hline
   $\MH[\UGeVZ]$ & $\sigma[\UpbZ]$ & Scale [\%] & PDF+$\alphas$ [\%] & \small{PDF4LHC} [\%]\\
   \hline
 $ 90 $&$ 91.49 $& $ {+10.5}  \;{-14.0} $ & $ {+ 4.1}  \;{- 3.1} $ & $ {+ 7.5}  \;{- 6.2} $ \\
 $ 95 $&$ 83.22 $& $ {+10.1}  \;{-13.5} $ & $ {+ 4.0}  \;{- 3.1} $ & $ {+ 7.4}  \;{- 6.1} $ \\
 $100 $&$ 76.07$ & $ {+ 9.9}  \;{-13.1} $ & $ {+ 4.0}  \;{- 3.1} $ & $ {+ 7.4}  \;{- 6.1} $ \\
 $105 $&$ 69.84$ & $ {+ 9.6}  \;{-12.7} $ & $ {+ 4.0}  \;{- 3.0} $ & $ {+ 7.4}  \;{- 6.1} $ \\
 $110 $&$ 64.38 $& $ {+ 9.3}  \;{-12.3} $ & $ {+ 3.9}  \;{- 3.0} $ & $ {+ 7.3}  \;{- 6.1} $ \\
 $115 $&$ 59.56 $& $ {+ 9.1}  \;{-11.9} $ & $ {+ 3.9}  \;{- 3.0} $ & $ {+ 7.3}  \;{- 6.1} $ \\
 $120 $&$ 55.29 $& $ {+ 8.9}  \;{-11.6} $ & $ {+ 3.9}  \;{- 2.9} $ & $ {+ 7.3}  \;{- 6.1} $ \\
 $125 $&$ 51.47 $& $ {+ 8.7}  \;{-11.3} $ & $ {+ 3.9}  \;{- 2.9} $ & $ {+ 7.3}  \;{- 6.1} $ \\
 $130 $&$ 48.06 $& $ {+ 8.6}  \;{-11.1} $ & $ {+ 3.8}  \;{- 2.9} $ & $ {+ 7.3}  \;{- 6.1} $ \\
 $135 $&$ 44.98 $& $ {+ 8.4}  \;{-10.8} $ & $ {+ 3.8}  \;{- 2.9} $ & $ {+ 7.3}  \;{- 6.1} $ \\
 $140 $&$ 42.21 $& $ {+ 8.2}  \;{-10.5} $ & $ {+ 3.8}  \;{- 2.9} $ & $ {+ 7.3}  \;{- 6.2} $ \\
 $145 $&$ 39.71 $& $ {+ 8.1}  \;{-10.3} $ & $ {+ 3.8}  \;{- 2.9} $ & $ {+ 7.3}  \;{- 6.2} $ \\
 $150 $&$ 37.43 $& $ {+ 8.0}  \;{-10.1} $ & $ {+ 3.8}  \;{- 2.8} $ & $ {+ 7.2}  \;{- 6.2} $ \\
 $155 $&$ 35.34 $& $ {+ 7.8}  \;{- 9.9} $ & $ {+ 3.8}  \;{- 2.8} $ & $ {+ 7.2}  \;{- 6.2} $ \\
 $160 $&$ 33.19 $& $ {+ 7.7}  \;{- 9.7} $ & $ {+ 3.7}  \;{- 2.8} $ & $ {+ 7.2}  \;{- 6.2} $ \\
 $165 $&$ 30.60 $& $ {+ 7.6}  \;{- 9.5} $ & $ {+ 3.7}  \;{- 2.8} $ & $ {+ 7.2}  \;{- 6.2} $ \\
 $170 $&$ 28.69 $& $ {+ 7.5}  \;{- 9.4} $ & $ {+ 3.7}  \;{- 2.8} $ & $ {+ 7.2}  \;{- 6.3} $ \\
 $175 $&$ 27.09 $& $ {+ 7.5}  \;{- 9.2} $ & $ {+ 3.7}  \;{- 2.8} $ & $ {+ 7.2}  \;{- 6.3} $ \\
 $180 $&$ 25.65 $& $ {+ 7.4}  \;{- 9.1} $ & $ {+ 3.7}  \;{- 2.8} $ & $ {+ 7.2}  \;{- 6.3} $ \\
 $185 $&$ 24.09 $& $ {+ 7.3}  \;{- 8.9} $ & $ {+ 3.7}  \;{- 2.8} $ & $ {+ 7.1}  \;{- 6.4} $ \\
 $190 $&$ 22.75 $& $ {+ 7.3}  \;{- 8.8} $ & $ {+ 3.7}  \;{- 2.8} $ & $ {+ 7.1}  \;{- 6.4} $ \\
 $195 $&$ 21.63 $& $ {+ 7.2}  \;{- 8.7} $ & $ {+ 3.7}  \;{- 2.8} $ & $ {+ 7.1}  \;{- 6.4} $ \\
 $200 $&$ 20.64 $& $ {+ 7.1}  \;{- 8.5} $ & $ {+ 3.7}  \;{- 2.8} $ & $ {+ 7.1}  \;{- 6.4} $ \\
 $210 $&$ 18.92 $& $ {+ 7.0}  \;{- 8.3} $ & $ {+ 3.6}  \;{- 2.8} $ & $ {+ 7.1}  \;{- 6.5} $ \\
 $220 $&$ 17.47 $& $ {+ 6.9}  \;{- 8.1} $ & $ {+ 3.6}  \;{- 2.8} $ & $ {+ 7.1}  \;{- 6.6} $ \\
 $230 $&$ 16.22 $& $ {+ 6.8}  \;{- 8.0} $ & $ {+ 3.6}  \;{- 2.8} $ & $ {+ 7.0}  \;{- 6.6} $ \\
 $240 $&$ 15.15 $& $ {+ 6.7}  \;{- 7.9} $ & $ {+ 3.6}  \;{- 2.8} $ & $ {+ 7.0}  \;{- 6.7} $ \\
 $250 $&$ 14.23 $& $ {+ 6.6}  \;{- 7.9} $ & $ {+ 3.6}  \;{- 2.8} $ & $ {+ 7.0}  \;{- 6.8} $ \\
 $260 $&$ 13.43 $& $ {+ 6.5}  \;{- 7.8} $ & $ {+ 3.6}  \;{- 2.8} $ & $ {+ 7.0}  \;{- 6.9} $ \\
 $270 $&$ 12.74 $& $ {+ 6.4}  \;{- 7.8} $ & $ {+ 3.6}  \;{- 2.8} $ & $ {+ 7.0}  \;{- 6.9} $ \\
 $280 $&$ 12.15 $& $ {+ 6.4}  \;{- 7.8} $ & $ {+ 3.6}  \;{- 2.8} $ & $ {+ 7.0}  \;{- 7.0} $ \\
 $290 $&$ 11.67 $& $ {+ 6.3}  \;{- 7.7} $ & $ {+ 3.6}  \;{- 2.9} $ & $ {+ 6.9}  \;{- 7.0} $ \\
 $300 $&$ 11.28 $& $ {+ 6.2}  \;{- 7.7} $ & $ {+ 3.6}  \;{- 2.9} $ & $ {+ 6.9}  \;{- 7.0} $ \\
 $320 $&$ 10.81 $& $ {+ 6.2}  \;{- 7.7} $ & $ {+ 3.6}  \;{- 2.9} $ & $ {+ 6.9}  \;{- 7.0} $ \\
 $340 $&$ 11.00 $& $ {+ 6.2}  \;{- 7.7} $ & $ {+ 3.6}  \;{- 2.9} $ & $ {+ 6.9}  \;{- 7.1} $ \\
 $360 $&$ 12.30 $& $ {+ 6.1}  \;{- 7.7} $ & $ {+ 3.6}  \;{- 3.0} $ & $ {+ 6.9}  \;{- 7.1} $ \\
 $380 $&$ 12.01 $& $ {+ 5.7}  \;{- 7.4} $ & $ {+ 3.6}  \;{- 3.0} $ & $ {+ 6.9}  \;{- 7.1} $ \\
 $400 $&$ 10.98 $& $ {+ 5.3}  \;{- 7.1} $ & $ {+ 3.6}  \;{- 3.1} $ & $ {+ 6.9}  \;{- 7.2} $ \\
 $450 $&$  7.81 $& $ {+ 4.7}  \;{- 6.7} $ & $ {+ 3.7}  \;{- 3.2} $ & $ {+ 7.0}  \;{- 7.2} $ \\
 $500 $&$  5.24 $& $ {+ 4.3}  \;{- 6.4} $ & $ {+ 3.7}  \;{- 3.3} $ & $ {+ 7.1}  \;{- 7.3} $ \\
 $550 $&$  3.48 $& $ {+ 4.0}  \;{- 6.2} $ & $ {+ 3.8}  \;{- 3.4} $ & $ {+ 7.3}  \;{- 7.5} $ \\
 $600 $&$  2.32 $& $ {+ 3.8}  \;{- 6.0} $ & $ {+ 3.9}  \;{- 3.5} $ & $ {+ 7.4}  \;{- 7.5} $ \\
 $650 $&$  1.57 $& $ {+ 3.6}  \;{- 5.9} $ & $ {+ 4.0}  \;{- 3.6} $ & $ {+ 7.5}  \;{- 7.5} $ \\
 $700 $&$  1.07 $& $ {+ 3.5}  \;{- 5.8} $ & $ {+ 4.1}  \;{- 3.8} $ & $ {+ 7.7}  \;{- 7.6} $ \\
 $750 $&$  0.746 $& $ {+ 3.3}  \;{- 5.7} $ & $ {+ 4.3}  \;{- 3.9} $ & $ {+ 7.8}  \;{- 7.7} $ \\
 $800 $&$  0.525 $& $ {+ 3.2}  \;{- 5.7} $ & $ {+ 4.4}  \;{- 4.0} $ & $ {+ 7.9}  \;{- 7.8} $ \\
 $850 $&$  0.374 $& $ {+ 3.2}  \;{- 5.6} $ & $ {+ 4.5}  \;{- 4.1} $ & $ {+ 8.0}  \;{- 7.9} $ \\
 $900 $&$  0.270 $& $ {+ 3.1}  \;{- 5.6} $ & $ {+ 4.6}  \;{- 4.3} $ & $ {+ 8.1}  \;{- 8.0} $ \\
 $950 $&$  0.197 $& $ {+ 3.0}  \;{- 5.5} $ & $ {+ 4.8}  \;{- 4.4} $ & $ {+ 8.2}  \;{- 8.1} $ \\
$1000 $&$  0.146 $& $ {+ 3.0}  \;{- 5.5} $ & $ {+ 4.9}  \;{- 4.5} $ & $ {+ 8.3}  \;{- 8.3} $ \\

\hline
\end{tabular}
\end{center}
\end{table}


\subsection{Uncertainties}

We now discuss the various sources of uncertainty affecting the 
cross sections presented in \Trefs{tab:dFG7}--\ref{tab:ABPS14}. The uncertainty has two primary origins:
From missing terms in the partonic cross sections and from our limited
knowledge of the PDFs.

\begin{itemize}

\item[$\bullet$] Uncalculated higher-order QCD radiative corrections are one of the most important sources of uncertainty
on the partonic cross section. The customary method used in perturbative QCD calculations to 
estimate their size is to vary the renormalization and factorization scales around
a central value $\mu_0$, which is chosen to be of the order of the hard scale of the process.  The 
uncertainty of the ABPS and dFG calculations is quantified in this way.
The factorization and renormalization scales $\mu_F$ and $\mu_R$ are varied 
in the range $0.5 \mu_0< \mu_F,\mu_R< 2 \mu_0$, with the constraint
$0.5 < \mu_F/\mu_R < 2$. The choice of the central scale $\mu_0$ is instead different: dFG choose $\mu_0=\MH$, whereas ABPS 
choose $\mu_0=\MH/2$.  The structure of the scale dependent logarithmic contributions in the fixed-order calculation of ABPS suggests that the central value 
of the scale should be chosen parametrically smaller than $\MH$.  This is supported by the better convergence of the cross section 
through NNLO and also after including the leading N$^3$LO terms~\cite{Moch:2005ky}.  The resummation implemented in the NNLL result of dFG 
minimizes the sensitivity to the choice of central scale. 
This is clearly shown in \Fref{fig:compband}, where the scale dependent bands for different values
of the reference scale $\mu_0$ are shown.
The results of dFG show a remarkable stability with respect to the choice of $\mu_0$ both at $7\UTeV$ and at $14\UTeV$.

\begin{figure}[htb]
\vspace{5pt}
\begin{center}
\begin{tabular}{cc}
\includegraphics[width=.42\linewidth]{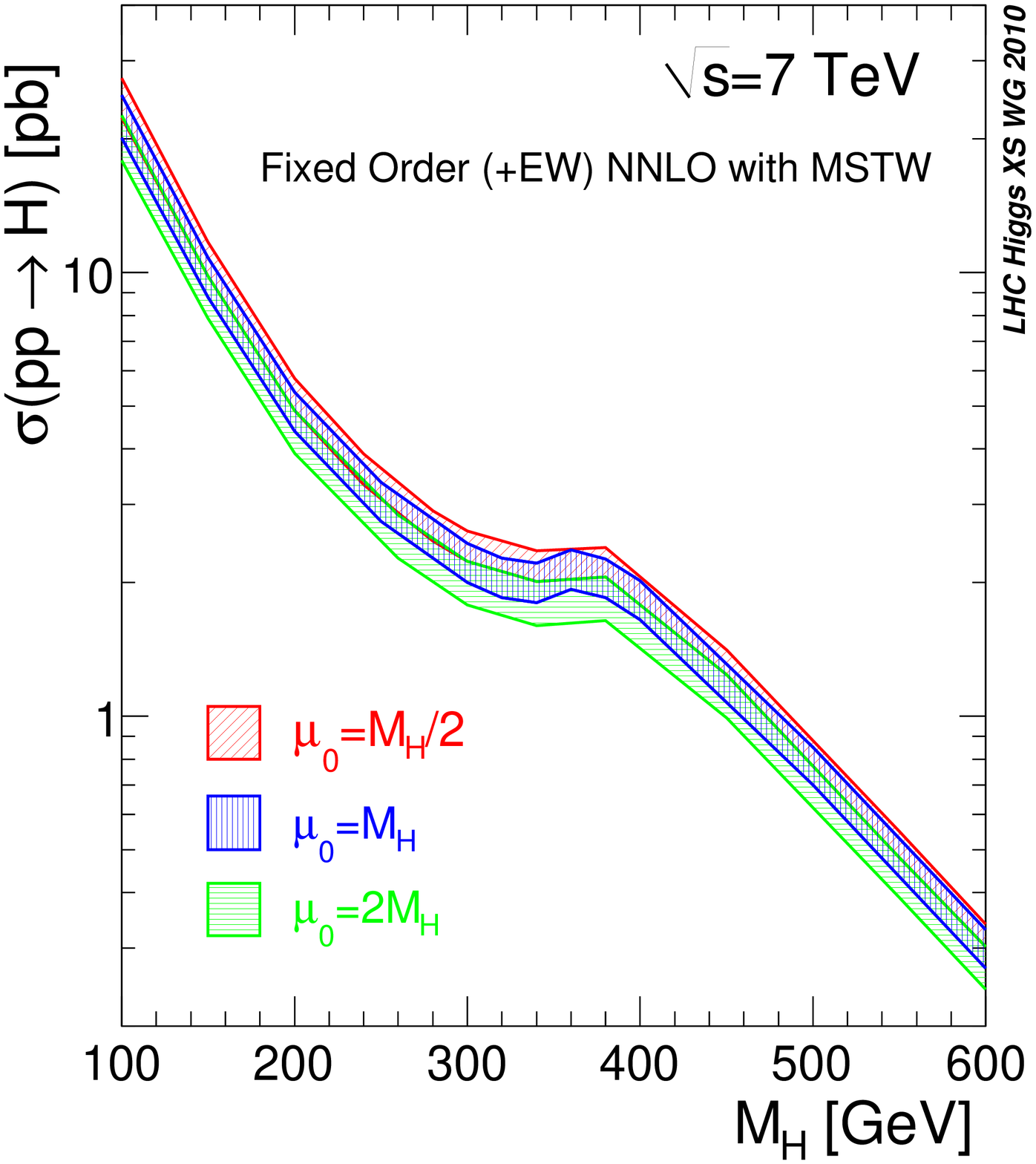} &
\includegraphics[width=.42\linewidth]{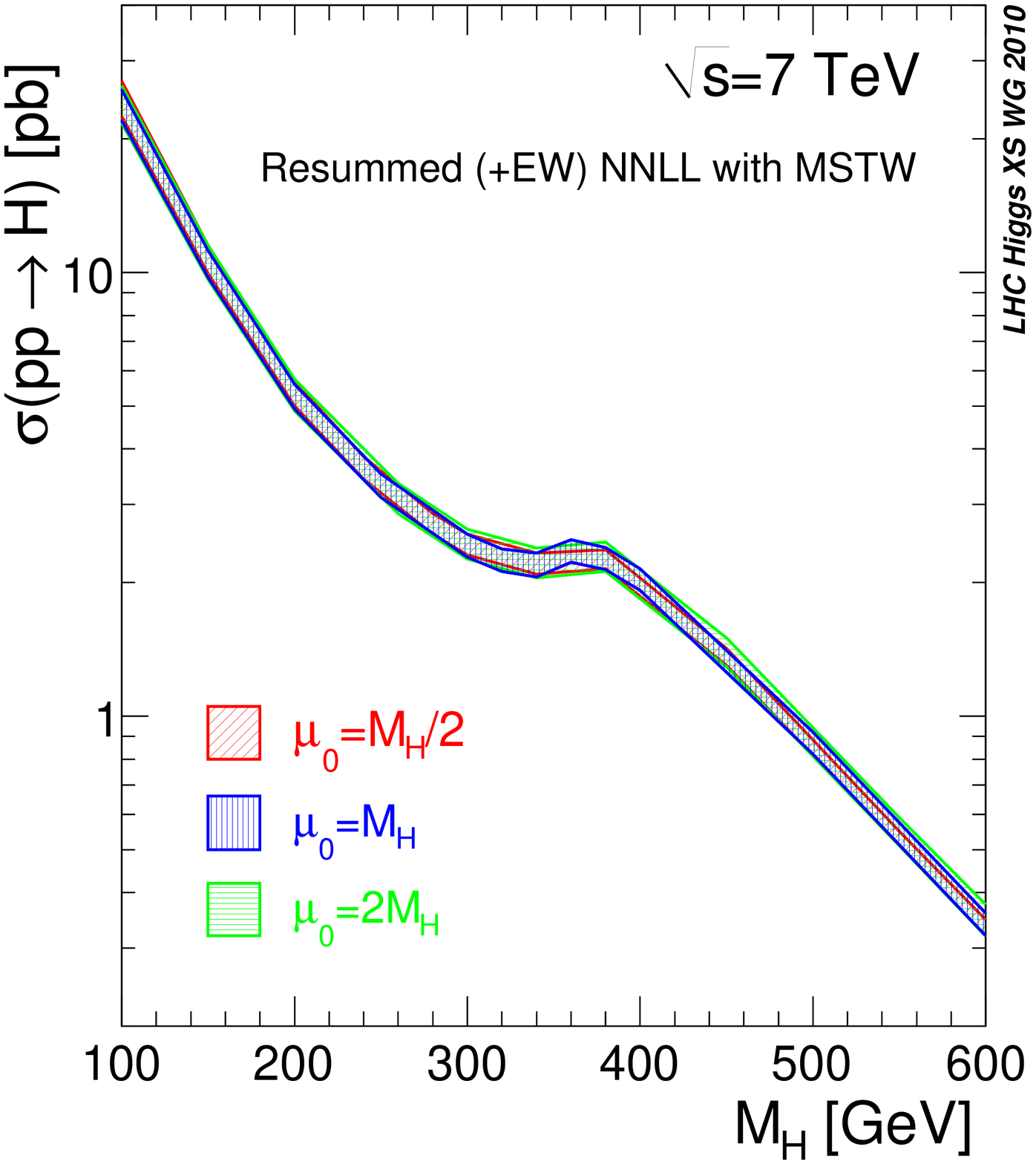} \\
\includegraphics[width=.42\linewidth]{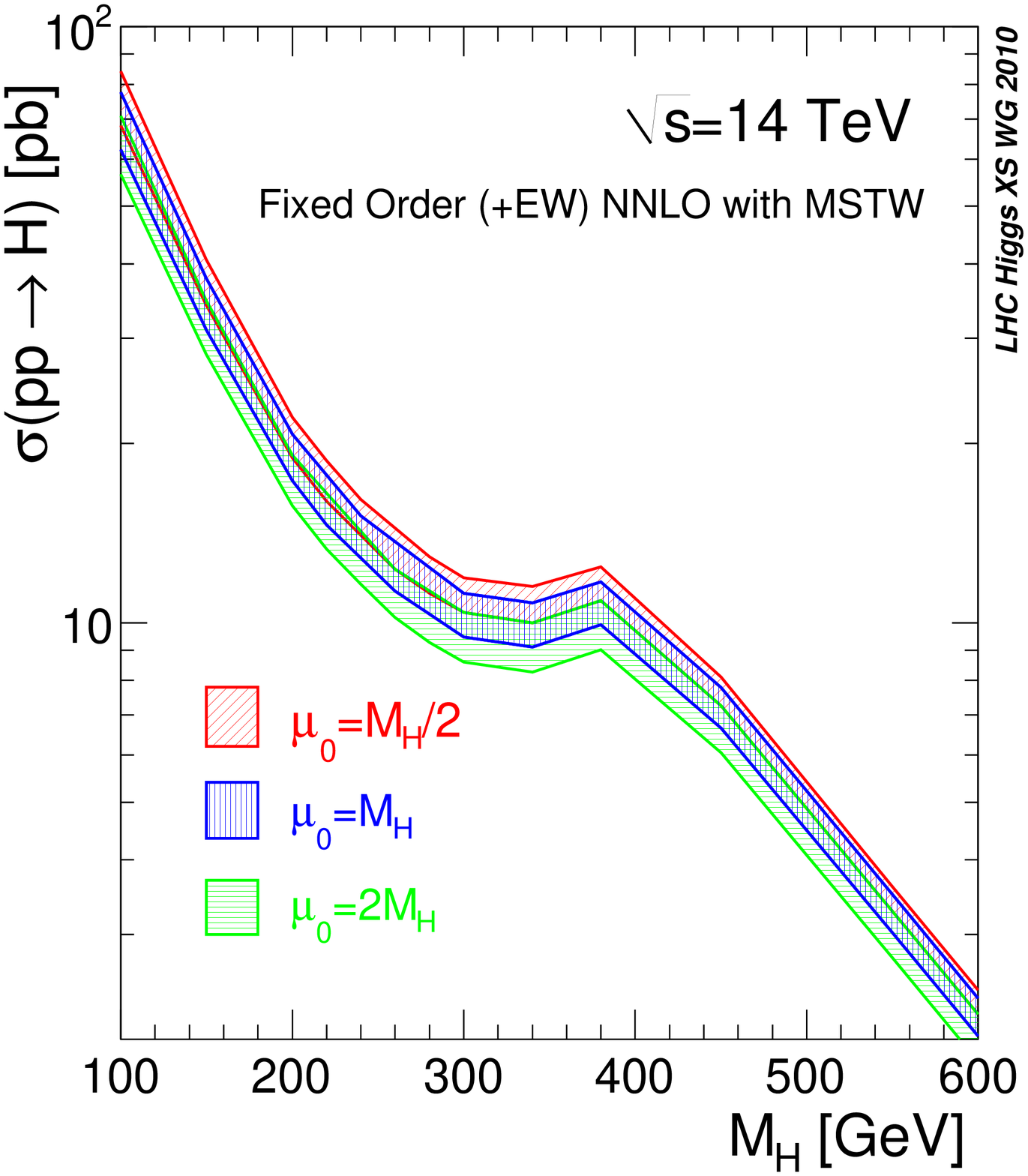} &
\includegraphics[width=.42\linewidth]{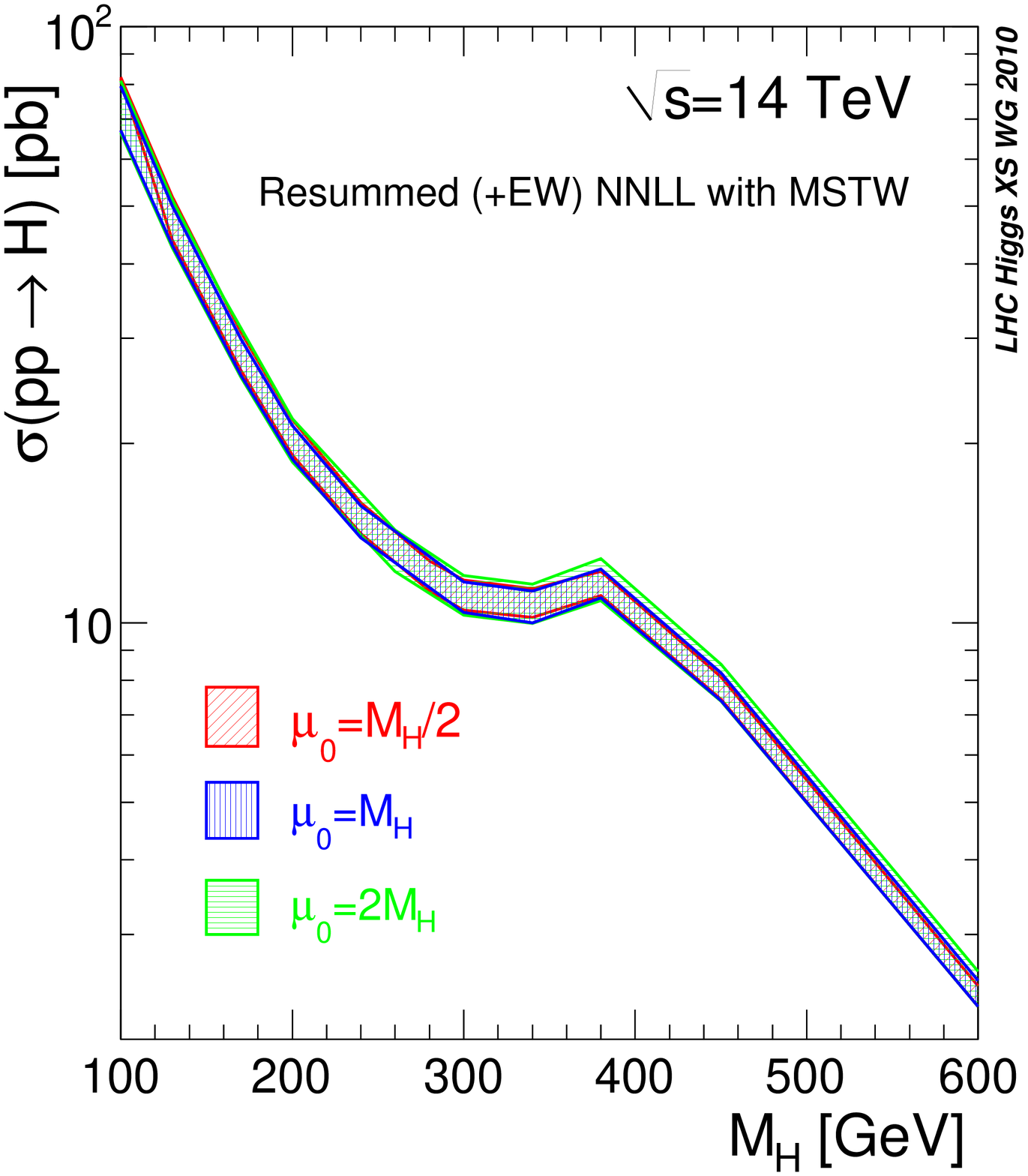} \\
\end{tabular}
\end{center}
\caption{Comparison of NNLO and NNLL bands with different choice of the central scale.}
\label{fig:compband}
\end{figure}

In principle, the uncertainty obtained through scale variations
can only give a lower limit on the {\it true} uncertainty.
Nonetheless, we point out that the results of ABPS and dFG are consistent with
those obtained at the previous order (i.e., dFG NNLL bands overlap with the NNLO band, and ABPS NNLO band overlap
with the NLO band), thus suggesting that the uncertainty obtained with this procedure provides a
reasonable estimate of the true perturbative uncertainty.  At $\sqrt{s}=7$ $(14)\UTeV$ the scale uncertainty of the ABPS result is about $\pm 9{-}10\%$ ($\pm 8{-}13\%$) in the range 
$\MH=100{-}300$\UGeV, and it decreases to about $\pm 7\%$ ($\pm 5\%$) as $\MH$ increases.
At $\sqrt{s}=7$ $(14)\UTeV$ the scale uncertainty of the dFG result is about $\pm 6{-}8\%$ ($\pm 6{-}9\%$) in the range $\MH=100{-}300$\UGeV, and it decreases slightly to about $\pm 5{-}7\%$ ($\pm 5\%$) as $\MH$ increases.

\item[$\bullet$] Another source of perturbative uncertainty on the partonic cross sections comes from the implementation of
the EW corrections. Both ABPS and dFG results are obtained in the complete factorization scheme discussed above.
The partial factorization scheme would lead to a change of the results ranging
from about $-3 \%$ ($\MH=110$\UGeV) to $+1\%$ ($\MH=300$\UGeV).  We note that the effective-theory calculation of \Bref{Anastasiou:2008tj} 
supports the use of the complete factorization scheme.  When the three-loop mixed QCD--EW correction derived there is normalized with the 
exact two-loop light-quark terms derived in \Brefs{Actis:2008ug,Actis:2008ts}, the dominant parts of the exact QCD corrections to 
the EW contributions are properly included.  This is the same reason that the NLO correction found using the large-$\Mt$ approximation 
only differs from the exact result by $10{-}15\%$ even for $\MH \sim 1$\UTeV, well outside the expected range of validity $\MH < 2 \Mt$. We expect that the 
exact three-loop mixed QCD--EW correction is estimated with a similar $\pm 10\%$ uncertainty using the effective-theory calculation of 
\Bref{Anastasiou:2008tj}.  As the two-loop EW contribution to the cross section reaches a 
maximum of only $+5\%$, we estimate an uncertainty of $\pm 1\%$ coming from missing EW corrections for $\MH \lsim 300$\UGeV.

\item[$\bullet$]  The use of the large-$\Mt$ approximation induces another source of uncertainty.
The ABPS and dFG calculations both include the exact NLO corrections with full dependence on the masses of the top and bottom quarks. 
The NNLO (NNLL) top-quark contributions are instead evaluated in the large-$\Mt$ limit. In \Brefs{Marzani:2008az,Harlander:2009bw,Harlander:2009mq,Harlander:2009my,Pak:2009bx,Pak:2009dg}
subleading corrections to the large-$\Mt$ limit have been computed. These
works have shown that for a relatively light Higgs boson ($\MH\lsim 300$\UGeV),
the approximation works to better than $1\%$. For a heavier Higgs boson
($\MH\gsim 300$\UGeV), the accuracy of the large-$\Mt$ approximation is expected to be worse, but still within a few percent.

\item[$\bullet$] Different choices of the input quark masses $\Mt$ and $\Mb$ lead to
a scheme dependence in the cross section.  We have 
checked that different values of $\Mt$ produce a negligible effect on the final cross section.  Although the contribution of the bottom quark to the 
production rate is much smaller than that of the top quark, large logarithms of the form ${\rm ln}(\MH / \Mb)$ lead to a non-negligible shift in the 
cross section. We estimate this by evaluating the cross section using both the pole mass and the $\MSbar$ mass for the \PQb\ quark, and interpreting 
the difference as a measure of uncertainty.  We use the $\MSbar$ mass evaluated at the renormalization scale, $\Mb(\mu_R)$. This leads to an uncertainty estimate of approximately $\pm 1{-}2\%$ on the final result.

\item[$\bullet$] The other important source of uncertainty in the cross section is
the one coming from PDFs.
Modern PDF sets let the user estimate the experimental uncertainty
originating from the accuracy of the data points used to perform the fit.
The MSTW2008 NNLO set \cite{Martin:2009iq} provides 40 different grids that allow evaluation of the experimental uncertainties according
to the procedure discussed in \Bref{Martin:2002aw}.
A related and important uncertainty is the one coming from the value of the QCD coupling.
Higgs production through gluon fusion starts at ${\cal O}(\alphas^2)$ and thus
this uncertainty is expected to have a sizeable effect on the production rate.  
Recently, the MSTW collaboration has studied the combined effect of PDF+$\alphas$
uncertainties~\cite{Martin:2009bu}.
The PDF+$\alphas$ uncertainties at $68\%$ confidence limit (CL) of the ABPS and dFG calculations are
reported in \Trefs{tab:dFG7}--\ref{tab:ABPS14}.
The uncertainties turn out to be quite similar,
being about $\pm 3{-}4\%$ in the range $\MH=100{-}300$\UGeV\ both at $\sqrt{s}=7\UTeV$ and $14$\UTeV. At $\sqrt{s}=7$ $(14)\UTeV$ they
increase to about $\pm 8{-}9\%$ ($\pm 5\%$) at high Higgs-boson masses.
In \Trefs{tab:dFG7}--\ref{tab:ABPS14} we also report the
uncertainties (see \refS{sec:pdf4lhcreco}) obtained through the PDF4LHC recommendation\footnote{We thank A. Vicini for providing us with the PDF4LHC correction factors.} \cite{PDF4LHCwebpage}.
At $7$ $(14)\UTeV$ the uncertainties are about $\pm 7{-}8\%$ ($\pm 6{-}7\%$) in the range $\MH=100{-}300$\UGeV,
and increase at high Higgs-boson masses. This is not completely unexpected: as the Higgs mass increases,
larger values of $x$ are probed, where the gluon distribution is more uncertain.

We finally point out that, besides MSTW, we have at present three other
NNLO parton analyses: ABKM09 \cite{Alekhin:2009ni}, JR09VFNNLO
\cite{JimenezDelgado:2009tv}, and HERAPDF \cite{CooperSarkar:2010ik}. These PDF sets tend to give smaller cross sections
both at $7\UTeV$ and $14\UTeV$ with respect to MSTW.
For example, at $14\UTeV$
the ABKM09 (JR09) result is smaller than the MSTW result by about $6{-}10\%$ ($13{-}8\%$)
in the range $\MH=100{-}300$\UGeV.
At $7\UTeV$ the ABKM09 (JR09) cross section is smaller than the MSTW cross section
by $9{-}16\%$ ($12{-}4\%$) in the same range of Higgs-boson masses.
HERAPDF has released two NNLO PDF sets corresponding to $\alphas(\MZ)=0.1145$ and $\alphas(\MZ)=0.1176$.
At $14\UTeV$ the result corresponding to $\alphas(\MZ)=0.1145$ ($\alphas(\MZ)=0.1176$)
is smaller than the MSTW result by about $8{-}10\%$ ($4{-}5\%$).
At $7\UTeV$ the cross section corresponding to $\alphas(\MZ)=0.1145$ ($\alphas(\MZ)=0.1176$)
is smaller than the MSTW result by about $10{-}14\%$ ($5{-}7\%$).

\end{itemize}

\subsection{Cross-section predictions II}
\label{se:ggfXS2}

\noindent


	 A study of both the central value and uncertainty of Higgs production
cross sections at the Tevatron was performed in \Bref{Baglio:2010um}. We refer to this
analysis with the acronym BD. The BD study was later extended to cover LHC
production \cite{Baglio:2010ae}, and the results are reported in
\Trefs{tab:BD7} and \ref{tab:BD14}. BD use  a
fixed-order calculation with the exact top- and bottom-quark mass effects at
NLO, and then add on the NNLO top contributions in the large-$m_t$ limit as well
as the electroweak corrections at NLO and NNLO, as done in ABPS. They assume a
central scale value $\mu_0=  \MH/2$, as also do ABPS. This leads to an
excellent agreement in central value and relatively good agreement in the
estimated scale variation error with the dFG and ABPS results. BD estimate the
error arising from the PDFs and $\alphas$ differently than do dFG and ABPS.
They first choose to consider the 90\%\ CL PDF+$\Delta^{\rm exp}\alphas$
uncertainty and then define an additional theoretical error of $\Delta^{\rm
th}\alphas = 0.002$ on the strong coupling constant and use PDF grids with a
fixed $\alphas$ provided by MSTW to define a resulting uncertainty on the
production cross section. The resulting BD uncertainty is then added in quadrature
with the combined PDF+$\Delta^{\rm exp}\alphas$  uncertainty (at 90\%\ CL)
obtained by using the MSTW procedure ~\cite{Martin:2009bu}, giving a combined PDF+$\Delta^{\rm 
exp+th}\alphas$ uncertainty estimate of $\pm 10\%$ for Higgs masses below 350
GeV. The BD procedure is motivated by having the PDF+$\alphas$ uncertainty
bands obtained using MSTW to be consistent with those obtained with the PDF set
of \Bref{Alekhin:2009ni}. The ensuing BD uncertainty is only slightly larger than the one
obtained by following the PDF4LHC recommendation. BD finally combine the
uncertainties as follows: the PDF+$\alphas$ uncertainties are evaluated
directly on the maximum and minimum cross sections that arise from scale
variation. This gives a combined BD uncertainty that is comparable to that obtained
with a linear sum of the scale and PDF+$\alphas$ uncertainties. 

The major difference between the BD estimate for the theory uncertainty compared
to dFG and ABPS, is that an additional uncertainty, which is mainly due to the
use of the effective-field-theory approach beyond NLO,  is considered. It
consists of three main components: $i)$ the difference between the partial and
complete factorisation schemes in the NLO electroweak 
corrections~\cite{Actis:2008ug} which
approximately is equivalent to the contributions of the mixed NNLO
QCD--electroweak corrections obtained in the limit  $\MH\muchless \MW$~\cite{Anastasiou:2008tj}; $ii)$ 
the missing \PQb-quark loop contribution at NNLO (and its interference with the
top-quark loop) and the scheme dependence in the renormalisation of the
\PQb-quark mass in the NLO QCD contributions; $iii)$ the use of the $\Mt \to
\infty$ effective approximation  for Higgs masses beyond the $2\Mt$ threshold in
the NNLO QCD contribution. The (linear) sum of these three uncertainties turns
out to be quite large: it is at the  level of about $6{-}7\%$ in the mass range
$\MH \lsim 160\UGeV$ where the  difference between the partial and complete
factorisation approaches is significant and becomes even larger for $\MH\!
\gsim 600\UGeV$ where the $\Mt \to \infty$ approximation starts to fail
badly. 

When the EFT uncertainty is added linearly with the combined scale and
PDF+$\alphas$ uncertainty, the total BD theoretical uncertainties
become definitely large, being at $\sqrt s=7$ TeV, about $\pm 25{-}30\%$
in the low- and high-Higgs mass  ranges.

\begin{table}[!h]


   \begin{center}
   \ccaption{}{\label{tab:BD7}{Results on $\Pp\Pp(\Pg\Pg)\to \PH+X$ cross
    sections with $\sqrt{s}=7$\UTeV\ based on BD calculation with MSTW
    PDFs.}}
    \small
    \begin{tabular}{cccccc}\hline
    $\MH[\UGeVZ]$ & $\sigma[\UpbZ]$  & Scale [\%] &  PDF+$\Delta^{exp+th}_{\alphas}$  [\%]  & ${\rm EFT}$  [\%]  \\ \hline
$90$  & $29.79$ & ${+10.4 \; -12.1}$ & ${+9.3}\;{ -8.9}$ & ${\pm 7.8} $  \\
$95$  & $26.77$ & ${+10.1 \; -11. }$ & ${+9.2}\;{ -8.9}$ & ${\pm 7.7} $  \\  
$100$ & $24.25$ & ${+9.9 \; -10.7 }$ & ${+9.2}\;{ -8.8}$ & ${\pm 7.6} $  \\  
$105$ & $22.01$ & ${+9.6 \; -10.5 }$ & ${+9.2}\;{ -8.8}$ & ${\pm 7.5} $  \\  
$110$ & $20.06$ & ${+9.4 \; -10.3 }$ & ${+9.1}\;{ -8.8}$ & ${\pm 7.4} $  \\  
$115$ & $18.35$ & ${+9.1 \; -10.2 }$ & ${+9.1}\;{ -8.8}$ & ${\pm 7.3} $  \\  
$120$ & $16.84$ & ${+8.9 \; -10.2 }$ & ${+9.1}\;{ -8.8}$ & ${\pm 7.3} $  \\  
$125$ & $15.51$ & ${ +8.8 \; -9.9 }$ & ${+9.1}\;{ -8.8}$ & ${\pm 7.2} $  \\  
$130$ & $14.32$ & ${ +8.5 \; -9.8 }$ & ${+9.1}\;{ -8.8}$ & ${\pm 7.1} $  \\  
$135$ & $13.26$ & ${ +8.4 \; -9.6 }$ & ${+9.1}\;{ -8.8}$ & ${\pm 7.0} $  \\  
$140$ & $12.31$ & ${ +8.3 \; -9.5 }$ & ${+9.1}\;{ -8.8}$ & ${\pm 7.0} $  \\  
$145$ & $11.45$ & ${ +8.2 \; -9.5 }$ & ${+9.1}\;{ -8.8}$ & ${\pm 6.9} $  \\  
$150$ & $10.67$ & ${ +8.1 \; -9.5 }$ & ${+9.1}\;{ -8.8}$ & ${\pm 6.8} $  \\  
$155$ & $9.94$  & ${ +7.9 \; -9.4 }$ & ${+9.1}\;{ -8.8}$ & ${\pm 6.6} $  \\  
$160$ & $9.21$  & ${ +7.8 \; -9.4 }$ & ${+9.1}\;{ -8.8}$ & ${\pm 5.9} $  \\  
$165$ & $8.47$  & ${ +7.7 \; -9.4 }$ & ${+9.1}\;{ -8.8}$ & ${\pm 4.9} $  \\  
$170$ & $7.87$  & ${ +7.7 \; -9.4 }$ & ${+9.1}\;{ -8.8}$ & ${\pm 4.2} $  \\  
$175$ & $7.35$  & ${ +7.6 \; -9.4 }$ & ${+9.1}\;{ -8.9}$ & ${\pm 3.7} $  \\  
$180$ & $6.86$  & ${ +7.5 \; -9.3 }$ & ${+9.2}\;{ -8.9}$ & ${\pm 3.1} $  \\  
$185$ & $6.42$  & ${ +7.4 \; -9.3 }$ & ${+9.2}\;{ -8.9}$ & ${\pm 3.0} $  \\  
$190$ & $6.01$  & ${ +7.4 \; -9.3 }$ & ${+9.2}\;{ -8.9}$ & ${\pm 3.4} $  \\  
$195$ & $5.65$  & ${ +7.4 \; -9.3 }$ & ${+9.2}\;{ -8.9}$ & ${\pm 3.6} $  \\  
$200$ & $5.34$  & ${ +7.3 \; -9.3 }$ & ${+9.3}\;{ -9.0}$ & ${\pm 3.7} $  \\  
$210$ & $4.81$  & ${ +7.2 \; -9.3 }$ & ${+9.3}\;{ -9.0}$ & ${\pm 3.7} $  \\  
$220$ & $4.36$  & ${ +7.2 \; -9.2 }$ & ${+9.3}\;{ -9.1}$ & ${\pm 3.6} $  \\  
$230$ & $3.97$  & ${ +7.0 \; -9.2 }$ & ${+9.4}\;{ -9.2}$ & ${\pm 3.5} $  \\  
$240$ & $3.65$  & ${ +7.0 \; -9.2 }$ & ${+9.5}\;{ -9.2}$ & ${\pm 3.3} $  \\  
$250$ & $3.37$  & ${ +6.9 \; -9.2 }$ & ${+9.5}\;{ -9.3}$ & ${\pm 3.1} $  \\  
$260$ & $3.11$  & ${ +6.8 \; -9.2 }$ & ${+9.6}\;{ -9.4}$ & ${\pm 3.0} $  \\  
$270$ & $2.89$  & ${ +6.7 \; -9.2 }$ & ${+9.7}\;{ -9.5}$ & ${\pm 2.8} $  \\  
$280$ & $2.71$  & ${ +6.8 \; -9.2 }$ & ${+9.8}\;{ -9.5}$ & ${\pm 2.6} $  \\  
$290$ & $2.55$  & ${ +6.8 \; -9.1 }$ & ${+9.8}\;{ -9.6}$ & ${\pm 2.4} $  \\  
$300$ & $2.42$  & ${ +6.7 \; -9.1 }$ & ${+9.9}\;{ -9.7}$ & ${\pm 2.3} $  \\  
$320$ & $2.23$  & ${ +6.7 \; -9.2 }$ & ${+10.1}\;{-9.9}$ & ${\pm 2.3} $  \\  
$340$ & $2.19$  & ${ +6.9 \; -9.2 }$ & ${+10.3}\;{-10.1}$& ${\pm 3.0} $ \\  
$360$ & $2.31$  & ${ +7.0 \; -9.2 }$ & ${+10.5}\;{-10.3}$& ${\pm 4.1} $ \\  
$380$ & $2.18$  & ${ +6.3 \; -9.1 }$ & ${+10.7}\;{-10.5}$& ${\pm 2.5} $  \\  
$400$ & $1.93$  & ${ +5.9 \; -8.8 }$ & ${+11.0}\;{-10.7}$& ${\pm 3.1} $  \\  
$450$ & $1.27$  & ${ +5.0 \; -8.4 }$ & ${+11.6}\;{-11.3}$& ${\pm 4.0} $  \\  
$500$ & $0.79$  & ${ +4.4 \; -8.1 }$ & ${+12.2}\;{-11.9}$& ${\pm 4.5} $  \\  
$550$ & $0.49$  & ${ +4.0 \; -7.9 }$ & ${+12.7}\;{-12.4}$& ${\pm 5.5} $  \\  
$600$ & $0.31$  & ${ +3.7 \; -7.7 }$ & ${+13.3}\;{-13.0}$& ${\pm 6.6} $  \\  
$650$ & $0.20$  & ${ +3.5 \; -7.6 }$ & ${+14.0}\;{-13.5}$& ${\pm 7.5} $  \\  
$700$ & $0.13$  & ${ +3.4 \; -7.5 }$ & ${+14.7}\;{-14.1}$& ${\pm 8.3} $  \\  
$750$ & $0.08$  & ${ +3.3 \; -7.4 }$ & ${+15.4}\;{-14.6}$& ${\pm 9.0} $  \\  
$800$ & $0.06$  & ${ +3.1 \; -7.4 }$ & ${+16.2}\;{-15.1}$& ${\pm 9.7} $  \\  
$850$ & $0.04$  & ${ +3.1 \; -7.3 }$ & ${+17.1}\;{-15.7}$& ${\pm 10.2}$  \\  
$900$ & $0.03$  & ${ +3.0 \; -7.3 }$ & ${+18.0}\;{-16.2}$& ${\pm 10.8}$  \\  
$950$ & $0.02$  & ${ +3.0 \; -7.3 }$ & ${+18.9}\;{-16.8}$& ${\pm 11.3}$  \\  
$1000$& $0.01$  & ${ +2.9 \; -7.2 }$ & ${+19.9}\;{-17.3}$& ${\pm 11.8}$  \\ \hline
\end{tabular} 
\end{center} 
\end{table}

\begin{table}[!h]

   \begin{center}
   \ccaption{}{\label{tab:BD14}{Results on $\Pp\Pp(\Pg\Pg)\to \PH+X$
    cross sections with $\sqrt{s}=14$\UTeV\ based on BD calculation with MSTW PDFs.}}
    \small
    \begin{tabular}{cccccc}\hline
    $\MH[\UGeVZ]$ & $\sigma[\UpbZ]$  & Scale [\%] &  PDF+$\Delta^{exp+th}_{\alphas}$  [\%]  & ${\rm EFT}$  [\%] \\ \hline
$90$  & $90.02$ & ${+10.8 \; -14.3}$ & ${ +9.1\; -8.9}$ & ${\pm 8.3} $  \\  
$95$  & $82.09$ & ${+10.4 \; -13.8}$ & ${ +9.0\; -8.8}$ & ${\pm 8.2} $  \\  
$100$ & $75.41$ & ${+10.2 \; -13.4}$ & ${ +8.9\; -8.7}$ & ${\pm 8.1} $  \\  
$105$ & $69.38$ & ${+9.9 \; -13.0 }$ & ${ +8.8\; -8.7}$ & ${\pm 8.0} $  \\  
$110$ & $64.07$ & ${+9.6 \; -12.6 }$ & ${ +8.7\; -8.6}$ & ${\pm 7.9} $  \\  
$115$ & $59.37$ & ${+9.4 \; -12.2 }$ & ${ +8.7\; -8.5}$ & ${\pm 7.8} $  \\  
$120$ & $55.20$ & ${+9.2 \; -11.9 }$ & ${ +8.6\; -8.4}$ & ${\pm 7.7} $  \\  
$125$ & $51.45$ & ${+9.0 \; -11.6 }$ & ${ +8.5\; -8.4}$ & ${\pm 7.6 } $  \\  
$130$ & $48.09$ & ${+8.9 \; -11.4 }$ & ${ +8.5\; -8.3}$ & ${\pm 7.5 } $  \\  
$135$ & $45.06$ & ${+8.7 \; -11.1 }$ & ${ +8.4\; -8.2}$ & ${\pm 7.5 } $  \\  
$140$ & $42.30$ & ${+8.5 \; -10.8 }$ & ${ +8.4\; -8.2}$ & ${\pm 7.4 } $  \\  
$145$ & $39.80$ & ${+8.4 \; -10.6 }$ & ${ +8.3\; -8.1}$ & ${\pm 7.3 } $ \\  
$150$ & $37.50$ & ${+8.3 \; -10.4 }$ & ${ +8.3\; -8.1}$ & ${\pm 7.2 } $  \\  
$155$ & $35.32$ & ${+8.1 \; -10.2 }$ & ${ +8.3\; -8.1}$ & ${\pm 7.0 } $ \\  
$160$ & $33.08$ & ${+8.0 \; -10.0 }$ & ${ +8.2\; -8.0}$ & ${\pm 6.3 } $  \\  
$165$ & $30.77$ & ${ +7.9 \; -9.8 }$ & ${ +8.2\; -8.0}$ & ${\pm 5.3 } $ \\  
$170$ & $28.89$ & ${ +7.8 \; -9.7 }$ & ${ +8.2\; -8.0}$ & ${\pm 4.5 } $  \\  
$175$ & $27.24$ & ${ +7.8 \; -9.5 }$ & ${ +8.2\; -7.9}$ & ${\pm 4.0 } $  \\  
$180$ & $25.71$ & ${ +7.7 \; -9.4 }$ & ${ +8.2\; -7.9}$ & ${\pm 3.5 } $  \\  
$185$ & $24.28$ & ${ +7.6 \; -9.1 }$ & ${ +8.1\; -7.9}$ & ${\pm 3.3 } $  \\  
$190$ & $22.97$ & ${ +7.6 \; -9.1 }$ & ${ +8.1\; -7.9}$ & ${\pm 3.8 } $ \\  
$195$ & $21.83$ & ${ +7.5 \; -9.0 }$ & ${ +8.1\; -7.9}$ & ${\pm 4.0 } $ \\  
$200$ & $20.83$ & ${ +7.4 \; -8.8 }$ & ${ +8.1\; -7.9}$ & ${\pm 4.1 } $ \\  
$210$ & $19.10$ & ${ +7.3 \; -8.6 }$ & ${ +8.1\; -7.8}$ & ${\pm 4.1 } $  \\  
$220$ & $17.64$ & ${ +7.2 \; -8.4 }$ & ${ +8.1\; -7.8}$ & ${\pm 4.0 } $  \\  
$230$ & $16.38$ & ${ +7.1 \; -8.3 }$ & ${ +8.0\; -7.8}$ & ${\pm 3.8 } $ \\  
$240$ & $15.30$ & ${ +7.0 \; -8.2 }$ & ${ +8.0\; -7.8}$ & ${\pm 3.7 } $  \\  
$250$ & $14.38$ & ${ +6.9 \; -8.2 }$ & ${ +8.0\; -7.8}$ & ${\pm 3.5 } $  \\  
$260$ & $13.52$ & ${ +6.8 \; -8.1 }$ & ${ +8.0\; -7.8}$ & ${\pm 3.3 } $ \\  
$270$ & $12.79$ & ${ +6.7 \; -8.1 }$ & ${ +8.0\; -7.8}$ & ${\pm 3.1 } $  \\  
$280$ & $12.17$ & ${ +6.7 \; -8.1 }$ & ${ +8.0\; -7.8}$ & ${\pm 2.9 } $ \\  
$290$ & $11.65$ & ${ +6.6 \; -8.0 }$ & ${ +8.0\; -7.8}$ & ${\pm 2.8 } $  \\  
$300$ & $11.22$ & ${ +6.5 \; -8.0 }$ & ${ +8.0\; -7.8}$ & ${\pm 3.4 } $  \\  
$320$ & $10.70$ & ${ +6.5 \; -8.0 }$ & ${ +8.1\; -7.9}$ & ${\pm 3.1 } $ \\  
$340$ & $10.83$ & ${ +6.5 \; -8.0 }$ & ${ +8.1\; -7.9}$ & ${\pm 2.8 } $ \\  
$360$ & $11.77$ & ${ +6.4 \; -8.0 }$ & ${ +8.1\; -8.0}$ & ${\pm 3.5 } $  \\  
$380$ & $11.46$ & ${ +6.0 \; -7.7 }$ & ${ +8.2\; -8.1}$ & ${\pm 4.4 } $ \\  
$400$ & $10.46$ & ${ +5.6 \; -7.4 }$ & ${ +8.2\; -8.1}$ & ${\pm 5.0 } $ \\  
$450$ & $7.42$  & ${ +5.0 \; -7.0 }$ & ${ +8.4\; -8.3}$ & ${\pm 6.0 } $  \\  
$500$ & $4.97$  & ${ +4.6 \; -6.7 }$ & ${ +8.6\; -8.6}$ & ${\pm 6.4 } $ \\  
$550$ & $3.32$  & ${ +4.3 \; -6.5 }$ & ${ +8.9\; -8.8}$ & ${\pm 7.4 } $ \\  
$600$ & $2.24$  & ${ +4.1 \; -6.3 }$ & ${ +9.2\; -9.1}$ & ${\pm 8.3 } $  \\  
$650$ & $1.53$  & ${ +3.9 \; -6.2 }$ & ${ +9.5\; -9.4}$ & ${\pm 9.0 } $  \\  
$700$ & $1.05$  & ${ +3.8 \; -6.1 }$ & ${ +9.8\; -9.6}$ & ${\pm 9.6 } $  \\  
$750$ & $0.74$  & ${ +3.6 \; -6.0 }$ & ${+10.1\; -9.9}$ & ${\pm 10.1 } $  \\  
$800$ & $0.52$  & ${ +3.5 \; -6.0 }$ & ${+10.4\;-10.2}$ & ${\pm 10.5 } $ \\  
$850$ & $0.38$  & ${ +3.5 \; -5.9 }$ & ${+10.7\;-10.5}$ & ${\pm 11.0 } $  \\  
$900$ & $0.27$  & ${ +3.4 \; -5.9 }$ & ${+11.0\;-10.7}$ & ${\pm 11.3 } $  \\  
$950$ & $0.20$  & ${ +3.3 \; -5.8 }$ & ${+11.3\;-11.0}$ & ${\pm 11.7 } $ \\  
$1000$ & $0.15$ & ${ +3.3 \; -5.8 }$ & ${+11.5\;-11.3}$ & ${\pm 12.0 } $  \\ \hline
\end{tabular} 
\end{center} 
\end{table}

\subsection{An alternative cross-section calculation based on an
effective field theory}

In \Bref{Ahrens:2010rs} updated predictions for Higgs-boson production at the Tevatron and the LHC were presented.
The results of \Bref{Ahrens:2010rs} are based on the work of \Brefs{Ahrens:2008qu,Ahrens:2008nc},
where a new calculation of the Higgs production cross section was presented.
This calculation supplements the NNLO result, obtained in the large-$\Mt$ approximation, 
with soft-gluon resummation done in the framework of an effective field theory (EFT) approach, and with the resummation of
some ``$\pi^2$-terms'' originating from the analytic continuation of the gluon form factor.  These additional terms are obtained 
in the EFT formalism by choosing an imaginary matching scale, and are included by the authors to improve the convergence of the perturbative series.  
The update of \Bref{Ahrens:2010rs} treats both top- and bottom-quark loops in the heavy-quark approximation, and includes EW corrections 
assuming complete factorization.  In the range $\MH=115{-}200$\UGeV\ the central values of \Bref{Ahrens:2010rs} are in good agreement with those of the 
ABPS and dFG calculations (for example, the difference with the dFG results is at $1{-}2\%$ level).  However, we note that the reliability of $\pi^2$ resummation 
has been questioned, and that there are puzzling differences between this approach and the standard soft-gluon resummation.  The effect of 
resummation in \Bref{Ahrens:2010rs} is driven by the $\pi^2$ terms; without them, the effect of resummation is much smaller than the one obtained using the 
standard approach~\cite{Ahrens:2008nc}.  The numerical agreement between central values therefore appears accidental.  
Soft-gluon resummations typically deal with logarithmically terms that are enhanced in some region of the phase space.
As an example, in the soft-gluon resummation of \Bref{Catani:2003zt} the logarithmic terms are $\log^n(1-z)$ where
$1-z=1-\MH^2/{\hat s}$ is the distance from the partonic threshold. These logarithmic terms can be precisely traced back and identified at each 
perturbative order. On the contrary, $\pi^2$ terms are just numbers, and there is no limit in which they can dominate. Moreover, only those 
$\pi^2$ terms coming from the analytic continuation of the gluon form factor can actually be controlled in this way.  Other $\pi^2$ terms are
present at each order in perturbation theory, and they can be obtained only through an explicit computation.
We add a final comment on the perturbative uncertainties quoted in the calculation of \Bref{Ahrens:2010rs}. The scale uncertainty of the 
results are of the order of $\pm 3\%$ or smaller.
This should be contrasted with the uncertainties of the ABPS and dFG calculations, which are a factor of $2{-}3$ larger. Since the calculation of 
\Bref{Ahrens:2010rs} does not contain new information beyond NNLO with respect to those of ABPS, dFG, and BD, we feel uncomfortable with such a 
small uncertainty and believe it is underestimated. For comparison, it should be noticed that the perturbative uncertainty of a full N$^3$LO
calculation, estimated through scale variations, would be of the order of about $\pm 5\%$~\cite{Moch:2005ky}.

\clearpage

\section{Vector-Boson-Fusion process\footnote{A.~Denner, S.~Farrington,
  C.~Hackstein, C.~Oleari, D.~Rebuzzi (eds.); P.~Bolzoni, S.~Dittmaier, F.~Maltoni,
  S.-O.~Moch, A.~M\"uck, S. Palmer and M.~Zaro.} }

\providecommand{\lsim}
{\;\raisebox{-.3em}{$\stackrel{\displaystyle <}{\sim}$}\;}
\providecommand{\gsim}
{\;\raisebox{-.3em}{$\stackrel{\displaystyle >}{\sim}$}\;}

\subsection{Higgs-boson production in vector-boson fusion}
\label{sec:intro}
The production of a Standard Model Higgs boson in association with two hard
jets in the forward and backward regions of the detector, frequently quoted
as the ``vector-boson fusion''~(VBF) channel, is a cornerstone in
the Higgs-boson search both in the ATLAS~\cite{Asai:2004ws} and
CMS~\cite{Abdullin:2005yn} experiments at the LHC.  Higgs-boson production in
the VBF channel plays also an important role in the determination of
Higgs-boson couplings at the LHC (see e.g.,\ \Bref{Duhrssen:2004cv}). Bounds on
non-standard couplings between Higgs and electroweak~(EW) gauge bosons can be
imposed from precision studies in this channel~\cite{Hankele:2006ma}.
In addition this channel contributes in a significant way to the inclusive Higgs 
production over the full Higgs-mass range.

The production of a Higgs boson~+~2~jets receives two contributions at hadron
colliders. The first type, where the Higgs boson couples to a weak boson that
links two quark lines, is dominated by $t$- and $u$-channel-like diagrams and
represents the genuine VBF channel.  The hard jet pairs have a strong
tendency to be forward--backward directed in contrast to other jet-production
mechanisms, offering a good background suppression (transverse-momentum and
rapidity cuts on jets, jet rapidity gap, central-jet veto, etc.). 

If one is interested in the measurement of the Higgs-boson couplings in VBF,
especially for the measurement of the $\PH\PW\PW$ and $\PH\PZ\PZ$ couplings, 
cuts should
be applied in order to suppress events from Higgs~+~2~jet production via
gluon fusion, which become a background to the signal VBF production.  In the
gluon-fusion channel, the Higgs boson is radiated off a heavy-quark loop that
couples to any parton of the incoming hadrons via
gluons~\cite{DelDuca:2001fn,Campbell:2006xx}. Although the final states are
similar, the kinematic distributions of jets are very different.  Applying
appropriate event selection criteria, called VBF cuts (see
e.g.,\ \Brefs{Barger:1994zq, Rainwater:1997dg, Rainwater:1998kj,
  Rainwater:1999sd, DelDuca:2006hk}), it is possible to sufficiently suppress
the gluon-fusion Higgs-boson production mechanism with respect to the VBF
one.  According to a recent estimate~\cite{Nikitenko:2007it}, gluon fusion
contributes about $4{-}5\%$ to the Higgs~+~2~jet events for a Higgs-boson
mass of $120$\UGeV, after applying VBF cuts.
A next-to-leading order~(NLO)
analysis of the gluon-fusion contribution~\cite{Campbell:2006xx} shows that its residual
scale dependence is still of the order of $35\%$.

Electroweak Higgs-boson production at leading order~(LO) involve only quark
and antiquark initial states, $\Pq \Pq\to \Pq \Pq\PH$.  The topologies of the
LO Feynman diagrams contributing to various partonic processes are shown
in~\Figure~\ref{fig:VBFLOtops}.  As $s$-channel diagrams and interferences
tend to be suppressed when imposing VBF cuts, the cross section can be
approximated by the contribution of squared $t$- and $u$-channel diagrams only
without their interference.  The
corresponding QCD corrections reduce to vertex corrections to the
weak-boson--quark coupling.  Explicit NLO QCD calculations in this
approximation~\cite{Spira:1997dg,Han:1992hr,Figy:2003nv,Figy:2004pt,Berger:2004pca}
confirm the expectation that these QCD corrections are small, because they
are shifted to the parton distribution functions (PDFs) via QCD factorization
to a large extent. The resulting QCD corrections are of the order of
$5{-}10\%$ and reduce the remaining factorization and renormalization scale
dependence of the NLO cross section to a few percent.
For the NLO QCD predictions from {\sc
  HAWK}~\cite{Ciccolini:2007jr,Ciccolini:2007ec,HAWK}, {\sc
  VBFNLO}~\cite{Figy:2003nv,Arnold:2008rz}, and {\sc VV2H}~\cite{VV2H} (this last
program calculates only total cross sections without cuts), a tuned
comparison has been performed in \Bref{HiggsLH:2008uu}, neglecting $s$-channel
diagrams and interferences. Recently, {\sc VBF@NNLO}~\cite{Bolzoni:2010xr}
was also run in the same setup.  The results of all four codes were found to
agree within the statistical errors at the level of $0.1\%$.

\newcommand\scalefac{0.9}
\newlength{\largfig}
\largfig=\scalefac pt

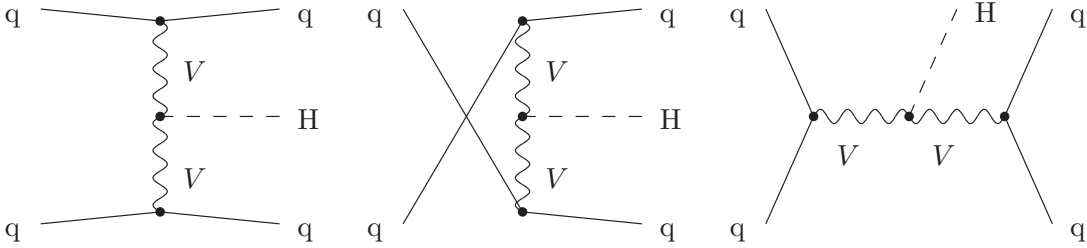
\begin{figure}
  \begin{center}
    {\unitlength \largfig 
      \begin{picture}(125,130)(0,0)
        \SetScale{\scalefac}
        \Line( 15,95)( 65,90)
        \Line( 15, 5)( 65,10)
        \Line(115, 5)( 65,10)
        \Line(115,95)( 65,90)
        \Photon( 65,10)( 65,90){3}{7}
        \DashLine(115,50)( 65,50){6}
        \Vertex(65,90){2}
        \Vertex(65,10){2}
        \Vertex(65,50){2}
        \put(  0,90){$\Pq$}
        \put(  0,0){$\Pq$}
        \put(123,90){{$\Pq$}}
        \put(123,0){{$\Pq$}}
        \put(123,45){{$\PH$}}
        \put( 75,65){{$V$}}
        \put( 75,20){{$V$}}
        \SetScale{1}
      \end{picture}
    }%
    \hspace*{2em}%
            {\unitlength  \largfig 
              \begin{picture}(125,100)(0,0)
                \SetScale{\scalefac}
                \Line( 15,95)( 65,10)
                \Line( 15, 5)( 65,90)
                \Line(115, 5)( 65,10)
                \Line(115,95)( 65,90)
                \Photon( 65,10)( 65,90){3}{7}
                \DashLine(115,50)( 65,50){6}
                \Vertex(65,90){2}
                \Vertex(65,10){2}
                \Vertex(65,50){2}
                \put(  0,90){$\Pq$}
                \put(  0,0){$\Pq$}
                \put(123,90){{$\Pq$}}
                \put(123,0){{$\Pq$}}
                \put(123,45){{$\PH$}}
                \put( 75,65){{$V$}}
                \put( 75,20){{$V$}}
                \SetScale{1}
              \end{picture}
            }%
            \hspace*{2em}%
                    {\unitlength  \largfig  
                      \begin{picture}(145,100)(0,0)
                        \SetScale{\scalefac}
                        \Line( 15,95)( 35,50)
                        \Line( 15, 5)( 35,50)
                        \Line(135,95)(115,50)
                        \Line(135, 5)(115,50)
                        \Photon( 35,50)(115,50){3}{7}
                        \DashLine(75,50)(95,95){6}
                        \Vertex( 35,50){2}
                        \Vertex(115,50){2}
                        \Vertex( 75,50){2}
                        \put(  0,90){$\Pq$}
                        \put(  0,0){$\Pq$}
                        \put(143,90){{$\Pq$}}
                        \put(143,0){{$\Pq$}}
                        \put(103,90){{$\PH$}}
                        \put( 45,30){{$V$}}
                        \put( 85,30){{$V$}}
                        \SetScale{1}
                      \end{picture}
                    }                      
  \end{center}
  \caption{Topologies of $t$-, $u$-, and $s$-channel contributions
for electroweak Higgs-boson production, $\Pq
    \Pq\to \Pq \Pq\PH$ at LO, where $\Pq$ denotes any quark or antiquark and
    $V$ stands for \PW and \PZ~boson.}
  \label{fig:VBFLOtops}
\end{figure}

In \Brefs{Ciccolini:2007jr,Ciccolini:2007ec} the full NLO EW +
QCD corrections have been computed with {\sc HAWK}, including the complete set of
$t$-, $u$-, and $s$-channel Feynman diagrams and taking into account
real corrections induced by photons in the initial state and QED
corrections implicitly contained in the DGLAP evolution of PDFs.  The
size of the electroweak corrections sensitively depends on the chosen
renormalization scheme to define the weak couplings, most notably on
the chosen value for the electromagnetic coupling $\alpha$. The
preferred choice, which should be most robust with respect to
higher-order corrections, is the so-called $\GF$ scheme, where
$\alpha$ is derived from Fermi's constant $\GF$.  The impact of EW and
QCD corrections in the favoured Higgs-mass range between $100$ and
$200$\UGeV{} are of order $5\%$ and negative, and thus as important as
the QCD corrections.  Photon-induced processes lead to corrections at
the percent level.

Approximate next-to-next-to-leading order~(NNLO) QCD corrections to the total
inclusive cross section for VBF have been presented in \Bref{Bolzoni:2010xr}.
The theoretical predictions are obtained using the structure-function
approach~\cite{Han:1992hr}. Upon including the NNLO corrections in QCD for
the VBF production mechanism via the structure-function approach the
theoretical uncertainty for this channel, i.e. the scale dependence, reduces from the $5{-}10\%$ of the
NLO QCD and electroweak combined
computations~\cite{Han:1992hr,Ciccolini:2007ec} down to $1{-}2\%$. The
uncertainties due to parton distributions are estimated to be at the same
level.

\subsection{Higher-order calculations}
\label{sec:NLO_calculations}
In order to study the NLO corrections to Higgs-boson production in VBF, we
have used two existing partonic Monte Carlo programs: {\sc HAWK} and {\sc
  VBFNLO}, which we now present. Furthermore we also give results of the
NNLO QCD calculation based on {\sc VBF@NNLO} and combine them with the
electroweak corrections obtained from {\sc HAWK}.

\subsubsection{HAWK -- NLO QCD and EW corrections}
\label{sec:HAWK}
{\sc HAWK}~\cite{Ciccolini:2007jr,Ciccolini:2007ec,HAWK} is a Monte Carlo
event generator for $\Pp\Pp\to\PH + 2\,\mathrm{jets}$.  It includes the
complete NLO QCD and electroweak corrections and all weak-boson fusion and
quark--antiquark annihilation diagrams, i.e.~$t$-channel and $u$-channel
diagrams with VBF-like vector-boson exchange and $s$-channel Higgs-strahlung
diagrams with hadronic weak-boson decay.  Also, all interferences at LO and
NLO are included. If it is supported by the PDF set, contributions from
incoming photons, which are at the level of $1{-}2\%$, can be taken into
account.  Leading heavy-Higgs-boson effects at two-loop order proportional to
$\GF^2 \MH^4$ are included according to
\Brefs{Ghinculov:1995bz,Frink:1996sv}. While these contributions are
negligible for small Higgs-boson masses, they become important for
Higgs-boson masses above $400$\UGeV.  For $\MH=700$\UGeV{} they yield
$+4\%$, i.e.~about half of the total EW corrections. This signals a breakdown
of the perturbative expansion, and these contributions can be viewed as an
estimate of the theoretical uncertainty.  Contributions of $\PQb$-quark PDFs
and final-state $\PQb$ quarks can be taken into account at LO. While the
effect of only initial $\PQb$ quarks is negligible, final-state $\PQb$ quarks can increase the cross section by up to $4\%$.
While $s$-channel diagrams can contribute up to $25\%$ for small
Higgs-boson masses in the total cross section without cuts, their contribution is below
 $1\%$ once VBF cuts are applied. Since the $s$-channel diagrams are
actually a contribution to $\PW\PH$ and $\PZ\PH$ production, they
are switched off in the following.

The code is interfaced to LHAPDF and allows to evaluate
the PDF uncertainties in a single run.  The calculation can be performed for an
on-shell Higgs boson or for an off-shell Higgs boson decaying into a pair of
gauge singlets, thus mimicking an off-shell Higgs boson. While the effects of
the off-shellness are negligible for small Higgs-boson masses, they should be
taken into account for $\MH \gsim 400$\UGeV. As a flexible partonic Monte
Carlo generator, {\sc HAWK} allows to apply phase-space cuts on the jets and
the Higgs-boson decay products and to switch off certain contributions.

\subsubsection{VBFNLO -- NLO QCD and EW corrections}
\label{sec:VBFNLO}
{\sc VBFNLO}~\cite{webVBFNLO} is a fully flexible partonic Monte Carlo
program for VBF, double and triple vector-boson production processes at NLO
QCD accuracy.  Arbitrary cuts can be specified as well as various scale
choices: in fact, {\sc VBFNLO} can use fixed or dynamical renormalization and
factorization scales. Any currently available parton distribution function
set can be used through the LHAPDF library.  For processes implemented at
leading order, the program is capable of generating event files in the Les
Houches Accord (LHA) format~\cite{Alwall:2006yp}.

Since, in the phase-space regions which are accessible at hadron colliders,
VBF reactions are dominated by $t$-channel electroweak gauge-boson exchange,
in {\sc VBFNLO}, $s$-channel exchange contributions and
kinematically-suppressed fermion-interference
contributions~\cite{Andersen:2007mp,Bredenstein:2008tm} are disregarded. 
While the interference
effects are always well below $1\%$, they are entirely negligible once VBF
cuts are applied. Here, even the $s$-channel contributions which, with
excellent accuracy, can be regarded as a separate "Higgs-strahlung" process, drop below 1\%.
The subsequent decay of the Higgs boson is simulated in the narrow-width
approximation.  For the $\PH\to \PW^+\PW^- $ and the $\PH\to \PZ\PZ$ modes,
full off-shell effects and spin correlations of the decay leptons are
included.  Details of the calculation can be found in
\Bref{Figy:2003nv}.  Very recently, the EW corrections to VBF
Higgs-boson production have been added to the code~\cite{Figy:2010ct}.


\subsubsection{VBF@NNLO -- NNLO QCD corrections}
\label{sec:VBF-NNLO}

{\sc VBF@NNLO}~\cite{Bolzoni:2010xr} computes VBF Higgs cross sections 
at LO, NLO, and NNLO in QCD via the structure-function approach.  This
approach~\cite{Han:1992hr} consists basically in viewing the VBF process as a
double deep-inelastic scattering (DIS) attached to the colourless pure
electroweak vector-boson fusion into a Higgs boson.  According to this
approach one can include NLO QCD corrections to the VBF process employing the
standard DIS structure functions $F_i(x,Q^2);\,i=1,2,3$ at
NLO~\cite{Bardeen:1978yd} or similarly the corresponding structure
functions~\cite{Kazakov:1990fu,Zijlstra:1992kj,Zijlstra:1992qd,Moch:1999eb}.

The structure-function approach does not include all types of contributions.
At LO a structure-function-violating contribution comes from the
interferences between identical final-state quarks (e.g.,\
$\PQu\PQu\rightarrow \PH\PQu\PQu$) or between processes where either a $\PW$
or a $\PZ$ can be exchanged (e.g.,\ $\PQu\PQd\rightarrow \PH\PQu\PQd$).  These
LI contributions have been included in the NNLO results. Apart from such
contributions, the structure-function approach represents an exact approach
also at NLO.  At NNLO, however, several types of diagrams violate it.  Some
are colour suppressed and kinematically
suppressed~\cite{vanNeerven:1984ak,Blumlein:1992eh,Figy:2007kv}, others have
been shown in \Bref{Harlander:2008xn} to be small enough not to produce
a significant deterioration of the VBF signal.  A first rough estimation for
a third set showed that their contribution is small and can be safely
neglected.  At NNLO in QCD, the theoretical uncertainty is reduced to be less
than $2\%$.

\subsection{Results}
In the following, we present VBF results for LHC at $7\UTeV$ and $14\UTeV$
calculated at NLO, from {\sc HAWK} and {\sc VBFNLO}~\cite{webVBFNLO}, and at
NNLO, from {\sc VBF@NNLO}~\cite{WebInterface:2010}.

All results have been computed using the values of the electroweak parameters
given in~\refA{sminput}. The renormalization and factorization scales have
been fixed to $\MW$, and both the scales varied in the range
$\MW/2<\mu<2\MW$. The Higgs boson has been treated as stable on on-shell, and
the contributions from $s$-channel diagrams have been neglected.

\begin{figure}
  \includegraphics[width=0.5\textwidth]{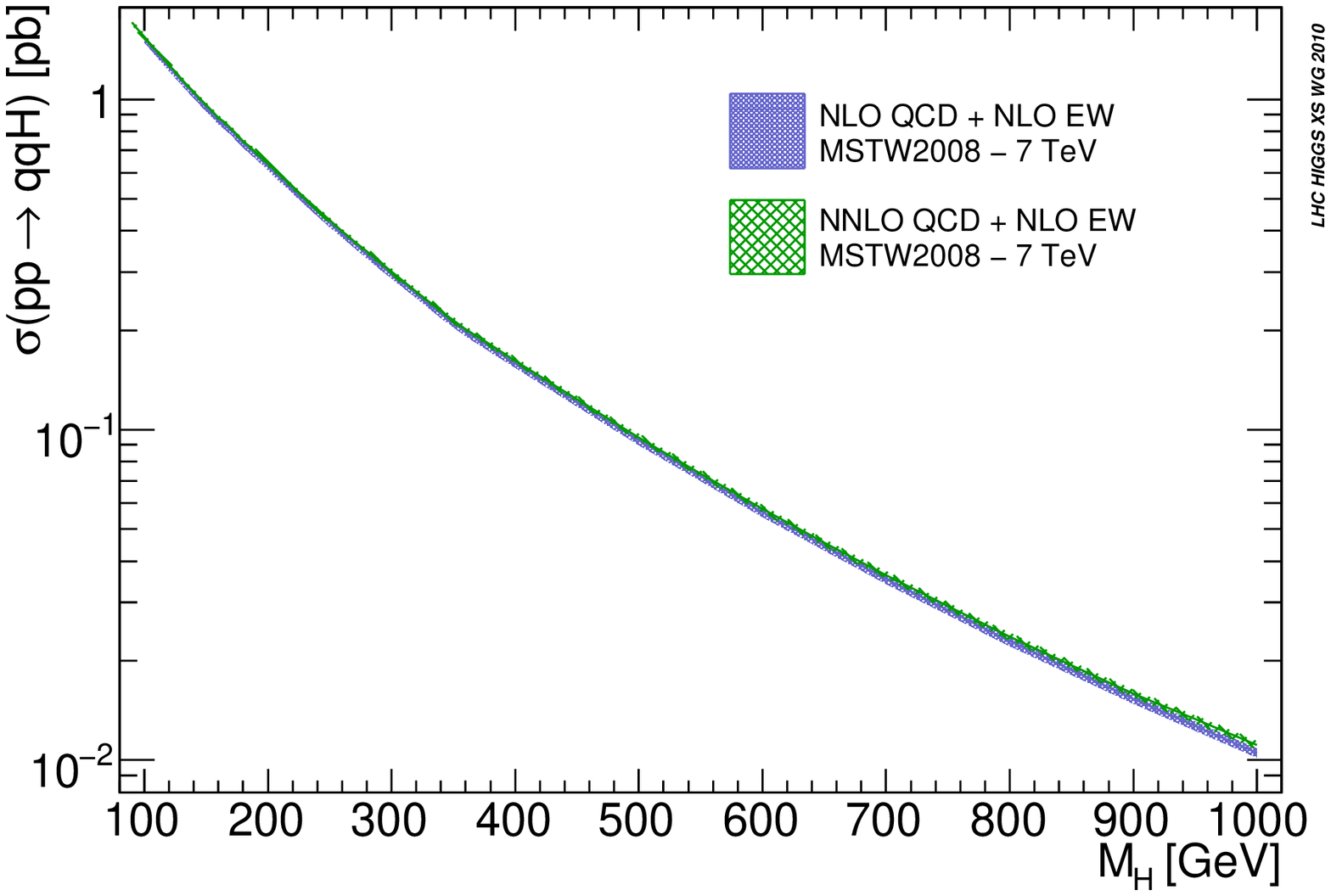}
  \includegraphics[width=0.5\textwidth]{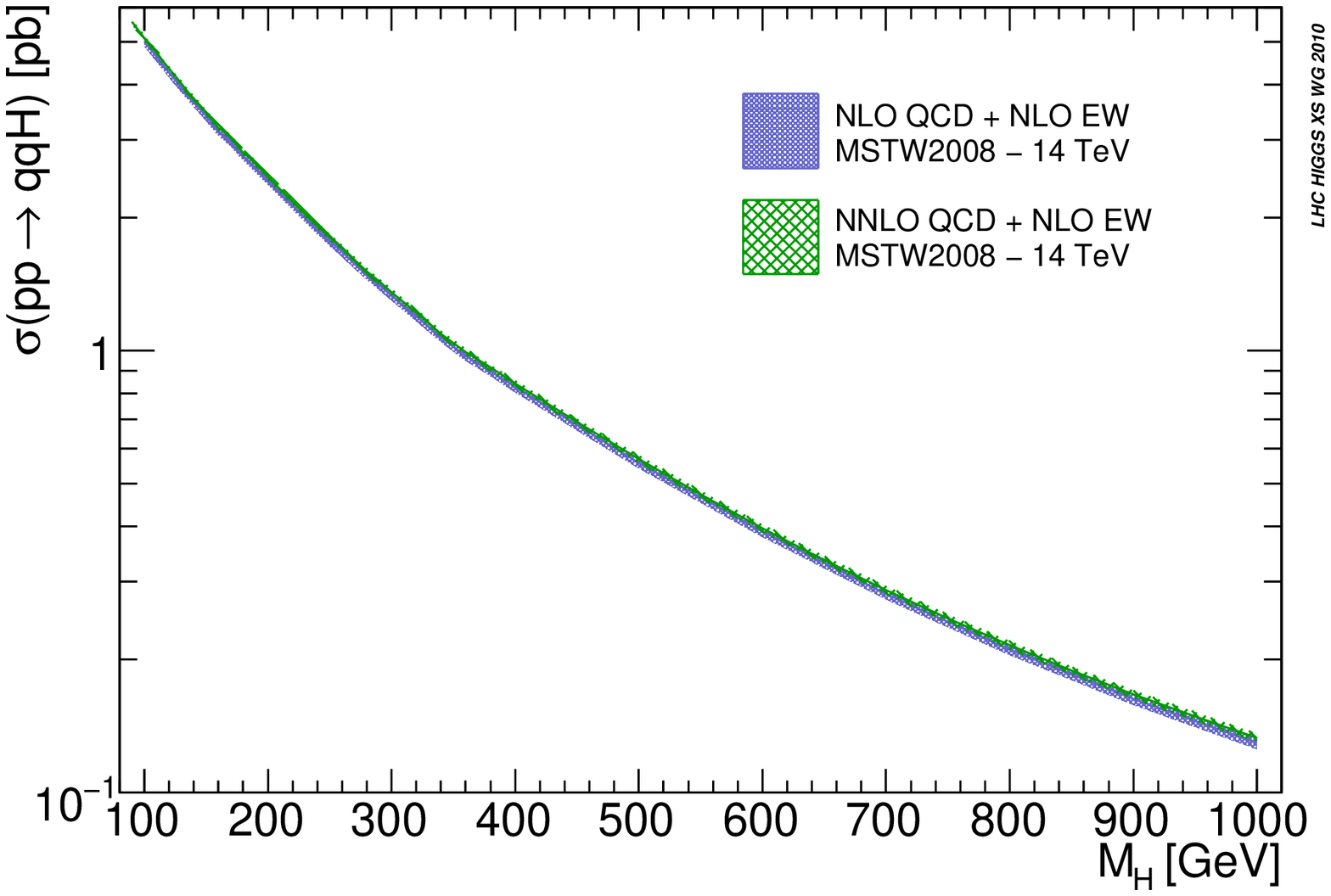}
  \caption{VBF cross sections at the LHC at $7\UTeV$~(left) and $14\UTeV$ (right)
    estimated with MSTW2008 PDF set.  NLO QCD results and NNLO QCD
results are shown both with the
    EW corrections. The bands represent the
    PDF + $\alphas$ 68\% CL uncertainty.}
  \label{fig:totXsecMSTW2008}
\end{figure}

\begin{figure}
  \includegraphics[width=0.5\textwidth]{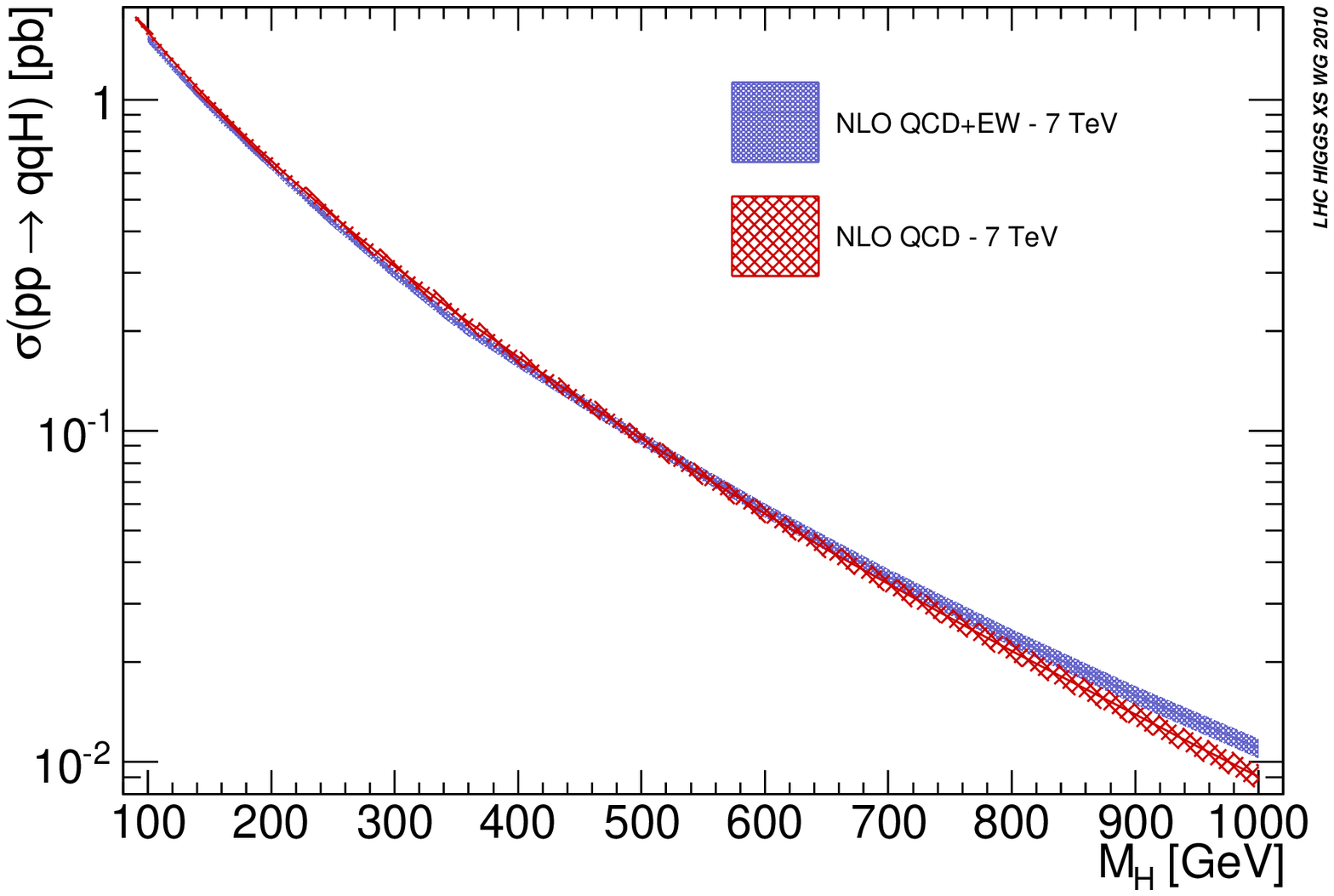}
  \includegraphics[width=0.5\textwidth]{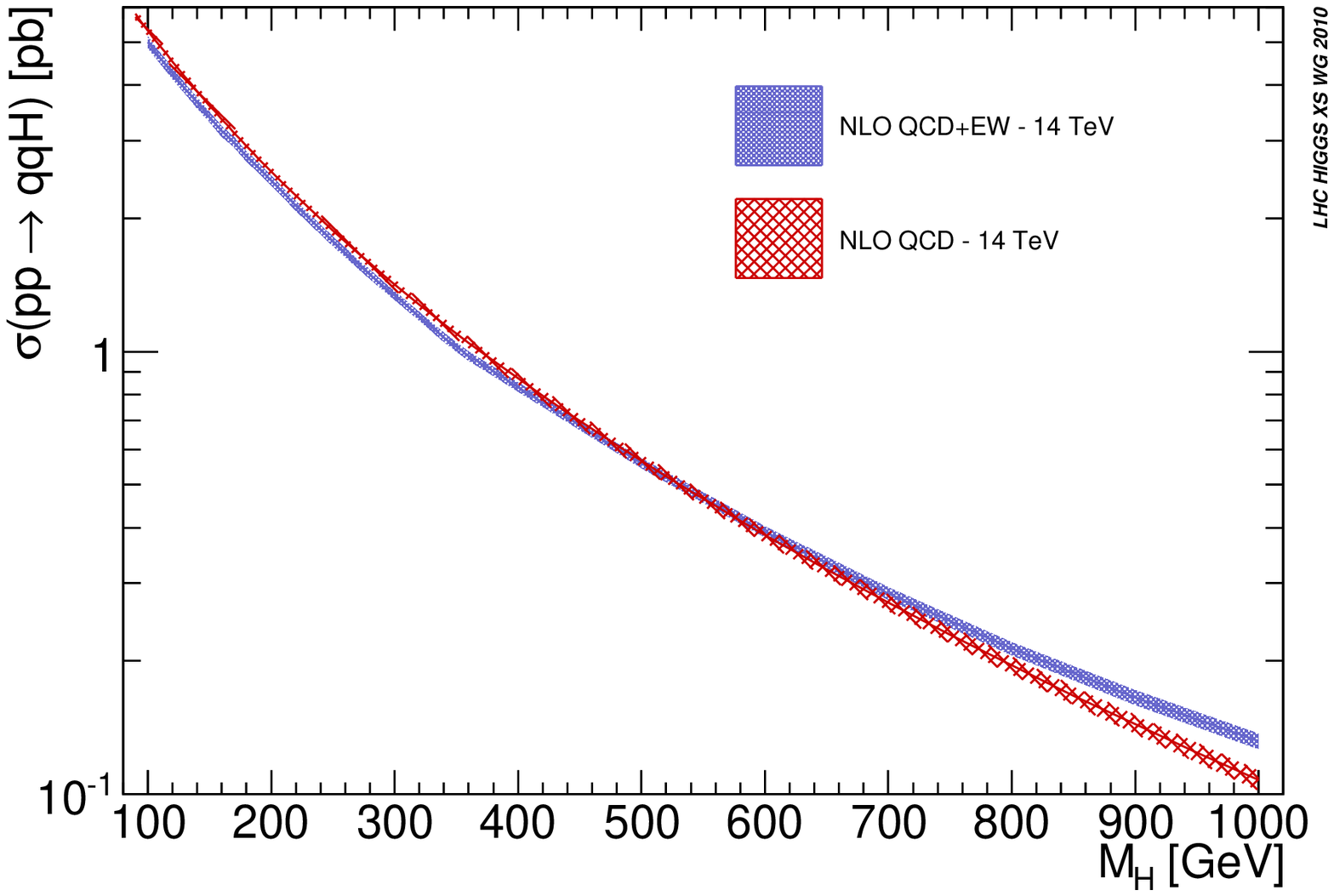}
  \caption{NLO VBF cross sections at the LHC at $7\UTeV$~(left) and
    $14\UTeV$ (right). Results with and without the EW corrections
    are plotted. The bands represent the PDF + $\alphas$ 68\% CL uncertainty coming from
	 the {\em envelope} of three PDF sets (see text for details).}
  \label{fig:totXsecEnvelope}
\end{figure}

Figures~\ref{fig:totXsecMSTW2008} and \ref{fig:totXsecEnvelope}
summarize the VBF results at the LHC at $7\UTeV$ and $14\UTeV$.  In
\Figure~\ref{fig:totXsecMSTW2008}, the cross section results at NLO
QCD and NNLO QCD both with EW corrections are shown
as a function of the Higgs-boson mass.
Calculations are performed with the
MSTW2008 68\% CL PDF set.  In \Figure~\ref{fig:totXsecEnvelope}, the NLO and
NNLO results, with and without the EW corrections, are shown as a function of
the Higgs-boson mass. For these calculations, the full estimation of central
values and $\alphas$ + PDF uncertainty over three PDF sets (namely MSTW2008,
CTEQ6.6, and NNPDF2.0, combined according to the PDF4LHC prescription) is
available and represented in the plots by the error bands.


In \Tables~\ref{tab:yesEWnoSch7TeV} and~\ref{tab:yesEWnoSch14TeV}, we collect
the NLO QCD + EW results, for the LHC at $7\UTeV$ and $14\UTeV$, respectively.
Numbers have been obtained with {\sc HAWK}. {\sc VBFNLO} results (obtained
with CTEQ6.6 PDF set) are listed in the rightmost column, for the sake of
comparison. For some of the mass points, a full PDF + $\alphas$ uncertainty
estimation has been performed according to the PDF4LHC prescription. In this
case, the uncertainty comes from the {\em envelope} among three PDF sets
(namely CTEQ6.6, MSTW2008NLO, and NNPDF2.0), and the central cross section
values are taken from the mid-point of the envelope width.  Integration
errors, affecting the last shown digit, are below $0.1\%$.  The integration
error for the {\sc VBFNLO} results is of order $0.3\%$.

In \Tables~\ref{tab:noEWnoSch7TeV} and~\ref{tab:noEWnoSch14TeV} we collect the
results on NLO QCD correction for the LHC at $7\UTeV$ and $14\UTeV$, respectively.
Numbers have been obtained with {\sc VBFNLO}. In
\Table~\ref{tab:noEWnoSch7TeV}, {\sc HAWK} results (obtained with MSTW2008 PDF
set) are listed in the rightmost column, for the sake of comparison.

\begin{table}[h!]
  \caption{NLO QCD + EW results on VBF cross sections at $\sqrt{s} = 7$\UTeV: central values  and relative uncertainties  from {\sc HAWK}. Integration errors,
  affecting the last shown digit, are below $0.1\%$. In the last column, {\sc
    VBFNLO} results obtained with CTEQ6.6, for the sake of comparison
  (integration errors at the $0.3\%$ level).}
  \centering
  \small
  \begin{tabular}{ccccc}\hline
$\MH [\UGeVZ]$ & $\sigma[\UfbZ]$ & Scale uncert. [\%] & PDF4LHC [\%] &  {\sc VBFNLO} $[\UfbZ]$ \\  
\hline
$90  $&$ 1682  $&$ +0.8 \; -\!0.2 $&$          $&$ 1706  $ \\
$95  $&$ 1598  $&$ +0.8 \; -\!0.3 $&$          $&$ 1613  $ \\
$100 $&$ 1530  $&$ +0.8 \; -\!0.1 $&$ \pm 2.2  $&$ 1531  $ \\
$105 $&$ 1445  $&$ +0.7 \; -\!0.2 $&$          $&$ 1450  $ \\
$110 $&$ 1385  $&$ +0.7 \; -\!0.1 $&$ \pm 2.2  $&$ 1385  $ \\
$115 $&$ 1312  $&$ +0.7 \; -\!0.1 $&$          $&$ 1314  $ \\
$120 $&$ 1257  $&$ +0.7 \; -\!0.0 $&$ \pm 2.1  $&$ 1253  $ \\
$125 $&$ 1193  $&$ +0.6 \; -\!0.0 $&$          $&$ 1193  $ \\
$130 $&$ 1144  $&$ +0.6 \; -\!0.0 $&$ \pm 2.1  $&$ 1138  $ \\
$135 $&$ 1087  $&$ +0.6 \; -\!0.1 $&$          $&$ 1085  $ \\
$140 $&$ 1042  $&$ +0.6 \; -\!0.0 $&$ \pm 2.1  $&$ 1037  $ \\
$145 $&$  992  $&$ +0.6 \; -\!0.1 $&$          $&$ 989 $ \\
$150 $&$  951  $&$ +0.6 \; -\!0.1 $&$ \pm 2.1  $&$ 946 $ \\
$155 $&$  907  $&$ +0.5 \; -\!0.1 $&$          $&$ 903 $ \\
$160 $&$  869  $&$ +0.5 \; -\!0.1 $&$ \pm 2.2  $&$ 864 $ \\
$165 $&$  842  $&$ +0.5 \; -\!0.1 $&$          $&$ 836 $ \\
$170 $&$  808  $&$ +0.4 \; -\!0.1 $&$ \pm 2.2  $&$ 802 $ \\
$175 $&$  772  $&$ +0.4 \; -\!0.1 $&$          $&$ 767 $ \\
$180 $&$  738  $&$ +0.4 \; -\!0.1 $&$ \pm 2.2  $&$ 735 $ \\
$185 $&$  713  $&$ +0.3 \; -\!0.1 $&$          $&$ 709 $ \\
$190 $&$  684  $&$ +0.3 \; -\!0.1 $&$ \pm 2.2  $&$ 680 $ \\
$195 $&$  658  $&$ +0.3 \; -\!0.1 $&$          $&$ 652 $ \\
$200 $&$  630  $&$ +0.3 \; -\!0.1 $&$ \pm 2.2  $&$ 625 $ \\
$210 $&$  580  $&$ +0.3 \; -\!0.0 $&$ \pm 2.2  $&$ 576 $ \\
$220 $&$  535  $&$ +0.4 \; -\!0.0 $&$ \pm 2.3  $&$ 531 $ \\
$230 $&$  495  $&$ +0.3 \; -\!0.0 $&$ \pm 2.3  $&$ 490 $ \\
$240 $&$  458  $&$ +0.3 \; -\!0.0 $&$ \pm 2.4  $&$ 453 $ \\
$250 $&$  425  $&$ +0.3 \; -\!0.0 $&$ \pm 2.4  $&$ 422 $ \\
$260 $&$  395  $&$ +0.3 \; -\!0.0 $&$ \pm 2.5  $&$ 392 $ \\
$270 $&$  368  $&$ +0.4 \; -\!0.0 $&$ \pm 2.6  $&$ 364 $ \\
$280 $&$  343  $&$ +0.4 \; -\!0.0 $&$ \pm 2.7  $&$ 340 $ \\
$290 $&$  320  $&$ +0.4 \; -\!0.0 $&$ \pm 2.7  $&$ 316 $ \\
$300 $&$  298  $&$ +0.5 \; -\!0.0 $&$ \pm 2.8  $&$ 296 $ \\
$320 $&$  260  $&$ +0.4 \; -\!0.1 $&$ \pm 2.9  $&$ 257 $ \\
$340 $&$  227  $&$ +0.4 \; -\!0.1 $&$ \pm 3.0  $&$ 225 $ \\
$360 $&$  200  $&$ +0.4 \; -\!0.0 $&$ \pm 3.1  $&$ 198 $ \\
$380 $&$  180  $&$ +0.6 \; -\!0.1 $&$ \pm 3.3  $&$ 178 $ \\
$400 $&$  161  $&$ +0.8 \; -\!0.1 $&$ \pm 3.4  $&$ 159 $ \\
$450 $&$  125  $&$ +1.1 \; -\!0.2 $&$          $&$ 122 $ \\
$500 $&$  94.6 $&$ +1.4 \; -\!0.2 $&$ \pm 4.0  $&$ 93.4 $ \\
$550 $&$  74.8 $&$ +1.7 \; -\!0.2 $&$          $&$ 72.8 $ \\
$600 $&$  57.6 $&$ +2.0 \; -\!0.3 $&$ \pm 4.5  $&$ 56.9 $ \\
$650 $&$  46.6 $&$ +2.3 \; -\!0.3 $&$          $&$ 44.7 $ \\
$700 $&$  36.4 $&$ +2.6 \; -\!0.3 $&$ \pm 5.1  $&$ 35.7 $ \\
$750 $&$  30.0 $&$ +2.9 \; -\!0.4 $&$          $&$ 28.6 $ \\
$800 $&$  23.7 $&$ +3.3 \; -\!0.4 $&$ \pm 5.6  $&$ 23.5 $ \\
$850 $&$  19.9 $&$ +3.9 \; -\!0.4 $&$          $&$ 18.9 $ \\
$900 $&$  15.9 $&$ +4.3 \; -\!0.4 $&$ \pm 6.1  $&$ 15.5 $ \\
$950 $&$  13.6 $&$ +4.9 \; -\!0.5 $&$          $&$ 13.0 $ \\
$1000$&$  11.0 $&$ +5.6 \; -\!0.5 $&$ \pm 6.6  $&$ 10.6 $ \\
\hline
\end{tabular}   
\label{tab:yesEWnoSch7TeV}
\end{table}

\begin{table}
  \caption{NLO QCD results on VBF cross sections (NLO EW corrections not included) at $\sqrt{s} =
  7$\UTeV: central values  and relative uncertainties  from
  {\sc VBFNLO}. Integration errors, affecting the last shown digit, are below
  $0.3\%$. In the last column, {\sc HAWK} results obtained with MSTW2008NLO,
  for the sake of comparison (integration errors at the $0.1\%$ level).}
  \centering
  \small
  \begin{tabular}{ccccc}\hline
$\MH [\UGeVZ]$ & $\sigma[\UfbZ]$ & Scale uncert. [\%] & PDF4LHC [\%] & {\sc HAWK}$[\UfbZ]$ \\  
\hline
$90  $&$ 1776 $&$ +0.0 \; -\!0.5 $&$ \pm 2.4 $&$ 1772  $ \\
$95  $&$ 1685 $&$ +0.1 \; -\!0.3 $&$ \pm 2.5 $&$ 1682  $ \\
$100 $&$ 1601 $&$ +0.1 \; -\!0.4 $&$ \pm 2.5 $&$ 1597  $ \\
$105 $&$ 1522 $&$ +0.1 \; -\!0.4 $&$ \pm 2.5 $&$ 1519  $ \\
$110 $&$ 1448 $&$ +0.2 \; -\!0.4 $&$ \pm 2.5 $&$ 1445  $ \\
$115 $&$ 1377 $&$ +0.1 \; -\!0.3 $&$ \pm 2.6 $&$ 1375  $ \\
$120 $&$ 1312 $&$ +0.2 \; -\!0.3 $&$ \pm 2.6 $&$ 1310  $ \\
$125 $&$ 1251 $&$ +0.2 \; -\!0.3 $&$ \pm 2.6 $&$ 1249  $ \\
$130 $&$ 1193 $&$ +0.3 \; -\!0.3 $&$ \pm 2.6 $&$ 1190  $ \\
$135 $&$ 1139 $&$ +0.3 \; -\!0.2 $&$ \pm 2.6 $&$ 1136  $ \\
$140 $&$ 1088 $&$ +0.4 \; -\!0.2 $&$ \pm 2.7 $&$ 1084  $ \\
$145 $&$ 1040 $&$ +0.4 \; -\!0.2 $&$ \pm 2.7 $&$ 1036  $ \\
$150 $&$  994 $&$ +0.4 \; -\!0.3 $&$ \pm 2.8 $&$ 990   $ \\
$155 $&$  951 $&$ +0.5 \; -\!0.2 $&$ \pm 2.8 $&$ 947   $ \\
$160 $&$  910 $&$ +0.5 \; -\!0.1 $&$ \pm 2.9 $&$ 906   $ \\
$165 $&$  872 $&$ +0.6 \; -\!0.1 $&$ \pm 3.0 $&$ 867   $ \\
$170 $&$  836 $&$ +0.6 \; -\!0.2 $&$ \pm 3.0 $&$ 831   $ \\
$175 $&$  801 $&$ +0.7 \; -\!0.1 $&$ \pm 3.0 $&$ 796   $ \\
$180 $&$  768 $&$ +0.6 \; -\!0.0 $&$ \pm 3.1 $&$ 763   $ \\
$185 $&$  737 $&$ +0.7 \; -\!0.1 $&$ \pm 3.1 $&$ 732   $ \\
$190 $&$  707 $&$ +0.7 \; -\!0.1 $&$ \pm 3.1 $&$ 702   $ \\
$195 $&$  679 $&$ +0.6 \; -\!0.0 $&$ \pm 3.2 $&$ 674   $ \\
$200 $&$  653 $&$ +0.7 \; -\!0.0 $&$ \pm 3.2 $&$ 648   $ \\
$210 $&$  603 $&$ +0.8 \; -\!0.1 $&$ \pm 3.3 $&$ 598   $ \\
$220 $&$  558 $&$ +0.9 \; -\!0.0 $&$ \pm 3.4 $&$ 553   $ \\
$230 $&$  517 $&$ +1.0 \; -\!0.0 $&$ \pm 3.5 $&$ 512   $ \\
$240 $&$  480 $&$ +1.0 \; -\!0.0 $&$ \pm 3.6 $&$ 475   $ \\
$250 $&$  446 $&$ +1.2 \; -\!0.0 $&$ \pm 3.6 $&$ 440   $ \\
$260 $&$  415 $&$ +1.1 \; -\!0.1 $&$ \pm 3.7 $&$ 410   $ \\
$270 $&$  386 $&$ +1.1 \; -\!0.1 $&$ \pm 3.8 $&$ 382   $ \\
$280 $&$  360 $&$ +1.2 \; -\!0.1 $&$ \pm 3.9 $&$ 356   $ \\
$290 $&$  336 $&$ +1.3 \; -\!0.1 $&$ \pm 3.9 $&$ 332   $ \\
$300 $&$  314 $&$ +1.4 \; -\!0.1 $&$ \pm 4.0 $&$ 310   $ \\
$320 $&$  275 $&$ +1.4 \; -\!0.1 $&$ \pm 4.2 $&$ 271   $ \\
$340 $&$  242 $&$ +1.5 \; -\!0.2 $&$ \pm 4.3 $&$ 238   $ \\
$360 $&$  213 $&$ +1.5 \; -\!0.2 $&$ \pm 4.4 $&$ 209   $ \\
$380 $&$  189 $&$ +1.7 \; -\!0.2 $&$ \pm 4.5 $&$ 185   $ \\
$400 $&$  167 $&$ +1.7 \; -\!0.3 $&$ \pm 4.7 $&$ 163   $ \\
$500 $&$ 94.9 $&$ +2.2 \; -\!0.4 $&$ \pm 5.3 $&$ 92.0  $ \\
$600 $&$ 56.3 $&$ +2.5 \; -\!0.6 $&$ \pm 5.9 $&$ 54.3  $ \\
$650 $&$ 43.9 $&$ +2.7 \; -\!0.7 $&$ \pm 6.2 $&$ 42.2  $ \\
$700 $&$ 34.5 $&$ +2.9 \; -\!0.7 $&$ \pm 6.5 $&$ 33.1  $ \\
$750 $&$ 27.3 $&$ +3.0 \; -\!0.8 $&$ \pm 6.8 $&$ 26.1  $ \\
$800 $&$ 21.7 $&$ +3.1 \; -\!1.0 $&$ \pm 7.1 $&$ 20.7  $ \\
$850 $&$ 17.3 $&$ +3.3 \; -\!1.1 $&$ \pm 7.4 $&$ 16.5  $ \\
$900 $&$ 13.9 $&$ +3.5 \; -\!1.2 $&$ \pm 7.7 $&$ 13.2  $ \\
$950 $&$ 11.2 $&$ +3.7 \; -\!1.2 $&$ \pm 8.0 $&$ 10.6  $ \\
$1000$&$ 9.03 $&$ +3.9 \; -\!1.2 $&$ \pm 8.3 $&$ 8.51  $ \\
\hline
\end{tabular}
\label{tab:noEWnoSch7TeV}
\end{table}

\begin{table}
  \caption{NLO QCD + EW results on VBF cross sections at $\sqrt{s} = 14$\UTeV: central values  and relative uncertainties  for {\sc HAWK}. Integration errors,
  affecting the last shown digit, are below $0.1\%$. In the last column, the
  {\sc VBFNLO} results obtained with CTEQ6.6, for the sake of comparison
  (integration errors at the $0.3\%$ level).}
  \label{tab:siEWnoSch14TeV}
  \centering
  \small
  \begin{tabular}{ccccc}\hline
$\MH [\UGeVZ]$ & $\sigma[\UfbZ]$ & Scale uncert. [\%] & PDF4LHC [\%] & {\sc VBFNLO} $[\UfbZ]$\\
  \hline
$90  $&$ 5375  $&$ +1.0 \; -\!0.5 $&$          $&$ 5517  $ \\
$95  $&$ 5156  $&$ +0.9 \; -\!0.5 $&$          $&$ 5272  $ \\
$100 $&$ 5004  $&$ +1.0 \; -\!0.4 $&$ \pm 2.6  $&$ 5057  $ \\
$105 $&$ 4746  $&$ +1.0 \; -\!0.4 $&$          $&$ 4839  $ \\
$110 $&$ 4607  $&$ +1.0 \; -\!0.5 $&$ \pm 2.6  $&$ 4642  $ \\
$115 $&$ 4373  $&$ +0.9 \; -\!0.5 $&$          $&$ 4455  $ \\
$120 $&$ 4254  $&$ +0.9 \; -\!0.4 $&$ \pm 2.6  $&$ 4272  $ \\
$125 $&$ 4048  $&$ +0.8 \; -\!0.4 $&$          $&$ 4109  $ \\
$130 $&$ 3938  $&$ +1.0 \; -\!0.3 $&$ \pm 2.5  $&$ 3952  $ \\
$135 $&$ 3754  $&$ +0.9 \; -\!0.4 $&$          $&$ 3807  $ \\
$140 $&$ 3651  $&$ +0.8 \; -\!0.3 $&$ \pm 2.5  $&$ 3666  $ \\
$145 $&$ 3485  $&$ +0.8 \; -\!0.3 $&$          $&$ 3431  $ \\
$150 $&$ 3394  $&$ +0.7 \; -\!0.3 $&$ \pm 2.5  $&$ 3403  $ \\
$155 $&$ 3237  $&$ +0.8 \; -\!0.3 $&$          $&$ 3277  $ \\
$160 $&$ 3147  $&$ +1.0 \; -\!0.2 $&$ \pm 2.4  $&$ 3156  $ \\
$165 $&$ 3047  $&$ +0.8 \; -\!0.3 $&$          $&$ 3083  $ \\
$170 $&$ 2975  $&$ +0.8 \; -\!0.3 $&$ \pm 2.4  $&$ 2978  $ \\
$175 $&$ 2842  $&$ +0.8 \; -\!0.3 $&$          $&$ 2866  $ \\
$180 $&$ 2765  $&$ +0.9 \; -\!0.3 $&$ \pm 2.3  $&$ 2764  $ \\
$185 $&$ 2667  $&$ +0.9 \; -\!0.3 $&$          $&$ 2679  $ \\
$190 $&$ 2601  $&$ +1.0 \; -\!0.0 $&$ \pm 2.3  $&$ 2595  $ \\
$195 $&$ 2494  $&$ +0.8 \; -\!0.2 $&$          $&$ 2512  $ \\
$200 $&$ 2432  $&$ +0.8 \; -\!0.0 $&$ \pm 2.3  $&$ 2437  $ \\
$210 $&$ 2279  $&$ +0.8 \; -\!0.0 $&$ \pm 2.2  $&$ 2274  $ \\
$220 $&$ 2135  $&$ +0.6 \; -\!0.2 $&$ \pm 2.3  $&$ 2135  $ \\
$230 $&$ 2006  $&$ +0.7 \; -\!0.3 $&$ \pm 2.2  $&$ 1999  $ \\
$240 $&$ 1885  $&$ +0.7 \; -\!0.2 $&$ \pm 2.3  $&$ 1883  $ \\
$250 $&$ 1777  $&$ +0.6 \; -\!0.1 $&$ \pm 2.2  $&$ 1770  $ \\
$260 $&$ 1675  $&$ +0.7 \; -\!0.1 $&$ \pm 2.1  $&$ 1668  $ \\
$270 $&$ 1581  $&$ +0.7 \; -\!0.1 $&$ \pm 2.1  $&$ 1575  $ \\
$280 $&$ 1494  $&$ +0.7 \; -\!0.0 $&$ \pm 2.1  $&$ 1488  $ \\
$290 $&$ 1413  $&$ +0.8 \; -\!0.0 $&$ \pm 2.1  $&$ 1407  $ \\
$300 $&$ 1338  $&$ +0.7 \; -\!0.0 $&$ \pm 2.1  $&$ 1329  $ \\
$320 $&$ 1202  $&$ +0.6 \; -\!0.1 $&$ \pm 2.1  $&$ 1195  $ \\
$340 $&$ 1077  $&$ +0.6 \; -\!0.1 $&$ \pm 2.1  $&$ 1069  $ \\
$360 $&$  977  $&$ +0.6 \; -\!0.2 $&$ \pm 2.1  $&$  973  $ \\
$380 $&$  901  $&$ +0.5 \; -\!0.0 $&$ \pm 2.1  $&$  893  $ \\
$400 $&$  830  $&$ +0.4 \; -\!0.2 $&$ \pm 2.2  $&$  826  $ \\
$450 $&$  681  $&$ +0.5 \; -\!0.2 $&$          $&$  673  $ \\
$500 $&$  560  $&$ +0.6 \; -\!0.0 $&$ \pm 2.3  $&$  561  $ \\
$550 $&$  469  $&$ +0.6 \; -\!0.1 $&$          $&$  463  $ \\
$600 $&$  391  $&$ +0.8 \; -\!0.1 $&$ \pm 2.6  $&$  388  $ \\
$650 $&$  335  $&$ +1.2 \; -\!0.0 $&$          $&$  330  $ \\
$700 $&$  284  $&$ +1.4 \; -\!0.0 $&$ \pm 3.0  $&$  282  $ \\
$750 $&$  248  $&$ +1.8 \; -\!0.0 $&$          $&$  242  $ \\
$800 $&$  213  $&$ +1.9 \; -\!0.0 $&$ \pm 3.2  $&$  212  $ \\
$850 $&$  189  $&$ +2.4 \; -\!0.0 $&$          $&$  185  $ \\
$900 $&$  165  $&$ +2.6 \; -\!0.1 $&$ \pm 3.6  $&$  164  $ \\
$950 $&$  149  $&$ +3.0 \; -\!0.0 $&$          $&$  146  $ \\
$1000$&$  132  $&$ +3.6 \; -\!0.1 $&$ \pm 3.9  $&$  130  $ \\
\hline
\end{tabular}
\label{tab:yesEWnoSch14TeV}
\end{table}
                  
\begin{table}
  \caption{NLO QCD results on VBF cross sections (NLO EW corrections not included) at $\sqrt{s} = 14$\UTeV: central values 
  and relative uncertainties  from {\sc VBFNLO}. Integration errors, affecting 
  the last shown digit, are below $0.3\%$.
  }
  \centering
  \small
  \begin{tabular}{cccc}\hline
$\MH [\UGeVZ]$ & $\sigma[\UfbZ]$ & Scale uncert. [\%] & PDF4LHC [\%] \\
  \hline
$90  $&$ 5792   $&$ +1.0  \; -\!0.9  $&$ \pm 3.0 $ \\
$95  $&$ 5550   $&$ +0.8  \; -\!0.9  $&$ \pm 3.0 $ \\
$100 $&$ 5320   $&$ +0.8  \; -\!0.7  $&$ \pm 2.9 $ \\
$105 $&$ 5104   $&$ +0.7  \; -\!0.9  $&$ \pm 2.9 $ \\
$110 $&$ 4898   $&$ +0.7  \; -\!0.7  $&$ \pm 2.8 $ \\
$115 $&$ 4702   $&$ +0.8  \; -\!0.6  $&$ \pm 2.8 $ \\
$120 $&$ 4521   $&$ +0.7  \; -\!0.8  $&$ \pm 2.8 $ \\
$125 $&$ 4344   $&$ +0.7  \; -\!0.6  $&$ \pm 2.7 $ \\
$130 $&$ 4182   $&$ +0.5  \; -\!0.8  $&$ \pm 2.7 $ \\
$135 $&$ 4025   $&$ +0.5  \; -\!0.8  $&$ \pm 2.7 $ \\
$140 $&$ 3874   $&$ +0.5  \; -\!0.7  $&$ \pm 2.6 $ \\
$145 $&$ 3734   $&$ +0.4  \; -\!0.8  $&$ \pm 2.6 $ \\
$150 $&$ 3599   $&$ +0.5  \; -\!0.6  $&$ \pm 2.6 $ \\
$155 $&$ 3472   $&$ +0.4  \; -\!0.7  $&$ \pm 2.6 $ \\
$160 $&$ 3349   $&$ +0.4  \; -\!0.7  $&$ \pm 2.5 $ \\
$165 $&$ 3234   $&$ +0.3  \; -\!0.6  $&$ \pm 2.5 $ \\
$170 $&$ 3124   $&$ +0.3  \; -\!0.6  $&$ \pm 2.5 $ \\
$175 $&$ 3017   $&$ +0.3  \; -\!0.6  $&$ \pm 2.4 $ \\
$180 $&$ 2917   $&$ +0.4  \; -\!0.6  $&$ \pm 2.4 $ \\
$185 $&$ 2819   $&$ +0.3  \; -\!0.5  $&$ \pm 2.4 $ \\
$190 $&$ 2726   $&$ +0.3  \; -\!0.5  $&$ \pm 2.4 $ \\
$195 $&$ 2639   $&$ +0.2  \; -\!0.5  $&$ \pm 2.4 $ \\
$200 $&$ 2553   $&$ +0.2  \; -\!0.5  $&$ \pm 2.4 $ \\
$210 $&$ 2395   $&$ +0.1  \; -\!0.5  $&$ \pm 2.4 $ \\
$220 $&$ 2248   $&$ +0.1  \; -\!0.4  $&$ \pm 2.5 $ \\
$230 $&$ 2115   $&$ +0.1  \; -\!0.4  $&$ \pm 2.5 $ \\
$240 $&$ 1991   $&$ +0.0  \; -\!0.4  $&$ \pm 2.5 $ \\
$250 $&$ 1877   $&$ +0.1  \; -\!0.5  $&$ \pm 2.5 $ \\
$260 $&$ 1771   $&$ +0.1  \; -\!0.4  $&$ \pm 2.5 $ \\
$270 $&$ 1673   $&$ +0.2  \; -\!0.4  $&$ \pm 2.5 $ \\
$280 $&$ 1583   $&$ +0.2  \; -\!0.3  $&$ \pm 2.5 $ \\
$290 $&$ 1498   $&$ +0.1  \; -\!0.3  $&$ \pm 2.6 $ \\
$300 $&$ 1419   $&$ +0.2  \; -\!0.2  $&$ \pm 2.5 $ \\
$320 $&$ 1279   $&$ +0.3  \; -\!0.3  $&$ \pm 2.7 $ \\
$340 $&$ 1156   $&$ +0.4  \; -\!0.4  $&$ \pm 2.7 $ \\
$360 $&$ 1048   $&$ +0.5  \; -\!0.3  $&$ \pm 2.8 $ \\
$380 $&$  953   $&$ +0.5  \; -\!0.1  $&$ \pm 3.0 $ \\
$400 $&$  869   $&$ +0.6  \; -\!0.2  $&$ \pm 3.0 $ \\
$500 $&$  566   $&$ +0.9  \; -\!0.2  $&$ \pm 3.4 $ \\
$600 $&$  385   $&$ +1.2  \; -\!0.1  $&$ \pm 3.8 $ \\
$650 $&$  322   $&$ +1.4  \; -\!0.0  $&$ \pm 4.0 $ \\
$700 $&$  271   $&$ +1.4  \; -\!0.1  $&$ \pm 4.2 $ \\
$750 $&$  229   $&$ +1.5  \; -\!0.1  $&$ \pm 4.4 $ \\
$800 $&$  195   $&$ +1.6  \; -\!0.1  $&$ \pm 4.5 $ \\
$850 $&$  167   $&$ +1.7  \; -\!0.2  $&$ \pm 4.7 $ \\
$900 $&$  144   $&$ +1.8  \; -\!0.1  $&$ \pm 4.9 $ \\
$950 $&$  124   $&$ +1.9  \; -\!0.2  $&$ \pm 5.0 $ \\
$1000$&$  108   $&$ +2.0  \; -\!0.2  $&$ \pm 5.1 $ \\
\hline
\end{tabular} 
\label{tab:noEWnoSch14TeV}
\end{table}

\begin{table}
  \caption{NNLO QCD results on VBF cross sections at $\sqrt{s} = 7$\UTeV: central values 
  and relative uncertainties. PDF uncertainties are
evaluated with MSTW2008NNLO PDF set.  Integration errors
  are below the $0.1\%$ level.}  
  \centering
  \small
  \begin{tabular}{ccccccc}\hline
$\MH [\UGeV]$ & $\sigma[\UfbZ]$ &  $\left(1+\delta_{\rm
EW}\right)\sigma[\UfbZ]$&  Scale uncert. [\%] & PDF + \alphas [\%] & PDF4LHC [\%]\\
\hline
 $  90$ & $  1788 $ & $  1710 $ & $ +0.6 \; -\!0.2$ & $ +1.8 \; -\!1.8$ & $ +2.1 \; -\!2.1$ \\
 $  95$ & $  1703 $ & $  1628 $ & $ +0.4 \; -\!0.4$ & $ +1.8 \; -\!1.8$ & $ +2.1 \; -\!2.1$ \\
 $ 100$ & $  1616 $ & $  1546 $ & $ +0.4 \; -\!0.3$ & $ +1.8 \; -\!1.8$ & $ +2.2 \; -\!2.1$ \\
 $ 105$ & $  1539 $ & $  1472 $ & $ +0.3 \; -\!0.3$ & $ +1.8 \; -\!1.8$ & $ +2.2 \; -\!2.1$ \\
 $ 110$ & $  1461 $ & $  1398 $ & $ +0.5 \; -\!0.2$ & $ +1.8 \; -\!1.8$ & $ +2.3 \; -\!2.1$ \\
 $ 115$ & $  1393 $ & $  1332 $ & $ +0.2 \; -\!0.2$ & $ +1.8 \; -\!1.8$ & $ +2.3 \; -\!2.1$ \\
 $ 120$ & $  1326 $ & $  1269 $ & $ +0.3 \; -\!0.4$ & $ +1.8 \; -\!1.8$ & $ +2.4 \; -\!2.1$ \\
 $ 125$ & $  1265 $ & $  1211 $ & $ +0.3 \; -\!0.3$ & $ +1.8 \; -\!1.8$ & $ +2.5 \; -\!2.1$ \\
 $ 130$ & $  1205 $ & $  1154 $ & $ +0.3 \; -\!0.2$ & $ +1.8 \; -\!1.8$ & $ +2.5 \; -\!2.1$ \\
 $ 135$ & $  1148 $ & $  1100 $ & $ +0.5 \; -\!0.1$ & $ +1.8 \; -\!1.8$ & $ +2.6 \; -\!2.1$ \\
 $ 140$ & $  1099 $ & $  1052 $ & $ +0.2 \; -\!0.2$ & $ +1.8 \; -\!1.8$ & $ +2.6 \; -\!2.1$ \\
 $ 145$ & $  1048 $ & $  1004 $ & $ +0.4 \; -\!0.0$ & $ +1.9 \; -\!1.9$ & $ +2.7 \; -\!2.1$ \\
 $ 150$ & $  1004 $ & $  961.7$ & $ +0.2 \; -\!0.1$ & $ +1.9 \; -\!1.9$ & $ +2.7 \; -\!2.1$ \\
 $ 155$ & $  959.6$ & $  918.0$ & $ +0.3 \; -\!0.0$ & $ +1.9 \; -\!1.9$ & $ +2.8 \; -\!2.1$ \\
 $ 160$ & $  920.0$ & $  878.7$ & $ +0.1 \; -\!0.2$ & $ +1.9 \; -\!1.9$ & $ +2.8 \; -\!2.1$ \\
 $ 165$ & $  880.0$ & $  851.7$ & $ +0.2 \; -\!0.1$ & $ +1.9 \; -\!1.9$ & $ +2.9 \; -\!2.1$ \\
 $ 170$ & $  843.9$ & $  817.3$ & $ +0.2 \; -\!0.2$ & $ +1.9 \; -\!1.9$ & $ +3.0 \; -\!2.1$ \\
 $ 175$ & $  808.2$ & $  781.4$ & $ +0.2 \; -\!0.1$ & $ +1.9 \; -\!1.9$ & $ +3.0 \; -\!2.1$ \\
 $ 180$ & $  776.0$ & $  748.0$ & $ +0.0 \; -\!0.3$ & $ +1.9 \; -\!1.9$ & $ +3.1 \; -\!2.1$ \\
 $ 185$ & $  742.1$ & $  719.3$ & $ +0.3 \; -\!0.1$ & $ +1.9 \; -\!1.9$ & $ +3.1 \; -\!2.0$ \\
 $ 190$ & $  713.5$ & $  692.5$ & $ +0.1 \; -\!0.2$ & $ +1.9 \; -\!1.9$ & $ +3.2 \; -\!2.0$ \\
 $ 195$ & $  685.0$ & $  664.3$ & $ +0.2 \; -\!0.4$ & $ +1.9 \; -\!1.9$ & $ +3.2 \; -\!2.0$ \\
 $ 200$ & $  657.9$ & $  637.1$ & $ +0.1 \; -\!0.2$ & $ +1.9 \; -\!1.9$ & $ +3.3 \; -\!2.0$ \\
 $ 210$ & $  607.6$ & $  586.9$ & $ +0.1 \; -\!0.3$ & $ +2.0 \; -\!2.0$ & $ +3.4 \; -\!2.0$ \\
 $ 220$ & $  562.3$ & $  542.0$ & $ +0.0 \; -\!0.4$ & $ +2.0 \; -\!2.0$ & $ +3.5 \; -\!2.0$ \\
 $ 230$ & $  520.8$ & $  501.1$ & $ +0.1 \; -\!0.4$ & $ +2.0 \; -\!2.0$ & $ +3.6 \; -\!2.0$ \\
 $ 240$ & $  483.2$ & $  464.1$ & $ +0.1 \; -\!0.5$ & $ +2.0 \; -\!2.0$ & $ +3.7 \; -\!2.0$ \\
 $ 250$ & $  448.7$ & $  430.4$ & $ +0.1 \; -\!0.6$ & $ +2.0 \; -\!2.0$ & $ +3.8 \; -\!2.0$ \\
 $ 260$ & $  416.2$ & $  398.8$ & $ +0.3 \; -\!0.4$ & $ +2.0 \; -\!2.0$ & $ +3.9 \; -\!2.0$ \\
 $ 270$ & $  388.1$ & $  371.5$ & $ +0.1 \; -\!0.6$ & $ +2.0 \; -\!2.0$ & $ +4.0 \; -\!2.0$ \\
 $ 280$ & $  361.9$ & $  346.1$ & $ +0.2 \; -\!0.7$ & $ +2.0 \; -\!2.0$ & $ +4.2 \; -\!2.0$ \\
 $ 290$ & $  337.7$ & $  322.6$ & $ +0.2 \; -\!0.7$ & $ +2.1 \; -\!2.1$ & $ +4.3 \; -\!2.0$ \\
 $ 300$ & $  315.4$ & $  301.0$ & $ +0.2 \; -\!0.8$ & $ +2.1 \; -\!2.1$ & $ +4.4 \; -\!2.0$ \\
 $ 320$ & $  275.4$ & $  262.2$ & $ +0.3 \; -\!0.7$ & $ +2.1 \; -\!2.1$ & $ +4.6 \; -\!1.9$ \\
 $ 340$ & $  241.9$ & $  228.6$ & $ +0.3 \; -\!0.9$ & $ +2.1 \; -\!2.1$ & $ +4.8 \; -\!1.9$ \\
 $ 360$ & $  213.2$ & $  201.8$ & $ +0.3 \; -\!1.1$ & $ +2.2 \; -\!2.2$ & $ +5.0 \; -\!1.9$ \\
 $ 380$ & $  188.2$ & $  180.7$ & $ +0.4 \; -\!1.1$ & $ +2.2 \; -\!2.2$ & $ +5.2 \; -\!1.9$ \\
 $ 400$ & $  166.6$ & $  161.9$ & $ +0.4 \; -\!1.2$ & $ +2.2 \; -\!2.2$ & $ +5.5 \; -\!1.9$ \\
 $ 450$ & $  124.4$ & $  123.5$ & $ +0.6 \; -\!1.3$ & $ +2.2 \; -\!2.2$ & $ +6.0 \; -\!1.8$ \\
 $ 500$ & $  94.07$ & $  94.91$ & $ +0.7 \; -\!1.6$ & $ +2.3 \; -\!2.3$ & $ +6.6 \; -\!1.8$ \\
 $ 550$ & $  71.90$ & $  73.56$ & $ +0.8 \; -\!1.7$ & $ +2.3 \; -\!2.3$ & $ +7.1 \; -\!1.8$ \\
 $ 600$ & $  55.52$ & $  57.63$ & $ +1.0 \; -\!2.0$ & $ +2.4 \; -\!2.4$ & $ +7.6 \; -\!1.7$ \\
 $ 650$ & $  43.22$ & $  45.56$ & $ +1.1 \; -\!2.2$ & $ +2.4 \; -\!2.4$ & $ +8.2 \; -\!1.7$ \\
 $ 700$ & $  33.89$ & $  36.35$ & $ +1.2 \; -\!2.4$ & $ +2.5 \; -\!2.5$ & $ +8.7 \; -\!1.6$ \\
 $ 750$ & $  26.74$ & $  29.24$ & $ +1.4 \; -\!2.6$ & $ +2.5 \; -\!2.5$ & $ +9.3 \; -\!1.6$ \\
 $ 800$ & $  21.21$ & $  23.71$ & $ +1.5 \; -\!2.8$ & $ +2.6 \; -\!2.6$ & $ +9.8 \; -\!1.6$ \\
 $ 850$ & $  16.90$ & $  19.37$ & $ +1.6 \; -\!3.0$ & $ +2.6 \; -\!2.6$ & $ +10.4 \; -\!1.5$ \\
 $ 900$ & $  13.52$ & $  15.95$ & $ +1.7 \; -\!3.2$ & $ +2.7 \; -\!2.7$ & $ +10.9 \; -\!1.5$ \\
 $ 950$ & $  10.86$ & $  13.21$ & $ +2.0 \; -\!3.3$ & $ +2.7 \; -\!2.7$ & $ +11.5 \; -\!1.4$ \\
 $1000$ & $  8.752$ & $  11.03$ & $ +2.2 \; -\!3.5$ & $ +2.8 \; -\!2.8$ & $ +12.0 \; -\!1.4$ \\
\hline
\end{tabular}
\label{tab:NNLO7TeV}
\end{table}

\begin{table}
  \caption{NNLO QCD results on VBF cross sections at $\sqrt{s} = 14$\UTeV: central values  and relative uncertainties. PDF uncertainties are evaluated
  with MSTW2008NNLO PDF set.  Integration errors are below the $0.1\%$ level.}
  \centering
  \small
  \begin{tabular}{ccccccc}\hline
$\MH [\UGeVZ]$ & $\sigma[\UfbZ]$ &  $\left(1+\delta_{\rm
EW}\right)\sigma[\UfbZ]$&  Scale uncert. [\%] & PDF + \alphas [\%] & PDF4LHC [\%]\\
\hline
 $  90$ & $  5879 $ & $  5569 $ & $ +1.0 \; -\!0.4$ & $ +1.6 \; -\!1.6$ & $ +1.9 \; -\!2.6$ \\
 $  95$ & $  5637 $ & $  5338 $ & $ +1.0 \; -\!0.5$ & $ +1.6 \; -\!1.6$ & $ +2.0 \; -\!2.6$ \\
 $ 100$ & $  5401 $ & $  5114 $ & $ +0.8 \; -\!0.5$ & $ +1.6 \; -\!1.6$ & $ +2.0 \; -\!2.6$ \\
 $ 105$ & $  5175 $ & $  4900 $ & $ +1.2 \; -\!0.3$ & $ +1.6 \; -\!1.6$ & $ +2.0 \; -\!2.6$ \\
 $ 110$ & $  5015 $ & $  4750 $ & $ +0.2 \; -\!1.3$ & $ +1.6 \; -\!1.6$ & $ +2.0 \; -\!2.6$ \\
 $ 115$ & $  4771 $ & $  4520 $ & $ +0.9 \; -\!0.4$ & $ +1.6 \; -\!1.6$ & $ +2.0 \; -\!2.6$ \\
 $ 120$ & $  4603 $ & $  4361 $ & $ +0.4 \; -\!0.9$ & $ +1.6 \; -\!1.6$ & $ +2.1 \; -\!2.6$ \\
 $ 125$ & $  4412 $ & $  4180 $ & $ +0.7 \; -\!0.4$ & $ +1.6 \; -\!1.6$ & $ +2.1 \; -\!2.6$ \\
 $ 130$ & $  4252 $ & $  4029 $ & $ +0.4 \; -\!0.5$ & $ +1.6 \; -\!1.6$ & $ +2.1 \; -\!2.6$ \\
 $ 135$ & $  4076 $ & $  3862 $ & $ +0.9 \; -\!0.2$ & $ +1.6 \; -\!1.6$ & $ +2.2 \; -\!2.6$ \\
 $ 140$ & $  3938 $ & $  3732 $ & $ +0.5 \; -\!0.8$ & $ +1.6 \; -\!1.6$ & $ +2.2 \; -\!2.6$ \\
 $ 145$ & $  3789 $ & $  3590 $ & $ +0.8 \; -\!0.4$ & $ +1.6 \; -\!1.6$ & $ +2.2 \; -\!2.6$ \\
 $ 150$ & $  3653 $ & $  3460 $ & $ +0.6 \; -\!0.4$ & $ +1.6 \; -\!1.6$ & $ +2.2 \; -\!2.6$ \\
 $ 155$ & $  3522 $ & $  3332 $ & $ +0.7 \; -\!0.4$ & $ +1.6 \; -\!1.6$ & $ +2.2 \; -\!2.6$ \\
 $ 160$ & $  3386 $ & $  3198 $ & $ +0.9 \; -\!0.2$ & $ +1.6 \; -\!1.6$ & $ +2.3 \; -\!2.6$ \\
 $ 165$ & $  3278 $ & $  3137 $ & $ +0.7 \; -\!0.3$ & $ +1.7 \; -\!1.7$ & $ +2.3 \; -\!2.6$ \\
 $ 170$ & $  3168 $ & $  3033 $ & $ +0.5 \; -\!0.4$ & $ +1.7 \; -\!1.7$ & $ +2.3 \; -\!2.6$ \\
 $ 175$ & $  3058 $ & $  2922 $ & $ +1.1 \; -\!0.2$ & $ +1.7 \; -\!1.7$ & $ +2.3 \; -\!2.6$ \\
 $ 180$ & $  2945 $ & $  2805 $ & $ +0.9 \; -\!0.2$ & $ +1.7 \; -\!1.7$ & $ +2.4 \; -\!2.6$ \\
 $ 185$ & $  2860 $ & $  2740 $ & $ +0.4 \; -\!0.3$ & $ +1.7 \; -\!1.7$ & $ +2.4 \; -\!2.6$ \\
 $ 190$ & $  2766 $ & $  2652 $ & $ +0.3 \; -\!0.3$ & $ +1.7 \; -\!1.7$ & $ +2.4 \; -\!2.6$ \\
 $ 195$ & $  2678 $ & $  2566 $ & $ +0.4 \; -\!0.3$ & $ +1.7 \; -\!1.7$ & $ +2.4 \; -\!2.6$ \\
 $ 200$ & $  2583 $ & $  2472 $ & $ +0.7 \; -\!0.1$ & $ +1.7 \; -\!1.7$ & $ +2.5 \; -\!2.6$ \\
 $ 210$ & $  2425 $ & $  2315 $ & $ +0.7 \; -\!0.1$ & $ +1.7 \; -\!1.7$ & $ +2.5 \; -\!2.6$ \\
 $ 220$ & $  2280 $ & $  2171 $ & $ +0.4 \; -\!0.5$ & $ +1.7 \; -\!1.7$ & $ +2.6 \; -\!2.6$ \\
 $ 230$ & $  2142 $ & $  2036 $ & $ +0.6 \; -\!0.2$ & $ +1.7 \; -\!1.7$ & $ +2.6 \; -\!2.6$ \\
 $ 240$ & $  2021 $ & $  1918 $ & $ +0.4 \; -\!0.1$ & $ +1.7 \; -\!1.7$ & $ +2.7 \; -\!2.6$ \\
 $ 250$ & $  1908 $ & $  1807 $ & $ +0.2 \; -\!0.4$ & $ +1.7 \; -\!1.7$ & $ +2.7 \; -\!2.6$ \\
 $ 260$ & $  1809 $ & $  1711 $ & $ +0.2 \; -\!1.1$ & $ +1.8 \; -\!1.8$ & $ +2.8 \; -\!2.6$ \\
 $ 270$ & $  1699 $ & $  1606 $ & $ +0.2 \; -\!0.3$ & $ +1.8 \; -\!1.8$ & $ +2.8 \; -\!2.6$ \\
 $ 280$ & $  1603 $ & $  1514 $ & $ +0.4 \; -\!0.1$ & $ +1.8 \; -\!1.8$ & $ +2.8 \; -\!2.6$ \\
 $ 290$ & $  1522 $ & $  1436 $ & $ +0.3 \; -\!0.2$ & $ +1.8 \; -\!1.8$ & $ +2.9 \; -\!2.6$ \\
 $ 300$ & $  1441 $ & $  1358 $ & $ +0.2 \; -\!0.3$ & $ +1.8 \; -\!1.8$ & $ +2.9 \; -\!2.6$ \\
 $ 320$ & $  1298 $ & $  1220 $ & $ +0.2 \; -\!0.2$ & $ +1.8 \; -\!1.8$ & $ +3.0 \; -\!2.6$ \\
 $ 340$ & $  1173 $ & $  1094 $ & $ +0.2 \; -\!0.2$ & $ +1.8 \; -\!1.8$ & $ +3.1 \; -\!2.6$ \\
 $ 360$ & $  1063 $ & $  993.0$ & $ +0.1 \; -\!0.2$ & $ +1.9 \; -\!1.9$ & $ +3.2 \; -\!2.6$ \\
 $ 380$ & $  965.3$ & $  914.8$ & $ +0.1 \; -\!0.1$ & $ +1.9 \; -\!1.9$ & $ +3.3 \; -\!2.6$ \\
 $ 400$ & $  878.6$ & $  842.2$ & $ +0.2 \; -\!0.1$ & $ +1.9 \; -\!1.9$ & $ +3.4 \; -\!2.6$ \\
 $ 450$ & $  703.6$ & $  689.3$ & $ +0.2 \; -\!0.4$ & $ +1.9 \; -\!1.9$ & $ +3.7 \; -\!2.6$ \\
 $ 500$ & $  570.7$ & $  568.4$ & $ +0.1 \; -\!0.3$ & $ +2.0 \; -\!2.0$ & $ +3.9 \; -\!2.6$ \\
 $ 550$ & $  467.6$ & $  472.4$ & $ +0.3 \; -\!0.4$ & $ +2.0 \; -\!2.0$ & $ +4.1 \; -\!2.6$ \\
 $ 600$ & $  386.9$ & $  396.5$ & $ +0.3 \; -\!0.5$ & $ +2.0 \; -\!2.0$ & $ +4.4 \; -\!2.6$ \\
 $ 650$ & $  322.8$ & $  336.0$ & $ +0.3 \; -\!0.6$ & $ +2.1 \; -\!2.1$ & $ +4.6 \; -\!2.6$ \\
 $ 700$ & $  271.3$ & $  287.2$ & $ +0.4 \; -\!0.8$ & $ +2.1 \; -\!2.1$ & $ +4.9 \; -\!2.6$ \\
 $ 750$ & $  229.3$ & $  247.6$ & $ +0.5 \; -\!0.9$ & $ +2.1 \; -\!2.1$ & $ +5.1 \; -\!2.6$ \\
 $ 800$ & $  195.1$ & $  215.5$ & $ +0.5 \; -\!1.1$ & $ +2.2 \; -\!2.2$ & $ +5.3 \; -\!2.6$ \\
 $ 850$ & $  166.5$ & $  188.5$ & $ +0.7 \; -\!1.0$ & $ +2.2 \; -\!2.2$ & $ +5.6 \; -\!2.6$ \\
 $ 900$ & $  143.0$ & $  166.6$ & $ +0.6 \; -\!1.2$ & $ +2.2 \; -\!2.2$ & $ +5.8 \; -\!2.6$ \\
 $ 950$ & $  123.4$ & $  148.4$ & $ +0.5 \; -\!1.4$ & $ +2.2 \; -\!2.2$ & $ +6.1 \; -\!2.6$ \\
 $1000$ & $  106.7$ & $  133.0$ & $ +0.7 \; -\!1.4$ & $ +2.3 \; -\!2.3$ & $ +6.3 \; -\!2.6$ \\
\hline
\end{tabular}
\label{tab:NNLO14TeV}
\end{table}


\begin{sloppy}
In Tables~\ref{tab:NNLO7TeV} and~\ref{tab:NNLO14TeV} we show the NNLO QCD
results (second column), obtained with {\sc VBF@NNLO}, and the combination of
NNLO QCD and NLO EW corrections (third column).  The combination has been
performed under the assumption that QCD and EW corrections factorize
completely, i.e. the cross section has been obtained as
\begin{equation}
\sigma = \sigma_{\rm NNLO}\times (1 + \delta_{\rm EW})\,,
\end{equation} 
where $\sigma_{\mathrm{NNLO}}$ is the NNLO QCD result and
$\delta_{\mathrm{EW}}$ the relative EW correction determined in the limit
$\alphas=0$.
To estimate the uncertainties coming from the parton distributions, we have
employed the MSTW $68\%$ confidence level PDF sets~\cite{Martin:2009iq} and
compared with other NNLO PDF sets, i.e.\ ABKM09~\cite{Alekhin:2009ni} and
JR09VF~\cite{JimenezDelgado:2009tv}. The results show that an almost constant
$2\%$ PDF uncertainty can be associated to the cross section for the LHC.
The above discussed NNLO results calculated with MSTW2008 PDFs are similar
to the ones based on ABKM09, both in 
central values and PDF uncertainties of $\cal{O}$(2\%), over the whole mass
range.  
JR09 is in agreement with this for small Higgs masses ($100{-}200$\UGeV)
and predicts $\cal{O}$(10\%) larger cross sections at high masses (1\UTeV).
The numbers of the NNLO calculation presented here can also be obtained via
the web interface~\cite{WebInterface:2010}, where the code {\sc VBF@NNLO} can
be run online.
\end{sloppy}

\clearpage


\section{$\PW\PH$/$\PZ\PH$ production mode\footnote{%
    S.~Dittmaier, R.V.~Harlander, J.~Olsen, G.~Piacquadio (eds.);
    O.~Brein, M.~Kr\"amer and T.~Zirke.}}

\subsection{Experimental overview}

Searches for the Higgs boson in the $\PW\PH$ and $\PZ\PH$ production modes,
usually defined as Higgs-strahlung processes, have been considered
mainly by exploiting two decay modes, $\PH \rightarrow \PW^{+}\PW^{-}$ and $\PH
\rightarrow \PQb\bar{\PQb}$. While the former is looked for mainly because it
could contribute to the measurement of the Higgs-boson coupling to $\PW$
bosons, the latter decay mode might contribute to the discovery of a 
low-mass Higgs boson and later allow to measure the coupling of the Higgs
boson to $\PQb$~quarks. The experimental sensitivity to $\PH \rightarrow \PW^{+}\PW^{-}$ 
is highest for Higgs-boson masses above about $160$\UGeV, while the 
$\PH \rightarrow \PQb\bar{\PQb}$ decay modes are investigated for 
the low Higgs-boson mass region, below about $130$\UGeV.

The $\PW\PH \rightarrow \PW\PW\PW$ channel in the tri-lepton mode was explored
with a parton-level study in \Bref{Baer:1998cm}, while a first
estimate of the discovery sensitivity at the LHC was presented in
\Brefs{atlasphystdr,ATL-PHYS-2000-008}, based on
a fast simulation of the ATLAS detector only. In the latter document the
statistical discovery significance of the ATLAS detector with an
integrated luminosity of $30$\Ufb$^{-1}$ was estimated to be above
$3\sigma$ for Higgs-boson masses in the range $160{-}170$\UGeV. However, a
more realistic study based on samples of fully simulated Monte Carlo
events, presented in \Bref{CSC}, shows that the extraction of this
signal might be significantly harder than previously thought, in
particular due to the very high $\PQt\bar{\PQt}$ background, although a
precise quantitative estimate of the discovery significance suffers from
the limited available statistics of the samples and from the fact that
the continuum $\PW\PW\PW$ background was not considered in the study.

The decay channel $\PH \rightarrow \PQb\bar{\PQb}$ is dominant at low
Higgs-boson masses, below about $130$\UGeV. Given the large $\PQb\bar{\PQb}$ 
backgrounds from pure QCD-driven processes, this decay mode is not accessible 
in gluon-fusion production mode and is only marginally accessible 
in combination with the vector-boson fusion. The most promising 
sensitivity studies rely on the associated production of a Higgs boson either with a
$\PZ$ or $\PW$ boson ($\PW\PH$ or $\PZ\PH$) or with a $\PQt\bar{\PQt}$
pair. The $\PW\PH$ and $\PZ\PH$ channels with $\PH \to \PQb\bar{\PQb}$
are the main search channels at the Tevatron for a Higgs boson with low
mass, but at the LHC it is significantly more challenging to extract
these signals from the backgrounds. A first study of the sensitivity to
a Higgs boson in the $\PW\PH$ and $\PZ\PH$ channels was presented in the
ATLAS TDR~\cite{atlasphystdr} and one year later in
\Brefs{ATL-PHYS-2000-023,ATL-PHYS-2000-024}. The channel with the
most significant predicted signal is $\PW\PH$, which however results in
a predicted discovery significance of about $2$ after30\Ufb$^{-1}$ and a signal to background ratio of about $2$\%. Under these conditions, the extraction of the signal is extremely
challenging, since the significance is low and the normalization of the
backgrounds in the signal region must be controlled at the percent level.

More recently, in \Bref{Butterworth:2008iy}, it has been proposed
to focus the search for a Higgs boson in the $\PW\PH$ and $\PZ\PH$
channels with the decay $\PH \rightarrow \PQb\bar{\PQb}$ into the very
specific kinematic region where both the Higgs boson and the $\PW$ or
$\PZ$ boson produced in association with it are emitted at high $\pT$
(e.g.\ $\pT>200$\UGeV), i.e.\ in a topological configuration where they
are back-to-back in the transverse plane and highly boosted. As a first
consequence, the intermediate virtual $\PW$ or $\PZ$ boson producing the
Higgs boson and the associated vector boson must be very massive, thus
even with the LHC center-of-mass energy it will produced quite
centrally, so that the kinematic acceptance of its decay products, the
Higgs and the $\PW$ bosons, will be significantly improved. In addition,
for various reasons, the signal-to-background ratio is significantly
improved, reducing the impact of background uncertainties onto the
discovery significance. A first study based on a realistic simulation of
the ATLAS detector, but based only on $LO$ Monte Carlo generators, was
performed in \Bref{ATL-PHYS-PUB-2009-088}, where it was found that
after 30\Ufb$^{-1}$ of data collected at a center-of-mass energy of
$14\UTeV$ a discovery significance above $3$ should be achievable and that
these channels might contribute, in combination with others, to the
discovery of a low-mass Higgs boson with around 10\Ufb$^{-1}$ of
integrated luminosity.

In the past months the expected sensitivity in the $\PW\PH$ and $\PZ\PH$
channels has been re-evaluated for lower center-of-mass 
energies, by both the ATLAS and CMS Collaborations. With $1\Ufb^{-1}$ of data and 
$\sqrt{s}=7$\UTeV\ ATLAS expects to exclude a Higgs boson at $95\%$ CL 
with a cross section equivalent to about $6$ times the 
SM one~\cite{ATL-PHYS-PUB-2010-015}, 
while with $5\Ufb^{-1}$ of data and $\sqrt{s}=8$\UTeV\ CMS expects to exclude a 
Higgs boson at $95\%$ CL with a cross section equivalent to about
$2$ times the SM one~\cite{CMSPublicPhysicsResultsHIGStandardModelProjectionsWebPage}. 
These results are very preliminary and partially rely on analyses which have not been 
re-optimized for the lower center-of-mass energy.

One of the main challenges of these searches is to control the backgrounds down to
a precision of about $10\%$ or better in the very specific kinematic
region where the signal is expected. Precise differential predictions 
for these backgrounds as provided by theoretical perturbative calculations and parton-shower 
Monte Carlo generators are therefore crucial. Further studies (e.g.\ in
\Bref{CERN-THESIS-2010-027}) suggest that with data corresponding to 
an integrated luminosity of the order of 30\Ufb$^{-1}$ the $\PQt\bar{\PQt}$
background might be extracted from data in a signal-free control region, 
while this seems to be significantly harder for the $\PW\PQb\bar{\PQb}$ or
$\PZ\PQb\bar{\PQb}$ irreducible backgrounds, even in the presence of such a 
large amount of data.

For all search channels previously mentioned, a precise prediction of the
signal cross section and of the kinematic properties of the produced
final-state particles is of utmost importance, together with a possibly
accurate estimation of the connected systematic uncertainties.  The
scope of this section is to present the state-of-the-art inclusive cross
sections for the $\PW\PH$ and $\PZ\PH$ Higgs-boson production modes at different
LHC center-of-mass energies and for different possible values of the
Higgs-boson mass and their connected uncertainties.

\subsection{Theoretical framework}

The inclusive partonic cross section for associated production of a
Higgs boson ($\PH$) and a weak gauge boson ($V$) can be written as
\begin{equation}
\begin{split}
\hat \sigma(\hat s) = \int_0^{\hat s} {\rm d}k^2 \,
\sigma(V^\ast(k))\,\frac{{\rm d}\Gamma}{{\rm d}k^2} (V^\ast(k)\to \PH V) +
\Delta\sigma\,,
\label{eq:sigpart}
\end{split}
\end{equation}
where $\sqrt{\hat s}$ is the partonic center-of-mass energy.
The first term on the r.h.s.\ arises from terms where a virtual gauge
boson $V^\ast$ with momentum $k$ is produced in a Drell--Yan-like
process, which then radiates a Higgs boson. The factor $\sigma(V^\ast)$
is the total cross section for producing the intermediate vector boson
and is fully analogous to the Drell--Yan expression. The second term on
the r.h.s., $\Delta\sigma$, comprises all remaining contributions.
The hadronic cross section is obtained from the partonic expression of
Eq.\,(\ref{eq:sigpart}) by convoluting it with the parton densities in
the usual way.

The LO prediction for $pp\to V\PH$ ($V=\PW,\PZ$) is based on the Feynman
diagrams shown in \Fref{fig:ppVH-LO-diags}\,(a),(b) and leads to a
LO cross section of ${\cal O}(\GF^2)$.
\begin{figure}
\begin{center}
\begin{tabular}{ccc}
{\unitlength 1pt \SetScale{1}
\begin{picture}(150,100)(-5,-10)
\ArrowLine(20, 5)(50,40)
\ArrowLine(50,40)(20,75)
\Photon(50,40)(90,40){2}{5}
\Photon(120,20)(90,40){2}{5}
\DashLine(90,40)(120,60){5}
\Vertex(50,40){2}
\Vertex(90,40){2}
\put( -4, 2){${\PQu/\PQd}$}
\put( -4,72){$\PAQd/\PAQu$}
\put(125,56){${\PH}$}
\put( 62,25){${\PW}$}
\put(125,12){${\PW^\pm}$}
\end{picture}
}\hspace*{-2em} &
{\unitlength 1pt \SetScale{1}
\begin{picture}(150,100)(-5,-10)
\ArrowLine(20, 5)(50,40)
\ArrowLine(50,40)(20,75)
\Photon(50,40)(90,40){2}{5}
\Photon(120,20)(90,40){2}{5}
\DashLine(90,40)(120,60){5}
\Vertex(50,40){2}
\Vertex(90,40){2}
\put(  8, 2){${\PQq}$}
\put(  8,72){${\PAQq}$}
\put(125,56){${\PH}$}
\put( 65,25){${\PZ}$}
\put(125,12){${\PZ}$}
\end{picture}
}\hspace*{-2em} &
{\unitlength 1pt \SetScale{1}
\begin{picture}(150,100)(-5,-10)
\Gluon(50,15)(20, 5){4}{3}
\Gluon(20,75)(50,65){4}{3}
\ArrowLine(50,65)(50,15)
\ArrowLine(90,65)(50,65)
\ArrowLine(50,15)(90,15)
\ArrowLine(90,15)(90,65)
\Photon(120,10)(90,15){2}{5}
\DashLine(90,65)(120,70){5}
\Vertex(90,65){2}
\Vertex(90,15){2}
\Vertex(50,15){2}
\Vertex(50,65){2}
\put(  55, 35){$\PQt$}
\put(  8, 2){$\Pg$}
\put(  8,72){$\Pg$}
\put(125,68){$\PH$}
\put(125,7){$\PZ$}
\end{picture}
}\\
(a) & (b) & (c)
\end{tabular}
\end{center}
\vspace*{-1em}
\caption{(a), (b) LO diagrams for the partonic processes $\Pp\Pp\to V\PH$
  ($V=\PW,\PZ$); (c) diagram contributing to the $\Pg\Pg\to \PH\PZ$ channel.}
\label{fig:ppVH-LO-diags}
\end{figure}
Through NLO, the QCD
corrections are fully given by the NLO QCD corrections to the Drell--Yan
cross section $\hat
\sigma(V^\ast)$~\cite{Han:1991ia,Baer:1992vx,Ohnemus:1992bd}. For $V=\PW$,
this observation carries over to NNLO%
\footnote{\samepage This statement holds up to two-loop diagrams where the Higgs
  boson is attached to a one-loop Drell--Yan diagram via the loop-induced
  $\Pg\Pg\PH$ coupling. Such diagrams, which are neglected so far, are believed 
  to have only a small impact; their calculation is in progress.},
so that the corresponding QCD
corrections can be easily derived by integrating the classic Drell--Yan
result~\cite{Hamberg:1990np,Harlander:2002wh} over the virtuality of the
intermediate gauge boson. For that purpose, the program {\sc VH@NNLO}
has been developed~\cite{Brein:2003wg}, building on the publicly
available code {\sc zwprod.f}~\cite{Hamberg:1990np}.

The Drell--Yan-like corrections that determine the NNLO result for
$\PW\PH$ production also give the bulk of the $\PZ\PH$
contribution. However, in that case, there are gluon--gluon-induced
terms that do not involve a virtual weak gauge boson; both $\PZ$ and
$\PH$ couple to the gluons via a top-quark loop in this case, see
\Fref{fig:ppVH-LO-diags}\,(c). This class of diagrams is not taken
into account in {\sc VH@NNLO}; it was computed in
\Bref{Brein:2003wg}, and the numbers included in the results below are based on
the corresponding numerical code.

As every hadron collider observable that is evaluated at fixed order
perturbation theory, the cross section depends on the unphysical
renormalization and factorization scales $\mu_R$ and $\mu_F$.  Since the
QCD corrections mostly affect the production of the intermediate gauge
boson, a natural choice for the central value of $\mu_F$ and $\mu_R$ is
the virtuality $k^2$ of this gauge boson.

NLO electroweak (EW) corrections have been evaluated in
\Bref{Ciccolini:2003jy}.  In contrast to the NLO QCD corrections,
EW corrections do not respect a factorization into
Drell--Yan-like production and decay, since there are irreducible (box)
corrections to $\PQq\PQq^{(\prime)}\to V\PH$ already at one loop.  Note also
that the size of the EW corrections (as usual) sensitively
depend on the chosen renormalization scheme to define the weak
couplings, most notably on the choice for the electromagnetic couplings
$\alpha$. The preferred choice, which should be most robust with
respect to higher-order corrections, is the so-called $\GF$ scheme,
where $\alpha$ is derived from Fermi's constant $\GF$.

The combination of QCD and EW corrections poses
the question on whether factorization of the
EW and QCD effects is a valid approximation to the actual
mixed ${\cal O}(\GF\alphas)$ corrections. Following \Bref{Brein:2004ue},
we present our result based
on the assumption that full factorization of the two effects is valid,
i.e., the cross section is determined as
\begin{equation}
\begin{split}
\sigma_{\PW\PH} = \sigma_{\PW\PH}^{\mbox{\footnotesize\sc VH@NNLO}} 
\times (1 + \delta_{\PW\PH,\rm EW})\,, \qquad
\sigma_{\PZ\PH} = \sigma_{\PZ\PH}^{\mbox{\footnotesize\sc VH@NNLO}} 
\times (1 + \delta_{\PZ\PH,\rm EW})
+\sigma_{\Pg\Pg\to\PZ\PH} \,,
\end{split}
\end{equation}
where $\sigma_{V\PH}^{\mbox{\footnotesize\sc VH@NNLO}}$ 
is the NNLO QCD result of {\sc VH@NNLO} 
through ${\cal O}(\alphas^2)$, $\delta_{V\PH,\rm EW}$ is the relative 
EW correction factor determined in the limit $\alphas=0$, and
$\sigma_{\Pg\Pg\to\PZ\PH}$ is the NNLO contribution 
to $\PZ\PH$ production induced by $\Pg\Pg$ fusion.

The PDF+$\alphas$ uncertainties are evaluated according to the recipe
proposed in Section~\ref{sec:pdf4lhcreco} of this report. The
uncertainties due to the residual dependence on the renormalization and
factorization scales are determined by considering the cross section
when mutually fixing one of $\mu_R$ and $\mu_F$ at the central scale
$\sqrt{k^2}$ (the mass of the intermediate gauge boson, see above), and varying
the other scale between $\sqrt{k^2}/3$ and $3\sqrt{k^2}$.
The EW factor $\delta_{V\PH,\rm EW}$ is always calculated in the same way as
the central value of the cross section, because the relative EW correction
is insensitive to the PDF and/or scale choices.

In principle there are also real NLO EW corrections induced by
initial-state photons, which are ignored, since current PDF sets do not
deliver a photon PDF.  The photon PDF is, however, strongly suppressed,
so that an uncertainty of not more than 1\% should arise from this
approximation.  This estimated percent uncertainty, which rests on the
comparison with other cross sections such as vector-boson
fusion~\cite{Ciccolini:2007ec,Ciccolini:2007jr} where these effects have
been calculated, also includes the neglect of NLO EW corrections in the
evolution of current PDFs.

\subsection{Numerical results}

The results for the NLO and the NNLO QCD cross sections for $\PW\PH$
production, including NLO EW effects, are shown in
\Fref{fig:wh-xsec}, both at $7\UTeV$ and $14\UTeV$. The numbers are obtained
by summing over $\PWp\PH$ and $\PWm\PH$ production.
The corresponding $K$-factors, obtained by normalizing the cross section
to the LO value (at central scales and PDFs), are shown in
\Fref{fig:wh-k}. The little kinks at around 160\UGeV\ and, somewhat
smaller, 180\UGeV\ are due to the $\PW\PW$ and $\PZ\PZ$ thresholds that occur
in the EW radiative corrections (see also \Bref{Ciccolini:2003jy}).
The present prediction does not properly describe the threshold
behaviour, which is in fact singular on threshold. Therefore, 
in practice, Higgs mass windows of $\sim \pm 5\UGeV$ around the 
thresholds should be obtained upon
interpolation unless the threshold regions are properly described
(e.g.\ by complex masses), a task which is in progress.
The uncertainty of the threshold interpolation is about 1\%.

The plots for $\PZ\PH$ production are shown in \Frefs{fig:zh-xsec}
and \ref{fig:zh-k}. The fact that the uncertainty bands at NNLO are of
the same order of magnitude as at NLO is due to the $\Pg\Pg$ channel
that occurs only at NNLO and is absent in the $\PW\PH$ case.
In more detail, for the centre-of-mass energy of $7\UTeV$ ($14\UTeV$) the
$\Pg\Pg$ channel contributes to $\PZ\PH$ production by
$2{-}6\%$ ($4{-}12\%$) with an uncertainty of 
$20{-}30\%$ from scale variation and of $4\%$ ($2\%$) from PDF, 
translating roughly into a $0.5{-}1.5\%$
($1{-}3\%$) uncertainty on the full result.

We have checked the NLO numbers against {\sc V2HV}~\cite{V2HV} and find
agreement at the permille level, once CKM mixing is included in {\sc
  V2HV}. Also, we find satisfactory agreement of the NLO result when
comparing to {\sc MCFM}~\cite{MCFM}. However, the comparison is less strict in
this case as {\sc MCFM} does not allow the same scale choice as used here.

The results for the central values of the cross section and the
corresponding theoretical uncertainties are shown in
\Trefs{tab:wzh7} and \ref{tab:wzh14} for $7\UTeV$ and $14\UTeV$,
respectively. Notice that the scale uncertainties for $\PZ\PH$ production
are consistently larger than for $\PW\PH$ production, because they are
dominated by the uncertainties of the $\Pg\Pg$ channel.

\begin{figure}
\vspace{0pt}
\begin{center}
\begin{tabular}{cc}
\includegraphics[
angle=0,width=.46\linewidth]{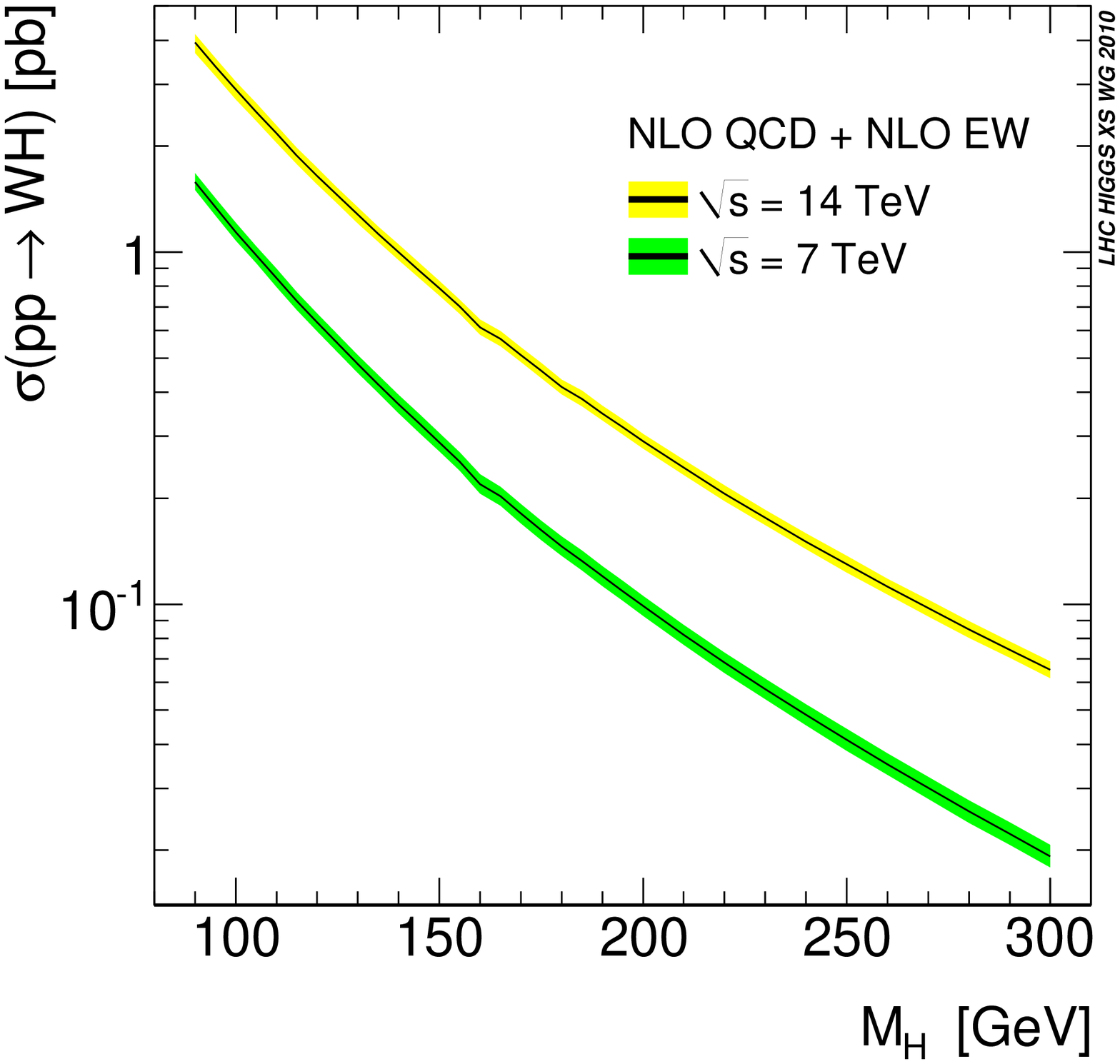} &
\includegraphics[
angle=0,width=.46\linewidth]{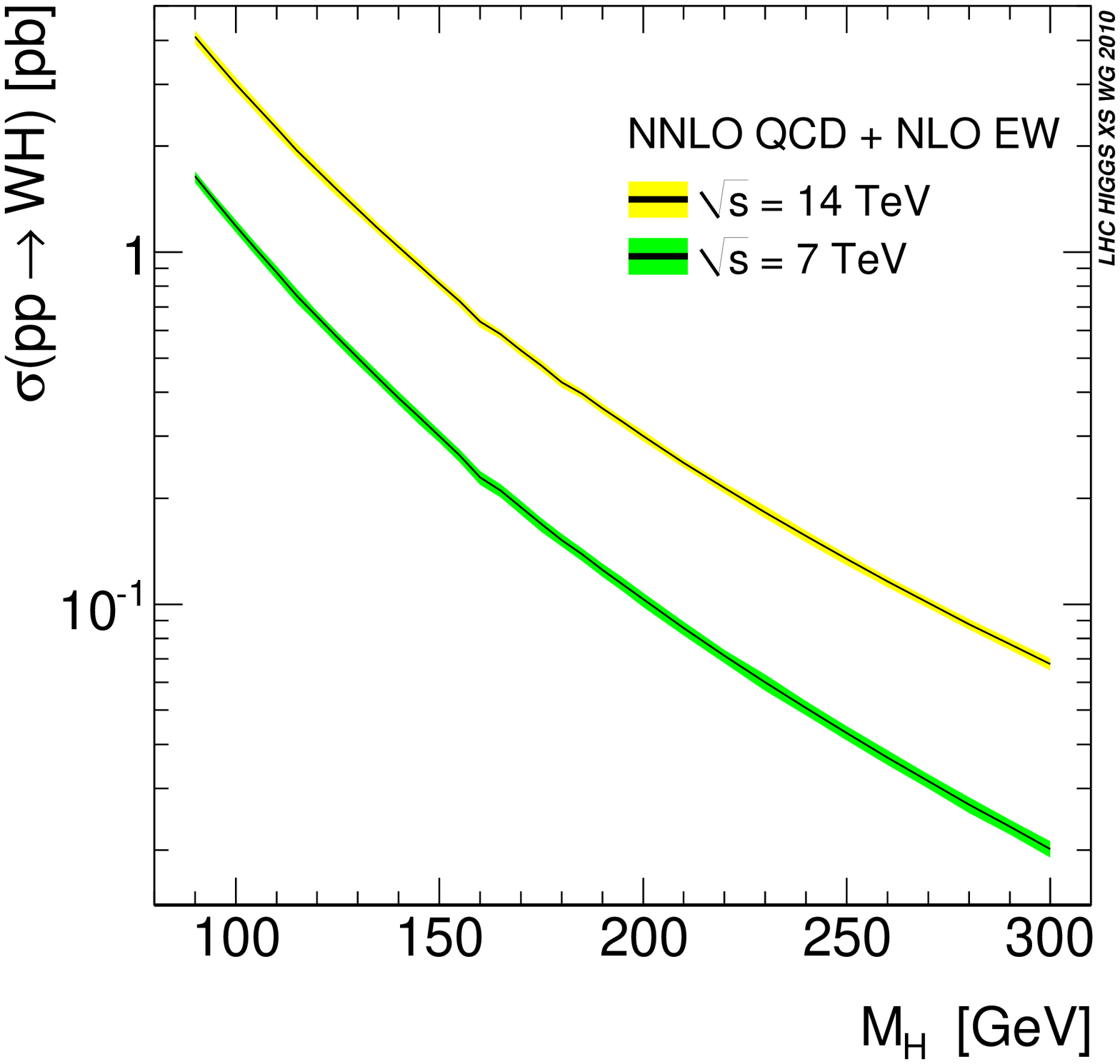}\\[-1.5em] 
(a) & (b)
\end{tabular}
\end{center}
\vspace*{-1em}
\caption[]{\label{fig:wh-xsec} Cross section for the sum of $\PWp\PH$
  and $\PWm\PH$ production for $7\UTeV$ and $14\UTeV$ at (a) NLO and (b) NNLO
  QCD, including NLO EW effects in both cases.}
%
\vspace{0pt}
\begin{center}
\begin{tabular}{cc}
\includegraphics[
angle=0,width=.46\linewidth]{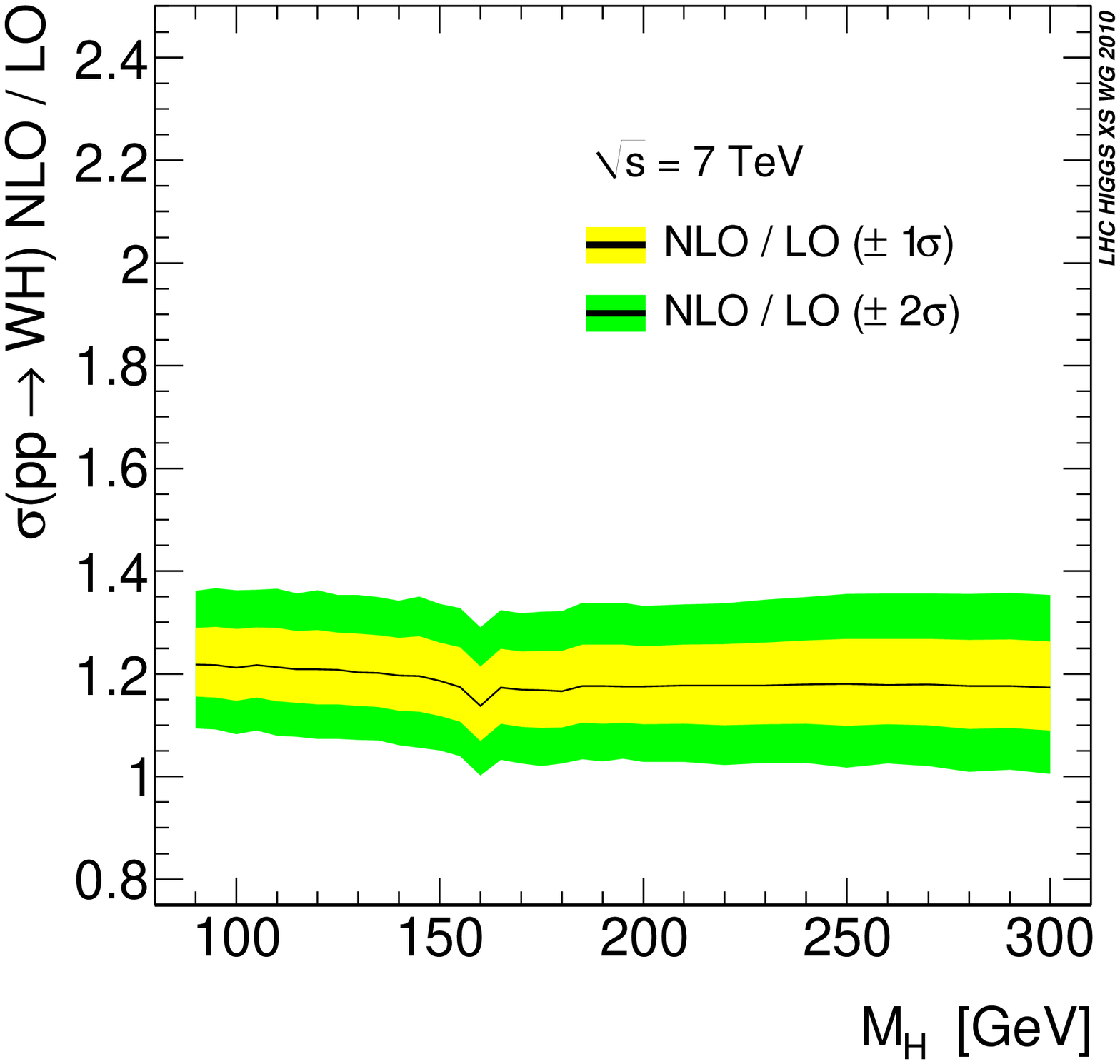} &
\includegraphics[
angle=0,width=.46\linewidth]{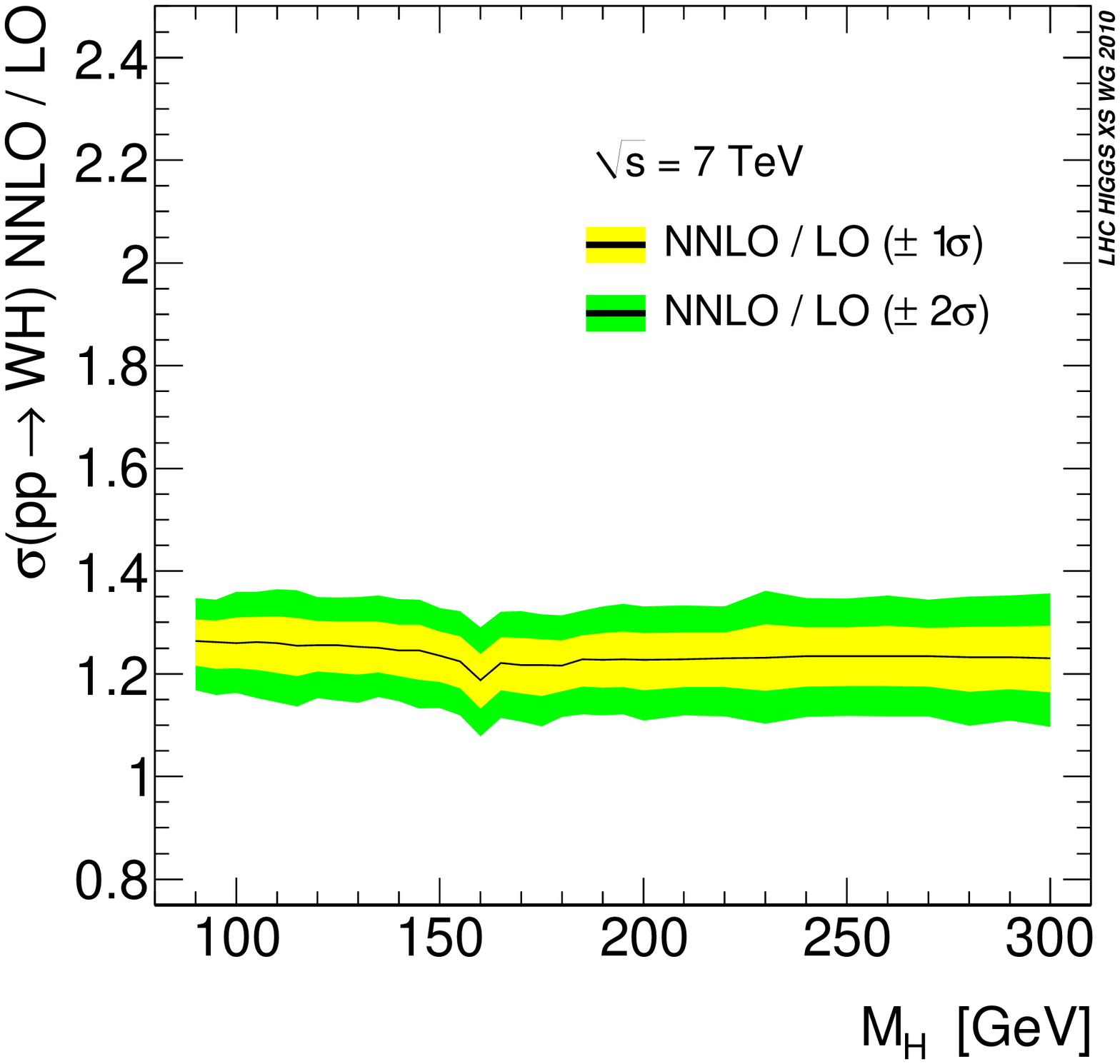}\\[-1.5em] 
(a) & (b)\\[.5em]
\includegraphics[
angle=0,width=.46\linewidth]{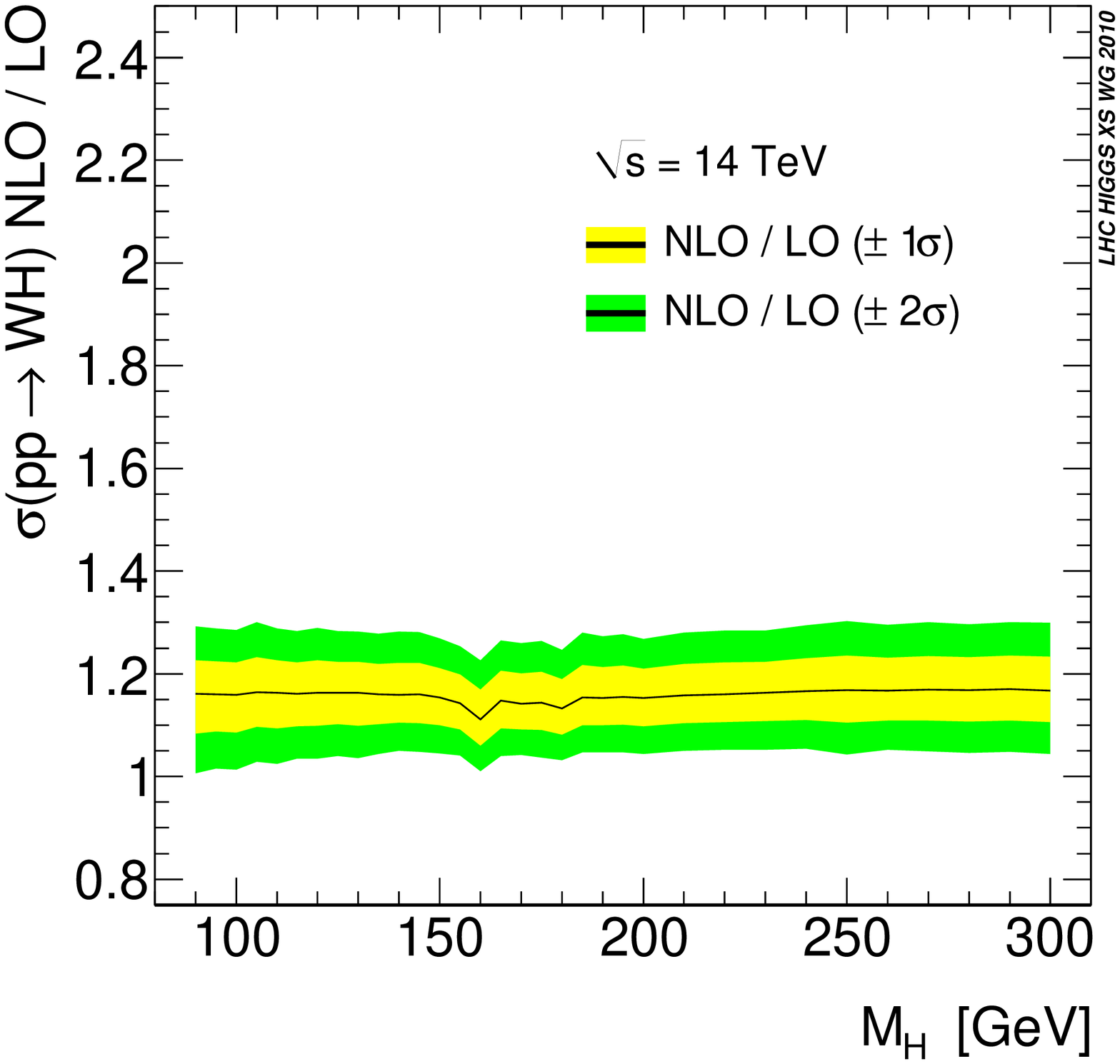} &
\includegraphics[
angle=0,width=.46\linewidth]{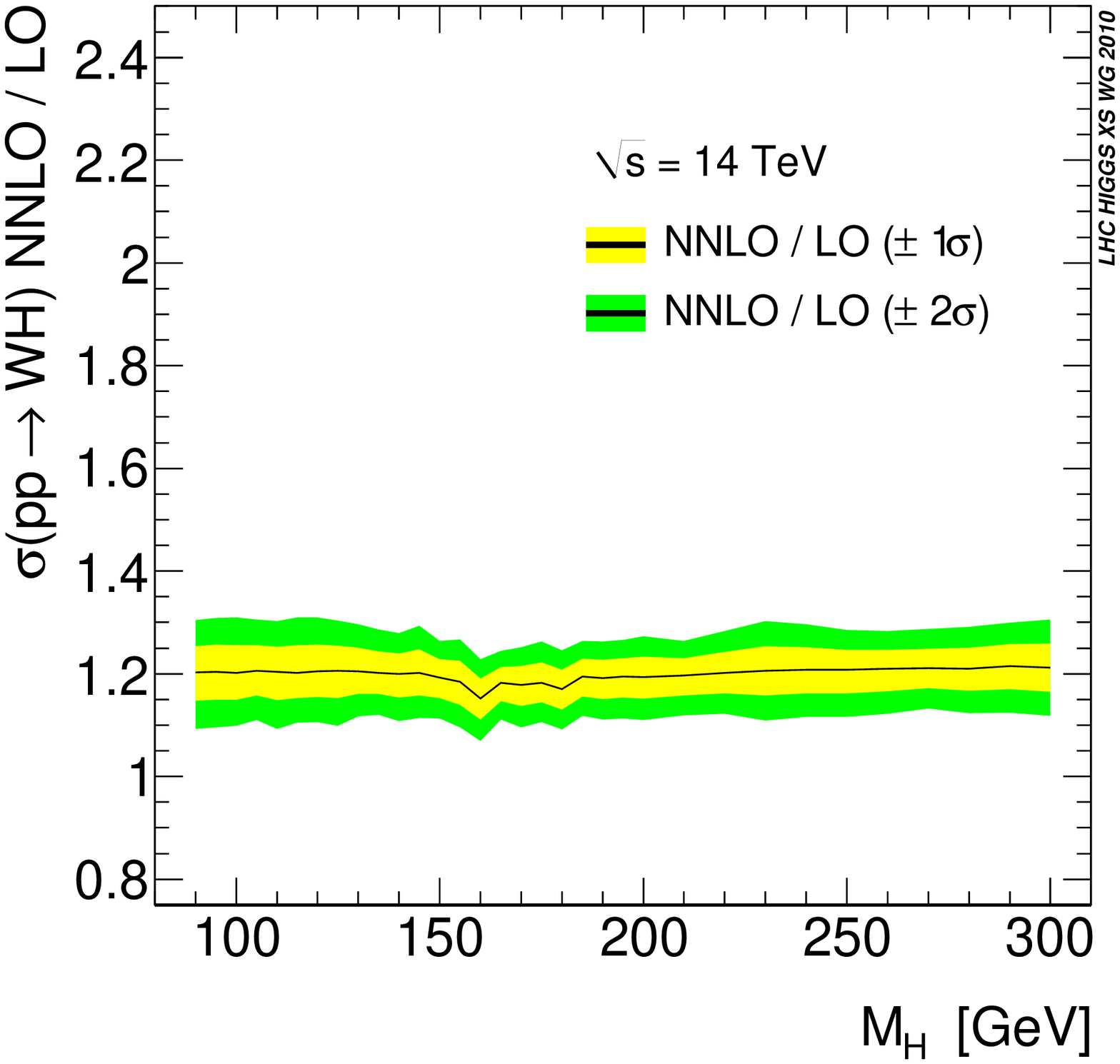}\\[-1.5em] 
(c) & (d)
\end{tabular}
\end{center}
\vspace*{-1em}
\caption[]{\label{fig:wh-k}$K$-factors (ratio to LO prediction) for the
  NLO and NNLO cross sections of \Fref{fig:wh-xsec}.}
\end{figure}

\begin{figure}
\vspace{0pt}
\begin{center}
\begin{tabular}{cc}
\includegraphics[
angle=0,width=.46\linewidth]{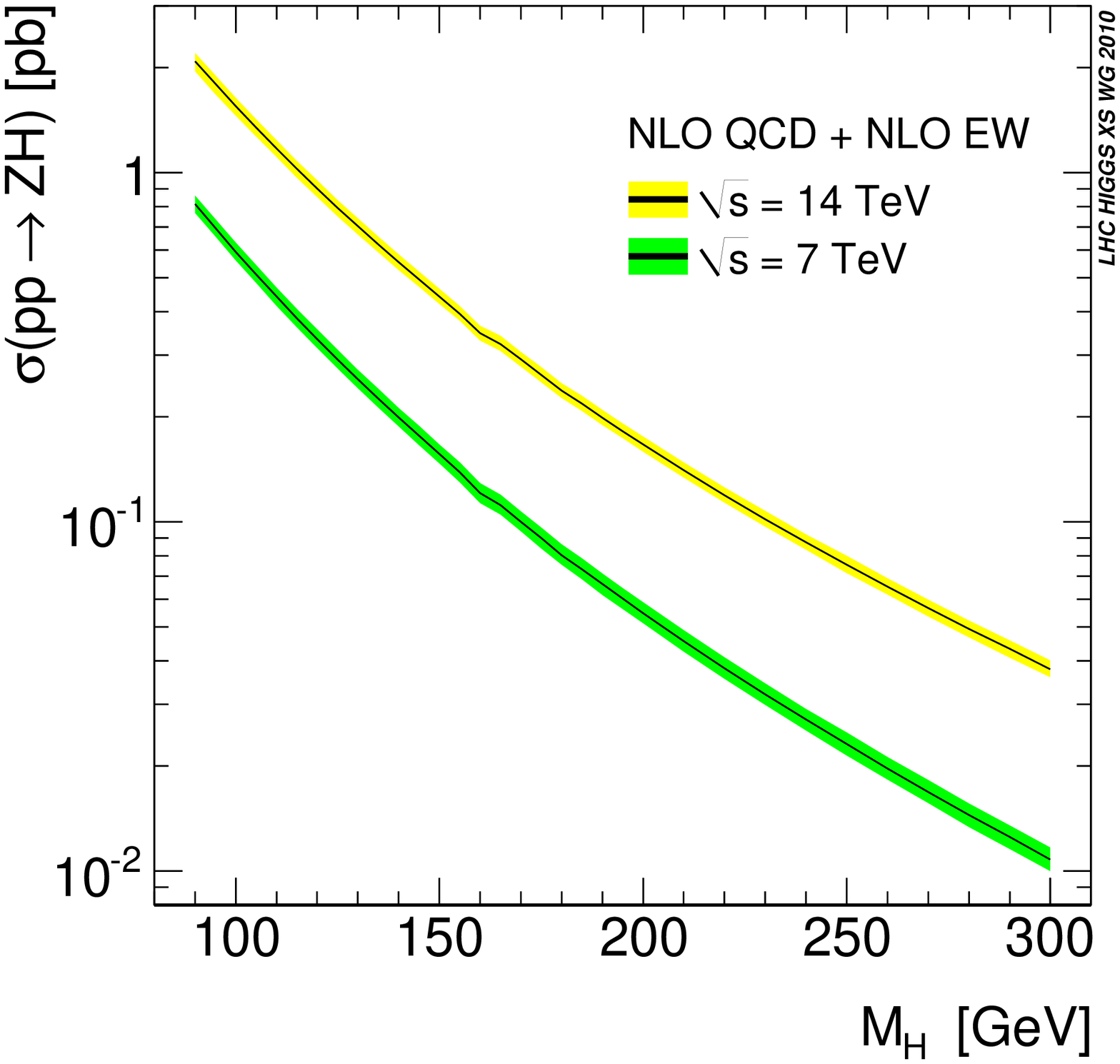} &
\includegraphics[
angle=0,width=.46\linewidth]{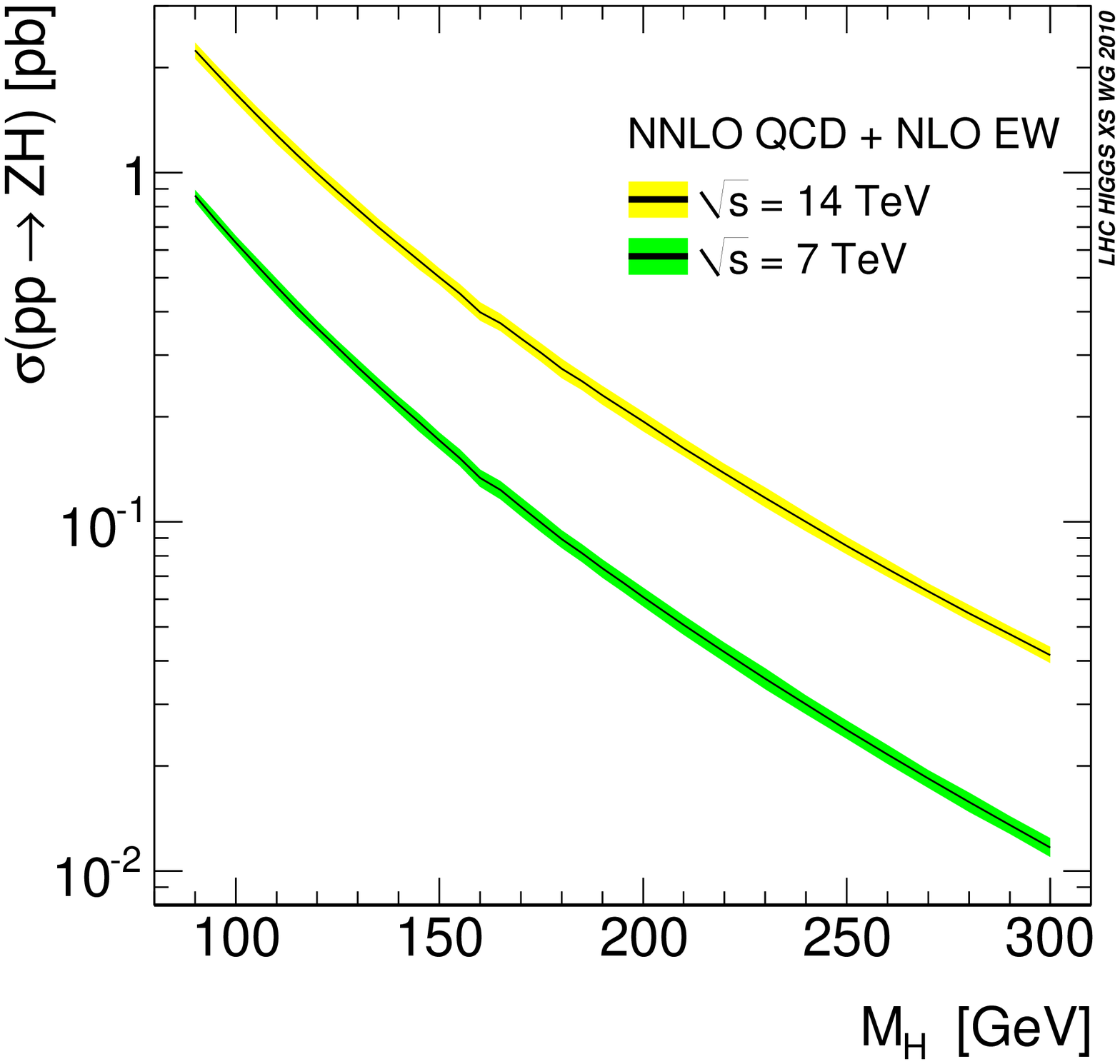}\\[-1.5em]  
(a) & (b)
\end{tabular}
\end{center}
\vspace*{-1em}
\caption[]{\label{fig:zh-xsec} Cross section for $\PZ\PH$
  production for $7\UTeV$ and $14\UTeV$ at (a) NLO and (b) NNLO
  QCD, including NLO EW effects in both cases.}
%
\vspace{0pt}
\begin{center}
\begin{tabular}{cc}
\includegraphics[
angle=0,width=.46\linewidth]{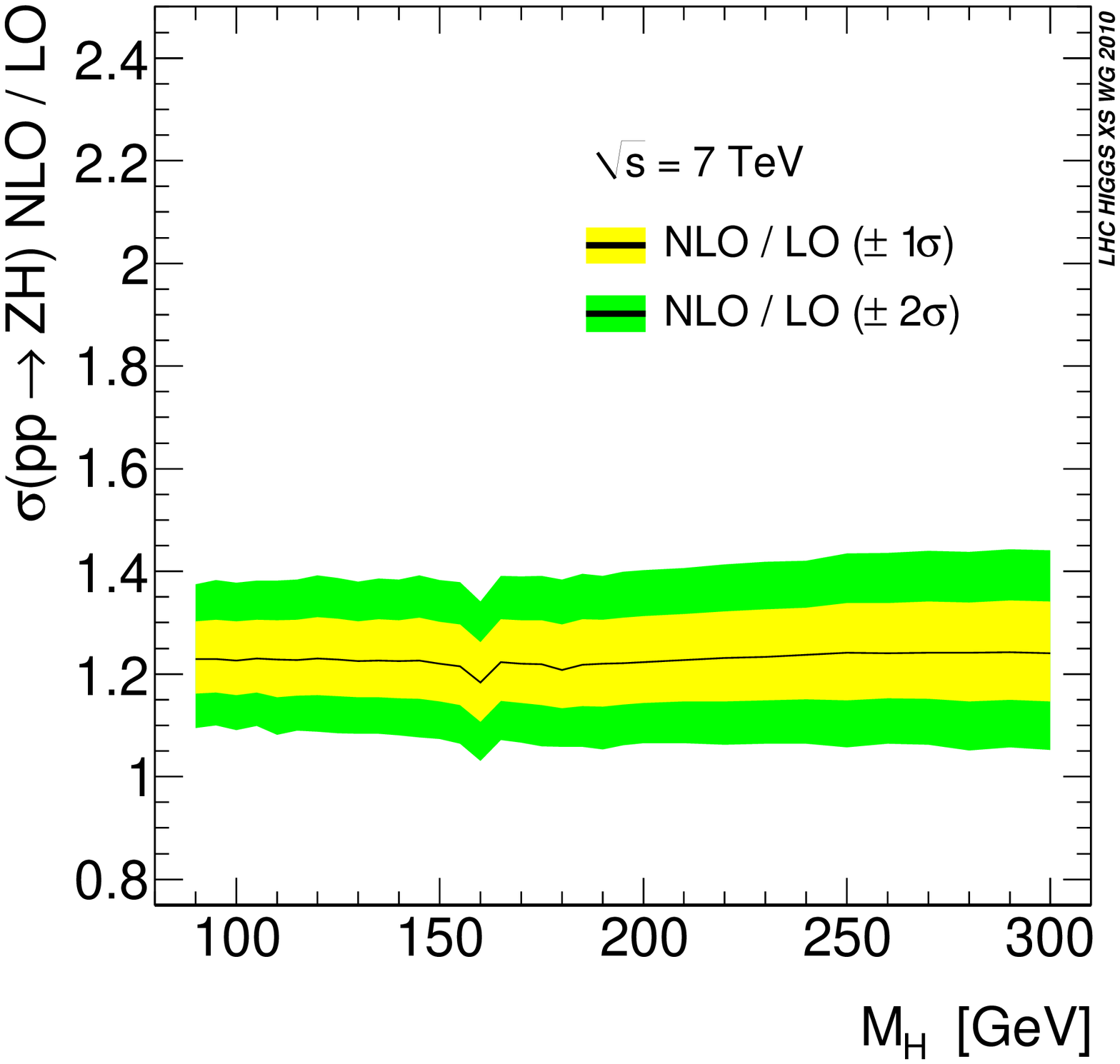} &
\includegraphics[
angle=0,width=.46\linewidth]{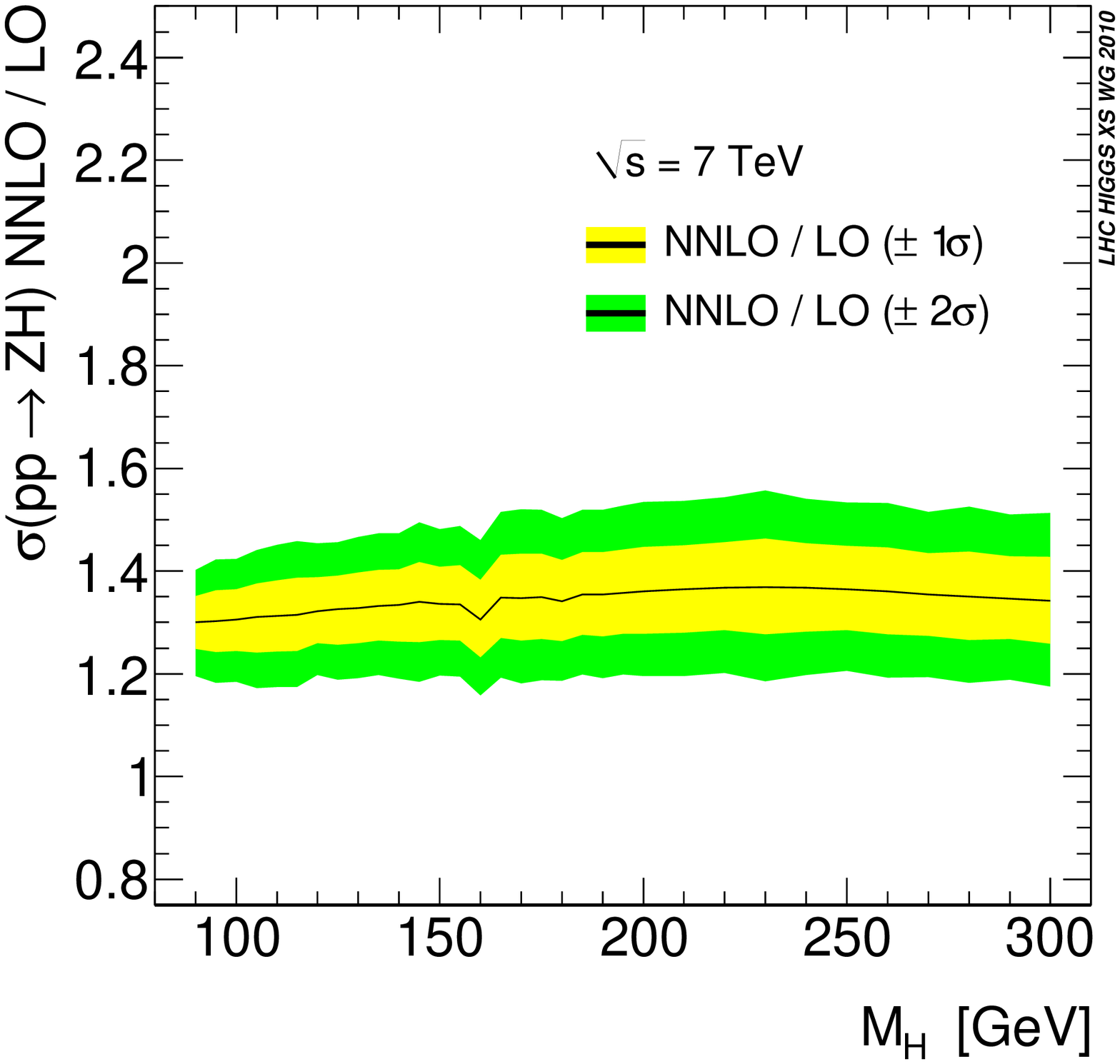}\\[-1.5em]  
(a) & (b)\\[.5em]
\includegraphics[
angle=0,width=.46\linewidth]{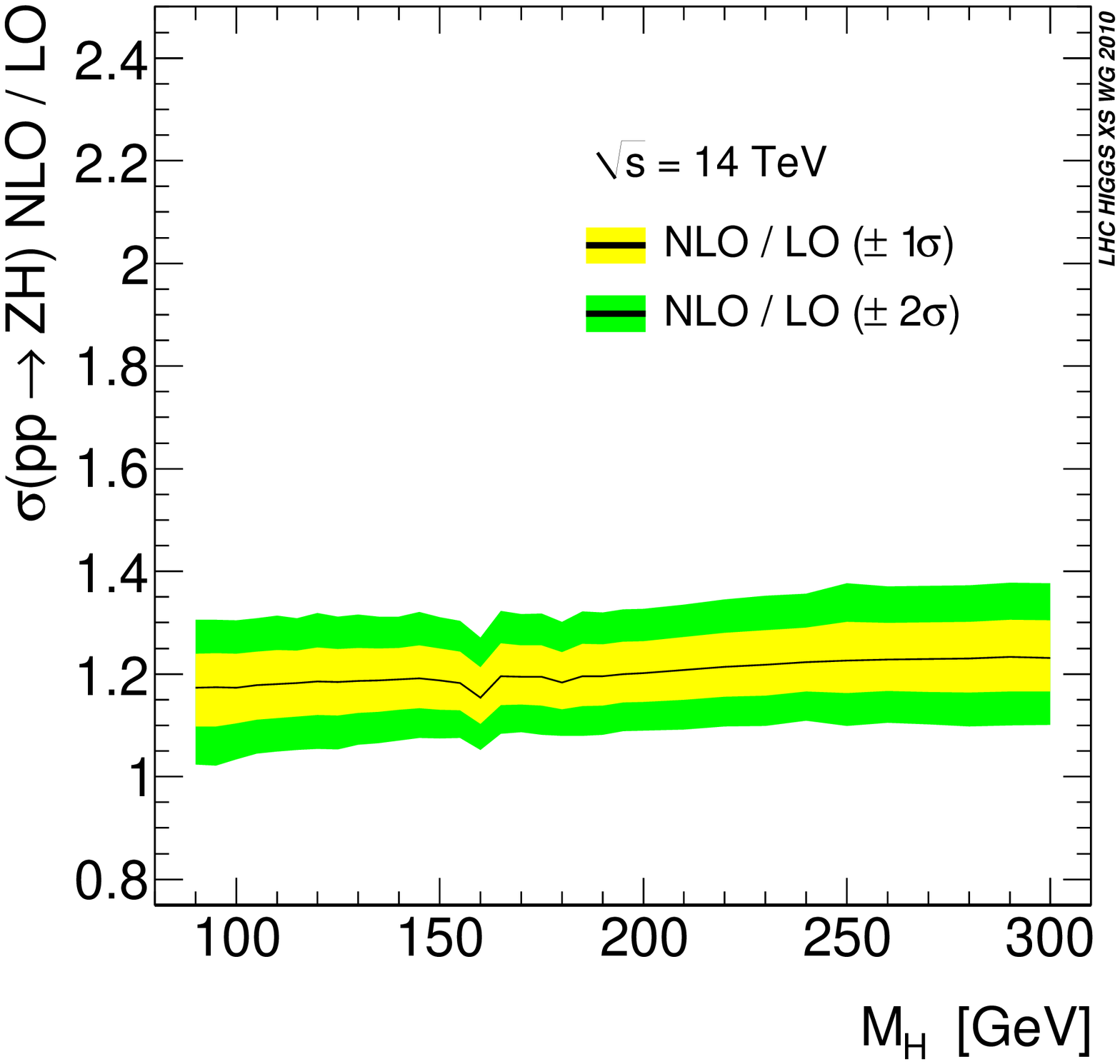} &
\includegraphics[
angle=0,width=.46\linewidth]{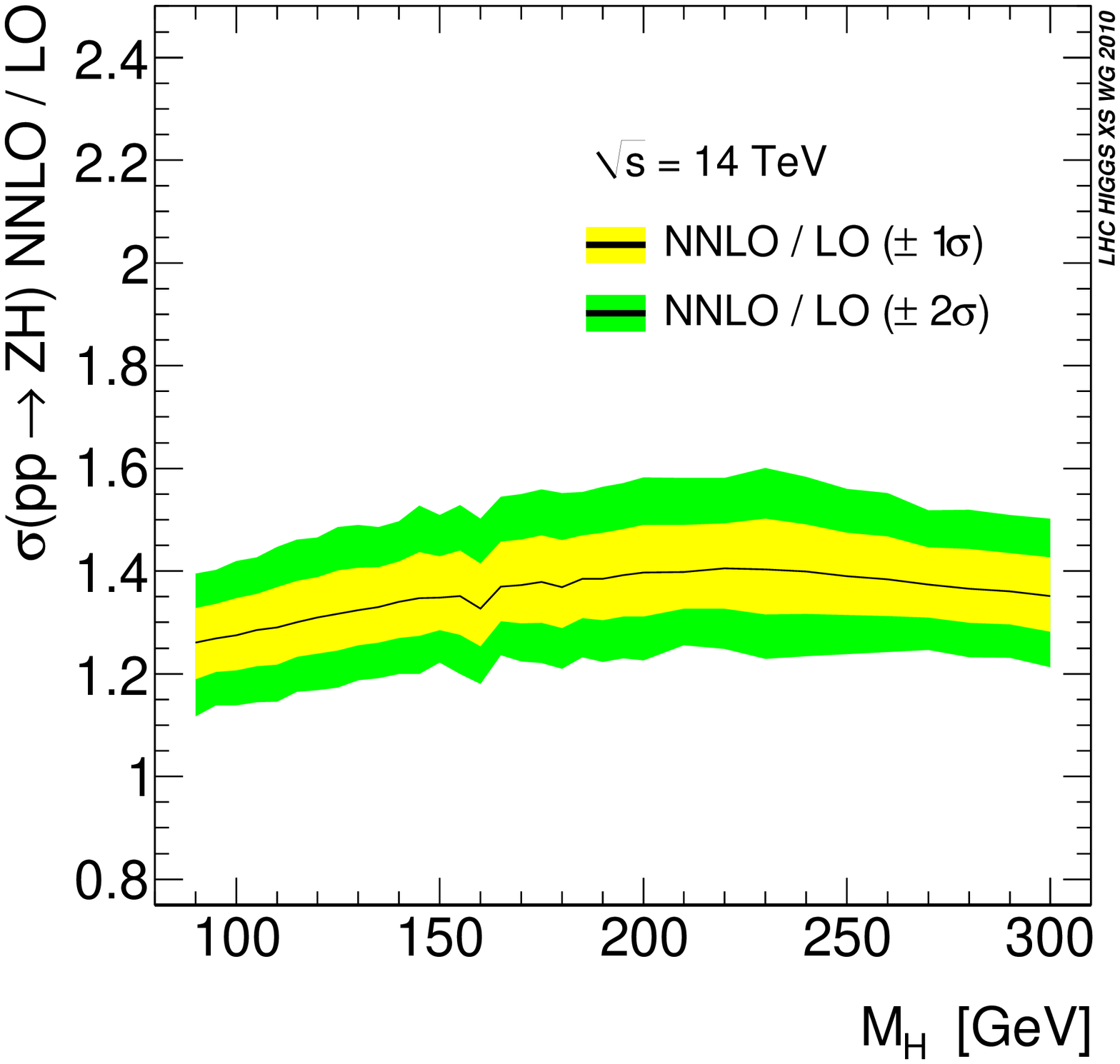}\\[-1.5em]  
(c) & (d)
\end{tabular}
\end{center}
\vspace*{-1em}
\caption[]{\label{fig:zh-k}$K$-factors (ratio to LO prediction) for the
  NLO and NNLO cross sections of \Fref{fig:zh-xsec}.}
\end{figure}

\begin{table}
   \begin{center}
   \caption[]{\label{tab:wzh7}Total inclusive cross section at $7\UTeV$ for
  $\PW\PH$ and $\PZ\PH$ production at NNLO QCD + NLO EW. The first error
  indicates the uncertainty from the renormalization and factorization
  scale variation, the second from the PDF+$\alphas$ variation.}
   \small
   \begin{tabular}{ccccccc}
   \hline
$\MH[\UGeVZ]$ & $\sigma(\PW\PH)[\UpbZ]$ & Scale [\%] & PDF4LHC [\%] &
$\sigma(\PZ\PH)[\UpbZ]$ & Scale [\%] & PDF4LHC [\%]
\\ \hline
$ 90 $ & $    1.640 $ & $ +0.3   -\!0.8  $ & 
$ \pm 3.0 $
 & $   0.8597 $ & $ +0.9   -\!1.0  $ &
$ \pm 3.0 $ \\[.2em]

$ 95 $ & $    1.392 $ & $ +0.1   -\!0.9  $ & 
$ \pm 3.2 $
 & $   0.7348 $ & $ +1.0   -\!1.1  $ &
$ \pm 3.6 $ \\[.2em]

$ 100 $ & $    1.186 $ & $ +0.6   -\!0.5  $ & 
$ \pm 3.4 $
 & $   0.6313 $ & $ +1.1   -\!1.2  $ &
$ \pm 3.4 $ \\[.2em]

$ 105 $ & $    1.018 $ & $ +0.3   -\!0.8  $ & 
$ \pm 3.5 $
 & $   0.5449 $ & $ +1.3   -\!1.6  $ &
$ \pm 3.7 $ \\[.2em]

$ 110 $ & $   0.8754 $ & $ +0.3   -\!0.7  $ & 
$ \pm 3.8 $
 & $   0.4721 $ & $ +1.2   -\!1.2  $ &
$ \pm 4.1 $ \\[.2em]

$ 115 $ & $   0.7546 $ & $ +0.4   -\!0.8  $ & 
$ \pm 3.9 $
 & $   0.4107 $ & $ +1.3   -\!1.2  $ &
$ \pm 4.2 $ \\[.2em]

$ 120 $ & $   0.6561 $ & $ +0.4   -\!0.7  $ & 
$ \pm 3.4 $
 & $   0.3598 $ & $ +1.5   -\!1.2  $ &
$ \pm 3.5 $ \\[.2em]

$ 125 $ & $   0.5729 $ & $ +0.2   -\!0.8  $ & 
$ \pm 3.5 $
 & $   0.3158 $ & $ +1.4   -\!1.6  $ &
$ \pm 3.5 $ \\[.2em]

$ 130 $ & $   0.5008 $ & $ +0.3   -\!0.8  $ & 
$ \pm 3.5 $
 & $   0.2778 $ & $ +1.5   -\!1.4  $ &
$ \pm 3.7 $ \\[.2em]

$ 135 $ & $   0.4390 $ & $ +0.7   -\!0.4  $ & 
$ \pm 3.4 $
 & $   0.2453 $ & $ +1.7   -\!1.4  $ &
$ \pm 3.6 $ \\[.2em]

$ 140 $ & $   0.3857 $ & $ +0.5   -\!0.5  $ & 
$ \pm 3.5 $
 & $   0.2172 $ & $ +1.5   -\!1.6  $ &
$ \pm 3.7 $ \\[.2em]

$ 145 $ & $   0.3406 $ & $ +0.2   -\!0.8  $ & 
$ \pm 3.8 $
 & $   0.1930 $ & $ +1.8   -\!1.8  $ &
$ \pm 4.0 $ \\[.2em]

$ 150 $ & $   0.3001 $ & $ +0.4   -\!0.8  $ & 
$ \pm 3.3 $
 & $   0.1713 $ & $ +1.8   -\!1.6  $ &
$ \pm 3.6 $ \\[.2em]

$ 155 $ & $   0.2646 $ & $ +0.5   -\!0.8  $ & 
$ \pm 3.5 $
 & $   0.1525 $ & $ +2.1   -\!1.6  $ &
$ \pm 3.6 $ \\[.2em]

$ 160 $ & $   0.2291 $ & $ +0.5   -\!0.7  $ & 
$ \pm 3.8 $
 & $   0.1334 $ & $ +2.0   -\!1.7  $ &
$ \pm 4.0 $ \\[.2em]

$ 165 $ & $   0.2107 $ & $ +0.5   -\!0.7  $ & 
$ \pm 3.6 $
 & $   0.1233 $ & $ +2.1   -\!1.7  $ &
$ \pm 4.1 $ \\[.2em]

$ 170 $ & $   0.1883 $ & $ +0.5   -\!0.7  $ & 
$ \pm 3.8 $
 & $   0.1106 $ & $ +2.2   -\!1.9  $ &
$ \pm 4.2 $ \\[.2em]

$ 175 $ & $   0.1689 $ & $ +0.3   -\!1.1  $ & 
$ \pm 3.8 $
 & $  0.09950 $ & $ +2.1   -\!1.9  $ &
$ \pm 4.1 $ \\[.2em]

$ 180 $ & $   0.1521 $ & $ +0.6   -\!0.6  $ & 
$ \pm 3.5 $
 & $  0.08917 $ & $ +2.2   -\!1.9  $ &
$ \pm 3.8 $ \\[.2em]

$ 185 $ & $   0.1387 $ & $ +0.4   -\!0.9  $ & 
$ \pm 3.5 $
 & $  0.08139 $ & $ +2.3   -\!2.0  $ &
$ \pm 3.8 $ \\[.2em]

$ 190 $ & $   0.1253 $ & $ +0.5   -\!0.7  $ & 
$ \pm 3.7 $
 & $  0.07366 $ & $ +2.2   -\!2.1  $ &
$ \pm 3.9 $ \\[.2em]

$ 195 $ & $   0.1138 $ & $ +0.7   -\!0.6  $ & 
$ \pm 3.7 $
 & $  0.06699 $ & $ +2.3   -\!1.9  $ &
$ \pm 4.0 $ \\[.2em]

$ 200 $ & $   0.1032 $ & $ +0.4   -\!1.0  $ & 
$ \pm 3.8 $
 & $  0.06096 $ & $ +2.3   -\!1.9  $ &
$ \pm 4.1 $ \\[.2em]

$ 210 $ & $  0.08557 $ & $ +0.5   -\!0.7  $ & 
$ \pm 3.7 $
 & $  0.05068 $ & $ +2.1   -\!2.0  $ &
$ \pm 4.2 $ \\[.2em]

$ 220 $ & $  0.07142 $ & $ +0.3   -\!0.9  $ & 
$ \pm 3.7 $
 & $  0.04235 $ & $ +2.2   -\!1.9  $ &
$ \pm 4.2 $ \\[.2em]

$ 230 $ & $  0.06006 $ & $ +0.7   -\!0.7  $ & 
$ \pm 4.5 $
 & $  0.03560 $ & $ +2.1   -\!1.9  $ &
$ \pm 4.8 $ \\[.2em]

$ 240 $ & $  0.05075 $ & $ +0.5   -\!0.7  $ & 
$ \pm 4.0 $
 & $  0.02999 $ & $ +1.9   -\!1.8  $ &
$ \pm 4.4 $ \\[.2em]

$ 250 $ & $  0.04308 $ & $ +0.5   -\!0.7  $ & 
$ \pm 4.0 $
 & $  0.02540 $ & $ +2.0   -\!1.6  $ &
$ \pm 4.2 $ \\[.2em]

$ 260 $ & $  0.03674 $ & $ +0.8   -\!0.7  $ & 
$ \pm 4.0 $
 & $  0.02158 $ & $ +1.8   -\!1.7  $ &
$ \pm 4.5 $ \\[.2em]

$ 270 $ & $  0.03146 $ & $ +0.6   -\!0.9  $ & 
$ \pm 3.8 $
 & $  0.01839 $ & $ +1.7   -\!1.7  $ &
$ \pm 4.3 $ \\[.2em]

$ 280 $ & $  0.02700 $ & $ +0.4   -\!1.0  $ & 
$ \pm 4.4 $
 & $  0.01575 $ & $ +1.6   -\!1.3  $ &
$ \pm 4.9 $ \\[.2em]

$ 290 $ & $  0.02333 $ & $ +0.7   -\!0.8  $ & 
$ \pm 4.2 $
 & $  0.01355 $ & $ +1.5   -\!1.3  $ &
$ \pm 4.5 $ \\[.2em]

$ 300 $ & $  0.02018 $ & $ +0.6   -\!0.9  $ & 
$ \pm 4.5 $
 & $  0.01169 $ & $ +1.4   -\!1.2  $ &
$ \pm 5.0 $ \\[.2em]

   \hline
   \end{tabular}
   \end{center}
\end{table}

\begin{table}
    \begin{center}
      \caption[]{\label{tab:wzh14}Total inclusive cross section at
  14\UTeV\ for $\PW\PH$ and $\PZ\PH$ production at NNLO QCD + NLO
  EW. The first error indicates the uncertainty from the renormalization
  and factorization scale variation, the second from the PDF+$\alphas$
  variation.}
   \small
   \begin{tabular}{ccccccc}
   \hline
$\MH[\UGeVZ]$ & $\sigma(\PW\PH)[\UpbZ]$ & Scale [\%] & PDF4LHC [\%] &
$\sigma(\PZ\PH)[\UpbZ]$ & Scale [\%] & PDF4LHC [\%]
\\ \hline
$ 90 $ & $    4.090 $ & $   +0.4   -\!0.7  $ & 
$ \pm 3.9 $
 & $    2.245 $ & $  +1.3   -\!1.7  $ &
$ \pm 4.0 $ \\[.2em]

$ 95 $ & $    3.499 $ & $   +0.6   -\!0.7  $ & 
$ \pm 3.8 $
 & $    1.941 $ & $   +1.6   -\!1.6  $ &
$ \pm 3.6 $ \\[.2em]

$ 100 $ & $    3.002 $ & $   +0.8   -\!0.6  $ & 
$ \pm 3.7 $
 & $    1.683 $ & $   +1.9   -\!1.5  $ &
$ \pm 3.8 $ \\[.2em]

$ 105 $ & $    2.596 $ & $   +0.6   -\!0.5  $ & 
$ \pm 3.5 $
 & $    1.468 $ & $   +1.7   -\!1.7  $ &
$ \pm 3.7 $ \\[.2em]

$ 110 $ & $    2.246 $ & $   +0.3   -\!0.8  $ & 
$ \pm 3.8 $
 & $    1.283 $ & $   +2.1   -\!1.6  $ &
$ \pm 4.0 $ \\[.2em]

$ 115 $ & $    1.952 $ & $   +0.7   -\!0.2  $ & 
$ \pm 3.8 $
 & $    1.130 $ & $   +2.5   -\!1.5  $ &
$ \pm 3.7 $ \\[.2em]

$ 120 $ & $    1.710 $ & $   +0.6   -\!0.3  $ & 
$ \pm 3.8 $
 & $   0.9967 $ & $   +2.4   -\!1.8  $ &
$ \pm 3.6 $ \\[.2em]

$ 125 $ & $    1.504 $ & $   +0.3   -\!0.6  $ & 
$ \pm 3.8 $
 & $   0.8830 $ & $   +2.7   -\!1.8  $ &
$ \pm 3.7 $ \\[.2em]

$ 130 $ & $    1.324 $ & $   +0.5   -\!0.4  $ & 
$ \pm 3.3 $
 & $   0.7846 $ & $   +2.9   -\!1.8  $ &
$ \pm 3.4 $ \\[.2em]

$ 135 $ & $    1.167 $ & $   +0.6   -\!0.5  $ & 
$ \pm 2.9 $
 & $   0.6981 $ & $   +2.9   -\!2.2  $ &
$ \pm 3.0 $ \\[.2em]

$ 140 $ & $    1.034 $ & $   +0.2   -\!0.7  $ & 
$ \pm 3.1 $
 & $   0.6256 $ & $   +2.8   -\!2.2  $ &
$ \pm 3.0 $ \\[.2em]

$ 145 $ & $   0.9200 $ & $   +0.5   -\!0.4  $ & 
$ \pm 3.3 $
 & $   0.5601 $ & $   +3.3   -\!2.1  $ &
$ \pm 3.4 $ \\[.2em]

$ 150 $ & $   0.8156 $ &  $  +0.3   -\!0.6  $ & 
$ \pm 2.7 $
 & $   0.5016 $ & $   +3.3   -\!2.0  $ &
$ \pm 2.7 $ \\[.2em]

$ 155 $ & $   0.7255 $ &  $  +0.4   -\!0.6  $ & 
$ \pm 3.1 $
 & $   0.4513 $ & $   +3.3   -\!2.4  $ &
$ \pm 3.2 $ \\[.2em]

$ 160 $ & $   0.6341 $ &  $  +0.2   -\!0.5  $ & 
$ \pm 3.1 $
 & $   0.3986 $ & $   +3.5   -\!2.4  $ &
$ \pm 3.1 $ \\[.2em]

$ 165 $ & $   0.5850 $ &  $  +0.2   -\!0.6  $ & 
$ \pm 2.4 $
 & $   0.3705 $ & $   +3.8   -\!2.3  $ &
$ \pm 2.6 $ \\[.2em]

$ 170 $ & $   0.5260 $ &  $  +0.3   -\!0.7  $ & 
$ \pm 2.8 $
 & $   0.3355 $ & $   +3.5   -\!2.4  $ &
$ \pm 3.0 $ \\[.2em]

$ 175 $ & $   0.4763 $ &  $  +0.5   -\!0.3  $ & 
$ \pm 2.9 $
 & $   0.3044 $ & $   +3.5   -\!2.6  $ &
$ \pm 3.1 $ \\[.2em]

$ 180 $ & $   0.4274 $ & $   +0.4   -\!0.6  $ & 
$ \pm 2.8 $
 & $   0.2744 $ & $   +3.7   -\!2.8  $ &
$ \pm 3.0 $ \\[.2em]

$ 185 $ & $   0.3963 $ & $   +0.4   -\!0.7  $ & 
$ \pm 2.5 $
 & $   0.2524 $ & $   +3.5   -\!2.9  $ &
$ \pm 2.6 $ \\[.2em]

$ 190 $ & $   0.3600 $ & $   +0.2   -\!0.6  $ & 
$ \pm 2.8 $
 & $   0.2301 $ & $  +3.5   -\!2.9  $ &
$ \pm 3.0 $ \\[.2em]

$ 195 $ & $   0.3291 $ & $   +0.3   -\!0.7  $ & 
$ \pm 2.7 $
 & $   0.2112 $ & $  +3.5   -\!2.9  $ &
$ \pm 2.9 $ \\[.2em]

$ 200 $ & $   0.3004 $ & $   +0.4   -\!0.5  $ & 
$ \pm 3.0 $
 & $   0.1936 $ & $  +3.6   -\!3.0  $ &
$ \pm 3.1 $ \\[.2em]

$ 210 $ & $   0.2526 $ & $   +0.2   -\!0.7  $ & 
$ \pm 2.6 $
 & $   0.1628 $ & $  +3.9   -\!2.5  $ &
$ \pm 2.6 $ \\[.2em]

$ 220 $ & $   0.2138 $ & $   +0.6   -\!0.5  $ & 
$ \pm 2.8 $
 & $   0.1380 $ & $   +3.4   -\!2.7  $ &
$ \pm 2.9 $ \\[.2em]

$ 230 $ & $   0.1826 $ & $   +0.4   -\!0.5  $ & 
$ \pm 3.5 $
 & $   0.1173 $ & $   +3.4   -\!2.6  $ &
$ \pm 3.6 $ \\[.2em]

$ 240 $ & $   0.1561 $ & $   +0.4   -\!0.5  $ & 
$ \pm 3.3 $
 & $  0.09996 $ & $   +3.1   -\!2.5  $ &
$ \pm 3.4 $ \\[.2em]

$ 250 $ & $   0.1343 $ & $   +0.2   -\!0.7  $ & 
$ \pm 3.0 $
 & $  0.08540 $ & $  +3.0   -\!2.3  $ &
$ \pm 3.2 $ \\[.2em]

$ 260 $ & $   0.1161 $ & $   +0.2   -\!0.7  $ & 
$ \pm 2.8 $
 & $  0.07341 $ & $  +3.0   -\!2.1  $ &
$ \pm 3.1 $ \\[.2em]

$ 270 $ & $   0.1009 $ & $   +0.5   -\!0.6  $ & 
$ \pm 2.6 $
 & $  0.06325 $ & $  +2.5   -\!1.9  $ &
$ \pm 2.8 $ \\[.2em]

$ 280 $ & $  0.08781 $ & $   +0.4   -\!0.6  $ & 
$ \pm 3.0 $
 & $  0.05474 $ & $  +2.5   -\!1.8  $ &
$ \pm 3.2 $ \\[.2em]

$ 290 $ & $  0.07714 $ & $   +0.3   -\!0.6  $ & 
$ \pm 3.2 $
 & $  0.04769 $ & $  +2.2   -\!1.5  $ &
$ \pm 3.2 $ \\[.2em]

$ 300 $ & $  0.06755 $ & $   +0.6   -\!0.5  $ & 
$ \pm 3.3 $
 & $  0.04156 $ & $   +2.0   -\!1.6  $ &
$ \pm 3.6 $ \\[.2em]

   \hline
   \end{tabular}
   \end{center}
\end{table}

\clearpage



\section{ttH process\footnote{C.~Collins-Tooth, C.~Neu, L.~Reina,
M.~Spira (eds.); S.~Dawson, S.~Dean, S.~Dittmaier, M.~Kr\"amer,
C.T.~Potter and D.~Wackeroth.}}




%

\subsection{Higgs-boson production in association with $\PQt\PAQt$ pairs}
Higgs radiation off top quarks $\PQq\PAQq/\Pg\Pg\to \PH\PQt\PAQt$ (see
\Fref{fg:lodiatth}) plays a role for light Higgs masses below $\sim
150$ \UGeV\ at the LHC. The measurement of the $\PQt\PAQt\PH$ production
rate can provide relevant information on the top--Higgs Yukawa coupling.
The leading-order (LO) cross section was computed a long time ago
\cite{Raitio:1978pt,Ng:1983jm,Kunszt:1984ri,Gunion:1991kg,Marciano:1991qq}.
These LO results are plagued by large theoretical uncertainties due to
the strong dependence on the renormalization scale of the strong
coupling constant and on the factorization scales of the parton
density functions inside the proton, respectively. For the LO cross
section there are several public codes available, as e.g.\ {\sc HQQ}
\cite{Spira:1997dg,HQQ}, {\sc Madgraph/Madevent}
\cite{Stelzer:1994tk,Madgraph}, {\sc MCFM} \cite{MCFM}, or {\sc PYTHIA}
\cite{PYTHIA}.  The dominant background processes for this signal
process are $\PQt\PAQt\PQb\PAQb$, $\PQt\PAQt jj$, $\PQt\PAQt\PGg\PGg$,
$\PQt\PAQt\PZ$, and $\PQt\PAQt\PWp\PWm$
production depending on the final-state Higgs-boson decay.
\begin{figure}[htb]
\begin{center}
\SetScale{0.8}
\begin{picture}(130,90)(0,0)
\ArrowLine(0,100)(50,50)
\ArrowLine(50,50)(0,0)
\Gluon(50,50)(100,50){3}{5}
\ArrowLine(100,50)(120,70)
\ArrowLine(120,70)(150,100)
\ArrowLine(150,0)(100,50)
\DashLine(120,70)(150,70){5}
\Vertex(50,50){2}
\Vertex(100,50){2}
\Vertex(120,70){2}
\put(-12,78){$\PQq$}
\put(-12,-2){$\PAQq$}
\put(125,53){$\PH$}
\put(125,78){$\PQt$}
\put(125,-2){$\PAQt$}
\end{picture}
\begin{picture}(130,90)(-80,0)
\Gluon(0,0)(50,0){3}{5}
\Gluon(0,100)(50,100){3}{5}
\ArrowLine(100,0)(50,0)
\ArrowLine(50,0)(50,50)
\ArrowLine(50,50)(50,100)
\ArrowLine(50,100)(100,100)
\DashLine(50,50)(100,50){5}
\Vertex(50,100){2}
\Vertex(50,50){2}
\Vertex(50,0){2}
\put(85,35){$\PH$}
\put(-12,78){$\Pg$}
\put(-12,-2){$\Pg$}
\put(85,78){$\PQt$}
\put(85,-2){$\PAQt$}
\end{picture}
\end{center}
\caption[]{Examples of LO Feynman diagrams for the
partonic processes $\PQq\PAQq,\Pg\Pg\to\PQt\PAQt\PH$.}
\label{fg:lodiatth} 
\end{figure}

The full next-to-leading-order (NLO) QCD corrections to $\PQt\PAQt\PH$
production have been calculated
\cite{Beenakker:2001rj,Beenakker:2002nc,Reina:2001sf,Dawson:2002tg}
resulting in a moderate increase of the total cross section at the LHC
by at most $\sim 20\%$, depending on the value of $\MH$ and on the PDF
set used. Indeed, when using CTEQ6.6 the NLO corrections are always
positive and the $K$-factor varies between $1.14$ and $1.22$ for
$\MH=90,\ldots,300\UGeV$, while when using MSTW2008 the impact of NLO
corrections is much less uniform: NLO corrections can either increase
or decrease the LO cross section by a few percents and result in 
$K$-factors between $1.05$ and $0.98$ for $\MH=90,\ldots,300\UGeV$.  

The residual scale dependence has decreased from ${\cal O}(50\%)$ to a
level of ${\cal O}(10\%)$ at NLO, if the renormalization and
factorization scales are varied by a factor $2$ up- and downwards around
the central scale choice, thus signalling a significant improvement of
the theoretical prediction at NLO. The full NLO results confirm former
estimates based on an effective-Higgs approximation \cite{Dawson:1997im}
which approximates Higgs radiation as a fragmentation process in the
high-energy limit. The NLO effects on the relevant parts of final-state
particle distribution shapes are of moderate size, i.e.~${\cal
O}(10\%)$, so that former experimental analyses are not expected to
change much due to these results. There is no public NLO code for the
signal process available yet.

\subsection{Background processes}
Recently the NLO QCD corrections to the $\PQt\PAQt\PQb\PAQb$ production
background have been calculated
\cite{Bredenstein:2009aj,Bredenstein:2008zb,Bredenstein:2010rs,
Bevilacqua:2009zn,Binoth:2010ra}. By choosing
$\mu_R^2=\mu_F^2=\Mt\sqrt{p_{T\PQb}p_{T\PAQb}}$ as the central
renormalization and factorization scales the NLO corrections increase
the background cross section within the signal region by about $20$--$30\%$.
The scale dependence is significantly reduced to a level significantly
below $30\%$. The new predictions for the NLO QCD cross sections with the
new scale choice $\mu_R^2=\mu_F^2=\Mt\sqrt{p_{T\PQb}p_{T\PAQb}}$ are
larger than the old LO predictions with the old scale choice
$\mu_R=\mu_F=\Mt+m_{\PQb\PAQb}/2$ by more than $100\%$ within the typical
experimental cuts \cite{Bredenstein:2010rs}. In addition the signal
process $\Pp\Pp\to \PQt\PAQt\PH\to \PQt\PAQt\PQb\PAQb$ has been added to
these background calculations in the narrow-width approximation
\cite{Binoth:2010ra}.  This makes it possible to study the signal and
background processes including the final-state Higgs decay into
$\PQb\PAQb$ with cuts at the same time at NLO. However, it should be
noted that the final-state top decays have not been included at NLO so
that a full NLO signal and background analysis including all
experimental cuts is not possible yet. The top-quark decays are
expected to affect the final-state distributions more than the Higgs
decays into $\PQb\PAQb$ pairs. For highly boosted Higgs bosons the
shapes of the background distributions are affected by the QCD
corrections which thus have to be taken into account properly. The
effects of a jet veto for the boosted-Higgs regime require further
detailed investigations. Very recently the NLO QCD corrections to
$\PQt\PAQt jj$ production have been calculated \cite{Bevilacqua:2010ve}.
However, a full numerical analysis of these results has not been
performed so far.  As it is the case for the signal process, there is no
public code available for the NLO calculations of the background
processes $\Pp\Pp\to\PQt\PAQt\PQb\PAQb, \PQt\PAQt jj$.

\subsection{Numerical analysis and results}
In the following we provide results for the inclusive NLO signal
cross section for different values of Higgs masses. The central scale
has been chosen as $\mu_R=\mu_F=\mu_0=\Mt+\MH/2$. In addition, the
uncertainties due to scale variations of a factor of two around the
central scale $\mu_0$ as well as the 68\% CL uncertainties due to the
PDFs and the strong coupling $\alphas$ are given explicitly.  
In this study we used the on-shell top-quark mass and did not include
the parametric uncertainties due to the experimental error on the
top-quark mass. Loop diagrams with a bottom-quark loop were calculated
using the $\Pb$-quark pole mass. The top-quark Yukawa coupling was defined
in terms of the top pole mass. The values for the top and bottom masses
are chosen according to the parameters given in Appendix~\ref{sminput}.  
We have used the MSTW2008
\cite{Martin:2009iq,Martin:2009bu}, CTEQ6.6 \cite{Pumplin:2002vw}, and
NNPDF2.0 \cite{Ball:2010de} sets of parton density functions.  The
central values of the strong coupling constant have been implemented
according to the corresponding PDFs for the sake of consistency.
In \refT{tb:siglo} we show the LO cross sections for the signal
process and their respective scale and PDF uncertainties calculated with
MSTW2008 PDFs. For comparison we also show the central LO cross sections
obtained with CTEQ6L1 PDFs. It is remarkable that the numbers using the
LO PDFs of MSTW2008 and CTEQ6L1 differ by about $20\%$. The scale
uncertainties at LO are typically of the order of $30{-}40\%$, while the
PDF uncertainties amount to about $2{-}3$\%.

\begin{table}
  \begin{center}
  \caption{\label{tb:siglo} LO cross sections of
$\Pp\Pp\to\PQt\PAQt\PH$ for $\sqrt{s}=7$\UTeV\ using MSTW2008 and CTEQ6.6
PDFs. The scale dependence is given for the scale variation $\mu_0/2 <
\mu_R,\mu_F < 2\mu_0$ with $\mu_0=\Mt+\MH/2$. The PDF uncertainties are
defined at 68\% CL using MSTW2008.
}
  \small
  \begin{tabular}{ccccc} \hline
$\MH$ [\UGeVZ] & LO [\UfbZ], MSTW2008 & LO [\UfbZ], CTEQ6L1 & Scale
  [\%]& PDF[\%] \\ \hline
$90  $&$ 213.2  $&$ 174.2  $&$ +40.0 -\!26.3  $&$ +2.5 -\!2.6  $\\
$100 $&$ 162.7  $&$ 133.0  $&$ +39.9 -\!26.3  $&$ +2.5 -\!2.6  $\\
$110 $&$ 126.1  $&$ 102.8  $&$ +39.9 -\!26.2  $&$ +2.5 -\!2.5  $\\
$120 $&$ 98.66  $&$ 80.43  $&$ +39.8 -\!26.2  $&$ +2.5 -\!2.6  $\\
$130 $&$ 78.09  $&$ 63.62  $&$ +39.8 -\!26.2  $&$ +2.5 -\!2.5  $\\
$140 $&$ 62.43  $&$ 50.79  $&$ +39.9 -\!26.2  $&$ +2.5 -\!2.6  $\\
$150 $&$ 50.35  $&$ 40.94  $&$ +39.8 -\!26.2  $&$ +2.6 -\!2.6  $\\
$160 $&$ 40.98  $&$ 33.29  $&$ +39.8 -\!26.2  $&$ +2.6 -\!2.6  $\\
$170 $&$ 33.62  $&$ 27.30  $&$ +39.8 -\!26.2  $&$ +2.6 -\!2.6  $\\
$180 $&$ 27.83  $&$ 22.57  $&$ +39.8 -\!26.2  $&$ +2.6 -\!2.6  $\\
$190 $&$ 23.20  $&$ 18.80  $&$ +39.8 -\!26.2  $&$ +2.7 -\!2.6  $\\
$200 $&$ 19.48  $&$ 15.78  $&$ +39.9 -\!26.2  $&$ +2.7 -\!2.7  $\\ \hline

  \end{tabular}
  \end{center}
\end{table}

In \refT{tb:mstw} the NLO signal cross section is listed
including the scale, $\alphas$, and PDF uncertainties at 68\% CL for
MSTW2008 PDFs. It should be noted that the LO and NLO cross sections are
very similar so that the $K$-factor is about unity for the central scale
choice with MSTW2008 PDFs. The scale uncertainties amount to $5{-}10\%$ at
NLO typically, while the PDF uncertainties range at the level of $3{-}5\%$.
The uncertainties induced by the strong coupling $\alphas$ turn out to
be of ${\cal O}(2{-}3\%)$ for MSTW2008 PDFs, while the combined
PDF+$\alphas$ errors range at the level of $4{-}6\%$. In 
\refT{tb:cteq} we show the corresponding NLO numbers for the CTEQ6.6 PDFs
and in \refT{tb:nnpdf} for the NNPDF2.0 parton densities. The
difference of about $20\%$ between MSTW2008 and CTEQ6L1 at LO reduces to a
level of $7{-}8\%$ at NLO between MSTW2008 and CTEQ6.6. The PDF and
$\alphas$ uncertainties are larger with CTEQ6.6 PDFs than with
MSTW2008. For the NNPDF2.0 sets we obtain the smallest $\alphas$
uncertainties. The PDF uncertainties are comparable to MSTW2008.

\begin{table}[p]
  \begin{center}
  \caption{\label{tb:mstw} LO and NLO QCD cross sections of $\Pp\Pp\to\PQt
\PAQt\PH$ for $\sqrt{s}=7$\UTeV\ using MSTW2008 PDFs. The scale dependence
is given for the scale variation $\mu_0/2 < \mu_R,\mu_F < 2\mu_0$ with
$\mu_0=\Mt+\MH/2$. The $\alphas$ and PDF uncertainties are
defined at $68\%$ CL. The last column contains the combined
PDF+$\alphas$ uncertainties obtained with combined PDF sets.
  }
  \small
  \begin{tabular}{ccccccc} \hline
\!$\MH$[\UGeVZ]\!\! & \!LO[\UfbZ]\!\! & \!\!NLO QCD[\UfbZ]\!\! &
 Scale [\%] & $\alphas$ [\%] & PDF [\%] & PDF+$\alphas$ [\%] \\ \hline
$ 90 $&$ 213.2  $&$ 224.8 $&$ +4.1  -\!9.7  $&$ +2.2  -\!2.7 $&$ +2.9  -\!3.4 $&$ +4.2  -\!3.9 $\\
$ 95 $&$ 186.1  $&$ 195.6 $&$ +4.0  -\!9.6  $&$ +2.2  -\!2.7 $&$ +2.9  -\!3.4 $&$ +4.3  -\!3.9 $\\
$100 $&$ 162.7  $&$ 170.4 $&$ +3.9  -\!9.6  $&$ +2.2  -\!2.7 $&$ +2.9  -\!3.4 $&$ +4.3  -\!3.9 $\\
$105 $&$ 143.1  $&$ 149.0 $&$ +3.7  -\!9.5  $&$ +2.2  -\!2.7 $&$ +2.9  -\!3.4 $&$ +4.3  -\!3.9 $\\
$110 $&$ 126.1  $&$ 130.8 $&$ +3.6  -\!9.5  $&$ +2.2  -\!2.7 $&$ +2.9  -\!3.4 $&$ +4.3  -\!3.9 $\\
$115 $&$ 111.4  $&$ 115.0 $&$ +3.5  -\!9.4  $&$ +2.2  -\!2.7 $&$ +3.0  -\!3.4 $&$ +4.3  -\!3.9 $\\
$120 $&$  98.66 $&$ 101.4 $&$ +3.4  -\!9.4  $&$ +2.2  -\!2.7 $&$ +3.0  -\!3.4 $&$ +4.3  -\!3.9 $\\
$125 $&$  87.66 $&$  89.8 $&$ +3.3  -\!9.3  $&$ +2.2  -\!2.7 $&$ +3.0  -\!3.4 $&$ +4.3  -\!3.9 $\\
$130 $&$  78.09 $&$ 79.57 $&$ +3.2  -\!9.3  $&$ +2.2  -\!2.7 $&$ +3.0  -\!3.3 $&$ +4.3  -\!3.9 $\\
$135 $&$  69.71 $&$ 70.75 $&$ +3.1  -\!9.2  $&$ +2.2  -\!2.7 $&$ +3.0  -\!3.4 $&$ +4.3  -\!3.9 $\\
$140 $&$  62.43 $&$ 63.06 $&$ +3.0  -\!9.2  $&$ +2.2  -\!2.7 $&$ +3.0  -\!3.4 $&$ +4.4  -\!3.9 $\\
$145 $&$  55.96 $&$ 56.50 $&$ +2.9  -\!9.1  $&$ +2.2  -\!2.7 $&$ +3.1  -\!3.4 $&$ +4.4  -\!3.9 $\\
$150 $&$  50.35 $&$ 50.59 $&$ +2.9  -\!9.1  $&$ +2.2  -\!2.7 $&$ +3.1  -\!3.4 $&$ +4.4  -\!3.9 $\\
$155 $&$  45.37 $&$ 45.49 $&$ +2.8  -\!9.1  $&$ +2.2  -\!2.7 $&$ +3.1  -\!3.4 $&$ +4.4  -\!3.9 $\\
$160 $&$  40.98 $&$ 41.01 $&$ +2.8  -\!9.1  $&$ +2.2  -\!2.7 $&$ +3.1  -\!3.4 $&$ +4.4  -\!3.9 $\\
$165 $&$  37.09 $&$ 36.99 $&$ +2.7  -\!9.1  $&$ +2.2  -\!2.6 $&$ +3.2  -\!3.4 $&$ +4.5  -\!3.9 $\\
$170 $&$  33.62 $&$ 33.47 $&$ +2.7  -\!9.0  $&$ +2.2  -\!2.6 $&$ +3.2  -\!3.4 $&$ +4.5  -\!3.9 $\\
$175 $&$  30.56 $&$ 30.31 $&$ +2.6  -\!9.0  $&$ +2.2  -\!2.6 $&$ +3.2  -\!3.4 $&$ +4.5  -\!3.9 $\\
$180 $&$  27.83 $&$ 27.55 $&$ +2.6  -\!9.0  $&$ +2.2  -\!2.7 $&$ +3.2  -\!3.4 $&$ +4.6  -\!4.0 $\\
$185 $&$  25.38 $&$ 25.09 $&$ +2.6  -\!9.0  $&$ +2.2  -\!2.7 $&$ +3.3  -\!3.5 $&$ +4.6  -\!4.0 $\\
$190 $&$  23.20 $&$ 22.93 $&$ +2.6  -\!9.0  $&$ +2.2  -\!2.7 $&$ +3.3  -\!3.5 $&$ +4.6  -\!4.0 $\\
$195 $&$  21.25 $&$ 20.94 $&$ +2.6  -\!9.0  $&$ +2.2  -\!2.7 $&$ +3.4  -\!3.5 $&$ +4.7  -\!4.0 $\\
$200 $&$  19.48 $&$ 19.20 $&$ +2.6  -\!9.1  $&$ +2.2  -\!2.7 $&$ +3.4  -\!3.6 $&$ +4.7  -\!4.1 $\\
$210 $&$  16.49 $&$ 16.23 $&$ +2.8  -\!9.2  $&$ +2.2  -\!2.7 $&$ +3.5  -\!3.7 $&$ +4.8  -\!4.1 $\\
$220 $&$  14.04 $&$ 13.81 $&$ +2.9  -\!9.3  $&$ +2.2  -\!2.7 $&$ +3.6  -\!3.7 $&$ +4.9  -\!4.2 $\\
$230 $&$  12.04 $&$ 11.86 $&$ +3.2  -\!9.4  $&$ +2.3  -\!2.7 $&$ +3.7  -\!3.9 $&$ +5.0  -\!4.3 $\\
$240 $&$  10.38 $&$ 10.24 $&$ +3.2  -\!9.5  $&$ +2.3  -\!2.7 $&$ +3.8  -\!4.0 $&$ +5.2  -\!4.4 $\\
$250 $&$  9.011 $&$ 8.899 $&$ +3.5  -\!9.7  $&$ +2.3  -\!2.7 $&$ +4.0  -\!4.1 $&$ +5.3  -\!4.5 $\\
$260 $&$  7.850 $&$ 7.777 $&$ +3.9  -\!9.9  $&$ +2.3  -\!2.8 $&$ +4.1  -\!4.3 $&$ +5.5  -\!4.6 $\\
$270 $&$  6.888 $&$ 6.866 $&$ +4.3  -\!10.1 $&$ +2.4  -\!2.8 $&$ +4.2  -\!4.4 $&$ +5.6  -\!4.7 $\\
$280 $&$  6.075 $&$ 6.092 $&$ +4.7  -\!10.4 $&$ +2.4  -\!2.8 $&$ +4.4  -\!4.6 $&$ +5.8  -\!4.9 $\\
$290 $&$  5.376 $&$ 5.405 $&$ +5.2  -\!10.6 $&$ +2.4  -\!2.8 $&$ +4.6  -\!4.7 $&$ +6.0  -\!5.0 $\\
$300 $&$  4.780 $&$ 4.848 $&$ +5.6  -\!10.9 $&$ +2.5  -\!2.9 $&$ +4.7  -\!4.9 $&$ +6.2  -\!5.2 $\\ \hline
  \end{tabular}
   \end{center}
\end{table}

\begin{table}[p]
  \begin{center}
    \caption{\label{tb:cteq} LO and NLO QCD cross sections of $\Pp\Pp\to\PQt
\PAQt\PH$ for $\sqrt{s}=7$\UTeV\ using CTEQ6.6 PDFs. The scale dependence
is given for the scale variation $\mu_0/2 < \mu_R,\mu_F < 2\mu_0$ with
$\mu_0=\Mt+\MH/2$. The $\alphas$ and PDF uncertainties are
defined at 68\% CL.
    }
    \small
    \begin{tabular}{cccccc} \hline
$\MH$ [\UGeVZ] & LO [\UfbZ] & NLO QCD [\UfbZ] & Scale [\%] &
  $\alphas$ [\%]& PDF [\%]
\\ \hline
$ 90 $&$ 174.2   $&$ 210.0  $&$  +4.2  -\!9.4 $&$ +3.5  -\!2.5 $&$  +5.9  -\!5.1 $\\
$ 95 $&$ 151.9   $&$ 182.5 $&$  +4.1  -\!9.4 $&$ +3.5  -\!2.5 $&$  +5.9  -\!5.1 $\\
$100 $&$ 133.0   $&$ 159.1 $&$  +4.0  -\!9.3 $&$ +3.5  -\!2.5 $&$  +6.0  -\!5.1 $\\
$105 $&$ 116.7   $&$ 139.3 $&$  +3.8  -\!9.2 $&$ +3.5  -\!2.5 $&$  +6.1  -\!5.2 $\\
$110 $&$ 102.8   $&$ 122.1 $&$  +3.7  -\!9.2 $&$ +3.6  -\!2.5 $&$  +6.1  -\!5.2 $\\
$115 $&$  90.81  $&$ 107.5 $&$  +3.6  -\!9.2 $&$ +3.5  -\!2.5 $&$  +6.2  -\!5.2 $\\
$120 $&$  80.43  $&$  94.91 $&$  +3.5  -\!9.1 $&$ +3.5  -\!2.6 $&$  +6.2  -\!5.3 $\\
$125 $&$  71.44  $&$  83.94 $&$  +3.5  -\!9.1 $&$ +3.6  -\!2.5 $&$  +6.3  -\!5.3 $\\
$130 $&$  63.62  $&$  74.54 $&$  +3.4  -\!9.0 $&$ +3.6  -\!2.5 $&$  +6.4  -\!5.3 $\\
$135 $&$  56.77  $&$  66.32 $&$  +3.3  -\!9.0 $&$ +3.6  -\!2.5 $&$  +6.4  -\!5.4 $\\
$140 $&$  50.79  $&$  59.16 $&$  +3.2  -\!9.0 $&$ +3.6  -\!2.5 $&$  +6.5  -\!5.4 $\\
$145 $&$  45.55  $&$  52.92 $&$  +3.2  -\!8.9 $&$ +3.6  -\!2.5 $&$  +6.6  -\!5.5 $\\
$150 $&$  40.94  $&$  47.45 $&$  +3.1  -\!8.9 $&$ +3.6  -\!2.5 $&$  +6.6  -\!5.5 $\\
$155 $&$  36.88  $&$  42.60 $&$  +3.1  -\!8.9 $&$ +3.6  -\!2.5 $&$  +6.7  -\!5.6 $\\
$160 $&$  33.29  $&$  38.38 $&$  +3.0  -\!8.9 $&$ +3.6  -\!2.6 $&$  +6.8  -\!5.6 $\\
$165 $&$  30.12  $&$  34.68 $&$  +3.0  -\!8.9 $&$ +3.7  -\!2.6 $&$  +6.9  -\!5.7 $\\
$170 $&$  27.30  $&$  31.38 $&$  +3.0  -\!8.9 $&$ +3.7  -\!2.6 $&$  +7.0  -\!5.7 $\\
$175 $&$  24.81  $&$  28.47 $&$  +3.0  -\!8.9 $&$ +3.7  -\!2.6 $&$  +7.0  -\!5.8 $\\
$180 $&$  22.57  $&$  25.88 $&$  +3.0  -\!8.9 $&$ +3.7  -\!2.6 $&$  +7.1  -\!5.8 $\\
$185 $&$  20.58  $&$  23.56 $&$  +3.0  -\!8.9 $&$ +3.7  -\!2.6 $&$  +7.2  -\!5.9 $\\
$190 $&$  18.80  $&$  21.52 $&$  +3.0  -\!8.9 $&$ +3.8  -\!2.6 $&$  +7.3  -\!6.0 $\\
$195 $&$  17.20  $&$  19.70 $&$  +3.0  -\!8.9 $&$ +3.8  -\!2.6 $&$  +7.4  -\!6.0 $\\
$200 $&$  15.78  $&$  18.06 $&$  +3.1  -\!9.0 $&$ +3.8  -\!2.6 $&$  +7.5  -\!6.1 $\\
$210 $&$  13.33  $&$  15.27 $&$  +3.2  -\!9.1 $&$ +3.9  -\!2.6 $&$  +7.8  -\!6.3 $\\
$220 $&$  11.32  $&$  13.02 $&$  +3.3  -\!9.1 $&$ +3.9  -\!2.6 $&$  +8.0  -\!6.4 $\\
$230 $&$   9.696 $&$  11.20 $&$  +3.5  -\!9.3 $&$ +4.0  -\!2.7 $&$  +8.3  -\!6.6 $\\
$240 $&$   8.344 $&$   9.685 $&$ +3.6  -\!9.4 $&$ +4.1  -\!2.7 $&$  +8.5  -\!6.8 $\\
$250 $&$   7.227 $&$   8.450 $&$ +3.9  -\!9.6 $&$ +4.2  -\!2.7 $&$  +8.8  -\!7.0 $\\
$260 $&$   6.286 $&$   7.418 $&$ +4.1  -\!9.7 $&$ +4.3  -\!2.8 $&$  +9.1  -\!7.2 $\\
$270 $&$   5.501 $&$   6.541 $&$ +4.4  -\!9.9 $&$ +4.4  -\!2.8 $&$  +9.5  -\!7.4 $\\
$280 $&$   4.837 $&$   5.809 $&$ +4.6  -\!10.1 $&$ +4.5  -\!2.9 $&$  +9.8  -\!7.7 $\\
$290 $&$   4.267 $&$   5.186 $&$ +4.9  -\!10.3 $&$ +4.6  -\!2.9 $&$ +10.1  -\!7.9 $\\
$300 $&$   3.785 $&$   4.653 $&$ +5.2  -\!10.5 $&$ +4.7  -\!3.0 $&$ +10.5  -\!8.2 $\\ \hline
    \end{tabular}
    \end{center}
\end{table}

\begin{table}[p]
  \begin{center}
 \caption{\label{tb:nnpdf} NLO QCD cross sections of
$\Pp\Pp\to\PQt\PAQt\PH$ for $\sqrt{s}=7$\UTeV\ using NNPDF2.0 PDFs. The scale
dependence is given for the scale variation $\mu_0/2 < \mu_R,\mu_F <
2\mu_0$ with $\mu_0=\Mt+\MH/2$. The $\alphas$ and PDF
uncertainties are defined at 68\% CL.}
 \small
 \begin{tabular}{cccccc} \hline
$\MH$ [\UGeVZ] & NLO QCD [\UfbZ] & Scale [\%] & $\alphas$ [\%]& PDF [\%]
\\ \hline
$ 90 $&$ 221.3    $&$ +4.8  -\!10.7 $&$ +1.6  -\!2.3 $&$ \pm 4.1 $\\
$ 95 $&$ 192.0    $&$ +4.7  -\!10.6 $&$ +1.6  -\!2.3 $&$ \pm 4.1 $\\
$100 $&$ 167.1    $&$ +4.5  -\!10.6 $&$ +1.6  -\!2.2 $&$ \pm 4.1 $\\
$105 $&$ 145.9    $&$ +4.4  -\!10.5 $&$ +1.6  -\!2.2 $&$ \pm 4.1 $\\
$110 $&$ 127.8    $&$ +4.3  -\!10.4 $&$ +1.6  -\!2.2 $&$ \pm 4.2 $\\
$115 $&$ 112.3    $&$ +4.2  -\!10.4 $&$ +1.6  -\!2.2 $&$ \pm 4.2 $\\
$120 $&$ 99.01    $&$ +4.1  -\!10.3 $&$ +1.6  -\!2.2 $&$ \pm 4.2 $\\
$125 $&$ 87.50    $&$ +4.1  -\!10.2 $&$ +1.6  -\!2.2 $&$ \pm 4.2 $\\
$130 $&$ 77.54    $&$ +4.0  -\!10.2 $&$ +1.6  -\!2.2 $&$ \pm 4.2 $\\
$135 $&$ 68.89    $&$ +3.9  -\!10.1 $&$ +1.6  -\!2.1 $&$ \pm 4.2 $\\
$140 $&$ 61.37    $&$ +3.8  -\!10.1 $&$ +1.6  -\!2.1 $&$ \pm 4.3 $\\
$145 $&$ 54.81    $&$ +3.8  -\!10.0 $&$ +1.6  -\!2.1 $&$ \pm 4.3 $\\
$150 $&$ 49.07    $&$ +3.7  -\!10.0 $&$ +1.6  -\!2.1 $&$ \pm 4.3 $\\
$155 $&$ 44.03    $&$ +3.7  -\!9.9  $&$ +1.6  -\!2.1 $&$ \pm 4.3 $\\
$160 $&$ 39.61    $&$ +3.6  -\!9.9  $&$ +1.6  -\!2.1 $&$ \pm 4.4 $\\
$165 $&$ 35.72    $&$ +3.6  -\!9.9  $&$ +1.6  -\!2.1 $&$ \pm 4.4 $\\
$170 $&$ 32.28    $&$ +3.6  -\!9.9  $&$ +1.6  -\!2.1 $&$ \pm 4.5 $\\
$175 $&$ 29.24    $&$ +3.6  -\!9.9  $&$ +1.6  -\!2.1 $&$ \pm 4.5 $\\
$180 $&$ 26.55    $&$ +3.6  -\!9.8  $&$ +1.6  -\!2.1 $&$ \pm 4.5 $\\
$185 $&$ 24.16    $&$ +3.6  -\!9.8  $&$ +1.6  -\!2.1 $&$ \pm 4.6 $\\
$190 $&$ 22.03    $&$ +3.6  -\!9.9  $&$ +1.6  -\!2.0 $&$ \pm 4.6 $\\
$195 $&$ 20.13    $&$ +3.6  -\!9.9  $&$ +1.6  -\!2.0 $&$ \pm 4.7 $\\
$200 $&$ 18.44    $&$ +3.7  -\!9.9  $&$ +1.6  -\!2.0 $&$ \pm 4.7 $\\
$210 $&$ 15.56    $&$ +3.8  -\!10.0 $&$ +1.6  -\!2.0 $&$ \pm 4.9 $\\
$220 $&$ 13.24    $&$ +3.9  -\!10.0 $&$ +1.6  -\!2.0 $&$ \pm 5.0 $\\
$230 $&$ 11.35    $&$ +4.1  -\!10.1 $&$ +1.6  -\!2.0 $&$ \pm 5.2 $\\
$240 $&$ 9.805    $&$ +4.3  -\!10.3 $&$ +1.6  -\!2.0 $&$ \pm 5.3 $\\
$250 $&$ 8.527    $&$ +4.5  -\!10.4 $&$ +1.6  -\!2.0 $&$ \pm 5.5 $\\
$260 $&$ 7.465    $&$ +4.8  -\!10.6 $&$ +1.6  -\!2.1 $&$ \pm 5.7 $\\
$270 $&$ 6.575    $&$ +5.1  -\!10.7 $&$ +1.6  -\!2.1 $&$ \pm 5.9 $\\
$280 $&$ 5.824    $&$ +5.4  -\!10.9 $&$ +1.6  -\!2.1 $&$ \pm 6.1 $\\
$290 $&$ 5.187    $&$ +5.7  -\!11.1 $&$ +1.5  -\!2.1 $&$ \pm 6.4 $\\
$300 $&$ 4.642    $&$ +6.0  -\!11.3 $&$ +1.5  -\!2.1 $&$ \pm 6.6 $\\
\hline
 \end{tabular}
 \end{center}
\end{table}

\refTs{tb:sig7} and \ref{tb:sig14} contain our final results for
$\sqrt{s}=7\UTeV$ and $14\UTeV$, respectively. We exhibit the central
values and the PDF+$\alphas$ uncertainties according to the envelope
method of the PDF4LHC recommendation and the relative scale variations
using MSTW2008 PDFs (see \refT{tb:mstw} for $\sqrt{s}=7\UTeV$). The
last column displays the total uncertainties by adding the final errors
linearly. The cross sections for $\sqrt{s}=14\UTeV$ are $7{-}10$ times
larger than the corresponding values for $\sqrt{s}=7\UTeV$. The total
uncertainties amount to typically $10{-}15\%$ apart from Higgs masses
beyond 200\UGeV\ where they are slightly larger.

In \Fref{fg:cxn}a we show the LO and NLO QCD cross sections for
$\sqrt{s}=7\UTeV$ for the MSTW2008, CTEQ6.6, and NNPDF2.0 PDF sets
individually. It is clearly visible that the LO and NLO cross sections
nearly coincide for the central scale choice with MSTW2008 PDFs, while
there are corrections of ${\cal O}(20\%)$ with CTEQ6.6 PDFs. At NLO all
three PDF sets yield consistent values within less than $10\%$.

The final total cross sections for $\Pp\Pp\to \PQt\PAQt\PH + X$ are
shown in \Fref{fg:cxn}b for both energies $\sqrt{s}=7,14\UTeV$.
The error bands include the total uncertainties according to the PDF4LHC
recommendation as given in \refTs{tb:sig7} and \ref{tb:sig14}.

\begin{figure}[h]
   \begin{center}
     \begin{tabular}{cc}
       \includegraphics[
angle=0,width=.46\linewidth]{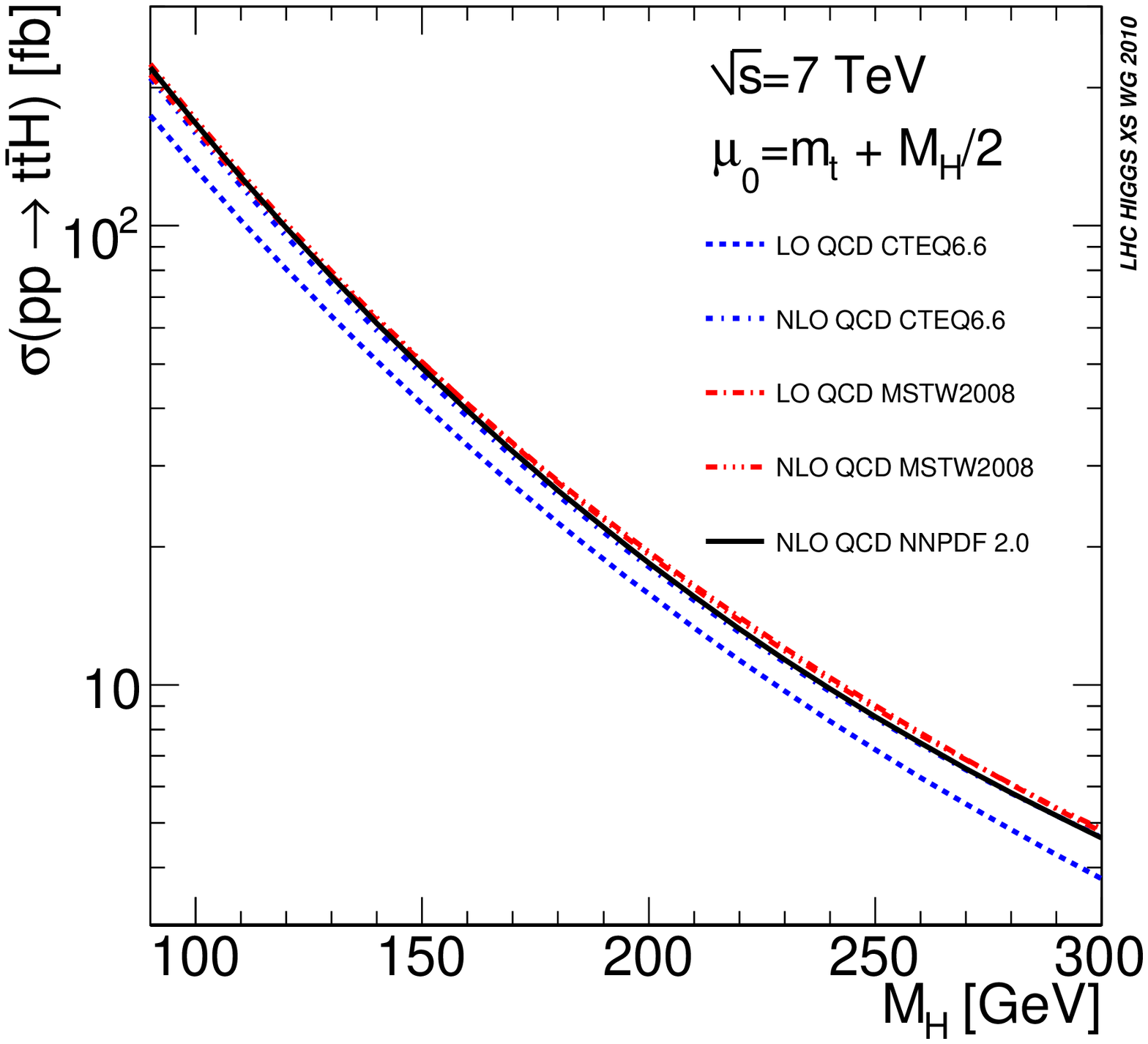} &
       \includegraphics[
angle=0,width=.46\linewidth]{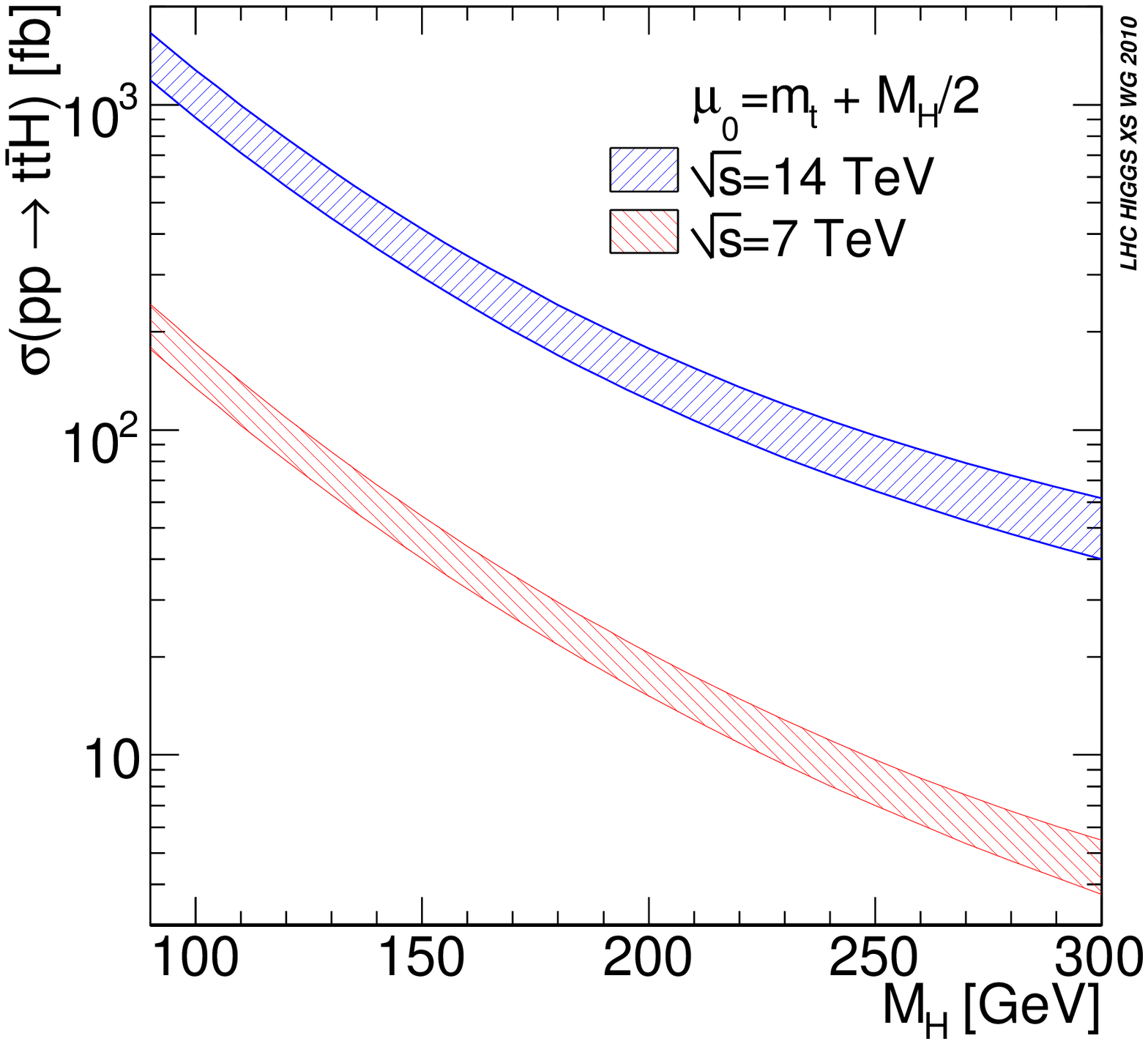}\\[-1.5em]
(a) & (b)
     \end{tabular}
   \end{center}
   \caption{\label{fg:cxn} (a) Total production cross sections of
$\Pp\Pp\to\PQt \PAQt\PH + X$ for $\sqrt{s}=7$\UTeV\ at LO and NLO QCD
for the different sets of PDFs. (b) Total production cross sections of
$\Pp\Pp\to\PQt \PAQt\PH + X$ for $\sqrt{s}=7\UTeV$ and $14\UTeV$ at NLO
QCD including the total uncertainties according to the PDF4LHC
recommendation.}
\end{figure}

\begin{table}
  \begin{center}
  \caption{\label{tb:sig7} NLO QCD cross sections of
$\Pp\Pp\to\PQt\PAQt\PH$ for $\sqrt{s}=7$\UTeV\ obtained according to the
envelope method of the PDF4LHC group. }
  \small
  \begin{tabular}{cccc} \hline
$\MH$ [\UGeVZ] & NLO QCD [\UfbZ] & Scale [\%] & PDF4LHC [\%]\\ \hline
$ 90 $&$ 216.2 $&$ +4.1  -\!9.7  $&$ \pm 8.4  $\\
$ 95 $&$ 188.0 $&$ +4.0  -\!9.6  $&$ \pm 8.4  $\\
$100 $&$ 163.8 $&$ +3.9  -\!9.6  $&$ \pm 8.4  $\\
$105 $&$ 143.3 $&$ +3.7  -\!9.5  $&$ \pm 8.4  $\\
$110 $&$ 125.7 $&$ +3.6  -\!9.5  $&$ \pm 8.5  $\\
$115 $&$ 110.6 $&$ +3.5  -\!9.4  $&$ \pm 8.4  $\\
$120 $&$ 97.56 $&$ +3.4  -\!9.4  $&$ \pm 8.4  $\\
$125 $&$ 86.34 $&$ +3.3  -\!9.3  $&$ \pm 8.5  $\\
$130 $&$ 76.58 $&$ +3.2  -\!9.3  $&$ \pm 8.4  $\\
$135 $&$ 68.10 $&$ +3.1  -\!9.2  $&$ \pm 8.4  $\\
$140 $&$ 60.72 $&$ +3.0  -\!9.2  $&$ \pm 8.4  $\\
$145 $&$ 54.35 $&$ +2.9  -\!9.1  $&$ \pm 8.5  $\\
$150 $&$ 48.69 $&$ +2.9  -\!9.1  $&$ \pm 8.4  $\\
$155 $&$ 43.74 $&$ +2.8  -\!9.1  $&$ \pm 8.6  $\\
$160 $&$ 39.42 $&$ +2.8  -\!9.1  $&$ \pm 8.6  $\\
$165 $&$ 35.59 $&$ +2.7  -\!9.1  $&$ \pm 8.6  $\\
$170 $&$ 32.19 $&$ +2.7  -\!9.0  $&$ \pm 8.6  $\\
$175 $&$ 29.18 $&$ +2.6  -\!9.0  $&$ \pm 8.6  $\\
$180 $&$ 26.52 $&$ +2.6  -\!9.0  $&$ \pm 8.6  $\\
$185 $&$ 24.14 $&$ +2.6  -\!9.0  $&$ \pm 8.7  $\\
$190 $&$ 22.06 $&$ +2.6  -\!9.0  $&$ \pm 8.7  $\\
$195 $&$ 20.16 $&$ +2.6  -\!9.0  $&$ \pm 8.7  $\\
$200 $&$ 18.49 $&$ +2.6  -\!9.1  $&$ \pm 8.7  $\\
$210 $&$ 15.62 $&$ +2.8  -\!9.2  $&$ \pm 8.9  $\\
$220 $&$ 13.30 $&$ +2.9  -\!9.3  $&$ \pm 8.9  $\\
$230 $&$ 11.43 $&$ +3.2  -\!9.4  $&$ \pm 9.0  $\\
$240 $&$ 9.873 $&$ +3.2  -\!9.5  $&$ \pm 9.1  $\\
$250 $&$ 8.593 $&$ +3.5  -\!9.7  $&$ \pm 9.1  $\\
$260 $&$ 7.524 $&$ +3.9  -\!9.9  $&$ \pm 9.0  $\\
$270 $&$ 6.636 $&$ +4.3  -\!10.1 $&$ \pm 9.3  $\\
$280 $&$ 5.889 $&$ +4.7  -\!10.4 $&$ \pm 9.5  $\\
$290 $&$ 5.256 $&$ +5.2  -\!10.6 $&$ \pm 9.7  $\\
$300 $&$ 4.719 $&$ +5.6  -\!10.9 $&$ \pm 10.0 $\\ \hline
   \end{tabular}
   \end{center}
\end{table}
  
\begin{table}
  \begin{center}
  \caption{\label{tb:sig14} NLO QCD cross sections of
$\Pp\Pp\to\PQt\PAQt\PH$ for $\sqrt{s}=14$\UTeV\ obtained according to the
envelope method of the PDF4LHC goup. }
  \small
  \begin{tabular}{cccc} \hline
$\MH$ [\UGeVZ] & NLO QCD [\UfbZ] & Scale [\%]& PDF4LHC [\%]  \\ \hline
$ 90 $&$ 1449  $&$ +6.2  -\!9.3  $&$ \pm 8.7  $\\
$ 95 $&$ 1268  $&$ +6.1  -\!9.3  $&$ \pm 8.7  $\\
$100 $&$ 1114  $&$ +6.1  -\!9.3  $&$ \pm 8.7  $\\
$105 $&$ 981.6 $&$ +6.0  -\!9.3  $&$ \pm 8.7  $\\
$110 $&$ 868.1 $&$ +6.0  -\!9.3  $&$ \pm 8.8  $\\
$115 $&$ 769.9 $&$ +6.0  -\!9.3  $&$ \pm 8.8  $\\
$120 $&$ 685.0 $&$ +5.9  -\!9.3  $&$ \pm 8.8  $\\
$125 $&$ 611.3 $&$ +5.9  -\!9.3  $&$ \pm 8.9  $\\
$130 $&$ 547.2 $&$ +5.9  -\!9.3  $&$ \pm 8.9  $\\
$135 $&$ 491.0 $&$ +5.9  -\!9.3  $&$ \pm 8.9  $\\
$140 $&$ 441.9 $&$ +5.9  -\!9.3  $&$ \pm 8.9  $\\
$145 $&$ 398.9 $&$ +5.9  -\!9.3  $&$ \pm 9.0  $\\
$150 $&$ 360.9 $&$ +5.9  -\!9.3  $&$ \pm 9.0  $\\
$155 $&$ 327.5 $&$ +5.9  -\!9.4  $&$ \pm 9.0  $\\
$160 $&$ 298.0 $&$ +5.9  -\!9.4  $&$ \pm 9.1  $\\
$165 $&$ 271.8 $&$ +6.0  -\!9.4  $&$ \pm 9.1  $\\
$170 $&$ 248.7 $&$ +6.5  -\!9.7  $&$ \pm 9.2  $\\
$175 $&$ 227.9 $&$ +6.6  -\!9.7  $&$ \pm 9.2  $\\
$180 $&$ 209.5 $&$ +6.6  -\!9.8  $&$ \pm 9.2  $\\
$185 $&$ 193.0 $&$ +6.6  -\!9.8  $&$ \pm 9.2  $\\
$190 $&$ 178.3 $&$ +6.7  -\!9.9  $&$ \pm 9.3  $\\
$195 $&$ 165.0 $&$ +6.7  -\!9.9  $&$ \pm 9.3  $\\
$200 $&$ 153.2 $&$ +6.8  -\!10.0 $&$ \pm 9.4  $\\
$210 $&$ 132.9 $&$ +7.0  -\!10.1 $&$ \pm 9.4  $\\
$220 $&$ 116.2 $&$ +7.2  -\!10.3 $&$ \pm 9.5  $\\
$230 $&$ 102.5 $&$ +7.5  -\!10.4 $&$ \pm 9.6  $\\
$240 $&$ 91.09 $&$ +7.6  -\!10.6 $&$ \pm 9.7  $\\
$250 $&$ 81.56 $&$ +8.0  -\!10.8 $&$ \pm 9.7  $\\
$260 $&$ 73.51 $&$ +8.3  -\!11.0 $&$ \pm 9.8  $\\
$270 $&$ 66.67 $&$ +8.6  -\!11.2 $&$ \pm 9.9  $\\
$280 $&$ 60.81 $&$ +9.0  -\!11.4 $&$ \pm 10.0 $\\
$290 $&$ 55.75 $&$ +9.3  -\!11.6 $&$ \pm 10.1 $\\
$300 $&$ 51.33 $&$ +9.7  -\!11.8 $&$ \pm 10.1 $\\ \hline
   \end{tabular}
   \end{center}
\end{table}

\clearpage

\section{MSSM neutral Higgs production processes\footnote{M.~Spira,
M.~Vazquez Acosta, M.~Warsinsky, G.~Weiglein (eds.); S.~Dittmaier,
R.~Harlander, S.~Heinemeyer, A.~Kalinowski, M.~M\"uhlleitner,
M.~Kr\"amer, H.~Rzehak, M.~Schumacher, P.~Slavich and T.~Vickey.}}

\label{sec:mssm_neutral}






\providecommand{\PA}{\mathrm{A}}
\newcommand{\orderx}[1]{\ensuremath{{\cal O}(#1)}}
\newcommand{\mhmaxx}{\ensuremath{m_h^{\rm max}}}
\newcommand{\cp}{\mathrm{CP}}
\newcommand{\MHp}{M_{\PSHpm}}
\providecommand{\lsim}
{\;\raisebox{-.3em}{$\stackrel{\displaystyle <}{\sim}$}\;}
\providecommand{\gsim}
{\;\raisebox{-.3em}{$\stackrel{\displaystyle >}{\sim}$}\;}
\providecommand{\gghnnlo}{{\sc ggH@NNLO}}
\providecommand{\bbhnnlo}{{\sc bbH@NNLO}}
\providecommand{\HDECAY}{{\sc HDECAY}}
\providecommand{\HIGLU}{{\sc HIGLU}}
\providecommand{\Prophecy}{{\sc Prophecy4f}}
\providecommand{\CPsuperH}{{\sc CPsuperH}}
\providecommand{\FeynHiggs}{{\sc FeynHiggs}}

\subsection{Higgs phenomenology in the MSSM}
\label{sec:mssmvssm}

The Higgs sector of the Minimal Supersymmetric Standard Model (MSSM)
with two scalar doublets accommodates five physical Higgs bosons. In
lowest order these are the light and heavy CP-even $\PSh$ and $\PH$, the
$\cp$-odd $\PSA$, and the charged Higgs bosons $\PSHpm$.  The MSSM Higgs
sector can be expressed at lowest order in terms of the gauge couplings
and two further input parameters, conventionally chosen as $\tanb \equiv
v_2/v_1$, the ratio of the two vacuum expectation values, and either
$\MA$ or $M_{\PSHpm}$. All other masses and mixing angles can therefore be
predicted. However, the Higgs sector of the MSSM is affected by large
higher-order corrections, which have to be taken into account for
reliable phenomenological predictions. In particular, owing to the large
top Yukawa coupling, loop contributions from the top and stop sector to
the Higgs masses and couplings can be numerically very important. For
large values of $\tanb$ also effects from the bottom/sbottom sector can
be large. The relation between the bottom-quark mass and the bottom
Yukawa coupling is affected by a $\tanb$-enhanced contribution
$\Delta_{\PQb}$~\cite{Hall:1993gn,Hempfling:1993kv,
Carena:1994bv,Pierce:1996zz,Carena:1999py,Guasch:2003cv,Noth:2008tw,Noth:2010jy,Mihaila:2010mp,Hofer:2009xb}, which is
non-vanishing even in the limit of asymptotically large values of the
SUSY mass parameters (an analogous contribution also exists for the
$\PGt$ lepton).  While the MSSM Higgs sector is CP-conserving at lowest
order, CP-violating effects can enter via the potentially large loop
corrections, giving rise to a mixing between all three neutral mass
eigenstates. In the following we will focus on the CP-conserving case
and use $\MA$ as input parameter. 

Higgs phenomenology in the MSSM can differ very significantly from the
SM case. The relevant couplings entering production and decay processes
of an MSSM Higgs boson can be much different from the corresponding
couplings in the SM case. The lower bound on the Higgs mass in the SM
from the searches at LEP cannot directly be applied to the MSSM
case~\cite{Barate:2003sz,Schael:2006cr}, and in fact much lighter Higgs 
masses are possible in the MSSM without being in conflict with the
present search limits. The presence of more than one Higgs boson in the
spectrum can give rise to overlapping signals in the Higgs searches, in 
particular in parameter regions where the Higgs-boson widths are large.
On the other hand, 
in the decoupling limit, $\MA \gg \MZ$ (in practice realised already for 
$\MA \gsim 2 \MZ$), the couplings of the light CP-even Higgs boson 
to gauge bosons and fermions become SM-like. In this parameter region 
the light CP-even Higgs boson of the MSSM resembles the 
Higgs boson of the SM. In addition to the production and decay processes 
present for a SM Higgs, further channels are possible in the MSSM case. 
In particular, MSSM Higgs bosons can be produced in association with or
in decays of SUSY particles, and decays of MSSM Higgs bosons into SUSY
particles, if kinematically allowed, can have a large impact on 
the Higgs branching ratios. In some parameter regions even decays of 
heavy MSSM Higgs bosons into lighter Higgs states can be relevant, 
which if detectable could be of great interest to gain information on 
the Higgs self-couplings. In the following we will mainly focus on the 
production processes that are expected to be most relevant for early
searches for MSSM Higgs bosons at the LHC, namely Higgs production in
gluon fusion and in association with bottom quarks.

It is customary to discuss searches for MSSM Higgs bosons in terms of 
benchmark scenarios where the lowest-order input parameters 
$\tanb$ and $\MA$ (or $\MHp$) are varied, while the other SUSY parameters
entering via radiative corrections are set to certain benchmark values. 
In the following we will focus on the $\mhmaxx$ benchmark 
scenario~\cite{Carena:2002qg}, which in the on-shell scheme is defined as
\begin{equation}
M_{\mathrm{SUSY}} = 1 \UTeV, \; X_{\PQt} = 2 M_{\mathrm{SUSY}}, \; \mu = 200 \UGeV, \;
M_{\PSg}=800 \UGeV, \; M_2 = 200 \UGeV, \; A_{\PQb} = A_{\PQt},
\label{YRHXS_MSSM_neutral_eq:mhmax}
\end{equation}
where $M_{SUSY}$ denotes the common soft-SUSY-breaking squark mass of
the third generation, $X_{\PQt}=A_{\PQt}-\mu/\tanb$ the stop mixing
parameter, $A_\mathrm{t}$ and $A_\mathrm{b}$ the stop and sbottom
trilinear couplings, respectively, $\mu$ the Higgsino mass parameter,
$M_{\PSg}$ the gluino mass, and $M_2$ the SU(2)-gaugino mass parameter.
$M_1$ is fixed via the GUT-relation $M_1 = 5/3 M_2 \sin \theta_w/\cos
\theta_w$.

In contrast to the SM case, where the Higgs mass is a free input
parameter, calculations of Higgs-boson production and decay processes in
the MSSM require as a first step the evaluation of the Higgs-boson
masses and mixing contributions in terms of $\MA$, $\tanb$, and all other
SUSY parameters that enter via radiative corrections. The mixing between
the CP-even states $\PSh$ and $\PSH$ (in the approximation where
CP-violating effects are neglected; in general mixing between $\PSh$,
$\PSH$, and $\PSA$ has to be considered) must be taken into account
correctly in order to ensure the correct on-shell properties of the
Higgs fields appearing in the $S$-matrix elements of Higgs-boson
production or decay processes. 

Two dedicated codes exist for calculating the Higgs-boson masses and
mixing contributions in terms of the MSSM input parameters,
\FeynHiggs~\cite{Heinemeyer:1998yj,Heinemeyer:1998np,Degrassi:2002fi,
Frank:2006yh} and \CPsuperH~\cite{Lee:2003nta,Lee:2007gn}, which
incorporate higher-order corrections in the MSSM Higgs sector up to the
two-loop level. In the case of real parameters a more complete set of
higher-order corrections is included in \FeynHiggs. We will therefore
use \FeynHiggs\ for evaluating the Higgs-boson masses and effective
couplings in the MSSM. We have performed a comparison between the
predictions of \FeynHiggs\ and \CPsuperH\ (using an appropriate
parameter transformation to take account of the different
renormalization schemes used in the calculations incorporated in the two
codes) in the $\mhmaxx$ and no-mixing benchmark
scenarios~\cite{Carena:2002qg,Carena:2000dp}. We have found in general
good agreement, with deviations in the prediction of the lightest MSSM
Higgs mass, $\Mh$, of \orderx{1}\UGeV, and deviations of up to $\sim
10\%$ in the effective mixing angle of the neutral $\cp$-even Higgs
sector for large values of $\tanb$.  The deviations can nevertheless be
relevant in the parameter regions that are tested first by the LHC:
relatively low $\MA$ and large $\tanb$.  A numerical comparison of
\FeynHiggs\ and \CPsuperH\ with the program
\HDECAY~\cite{Djouadi:1997yw,Spira:1997dg,hdecay2}, which
performs the calculation of Higgs-boson masses and mixings in the MSSM
using a less complete set of higher-order corrections, is in progress.

In making predictions for Higgs-boson production or decay processes in the MSSM 
one has to face the fact that certain types of higher-order 
corrections have only been calculated in the SM case up to now, while
their counterpart for the case of the MSSM is not yet available. Instead
of starting from dedicated MSSM calculations for Higgs cross sections or
decay widths,  
which treat higher-order corrections of SM-type and SUSY-type on the 
same footing but may be lacking the most up-to-date SM-type corrections,
it can be advantageous to start from SM-type processes including the 
relevant higher-order corrections and to dress suitable building blocks
with appropriate MSSM coupling factors (using also the MSSM predictions
for the Higgs masses). For the numerical results presented below 
on MSSM Higgs production in gluon fusion and in association with bottom
quarks we have followed the latter approach, as explained in more detail
below.


\subsection{Overview about the most relevant MSSM Higgs production processes}

The dominant neutral MSSM Higgs production mechanisms for small and
moderate values of $\tan\beta$ are the gluon-fusion processes (see
\Fref{YRHXS_MSSM_neutral_dia1})
\begin{displaymath}
\Pg\Pg \to \PSh,\PSH,\PSA
\end{displaymath}
\begin{figure}[hbt]
\begin{center}
\SetScale{0.8}
\begin{picture}(180,80)(0,0)
\Gluon(0,20)(50,20){-3}{5}
\Gluon(0,80)(50,80){3}{5}
\ArrowLine(50,20)(50,80)
\ArrowLine(50,80)(100,50)
\ArrowLine(100,50)(50,20)
\DashLine(100,50)(150,50){5}
\Vertex(50,20){2}
\Vertex(50,80){2}
\Vertex(100,50){2}
\put(126,37){$\PSh,\PSH,\PSA$}
\put(0,37){$\PQt,\PQb,\PSQt, \PSQb$}
\put(-12,14){$\Pg$}
\put(-12,62){$\Pg$}
\end{picture}  \\
\caption{\label{YRHXS_MSSM_neutral_dia1} Typical diagram
contributing to $\Pg\Pg\to \PSh,\PSH,\PSA$ at lowest order.}
\end{center}
\end{figure}
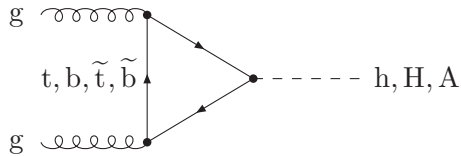
which are mediated predominantly by top and bottom loops as in the SM
case, but in addition by stop and sbottom loops for the scalar Higgs
bosons $\PSh,\PSH$, if the squarks are light \cite{Dawson:1996xz}. The
NLO QCD corrections to the quark loops are known in the heavy-quark
limit \cite{Djouadi:1991tka,Dawson:1991zj,Kauffman:1993nv,Dawson:1993qf}
as well as including the full quark mass dependence
\cite{Graudenz:1992pv,Spira:1993bb,Spira:1995rr,Harlander:2005rq,
Anastasiou:2006hc,Aglietti:2006tp,Bonciani:2007ex}. They increase the
cross sections by up to about $100\%$ for smaller $\tanb$ and up to about
$50\%$ for large $\tanb$, where the bottom loop contributions become
dominant due to the strongly enhanced bottom Yukawa couplings (for the
light CP-even Higgs this enhancement is only present away from the
decoupling limit, i.e.\ for relatively small $\MA$).  The limit of heavy
quarks is only applicable for $\tanb\lesssim 5$ within about $20{-}25\%$,
if the full mass dependence of the LO terms is taken into account
\cite{Kramer:1996iq,Spira:1997dg,Djouadi:2005gi,Djouadi:2005gj}. Thus
the available NNLO QCD corrections in the heavy-quark limit
\cite{Harlander:2002wh,Harlander:2002vv,
Anastasiou:2002yz,Anastasiou:2002wq,Ravindran:2003um} can only be used
for small and moderate $\tanb$, while for large $\tanb$ one has to rely
on the NLO results including the full mass dependence
\cite{Graudenz:1992pv,Spira:1993bb,Spira:1995rr,
Anastasiou:2006hc,Aglietti:2006tp,Bonciani:2007ex}. The QCD corrections
to the squark loops are known in the heavy-squark limit
\cite{Dawson:1996xz} and including the full mass dependence
\cite{Muhlleitner:2006wx,Anastasiou:2006hc,
Aglietti:2006tp,Bonciani:2007ex}. The full SUSY QCD corrections have
been obtained in the limit of heavy squarks and gluinos
\cite{Harlander:2003bb,Harlander:2003kf,Harlander:2004tp,
Harlander:2005if,Degrassi:2008zj,Muhlleitner:2008yw,Degrassi:2010eu} and
recently including the full mass dependences, too
\cite{Anastasiou:2008rm,Muhlleitner:2010nm}.  The pure QCD corrections
are of about the same size as those to the quark loops thus rendering
the total $K$-factor of similar size as for the quark loops alone with a
maximal deviation of about $10\%$ \cite{Dawson:1996xz}.  The pure
SUSY QCD corrections are small for small values of $\tanb$
\cite{Harlander:2003bb,Harlander:2003kf,Harlander:2004tp,
Harlander:2005if,Anastasiou:2008rm}. For large values of $\tanb$
sizable corrections arise due to $\tanb$-enhanced corrections
\cite{Muhlleitner:2010nm,Degrassi:2010eu}. The NNLL resummation of the
SM Higgs cross section \cite{Catani:2003zt,Ravindran:2005vv,Moch:2005ky}
can also be applied to the scalar MSSM Higgs cross sections in the
regions where the heavy-quark limit is valid. For the pseudoscalar 
Higgs-boson production the NNLL resummation has not been performed so far.

The vector-boson fusion processes
\cite{Cahn:1983ip,Hikasa:1985ee,Altarelli:1987ue} (see
\Fref{YRHXS_MSSM_neutral_dia2})
\begin{displaymath}
\PQq\PQq\to \PQq\PQq+\PW^*\PW^*/\PZ^*\PZ^*\to \PQq\PQq + \PSh/\PSH
\end{displaymath}
play an important role for the light CP-even Higgs boson $\PSh$ in the
decoupling limit, $\MA \gg \MZ$, where it becomes SM-like, and for the
heavy CP-even Higgs boson $\PSH$ for small $\MA$, where $\PSH$ becomes
SM-like. In the other regions the cross sections are suppressed by the
additional SUSY factors of the Higgs couplings. The NLO and approximate
NNLO QCD corrections to the total cross section and the distributions
can be taken from the SM Higgs case and are small
\cite{Han:1992hr,Figy:2003nv,Berger:2004pca,
Ciccolini:2007jr,Ciccolini:2007ec,Bolzoni:2010xr}. The SUSY QCD
corrections mediated by virtual gluino and squark exchange at the
vertices turned out to be small \cite{Djouadi:1999ht,Hollik:2008xn}. The
SUSY electroweak corrections are typically at the level of $1\%$ with
up to $2{-}4\%$ at the edge of the SUSY exclusion
limits~\cite{Hollik:2008xn}.
\begin{figure}[hbt] 
\begin{center}
\SetScale{0.8}
\begin{picture}(120,90)(0,0)
\ArrowLine(0,0)(50,0)
\ArrowLine(50,0)(100,0)
\ArrowLine(0,100)(50,100) 
\ArrowLine(50,100)(100,100)
\Photon(50,0)(50,50){3}{5}
\Photon(50,50)(50,100){3}{5}
\DashLine(50,50)(100,50){5}
\Vertex(50,100){2}
\Vertex(50,50){2}
\Vertex(50,0){2}
\put(87,37){$\PSh,\PSH$}
\put(-12,-2){$\PQq$}
\put(-12,78){$\PQq$}
\put(87,-2){$\PQq$}
\put(87,78){$\PQq$}
\put(44,17){$\PW,\PZ$}
\put(44,57){$\PW,\PZ$}
\end{picture}  \\
\caption{\label{YRHXS_MSSM_neutral_dia2} Diagram contributing to
$\PQq\PQq \to
\PQq\PQq V^*V^* \to \PQq\PQq + \PSh/\PSH$ ($V=\PW,\PZ$) at lowest order.} 
\end{center}
\end{figure}
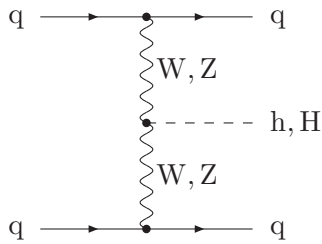

Higgs-strahlung off $\PW,\PZ$ gauge bosons
\cite{Glashow:1978ab,Kunszt:1991xk} (see
\Fref{YRHXS_MSSM_neutral_dia3})
\begin{displaymath}
\PQq\PAQq\to \PZ^*/\PW^* \to \PZ/\PW + \PSh/\PSH
\end{displaymath}
is most relevant for SM-like Higgs states. This class of processes
gained renewed attention at the LHC in the context of possible
improvements of jet reconstruction and decomposition
techniques~\cite{Butterworth:2008iy}.
The NLO \cite{Han:1991ia,Spira:1997dg} and NNLO \cite{Brein:2003wg} QCD
corrections can be translated from the SM to the MSSM case, and the SUSY QCD
corrections are small \cite{Djouadi:1999ht}. The SUSY electroweak
corrections are unknown.
\begin{figure}[hbt]
\begin{center}
\SetScale{0.8}
\begin{picture}(160,100)(0,-10)
\ArrowLine(0,100)(50,50)
\ArrowLine(50,50)(0,0)
\Photon(50,50)(100,50){3}{5}
\Photon(100,50)(150,100){3}{6}
\DashLine(100,50)(150,0){5}
\Vertex(50,50){2}
\Vertex(100,50){2}
\put(126,-4){$\PSh,\PSH$}
\put(-12,-2){$\PAQq$}
\put(-12,78){$\PQq$}
\put(50,50){$\PW,\PZ$}
\put(126,77){$\PW,\PZ$}
\end{picture}  \\
\caption{\label{YRHXS_MSSM_neutral_dia3} Diagram contributing to
$\PQq\PAQq \to V^* \to V + \PSh/\PSH$ ($V=\PW,\PZ$) at lowest order.}
\end{center}
\end{figure}
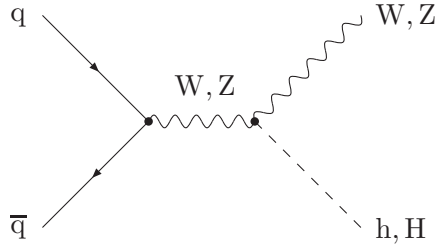

Higgs-boson radiation off top quarks
\cite{Raitio:1978pt,Ng:1983jm,Kunszt:1984ri,Gunion:1991kg,Marciano:1991qq}
(see \Fref{YRHXS_MSSM_neutral_dia4})
\begin{displaymath}
\PQq\PAQq/\Pg\Pg\to \PQt\PAQt + \PSh/\PSH/\PSA
\end{displaymath}
plays a significant role at the LHC for the light scalar Higgs particle
only. The NLO QCD corrections are the same as for the SM Higgs boson
with modified top and bottom Yukawa couplings and are thus of moderate
size \cite{Beenakker:2001rj,Beenakker:2002nc,
Reina:2001sf,Dawson:2002tg}. The SUSY QCD corrections have been
computed recently \cite{Peng:2005ti,Hollik:2006vn,
Hafliger:2006zz,Walser:2008zz}. They are of moderate size, too.
\begin{figure}[hbt]
\begin{center}
\SetScale{0.8}
\begin{picture}(360,100)(0,-10)
\ArrowLine(0,100)(50,50)
\ArrowLine(50,50)(0,0)
\Gluon(50,50)(100,50){3}{5}
\ArrowLine(100,50)(125,75)
\ArrowLine(125,75)(150,100)
\ArrowLine(150,0)(100,50)
\DashLine(125,75)(150,50){5}
\Vertex(50,50){2}
\Vertex(100,50){2}
\Vertex(125,75){2}
\put(126,37){$\PSh,\PSH,\PSA$}
\put(-12,78){$\PQq$}
\put(-12,-2){$\PAQq$}
\put(55,52){$\Pg$}
\put(126,78){$\PQt/\PQb$}
\put(126,-2){$\PAQt/\PAQb$}
\Gluon(250,0)(300,0){3}{5}
\Gluon(250,100)(300,100){3}{5}
\ArrowLine(350,0)(300,0)
\ArrowLine(300,0)(300,50)
\ArrowLine(300,50)(300,100)
\ArrowLine(300,100)(350,100)
\DashLine(300,50)(350,50){5}
\Vertex(300,100){2}
\Vertex(300,50){2}
\Vertex(300,0){2}
\put(286,37){$\PSh,\PSH,\PSA$}
\put(190,78){$\Pg$}
\put(190,-2){$\Pg$}
\put(286,78){$\PQt/\PQb$}
\put(286,-2){$\PAQt/\PAQb$}
\end{picture}  \\
\caption{\label{YRHXS_MSSM_neutral_dia4} Typical diagrams contributing to
$\PQq\PAQq/\Pg\Pg \to Q\bar Q + \PSh/\PSH/\PSA~~(Q=\PQt,\PQb)$ at lowest order.}
\end{center}
\end{figure}
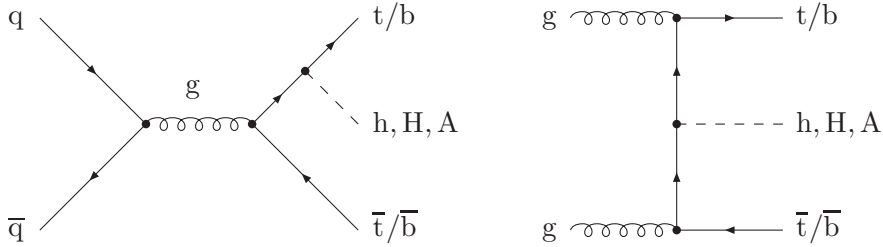

For large values of $\tanb$ Higgs-boson radiation off bottom quarks
\cite{Raitio:1978pt,Ng:1983jm,Kunszt:1984ri,Gunion:1991kg,Marciano:1991qq}
(see \Fref{YRHXS_MSSM_neutral_dia4})
\begin{displaymath}
\PQq\PAQq/\Pg\Pg\to \PQb\PAQb + \PSh/\PSH/\PSA
\end{displaymath}
constitute the dominant Higgs-boson production processes. The NLO QCD
corrections can be taken from the analogous calculation involving top
quarks. However, they turn out to be large
\cite{Dittmaier:2003ej,Dawson:2003kb}. The main
reason is that the integration over the transverse momenta of the final-state 
bottom quarks generates large logarithmic contributions. The
resummation of the latter can be established by the introduction of bottom-quark
densities in the proton, since the large logarithms correspond to the
DGLAP evolution of these densities. Their DGLAP evolution resums them.
This leads to an approximate approach starting from the process
\cite{Dicus:1988cx} (see \Fref{YRHXS_MSSM_neutral_dia5}a)
\begin{displaymath}
\PQb\PAQb\to \PSh/\PSH/\PSA
\end{displaymath}
at LO, where the transverse momenta of the incoming bottom quarks, their
masses and their off-shellness are neglected at LO. 
\begin{figure}[htbp]
\begin{center}
\SetScale{0.8}
\begin{picture}(130,90)(10,0)
\ArrowLine(0,0)(50,50)
\ArrowLine(50,50)(0,100)
\DashLine(50,50)(100,50){5}
\Vertex(50,50){2}
\put(-12,-2){$\PAQb$}
\put(-12,78){$\PQb$}
\put(88,36){$\PSh/\PSH/\PSA$}
\put(40,-30){{(a)}}
\end{picture}
\hspace*{2em}
\begin{picture}(130,90)(210,0)
\Gluon(250,0)(300,0){3}{5}
\Gluon(250,100)(300,100){3}{5}
\ArrowLine(350,0)(300,0)
\ArrowLine(300,0)(300,50)
\ArrowLine(300,50)(300,100)
\ArrowLine(300,100)(350,100)
\DashLine(300,50)(350,50){5}
\Vertex(300,100){2}
\Vertex(300,50){2}
\Vertex(300,0){2}
\put(286,36){$\PSh/\PSH/\PSA$}
\put(188,78){$\Pg$}
\put(188,-2){$\Pg$}
\put(286,78){$\PQb$}
\put(286,-2){$\PAQb$}
\put(240,-30){{(b)}}
\end{picture} \\[1.0cm]
\caption{\label{YRHXS_MSSM_neutral_dia5} Typical diagrams for the
Higgs-boson production mechanisms related to Higgs radiation off bottom
quarks in the 5FS and 4FS at leading order: {(a)} $\PQb\PAQb\to
\PSh/\PSH/\PSA$ (5FS) and {(b)} $\Pg\Pg\to \PQb\PAQb+\PSh/\PSH/\PSA$
(4FS).}
\end{center}
\end{figure}
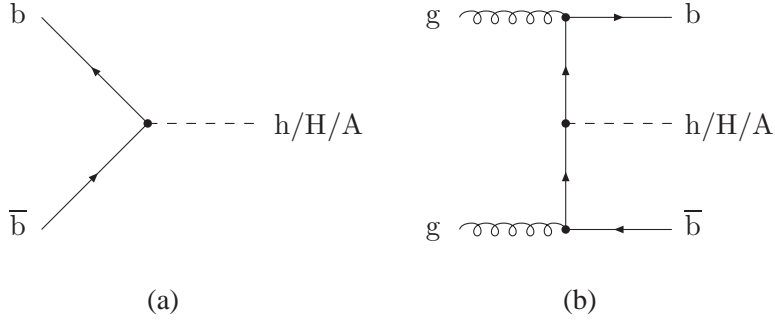
The NLO~\cite{Dicus:1998hs,Balazs:1998sb} and NNLO
\cite{Harlander:2003ai} QCD corrections to these bottom-initiated
processes are known and of moderate size, if the running bottom Yukawa
coupling is introduced at the scale of the corresponding Higgs-boson
mass. At NNLO the full process $\Pg\Pg\to \PQb\PAQb + \PSh/\PSH/\PSA$
(see \Fref{YRHXS_MSSM_neutral_dia5}b) contributes to the real
corrections for the first time.  The fully exclusive $\Pg\Pg\to
\PQb\PAQb + \PSh/\PSH/\PSA$ process, calculated with four active parton
flavours in a four-flavour scheme (4FS), and the result, calculated with
five active parton flavours in the five-flavour scheme (5FS), will converge
against the same value at higher perturbative orders. Reasonable
agreement between the NLO 4FS and NNLO 5FS is achieved if the
factorization scale of the bottom-quark densities is chosen as about a
quarter of the Higgs mass \cite{Campbell:2004pu,Dawson:2005vi}.  If both
bottom jets accompanying the Higgs boson in the final state are tagged,
one has to rely on the fully exclusive calculation for $\Pg\Pg\to
\PQb\PAQb + \PSh/\PSH/\PSA$. For the case of a single $\PQb$-tag in the
final state the corresponding calculation in the 5FS starts from the
process $\PQb\Pg\to \PQb+\PSh/\PSH/\PSA$ with the final-state bottom
quark carrying finite transverse momentum.
The NLO QCD and electroweak corrections to this process have been
calculated \cite{Campbell:2002zm,Dawson:2004sh,Beccaria:2010fg}
supplemented by the NLO SUSY QCD corrections recently \cite{Dawson:2007ur}.

In our study we concentrated on the gluon-fusion processes and neutral
Higgs-boson radiation off bottom quarks as the first step. We have
focused on the $\mhmaxx$ scenario \cite{Carena:2000dp,Carena:2002qg},
which is characterised by rather heavy SUSY particles. Genuine SUSY QCD
and SUSY electroweak corrections in this scenario are below the $10\%$
level for Higgs-boson radiation off bottom quarks as well as the 
gluon-fusion processes.  For the calculation of the MSSM Higgs-boson masses
and couplings we have used the program
\FeynHiggs\,2.7.4~\cite{Heinemeyer:1998yj,Heinemeyer:1998np,Degrassi:2002fi,
Frank:2006yh} which includes the most up-to-date radiative corrections
to the MSSM Higgs sector up to the two-loop level and the
$\Delta_{\PQb}$ terms as an approximation of the SUSY QCD and electroweak
corrections to the bottom Yukawa couplings. In further steps we will
have to include the full SUSY QCD and SUSY electroweak corrections
where available and in addition allow for complex MSSM parameters which
leads to additional complications of the Higgs sector, since the mass
eigenstates will no longer be CP-eigenstates.  Moreover, for this study
we have fixed the MSSM scenario, since otherwise general predictions as
in the SM case will not be possible due to the huge variety of the MSSM
parameter space.  However, the results in the $\mhmaxx$ scenario will
not be representative for all possible MSSM scenarios. In the further
progress of this work we will develop the machinery to be able to cover
as many aspects of the MSSM as possible.  This requires the combination
of the most advanced results and tools available in our HEP community
for neutral MSSM Higgs-boson production.

\subsection{Gluon fusion}
The gluon-fusion processes $\Pg\Pg\to\phi~(\phi = \PSh,\PSH,\PSA)$ have been
calculated by generating grids for the individual contributions of the
top and bottom-quark loops. Stop and sbottom loops have been neglected
in this first step but will be included in the next steps. We have
generated grids for the scalar and pseudoscalar Higgs bosons
individually with Yukawa couplings of SM-like strength. The MSSM cross
sections can then be obtained by rescaling the individual parts by the
corresponding MSSM Yukawa coupling factors,
\begin{eqnarray}
\sigma^{\mathrm{MSSM}}(\Pg\Pg\to\phi) & = & \left(\frac{g_{\PQt}^{\mathrm{MSSM}}}{g_{\PQt}^{\mathrm{SM}}}\right)^2
\sigma_{\PQt\PQt}(\Pg\Pg\to\phi) + \left(\frac{g_{\PQb}^{\mathrm{MSSM}}}{g_{\PQb}^{\mathrm{SM}}}\right)^2
\sigma_{\PQb\PQb}(\Pg\Pg\to\phi) \nonumber \\
& & +
\frac{g_{\PQt}^{\mathrm{MSSM}}}{g_{\PQt}^{\mathrm{SM}}}~\frac{g_{\PQb}^{\mathrm{MSSM}}}{g_{\PQb}^{\mathrm{SM}}}
\sigma_{\PQt\PQb}(\Pg\Pg\to\phi),
\label{YRHXS_MSSM_neutral_eq1}
\end{eqnarray}
where $\sigma_{\PQt\PQt}, \sigma_{\PQb\PQb}$, and $\sigma_{\PQt\PQb}$ denote the square of
the top contributions, the square of the bottom contributions, and the
top--bottom interference, respectively. For $\sigma_{\PQb\PQb}$ and
$\sigma_{\PQt\PQb}$ we have used the full NLO QCD calculation of \HIGLU
\cite{Spira:1995mt}. For $\sigma_{\PQt\PQt}$ we have used the full NLO QCD
result of \HIGLU~and added the NNLO corrections in the heavy-top-quark
limit by using the program \gghnnlo~\cite{Harlander:2002wh,Harlander:2002vv} in the following way:
$\sigma^0_{\LO}, \sigma^0_{\NLO}$, and $\sigma^0_{\mathrm{NNLO}}$ have been
calculated by \gghnnlo.  The additional part added to the full NLO
result of $\sigma_{\PQt\PQt}$ is then given by
\begin{eqnarray}
\Delta\sigma^{\mathrm{NNLO}}_{\PQt\PQt}(\Pg\Pg\to\phi) & = & \Delta K_{\mathrm{NNLO}}~
\sigma_{\PQt\PQt}^{\mathrm{LO}}(\Pg\Pg\to\phi), \nonumber \\
\Delta K_{\mathrm{NNLO}} & = & \frac{\sigma^0_{\mathrm{NNLO}}-\sigma^0_{\NLO}}{\sigma^0_{\LO}},
\end{eqnarray}
where the individual cross sections $\sigma^0_{\mathrm{LO}}, \sigma^0_{\NLO},
\sigma^0_{\mathrm{NNLO}}$ have been evaluated consistently with LO, NLO, and NNLO
PDFs, respectively.  Since top mass effects are small at NNLO
\cite{Marzani:2008az,Harlander:2009bw,Harlander:2009mq,Harlander:2009my,
Pak:2009bx,Pak:2009dg} this procedure provides a result that is
expected to be very close to full NNLO QCD accuracy for the
$\sigma_{\PQt\PQt}$ parts.  Electroweak corrections to MSSM Higgs-boson
production via gluon fusion have not been calculated. The corresponding
electroweak corrections in the SM case
\cite{Djouadi:1994ge,Aglietti:2004nj,Degrassi:2004mx,Actis:2008ug}
cannot be translated easily to the MSSM and have thus been neglected.
Moreover, we have neglected the NNLL resummation effects
\cite{Catani:2003zt,Ravindran:2005vv,Moch:2005ky} on the
$\sigma_{\PQt\PQt}$ part for two reasons: (i) The NNLL resummation has not
been calculated for the pseudoscalar Higgs boson so far so that in order
to treat the scalar and pseudoscalar Higgs bosons at the same level, the
NNLL effects should be neglected. (ii) For a completely consistent NNLL
prediction also NNLL PDFs would be needed which, however, are not
available. To use NNLO PDFs instead is not fully consistent.

The top and bottom-quark masses have been introduced as pole masses in
the calculation including the corresponding Yukawa couplings. The MSSM
Yukawa coupling ratios to the SM couplings in
Eq.~(\ref{YRHXS_MSSM_neutral_eq1}) have been taken from the program
\FeynHiggs\,2.7.4~\cite{Heinemeyer:1998yj,Heinemeyer:1998np,Degrassi:2002fi,
Frank:2006yh} . As mentioned above, for the numerical MSSM results we
have chosen the $\mhmaxx$ benchmark scenario as specified in
Eq.~(\ref{YRHXS_MSSM_neutral_eq:mhmax}).  As the central choices of the
renormalization and factorization scales we adopted the corresponding
Higgs-boson mass $M_\phi$. For the NLO pieces of the cross section we
used the NLO MSTW2008 PDFs, while for the NNLO contributions the NNLO
MSTW2008 PDFs have been used appropriately. The strong coupling constant
has been normalized according to the PDFs, i.e.~$\alphas(\MZ)=0.12018$
at NLO and $\alphas(\MZ)=0.11707$ at NNLO
\cite{Martin:2009iq,Martin:2009bu}. The scale uncertainty has been
determined by varying the renormalization and factorization scales
between $M_\phi/2$ and $2M_\phi$. It amounts to about $10{-}15\%$ for the whole
Higgs mass and $\tanb$ range although for large values of $\tanb$ the
results are dominated by the bottom-quark loops which are only known at
NLO, unless the light (heavy) scalar Higgs mass is close to its upper
(lower) bound, where the top loops are dominant for large values of
$\tanb$, too. However, the scale dependence of the bottom-quark
contributions is considerably smaller than that of the top quark ones
\cite{Spira:1993bb,Spira:1995rr}.  We have added the $68\%$
CL~PDF+$\alphas$ uncertainties of the MSTW2008 PDFs to the scale
uncertainties linearly. Since there are no NNLO PDF sets of CTEQ and
NNPDF we did not include those sets in this uncertainty.

We have generated grids of the three cross section parts
$\sigma_{\PQt\PQt}^{\mathrm{NNLO}}, \sigma_{\PQb\PQb}^{\NLO}$, and $\sigma_{\PQt\PQb}^{\NLO}$ for the
mass ranges from $70\UGeV$ up to $1\UTeV$ in steps of $1\UGeV$ for the
scalar and pseudoscalar Higgs bosons separately. These grids are then
used for interpolation and the resulting numbers rescaled and added
according to the coupling ratios of \FeynHiggs. For the $\mhmaxx$ scenario we
have included the $\tanb$-enhanced $\Delta_{\PQb}$ corrections in the
effective MSSM bottom Yukawa couplings, since we expect them to dominate
the full SUSY QCD corrections for squark and gluino masses much larger
than the Higgs masses \cite{Degrassi:2010eu}. The resulting cross
sections for the pseudoscalar Higgs boson are shown for various values
of $\tanb$ in \Fref{YRHXS_MSSM_neutral_fig1}, while
Figs.~\ref{YRHXS_MSSM_neutral_fig2} and \ref{YRHXS_MSSM_neutral_fig3}
display the corresponding results for the light and heavy CP-even MSSM
Higgs bosons. The overall scale and PDF+$\alphas$ uncertainties amount to
about $15\%$. It is visible that for small and moderate values of
$\tanb$ virtual $\PQt\PAQt$ thresholds develop for Higgs masses
$M_\phi=2\Mt$, while for large values of $\tanb$ this threshold
behaviour is strongly suppressed due to the dominance of the bottom-quark 
contributions. For the light CP-even Higgs boson most of the
displayed parameter region corresponds to rather low values of $\MA$ 
(which is the input parameter that has been varied in the plots), while
the decoupling region where $\MA \gg \MZ$ corresponds to the region of
the highest $\Mh$ values in the plots. It should be noted that in this
limit, i.e.\ for the upper bound of the light CP-even Higgs-boson mass
in the plots, the light scalar Higgs-boson 
production cross section approaches the NNLO SM result by
construction. Note that the full MSSM result including stop and
sbottom loop contributions does {\it not} approach the SM cross section
for the light scalar Higgs boson at its upper mass bound in general. The
additional contributions from the stops and sbottoms impose a mismatch
between the MSSM cross section in the decoupling limit and the
corresponding SM cross section. This differs from the results of
Section 2 which include the NNLL resummation effects by less than $10\%$,
i.e.~less than the residual scale uncertainties.
\vspace{1cm}

\begin{figure}[htb]
\includegraphics[width=0.5\textwidth]{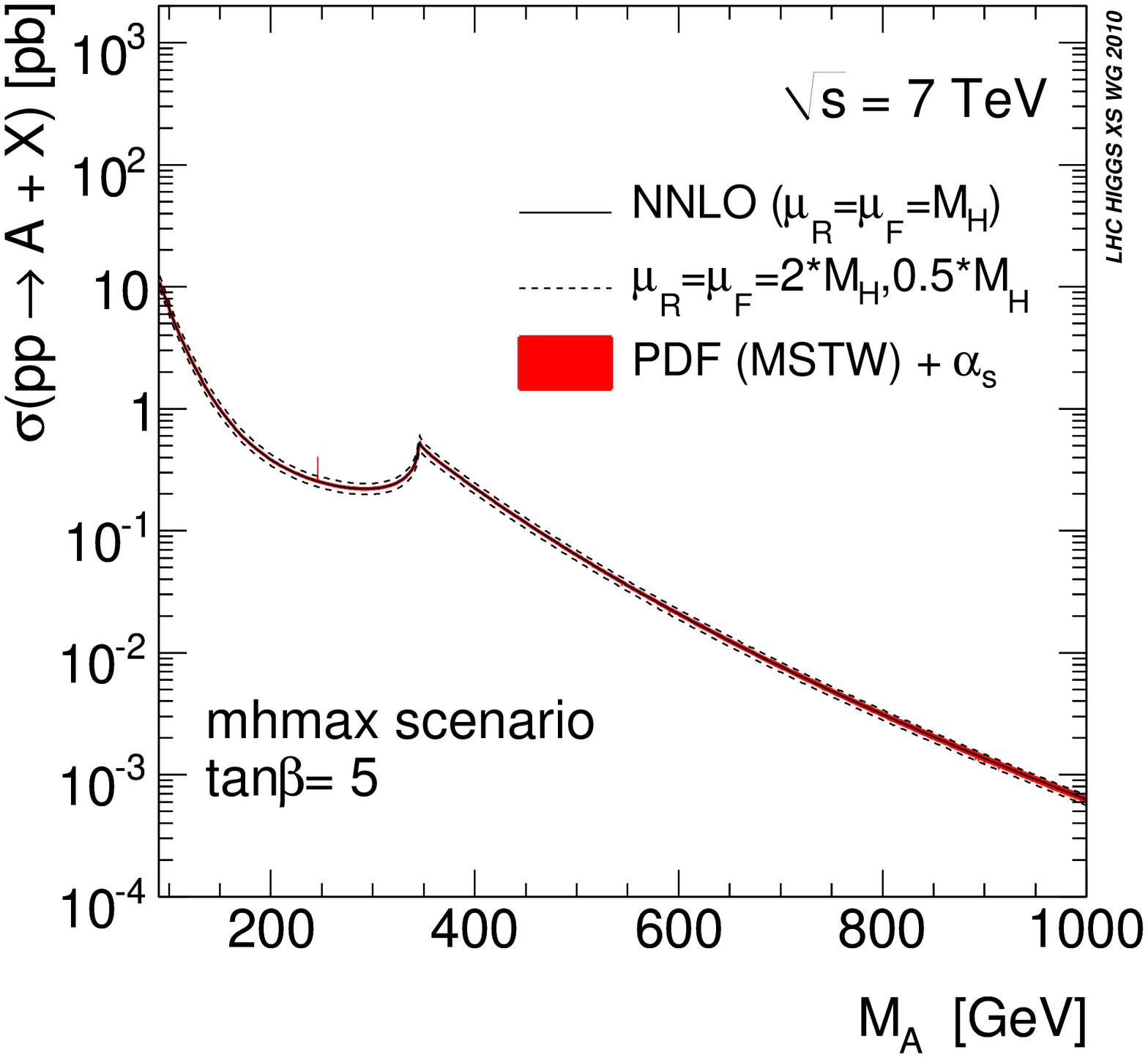}
\includegraphics[width=0.5\textwidth]{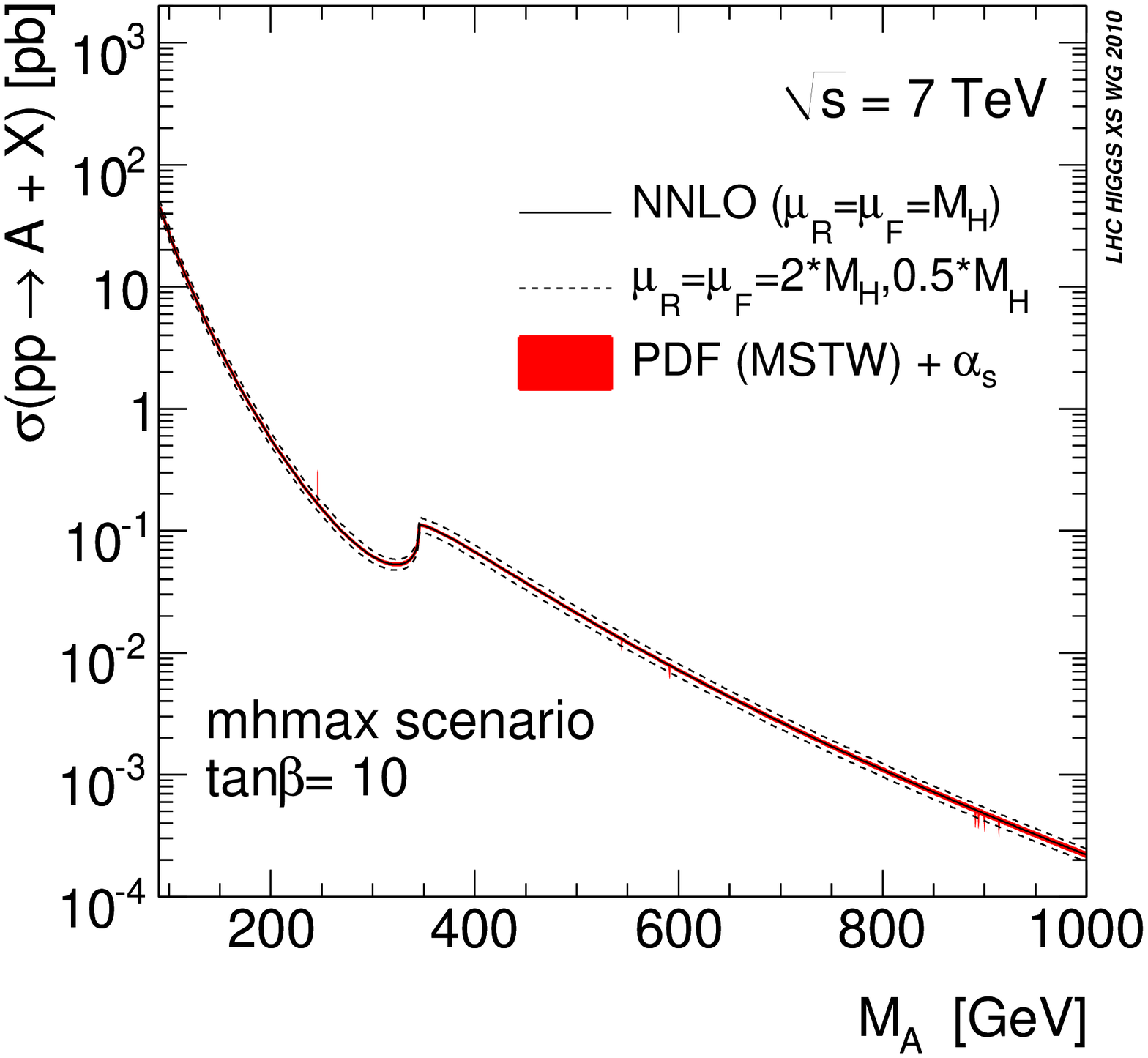}
\includegraphics[width=0.5\textwidth]{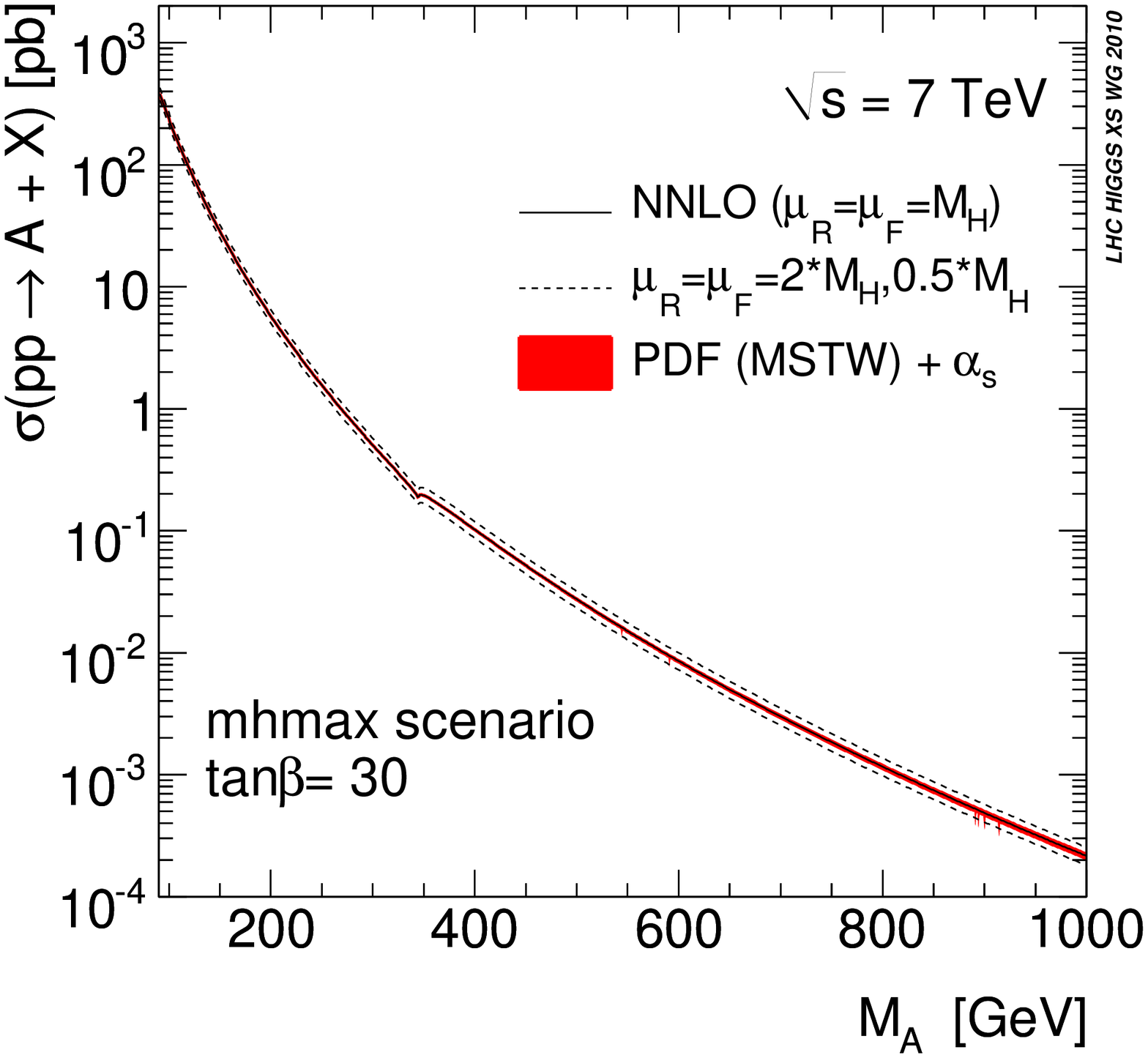}
\includegraphics[width=0.5\textwidth]{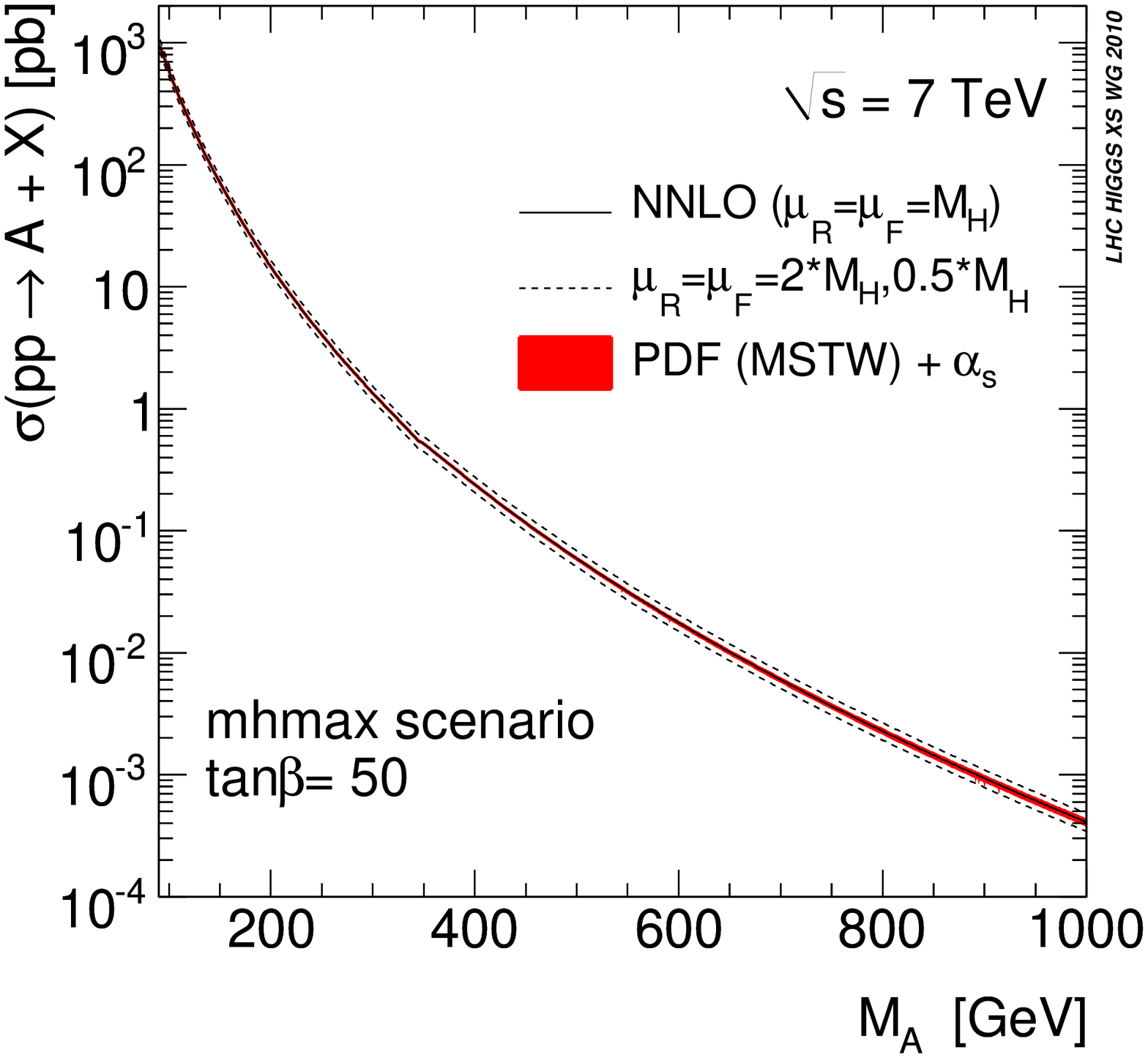}
\caption{\label{YRHXS_MSSM_neutral_fig1} Total gluon-fusion cross
sections of the pseudoscalar MSSM Higgs boson $\PSA$ for four values of
$\tanb$ within the $\mhmaxx$ scenario for $\sqrt{s}=7$\UTeV\ using
MSTW2008 PDFs \cite{Martin:2009iq,Martin:2009bu}. The NNLO results for
the SM-type contributions have been obtained from the programs
\HIGLU~and \gghnnlo, while the rescaling with MSSM coupling factors has
been done with \FeynHiggs.}
\end{figure}

\begin{figure}[htb]
\includegraphics[width=0.5\textwidth]{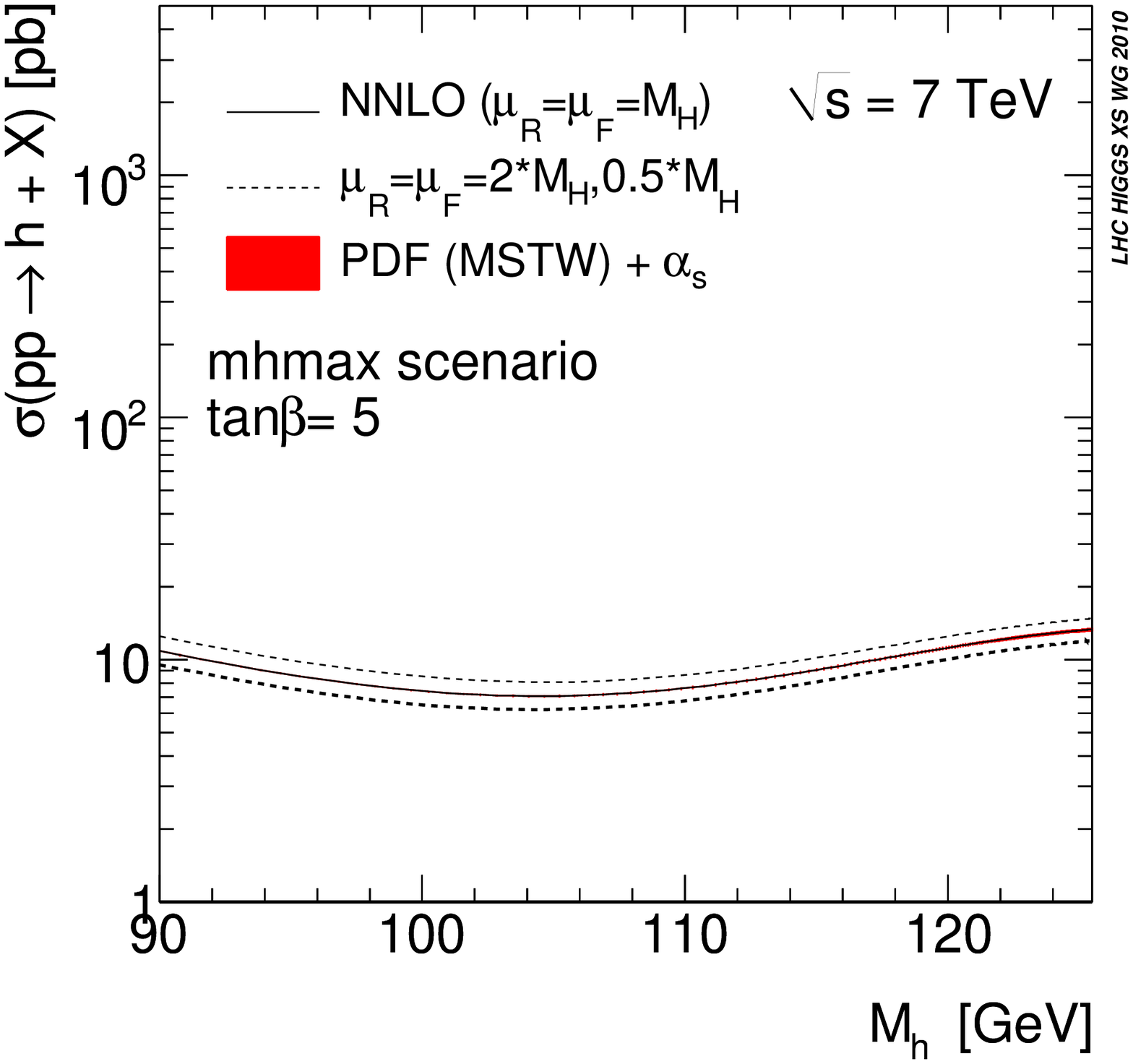}
\includegraphics[width=0.5\textwidth]{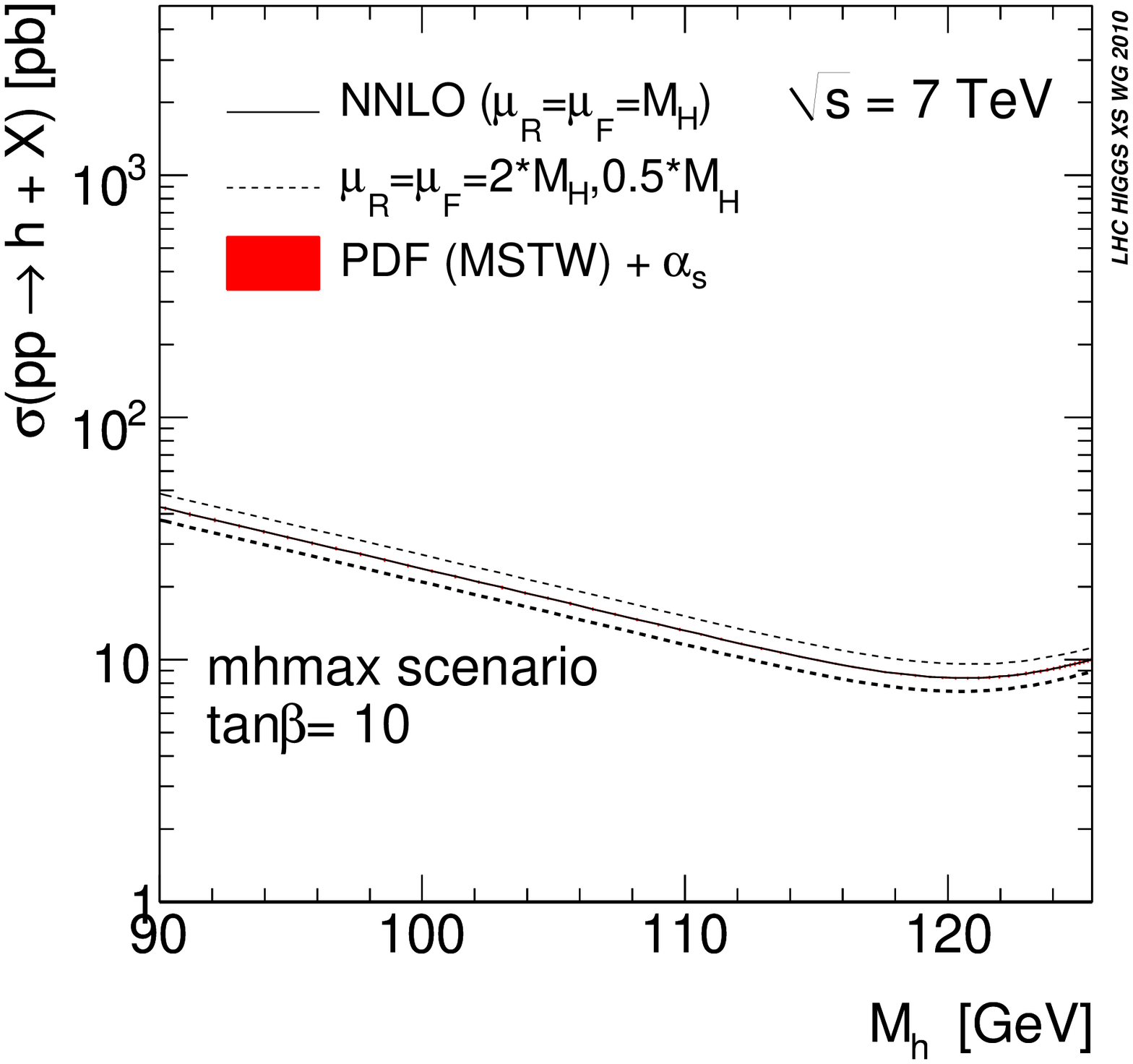}
\includegraphics[width=0.5\textwidth]{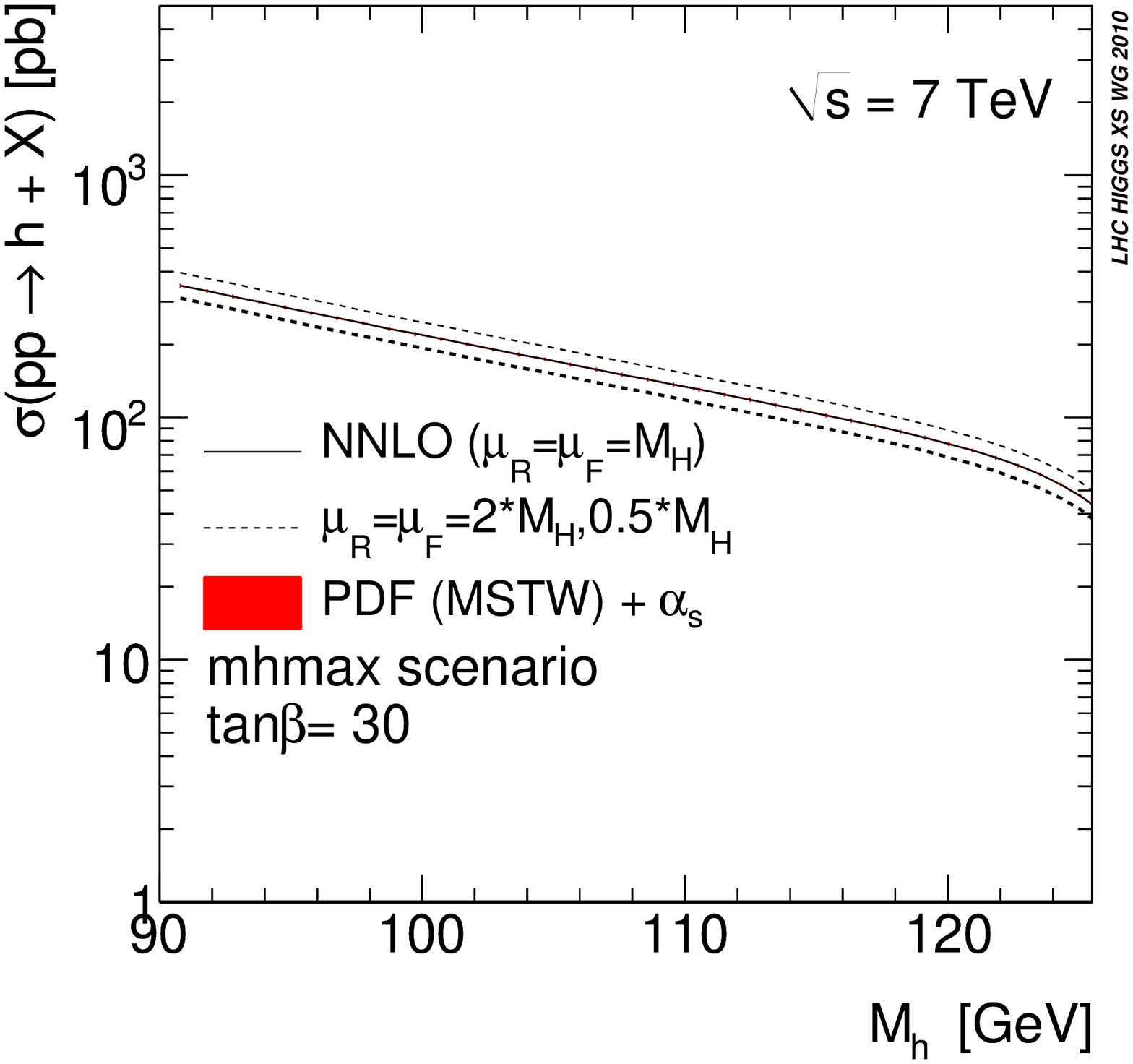}
\includegraphics[width=0.5\textwidth]{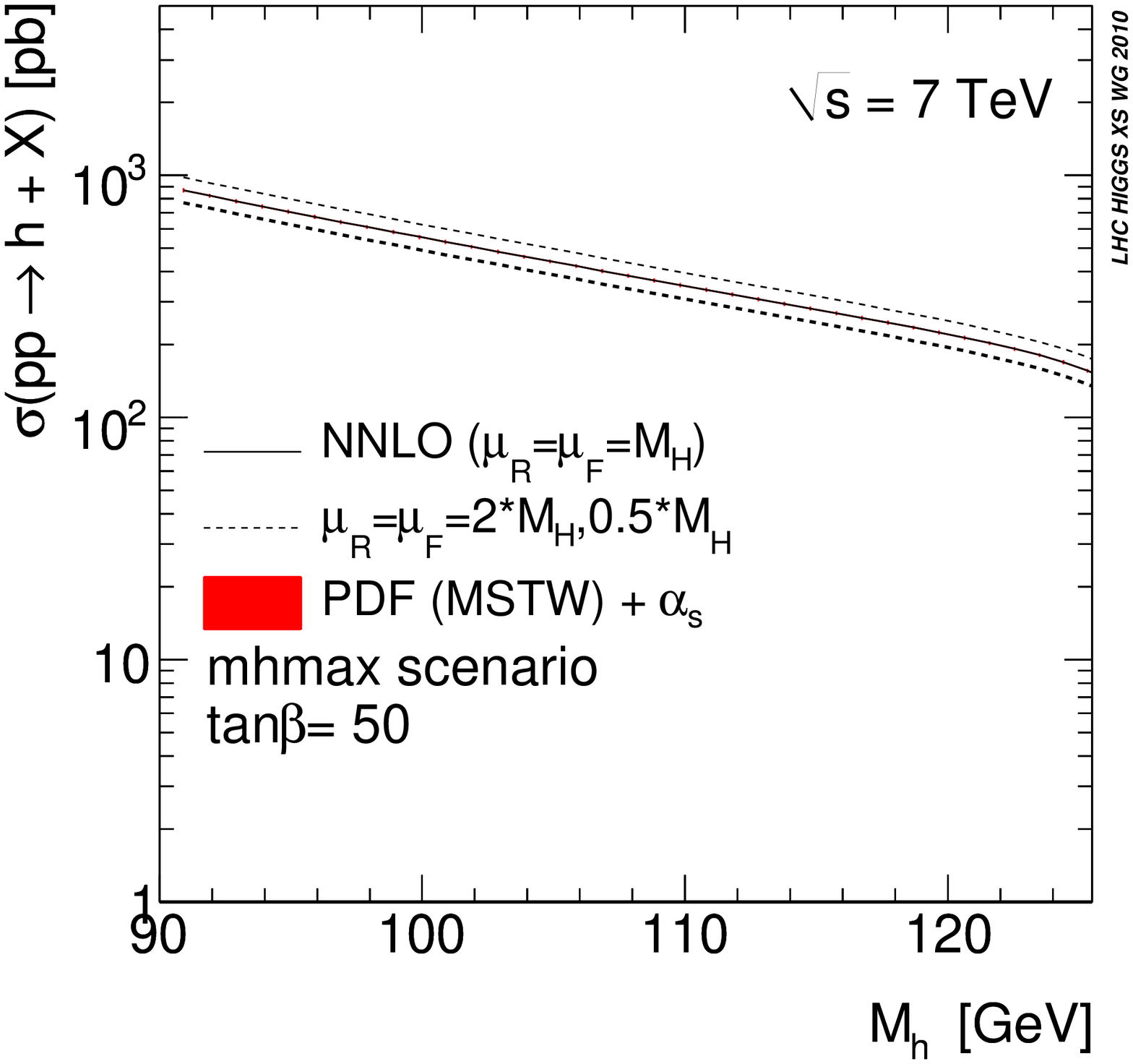}
\caption{\label{YRHXS_MSSM_neutral_fig2} Total gluon-fusion cross
sections of the light scalar (CP-even) MSSM Higgs boson $\PSh$ for four values of
$\tanb$ within the $\mhmaxx$ scenario for $\sqrt{s}=7$\UTeV\ using
MSTW2008 PDFs \cite{Martin:2009iq,Martin:2009bu}. The NNLO results for
the SM-type contributions have been obtained from the programs
\HIGLU~and \gghnnlo, while the rescaling with MSSM coupling factors has
been done with \FeynHiggs.}

\end{figure}
\begin{figure}[htb]
\includegraphics[width=0.5\textwidth]{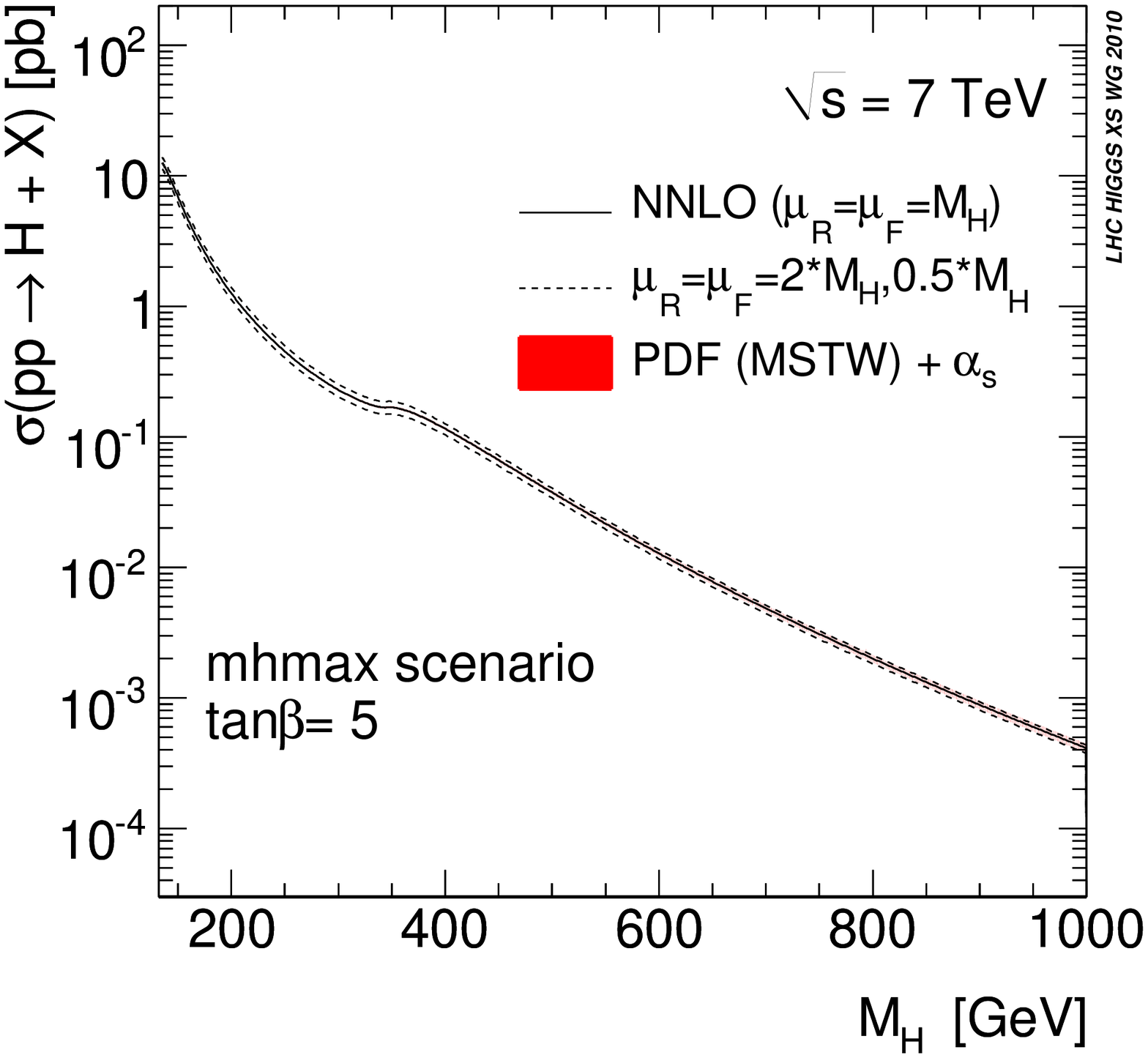}
\includegraphics[width=0.5\textwidth]{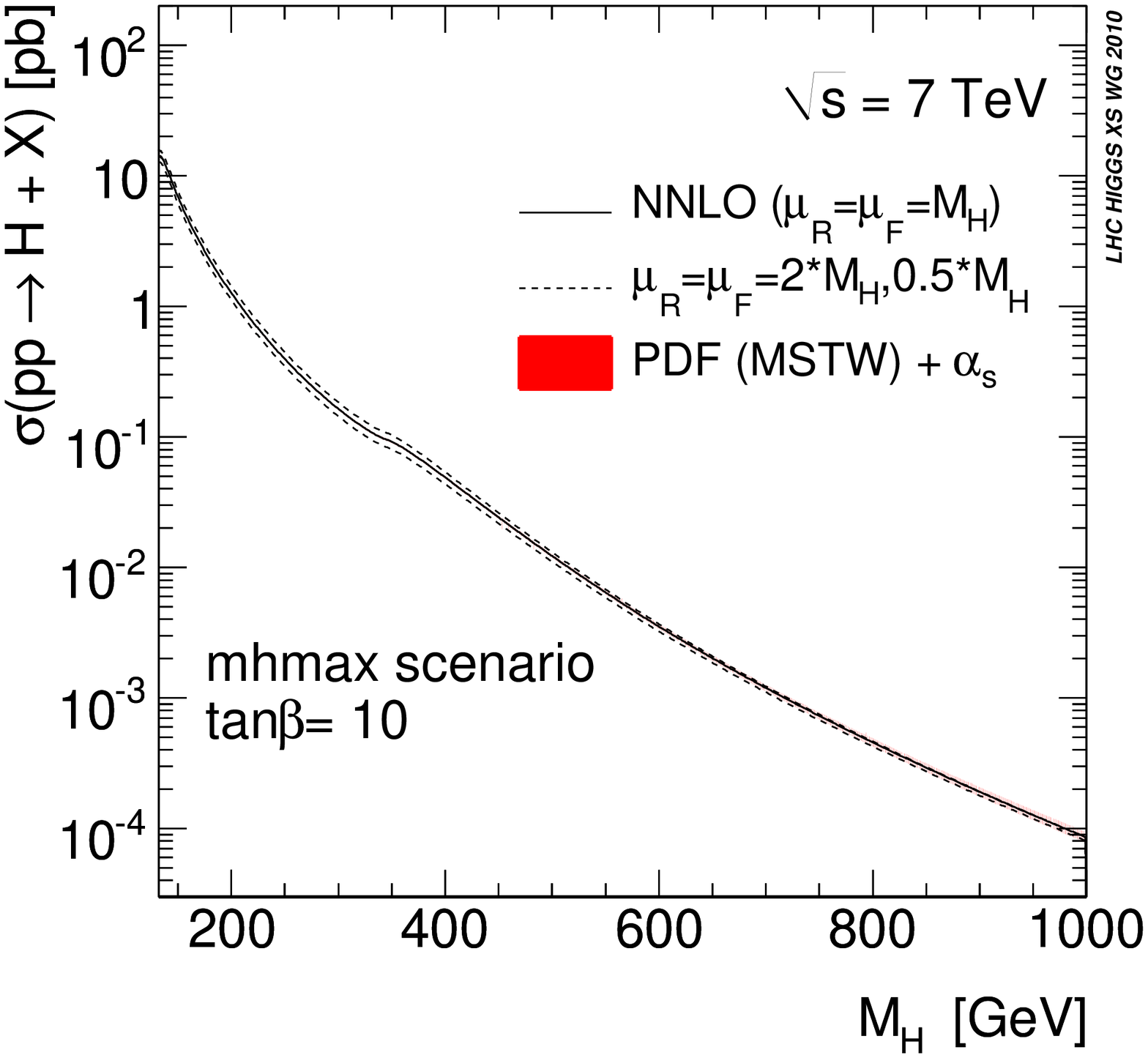}
\includegraphics[width=0.5\textwidth]{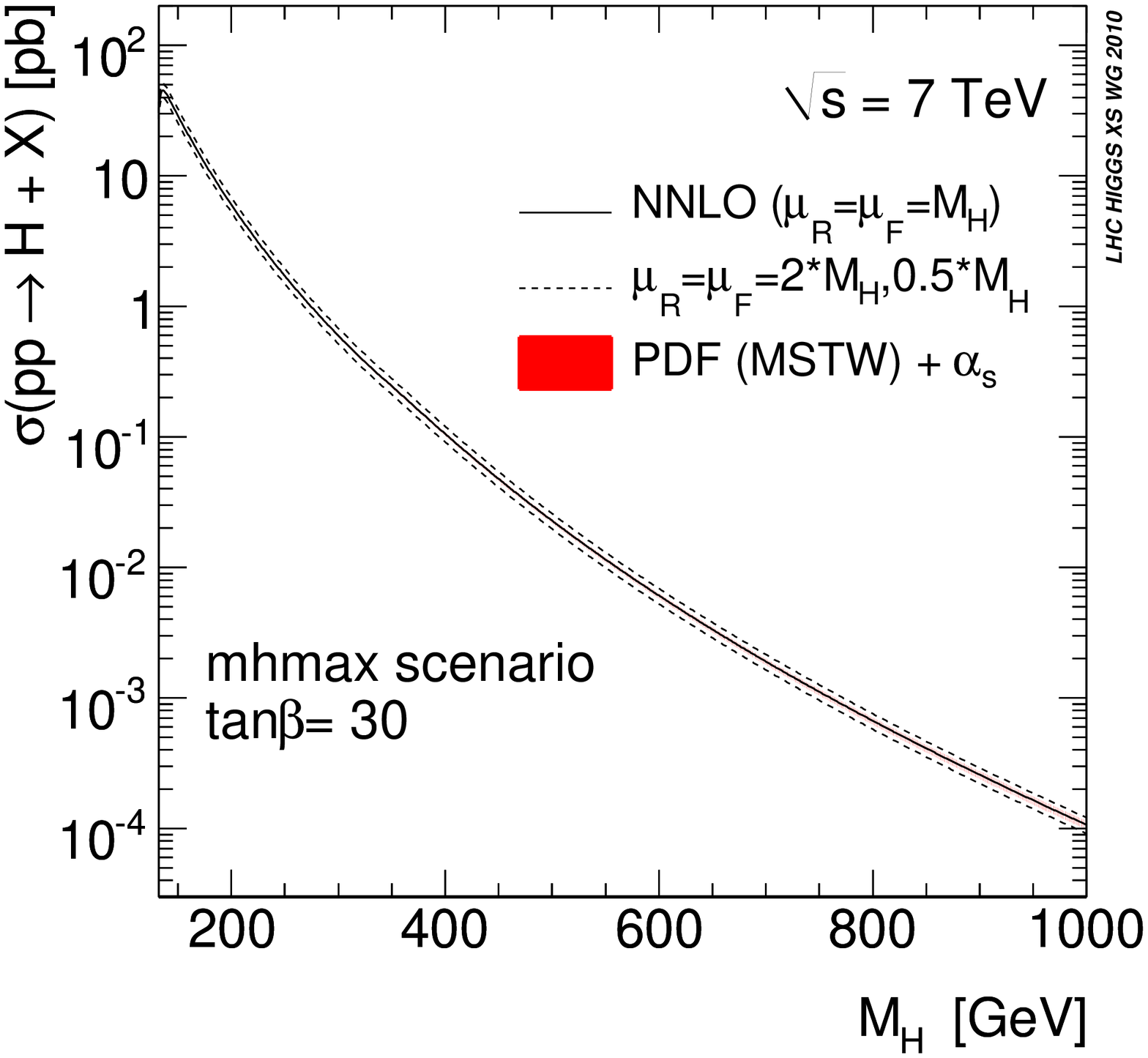}
\includegraphics[width=0.5\textwidth]{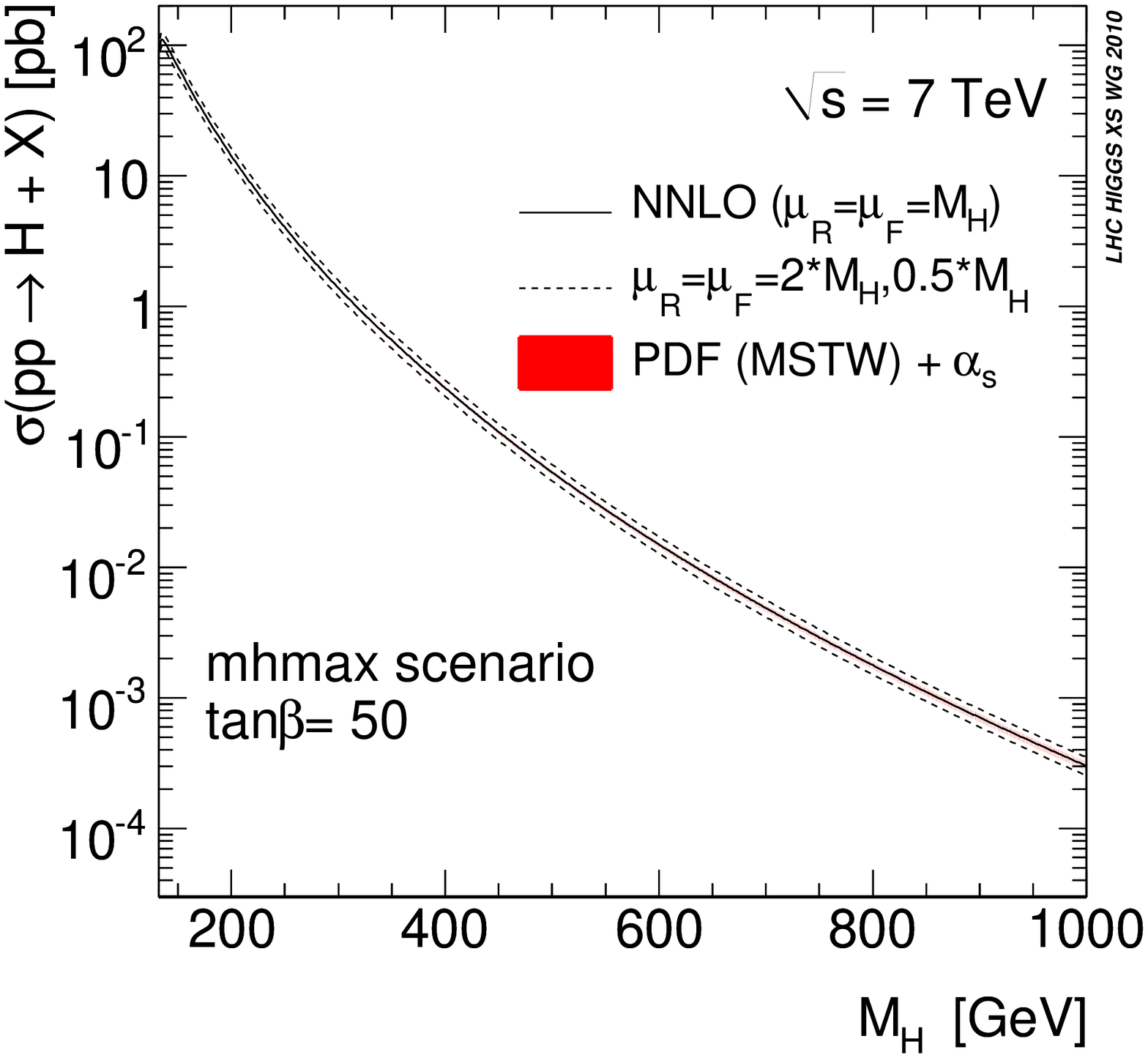}
\caption{\label{YRHXS_MSSM_neutral_fig3} Total gluon-fusion cross
sections of the heavy scalar (CP-even) MSSM Higgs boson $\PSH$ for four values of
$\tanb$ within the $\mhmaxx$ scenario for $\sqrt{s}=7$\UTeV\ using
MSTW2008 PDFs \cite{Martin:2009iq,Martin:2009bu}. The NNLO results for
the SM-type contributions have been obtained from the programs
\HIGLU~and \gghnnlo, while the rescaling with MSSM coupling factors has
been done with \FeynHiggs.}
\end{figure}

As the next step the inclusion of the full stop and sbottom loop
contributions has to be performed. This requires the generation of
multi-dimensional grids of the squark contributions including their
interference terms with the top and bottom contributions as well as
among each other along the same lines as in
Eq.~(\ref{YRHXS_MSSM_neutral_eq1}). This step, however, is beyond the
present write-up. The omission of the squark contributions as well as
the full SUSY QCD corrections to the gluon-fusion cross sections has to
be interpreted as an additional theoretical uncertainty on top of the
scale and PDF+$\alphas$ uncertainties. Since the corrections originating
from the $\Delta_{\PQb}$ terms are smaller than about $10\%$ in the
$\mhmaxx$ scenario, their impact on the overall uncertainties is of
moderate size. Since the full SUSY QCD corrections to the gluon-fusion
cross sections have not been included in our analysis, we take the
contribution of the $\Delta_{\PQb}$ terms as an estimate of the
uncertainties related to these corrections. The total uncertainties of
our gluon-fusion results can be estimated as $\sim 25{-}30\%$ within the
$\mhmaxx$ scenario.


\subsection{Higgs radiation off bottom quarks}
We have generated grids for the 5FS calculation of $\PQb\PAQb\to\phi$
and the 4FS calculation of $\Pg\Pg,\PQq\PAQq\to \PQb\PAQb\phi$. The
Higgs mass range $80{-}200$\UGeV\ has been covered with steps of $5\UGeV$ and
the range $200\UGeV{-}1\UTeV$ with steps of $20$\UGeV.

For the 5FS calculation we have used the program \bbhnnlo~\cite{Harlander:2003ai}. 
Scalar and pseudoscalar Higgs-boson production
are identical for the same masses and the same coupling factors due to
the chiral symmetry of massless bottom quarks.  The input value of the
$\overline{\rm MS}$ bottom mass has been chosen as
$\overline{m}_b(\overline{m}_b)=4.213$\UGeV~which corresponds to a NNLO
pole mass of $\Mb=4.75$\UGeV, i.e.~the bottom mass value of the MSTW2008
PDF sets \cite{Martin:2009iq,Martin:2009bu}. For the 5FS the NNLO PDFs
of MSTW2008 have been adopted with the strong coupling adjusted
accordingly, i.e.~$\alphas(\MZ)=0.11707$. As central scales we have
chosen $\mu_R=M_\phi$ for the renormalization scale and $\mu_F=M_\phi/4$
for the factorization scale, respectively. For the scale uncertainties
of the 5FS we have varied the scales in the intervals $M_\phi/5 < \mu_R
< 5 M_\phi$ and $M_\phi/10 < \mu_F <0.7 M_\phi$. These ranges
cover the maximal and minimal values of the cross sections within the
5FS. The central predictions of the 5FS calculation are shown in
\Fref{YRHXS_MSSM_neutral_fig4}a for SM-like couplings. These cross
sections have to be multiplied with the ratios
$\left(g_{\PQb}^{\mathrm{MSSM}}/g_{\PQb}^{\mathrm{SM}}\right)^2$ of the MSSM and SM Yukawa
couplings. The MSSM couplings $g_{\PQb}^{\mathrm{MSSM}}$ should contain the
$\Delta_{\PQb}$
terms \cite{Hall:1993gn,Hempfling:1993kv, Carena:1994bv,Pierce:1996zz,
Carena:1999py,Guasch:2003cv,Noth:2008tw,Noth:2010jy,Mihaila:2010mp}, 
since they approximate the full
genuine SUSY QCD \cite{Hafliger:2006zz,Walser:2008zz} and
SUSY electroweak \cite{Dittmaier:2006cz}
corrections within the percent level. The corresponding scale
uncertainties are shown in \Fref{YRHXS_MSSM_neutral_fig4}b. They
amount to less than $10\%$ for Higgs masses above about 200\UGeV, while for
smaller Higgs masses they can reach a level of $30{-}40\%$ as can be
inferred from \Fref{YRHXS_MSSM_neutral_fig4}b. The 68\% CL
PDF+$\alphas$ uncertainties are displayed in
\Fref{YRHXS_MSSM_neutral_fig4}c and the 90\% CL uncertainties in
\Fref{YRHXS_MSSM_neutral_fig4}d. The 68\% CL~uncertainties amount
to less than about $10\%$ in the relevant Higgs mass range below $\sim
500{-}600$\UGeV, while they are enhanced to a level below about $20\%$ at
90\% CL as shown in \Fref{YRHXS_MSSM_neutral_fig4}d. It is also
visible that these uncertainties are dominated by the pure PDF
uncertainties, while the $\alphas$ variation adds only a moderate
contribution.
\begin{figure}[htb]

\subfigure[]{\includegraphics[width=0.5\textwidth]{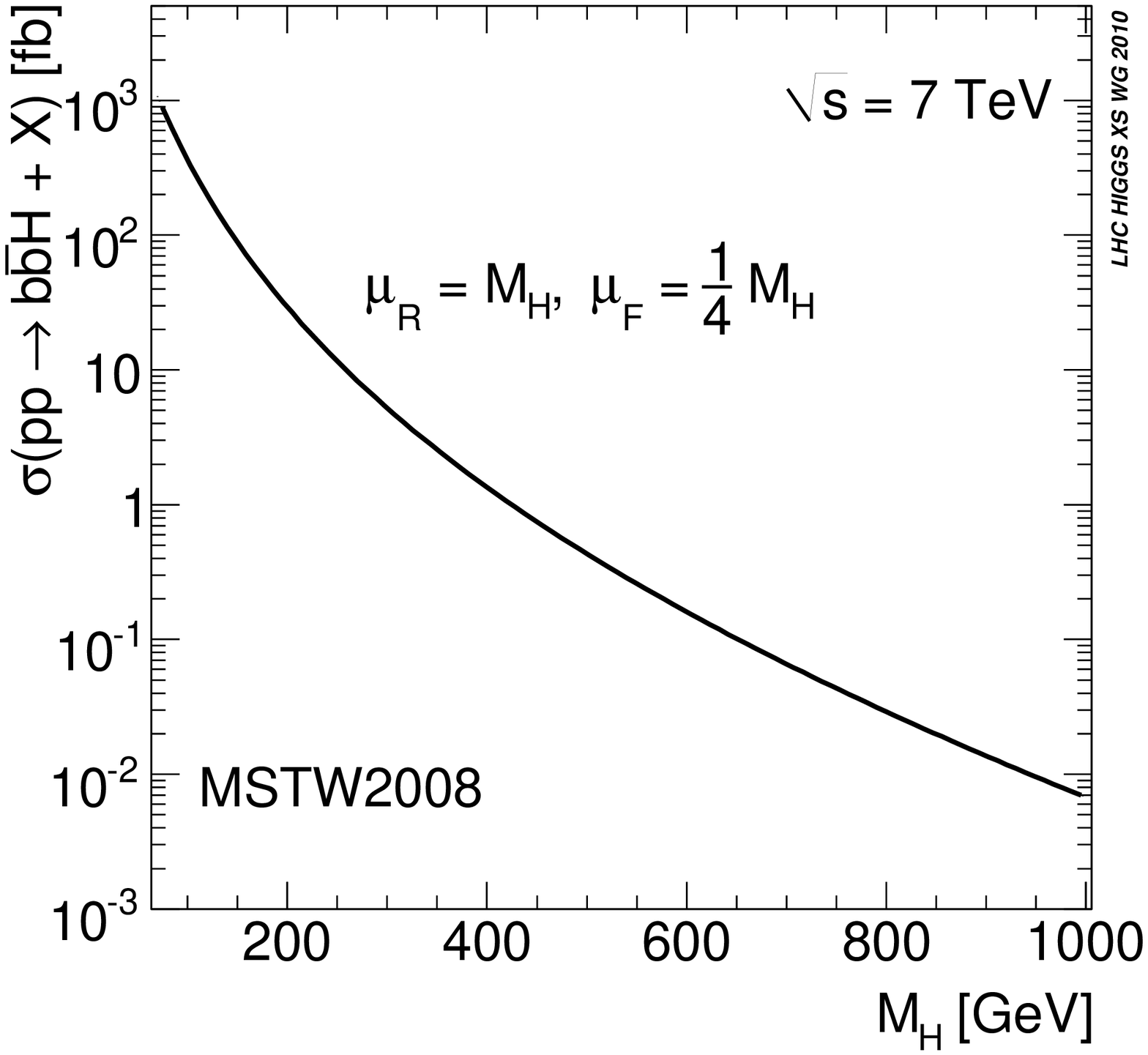}}
\subfigure[]{\includegraphics[width=0.5\textwidth]{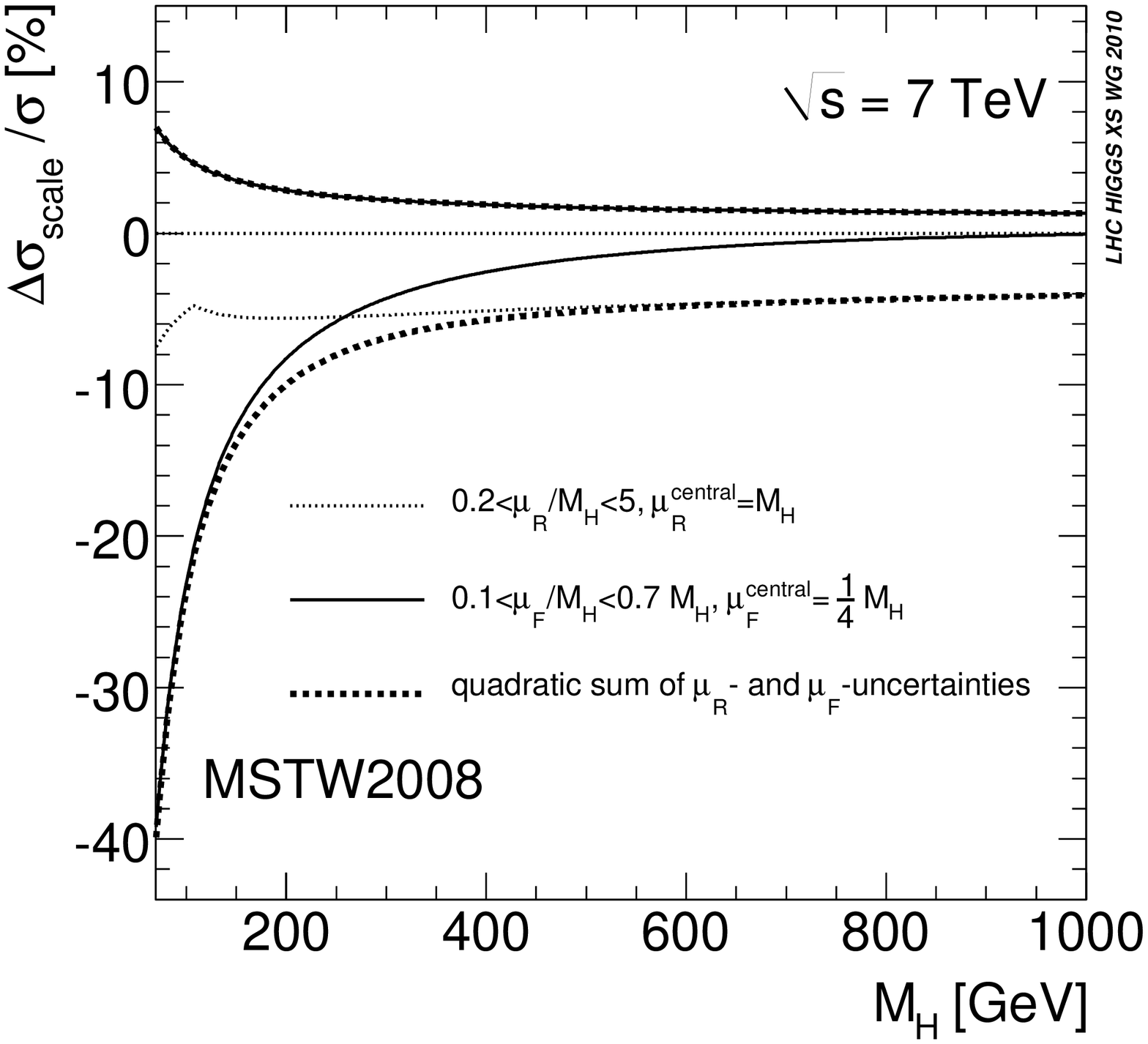}}
\subfigure[]{\includegraphics[width=0.5\textwidth]{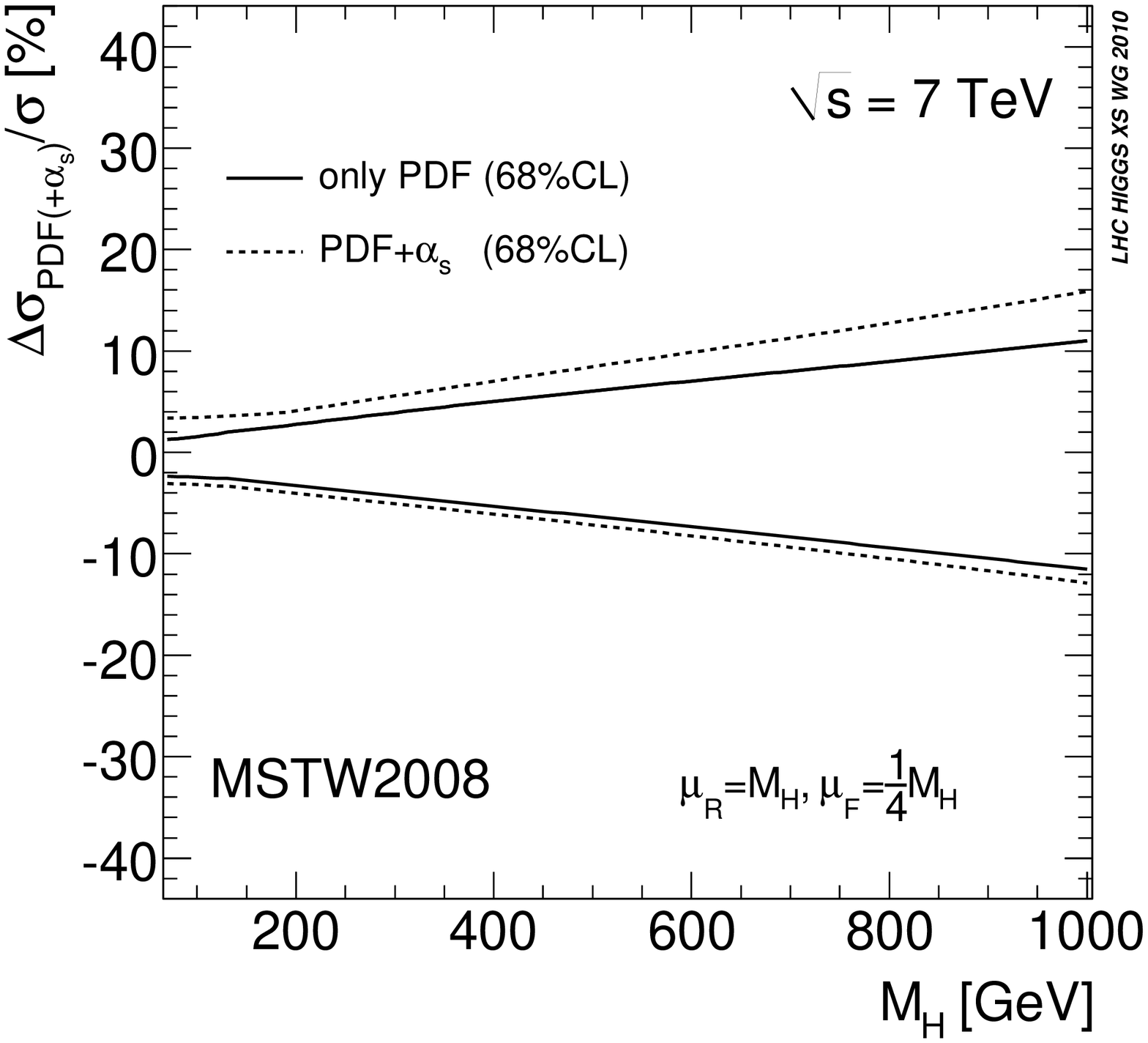}}
\subfigure[]{\includegraphics[width=0.5\textwidth]{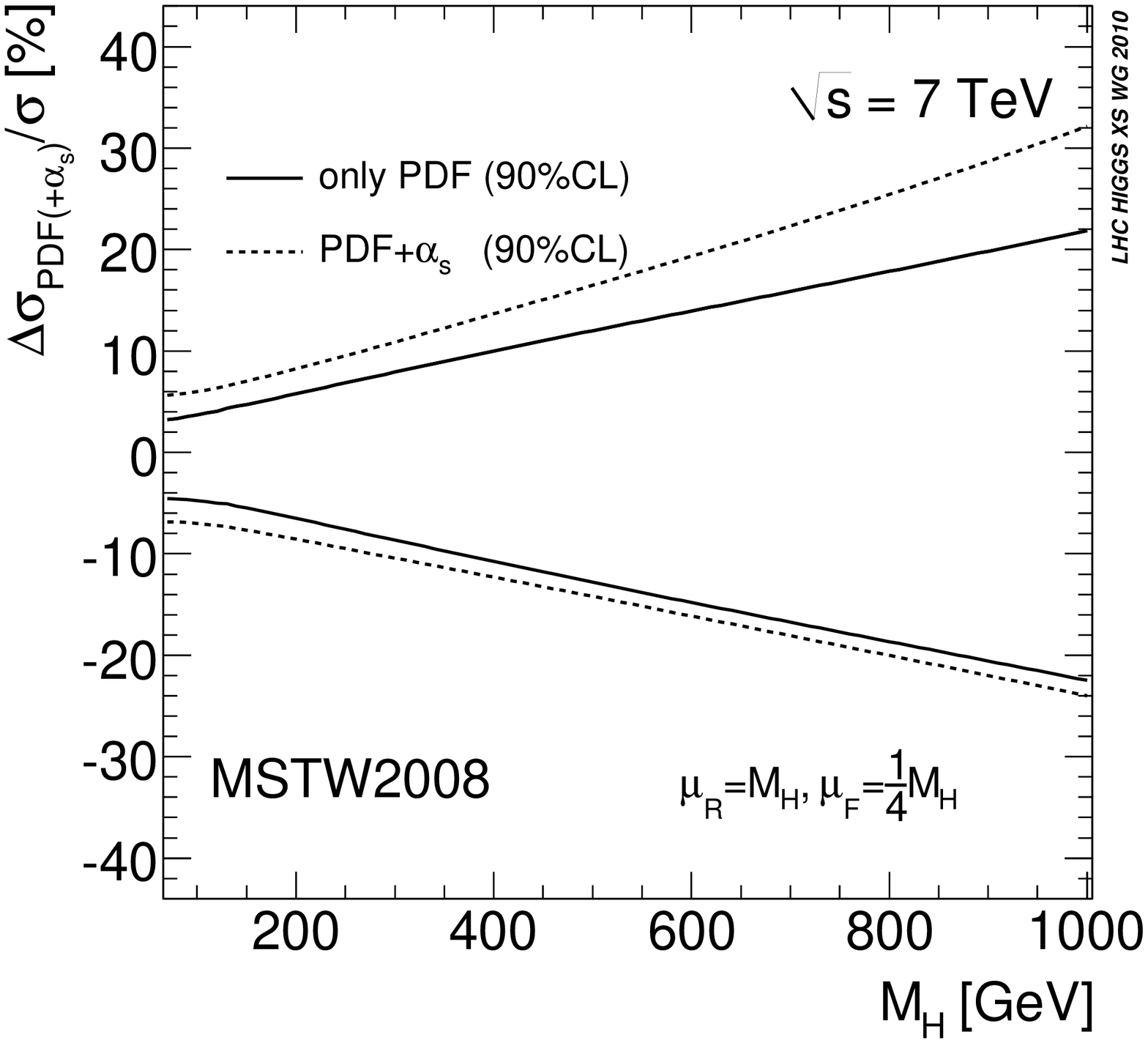}}
\caption{\label{YRHXS_MSSM_neutral_fig4} Total bottom-quark-fusion
cross sections of $\PQb\PAQb\to \PSH/\PSA+X$ within the 5FS for $\sqrt{s}=7$\UTeV\ with SM-like couplings using MSTW2008 PDFs
\cite{Martin:2009iq,Martin:2009bu}; (a) central prediction, (b) scale
uncertainties, (c) 68\% CL~PDF+$\alphas$ uncertainties, (d) 90\%
CL~PDF+$\alphas$ uncertainties.}
\end{figure}

In the corresponding 4FS calculation we have chosen the bottom-quark
pole mass as $\Mb=4.75$\UGeV~which corresponds to a NLO 
$\MSbar$
mass $\overline{m}_\PQb(\overline{m}_\PQb)=4.40$\UGeV. The closed top loop
contributions appearing in the virtual one-loop contributions have been
neglected for consistency, since for large values of $\tanb$ they
are strongly suppressed and in the 5FS calculation they vanish for
strictly massless bottom quarks. In the further progress of this study
we will generate separate grids for these top loop contributions so that
they can be included in the MSSM calculations consistently. The running
bottom-quark Yukawa coupling, expressed in terms of the $\overline{\rm
MS}$ bottom mass, has been chosen at the scale of the Higgs mass
$M_\phi$. The central scales $\mu_R=\mu_F=M_\phi/4$ have been adopted
for the renormalization and factorization scales, respectively. The
scale uncertainties have been obtained for scale variations
$M_\phi/8 < \mu_R,\mu_F < M_\phi/2$ where the choice $\mu_R=\mu_F =
M_\phi/8$ corresponds to the maximal cross sections and $\mu_R=\mu_F =
M_\phi/2$ to the minimal cross sections for all Higgs masses. The
four-flavour PDFs of MSTW2008 \cite{Martin:2010db} have been used for
the numerical analysis within the 4FS. Error PDFs within this scheme
have only been published very recently so that a full PDF uncertainty
analysis could not be performed for the 4FS yet.  However, the scale
uncertainties of $25{-}30\%$ are expected to dominate the overall
uncertainties of the 4FS calculation so that the additional
PDF+$\alphas$ uncertainties will be expected to modify the overall
uncertainties only mildly. The comparison of the 4FS and the 5FS for
SM-like couplings is shown in \Fref{YRHXS_MSSM_neutral_fig5} for
scalar and pseudoscalar Higgs-boson production in association with
bottom quarks.  The scalar and pseudoscalar cross sections for the same
mass differ by less than $2\%$ within the 4FS thanks to the approximate
chiral symmetry for the light bottom quarks compared to the Higgs-boson
masses.  \Fref{YRHXS_MSSM_neutral_fig5} shows good agreement of the
5FS and 4FS results for smaller Higgs masses while for large Higgs-boson
masses the 5FS cross sections are considerably larger than the
corresponding 4FS results. However, an overlap of both uncertainty bands
is visible for the whole mass range. This is the first completely
consistent comparison of both schemes resulting in a much better
agreement of both schemes than in all former studies
\cite{Campbell:2004pu}. The central values of the 4FS and 5FS differ by
up to $30\%$. In order to decide which of the two prescriptions is closer
to the experimentally relevant values of these production cross
sections, the comparison of the 4FS and 5FS calculations of $\PQb\PAQb{+}\PZ$
production with the forthcoming experimental data will be of big help.
\begin{figure}[htb]
\includegraphics[width=0.5\textwidth]{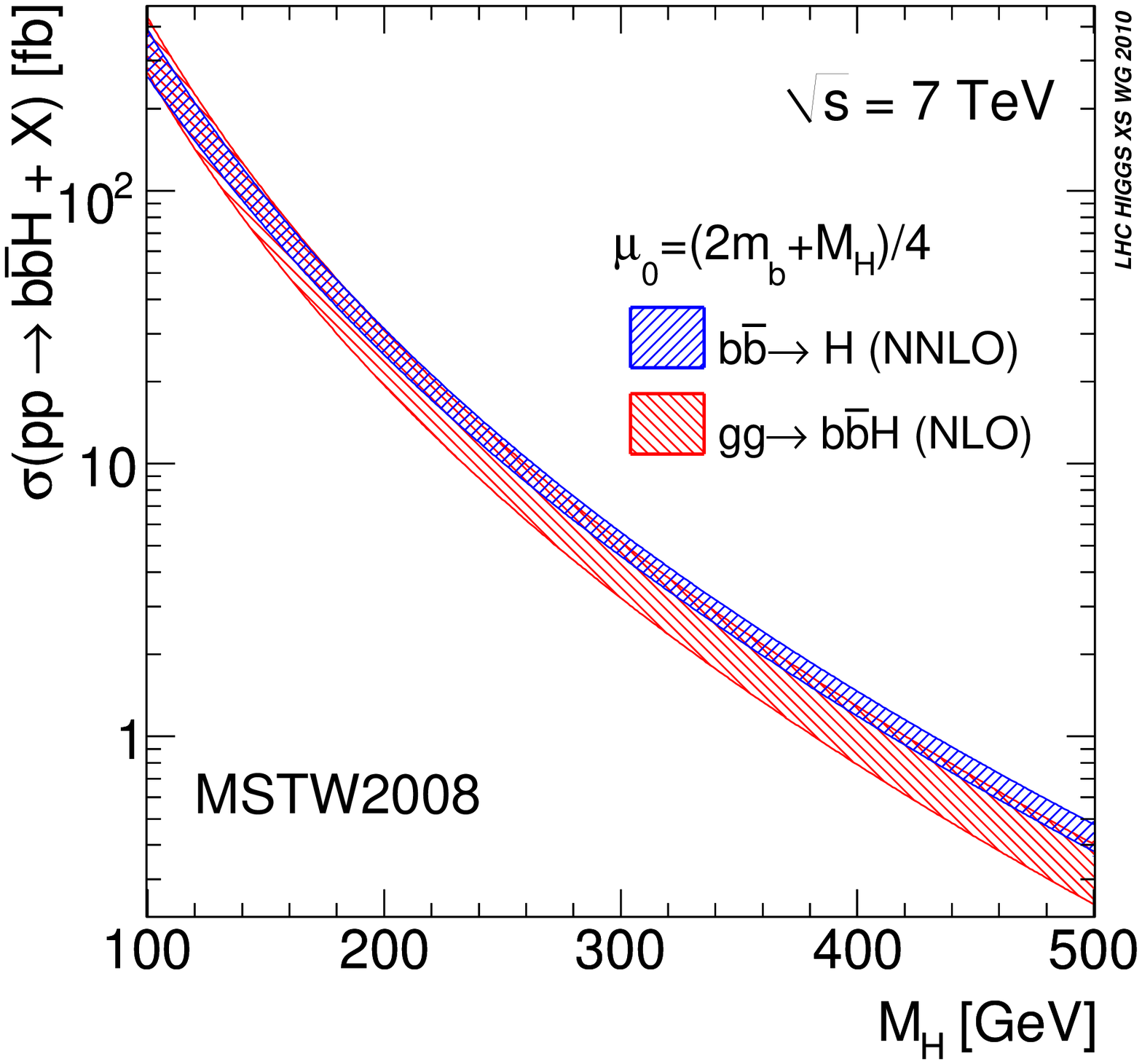}
\includegraphics[width=0.5\textwidth]{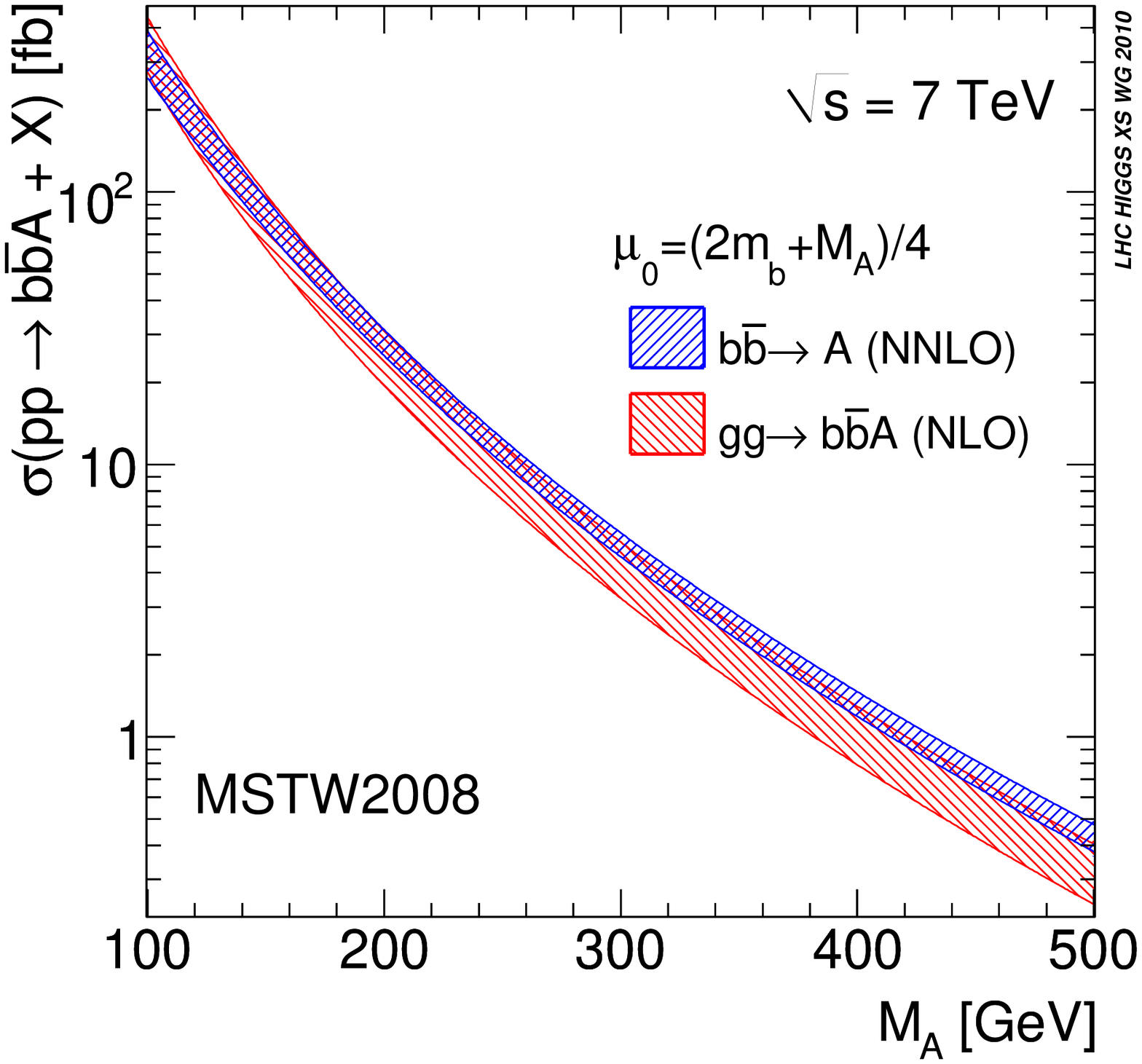}
\caption{\label{YRHXS_MSSM_neutral_fig5} Total production cross
sections of $\Pp\Pp\to \PQb\PAQb\PSH/\PSA+X$ for $\sqrt{s}=7$
\UTeV~within the 5FS and the 4FS using MSTW2008 PDFs
\cite{Martin:2009iq,Martin:2009bu}. The upper bands (blue bands) exhibit the combined
scale and 68\% CL~PDF+$\alphas$ uncertainties of the 5FS, while the
lower bands (red
bands) include the scale uncertainties of the 4FS only.} \end{figure}

In \Fref{YRHXS_MSSM_neutral_fig6} the central predictions for the
gluon-fusion processes $\Pg\Pg\to \PSh,\PSH,\PSA$ and neutral Higgs
radiation off bottom quarks within the 5FS are shown as a function of
the corresponding Higgs mass within the $\mhmaxx$ scenario for two
values of $\tanb=5,30$. These results have been obtained from the grids
generated by \gghnnlo~and \HIGLU~for the gluon-fusion process and
\bbhnnlo~for $\PQb\PAQb\to \phi$ and rescaling the corresponding Yukawa
couplings by the MSSM factors calculated with 
\FeynHiggs{}\footnote{Two complete scans of the ($M_{\PA},\tan\beta)$ 
plane for $\sqrt{s}=7$ and $14$\UTeV\ are available in electronic format 
on {\tt https://twiki.cern.ch/twiki/bin/view/LHCPhysics/MSSMNeutral} 
for the \mhmaxx scenario.}. It is clearly
visible that Higgs-boson radiation off bottom quarks plays the dominant
role for $\tanb=30$ while for $\tanb=5$ the gluon fusion is either
dominant or competitive.
\begin{figure}[htb]
\subfigure[]{\includegraphics[width=0.5\textwidth]{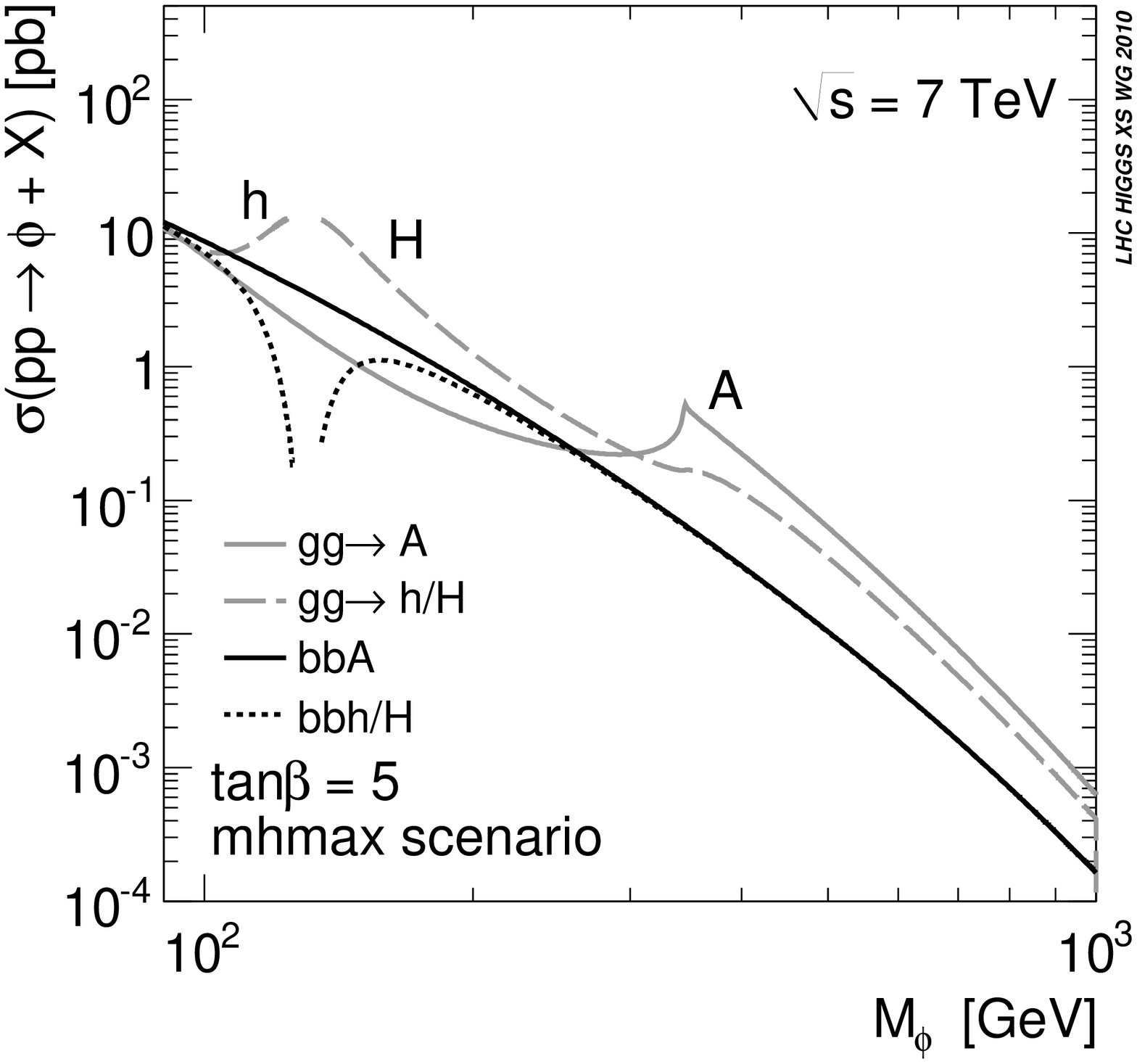}}
\subfigure[]{\includegraphics[width=0.5\textwidth]{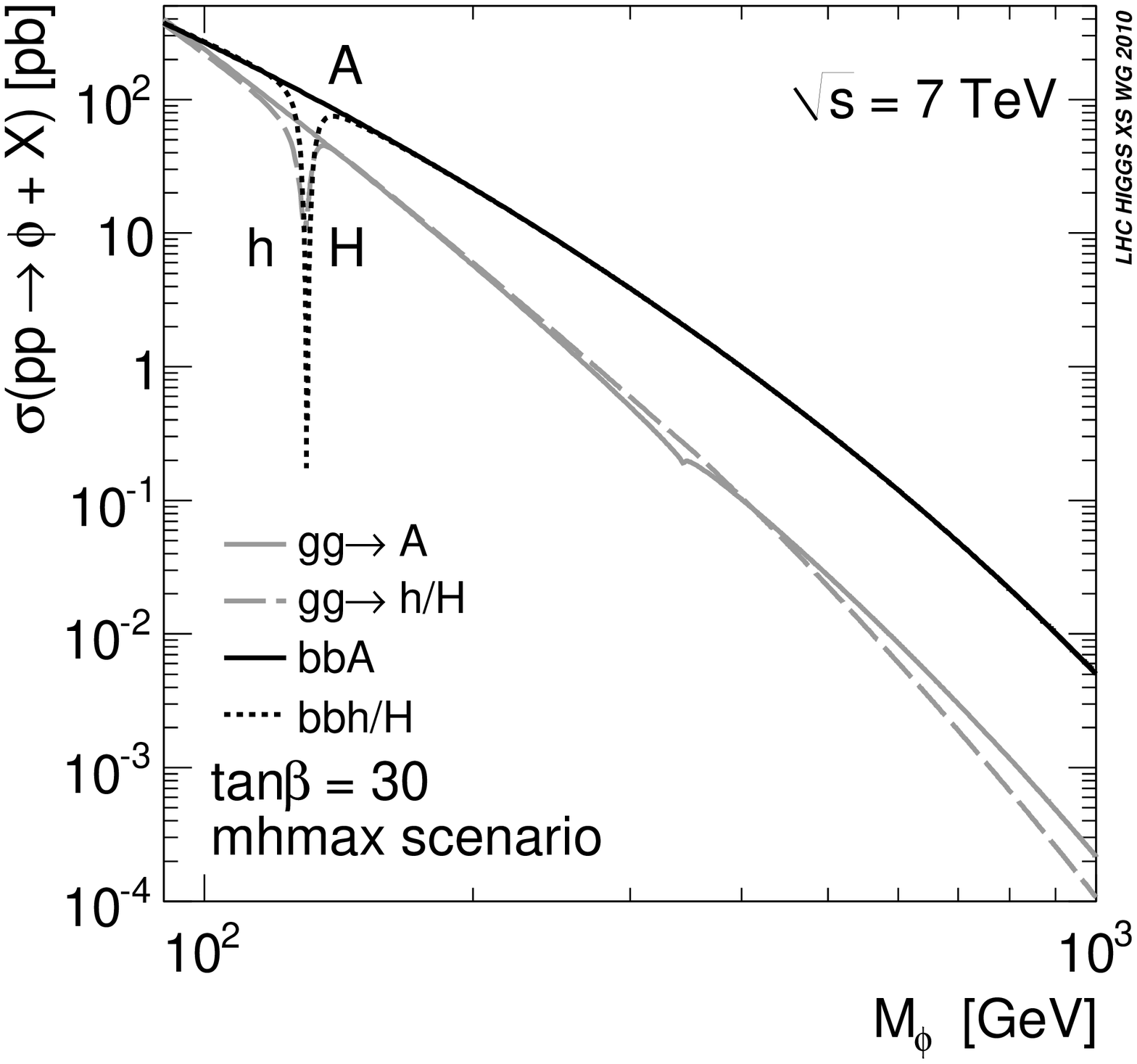}}
\caption{\label{YRHXS_MSSM_neutral_fig6} Central predictions for the
total MSSM production cross sections via gluon fusion and Higgs radiation off
bottom quarks within the 5FS for $\sqrt{s}=7$\UTeV\ using NNLO and NLO
MSTW2008 PDFs \cite{Martin:2009iq,Martin:2009bu} for the $\mhmaxx$
scenario; (a) $\tanb=5$, (b) $\tanb=30$.}
\end{figure}

\clearpage

\newcommand{\MHpm}{M_{\PSHpm}}

\section{MSSM charged Higgs production process\footnote{M.~Flechl,
    M.~Kr\"amer, S.~Lehti (eds.); S.~Dittmaier, T.~Hahn, 
    T.~Hartonen, S.~Heinemeyer, J.~S.~Lee, A.~Pilaftsis, 
    M.~Spira and C.~Weydert.}}
\label{sec:chiggs_intro}

Many extensions of the Standard Model, in particular supersymmetric
theories, require two Higgs doublets leading to five physical scalar
Higgs bosons, including two (mass-degenerate) charged particles $\PSHpm$.
The discovery of a charged Higgs boson would provide
unambiguous evidence for an extended Higgs sector beyond the Standard
Model. Searches at LEP have set a limit $\MHpm > 79.3$\UGeV\ 
on the mass of a charged Higgs boson in a general two-Higgs-doublet
model~\cite{Heister:2002ev}. Within the MSSM, the charged Higgs-boson
mass is constrained by the pseudoscalar Higgs mass and the W-boson
mass through $\MHpm^2 = \MA^2 + \MW^2$ at
tree level, with only moderate higher-order
corrections~\cite{Gunion:1988pc, Brignole:1991wp, Diaz:1991ki,
  Frank:2006yh}. A mass limit on the MSSM charged Higgs boson can thus
be derived from the limit on the pseudoscalar Higgs boson, 
$\MA > 93.4$\UGeV~\cite{Schael:2006cr}, resulting in 
$\MHpm \gsim 120$\UGeV. At the Tevatron, searches for light charged
Higgs bosons in top-quark decays $\Pt \to \Pb \PSHpm$~\cite{Aaltonen:2009ke,:2009zh} have placed some constraints
on the MSSM parameter space, but do not provide any further generic
bounds on $\MHpm$.

There are two main mechanisms for charged Higgs-boson production at
the LHC:
\begin{displaymath}
\begin{array}{lcl}
  \mbox{top-quark decay:} & \Pt \to \Pb \PSHpm{\rm + X} &\; {\rm if}\;\;
  \MHpm \lsim m_{\rm t}\,, \\
  \mbox{associated production:} & \Pp\Pp \to \Pt \Pb\PSHpm{\rm + X}
  &\; {\rm if}\;\; \MHpm \gsim \Mt\,.
\end{array}
\end{displaymath}
Alternative production mechanisms like quark--antiquark annihilation
$\Pq\bar \Pq'\to \PSHpm$ and $\PSHpm+\mathrm{jet}$ production~\cite{Dittmaier:2007uw},
associated $\PSHpm \PWmp$ production~\cite{Eriksson:2006yt}, or 
Higgs pair production~\cite{Alves:2005kr,Brein:1999sy} have
suppressed rates, and it is not yet clear whether a signal could be
established in any of those channels. 
Some of the above production processes may,
however, be enhanced in models with non-minimal flavour violation.

In this section we discuss charged Higgs-boson production in $\Pt \to
\Pb \PSHpm$ decays and compare the results of different software
packages for the calculation of this branching ratio. Furthermore, we
present NLO QCD predictions for the process $\Pp\Pp \to \Pt\Pb 
\PSHpm{\rm + X}$ in the four- and five-flavour schemes.

\subsection{Light charged Higgs production from top-quark decays}

If the charged Higgs boson is light, $\MHpm \lsim \Mt$, 
it is produced in top-quark decays. The branching ratio
calculation of the top quark to a light charged Higgs boson is
compared for two different programs, {\sc FeynHiggs}, version
2.7.3~\cite{Heinemeyer:1998yj,Heinemeyer:1998np,Degrassi:2002fi,Frank:2006yh},
and {\sc CPsuperH}, version 2.2~\cite{Lee:2003nta,Lee:2007gn}. 
We note that the decay  $\Pt \to \PSHp \Pb$ is also included in {\sc HDECAY}~\cite{Djouadi:1997yw}, 
which has however not been included in the comparison presented here. 
The
$m_h^{\rm max}$ benchmark scenario was used~\cite{Carena:2002qg},
which in the on-shell scheme is defined as described 
in \Eq~(\ref{YRHXS_MSSM_neutral_eq:mhmax}).
In addition to $\tan \beta$ and $\MHpm$, the $\mu$ parameter
was varied with values $\pm 1000, \pm 200$\UGeV~\cite{Carena:2005ek}. 
The Standard Model parameters were taken as given in the 
Appendix \Table~\ref{tab:SMinput}.

The calculation within {\sc FeynHiggs} is based on the evaluations of
$\Gamma(\Pt \to \PWp \Pb)$ and $\Gamma(\Pt \to \PSHp \Pb)$. The former is
calculated at NLO according to \Bref{Campbell:2004ch}. The decay
to the charged Higgs boson and the bottom quark uses $\Mt(\Mt)$ and 
$\Mb(\Mt)$ in the Yukawa coupling, where the
latter receives the additional correction factor $1/(1 +
\Delta_{\Pb})$. The leading contribution to $\Delta_{\Pb}$ is given
by~\cite{Carena:1999py}
\begin{equation}
\label{eq:qcd_tanb_enhanced}
\Delta_{\Pb} = \frac{C_F}{2}\frac{\alphas}{\pi} 
m_{\tilde{\Pg}} \mu \tan\beta  \,
I(m_{\tilde{\Pb}_1},m_{\tilde{\Pb}_2},m_{\tilde{\Pg}}) \, ,
\end{equation}
with $C_F = 4/3$ and the auxiliary function
\begin{equation}
\label{eq:I}
I(a,b,c) = \frac{1}{(a^2-b^2)(b^2-c^2)(a^2-c^2)} \left(
a^2 b^2 \ln \frac{a^2}{b^2} + 
b^2 c^2 \ln \frac{b^2}{c^2} + 
c^2 a^2 \ln \frac{c^2}{a^2}
\right) \, .
\end{equation}
Here, $\tilde{\Pb}_{1,2}$ are the sbottom mass eigenstates, and
$m_{\tilde{\Pg}}$ is the gluino mass. The numerical results presented
here have been based on the evaluation of $\Delta_{\Pb}$ in
\Bref{Hofer:2009xb}.  
Furthermore additional QCD corrections taken from
\Bref{Carena:1999py} are included.

The calculation within {\sc CPsuperH} is also based on the top-quark
decays $\Pt \to \PWp \Pb$ and $\Pt \to \PSHp \Pb$.  The decay width
$\Gamma(\Pt \to \PWp \Pb)$ is calculated by including ${\cal
  O}(\alphas)$ corrections~\cite{Chetyrkin:1999br}.
The partial decay width of the decay $\Pt \to \PSHp \Pb$ is given by
\begin{eqnarray}
\Gamma(\Pt\rightarrow \PH^+ b)&=&\frac{g_{\Pt\Pb}^2\Mt}{16\pi}
\left(|g^S_{_{{\PH}^+\bar{\Pt}{\rm b}}}|^2+|g^P_{_{\PH^+\bar{\Pt}{\rm b}}}|^2\right)
\left(1-\frac{\MHpm^2}{\Mt^2}\right)^2\,,
\end{eqnarray}
where $g_{\Pt\Pb} = - g \Mt/\sqrt{2} \MW$, 
      $g^S_{\PSHp\bar{\Pt}\Pb}=(g^L_{\PSHp\bar{\Pt}\Pb} + 
      \frac{\Mb}{\Mt}\, g^R_{\PSHp\bar{\Pt}\Pb})/2$, 
      and $g^P_{\PSHp\bar{\Pt}\Pb}=i(g^L_{\PSHp\bar{\Pt}\Pb} 
      - \frac{\Mb}{\Mt}\, g^R_{\PSHp\bar{\Pt}\Pb})/2$.  
In the couplings $g^{L,R}_{\PSHp\bar{\Pt}\Pb}$, all the
threshold corrections (both those enhanced and not enhanced by
$\tan\beta$) have been included as described in Appendix A of
\Bref{Lee:2003nta} and \Refs~\cite{Carena:2002bb,Ellis:2009di},
see also \Bref{Guasch:2003cv}. 
For $\Mt$ and $\Mb$ appearing in the couplings, we use the
quark masses evaluated at the scale $\MHpm$.

The comparison started by running {\sc FeynHiggs} with a selected set
of parameters. The {\sc FeynHiggs} output was used to set the values
for the CPsuperH input parameters. Due to differences in the parameter
definitions, the bottom-quark mass was changed from {\sc FeynHiggs}
$\Mb(\Mb) = 4.16$\UGeV\ to $\Mb(\Mt) = 2.64$\UGeV\ 
which is taken as input by {\sc CPsuperH}. The main result
from the comparison is shown in Figs.~\ref{fig:BRTop2HPlus} and
\ref{fig:BRTop2HPlusDiff}. A very good agreement within typically $0{-}2\%$ 
can be observed, except if simultaneously very small values of $\mu$, 
high $\tan\beta$, and relatively small $\MHpm$ are chosen.

\begin{figure}[ht]
  \centering
\begin{tabular}{cc}
  \includegraphics[width=0.46\textwidth]{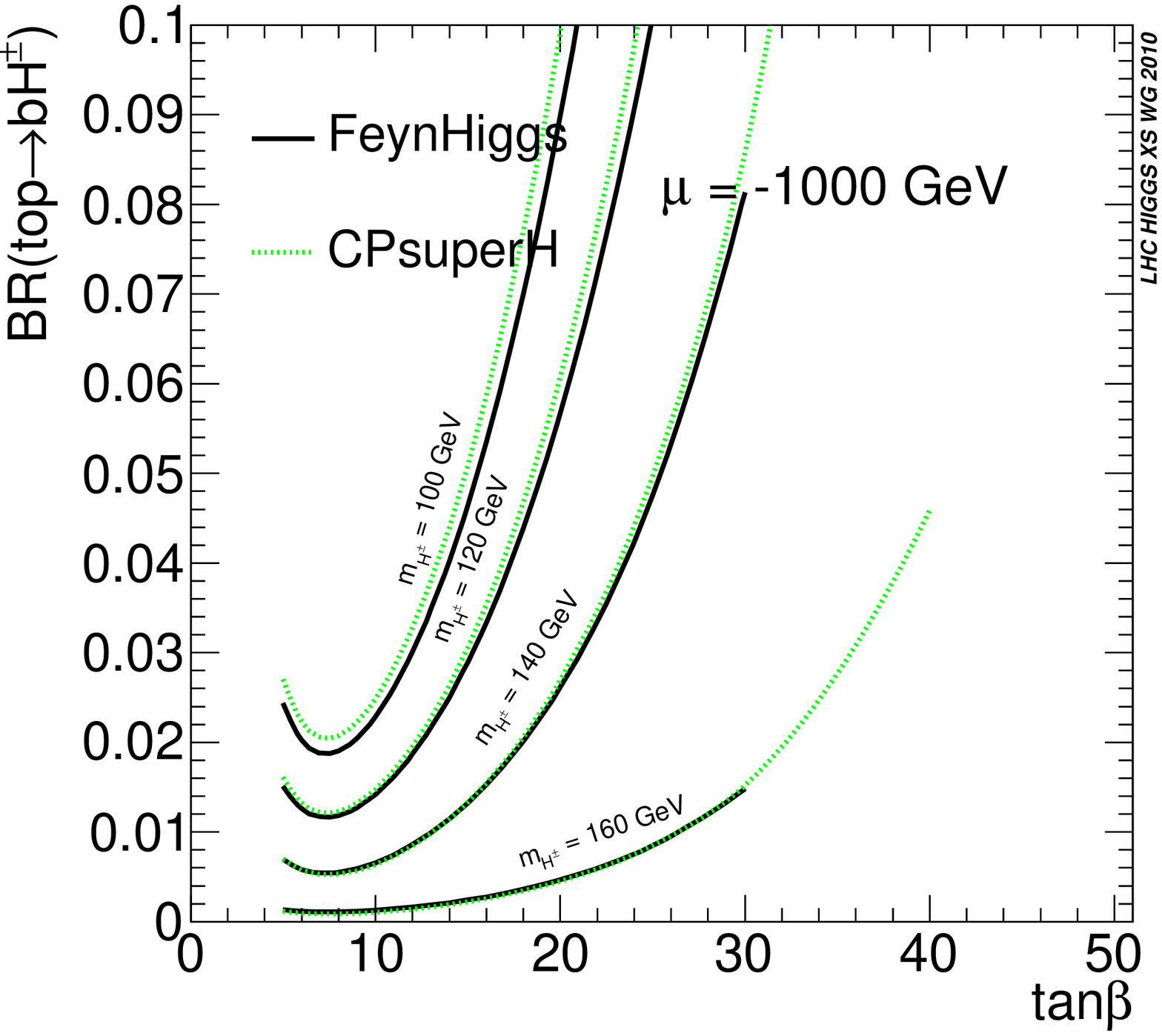} &
  \includegraphics[width=0.46\textwidth]{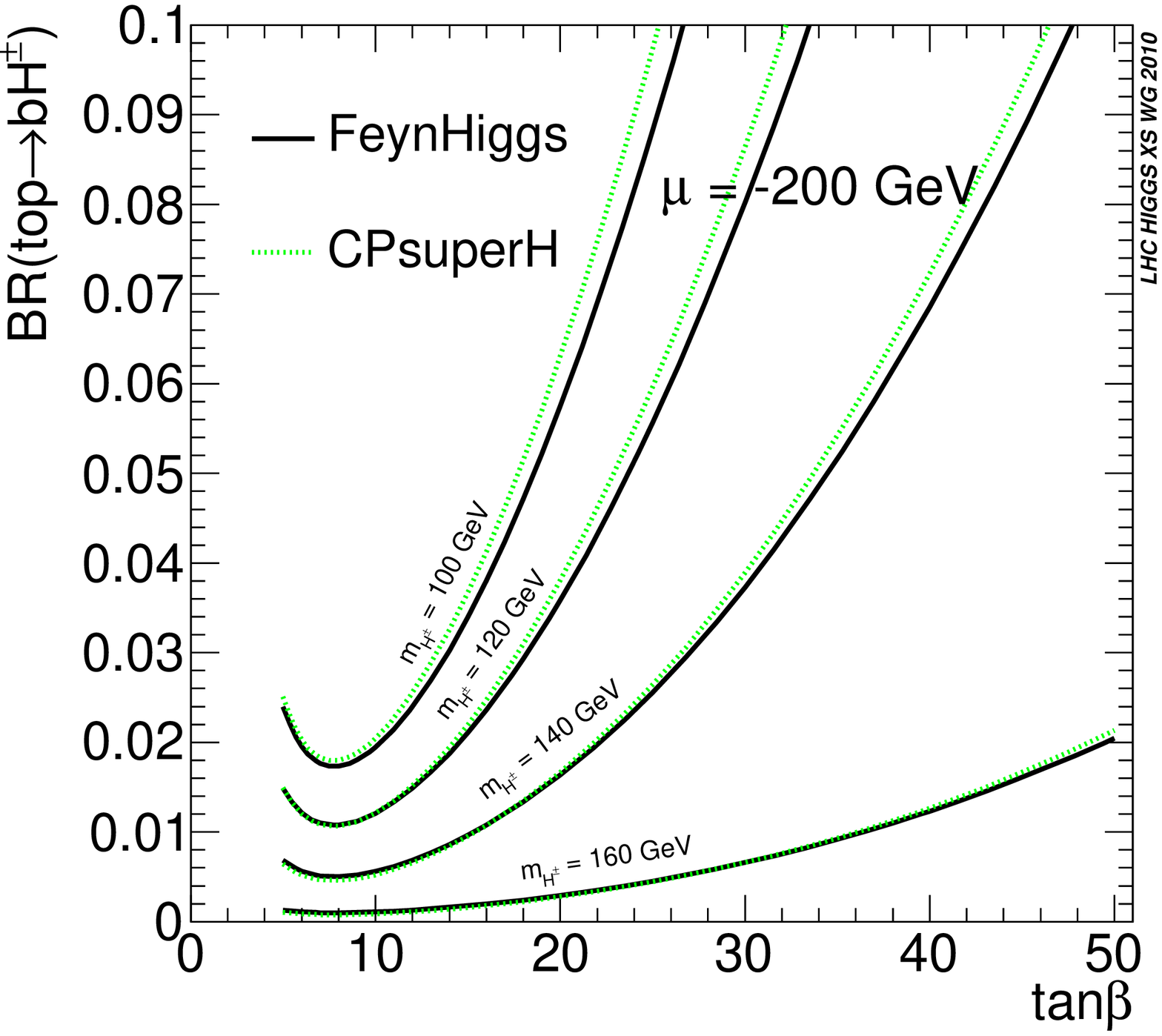} \\
  \includegraphics[width=0.46\textwidth]{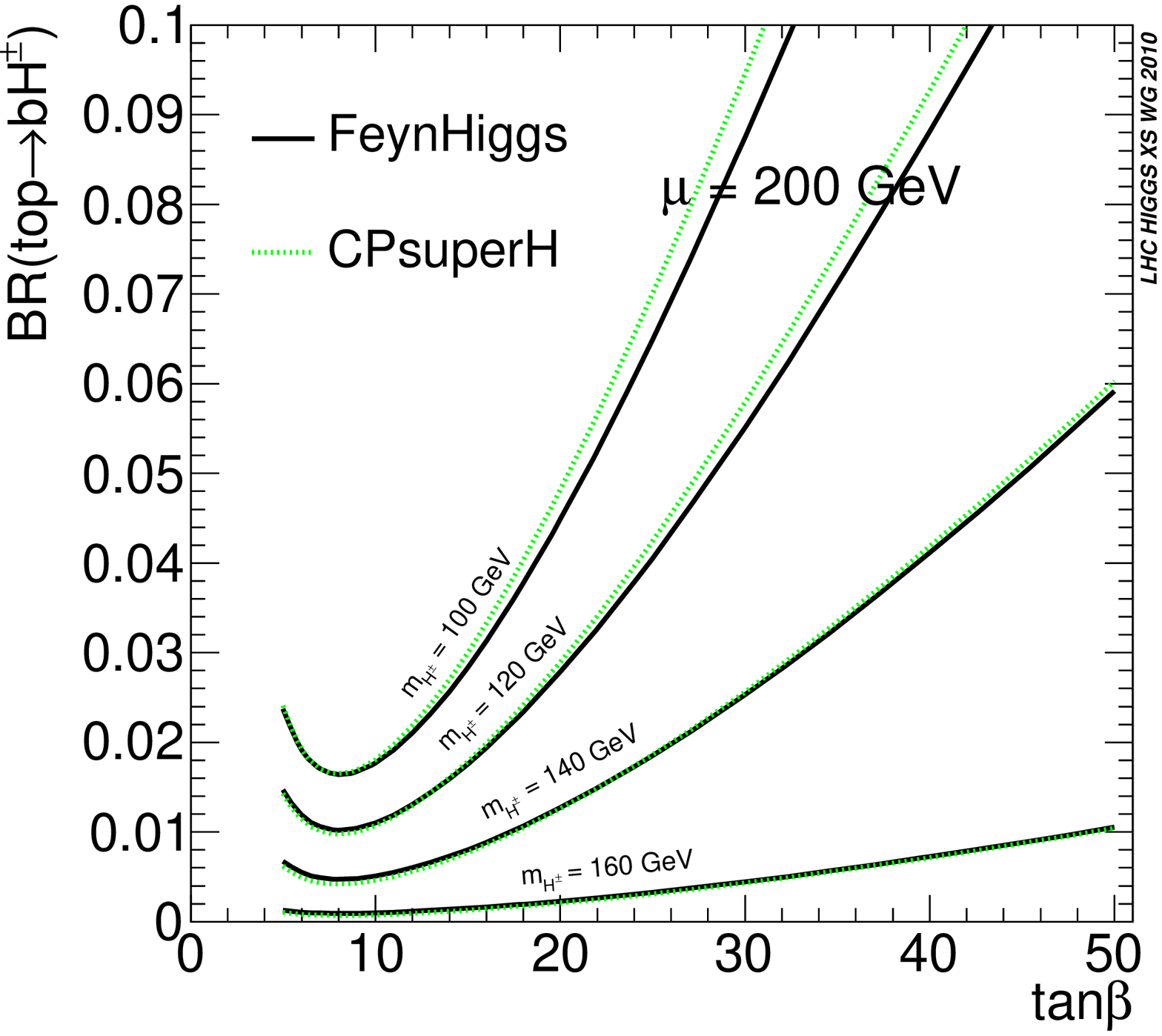} &
  \includegraphics[width=0.46\textwidth]{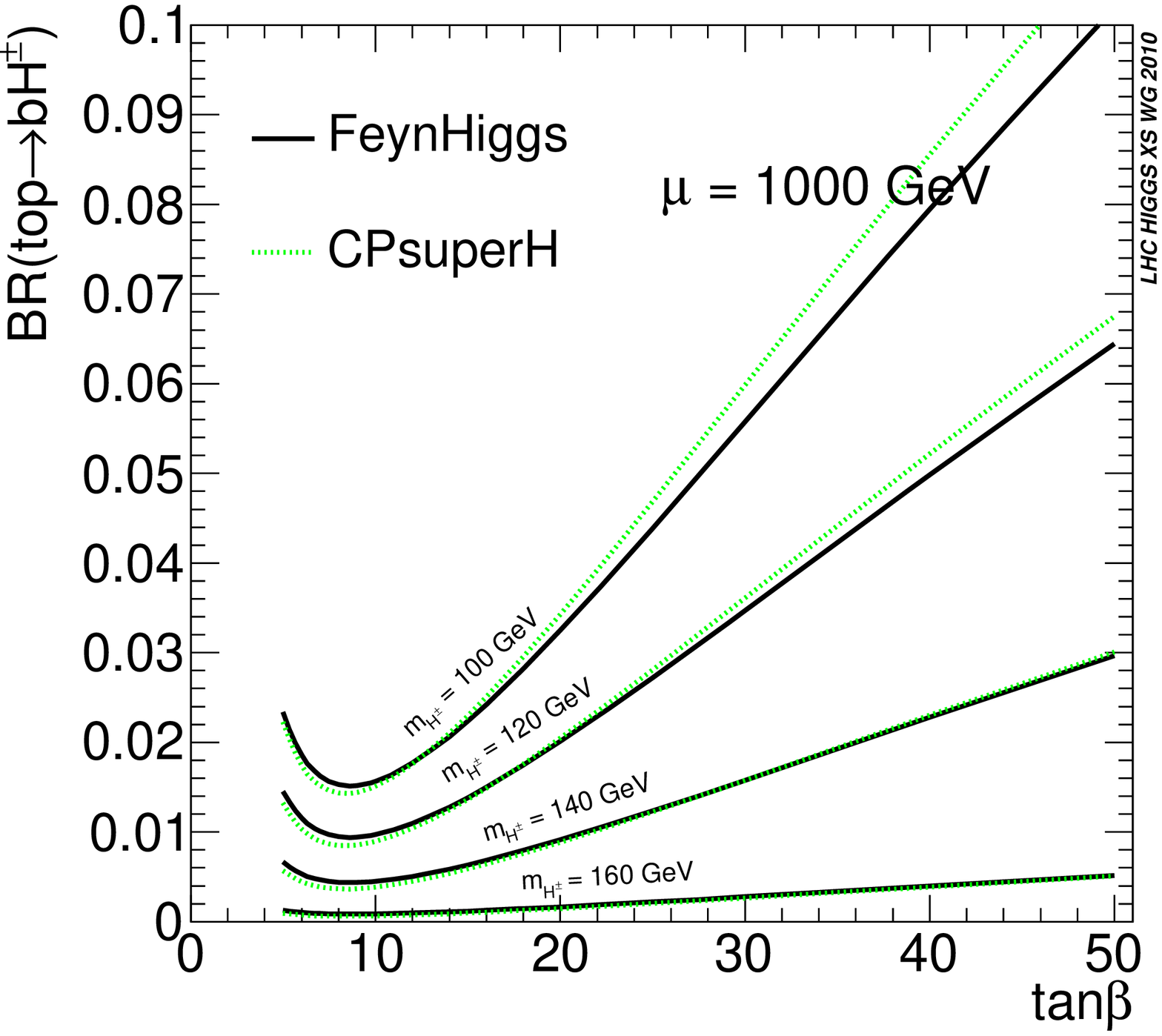} \\  
\end{tabular}
\caption{The branching fraction of $\Pt\rightarrow \Pb\PSHpm$ as a
  function of $\tan\beta$ for different values of $\mu$ and $\MHpm$. 
  The lines in the upper left plot terminate when the specific code
  reaches a negative light Higgs mass squared. Depending on the code this
  happens for slightly smaller or larger $\tan\beta$ values (in this
  extreme scenario).
}
\label{fig:BRTop2HPlus}
\end{figure}

\begin{figure}[ht]
  \centering
   \begin{tabular}{cc}
   \includegraphics[width=0.46\textwidth]{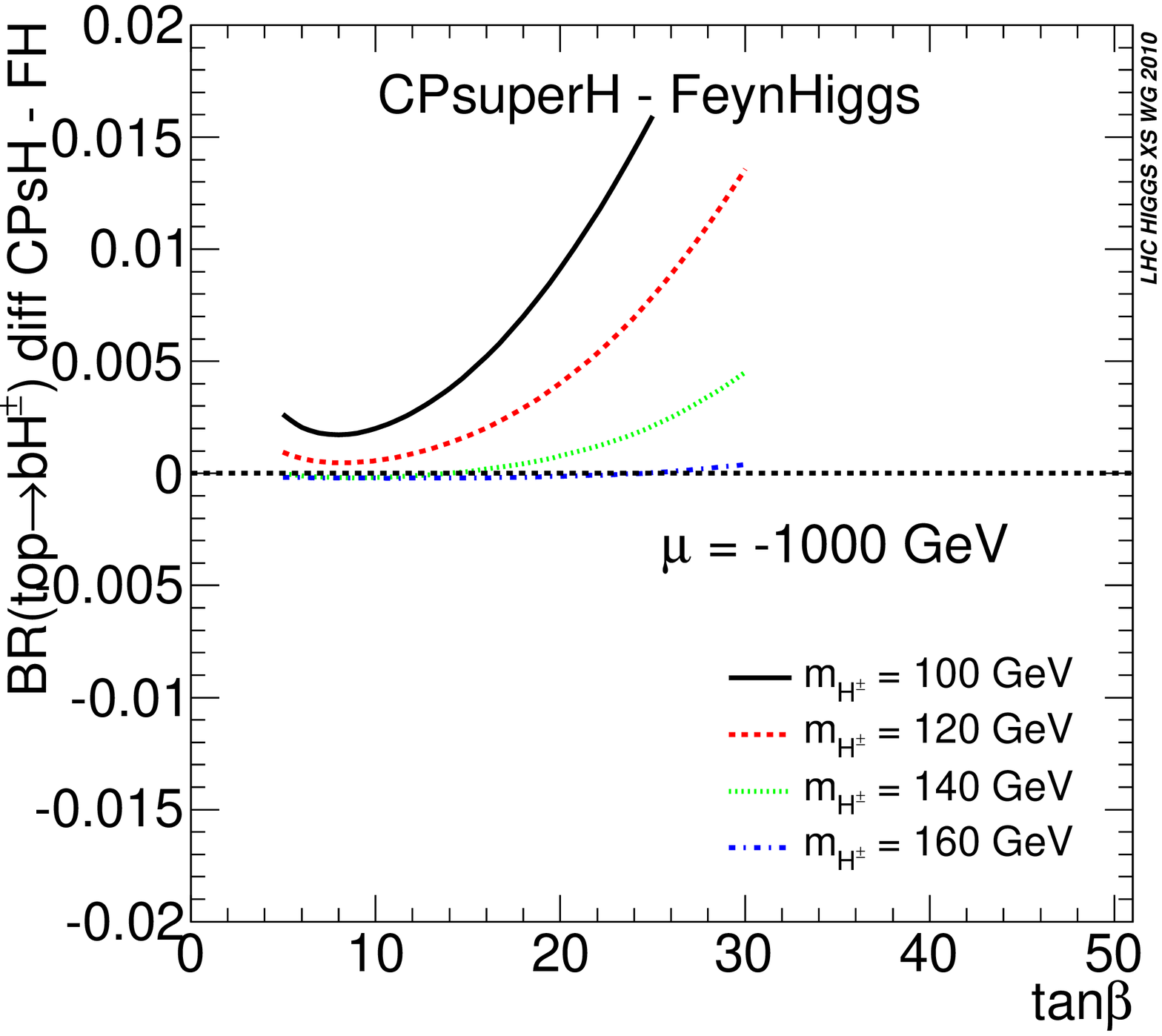} &
   \includegraphics[width=0.46\textwidth]{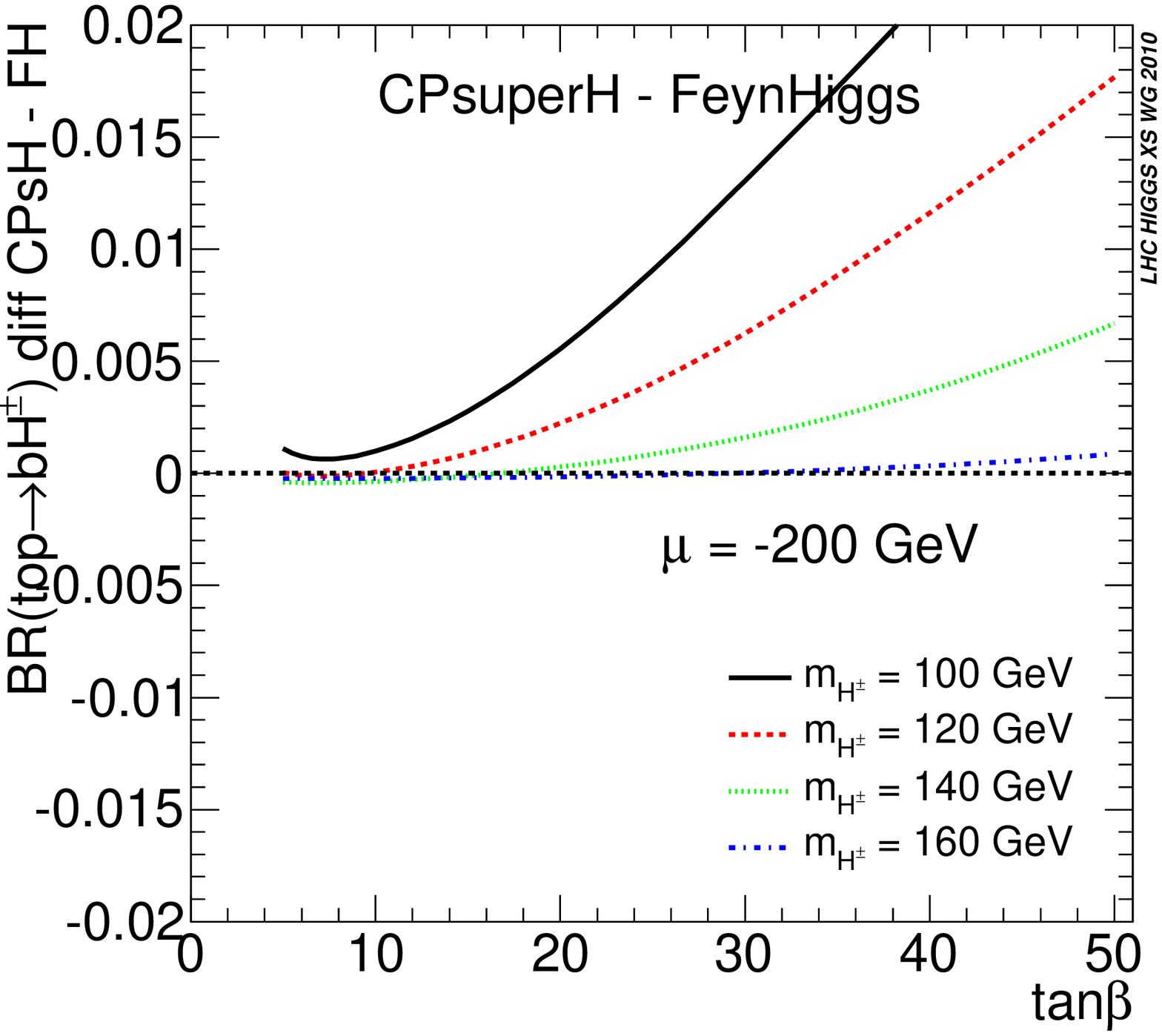} \\
   \includegraphics[width=0.46\textwidth]{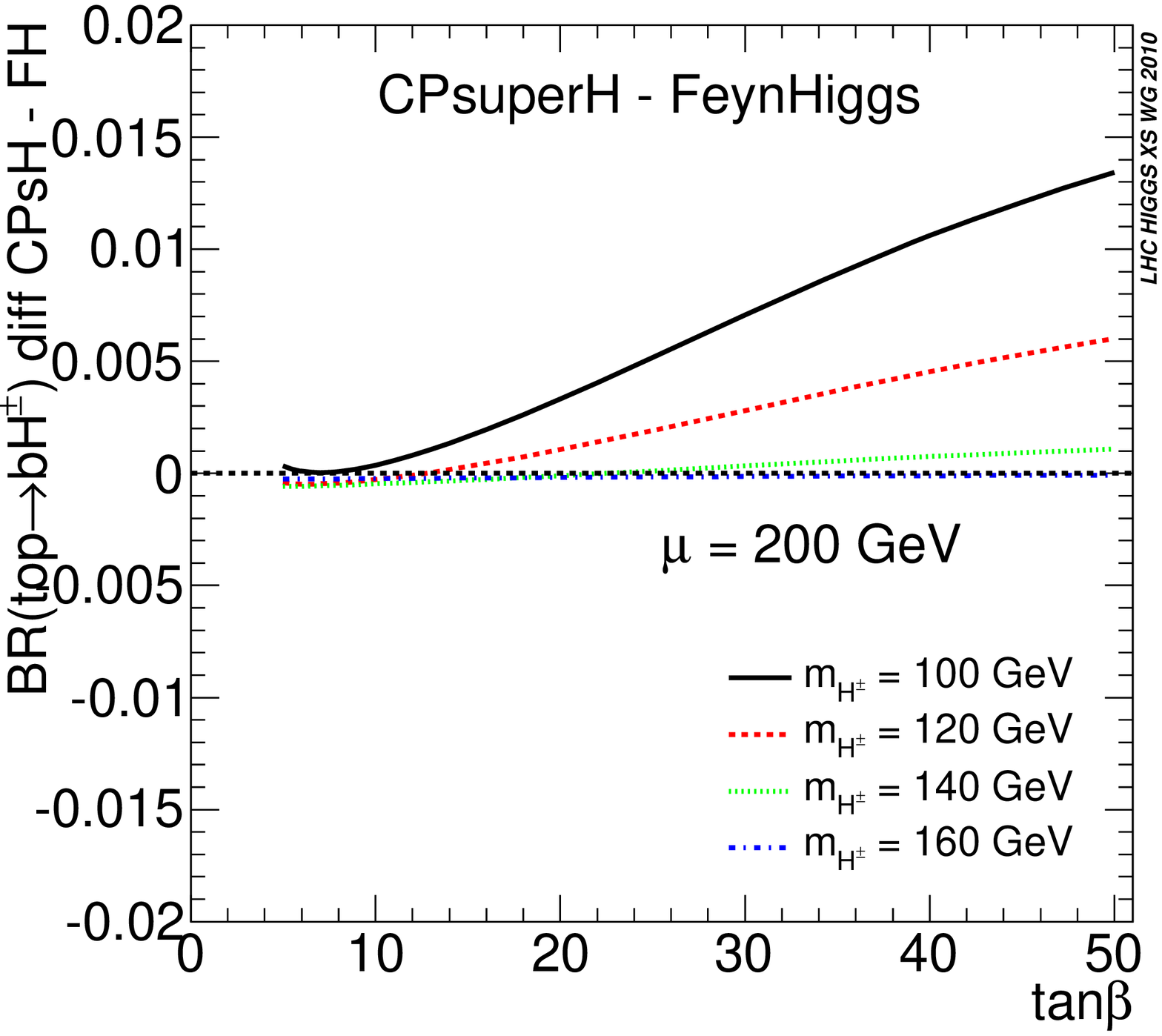} &
   \includegraphics[width=0.46\textwidth]{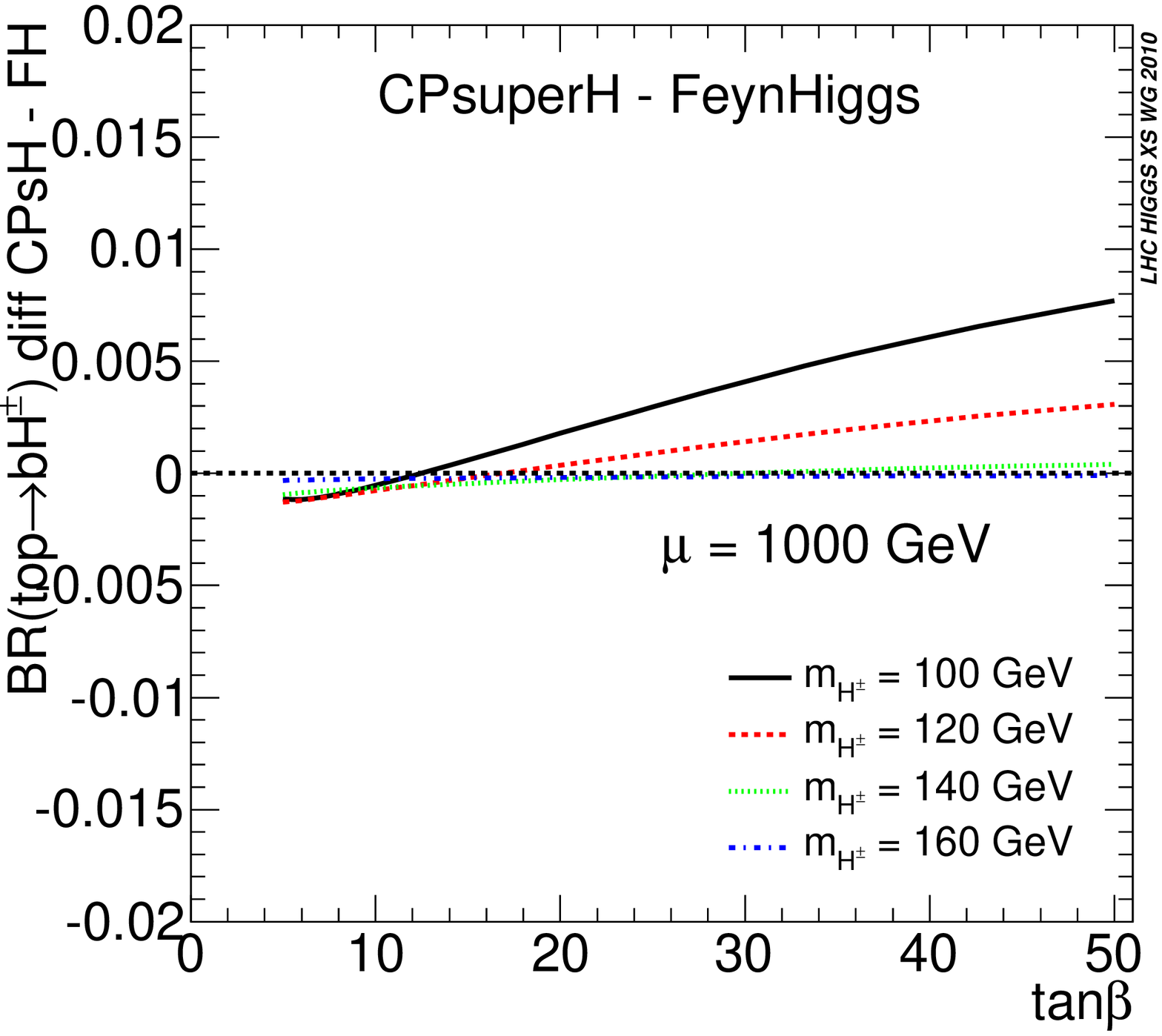} \\
  \end{tabular}
\caption{The difference of BR($\Pt\rightarrow \Pb\PSHpm$) calculated
  with {\sc CPsuperH} and {\sc FeynHiggs} as a function of $\tan\beta$
  for different values of $\mu$ and $\MHpm$.}
\label{fig:BRTop2HPlusDiff}
\end{figure}

\subsection{Heavy charged Higgs production with top and bottom quarks}
\label{subsec:CH_tbh}

For heavy charged Higgs bosons,  $\MHpm \gsim \Mt$,
associated production $\Pp\Pp \to {\Pt}\Pb \PSHpm{\rm + X}$ is the
dominant production mode. 
Two different formalisms can be employed to calculate the cross
section for associated ${\Pt}\Pb\PSHpm$ production.  In the
four-flavour scheme (4FS) with no $\Pb$ quarks in the initial state,
the lowest-order QCD production processes are gluon--gluon fusion and
quark--antiquark annihilation, $\Pg\Pg \to {\Pt}\Pb\PSHpm$ and
$\Pq\bar \Pq \to {\Pt}\Pb\PSHpm$, respectively. Potentially
large logarithms $\propto \ln(\mu_{\rm F}/\Mb)$, which arise from
the splitting of incoming gluons into nearly collinear $\Pb\bar \Pb$
pairs, can be summed to all orders in perturbation theory by
introducing bottom parton densities. This defines the five-flavour
scheme (5FS)~\cite{Barnett:1987jw}. The use of bottom distribution
functions is based on the approximation that the outgoing $\Pb$ quark
is at small transverse momentum and massless, and the virtual $\Pb$
quark is quasi on shell. In this scheme, the leading-order (LO)
process for the inclusive $\Pt\Pb\PSHpm$ cross section is
gluon--bottom fusion, $\Pg \Pb \to \Pt \PSHpm$.  The next-to-leading
order (NLO) cross section in the 5FS includes ${\cal
  O}(\alphas)$ corrections to $\Pg \Pb \to \Pt  \PSHpm$
and the tree-level processes $\Pg\Pg \to \Pt\Pb\PSHpm$ and $\Pq\bar
\Pq \to \Pt\Pb\PSHpm$. To all orders in perturbation theory the four-
and five-flavour schemes are identical, but the way of ordering the
perturbative expansion is different, and the results do not match
exactly at finite order. For the inclusive production of neutral Higgs 
bosons with bottom quarks, $\Pp\Pp \to \Pb\bar{\Pb}\PH{\rm + X}$, the
four- and five-flavour scheme calculations numerically agree within
their respective uncertainties, once higher-order QCD corrections are
taken into account~\cite{Dittmaier:2003ej, Campbell:2004pu,
  Dawson:2005vi, Buttar:2006zd}, see \Section~6 of this Report.
  
There has been considerable progress recently in improving the cross-section 
predictions for the associated production of charged Higgs
bosons with heavy quarks by calculating NLO SUSY QCD and electroweak
corrections in the four- and five-flavour schemes~\cite{Zhu:2001nt,
  Gao:2002is, Plehn:2002vy, Berger:2003sm, Kidonakis:2005hc,
  Peng:2006wv, Beccaria:2009my, Kidonakis:2010ux}, and the matching of
the NLO five-flavour scheme calculation with parton
showers~\cite{Weydert:2009vr}. Below, we shall present
state-of-the-art NLO QCD predictions in the 4FS
(\Section~\ref{subsubsec:CH_4FS}), in the 5FS
(\Section~\ref{subsubsec:CH_5FS}), and a first comparison of the two
schemes at NLO (\Section~\ref{subsubsec:CH_4and5FS}).

\subsubsection{NLO QCD predictions for $\Pp\Pp \to \Pt\Pb\PSHpm$ in the 4FS}
\label{subsubsec:CH_4FS}

In the 4FS the production of charged Higgs bosons in association with
top and bottom quarks proceeds at LO through the parton processes
$\Pg\Pg \to {\Pt}\bar{\Pb}\PSHm$ and $\Pq\bar{\Pq} \to \Pt\bar{\Pb}\PSHm$, 
and the charge-conjugate processes with the
$\bar{\Pt}\Pb\PSHp$ final state~\cite{DiazCruz:1992gg,
  Borzumati:1999th, Miller:1999bm}. Throughout this section we present
results for the ${\Pt}\bar{\Pb} \PSHm$ channels. Generic Feynman
diagrams that contribute at LO are displayed in \Fref{fig:diags}.
\begin{figure}
\begin{center}
\SetScale{0.8}
\begin{picture}(130,90)(0,0)
\ArrowLine(0,100)(50,50)
\ArrowLine(50,50)(0,0)
\Gluon(50,50)(100,50){3}{5}
\ArrowLine(100,50)(120,70)
\ArrowLine(120,70)(150,100)
\ArrowLine(150,0)(100,50)
\DashLine(120,70)(150,70){5}
\Vertex(50,50){2}
\Vertex(100,50){2}
\Vertex(120,70){2}
\put(-12,78){$\PQq$}
\put(-12,-2){$\PAQq$}
\put(125,53){$\PH^-$}
\put(125,78){$\PQt$}
\put(125,-2){$\PAQb$}
\end{picture}
\begin{picture}(130,90)(-80,0)
\Gluon(0,0)(50,0){3}{5}
\Gluon(0,100)(50,100){3}{5}
\ArrowLine(100,0)(50,0)
\ArrowLine(50,0)(50,50)
\ArrowLine(50,50)(50,100)
\ArrowLine(50,100)(100,100)
\DashLine(50,50)(100,50){5}
\Vertex(50,100){2}
\Vertex(50,50){2}
\Vertex(50,0){2}
\put(85,35){$\PH^-$}
\put(-12,78){$\Pg$}
\put(-12,-2){$\Pg$}
\put(85,78){$\PQt$}
\put(85,-2){$\PAQb$}
\end{picture}
\vspace*{1em}
\caption{Generic Feynman diagrams for 
          $\Pp\Pp \to {\Pt}\bar\Pb\PSHm{\rm + X}$ in the 4FS at LO.}
\label{fig:diags}
\end{center}
\end{figure}

The calculation of the NLO QCD corrections to charged Higgs production
in the 4FS has been discussed in detail in
\Bref{Dittmaier:2009np}, both within a two-Higgs-doublet model
with the SM particle content besides the extended Higgs sector, and
within the MSSM. The NLO QCD effects considerably enhance the cross
section and reduce the dependence on the renormalization and
factorization scales. In the MSSM, additional loop corrections from
squark and gluino exchange are sizable for large $\tan\beta$, but they
can be taken into account through the $\Delta_{\Pb}$ corrections
to
the bottom--Higgs-Yukawa coupling [cf.\ \Eq~(\ref{eq:qcd_tanb_enhanced})], i.e.\ through a rescaling of
the NLO QCD prediction according to $\Mb\tan\beta/v \to \Mb\tan\beta/v
\, (1 - \Delta_{\Pb}/\tan^2\beta) / (1 +
\Delta_{\Pb})$~\cite{Dittmaier:2009np}.

In \Tables~\ref{tab::charged_higgs_4fs_nlo_7tev} and
\ref{tab::charged_higgs_4fs_nlo_14tev} we present 4FS NLO QCD results
for the production of heavy charged Higgs bosons in a
two-Higgs-doublet model. Cross sections for MSSM scenarios with large
$\tan\beta$ can be obtained from the NLO QCD cross sections by the
rescaling defined above. Predictions are presented for LHC cross
sections at $7\UTeV$ and $14$\UTeV\ energy, with $\tan\beta = 30$ and the SM
input parameters according to \Table~\ref{tab:SMinput}.
\begin{table}[h!]
  \caption{\label{tab::charged_higgs_4fs_nlo_7tev}  
    NLO QCD cross sections for 
    $\Pp\Pp \to \Pt\bar{\Pb}\PSHm$ in the 4FS at the LHC with $7$\UTeV, $\tan \beta=30$.}
\centering
\small
\begin{tabular}{crcc}\hline
$\MHpm$~[GeV] & $\sigma$~[\UfbZ] & Scale uncert. [\%] & PDF + \alphas\; [\%]  \\
\hline
$200$ & $130$  &  $-33 \; +27$  & $-5.5 \; +4.5$ \\
$300$ & $45.9$ &  $-33 \; +34$  & $-6.7 \; +5.6$ \\
$400$ & $18.0$ &  $-34 \; +30$  & $-7.7 \; +6.6$ \\
$500$ & $7.59$ &  $-35 \; +32$  & $-8.6 \; +7.5$ \\
\hline
\end{tabular}
\end{table}
%
\begin{table}[h!]
  \caption{\label{tab::charged_higgs_4fs_nlo_14tev}  
    NLO QCD cross sections for 
    $\Pp\Pp \to \Pt\bar{\Pb}\PSHm$ in the 4FS at the LHC with $14$\UTeV, $\tan \beta=30$.}
\centering
\small
\begin{tabular}{crcc}\hline
$\MHpm$~[GeV] & $\sigma$~[\UfbZ] & Scale uncert. [\%] & PDF + \alphas\; [\%] \\
\hline
$200$ & $972$  &  $-30 \; +27$  & $-3.4 \; +2.7$ \\
$300$ & $405$  &  $-30 \; +26$  & $-4.0 \; +3.2$ \\
$400$ & $184$  &  $-30 \; +26$  & $-4.7 \; +3.7$ \\
$500$ & $92.6$ &  $-32 \; +29$  & $-5.1 \; +4.1$ \\
\hline
\end{tabular}
\end{table}

For a consistent evaluation of the hadronic cross sections in the 4FS
we adopt the recent MSTW four-flavour PDF~\cite{Martin:2010db} and the
corresponding four-flavour $\alphas$.  Note, however, that the
evaluation of the running \Pb-quark mass in the bottom--Higgs-Yukawa
coupling is based on a five-flavour $\alphas$ with 
$\alphas(\MZ) = 0.120$. The renormalization and factorization scales have
been identified and are set to $\mu = (\Mt + \Mb + M_{\PSHm})/3$ 
as our default choice. The NLO scale uncertainty has
been estimated from the variation of the renormalization and
factorization scales by a factor of three about the central scale
choice $\mu = (\Mt + \Mb + M_{\PSHm})/3$. As shown in \Bref{Dittmaier:2009np}, the variation of
the QCD scales by a factor three about the central scale provides
a more reliable estimate of the theory uncertainty than the usual
variation by a factor two, as the variation by a factor three encompasses 
the maximum of the NLO prediction. The residual NLO scale
uncertainty is then approximately $\pm 30$\%. While no four-flavour PDF
parametrization exists that would allow to estimate the combined PDF
and $\alphas$ error, the difference in the relative PDF error
as obtained from the MSTW four- and five-flavour sets is marginal.  We
have thus adopted the five-flavour MSTW PDF~\cite{Martin:2009iq} to
estimate the combined PDF and $\alphas$ uncertainty shown in
\Tables~\ref{tab::charged_higgs_4fs_nlo_7tev} and
\ref{tab::charged_higgs_4fs_nlo_14tev}. We find that the theoretical
uncertainty of the 4FS NLO QCD prediction for 
$\Pp\Pp \to \Pt \bar{\Pb}\PSHm$ at the LHC is by far 
dominated by the scale uncertainty.

The NLO QCD cross section for 
$\Pp\Pp \to {\Pt}\bar{\Pb}\PSHm$ at the LHC with $7\UTeV$ and $14$\UTeV\ is shown in
\Fref{fig:totalxs} as a function of the Higgs-boson mass. The
error band quantify the NLO scale uncertainty.


\begin{figure}
\begin{center}
\includegraphics[width=0.48\textwidth]{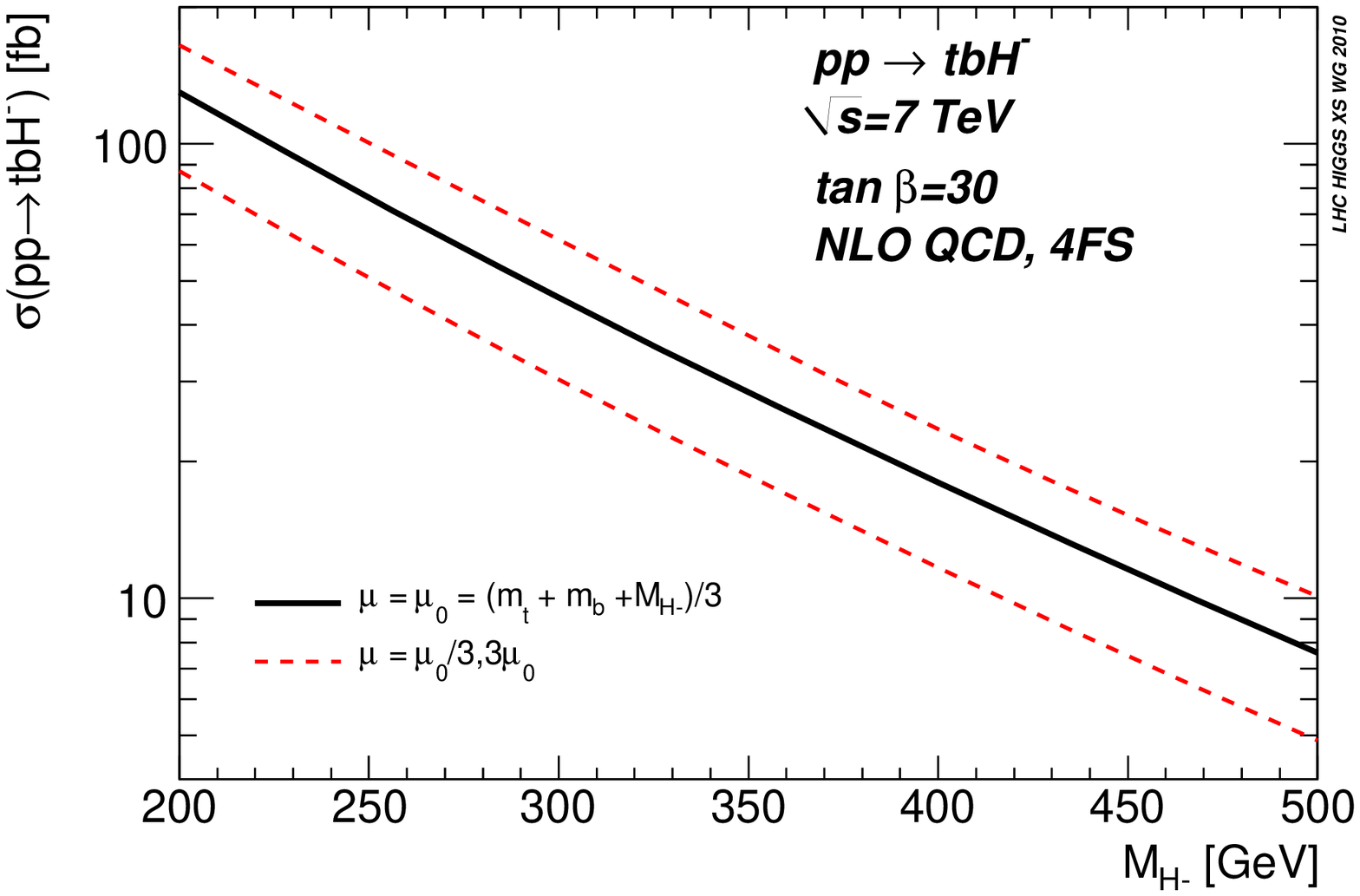}
\includegraphics[width=0.48\textwidth]{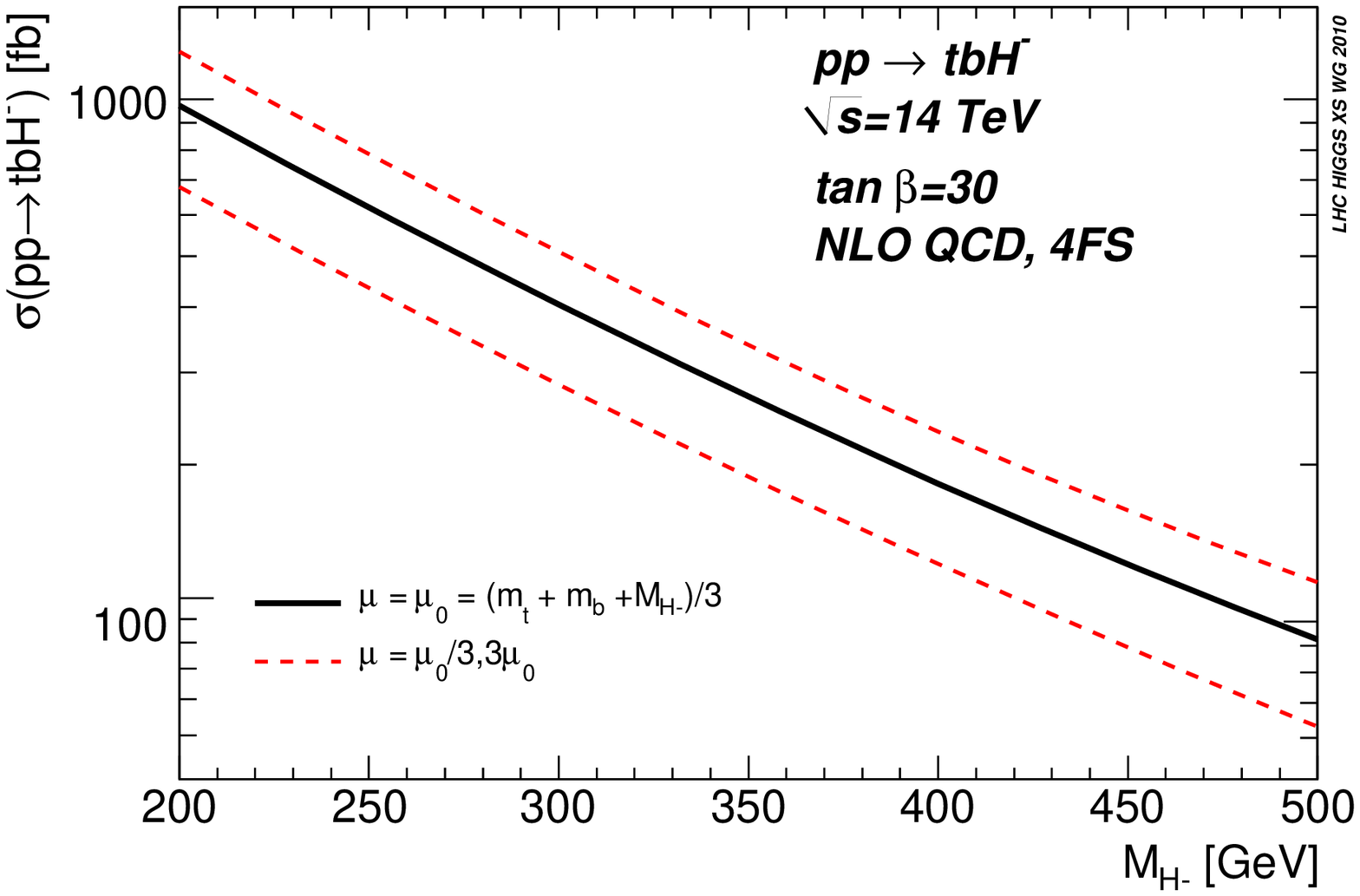}

 \caption{NLO QCD cross sections for 
     $\Pp\Pp \to {\Pt}\bar{\Pb}\PSHm$ in the 4FS at the LHC ($7$\UTeV\ and $14$\UTeV) as a function of
   the Higgs-boson mass. The error band includes the NLO scale
   uncertainty. 
   (Calculation from \Bref{Dittmaier:2009np}.)}
\label{fig:totalxs}
\end{center}
\end{figure}
        

\subsubsection{NLO QCD predictions for $\Pp\Pp \to \Pt \PSHpm$ in the 5FS}
\label{subsubsec:CH_5FS}

In the 5FS the LO process for the inclusive $\PSHpm$ cross
section is gluon--bottom fusion, $\Pg \Pb \to \Pt \PSHpm$. The NLO
cross section includes ${\cal O}(\alphas)$ corrections to
$\Pg \Pb \to \Pt \PSHpm$ and the tree-level processes $\Pg\Pg \to
\Pt\Pb\PSHpm$ and $\Pq\bar \Pq \to \Pt\Pb\PSHpm$, and has been
calculated in \Refs~\cite{Plehn:2002vy, Berger:2003sm,
  Weydert:2009vr}. In \Tables~\ref{tab::charged_higgs_5fs_nlo_7tev}
and \ref{tab::charged_higgs_5fs_nlo_14tev} we present NLO QCD results
for the production of heavy charged Higgs bosons in the 5FS, with
$\tan\beta = 30$ and the SM input parameters according to
\Table~\ref{tab:SMinput}. As in the 4FS calculation, cross sections
for MSSM scenarios with large $\tan\beta$ can be obtained from the
NLO QCD cross sections by rescaling the bottom--Higgs-Yukawa coupling.
\begin{table}[h!]
  \caption{\label{tab::charged_higgs_5fs_nlo_7tev}  
    NLO QCD cross sections for 
    $\Pp\Pp \to \Pt\PSHm$ in the 5FS at the LHC with $7$\UTeV, $\tan \beta=30$.}
  \centering
  \small
  \begin{tabular}{crr}\hline
$\MHpm$~[GeV] & $\sigma$~[\UfbZ] & Scale uncert. [\%] \\
\hline
$200$ & $178$  &  $-7.1  \; +9.4$~~  \\
$300$ & $62.7$ &  $-10.0 \; +4.7$~~  \\
$400$ & $24.7$ &  $-11.0 \; +2.7$~~  \\
$500$ & $10.5$ &  $-12.0 \; +1.1$~~  \\
\hline
\end{tabular}
\end{table}
\begin{table}[h!]
  \caption{\label{tab::charged_higgs_5fs_nlo_14tev}  
    NLO QCD cross sections for 
    $\Pp\Pp \to \Pt\PSHm$ in the 5FS at the LHC with $14$\UTeV, $\tan \beta=30$.}
\centering
\small
\begin{tabular}{crr}\hline
$\MHpm$~[GeV] & $\sigma$~[\UfbZ] & Scale uncert. [\%] \\
\hline
$200$ & $1237$ &  $-8.4  \; +13$\phantom{.}~~  \\
$300$ & $521$  &  $-9.0  \; +9.5$~~  \\
$400$ & $242$  &  $-9.8  \; +7.7$~~  \\
$500$ & $121$  &  $-10.0 \; +6.5$~~ \\
\hline
\end{tabular}
\end{table}
The NLO cross section values have been obtained using 
{\sc MC@NLO 4.0}~\cite{Frixione:2010wd}, with the option {\tt rflag} switched to 0 (for MSbar
Yukawa renormalization). The central scale has been set to
$\mu_0=(\Mt+M_{\PSHm})/4$, and the five-flavour MSTW
PDF~\cite{Martin:2009iq} has been adopted. We find a residual NLO
scale uncertainty of $10{-}20\%$. Since there are no direct experimental
constraints on the bottom PDF, the PDF uncertainty of the $\Pg \Pb \to
\Pt \PSHpm$ process is difficult to quantify. Thus, unfortunately,
no reliable estimates of the PDF and $\alphas$ uncertainty of
the 5FS calculation exist to date.

The total 5FS NLO QCD cross section for 
$\Pp\Pp \to \Pt\PSHm$ at the LHC with $7\UTeV$ and $14$\UTeV\ 
is shown in
\Fref{fig:totalxs_5fs} as a function of the Higgs-boson mass.
The error band includes the NLO scale uncertainty only.

\begin{figure}
\begin{center}
\includegraphics[width=0.48\textwidth]{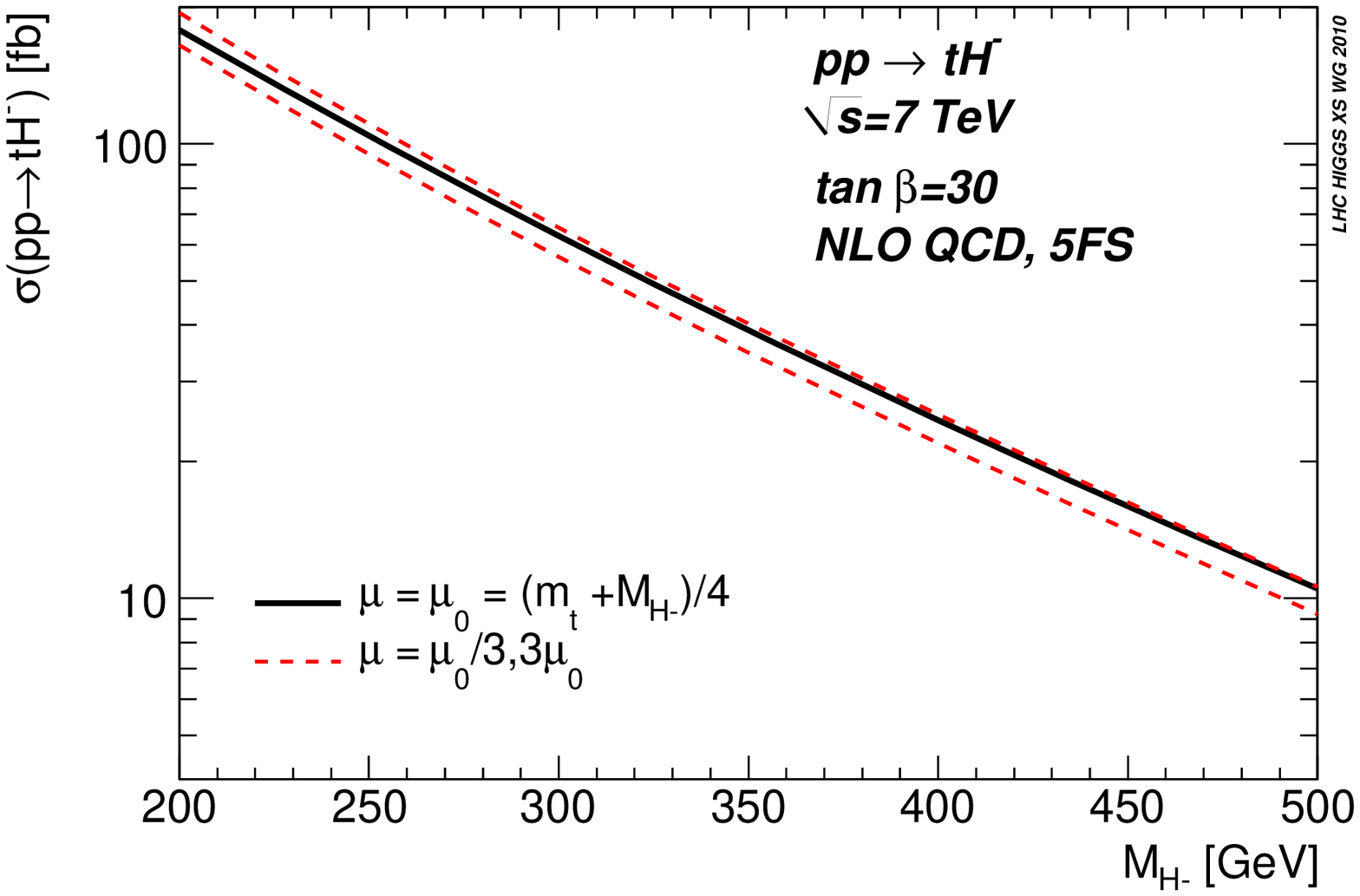}
\includegraphics[width=0.48\textwidth]{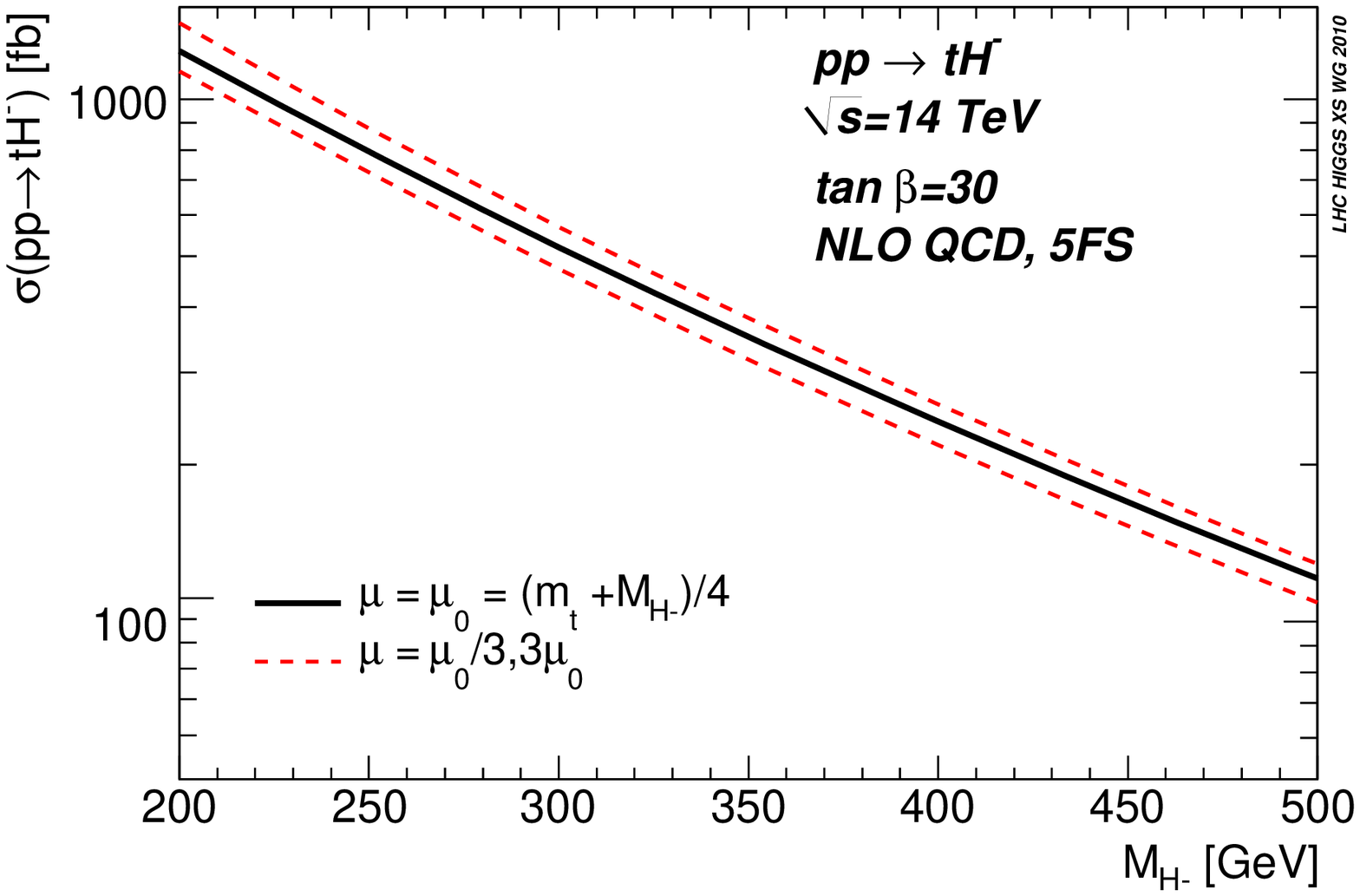}
\caption{NLO QCD cross sections for $\Pp\Pp \to {\Pt}\PSHm$ in the 5FS at the LHC ($7$\UTeV\ and $14$\UTeV) as a function of
  the Higgs-boson mass. The error band includes the NLO scale
  uncertainty. (Calculation from \Bref{Weydert:2009vr}.)}
\label{fig:totalxs_5fs}
\end{center}
\end{figure}
        
Note that supersymmetric electroweak ${\cal O(\alpha)}$ corrections to
charged Higgs-boson production in the five-flavour scheme have been
studied in \Bref{Beccaria:2009my}. These corrections depend
sensitively on the MSSM scenario and have thus not been included in
the numbers presented here.

\subsubsection{Comparison of the 4FS and 5FS calculations}
\label{subsubsec:CH_4and5FS}

The 4FS and 5FS calculations represent different ways of ordering the
perturbative expansion, and the results will not match exactly at
finite order. However, taking into account higher-order corrections,
the two predictions are expected to agree within their respective
uncertainties, see \Section~\ref{sec:mssm_neutral} of this Report for a similar comparison
for the inclusive production of neutral Higgs bosons with bottom
quarks.

In \Fref{fig:4fs_5fs} we present a comparison of the 4FS and
5FS calculations at NLO QCD for the inclusive 
$\Pp \Pp \to \Pt  \PSHm(\bar{\Pb})$ cross section at the LHC.
\begin{figure}
\begin{center}
\includegraphics[width=0.48\textwidth]{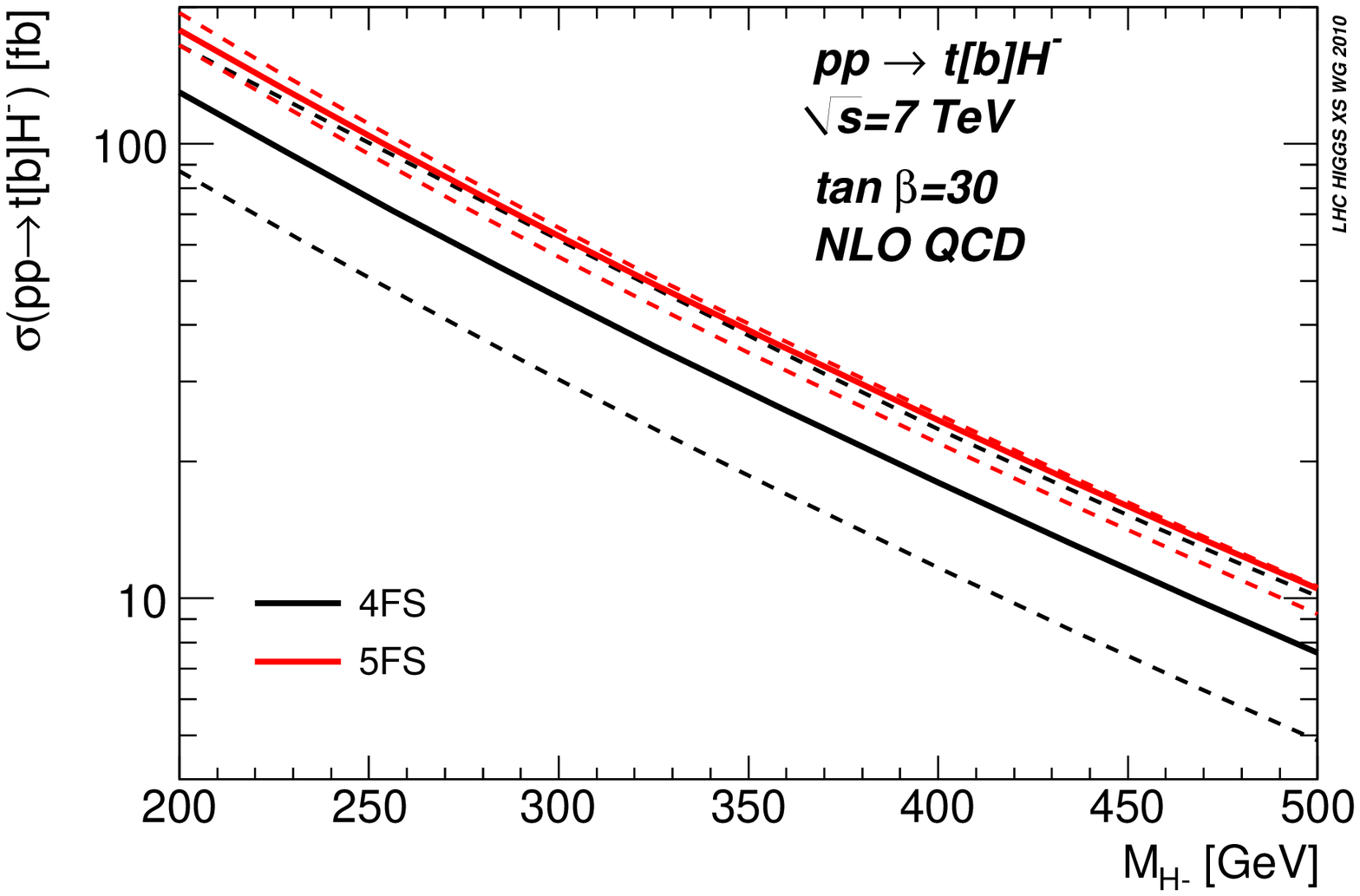}
\includegraphics[width=0.48\textwidth]{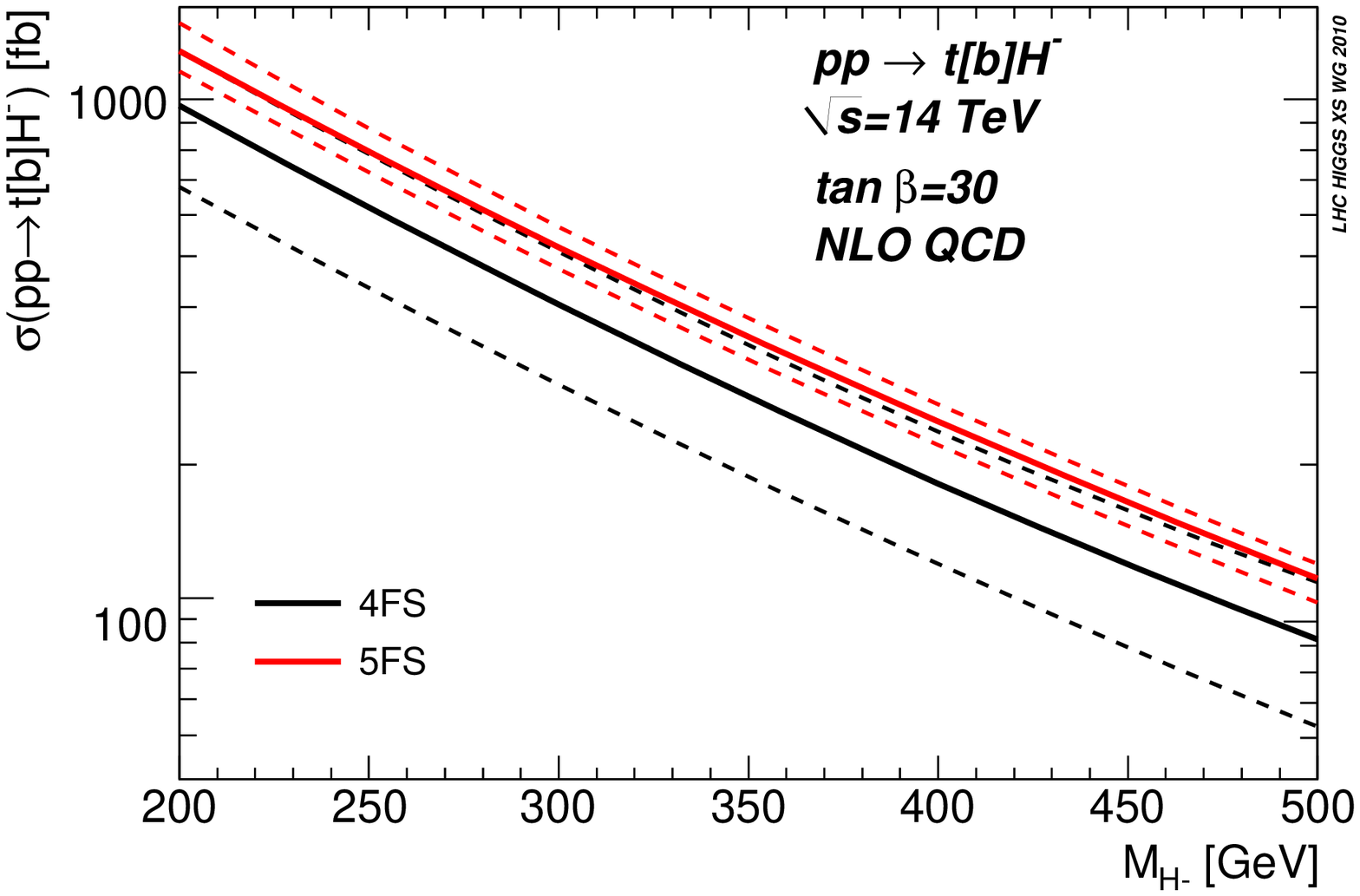}
\caption{NLO QCD cross sections for 
    $\Pp\Pp \to {\Pt}\PSHm(\bar{\Pb})$ in the 4FS and 5FS at the LHC 
($7$\UTeV\ and $14$\UTeV) as a
  function of the Higgs-boson mass. The error band includes the NLO
  scale uncertainty. (Calculations from \Refs~\cite{Dittmaier:2009np,
    Weydert:2009vr}.)}
\label{fig:4fs_5fs}
\end{center}
\end{figure}
The error band indicates the theoretical uncertainty when the
renormalization and factorization scales are varied between $\mu_0/3$
and $3\mu_0$,  with  $\mu_0 = (\Mt + \Mb + M_{\PSHm})/3$ 
and $\mu_0=(\Mt+M_{\PSHm})/4$ for the 4FS and 5FS 
calculations, respectively. 
Taking the scale uncertainty into account, the 4FS and
5FS cross sections at NLO are consistent, even though the predictions
in the 5FS at our choice of the central scales are larger than those of
the 4FS by approximately $30\%$, almost independently of the Higgs-boson
mass. Qualitatively similar results have been obtained from a
comparison of 4FS and 5FS NLO calculations for single-top production
at the LHC~\cite{Campbell:2009ss}. Note that the bottom PDF of the
recent five-flavour MSTW fit~\cite{Martin:2009iq} is considerably
smaller than that of previous fits~\cite{Martin:2004ir} and has lead
to a significant decrease in the 5FS cross section prediction.


\clearpage

\section{Parton distribution functions\protect\footnote{
  S. Forte, J. Huston, K. Mazumdar, R.S. Thorne and A. Vicini.}}
\label{pdfsection}

\subsection{Introduction}
\label{intro}

Parton distribution functions (PDFs) are crucial for the prediction
of any physical process to be measured at the LHC, hence PDFs and their
uncertainties 
 have very important significance, in particular for discovery and
 exclusion limits. 
At present these PDFs are obtained 
from fits to data from deep-inelastic scattering, Drell--Yan, and
jet production from a wide variety of different experiments. 
A number of groups have produced publicly available PDFs
using different data sets and analysis frameworks. Here we 
summarise our level of understanding as the first LHC cross sections
at $7$\UTeV\ are being determined.  There are many differences between
existing PDF analyses: different input data, different methodologies and 
criteria for  determining uncertainties, different ways of 
parametrizing the PDFs, different
number of parametrized PDFs, different treatments of heavy quarks, 
different perturbative orders,
different ways of treating  $\alphas$ (as an input or as a fit
parameter),
different values of physical parameters such as $\alphas$ and
heavy-quark masses, and more. 
Hence, we begin by summarizing the main features of
current PDF sets. We subsequently introduce various
theoretical uncertainties on PDFs, focusing on the uncertainty related to
the value of the strong coupling constant, and provide a presentation
of choices made by different groups. We then briefly 
summarise the computation of physical processes using various PDF sets. 
As an outcome of this, we motivate and 
describe the PDF4LHC interim recommendation~\cite{PDF4LHCwebpage} 
to obtain current combined predictions and uncertainties based on several 
global PDF sets, and illustrate it by showing its
application to the  Higgs production 
cross section via gluon--gluon fusion, both at
NLO and at NNLO. 

\begin{sloppypar}
We will discuss the following PDF sets (when several
releases are available, the reference release for our discussion below 
is given 
parenthesis in each case):
ABKM~\cite{Alekhin:2009ni}, CTEQ/CT (CTEQ6.6~\cite{Nadolsky:2008zw}),
GJR~\cite{Gluck:2007ck,Gluck:2008gs}, HERAPDF (HERAPDF1.0~\cite{:2009wt}), 
MSTW (MSTW08~\cite{Martin:2009iq,Martin:2009bu,Martin:2010db}) and 
NNPDF (NNPDF2.0~\cite{Ball:2010de}). ABKM, JR~\cite{JimenezDelgado:2008hf}
(for variable flavour see \Bref{JimenezDelgado:2009tv}), 
MSTW, and HERAPDF~\cite{CooperSarkar:2010ik} are 
available with NNLO evolution \cite{Moch:2004pa,Vogt:2004mw}.
A CTEQ/CT update is already
available (CT10~\cite{Lai:2010vv}),  while preliminary updates of
ABM~\cite{Alekhin:2010iu}, NNPDF~\cite{Rojo:2010gv},
HERAPDF~\cite{CooperSarkar:2010ik},
 and
MSTW~\cite{Thorne:2010kj} have also been presented.
\end{sloppypar}

\subsection{PDF determinations -- experimental uncertainties}
\label{sec:pdfdet1}

Experimental uncertainties on PDFs determined in global fits (usually 
called {\em PDF uncertainties}, for short) reflect the information available 
(or lack thereof) in the underlying data and the way it constrains PDFs; they
should be
interpreted as genuine statistical uncertainties, and indeed they are often
given in the form of confidence levels (CL). 
They may differ because of different choices made in the analysis
that extracts this information from the data, specifically in: 
(1) the choice of data sets; (2) the statistical
 treatment which is used to determine the uncertainties and which also
determines the way in which PDFs are delivered to the user; (3) the
form and size of parton parametrization.

\subsubsection{Data sets}
\label{data}
A wider data set contains more information, but data coming
from different experiments may be inconsistent to some extent. The
choices made by the various groups are the following:
\begin{itemize}
\item The data sets considered by CTEQ, MSTW, and NNPDF include data from both  
electroproduction 
and hadroproduction, in each case from both
  fixed-target and collider experiments. The electroproduction data
  include electron, muon and neutrino  
 deep-inelastic scattering data (both inclusive and charm
 production). The hadroproduction data include Drell--Yan (fixed-target
 virtual photon and collider $\PW$ and $\PZ$ production) and jet
 productions (Tevatron jets requiring some approximation for the MSTW NNLO
analysis). 
Details vary slightly among particular versions of CTEQ, MSTW, and NNPDF 
fits. 
\item For GJR (and JR) the data set consists of electroproduction data 
which include electron- and muon-inclusive
 deep-inelastic scattering data, and deep-inelastic charm 
 production from charged leptons and neutrinos from
  fixed-target and collider experiments, and a smaller set of 
hadroproduction data, i.e.\ fixed-target virtual photon Drell--Yan 
production and  Tevatron jet production.
\item The ABKM data set includes electroproduction  from
  fixed-target and collider experiments, including electron, 
muon, and neutrino deep-inelastic scattering data (both inclusive and charm
 production), and fixed-target hadroproduction data, i.e.\ virtual
 photon Drell--Yan production%
\footnote{An update is being prepared that includes the Tevatron jet data as well.}. 
\item The HERAPDF input contains all HERA deep-inelastic inclusive
data.   
\end{itemize}

\subsubsection{Statistical treatment}
\label{stat}

Available PDF determinations fall in two broad categories: those
based on a Hessian approach and those which use a Monte Carlo
approach. The output format for information on PDFs is different in 
each case. Here we  outline only the basic features. The precise manner 
in which to implement the PDFs 
is explained in more detail in the appropriate 
references for each group. 

Within the Hessian method, PDFs are determined by minimizing a
$\chi^2$ function defined as
  $\chi^2=\frac{1}{N_{\rm dat}}\sum_{i,j}
  (d_i-\bar d_i){\rm cov}_{ij}(d_j-\bar d_j)$, where $\bar d_i$ are data,
  $d_i$ theoretical predictions, ${N_{\rm dat}}$
  is the number of data points,
and ${\rm cov}_{ij}$ is
  the covariance matrix%
\footnote{Different groups use differing
  definitions of the covariance matrix -- including entirely or only
  partially correlated uncertainties -- see the papers for details. 
  Hence the values of the $\chi^2$
  quoted are only roughly comparable.}. 
The best fit is the point in parameter
space at which $\chi^2$ is minimum, while PDF uncertainties are found
by evaluating, and often diagonalizing the (Hessian) matrix of second
derivatives of the $\chi^2$ at the minimum, and then 
determining the range of parameter variation 
corresponding to a  prescribed
increase of the $\chi^2$ function with respect to the minimum.
In principle, the increase in $\chi^2$ which provides
$68\%$~CL  ($1 \,\sigma$) is $\Delta\chi^2=1$.  However, a larger variation
of $\Delta\chi^2=T^2$, with $T>1$  a suitable `tolerance'
 parameter~\cite{Stump:2001gu,Pumplin:2001ct,Martin:2002aw} may
turn out to be necessary for a more realistic error estimate for fits
containing a wide variety of input processes and data, and 
in particular, in order for each
 experiment which enters the global fit to be consistent
 with the global best fit within one sigma (or an alternative
 confidence level, 
e.g.\ $90\%$).  Possible reasons why this is necessary could be 
data inconsistencies or incompatibilities, 
underestimated experimental systematics, insufficiently flexible
parton parametrizations, theoretical uncertainties or approximations in
the PDF extraction. At present, HERAPDF and ABKM use $\Delta\chi^2=1$,
GJR uses $T\approx4.7$, CTEQ6.6
uses $T=10$ at $90\%$ ~CL (corresponding to $T\approx6$ at $68\%$~CL), while MSTW08 uses a dynamical tolerance~\cite{Martin:2009iq}, i.e.,\ a different value
of $T$ for each eigenvector, with values 
from $T\approx 1$ to
$T\approx 6.5$ and most values being in the range of $2<T<5$. 

Within a Monte Carlo method, PDFs are determined by first producing a
Monte Carlo sample of $N_{\rm rep}$ pseudo-data replicas. Each replica
contains a number of points equal to the number of original data
points.
 The sample is constructed
in such a way that, in the limit $N_{\rm rep}\to\infty$, 
the central value of the $i$-th
data point is equal to the mean over the $N_{\rm rep}$ values that the
$i$-th point takes in each replica,  the uncertainty of the
same point is  equal to the variance over the
replicas, and the correlations between any two original data
points is equal to their covariance over the replicas. From 
each pseudo-data replica, a PDF replica  is constructed by minimizing
a $\chi^2$ function. The PDF
central values, uncertainties and correlations are then computed by
taking means, variances, and covariances over this replica
sample. NNPDF uses a Monte Carlo method, with each PDF replica
obtained as the minimum $\chi^2$ which satisfies a cross-validation
criterion~\cite{Ball:2008by,Ball:2010de}, and is thus larger than the
absolute minimum of the $\chi^2$. This method has been used in all NNPDF sets
from NNPDF1.0 version onwards. 

\subsubsection{Parton parametrization}
\label{parm}
Existing PDF sets differ in the number and choice of linear combinations 
of PDFs which are independently parametrized and the functional form and 
number of parameters used.
For the functional form the most common choice is that
each PDF at some reference scale $Q_0$ is parametrized as
\beq
f_i(x,Q_0)=N x^{\alpha_i} (1-x)^{\beta_i} g_i(x)
\label{pdfparm}
\eeq
where $g_i(x)$ is a function which tends to a constant for both $x\to1$
and $x\to0$, for example $g_i(x)=1+ \epsilon_i \sqrt{x}+ D_i x+ E_i x^2$ (HERAPDF). The fit parameters are $\alpha_i$, $\beta_i$, and the
parameters in $g_i$. Some of these parameters may be chosen to take a
fixed value (including zero).
The general form (\ref{pdfparm})  is adopted in
all the PDF sets which we discuss here except for the case of NNPDF which, instead, defines
\beq
f_i(x,Q_0)= c_i(x) NN_i(x),
\label{pdfparmnn}
\eeq
where $NN_i(x)$ is a neural network (a feed-forward neural network
with two hidden layers, see
\Bref{Ball:2010de} for details)
and $c_i(x)$ is  a
{\em preprocessing} function. The fit parameters determine the shape 
of $NN_i(x)$. The function $c_i(x)$
is chosen randomly in a space of functions of
the form $x^{\alpha_i} (1-x)^{\beta_i}$, within some acceptable range of
the parameters $\alpha_i$ and $\beta_i$.
For each group the basis functions and number of parameters are the following.

\begin{itemize}
  
\item ABKM parametrizes the two lightest flavours, corresponding
  anti-flavours, the total strange-ness, and the gluon (six independent
  PDFs) with 21 free parameters.
\item CTEQ6.6 and CT10  parametrize  the two lightest flavours and
  anti-flavours, the total strangeness, and the gluon (six independent
  PDFs) with respectively $22$ and $26$ free parameters.
\item GJR parametrizes the two lightest flavours, their anti-flavours,
  and the gluon with $20$ free parameters (five independent PDFs); the
  strange distribution is assumed to be either proportional to the
  light sea or to vanish at a low scale $Q_0<1$\UGeV\ at which PDFs become valence-like.
\item HERAPDF parametrizes the two lightest flavours, $\PAQu$,
  the combination $\PAQd + \PAQs$, and the gluon with $10$ free
  parameters (five independent PDFs), strangeness is assumed to be
  proportional to the $\PAQd$ distribution; the
  effect of varying the form of the parametrization and the  
  size of the strange component is also studied.  
\item  MSTW parametrizes the three lightest flavours and anti-flavours, and
  the gluon with $28$ free parameters (seven independent PDFs) to find
  the best fit, but $8$ are held fixed in the determination of the uncertainty eigenvectors.
\item NNPDF parametrizes the three lightest flavours and
  anti-flavours, and the gluon with $259$ free parameters ($37$ for
  each of the seven independent PDFs).

\end{itemize}

\subsection{PDF determinations -- theoretical uncertainties}
\label{sec:pdfdet2}

The theoretical uncertainties of the PDFs
reflect the approximations in the theory
which is used in order to relate the PDFs to the measurable quantities.
The study of theoretical PDF uncertainties is currently less advanced
than that of  experimental uncertainties, and only some of
the theoretical
uncertainties have been explored till now. One might expect that one of 
the main theoretical uncertainties in PDF determinations
should be related to the treatment of the strong interaction: in
particular, to the values of the strong coupling constant ($\alphas$) 
and of the heavy-quark masses  ($\Mc$ and
$\Mb$), and the uncertainties related to the truncation of the perturbative
expansion (commonly estimated through the variation of renormalization
and factorization scales). The uncertainty on $\alphas$ has
been explored systematically by the PDF groups. The  effect of varying $\Mb$ and
$\Mc$ has been included by HERAPDF in  
model uncertainties, and these are parameters in the covariance matrix for 
ABKM~\cite{Alekhin:2009ni}. Sets with varying quark masses 
and implications have been 
made available by MSTW~\cite{Martin:2010db}, and 
preliminary studies of the effect of $\Mb$ and $\Mc$ have also been 
presented by NNPDF~\cite{Guffanti:2010yu}.
 Further uncertainties are related to the
treatment of the heavy-quark thresholds, which are handled in various ways
by different groups (see Section~22 of \Bref{Binoth:2010ra}), 
to numerical approximations,
and to the treatment of electroweak effects (such as QED PDF evolution~\cite{Martin:2004dh}).

\subsubsection{The value of $\alphas$ and its uncertainty}

The choice of value of $\alphas$ is clearly important because it
is strongly correlated to PDFs, especially the gluon distribution: this
correlation is studied in detail in \Bref{Demartin:2010er} using
CTEQ, MSTW, and NNPDF PDFs; 
$\alphas$ is a parameter in the covariance matrix for 
ABKM, GJR(JR). 
There are two separate issues related to the value of $\alphas$ in
PDF fits: first, the choice of $\alphas$ for which PDFs are made available, and second the choice of the preferred value of $\alphas$ to be
used when giving PDFs and their uncertainties. 
These are  two separate though related issues, and for each of them
two different basic philosophies 
may be adopted. In what concerns available  values, 
for some groups PDF fits 
are performed for a number of different values of
  $\alphas$. Though a  PDF set corresponding to some
  reference value of $\alphas$ is given, the user is free to
  choose any of the given sets. This philosophy is adopted by CTEQ ($0.118$),
  HERAPDF ($0.1177$), MSTW ($0.120$), and NNPDF ($0.119$), where, in parenthesis, the reference (NLO) value of $\alphas$ for each set is indicated.
  For others, $\alphas$ is treated as a fit parameter, and PDFs
  are given only for the best-fit value. This philosophy is adopted
  by ABKM ($0.11801$) and GJR ($0.1145$).
Concerning the preferred central value and uncertainty, 
for some groups the value of 
$\alphas(\MZ)$ is taken as an external parameter. 
  This philosophy is adopted by CTEQ,
  HERAPDF1.0, and NNPDF. In this case, there is no a-priori central value of $\alphas(\MZ)$, and the uncertainty on
  $\alphas(\MZ)$ is included by repeating the PDF determination as
  $\alphas$ is varied in a suitable range. 
  For others $\alphas$ is treated as a fit parameter, so its value
  and  uncertainty are
  determined along with the PDFs. This  philosophy is adopted by
  MSTW, ABKM, and GJR08.

When comparing results obtained using different PDF sets it should be
borne in mind that if different values of $\alphas$ are used,
 predictions for cross sections change both due to their dependence on 
$\alphas$ (which for some LHC processes,
such as top-pair production or Higgs 
production in $\Pg\Pg$ fusion may
be quite strong), as well as to the dependence of the PDFs on the
value of $\alphas$. Differences due to the PDFs alone can be isolated
only while performing comparisons at a common value of $\alphas$. The
different groups have different ways of calculating the total uncertainty 
due to both the PDFs and $\alphas$. This is explained in more detail 
in \Bref{Demartin:2010er}  and in publications from each of the groups, in 
particular, \Brefs{Martin:2009bu,Lai:2010nw,Alekhin:2009ni}.

\subsection{Comparison of results from different PDFs}
\label{sec:benchmarking}

To compare results from different PDF sets it is useful to introduce 
differential parton--parton luminosities, which, when multiplied by the 
dimensionless cross section $\hat{s}\hat{\sigma}$ for a process, 
provide an estimate  of 
the size of an event cross section at the LHC. 
The differential parton--parton luminosity $dL_{ij}/d\hat{s}$ is
\begin{equation}
\frac{d L_{ij}}{d\hat{s}\,dy} \equiv 
\frac{1}{s} \, \frac{1}{1+\delta_{ij}} \, 
[f_i(x_1,\mu) f_j(x_2,\mu) + (1\leftrightarrow 2)] \; ,
\label{eq1}
\end{equation}
where the prefactor avoids double-counting for identical partons. A 
generic cross section is written as
\begin{equation}
\sigma = \sum_{i,j} \int_0^1 dx_1 \, dx_2 \, 
f_i(x_1,\mu) \, f_j(x_2,\mu) \, \hat{\sigma}_{ij}
\equiv \sum_{i,j} \int \left(\frac{d\hat{s}}{\hat{s}} \right) 
\, \left(\frac{d L_{ij}}{d\hat{s}}\right) \, 
\left(\hat{s} \,\hat{\sigma}_{ij} \right) \; .
\label{eq:xseclum}
\end{equation}

\begin{figure}[ht]
\begin{center}
\includegraphics[width=0.48\textwidth]{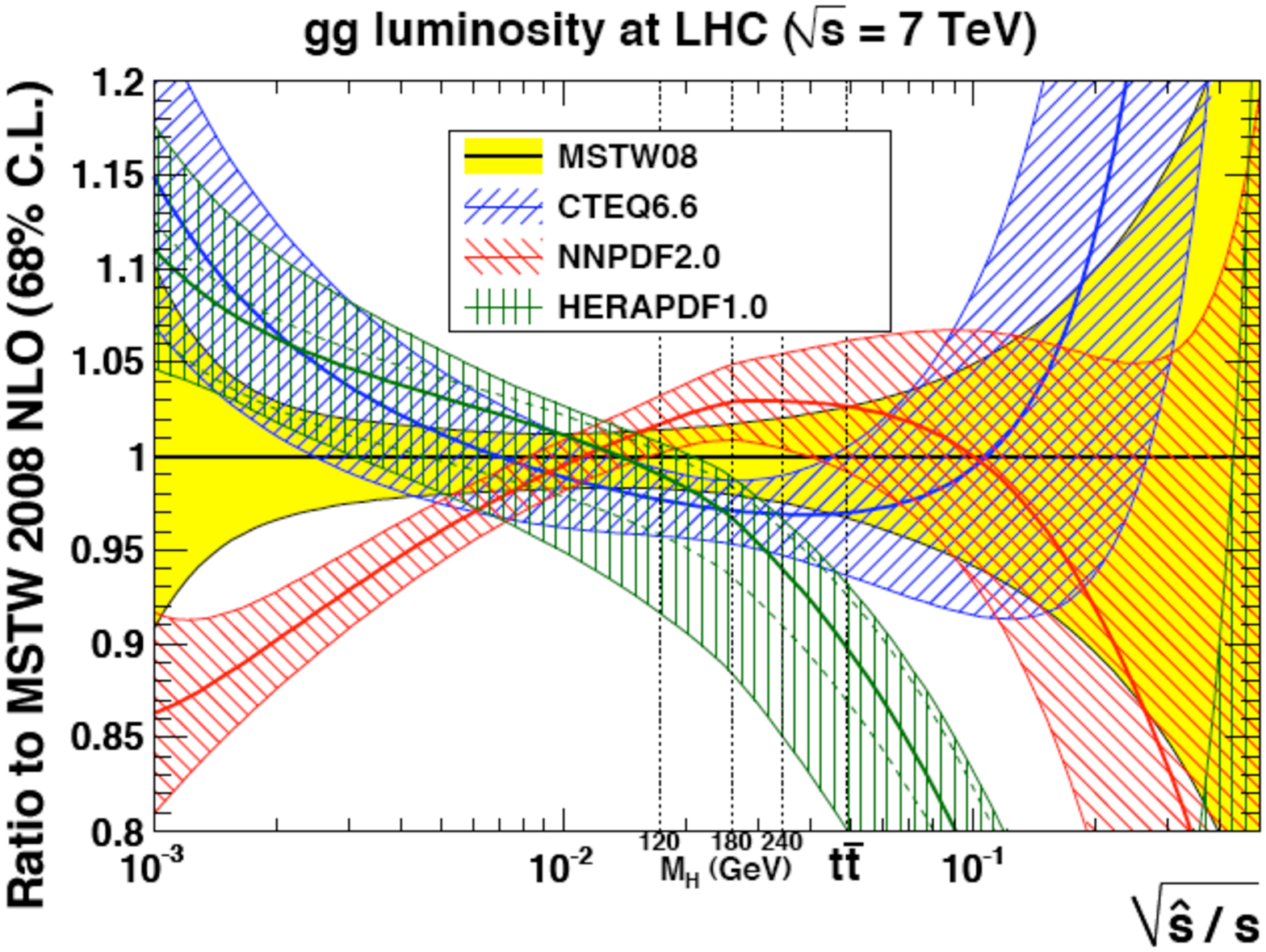}
\includegraphics[width=0.48\textwidth]{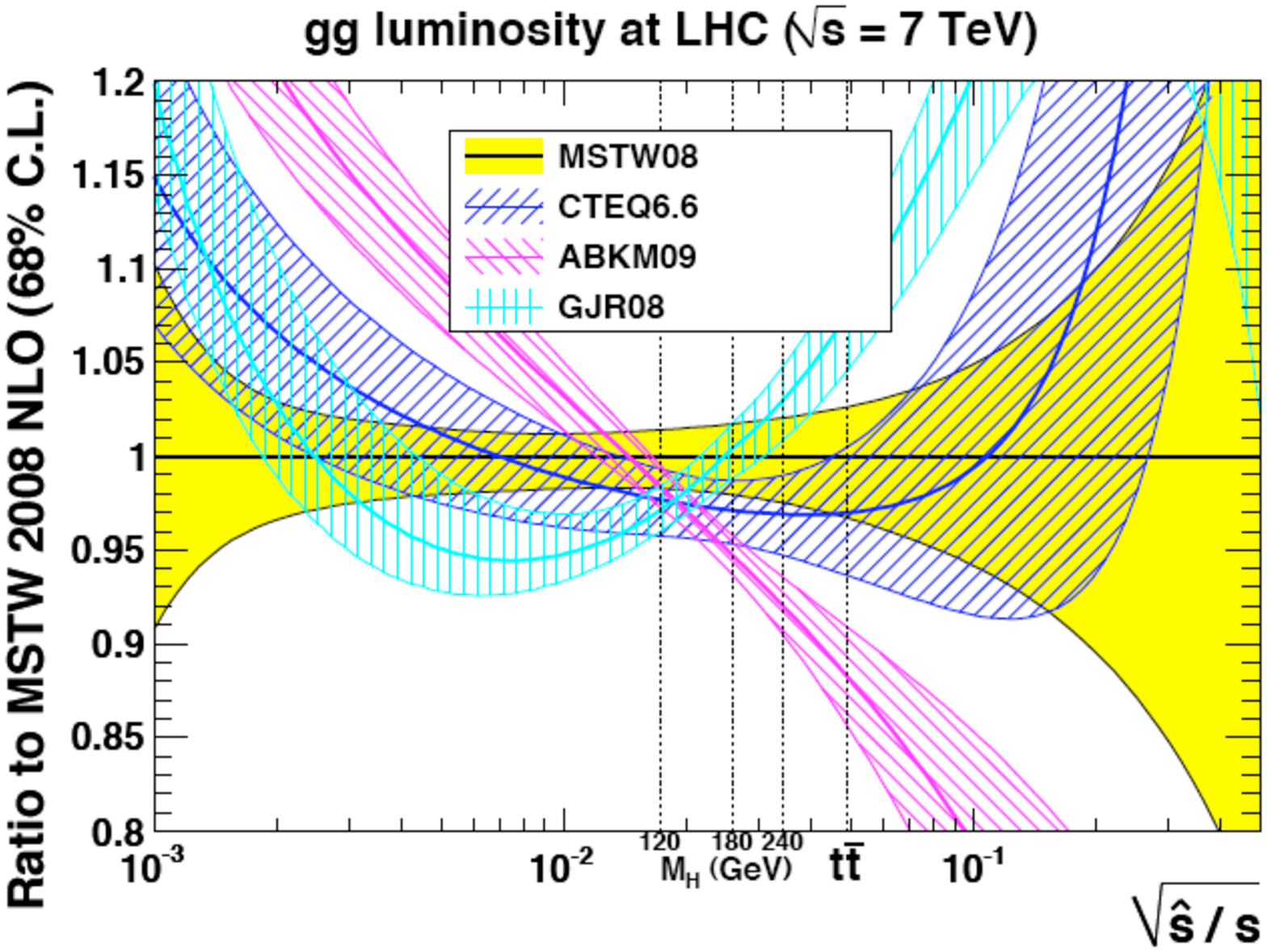}
\vspace{-0.8cm}
\caption{\label{fig:gglum} The $\Pg\Pg$  
luminosity functions and
uncertainties at $7$\UTeV, normalized to the MSTW08 result. 
(Plots by G.~Watt~\cite{Watt}.)}
\vspace{-0.6cm}
\end{center}
\end{figure}

The relative $\Pg\Pg$ PDF luminosities at NLO, along with their $68\%$~CL
error bands, are shown in \Fref{fig:gglum},
normalized to the MSTW08 central value~\cite{Watt}.
For HERAPDF1.0 the inner uncertainty bands (dashed lines) 
correspond to the experimental errors, and the
outer uncertainty bands (shaded regions) include errors due to model and
parametrization. The $\PQq\PAQq$ luminosity plots \cite{Watt} look 
similar, but turn upwards at high $\hat s/s$ for HERAPDF1.0. The error 
bands for each of the PDF
luminosities are of similar size. The luminosity  for the range of 
$\PQt\PAQt$ and Higgs production  are in
good agreement for CTEQ, MSTW, and NNPDF, while the agreement
with ABKM, HERAPDF, and GJR is less good at higher masses. 
It is notable that the PDF
luminosities tend to differ at low $x$ and high $x$. 
The CTEQ6.6 distributions, for
example, may be larger at low $x$ than MSTW2008, due to the
positive-definite parametrization of the gluon distribution; the MSTW
gluon starts off as negative at low $x$ and $Q^2$, and this results in an
impact for both the gluon and sea-quark distributions at larger $Q^2$
values. The NNPDF2.0 $\PQq\PAQq$ luminosity tends to be somewhat lower,
in the $\PW,\PZ$ region for example.  Part of this effect might come from
the use of a zero-mass heavy-quark scheme, although other differences 
might be relevant. However, there are other discrepancies of more than $20\%$
at high or low invariant masses. 

At small $x$ details of heavy-flavour 
treatment cause some deviation, and there is also an 
anticorrelation with the value of $\alphas$ which varies 
between groups (with the GJR value differing most). 
At high $x$ Tevatron jet data gives a constraint on the gluon 
(though there is some variation depending on the data set) 
and this data is not used in ABKM09 (investigations by ABM may be found at 
\Brefs{Trento,PDF4LHCnov10}) and 
HERAPDF1.0 fits. At high $x$, $\PW$ production data (not used by ABKM,
GJR, and HERAPDF) 
constrain the light-quark distributions, which are then correlated to
the gluon by the momentum sum rule.
 The high and low-$x$ gluon distributions are also 
anti-correlated by the momentum sum rule. All these factors, amongst
others, may influence the forms of the gluon luminosities and be
responsible for the discrepancies observed.
   
Benchmark computations of LHC total 
cross sections and rapidity distributions from various PDF groups
can be found in~\Bref{PDF4LHCwiki}
(see also \Bref{Alekhin:2010dd});  the degree of agreement and 
discrepancy between the groups is commensurate with the luminosity plots 
shown here. Differences between the luminosities and predictions 
for those sets which exist at NNLO are similar to NLO, showing that 
they are most likely due to choices of data sets in the fit or other assumptions 
rather than theoretical procedures, such as 
different schemes for the treatment heavy flavours, 
for which differences should become smaller at higher orders.

\begin{figure}
  \begin{center}
    \includegraphics[width=0.48\textwidth]{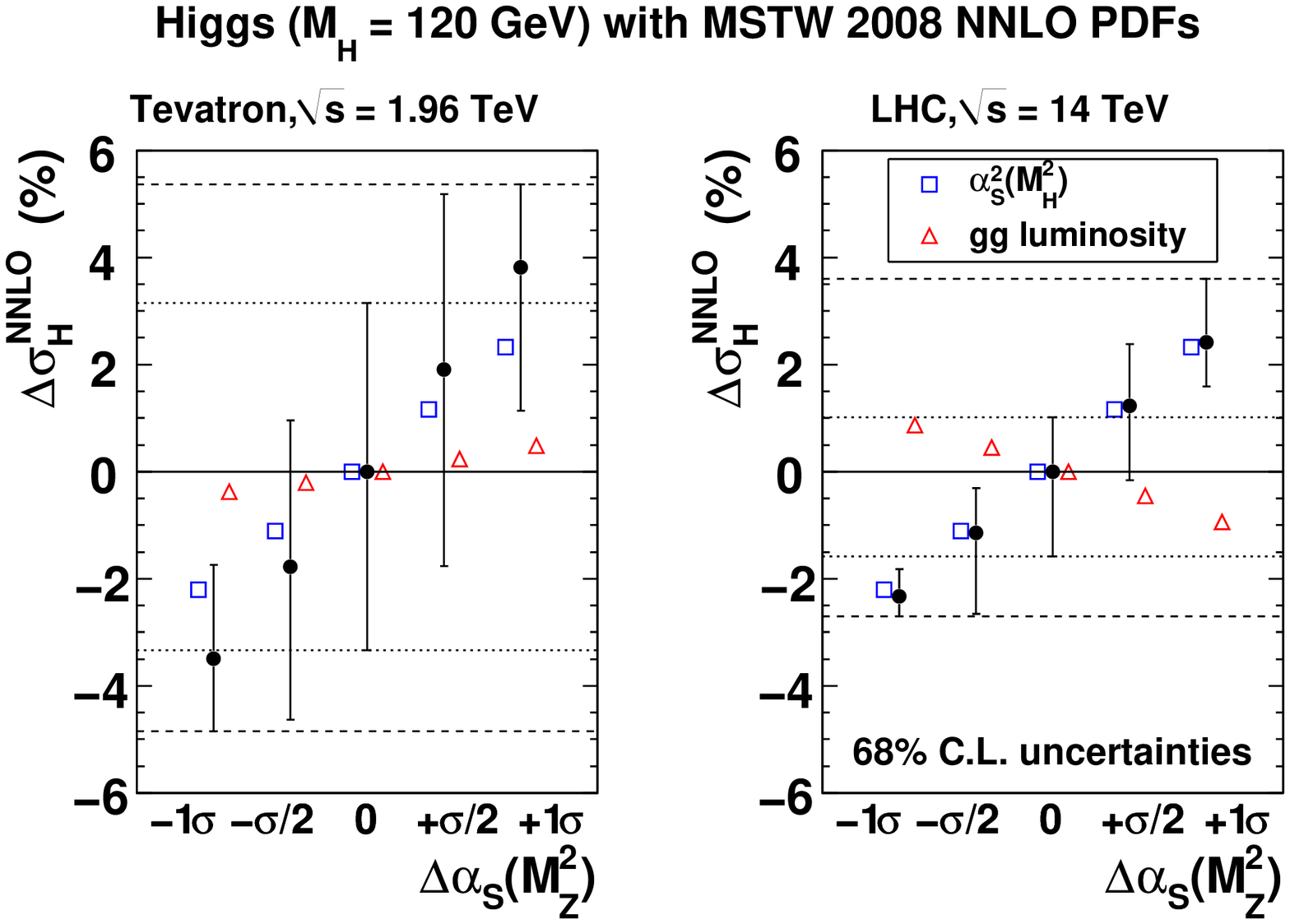}
    \includegraphics[width=0.48\textwidth]{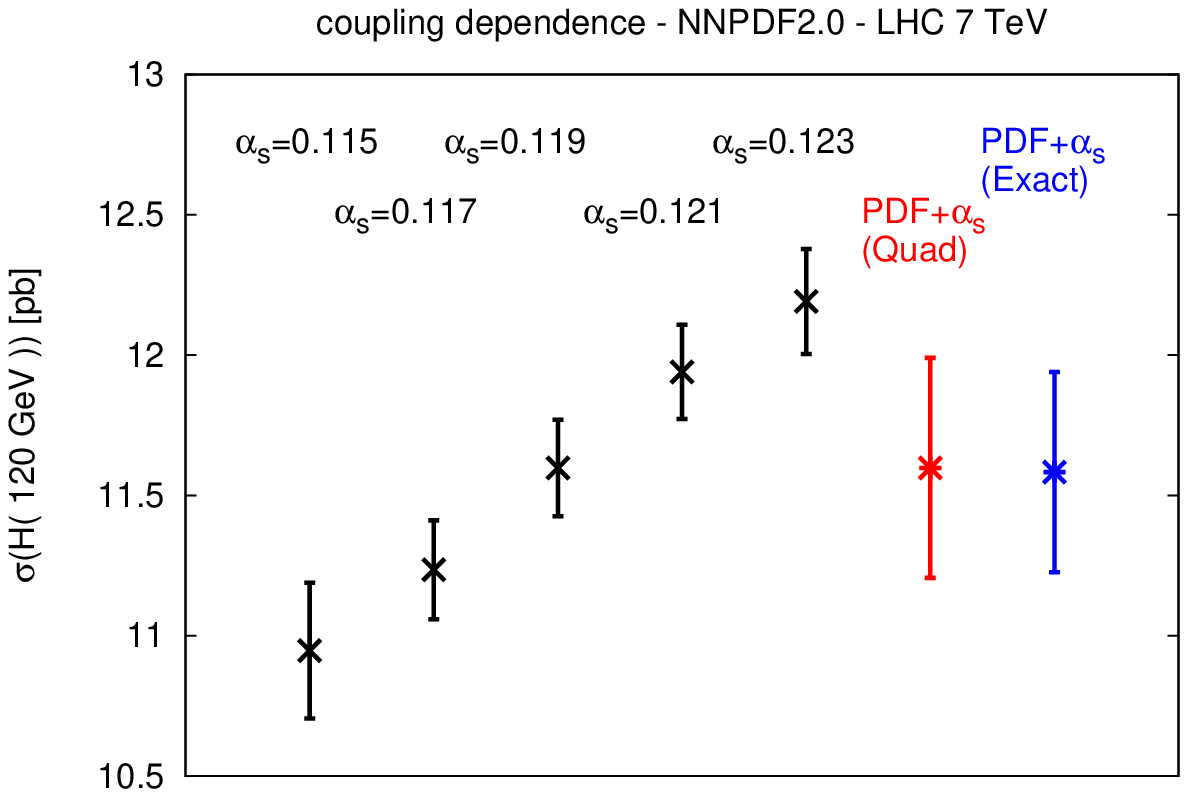}\\
\vspace{-0.2cm}
\includegraphics[width=0.48\textwidth]{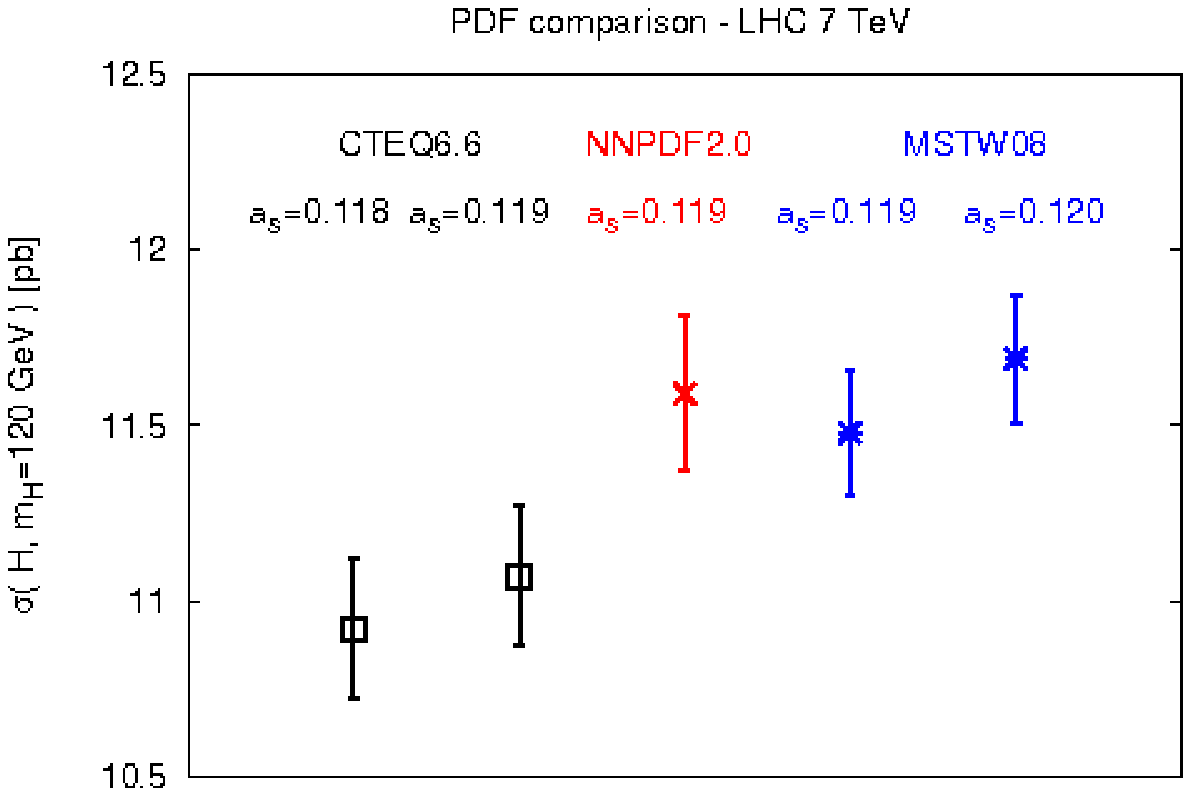}
\vspace{-0.6cm}
\caption{\label{fig:Higgscs} Cross-section predictions as a function
  of $\alphas$ for a Higgs 
($\Pg\Pg$ fusion) for a Higgs mass of $120$\UGeV\ at NNLO for the 
Tevatron and LHC at $14$\UTeV~\cite{Martin:2009bu} (top-left)  
and at NLO for the LHC at $7$\UTeV~\cite{Ubiali:2010xc} (top-right) and for 
various groups all at NLO~\cite{Ubiali:2010xc} (bottom).}
\vspace{-0.6cm}
\end{center}
\end{figure}

It is also very useful to show the cross sections as a
function of $\alphas$.
The predictions for Higgs production  from $\Pg\Pg$ fusion
(shown for MSTW08 and NNPDF2.0 in the top left and right plots of 
\Fref{fig:Higgscs}, respectively) depend strongly on the value
of $\alphas$: the anticorrelation (or correlation for the Tevatron) 
between the gluon distribution and
the value of $\alphas$ is not sufficient to offset the growth of the
cross section as seen from the top-left plot.
In the bottom plot one sees that 
CTEQ, MSTW, and NNPDF predictions are in moderate agreement
but CTEQ lies somewhat lower, to some extent due to
the lower choice of $\alphas$. Compared at the common value of 
$\alphas(\MZ^2)=0.119$, the CTEQ prediction and those from others have 
one-sigma PDF uncertainties which just about overlap  for
$\MH=120$\UGeV. This trend is similar up to about 
$\MH=180$\UGeV, and the agreement 
improves for higher masses, as seen in \Fref{nloenvelope} below. 
Hence, both the difference in PDFs and
in the dependence of the cross section on the value of $\alphas$ are
responsible for the differences observed. A useful measure of this 
is to note that the difference in the central values of the MSTW and 
CTEQ predictions for a common value of $\alphas(\MZ^2)=0.119$ and for a 
Higgs-boson  mass of $120$\UGeV\ (a typical discrepancy) is equivalent
 to a change in 
$\alphas(\MZ^2)$ of about $0.0025$. The worst discrepancy between 
CTEQ and either MSTW or NNPDF at any mass value of the Higgs 
is similar to a change in $\alphas$ of about $0.004$.
The predictions using some of the other PDF sets are rather 
lower~\cite{PDF4LHCwiki}, 
particularly at high masses, reflecting the
behaviour of the gluon luminosity of \Fref{fig:gglum}. 

\begin{figure}
\begin{center}
  \includegraphics[width=0.48\textwidth]{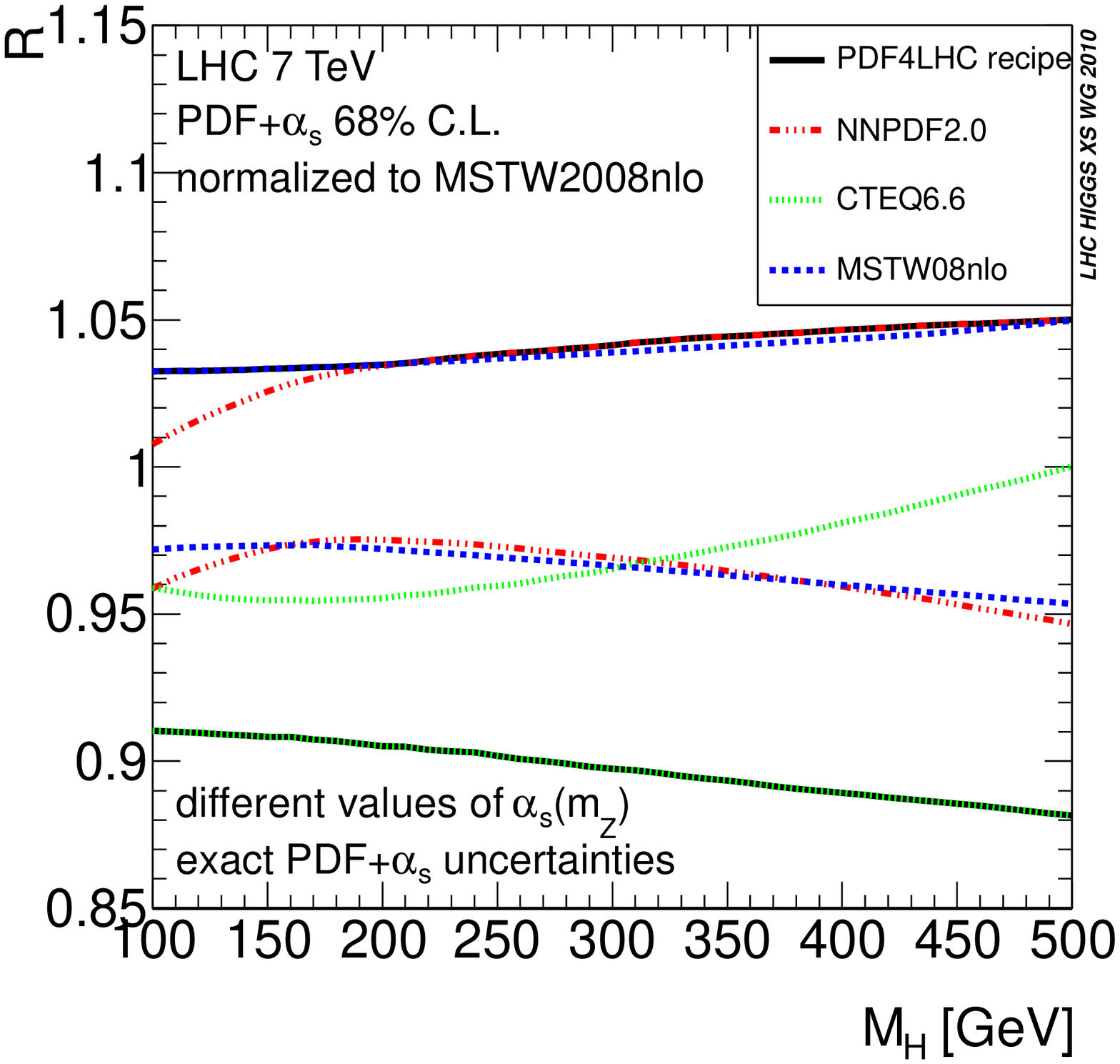}
  \includegraphics[width=0.48\textwidth]{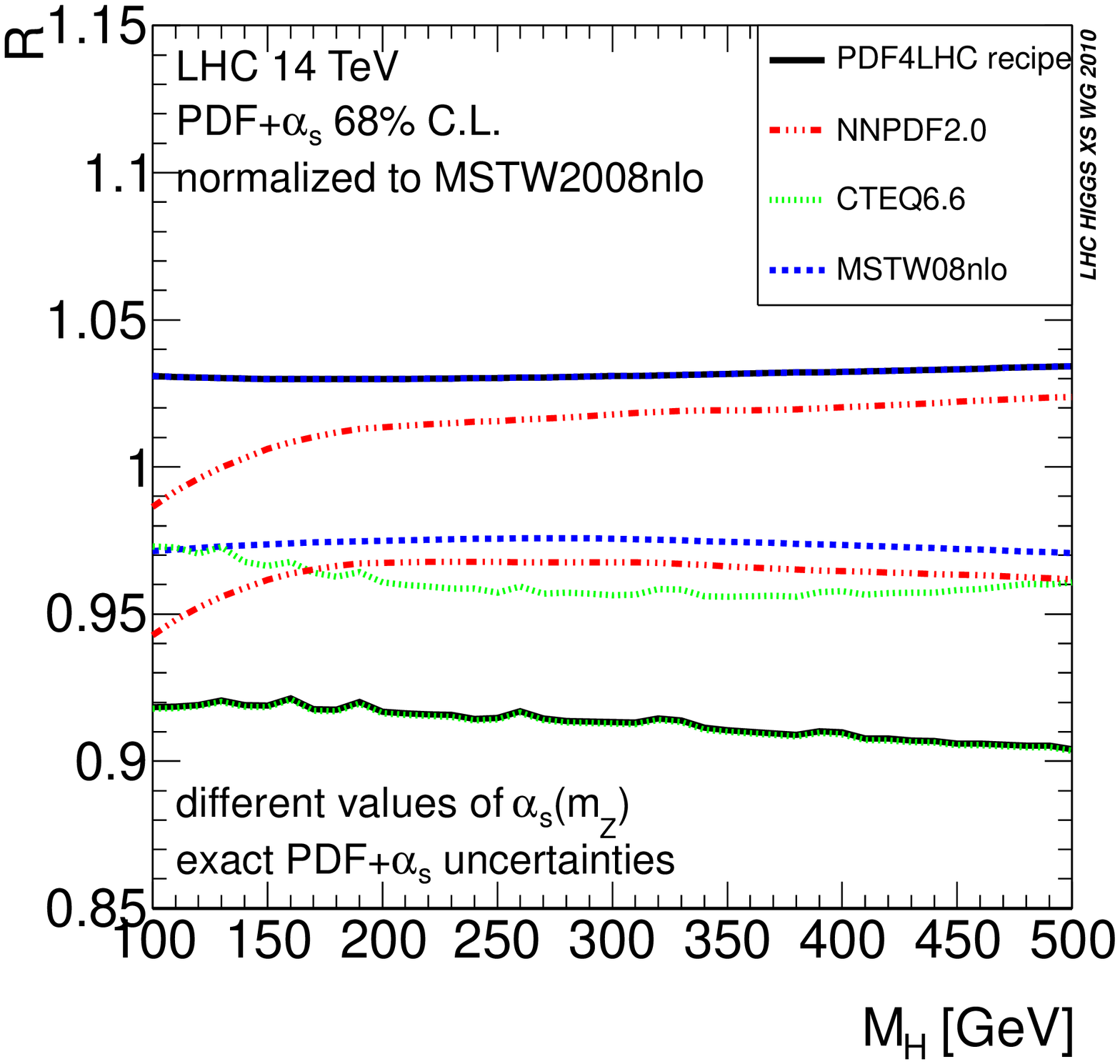}
\end{center}
\vspace{-0.6cm}
\caption{Combined PDF+$\alphas$ uncertainty band
for the total Higgs production cross section via gluon fusion,
at NLO,
evaluated according to the PDF4LHC recipe.
The bands are normalized to the central MSTW2008 NLO result.}
\vspace{-0.6cm}
\label{nloenvelope}
\end{figure}

\subsection{The PDF4LHC recommendation}
\label{sec:pdf4lhcreco}

Before we present our recommendation, we would like to highlight the
differences between two use cases: (1) cross sections which have not
yet been measured (such as, for example, Higgs production) and (2)
comparisons to existing cross sections. For the latter, the most
useful comparisons should be to the predictions using individual PDFs
(and their uncertainty bands) discussed above. Such cross sections
have the potential, for example, to provide information useful for
modification of those PDFs. For the former, in particular the 
cross-section predictions in this Report, we would like to provide a
reliable estimate of the true uncertainty, taking into account
possible differences between the central values of predictions using
different PDFs. From the results seen it is clear that this
uncertainty should be larger than that from any single PDF set; however
in order for the probabilistic interpretation of PDF uncertainties to
be preserved, it should not lose all connection to the individual PDF
uncertainties, which would inevitably happen for many processes if the full
spread of all PDFs were used. In order to do this,  
some compromise must be  reached.

As seen at NLO there is always reasonable agreement
between MSTW, CTEQ, and NNPDF, and potentially more deviation with the other 
sets. In some cases this deviation has at least one potential origin,
 e.g.\ the $\PQt\PAQt$ 
cross section at $7$\UTeV\ at the LHC probes similar PDFs as probed in
 the lower-$\pT$ 
jet production at the Tevatron, which has neither been fit nor validated 
against quantitatively by some groups (preliminary results for ABM may be 
found at \Brefs{Trento,PDF4LHCnov10}). As noted, large deviations 
in predictions between 
existing NNLO sets are similar to those between the same NLO sets.
Discrepancies in MSTW, CTEQ, and NNPDF
do not always have clear origin, or may be a matter of procedure (e.g.\ gluon 
parametrization) which is an ongoing debate between groups. Bearing this in 
mind and having been requested to provide a procedure to give a moderately 
conservative uncertainty, we adopt the following
PDF4LHC recommendation~\cite{PDF4LHCwebpage}. 

\subsubsection{NLO prescription}

At NLO the recommendation is to use (at least) 
predictions from the PDF fits from CTEQ, MSTW, and
NNPDF. These sets all use results from a hadron collider experiment, i.e.,\ the
Tevatron as well as fixed-target experiments and HERA,
and they make available specific sets for a
variety of values of $\alphas$. The PDF versions to be used are: CTEQ6.6,
MSTW2008, and NNPDF2.0. Neither the CTEQ6.6 nor the MSTW2008 
use the new combined very accurate HERA data sets, whereas NNPDF2.0 does
use this data (the CT10~\cite{Lai:2010vv}   update of 
the CTEQ PDFs does include them and future updates of MSTW~\cite{Thorne:2010kj} will
as well). It is to be noted 
that CTEQ6.6 and MSTW2008 are the PDF versions most commonly used by the LHC
experiments currently, hence it is these
versions that are suggested in the recommendation.
 The NNPDF2.0 set does not use a general-mass variable flavour
number scheme (the NNPDF2.1 PDF set, which does use a general-mass
variable flavour number scheme is currently being
finalized~\cite{Rojo:2010gv}), but the
alternative method which NNPDF use for determining PDF uncertainties
provides important independent information.  
Other PDF sets, GJR08, ABKM09, and HERAPDF1.0 are useful for more 
conservative or extensive evaluations of the uncertainty.  
For example a study of the 
theoretical uncertainties related to the charm-mass treatment is possible 
using HERAPDF1.0 and ABKM. 

The  $\alphas$ uncertainties can be evaluated by
taking a range of $\pm 0.0012$ for $68\%$~CL (or $\pm 0.002$ for $90\%$~CL) from the preferred central value for CTEQ and NNPDF. The total
PDF+$\alphas$ uncertainty can then be evaluated by adding the
variations in PDFs due to $\alphas$ uncertainty in quadrature with
the fixed $\alphas$ PDF uncertainty (shown~\cite{Lai:2010nw} 
to correctly incorporate
correlations in the quadratic error approximation) or, for NNPDF,
more efficiently taking a gaussian distribution of PDF replicas
corresponding to different values of $\alphas$. 
For MSTW the PDF+$\alphas$ uncertainties should be evaluated using
their prescription which better accounts for correlations between the
PDF and $\alphas$ uncertainties when using the MSTW dynamical
tolerance procedure for uncertainties. Adding the $\alphas$
uncertainty in quadrature for MSTW can be used as a simplification but
generally gives slightly smaller uncertainties.  

So the prescription for NLO is as follows:

\begin{itemize}

\item For the  calculation of uncertainties at the LHC, use  the
  envelope provided by the central values and PDF+$\alphas$ errors
  from the MSTW08, CTEQ6.6, and NNPDF2.0 PDFs, using each group's
  prescriptions for combining the two types of errors. We propose this
  definition of an envelope because the deviations between the
  predictions can sometimes be  as large as  their uncertainties. 
  As a central value, use the midpoint of this
  envelope. We follow the PDF4LHC prescription and
  recommend that a $68\%$~CL uncertainty envelope be
  calculated and the $\alphas$ variation suggested is consistent with
  this. Note that the CTEQ6.6 set has uncertainties and $\alphas$
  variations provided only at $90\%$~CL and thus their uncertainties
  should be reduced by a factor of $1.645$ for $68\%$~CL. Within the
  quadratic approximation, this procedure is exact.

\end{itemize}

\subsubsection{NNLO prescription}

For estimating uncertainties in cross section at NNLO, the 
recommendation is to use for base predictions 
the only NNLO set which currently includes a wide variety of hadron 
collider data sets, i.e.,\ MSTW2008. There seems to be
no reason to expect that the spread in predictions of the sets used in the 
NLO prescription, i.e.,\ MSTW, CTEQ, and NNPDF,  will diminish significantly 
at NNLO compared to NLO, where this spread was somewhat larger than the 
uncertainty from each single group. 
Hence, at NNLO the uncertainty obtained from MSTW alone should be expanded 
to some degree. It seems appropriate to do this by multiplying the MSTW 
uncertainty at NNLO by the factor obtained by dividing the full uncertainty 
obtained from the envelope of MSTW, CTEQ, and NNPDF results at NLO by the 
MSTW uncertainty at NLO. In all cases the $\alphas$ uncertainty should be 
included. We note that in several cases so far examined, for the LHC running at 
$7$\UTeV\ centre-of-mass energy, this factor of the envelope divided by the 
MSTW uncertainty is approximately constant, and 
quite close to $2$ for Higgs production as shown below: 
this constant factor can be used as a 
short-hand prescription. 

Since there are NNLO PDFs obtained from fits including fewer data sets by 
the ABKM, JR, and HERAPDF groups, these should ideally be compared with the 
above procedure, bearing in mind that it is possible there will be kinematic 
regions where the absence of data, or other reasons -- e.g.\ in the JR case 
a theoretical 
constraint is imposed on the input by the choice of assuming the form
Eq.~(\ref{pdfparm}) of PDFs at a very low, arguably non-perturbative,
starting scale (though data are fit only at higher scales) -- may lead to PDFs and predictions differing 
significantly from the central value and the extent of the uncertainty band.

So the prescription at NNLO is:

\begin{itemize}

\item As a central value, use the MSTW08 prediction. As an uncertainty, take the same percentage uncertainty on this NNLO prediction as found using the  NLO uncertainty prescription given above.

\end{itemize}

\subsubsection{Application to Higgs production via gluon fusion}
\label{sec:pdf4lchiggs}



\begin{figure}
  \begin{center}
    \includegraphics[width=0.48\textwidth]{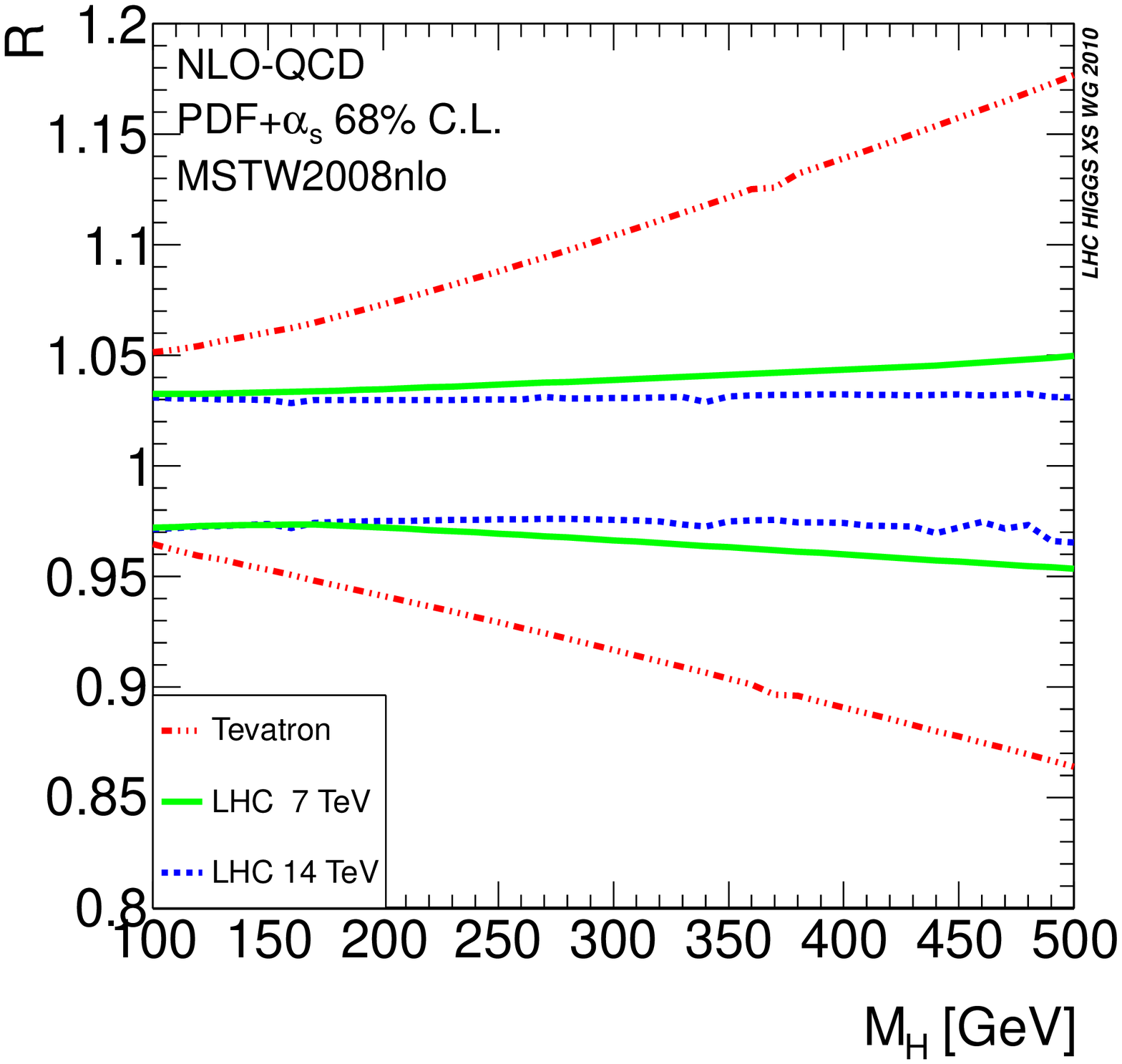}
    \includegraphics[width=0.48\textwidth]{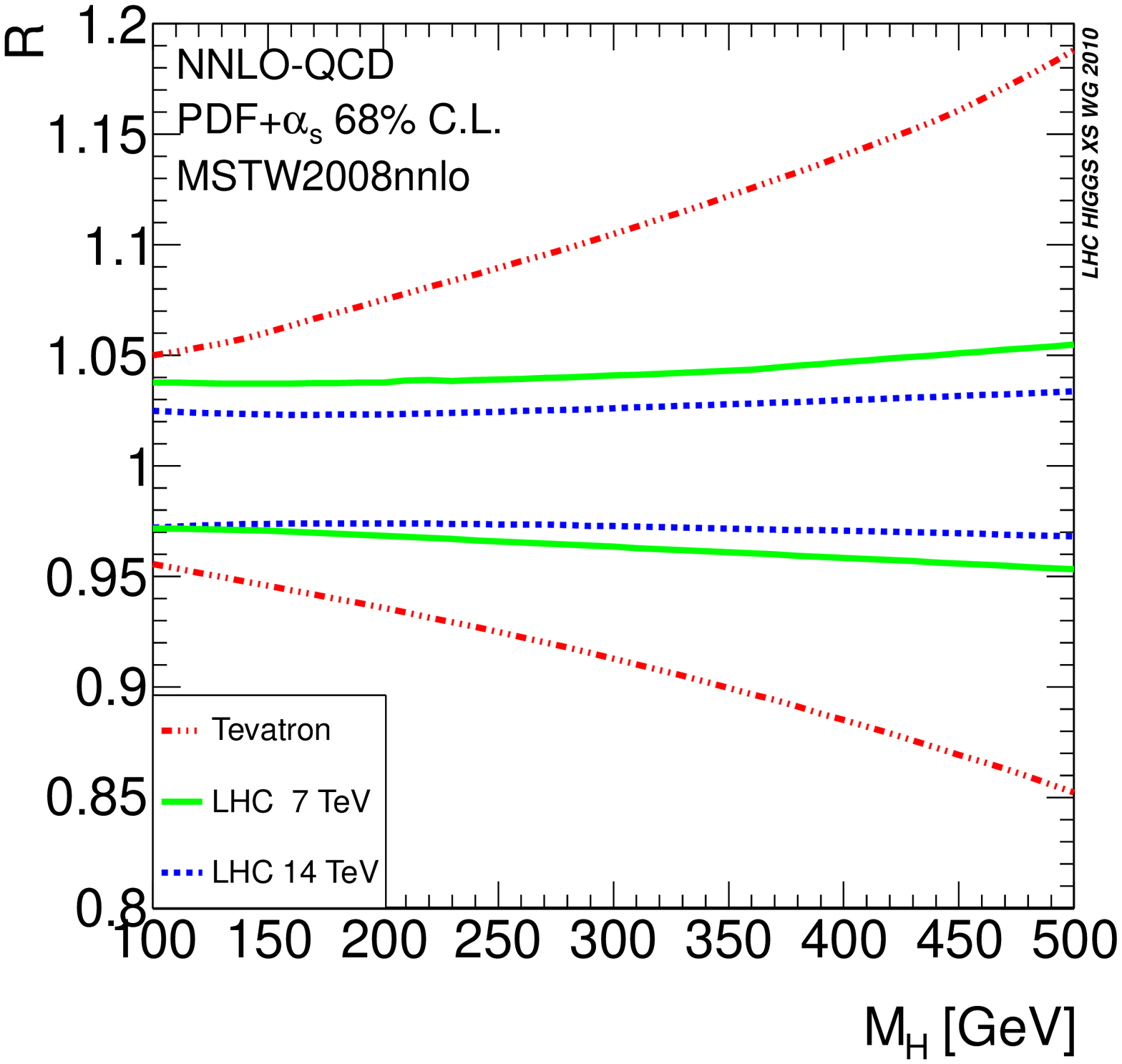}\\
    \includegraphics[width=0.48\textwidth]{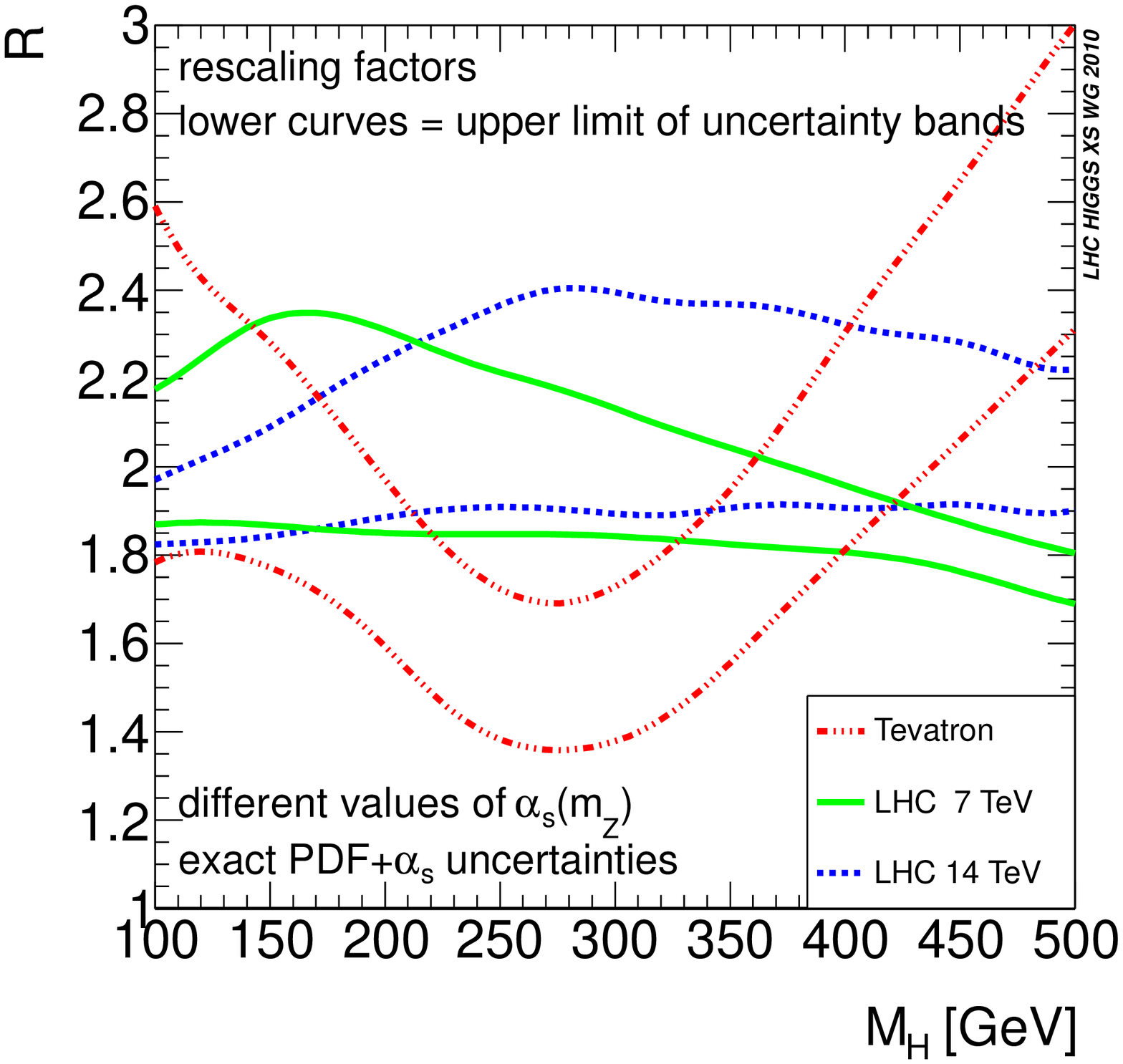}
    \includegraphics[width=0.48\textwidth]{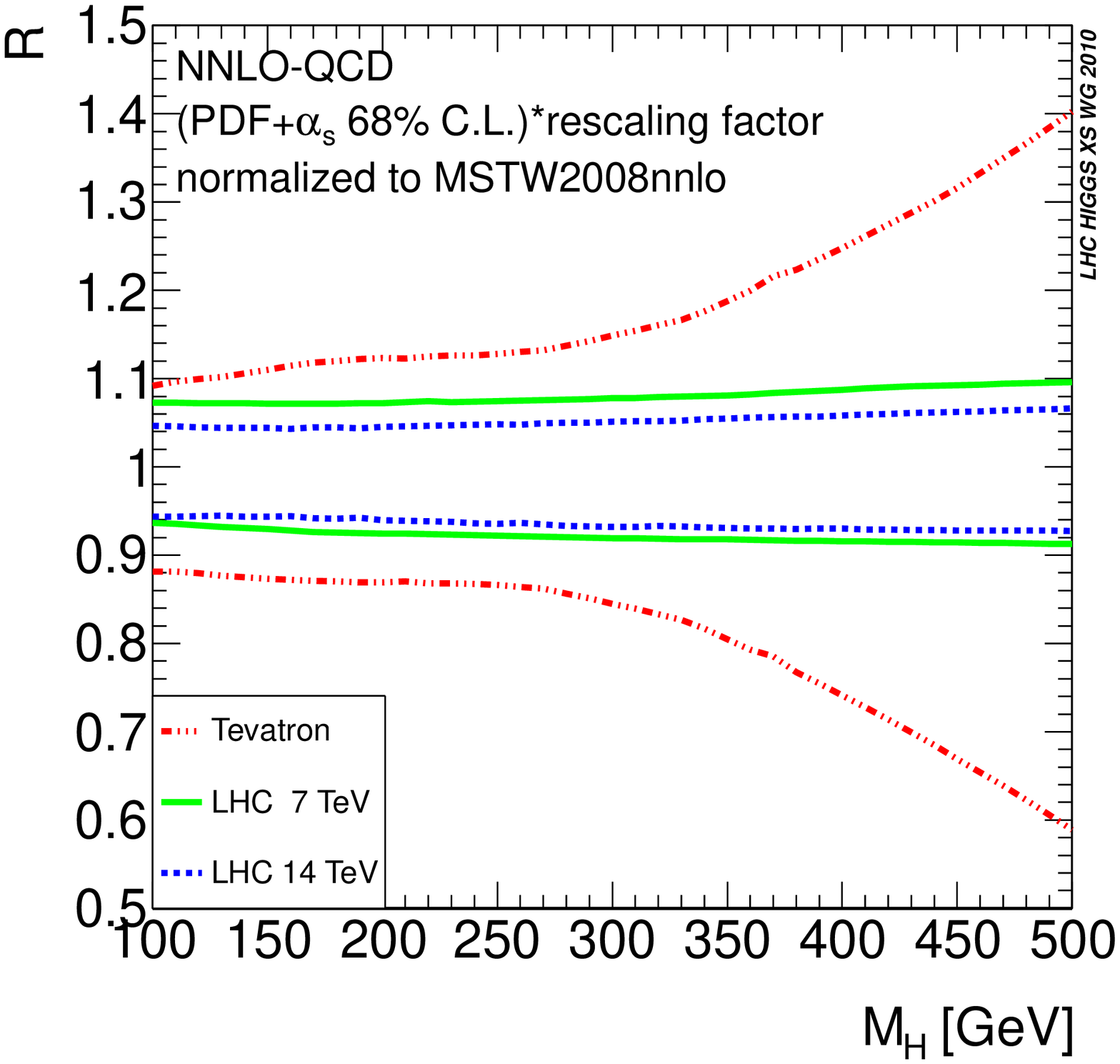}    
  \end{center}
  \vspace{-0.8cm}
\caption{
Top: Combined relative PDF+$\alphas$ uncertainty band
for the total  Higgs cross section from gluon fusion,
at NLO (left) and at NNLO (right) obtained using MSTW2008.
 Bottom:  rescaling factor for the NNLO uncertainty (left),
obtained as  the ratio of the percentage
  width of the NLO envelope with respect to its mid point
  to the percentage uncertainty of the MSTW2008 NLO band, and final
  NNLO uncertainty band obtained applying the rescaling to the MSTW08
  NNLO result (right).
}
\vspace{-0.5cm}
\label{nlonnlo}
\end{figure}
In accordance with the recommendation, we have considered the 
CTEQ6.6~\cite{Nadolsky:2008zw}, 
MSTW2008~\cite{Martin:2009iq, Martin:2009bu}, and
NNPDF2.0~\cite{Ball:2010de} PDF sets. Combined PDF+$\alphas$ uncertainties
for each of the three global sets are computed as discussed in
Sections~\ref{sec:pdfdet1} and~\ref{sec:pdfdet2} (more details are 
in \Bref{Demartin:2010er}). Computations have been performed 
using the code described in
\Brefs{Bonciani:2007ex, Aglietti:2006tp, Aglietti:2004nj, Aglietti:2004ki, Degrassi:2005mc, Degrassi:2004mx}, 
improved with the NNLO corrections~\cite{Harlander:2000mg, Catani:2001ic, Harlander:2001is, Harlander:2002wh, Anastasiou:2002yz, Ravindran:2003um}.

In order to obtain a meaningful comparison between
different PDF sets it is crucial to adopt the same
uncertainty range for the value of $\alphas$. Here we
assume the  same  range as 
for the PDF4LHC benchmarks of \Bref{PDF4LHCwebpage} 
namely
\beq
\delta^{(90)}\alphas=0.002 \quad\mbox{at}\quad 90\%~{\rm CL},\quad\quad
\delta^{(68)}\alphas=0.0012~=~0.002/C_{90}\quad\mbox{at}\quad 68\%~{\rm CL},
\eeq
where $C_{90}=1.64485$. 
In \Fref{nloenvelope} we show 
the combined PDF+$\alphas$ uncertainty bands obtained with
CTEQ6.6, MSTW2008, NNPDF2.0, for LHC at $7$\UTeV\ and $14$\UTeV,
all normalized to the central MSTW2008.
For different Higgs mass values the predictions show
partial agreement of different pairs of the three collaborations
in such a way that only an envelope (the black line)
of the three bands provides 
a conservative estimate of the uncertainty.
This black line corresponds to the NLO PDF4LHC prescription.


At NNLO, 
the PDF4LHC prescription amounts to multiplying the MSTW08 NNLO
percentage 
uncertainty by a factor obtained as the ratio of the MSTW08 NLO
percentage uncertainty to the
NLO envelope percentage uncertainty 
(all shown in \Fref{nlonnlo} along with the final result).  
Note that in this case
the MSTW2008 NLO and NNLO
PDF+$\alphas$ bands are very similar to each other. 
As can be observed in \Fref{nlonnlo},
the rescaling factor is of order $2$: it is approximately constant for
LHC at $7$\UTeV, but it displays a non-trivial Higgs mass dependence
at the Tevatron. Use of the full range of NNLO PDF sets would provide 
significantly more variation, e.g.\ in the above example for a Higgs mass
of $500$\UGeV\ the downwards error band for the LHC at $7$\UTeV\ would increase 
from $4.5\%$ for MSTW2008 to $27\%$, as opposed to $4.5\%$ to $8\%$ using the 
PDF4LHC prescription. 
Some updates on various sets were seen at \Bref{Trento}
with some signs of convergence evident.

\subsection{Summary}
\label{sec:summary}

We have summarized our understanding
of PDFs and the associated experimental and theoretical uncertainties.
The PDF4LHC recommendation is a pragmatic recommendation to be used when 
a prediction for the central value and a conservative estimate of the 
uncertainties is required, which acknowledges that the latter will be larger
than that from an individual set, but is still representative of this 
uncertainty. It has the feature that the uncertainty bands
are never too far from those PDF fits that include the largest number of 
data sets, in 
particular hadron collider data from the Tevatron which has the closest 
correlation to the measurements (particularly for high-mass final states) 
at the LHC. It is most likely expected to evolve when new experimental sets
and new PDF determinations become available.  In the near future 
some of the data used in the PDF determinations will be from the LHC, 
and this
will help to improve the PDFs from all groups. Comparison of current 
predictions, 
together with uncertainties, will help to determine which of the different choices
currently made by different groups are most successful. 

\clearpage

\section{Branching ratios\footnote{A.~Denner, S.~Heinemeyer,
    I.~Puljak, D.~Rebuzzi (eds.); S. Dittmaier, A. M\"uck, M. Spira,
    M. Weber and G. Weiglein.}}


\label{sec:BR}

\subsection{Standard Model (SM) Higgs branching ratios}
\label{sec:SM-BR}

\providecommand{\lsim}
{\;\raisebox{-.3em}{$\stackrel{\displaystyle <}{\sim}$}\;}
\providecommand{\gsim}
{\;\raisebox{-.3em}{$\stackrel{\displaystyle >}{\sim}$}\;}


\providecommand{\HDECAY}{{\sc HDECAY}}
\providecommand{\HIGLU}{{\sc HIGLU}}
\providecommand{\Prophecy}{{\sc Prophecy4f}}
\providecommand{\CPsuperH}{{\sc CPsuperH}}
\providecommand{\FeynHiggs}{{\sc FeynHiggs}}

The branching ratios of the Higgs boson in the Standard Model have
been determined using the programs {\sc HDECAY}
\cite{Djouadi:1997yw,Spira:1997dg,hdecay2}
and {\sc Prophecy4f}
\cite{Bredenstein:2006rh,Bredenstein:2006ha,Prophecy4f}. In a first
step, all partial widths have been calculated as accurately as
possible. Then the branching ratios have been derived from this full
set of partial widths. Since the widths are calculated for on-shell
Higgs bosons, the results have to be used with care for a heavy Higgs
boson ($\MH\gsim500\UGeV$).

The code {\sc HDECAY} calculates the decay widths and branching ratios
of the Higgs boson(s) in the SM and the MSSM. For the SM it includes
all kinematically allowed channels and all relevant higher-order
QCD corrections to decays into quark pairs and into gluons. More
details are given below.
The electroweak next-to-leading order (NLO)
corrections to the decays $\PH\to\PGg\PGg$ and
$\PH\to \Pg\Pg$ have been calculated in
\Brefs{Aglietti:2004ki,Aglietti:2004nj,Aglietti:2006ne,Degrassi:2004mx,Degrassi:2005mc,Actis:2008ug}.
They are implemented in {\sc HDECAY} in form of a grid based on the
calculation of \Bref{Actis:2008ug}.

{\sc Prophecy4f} is a Monte Carlo event generator for $\PH \to
\PW\PW/\PZ\PZ \to 4f$ (leptonic, semi-leptonic, and hadronic)
final states. It provides the leading-order (LO) and NLO partial widths for any
possible 4-fermion final state. It includes the complete NLO QCD and
electroweak corrections and all interferences at LO and NLO. In other
words, it takes into account both the corrections to the decays into
intermediate $\PW\PW$ and $\PZ\PZ$ states as well as their
interference for final states that allow for both. The dominant two-loop
contributions in the heavy-Higgs-mass limit proportional to $G_\mu^2
\MH^4$ are included according to \Brefs{Ghinculov:1995bz,Frink:1996sv}.
Since the calculation is consistently performed with off-shell gauge
bosons without any on-shell approximation, it is valid above, near,
and below the gauge-boson pair thresholds. Like all other light quarks
and leptons, bottom quarks are treated as massless.  Using the LO/NLO
gauge-boson widths in the LO/NLO calculation ensures that the
effective branching ratios of the $\PW$ and $\PZ$ bosons obtained by summing
over all decay channels add up to one.

The results presented below have been obtained as follows. The Higgs total
width resulting from {\sc HDECAY} has been modified according to the
prescription
\begin{equation}
\Gamma_{\PH}=\Gamma^{\mathrm{HD}}-\Gamma^{\mathrm{HD}}_{\PZ\PZ}-\Gamma^{\mathrm{HD}}_{\PW\PW}+\Gamma^{\mathrm{Proph.}}_{4f}
,
\end{equation}
where $\Gamma_{\PH}$ is the total Higgs width, $\Gamma^{\mathrm{HD}}$
the Higgs width obtained from {\sc HDECAY},
$\Gamma^{\mathrm{HD}}_{\PZ\PZ}$ and $\Gamma^{\mathrm{HD}}_{\PW\PW}$
stand for the partial widths to $\PZ\PZ$ and $\PW\PW$ calculated with
{\sc HDECAY}, while $\Gamma^{\mathrm{Proph.}}_{4f}$ represents the
partial width of $\PH\to 4f$ calculated with {\sc Prophecy4f}.  The
latter can be split into the decays into $\PZ\PZ$, $\PW\PW$, and the
interference,
\begin{equation}
\Gamma^{\mathrm{Proph.}}_{4f}=\Gamma_{{\mathrm{H}}\to \PW^*\PW^*\to 4f}
+ \Gamma_{{\mathrm{H}}\to \PZ^*\PZ^*\to 4f}
+ \Gamma_{\mathrm{\PW\PW/\PZ\PZ-int.}}.
\end{equation}

The relative theoretical uncertainties of the calculation resulting
from  missing higher-order corrections are summarized in \refT{tab:uncertainty}.

\begin{table}[ht]
   \caption{Estimate of theoretical uncertainties from missing higher orders.}
   \begin{center}
   \small  
   \begin{tabular}{lllll}
   \hline
   \textbf{Partial width} & \textbf{QCD} & \textbf{Electroweak} & \textbf{Total} \\
\hline
 $\PH \to \PQb\PQb/\PQc\PQc$ &    $\sim 0.1{-}0.2\%$
  &     $\sim 1$--$2\%$ for $\MH \lsim 135\UGeV$     &      $\sim1$--$2 \%$ \\
\hline
$\PH\to \PGt \PGt$ & & $\sim1$--$2\%$  for $\MH \lsim 135\UGeV$ &       $\sim1$--$2 \%$ \\
\hline
$\PH \to \PQt\PQt$ & $\sim 5\%$&
      ${\lsim  2}$--${5\%}$   for $\MH < 500\UGeV$     &       $\sim5\%$ \\
& &      $\sim 0.1 ({\MH}/{1\UTeV})^4$ for $\MH > 500\UGeV$  &       $\sim5$--$10\%$ \\
\hline
$\PH \to \Pg\Pg$ & ${\sim 10\%}$   &
$\sim 1\%$   &   $\sim10\%$\\
\hline
$\PH \to \PGg \PGg$  & ${<1\%}$ & $<1\%$    &  $\sim1\%$ \\
\hline
$\PH \to \PW\PW/\PZ\PZ\to4f$ & $<0.5\%$ &   $\sim 0.5\%$ for $\MH < 500\UGeV$ &  $\sim0.5\%$\\
$\phantom{\PH\to    \to 4f}$             && $\sim 0.17 ({\MH}/{1\UTeV})^4$ for $\MH > 500\UGeV$
&        $\sim0.5$--$15\%$                          \\
\hline
\end{tabular}
\end{center}
\label{tab:uncertainty}
\end{table}
For QCD corrections the uncertainties have been estimated by
the scale dependence of the widths resulting from a  variation of the scale
up and down by a factor $2$ or from the size of known omitted
corrections. For electroweak corrections the missing higher orders
have been estimated based on the known structure and size of the NLO
corrections. For cases where HDECAY takes into account the known NLO
corrections only approximately the accuracy of these approximations
has been used.
These theoretical uncertainties from missing higher-order corrections will
have to be combined with the parametric uncertainties (most notably from the
bottom-quark mass and $\alphas$) to arrive at the full theory uncertainties.

Specifically, the uncertainties of the results from \HDECAY\ are
obtained as follows: For the decays $\PH \to \PQb\PQb, \PQc\PQc$,
\HDECAY\ includes the complete massless QCD corrections up to and
including NNNLO, with a corresponding scale dependence of about
$0.1{-}0.2\%$. The NLO electroweak corrections
\cite{Fleischer:1980ub,Bardin:1990zj,Dabelstein:1991ky,Kniehl:1991ze}
are included in the approximation for small Higgs masses
\cite{Accomando:1997wt} which has an accuracy of about $1\%$ for $\MH <
135\UGeV$.  The same applies to the electroweak corrections to $\PH \to
\PGtp \PGtm$.  For Higgs decays into top quarks \HDECAY\ includes
the complete NLO QCD corrections for small Higgs masses
\cite{Braaten:1980yq,Sakai:1980fa,Inami:1980qp,Gorishnii:1983cu,Drees:1989du,Drees:1990dq,Drees:1991dq}
interpolated to the large-Higgs-mass results at NNNLO far above the
threshold
\cite{Gorishnii:1990zu,Gorishnii:1991zr,Kataev:1993be,Surguladze:1994gc,Larin:1995sq,Chetyrkin:1995pd,Chetyrkin:1996sr}.
The corresponding scale dependence is below $5\%$.  Only the NLO
electroweak corrections due to the self-interaction of the Higgs boson
are included, and the neglected electroweak corrections amount to
about $2{-}5\%$ for $\MH < 500\UGeV$, where $5\%$ refers to the region near
the $\PQt\bar\PQt$ threshold and $2\%$ to Higgs masses far above.  For
$\MH > 500$\UGeV\ higher-order heavy-Higgs corrections
\cite{Ghinculov:1994se,Ghinculov:1995err,Durand:1994pk,Durand:1994err,Durand:1994pw,Borodulin:1996br}
dominate the error, resulting in an uncertainty of
$0.1\times(\MH/1\UTeV)^4$ for $\MH > 500\UGeV$.  For $\PH \to \Pg\Pg$,
\HDECAY\ uses the NLO \cite{Inami:1982xt,Djouadi:1991tka,Spira:1995rr}
and NNLO \cite{Chetyrkin:1997iv} QCD corrections in the limit of heavy
top quarks, while NNNLO QCD corrections \cite{Baikov:2006ch} are
neglected.  The uncertainty from the scale dependence at NNLO is about
$10\%$ for $\MH < 135$\UGeV. The NLO electroweak corrections are included
via an interpolation based on a grid from \Bref{Actis:2008ug}; the
uncertainty from missing higher-order electroweak corrections is
estimated to be $1\%$.  For the decay $\PH\to\PGg \PGg$, \HDECAY\
includes the full NLO QCD corrections
\cite{Zheng:1990qa,Djouadi:1990aj,Dawson:1992cy,Djouadi:1993ji,Melnikov:1993tj,Inoue:1994jq,Spira:1995rr}
and a grid from \Bref{Actis:2008ug} for the NLO electroweak corrections.
Missing higher orders are estimated to be below $1\%$. The contribution of
the $\PH \rightarrow \PGg \Pe^{+}\Pe^{-}$ decay via virtual photon conversion,
evaluated in \Bref{Firan:2007tp} is not taken into account in the following results. Its correct treatment and its inclusion in HDECAY are in progress.

The decays $\PH \to \PW\PW/\PZ\PZ\to4f$ are based on \Prophecy, which
includes the complete NLO QCD and electroweak corrections with all
interferences and leading two-loop heavy-Higgs corrections.  For $\MH >
500$\UGeV\ higher-order heavy-Higgs corrections dominate the error
leading to an uncertainty of $0.17\times(\MH/1\UTeV)^4$ for $\MH >
500$\UGeV.

The assessment of parametric uncertainties of the Higgs branching ratios is still
work in progress. A thorough, but very conservative estimation has recently
been made in \Bref{Baglio:2010ae}.


\subsection{MSSM Higgs branching ratios: work in progress}
\label{sec:MSSM-BR}
\providecommand{\PA}{\mathrm{A}}
\providecommand{\MA}{M_{\PA}}
\newcommand{\tb}{\tan\beta}
\newcommand{\order}[1]{\ensuremath{{\cal O}(#1)}}
\newcommand{\mhmax}{\ensuremath{m_h^{\rm max}}}

The common issues of MSSM cross section and branching-ratio calculations
have been outlined in Section~\ref{sec:mssmvssm}.
It was stressed that {\em before} any branching-ratio calculation can be
performed in a first step the Higgs-boson masses, couplings, and mixings
have to be determined from the underlying set of (soft SUSY-breaking)
parameters. A brief comparison of the dedicated codes that provide this kind of
calculations
(\FeynHiggs~\cite{Heinemeyer:1998yj,Heinemeyer:1998np,Degrassi:2002fi,Frank:2006yh}
and \CPsuperH~\cite{Lee:2003nta,Lee:2007gn}) has been given, where in
the case of real parameters more corrections are included into
\FeynHiggs.

After the calculation of Higgs-boson masses and mixings from the
original SUSY input the branching-ratio calculation has to be
performed.  This can be done with the codes, \CPsuperH\ and
\FeynHiggs\ for real or complex parameters, or
\HDECAY~\cite{Djouadi:1997yw,Spira:1997dg,hdecay2} for real
parameters. The higher-order corrections included
in the calculation of the various decay channels differ in the three
codes. A detailed analysis of the accuracy of the different codes for
certain decay widths is currently performed.

As for MSSM Higgs-boson production cross sections (see
\Sref{sec:mssmvssm}) due to the
complexity of the MSSM parameter space, results can only be derived
in representative benchmark scenarios. In accordance with
\Sref{sec:mssm_neutral}
we show in \Tref{tab:BR-mssm} exemplary values for the
BR($\phi \to \PGtp\PGtm$) ($\phi = \Ph, \PH, \PA$),
in the $m_{\Ph}^{\rm max}$ scenario~\cite{Carena:2002qg}
(see Eq.~(\ref{YRHXS_MSSM_neutral_eq:mhmax}) for the definition of the SUSY
parameters) consistently derived
with \FeynHiggs\,2.7.4. In the further progress of this work a machinery will
be set up to evaluate MSSM Higgs-boson branching ratios (consistent with the
corresponding cross-section calculations) that will be valid for the full
MSSM parameter space.

\subsection{Results}

Final SM Higgs-boson branching ratios\footnote{Full listings can be found at
  {\sl https://twiki.cern.ch/twiki/bin/view/LHCPhysics/CERNYellowReportPageBR}}
for $2$-fermion final states,
gauge-boson pair and the total decay width are listed
in \Trefs{tab:BR-lm.part1}--\ref{tab:BR-hm.part2}.
In \Trefs{tab:PBR-lm}--\ref{tab:PBR-hm2} we list branching
ratios of the SM Higgs boson decaying into $4$-fermion final states,
where leptons $\Pl=\Pe,\PGm,\PGt,\PGn_{\Pe},\PGn_{\PGm},\PGn_{\PGt}$, 
and quarks $\PQq=\PQu,\PQd,\PQs,\PQc,\PQb$. 
All fermion masses are neglected, the branching ratios are therefore
identical for different flavours.
We display results for $4$-lepton final states ($\PH \to
\Pl\Pl\Pl\Pl$) in \Trefs{tab:PBR-lm}--\ref{tab:PBR-hm}. 
We also provide results for final states with $2$ arbitrary leptons
and quarks ($\PH \to \Pl\Pl\PQq\PQq$),
$4$ arbitrary quarks ($\PH \to \PQq\PQq\PQq\PQq$),
and for all possible $4$-fermion final states ($\PH \to \Pf\Pf\Pf\Pf$)
in \Trefs{tab:PBR-lm2}--\ref{tab:PBR-hm2}. 
For Higgs-boson masses below the
$\PZ\PZ$ threshold, interference contributions become relevant for $4$-fermion decays with identical
fermions like $\PH\to\PZ\PZ\to\Pe\Pe\Pe\Pe$ or
$\PH\to\PW\PW/\PZ\PZ\to\Pe\PGn_{\Pe}\Pe\PGn_{\Pe}$. These enhance
the branching ratios for $\PH \to
\Pe\Pe\Pe\Pe,\PGm\PGm\PGm\PGm$ by more than
$10\%$ and decrease those for
$\PH\to\Pe\PGn_{\Pe}\Pe\PGn_{\Pe}, \PGm\PGn_{\PGm}\PGm\PGn_{\PGm}$
by more than $5\%$ compared to those without identical fermions for
$\MH=120\UGeV$.
All partial widths are listed in Appendix~\ref{brappendix}.
Branching ratios as a function of the SM Higgs-boson mass up to $200\UGeV$ are shown in Fig.~\ref{fig:SMBR200}. The full mass range is displayed in Fig.~\ref{fig:SMBRs}. Figure~\ref{fig:SMWidth} shows the SM Higgs-boson total decay width as a function of its mass. 

MSSM Higgs-boson branching ratios to $\PGtp\PGtm$ final states in the
$m_{\Ph}^{\rm max}$ scenario as a function of $\MA$ [GeV] and $\tb$ are given
in Table~\ref{tab:BR-mssm} as an example of the MSSM results.

All results have been obtained using the values of the electroweak parameters as given in Appendix A. For the strong coupling constant we used $\alphas(\MZ^2) = 0.119$ with two-loop running.


\begin{figure}[h]
  \centering\includegraphics[width=.70\linewidth]{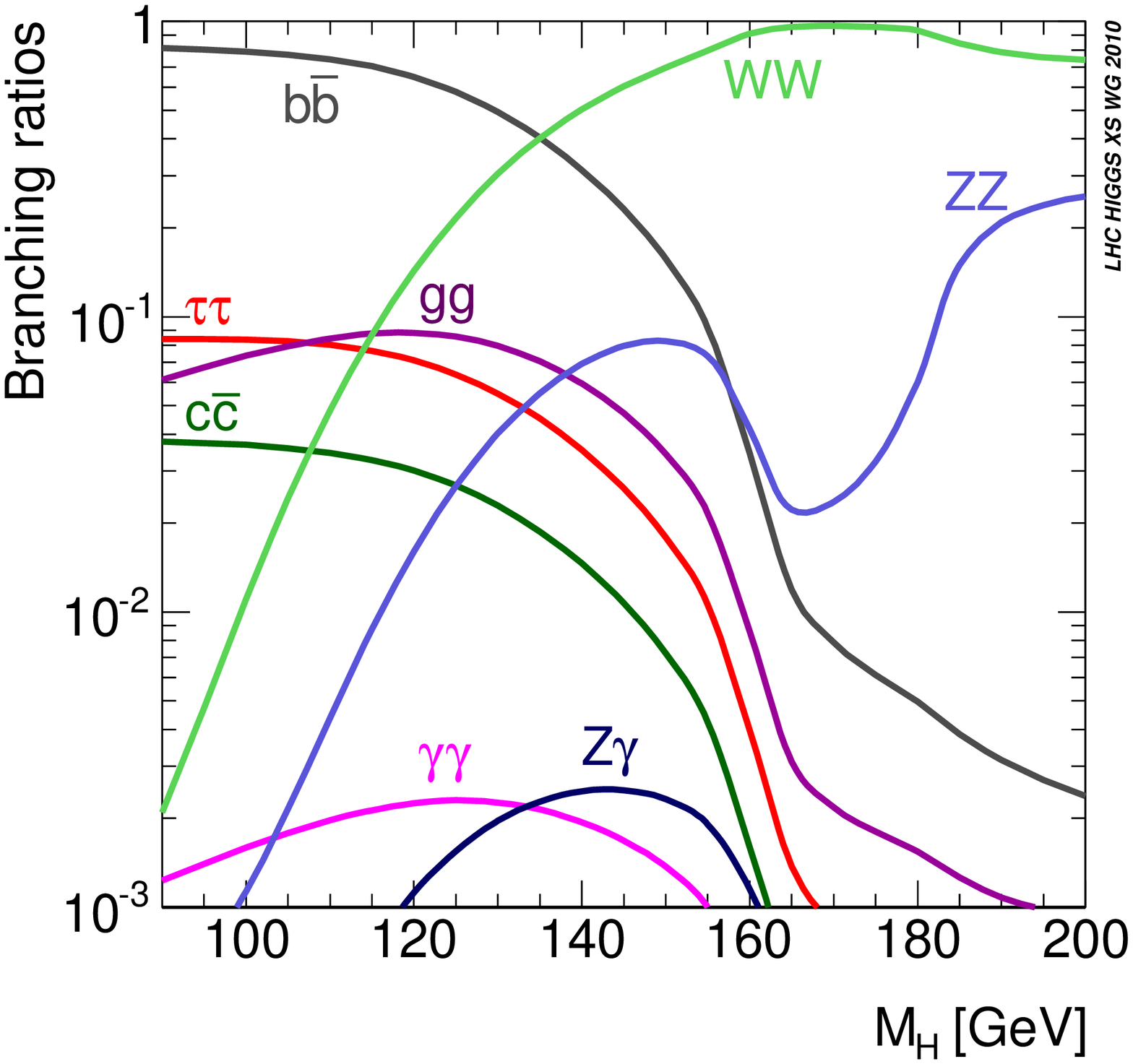}
  \caption{SM Higgs branching ratios as a function of the Higgs-boson mass.}
  \label{fig:SMBR200}
\end{figure}

\begin{figure}[h]
\centering\includegraphics[width=.70\linewidth]{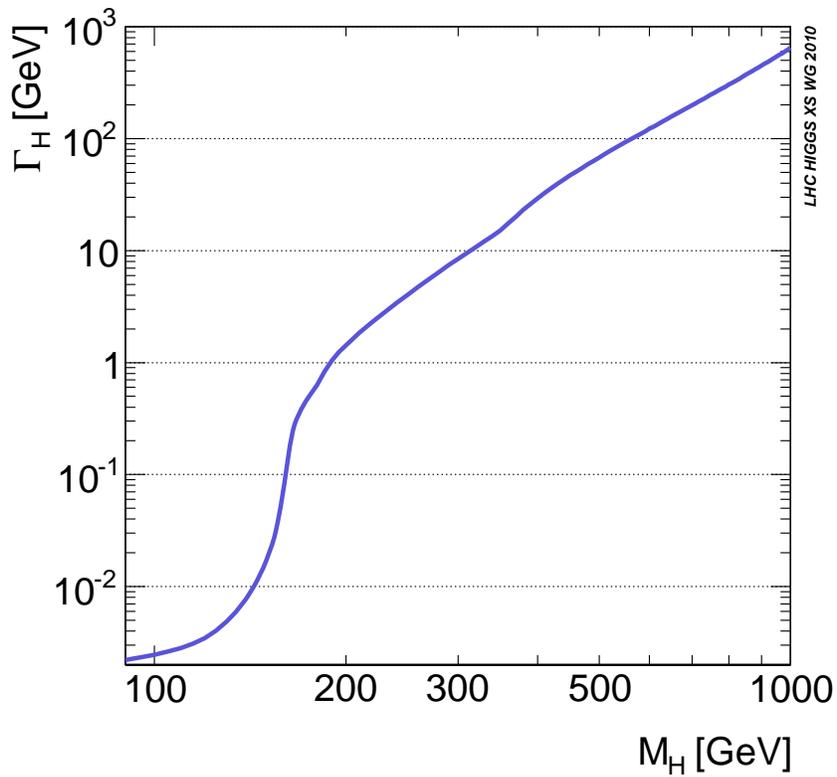}
\caption{SM Higgs total width  as a function of the Higgs-boson mass.}
\label{fig:SMWidth}
\end{figure}


\begin{table}
  \vspace{-\headsep}
  \caption{SM Higgs branching ratios in fermionic final states in the low-
  and intermediate-mass range.}
  \label{tab:BR-lm.part1}
  \centering
  \small
  \begin{tabular}{lcccccc}\hline
$\MH$ [GeV] & $\PH \rightarrow \PQb\PAQb$ & $\PH \rightarrow \PGt \PGt$ & $\PH \rightarrow
\PGm \PGm$ & $\PH \rightarrow \PQs \PAQs$ & $\PH \rightarrow \PQc \PAQc$
& $\PH \rightarrow \PQt \PAQt$ \\
\hline
$90 $&$ 8.12\cdot 10^{-1}  $&$ 8.41\cdot 10^{-2}  $&$ 2.92\cdot 10^{-4}  $&$ 6.20\cdot 10^{-4}  $&$ 3.78\cdot 10^{-2}  $&$ 0.00 $\\
$95 $&$ 8.04\cdot 10^{-1}  $&$ 8.41\cdot 10^{-2}  $&$ 2.92\cdot 10^{-4}  $&$ 6.13\cdot 10^{-4}  $&$ 3.73\cdot 10^{-2}  $&$ 0.00 $\\
$100 $&$ 7.91\cdot 10^{-1}  $&$ 8.36\cdot 10^{-2}  $&$ 2.90\cdot 10^{-4}  $&$ 6.03\cdot 10^{-4}  $&$ 3.68\cdot 10^{-2}  $&$ 0.00 $\\
$105 $&$ 7.73\cdot 10^{-1}  $&$ 8.25\cdot 10^{-2}  $&$ 2.86\cdot 10^{-4}  $&$ 5.89\cdot 10^{-4}  $&$ 3.59\cdot 10^{-2}  $&$ 0.00 $\\
$110 $&$ 7.45\cdot 10^{-1}  $&$ 8.03\cdot 10^{-2}  $&$ 2.79\cdot 10^{-4}  $&$ 5.68\cdot 10^{-4}  $&$ 3.46\cdot 10^{-2}  $&$ 0.00 $\\
$115 $&$ 7.05\cdot 10^{-1}  $&$ 7.65\cdot 10^{-2}  $&$ 2.66\cdot 10^{-4}  $&$ 5.37\cdot 10^{-4}  $&$ 3.27\cdot 10^{-2}  $&$ 0.00 $\\
$120 $&$ 6.49\cdot 10^{-1}  $&$ 7.11\cdot 10^{-2}  $&$ 2.47\cdot 10^{-4}  $&$ 4.94\cdot 10^{-4}  $&$ 3.01\cdot 10^{-2}  $&$ 0.00 $\\
$125 $&$ 5.78\cdot 10^{-1}  $&$ 6.37\cdot 10^{-2}  $&$ 2.21\cdot 10^{-4}  $&$ 4.40\cdot 10^{-4}  $&$ 2.68\cdot 10^{-2}  $&$ 0.00 $\\
$130 $&$ 4.94\cdot 10^{-1}  $&$ 5.49\cdot 10^{-2}  $&$ 1.91\cdot 10^{-4}  $&$ 3.76\cdot 10^{-4}  $&$ 2.29\cdot 10^{-2}  $&$ 0.00 $\\
$135 $&$ 4.04\cdot 10^{-1}  $&$ 4.52\cdot 10^{-2}  $&$ 1.57\cdot 10^{-4}  $&$ 3.07\cdot 10^{-4}  $&$ 1.87\cdot 10^{-2}  $&$ 0.00 $\\
$140 $&$ 3.14\cdot 10^{-1}  $&$ 3.54\cdot 10^{-2}  $&$ 1.23\cdot 10^{-4}  $&$ 2.39\cdot 10^{-4}  $&$ 1.46\cdot 10^{-2}  $&$ 0.00 $\\
$145 $&$ 2.31\cdot 10^{-1}  $&$ 2.62\cdot 10^{-2}  $&$ 9.09\cdot 10^{-5}  $&$ 1.76\cdot 10^{-4}  $&$ 1.07\cdot 10^{-2}  $&$ 0.00 $\\
$150 $&$ 1.57\cdot 10^{-1}  $&$ 1.79\cdot 10^{-2}  $&$ 6.20\cdot 10^{-5}  $&$ 1.19\cdot 10^{-4}  $&$ 7.25\cdot 10^{-3}  $&$ 0.00 $\\
$155 $&$ 9.18\cdot 10^{-2}  $&$ 1.06\cdot 10^{-2}  $&$ 3.66\cdot 10^{-5}  $&$ 6.98\cdot 10^{-5}  $&$ 4.25\cdot 10^{-3}  $&$ 0.00 $\\
$160 $&$ 3.44\cdot 10^{-2}  $&$ 3.97\cdot 10^{-3}  $&$ 1.38\cdot 10^{-5}  $&$ 2.61\cdot 10^{-5}  $&$ 1.59\cdot 10^{-3}  $&$ 0.00 $\\
$165 $&$ 1.19\cdot 10^{-2}  $&$ 1.38\cdot 10^{-3}  $&$ 4.78\cdot 10^{-6}  $&$ 9.02\cdot 10^{-6}  $&$ 5.49\cdot 10^{-4}  $&$ 0.00 $\\
$170 $&$ 7.87\cdot 10^{-3}  $&$ 9.20\cdot 10^{-4}  $&$ 3.19\cdot 10^{-6}  $&$ 5.99\cdot 10^{-6}  $&$ 3.64\cdot 10^{-4}  $&$ 0.00 $\\
$175 $&$ 6.12\cdot 10^{-3}  $&$ 7.19\cdot 10^{-4}  $&$ 2.49\cdot 10^{-6}  $&$ 4.65\cdot 10^{-6}  $&$ 2.83\cdot 10^{-4}  $&$ 0.00 $\\
$180 $&$ 4.97\cdot 10^{-3}  $&$ 5.87\cdot 10^{-4}  $&$ 2.04\cdot 10^{-6}  $&$ 3.78\cdot 10^{-6}  $&$ 2.30\cdot 10^{-4}  $&$ 0.00 $\\
$185 $&$ 3.85\cdot 10^{-3}  $&$ 4.57\cdot 10^{-4}  $&$ 1.59\cdot 10^{-6}  $&$ 2.93\cdot 10^{-6}  $&$ 1.78\cdot 10^{-4}  $&$ 0.00 $\\
$190 $&$ 3.15\cdot 10^{-3}  $&$ 3.76\cdot 10^{-4}  $&$ 1.30\cdot 10^{-6}  $&$ 2.39\cdot 10^{-6}  $&$ 1.46\cdot 10^{-4}  $&$ 0.00 $\\
$195 $&$ 2.70\cdot 10^{-3}  $&$ 3.24\cdot 10^{-4}  $&$ 1.13\cdot 10^{-6}  $&$ 2.06\cdot 10^{-6}  $&$ 1.25\cdot 10^{-4}  $&$ 0.00 $\\
$200 $&$ 2.38\cdot 10^{-3}  $&$ 2.87\cdot 10^{-4}  $&$ 9.96\cdot 10^{-7}  $&$ 1.81\cdot 10^{-6}  $&$ 1.10\cdot 10^{-4}  $&$ 0.00 $\\
$210 $&$ 1.92\cdot 10^{-3}  $&$ 2.34\cdot 10^{-4}  $&$ 8.11\cdot 10^{-7}  $&$ 1.46\cdot 10^{-6}  $&$ 8.89\cdot 10^{-5}  $&$ 0.00 $\\
$220 $&$ 1.60\cdot 10^{-3}  $&$ 1.96\cdot 10^{-4}  $&$ 6.81\cdot 10^{-7}  $&$ 1.22\cdot 10^{-6}  $&$ 7.40\cdot 10^{-5}  $&$ 0.00 $\\
$230 $&$ 1.36\cdot 10^{-3}  $&$ 1.68\cdot 10^{-4}  $&$ 5.82\cdot 10^{-7}  $&$ 1.03\cdot 10^{-6}  $&$ 6.27\cdot 10^{-5}  $&$ 0.00 $\\
$240 $&$ 1.17\cdot 10^{-3}  $&$ 1.45\cdot 10^{-4}  $&$ 5.04\cdot 10^{-7}  $&$ 8.86\cdot 10^{-7}  $&$ 5.39\cdot 10^{-5}  $&$ 0.00 $\\
$250 $&$ 1.01\cdot 10^{-3}  $&$ 1.27\cdot 10^{-4}  $&$ 4.42\cdot 10^{-7}  $&$ 7.70\cdot 10^{-7}  $&$ 4.68\cdot 10^{-5}  $&$ 0.00 $\\
$260 $&$ 8.89\cdot 10^{-4}  $&$ 1.12\cdot 10^{-4}  $&$ 3.90\cdot 10^{-7}  $&$ 6.75\cdot 10^{-7}  $&$ 4.11\cdot 10^{-5}  $&$ 5.14\cdot 10^{-8}  $\\
$270 $&$ 7.86\cdot 10^{-4}  $&$ 1.00\cdot 10^{-4}  $&$ 3.47\cdot 10^{-7}  $&$ 5.97\cdot 10^{-7}  $&$ 3.63\cdot 10^{-5}  $&$ 2.29\cdot 10^{-6}  $\\
$280 $&$ 7.00\cdot 10^{-4}  $&$ 8.98\cdot 10^{-5}  $&$ 3.11\cdot 10^{-7}  $&$ 5.31\cdot 10^{-7}  $&$ 3.23\cdot 10^{-5}  $&$ 1.09\cdot 10^{-5}  $\\
$290 $&$ 6.27\cdot 10^{-4}  $&$ 8.09\cdot 10^{-5}  $&$ 2.80\cdot 10^{-7}  $&$ 4.76\cdot 10^{-7}  $&$ 2.90\cdot 10^{-5}  $&$ 3.06\cdot 10^{-5}  $\\
$300 $&$ 5.65\cdot 10^{-4}  $&$ 7.33\cdot 10^{-5}  $&$ 2.54\cdot 10^{-7}  $&$ 4.29\cdot 10^{-7}  $&$ 2.61\cdot 10^{-5}  $&$ 6.87\cdot 10^{-5}  $\\
$310 $&$ 5.12\cdot 10^{-4}  $&$ 6.68\cdot 10^{-5}  $&$ 2.32\cdot 10^{-7}  $&$ 3.89\cdot 10^{-7}  $&$ 2.36\cdot 10^{-5}  $&$ 1.38\cdot 10^{-4}  $\\
$320 $&$ 4.66\cdot 10^{-4}  $&$ 6.12\cdot 10^{-5}  $&$ 2.12\cdot 10^{-7}  $&$ 3.54\cdot 10^{-7}  $&$ 2.15\cdot 10^{-5}  $&$ 2.66\cdot 10^{-4}  $\\
$330 $&$ 4.26\cdot 10^{-4}  $&$ 5.63\cdot 10^{-5}  $&$ 1.95\cdot 10^{-7}  $&$ 3.24\cdot 10^{-7}  $&$ 1.97\cdot 10^{-5}  $&$ 5.21\cdot 10^{-4}  $\\
$340 $&$ 3.92\cdot 10^{-4}  $&$ 5.20\cdot 10^{-5}  $&$ 1.80\cdot 10^{-7}  $&$ 2.98\cdot 10^{-7}  $&$ 1.81\cdot 10^{-5}  $&$ 1.20\cdot 10^{-3}  $\\
$350 $&$ 3.57\cdot 10^{-4}  $&$ 4.76\cdot 10^{-5}  $&$ 1.65\cdot 10^{-7}  $&$ 2.71\cdot 10^{-7}  $&$ 1.65\cdot 10^{-5}  $&$ 1.56\cdot 10^{-2}  $\\
$360 $&$ 3.16\cdot 10^{-4}  $&$ 4.23\cdot 10^{-5}  $&$ 1.47\cdot 10^{-7}  $&$ 2.40\cdot 10^{-7}  $&$ 1.46\cdot 10^{-5}  $&$ 5.15\cdot 10^{-2}  $\\
$370 $&$ 2.81\cdot 10^{-4}  $&$ 3.78\cdot 10^{-5}  $&$ 1.31\cdot 10^{-7}  $&$ 2.13\cdot 10^{-7}  $&$ 1.29\cdot 10^{-5}  $&$ 8.37\cdot 10^{-2}  $\\
$380 $&$ 2.52\cdot 10^{-4}  $&$ 3.40\cdot 10^{-5}  $&$ 1.18\cdot 10^{-7}  $&$ 1.91\cdot 10^{-7}  $&$ 1.16\cdot 10^{-5}  $&$ 1.10\cdot 10^{-1}  $\\
$390 $&$ 2.28\cdot 10^{-4}  $&$ 3.10\cdot 10^{-5}  $&$ 1.07\cdot 10^{-7}  $&$ 1.73\cdot 10^{-7}  $&$ 1.05\cdot 10^{-5}  $&$ 1.32\cdot 10^{-1}  $\\
$400 $&$ 2.08\cdot 10^{-4}  $&$ 2.84\cdot 10^{-5}  $&$ 9.83\cdot 10^{-8}  $&$ 1.58\cdot 10^{-7}  $&$ 9.59\cdot 10^{-6}  $&$ 1.48\cdot 10^{-1}  $\\
$410 $&$ 1.91\cdot 10^{-4}  $&$ 2.61\cdot 10^{-5}  $&$ 9.06\cdot 10^{-8}  $&$ 1.45\cdot 10^{-7}  $&$ 8.80\cdot 10^{-6}  $&$ 1.62\cdot 10^{-1}  $\\
$420 $&$ 1.76\cdot 10^{-4}  $&$ 2.43\cdot 10^{-5}  $&$ 8.41\cdot 10^{-8}  $&$ 1.34\cdot 10^{-7}  $&$ 8.13\cdot 10^{-6}  $&$ 1.72\cdot 10^{-1}  $\\
$430 $&$ 1.64\cdot 10^{-4}  $&$ 2.26\cdot 10^{-5}  $&$ 7.84\cdot 10^{-8}  $&$ 1.24\cdot 10^{-7}  $&$ 7.55\cdot 10^{-6}  $&$ 1.79\cdot 10^{-1}  $\\
$440 $&$ 1.53\cdot 10^{-4}  $&$ 2.12\cdot 10^{-5}  $&$ 7.34\cdot 10^{-8}  $&$ 1.16\cdot 10^{-7}  $&$ 7.05\cdot 10^{-6}  $&$ 1.85\cdot 10^{-1}  $\\
$450 $&$ 1.43\cdot 10^{-4}  $&$ 1.99\cdot 10^{-5}  $&$ 6.90\cdot 10^{-8}  $&$ 1.09\cdot 10^{-7}  $&$ 6.60\cdot 10^{-6}  $&$ 1.89\cdot 10^{-1}  $\\
$460 $&$ 1.35\cdot 10^{-4}  $&$ 1.88\cdot 10^{-5}  $&$ 6.51\cdot 10^{-8}  $&$ 1.02\cdot 10^{-7}  $&$ 6.21\cdot 10^{-6}  $&$ 1.91\cdot 10^{-1}  $\\
$470 $&$ 1.27\cdot 10^{-4}  $&$ 1.78\cdot 10^{-5}  $&$ 6.16\cdot 10^{-8}  $&$ 9.63\cdot 10^{-8}  $&$ 5.85\cdot 10^{-6}  $&$ 1.93\cdot 10^{-1}  $\\
$480 $&$ 1.20\cdot 10^{-4}  $&$ 1.69\cdot 10^{-5}  $&$ 5.85\cdot 10^{-8}  $&$ 9.10\cdot 10^{-8}  $&$ 5.53\cdot 10^{-6}  $&$ 1.94\cdot 10^{-1}  $\\
$490 $&$ 1.14\cdot 10^{-4}  $&$ 1.60\cdot 10^{-5}  $&$ 5.56\cdot 10^{-8}  $&$ 8.63\cdot 10^{-8}  $&$ 5.24\cdot 10^{-6}  $&$ 1.94\cdot 10^{-1}  $\\
\hline
  \end{tabular}
\end{table}

\begin{table}
  \vspace{-\headsep}
  \caption{SM Higgs branching ratios in fermionic final states in the high-mass range.}
  \label{tab:BR-hm.part1}
  \centering
  \small
  \begin{tabular}{lcccccc}\hline
$\MH$ [GeV] & $\PH \rightarrow \PQb\PAQb$ & $\PH \rightarrow \PGt \PGt$ & $\PH \rightarrow
\PGm \PGm$ & $\PH \rightarrow \PQs \PAQs$ & $\PH \rightarrow \PQc \PAQc$
& $\PH \rightarrow \PQt \PAQt$ \\
\hline
$500 $&$ 1.08\cdot 10^{-4}  $&$ 1.53\cdot 10^{-5}  $&$ 5.30\cdot 10^{-8}  $&$ 8.19\cdot 10^{-8}  $&$ 4.98\cdot 10^{-6}  $&$ 1.93\cdot 10^{-1}  $\\
$510 $&$ 1.03\cdot 10^{-4}  $&$ 1.46\cdot 10^{-5}  $&$ 5.06\cdot 10^{-8}  $&$ 7.80\cdot 10^{-8}  $&$ 4.74\cdot 10^{-6}  $&$ 1.92\cdot 10^{-1}  $\\
$520 $&$ 9.80\cdot 10^{-5}  $&$ 1.40\cdot 10^{-5}  $&$ 4.84\cdot 10^{-8}  $&$ 7.44\cdot 10^{-8}  $&$ 4.52\cdot 10^{-6}  $&$ 1.90\cdot 10^{-1}  $\\
$530 $&$ 9.36\cdot 10^{-5}  $&$ 1.34\cdot 10^{-5}  $&$ 4.64\cdot 10^{-8}  $&$ 7.10\cdot 10^{-8}  $&$ 4.31\cdot 10^{-6}  $&$ 1.88\cdot 10^{-1}  $\\
$540 $&$ 8.95\cdot 10^{-5}  $&$ 1.28\cdot 10^{-5}  $&$ 4.45\cdot 10^{-8}  $&$ 6.79\cdot 10^{-8}  $&$ 4.12\cdot 10^{-6}  $&$ 1.86\cdot 10^{-1}  $\\
$550 $&$ 8.57\cdot 10^{-5}  $&$ 1.23\cdot 10^{-5}  $&$ 4.27\cdot 10^{-8}  $&$ 6.50\cdot 10^{-8}  $&$ 3.95\cdot 10^{-6}  $&$ 1.84\cdot 10^{-1}  $\\
$560 $&$ 8.21\cdot 10^{-5}  $&$ 1.18\cdot 10^{-5}  $&$ 4.10\cdot 10^{-8}  $&$ 6.23\cdot 10^{-8}  $&$ 3.79\cdot 10^{-6}  $&$ 1.81\cdot 10^{-1}  $\\
$570 $&$ 7.88\cdot 10^{-5}  $&$ 1.14\cdot 10^{-5}  $&$ 3.95\cdot 10^{-8}  $&$ 5.98\cdot 10^{-8}  $&$ 3.63\cdot 10^{-6}  $&$ 1.78\cdot 10^{-1}  $\\
$580 $&$ 7.57\cdot 10^{-5}  $&$ 1.10\cdot 10^{-5}  $&$ 3.80\cdot 10^{-8}  $&$ 5.74\cdot 10^{-8}  $&$ 3.49\cdot 10^{-6}  $&$ 1.75\cdot 10^{-1}  $\\
$590 $&$ 7.28\cdot 10^{-5}  $&$ 1.06\cdot 10^{-5}  $&$ 3.67\cdot 10^{-8}  $&$ 5.52\cdot 10^{-8}  $&$ 3.35\cdot 10^{-6}  $&$ 1.72\cdot 10^{-1}  $\\
$600 $&$ 7.00\cdot 10^{-5}  $&$ 1.02\cdot 10^{-5}  $&$ 3.54\cdot 10^{-8}  $&$ 5.31\cdot 10^{-8}  $&$ 3.23\cdot 10^{-6}  $&$ 1.69\cdot 10^{-1}  $\\
$610 $&$ 6.74\cdot 10^{-5}  $&$ 9.86\cdot 10^{-6}  $&$ 3.42\cdot 10^{-8}  $&$ 5.12\cdot 10^{-8}  $&$ 3.11\cdot 10^{-6}  $&$ 1.66\cdot 10^{-1}  $\\
$620 $&$ 6.50\cdot 10^{-5}  $&$ 9.53\cdot 10^{-6}  $&$ 3.30\cdot 10^{-8}  $&$ 4.93\cdot 10^{-8}  $&$ 2.99\cdot 10^{-6}  $&$ 1.63\cdot 10^{-1}  $\\
$630 $&$ 6.27\cdot 10^{-5}  $&$ 9.21\cdot 10^{-6}  $&$ 3.19\cdot 10^{-8}  $&$ 4.76\cdot 10^{-8}  $&$ 2.89\cdot 10^{-6}  $&$ 1.60\cdot 10^{-1}  $\\
$640 $&$ 6.05\cdot 10^{-5}  $&$ 8.91\cdot 10^{-6}  $&$ 3.09\cdot 10^{-8}  $&$ 4.59\cdot 10^{-8}  $&$ 2.79\cdot 10^{-6}  $&$ 1.57\cdot 10^{-1}  $\\
$650 $&$ 5.84\cdot 10^{-5}  $&$ 8.63\cdot 10^{-6}  $&$ 2.99\cdot 10^{-8}  $&$ 4.43\cdot 10^{-8}  $&$ 2.69\cdot 10^{-6}  $&$ 1.54\cdot 10^{-1}  $\\
$660 $&$ 5.64\cdot 10^{-5}  $&$ 8.35\cdot 10^{-6}  $&$ 2.89\cdot 10^{-8}  $&$ 4.28\cdot 10^{-8}  $&$ 2.60\cdot 10^{-6}  $&$ 1.50\cdot 10^{-1}  $\\
$670 $&$ 5.45\cdot 10^{-5}  $&$ 8.09\cdot 10^{-6}  $&$ 2.80\cdot 10^{-8}  $&$ 4.14\cdot 10^{-8}  $&$ 2.51\cdot 10^{-6}  $&$ 1.47\cdot 10^{-1}  $\\
$680 $&$ 5.27\cdot 10^{-5}  $&$ 7.84\cdot 10^{-6}  $&$ 2.72\cdot 10^{-8}  $&$ 4.00\cdot 10^{-8}  $&$ 2.43\cdot 10^{-6}  $&$ 1.44\cdot 10^{-1}  $\\
$690 $&$ 5.10\cdot 10^{-5}  $&$ 7.60\cdot 10^{-6}  $&$ 2.64\cdot 10^{-8}  $&$ 3.87\cdot 10^{-8}  $&$ 2.35\cdot 10^{-6}  $&$ 1.41\cdot 10^{-1}  $\\
$700 $&$ 4.94\cdot 10^{-5}  $&$ 7.37\cdot 10^{-6}  $&$ 2.56\cdot 10^{-8}  $&$ 3.74\cdot 10^{-8}  $&$ 2.27\cdot 10^{-6}  $&$ 1.38\cdot 10^{-1}  $\\
$710 $&$ 4.78\cdot 10^{-5}  $&$ 7.16\cdot 10^{-6}  $&$ 2.48\cdot 10^{-8}  $&$ 3.62\cdot 10^{-8}  $&$ 2.20\cdot 10^{-6}  $&$ 1.35\cdot 10^{-1}  $\\
$720 $&$ 4.63\cdot 10^{-5}  $&$ 6.94\cdot 10^{-6}  $&$ 2.41\cdot 10^{-8}  $&$ 3.51\cdot 10^{-8}  $&$ 2.13\cdot 10^{-6}  $&$ 1.32\cdot 10^{-1}  $\\
$730 $&$ 4.48\cdot 10^{-5}  $&$ 6.74\cdot 10^{-6}  $&$ 2.34\cdot 10^{-8}  $&$ 3.40\cdot 10^{-8}  $&$ 2.07\cdot 10^{-6}  $&$ 1.29\cdot 10^{-1}  $\\
$740 $&$ 4.34\cdot 10^{-5}  $&$ 6.55\cdot 10^{-6}  $&$ 2.27\cdot 10^{-8}  $&$ 3.30\cdot 10^{-8}  $&$ 2.00\cdot 10^{-6}  $&$ 1.26\cdot 10^{-1}  $\\
$750 $&$ 4.21\cdot 10^{-5}  $&$ 6.36\cdot 10^{-6}  $&$ 2.20\cdot 10^{-8}  $&$ 3.19\cdot 10^{-8}  $&$ 1.94\cdot 10^{-6}  $&$ 1.23\cdot 10^{-1}  $\\
$760 $&$ 4.08\cdot 10^{-5}  $&$ 6.18\cdot 10^{-6}  $&$ 2.14\cdot 10^{-8}  $&$ 3.10\cdot 10^{-8}  $&$ 1.88\cdot 10^{-6}  $&$ 1.21\cdot 10^{-1}  $\\
$770 $&$ 3.96\cdot 10^{-5}  $&$ 6.00\cdot 10^{-6}  $&$ 2.08\cdot 10^{-8}  $&$ 3.00\cdot 10^{-8}  $&$ 1.82\cdot 10^{-6}  $&$ 1.18\cdot 10^{-1}  $\\
$780 $&$ 3.84\cdot 10^{-5}  $&$ 5.83\cdot 10^{-6}  $&$ 2.02\cdot 10^{-8}  $&$ 2.91\cdot 10^{-8}  $&$ 1.77\cdot 10^{-6}  $&$ 1.15\cdot 10^{-1}  $\\
$790 $&$ 3.73\cdot 10^{-5}  $&$ 5.67\cdot 10^{-6}  $&$ 1.97\cdot 10^{-8}  $&$ 2.83\cdot 10^{-8}  $&$ 1.72\cdot 10^{-6}  $&$ 1.13\cdot 10^{-1}  $\\
$800 $&$ 3.62\cdot 10^{-5}  $&$ 5.52\cdot 10^{-6}  $&$ 1.91\cdot 10^{-8}  $&$ 2.74\cdot 10^{-8}  $&$ 1.67\cdot 10^{-6}  $&$ 1.10\cdot 10^{-1}  $\\
$810 $&$ 3.51\cdot 10^{-5}  $&$ 5.36\cdot 10^{-6}  $&$ 1.86\cdot 10^{-8}  $&$ 2.66\cdot 10^{-8}  $&$ 1.62\cdot 10^{-6}  $&$ 1.07\cdot 10^{-1}  $\\
$820 $&$ 3.41\cdot 10^{-5}  $&$ 5.22\cdot 10^{-6}  $&$ 1.81\cdot 10^{-8}  $&$ 2.58\cdot 10^{-8}  $&$ 1.57\cdot 10^{-6}  $&$ 1.05\cdot 10^{-1}  $\\
$830 $&$ 3.31\cdot 10^{-5}  $&$ 5.07\cdot 10^{-6}  $&$ 1.76\cdot 10^{-8}  $&$ 2.51\cdot 10^{-8}  $&$ 1.52\cdot 10^{-6}  $&$ 1.02\cdot 10^{-1}  $\\
$840 $&$ 3.21\cdot 10^{-5}  $&$ 4.93\cdot 10^{-6}  $&$ 1.71\cdot 10^{-8}  $&$ 2.44\cdot 10^{-8}  $&$ 1.48\cdot 10^{-6}  $&$ 1.00\cdot 10^{-1}  $\\
$850 $&$ 3.12\cdot 10^{-5}  $&$ 4.80\cdot 10^{-6}  $&$ 1.66\cdot 10^{-8}  $&$ 2.37\cdot 10^{-8}  $&$ 1.44\cdot 10^{-6}  $&$ 9.77\cdot 10^{-2}  $\\
$860 $&$ 3.03\cdot 10^{-5}  $&$ 4.67\cdot 10^{-6}  $&$ 1.62\cdot 10^{-8}  $&$ 2.30\cdot 10^{-8}  $&$ 1.40\cdot 10^{-6}  $&$ 9.54\cdot 10^{-2}  $\\
$870 $&$ 2.94\cdot 10^{-5}  $&$ 4.55\cdot 10^{-6}  $&$ 1.58\cdot 10^{-8}  $&$ 2.23\cdot 10^{-8}  $&$ 1.36\cdot 10^{-6}  $&$ 9.31\cdot 10^{-2}  $\\
$880 $&$ 2.86\cdot 10^{-5}  $&$ 4.42\cdot 10^{-6}  $&$ 1.53\cdot 10^{-8}  $&$ 2.17\cdot 10^{-8}  $&$ 1.32\cdot 10^{-6}  $&$ 9.09\cdot 10^{-2}  $\\
$890 $&$ 2.78\cdot 10^{-5}  $&$ 4.31\cdot 10^{-6}  $&$ 1.49\cdot 10^{-8}  $&$ 2.11\cdot 10^{-8}  $&$ 1.28\cdot 10^{-6}  $&$ 8.87\cdot 10^{-2}  $\\
$900 $&$ 2.70\cdot 10^{-5}  $&$ 4.19\cdot 10^{-6}  $&$ 1.45\cdot 10^{-8}  $&$ 2.05\cdot 10^{-8}  $&$ 1.24\cdot 10^{-6}  $&$ 8.66\cdot 10^{-2}  $\\
$910 $&$ 2.62\cdot 10^{-5}  $&$ 4.08\cdot 10^{-6}  $&$ 1.41\cdot 10^{-8}  $&$ 1.99\cdot 10^{-8}  $&$ 1.21\cdot 10^{-6}  $&$ 8.45\cdot 10^{-2}  $\\
$920 $&$ 2.55\cdot 10^{-5}  $&$ 3.97\cdot 10^{-6}  $&$ 1.38\cdot 10^{-8}  $&$ 1.93\cdot 10^{-8}  $&$ 1.17\cdot 10^{-6}  $&$ 8.24\cdot 10^{-2}  $\\
$930 $&$ 2.48\cdot 10^{-5}  $&$ 3.86\cdot 10^{-6}  $&$ 1.34\cdot 10^{-8}  $&$ 1.88\cdot 10^{-8}  $&$ 1.14\cdot 10^{-6}  $&$ 8.04\cdot 10^{-2}  $\\
$940 $&$ 2.41\cdot 10^{-5}  $&$ 3.76\cdot 10^{-6}  $&$ 1.30\cdot 10^{-8}  $&$ 1.83\cdot 10^{-8}  $&$ 1.11\cdot 10^{-6}  $&$ 7.84\cdot 10^{-2}  $\\
$950 $&$ 2.34\cdot 10^{-5}  $&$ 3.66\cdot 10^{-6}  $&$ 1.27\cdot 10^{-8}  $&$ 1.77\cdot 10^{-8}  $&$ 1.08\cdot 10^{-6}  $&$ 7.65\cdot 10^{-2}  $\\
$960 $&$ 2.27\cdot 10^{-5}  $&$ 3.56\cdot 10^{-6}  $&$ 1.23\cdot 10^{-8}  $&$ 1.72\cdot 10^{-8}  $&$ 1.05\cdot 10^{-6}  $&$ 7.46\cdot 10^{-2}  $\\
$970 $&$ 2.21\cdot 10^{-5}  $&$ 3.47\cdot 10^{-6}  $&$ 1.20\cdot 10^{-8}  $&$ 1.68\cdot 10^{-8}  $&$ 1.02\cdot 10^{-6}  $&$ 7.27\cdot 10^{-2}  $\\
$980 $&$ 2.15\cdot 10^{-5}  $&$ 3.38\cdot 10^{-6}  $&$ 1.17\cdot 10^{-8}  $&$ 1.63\cdot 10^{-8}  $&$ 9.88\cdot 10^{-7}  $&$ 7.09\cdot 10^{-2}  $\\
$990 $&$ 2.09\cdot 10^{-5}  $&$ 3.29\cdot 10^{-6}  $&$ 1.14\cdot 10^{-8}  $&$ 1.58\cdot 10^{-8}  $&$ 9.61\cdot 10^{-7}  $&$ 6.91\cdot 10^{-2}  $\\
$1000 $&$ 2.03\cdot 10^{-5}  $&$ 3.20\cdot 10^{-6}  $&$ 1.11\cdot 10^{-8}  $&$ 1.54\cdot 10^{-8}  $&$ 9.34\cdot 10^{-7}  $&$ 6.74\cdot 10^{-2}  $\\
\hline
  \end{tabular}
\end{table}

\begin{table}
  \vspace{-\headsep}
  \caption{SM Higgs branching ratios in bosonic final states and Higgs total widths in the low- and intermediate-mass range.}
  \label{tab:BR-lm.part2}
  \centering
  \small
  \begin{tabular}{ccccccc}\hline
$\MH$ [GeV] & $\PH \rightarrow \Pg\Pg$ & $\PH \rightarrow \PGg \PGg$ & $\PH \rightarrow
\PZ\PGg$ & $\PH \rightarrow \PW\PW$
& $\PH \rightarrow \PZ\PZ$ & Total $\Gamma_{\PH}$ [GeV]\\
\hline
$90 $&$ 6.12\cdot 10^{-2}  $&$ 1.23\cdot 10^{-3}  $&$ 0.00 $&$ 2.09\cdot 10^{-3}  $&$ 4.21\cdot 10^{-4}  $&$ 2.20\cdot 10^{-3}  $\\
$95 $&$ 6.74\cdot 10^{-2}  $&$ 1.40\cdot 10^{-3}  $&$ 4.52\cdot 10^{-6}  $&$ 4.72\cdot 10^{-3}  $&$ 6.72\cdot 10^{-4}  $&$ 2.32\cdot 10^{-3}  $\\
$100 $&$ 7.37\cdot 10^{-2}  $&$ 1.59\cdot 10^{-3}  $&$ 4.98\cdot 10^{-5}  $&$ 1.11\cdot 10^{-2}  $&$ 1.13\cdot 10^{-3}  $&$ 2.46\cdot 10^{-3}  $\\
$105 $&$ 7.95\cdot 10^{-2}  $&$ 1.78\cdot 10^{-3}  $&$ 1.73\cdot 10^{-4}  $&$ 2.43\cdot 10^{-2}  $&$ 2.15\cdot 10^{-3}  $&$ 2.62\cdot 10^{-3}  $\\
$110 $&$ 8.44\cdot 10^{-2}  $&$ 1.97\cdot 10^{-3}  $&$ 3.95\cdot 10^{-4}  $&$ 4.82\cdot 10^{-2}  $&$ 4.39\cdot 10^{-3}  $&$ 2.82\cdot 10^{-3}  $\\
$115 $&$ 8.76\cdot 10^{-2}  $&$ 2.13\cdot 10^{-3}  $&$ 7.16\cdot 10^{-4}  $&$ 8.67\cdot 10^{-2}  $&$ 8.73\cdot 10^{-3}  $&$ 3.09\cdot 10^{-3}  $\\
$120 $&$ 8.82\cdot 10^{-2}  $&$ 2.25\cdot 10^{-3}  $&$ 1.12\cdot 10^{-3}  $&$ 1.43\cdot 10^{-1}  $&$ 1.60\cdot 10^{-2}  $&$ 3.47\cdot 10^{-3}  $\\
$125 $&$ 8.56\cdot 10^{-2}  $&$ 2.30\cdot 10^{-3}  $&$ 1.55\cdot 10^{-3}  $&$ 2.16\cdot 10^{-1}  $&$ 2.67\cdot 10^{-2}  $&$ 4.03\cdot 10^{-3}  $\\
$130 $&$ 7.96\cdot 10^{-2}  $&$ 2.26\cdot 10^{-3}  $&$ 1.96\cdot 10^{-3}  $&$ 3.05\cdot 10^{-1}  $&$ 4.02\cdot 10^{-2}  $&$ 4.87\cdot 10^{-3}  $\\
$135 $&$ 7.06\cdot 10^{-2}  $&$ 2.14\cdot 10^{-3}  $&$ 2.28\cdot 10^{-3}  $&$ 4.03\cdot 10^{-1}  $&$ 5.51\cdot 10^{-2}  $&$ 6.14\cdot 10^{-3}  $\\
$140 $&$ 5.94\cdot 10^{-2}  $&$ 1.94\cdot 10^{-3}  $&$ 2.47\cdot 10^{-3}  $&$ 5.04\cdot 10^{-1}  $&$ 6.92\cdot 10^{-2}  $&$ 8.12\cdot 10^{-3}  $\\
$145 $&$ 4.70\cdot 10^{-2}  $&$ 1.68\cdot 10^{-3}  $&$ 2.49\cdot 10^{-3}  $&$ 6.03\cdot 10^{-1}  $&$ 7.96\cdot 10^{-2}  $&$ 1.14\cdot 10^{-2}  $\\
$150 $&$ 3.43\cdot 10^{-2}  $&$ 1.37\cdot 10^{-3}  $&$ 2.32\cdot 10^{-3}  $&$ 6.99\cdot 10^{-1}  $&$ 8.28\cdot 10^{-2}  $&$ 1.73\cdot 10^{-2}  $\\
$155 $&$ 2.16\cdot 10^{-2}  $&$ 1.00\cdot 10^{-3}  $&$ 1.91\cdot 10^{-3}  $&$ 7.96\cdot 10^{-1}  $&$ 7.36\cdot 10^{-2}  $&$ 3.02\cdot 10^{-2}  $\\
$160 $&$ 8.57\cdot 10^{-3}  $&$ 5.33\cdot 10^{-4}  $&$ 1.15\cdot 10^{-3}  $&$ 9.09\cdot 10^{-1}  $&$ 4.16\cdot 10^{-2}  $&$ 8.29\cdot 10^{-2}  $\\
$165 $&$ 3.11\cdot 10^{-3}  $&$ 2.30\cdot 10^{-4}  $&$ 5.45\cdot 10^{-4}  $&$ 9.60\cdot 10^{-1}  $&$ 2.22\cdot 10^{-2}  $&$ 2.46\cdot 10^{-1}  $\\
$170 $&$ 2.18\cdot 10^{-3}  $&$ 1.58\cdot 10^{-4}  $&$ 4.00\cdot 10^{-4}  $&$ 9.65\cdot 10^{-1}  $&$ 2.36\cdot 10^{-2}  $&$ 3.80\cdot 10^{-1}  $\\
$175 $&$ 1.80\cdot 10^{-3}  $&$ 1.23\cdot 10^{-4}  $&$ 3.38\cdot 10^{-4}  $&$ 9.58\cdot 10^{-1}  $&$ 3.23\cdot 10^{-2}  $&$ 5.00\cdot 10^{-1}  $\\
$180 $&$ 1.54\cdot 10^{-3}  $&$ 1.02\cdot 10^{-4}  $&$ 2.96\cdot 10^{-4}  $&$ 9.32\cdot 10^{-1}  $&$ 6.02\cdot 10^{-2}  $&$ 6.31\cdot 10^{-1}  $\\
$185 $&$ 1.26\cdot 10^{-3}  $&$ 8.09\cdot 10^{-5}  $&$ 2.44\cdot 10^{-4}  $&$ 8.44\cdot 10^{-1}  $&$ 1.50\cdot 10^{-1}  $&$ 8.32\cdot 10^{-1}  $\\
$190 $&$ 1.08\cdot 10^{-3}  $&$ 6.74\cdot 10^{-5}  $&$ 2.11\cdot 10^{-4}  $&$ 7.86\cdot 10^{-1}  $&$ 2.09\cdot 10^{-1}  $&$ 1.04 $\\
$195 $&$ 9.84\cdot 10^{-4}  $&$ 5.89\cdot 10^{-5}  $&$ 1.91\cdot 10^{-4}  $&$ 7.57\cdot 10^{-1}  $&$ 2.39\cdot 10^{-1}  $&$ 1.24 $\\
$200 $&$ 9.16\cdot 10^{-4}  $&$ 5.26\cdot 10^{-5}  $&$ 1.75\cdot 10^{-4}  $&$ 7.41\cdot 10^{-1}  $&$ 2.56\cdot 10^{-1}  $&$ 1.43 $\\
$210 $&$ 8.27\cdot 10^{-4}  $&$ 4.34\cdot 10^{-5}  $&$ 1.52\cdot 10^{-4}  $&$ 7.23\cdot 10^{-1}  $&$ 2.74\cdot 10^{-1}  $&$ 1.85 $\\
$220 $&$ 7.69\cdot 10^{-4}  $&$ 3.67\cdot 10^{-5}  $&$ 1.34\cdot 10^{-4}  $&$ 7.14\cdot 10^{-1}  $&$ 2.84\cdot 10^{-1}  $&$ 2.31 $\\
$230 $&$ 7.27\cdot 10^{-4}  $&$ 3.14\cdot 10^{-5}  $&$ 1.19\cdot 10^{-4}  $&$ 7.08\cdot 10^{-1}  $&$ 2.89\cdot 10^{-1}  $&$ 2.82 $\\
$240 $&$ 6.97\cdot 10^{-4}  $&$ 2.72\cdot 10^{-5}  $&$ 1.07\cdot 10^{-4}  $&$ 7.04\cdot 10^{-1}  $&$ 2.94\cdot 10^{-1}  $&$ 3.40 $\\
$250 $&$ 6.75\cdot 10^{-4}  $&$ 2.37\cdot 10^{-5}  $&$ 9.54\cdot 10^{-5}  $&$ 7.01\cdot 10^{-1}  $&$ 2.97\cdot 10^{-1}  $&$ 4.04 $\\
$260 $&$ 6.59\cdot 10^{-4}  $&$ 2.08\cdot 10^{-5}  $&$ 8.57\cdot 10^{-5}  $&$ 6.99\cdot 10^{-1}  $&$ 2.99\cdot 10^{-1}  $&$ 4.76 $\\
$270 $&$ 6.48\cdot 10^{-4}  $&$ 1.84\cdot 10^{-5}  $&$ 7.72\cdot 10^{-5}  $&$ 6.97\cdot 10^{-1}  $&$ 3.02\cdot 10^{-1}  $&$ 5.55 $\\
$280 $&$ 6.42\cdot 10^{-4}  $&$ 1.63\cdot 10^{-5}  $&$ 6.98\cdot 10^{-5}  $&$ 6.95\cdot 10^{-1}  $&$ 3.04\cdot 10^{-1}  $&$ 6.43 $\\
$290 $&$ 6.42\cdot 10^{-4}  $&$ 1.45\cdot 10^{-5}  $&$ 6.32\cdot 10^{-5}  $&$ 6.93\cdot 10^{-1}  $&$ 3.05\cdot 10^{-1}  $&$ 7.39 $\\
$300 $&$ 6.46\cdot 10^{-4}  $&$ 1.30\cdot 10^{-5}  $&$ 5.75\cdot 10^{-5}  $&$ 6.92\cdot 10^{-1}  $&$ 3.07\cdot 10^{-1}  $&$ 8.43 $\\
$310 $&$ 6.56\cdot 10^{-4}  $&$ 1.17\cdot 10^{-5}  $&$ 5.24\cdot 10^{-5}  $&$ 6.90\cdot 10^{-1}  $&$ 3.08\cdot 10^{-1}  $&$ 9.57 $\\
$320 $&$ 6.73\cdot 10^{-4}  $&$ 1.05\cdot 10^{-5}  $&$ 4.79\cdot 10^{-5}  $&$ 6.89\cdot 10^{-1}  $&$ 3.09\cdot 10^{-1}  $&$ 10.8 $\\
$330 $&$ 6.99\cdot 10^{-4}  $&$ 9.56\cdot 10^{-6}  $&$ 4.39\cdot 10^{-5}  $&$ 6.88\cdot 10^{-1}  $&$ 3.10\cdot 10^{-1}  $&$ 12.1 $\\
$340 $&$ 7.42\cdot 10^{-4}  $&$ 8.73\cdot 10^{-6}  $&$ 4.04\cdot 10^{-5}  $&$ 6.87\cdot 10^{-1}  $&$ 3.11\cdot 10^{-1}  $&$ 13.5 $\\
$350 $&$ 8.05\cdot 10^{-4}  $&$ 7.62\cdot 10^{-6}  $&$ 3.65\cdot 10^{-5}  $&$ 6.76\cdot 10^{-1}  $&$ 3.07\cdot 10^{-1}  $&$ 15.2 $\\
$360 $&$ 8.42\cdot 10^{-4}  $&$ 6.10\cdot 10^{-6}  $&$ 3.17\cdot 10^{-5}  $&$ 6.51\cdot 10^{-1}  $&$ 2.97\cdot 10^{-1}  $&$ 17.6 $\\
$370 $&$ 8.54\cdot 10^{-4}  $&$ 4.85\cdot 10^{-6}  $&$ 2.76\cdot 10^{-5}  $&$ 6.28\cdot 10^{-1}  $&$ 2.87\cdot 10^{-1}  $&$ 20.2 $\\
$380 $&$ 8.51\cdot 10^{-4}  $&$ 3.86\cdot 10^{-6}  $&$ 2.42\cdot 10^{-5}  $&$ 6.09\cdot 10^{-1}  $&$ 2.79\cdot 10^{-1}  $&$ 23.1 $\\
$390 $&$ 8.40\cdot 10^{-4}  $&$ 3.09\cdot 10^{-6}  $&$ 2.14\cdot 10^{-5}  $&$ 5.94\cdot 10^{-1}  $&$ 2.73\cdot 10^{-1}  $&$ 26.1 $\\
$400 $&$ 8.22\cdot 10^{-4}  $&$ 2.47\cdot 10^{-6}  $&$ 1.90\cdot 10^{-5}  $&$ 5.82\cdot 10^{-1}  $&$ 2.69\cdot 10^{-1}  $&$ 29.2 $\\
$410 $&$ 8.02\cdot 10^{-4}  $&$ 1.98\cdot 10^{-6}  $&$ 1.70\cdot 10^{-5}  $&$ 5.72\cdot 10^{-1}  $&$ 2.65\cdot 10^{-1}  $&$ 32.5 $\\
$420 $&$ 7.80\cdot 10^{-4}  $&$ 1.60\cdot 10^{-6}  $&$ 1.53\cdot 10^{-5}  $&$ 5.64\cdot 10^{-1}  $&$ 2.63\cdot 10^{-1}  $&$ 35.9 $\\
$430 $&$ 7.56\cdot 10^{-4}  $&$ 1.28\cdot 10^{-6}  $&$ 1.38\cdot 10^{-5}  $&$ 5.59\cdot 10^{-1}  $&$ 2.61\cdot 10^{-1}  $&$ 39.4 $\\
$440 $&$ 7.33\cdot 10^{-4}  $&$ 1.03\cdot 10^{-6}  $&$ 1.26\cdot 10^{-5}  $&$ 5.54\cdot 10^{-1}  $&$ 2.60\cdot 10^{-1}  $&$ 43.1 $\\
$450 $&$ 7.09\cdot 10^{-4}  $&$ 8.27\cdot 10^{-7}  $&$ 1.15\cdot 10^{-5}  $&$ 5.51\cdot 10^{-1}  $&$ 2.59\cdot 10^{-1}  $&$ 46.9 $\\
$460 $&$ 6.85\cdot 10^{-4}  $&$ 6.62\cdot 10^{-7}  $&$ 1.05\cdot10^{-5}  $&$ 5.49\cdot 10^{-1}  $&$ 2.59\cdot 10^{-1}  $&$ 50.8 $\\
$470 $&$ 6.62\cdot 10^{-4}  $&$ 5.29\cdot 10^{-7}  $&$ 9.64\cdot 10^{-6}  $&$ 5.47\cdot 10^{-1}  $&$ 2.59\cdot 10^{-1}  $&$ 54.9 $\\
$480 $&$ 6.39\cdot 10^{-4}  $&$ 4.21\cdot 10^{-7}  $&$ 8.87\cdot 10^{-6}  $&$ 5.46\cdot 10^{-1}  $&$ 2.59\cdot 10^{-1}  $&$ 59.1 $\\
$490 $&$ 6.17\cdot 10^{-4}  $&$ 3.34\cdot 10^{-7}  $&$ 8.19\cdot 10^{-6}  $&$ 5.46\cdot 10^{-1}  $&$ 2.60\cdot 10^{-1}  $&$ 63.5 $\\
\hline
  \end{tabular}
\end{table}

\begin{table}
  \vspace{-\headsep}
  \caption{SM Higgs branching ratios in bosonic final states and Higgs total widths in the high-mass range.}
  \label{tab:BR-hm.part2}
  \centering
  \small
  \begin{tabular}{ccccccc}\hline
$\MH$ [GeV] & $\PH \rightarrow \Pg\Pg$ & $\PH \rightarrow \PGg \PGg$ & $\PH \rightarrow
\PZ\PGg$ & $\PH \rightarrow \PW\PW$
& $\PH \rightarrow \PZ\PZ$ & Total $\Gamma_{\PH}$ [GeV]\\
\hline
$500 $&$ 5.96\cdot 10^{-4}  $&$ 2.64\cdot 10^{-7}  $&$ 7.58\cdot 10^{-6}  $&$ 5.46\cdot 10^{-1}  $&$ 2.61\cdot 10^{-1}  $&$ 68.0  $\\
$510 $&$ 5.75\cdot 10^{-4}  $&$ 2.09\cdot 10^{-7}  $&$ 7.03\cdot 10^{-6}  $&$ 5.46\cdot 10^{-1}  $&$ 2.61\cdot 10^{-1}  $&$ 72.7  $\\
$520 $&$ 5.55\cdot 10^{-4}  $&$ 1.65\cdot 10^{-7}  $&$ 6.53\cdot 10^{-6}  $&$ 5.47\cdot 10^{-1}  $&$ 2.62\cdot 10^{-1}  $&$ 77.6  $\\
$530 $&$ 5.36\cdot 10^{-4}  $&$ 1.30\cdot 10^{-7}  $&$ 6.08\cdot 10^{-6}  $&$ 5.48\cdot 10^{-1}  $&$ 2.63\cdot 10^{-1}  $&$ 82.6  $\\
$540 $&$ 5.17\cdot 10^{-4}  $&$ 1.04\cdot 10^{-7}  $&$ 5.67\cdot 10^{-6}  $&$ 5.49\cdot 10^{-1}  $&$ 2.65\cdot 10^{-1}  $&$ 87.7  $\\
$550 $&$ 4.99\cdot 10^{-4}  $&$ 8.52\cdot 10^{-8}  $&$ 5.30\cdot 10^{-6}  $&$ 5.50\cdot 10^{-1}  $&$ 2.66\cdot 10^{-1}  $&$ 93.1  $\\
$560 $&$ 4.82\cdot 10^{-4}  $&$ 7.16\cdot 10^{-8}  $&$ 4.95\cdot 10^{-6}  $&$ 5.51\cdot 10^{-1}  $&$ 2.67\cdot 10^{-1}  $&$ 98.7 $\\
$570 $&$ 4.65\cdot 10^{-4}  $&$ 6.28\cdot 10^{-8}  $&$ 4.64\cdot 10^{-6}  $&$ 5.53\cdot 10^{-1}  $&$ 2.68\cdot 10^{-1}  $&$ 104 $\\
$580 $&$ 4.49\cdot 10^{-4}  $&$ 5.80\cdot 10^{-8}  $&$ 4.35\cdot 10^{-6}  $&$ 5.55\cdot 10^{-1}  $&$ 2.70\cdot 10^{-1}  $&$ 110 $\\
$590 $&$ 4.34\cdot 10^{-4}  $&$ 5.64\cdot 10^{-8}  $&$ 4.08\cdot 10^{-6}  $&$ 5.56\cdot 10^{-1}  $&$ 2.71\cdot 10^{-1}  $&$ 116 $\\
$600 $&$ 4.19\cdot 10^{-4}  $&$ 5.77\cdot 10^{-8}  $&$ 3.84\cdot 10^{-6}  $&$ 5.58\cdot 10^{-1}  $&$ 2.72\cdot 10^{-1}  $&$ 123 $\\
$610 $&$ 4.04\cdot 10^{-4}  $&$ 6.12\cdot 10^{-8}  $&$ 3.61\cdot 10^{-6}  $&$ 5.60\cdot 10^{-1}  $&$ 2.73\cdot 10^{-1}  $&$ 129 $\\
$620 $&$ 3.90\cdot 10^{-4}  $&$ 6.66\cdot 10^{-8}  $&$ 3.40\cdot 10^{-6}  $&$ 5.62\cdot 10^{-1}  $&$ 2.75\cdot 10^{-1}  $&$ 136 $\\
$630 $&$ 3.77\cdot 10^{-4}  $&$ 7.36\cdot 10^{-8}  $&$ 3.21\cdot 10^{-6}  $&$ 5.64\cdot 10^{-1}  $&$ 2.76\cdot 10^{-1}  $&$ 143 $\\
$640 $&$ 3.65\cdot 10^{-4}  $&$ 8.19\cdot 10^{-8}  $&$ 3.03\cdot 10^{-6}  $&$ 5.66\cdot 10^{-1}  $&$ 2.77\cdot 10^{-1}  $&$ 150 $\\
$650 $&$ 3.52\cdot 10^{-4}  $&$ 9.12\cdot 10^{-8}  $&$ 2.86\cdot 10^{-6}  $&$ 5.67\cdot 10^{-1}  $&$ 2.79\cdot 10^{-1}  $&$ 158 $\\
$660 $&$ 3.40\cdot 10^{-4}  $&$ 1.01\cdot 10^{-7}  $&$ 2.70\cdot 10^{-6}  $&$ 5.69\cdot 10^{-1}  $&$ 2.80\cdot 10^{-1}  $&$ 166 $\\
$670 $&$ 3.29\cdot 10^{-4}  $&$ 1.12\cdot 10^{-7}  $&$ 2.56\cdot 10^{-6}  $&$ 5.71\cdot 10^{-1}  $&$ 2.81\cdot 10^{-1}  $&$ 174 $\\
$680 $&$ 3.18\cdot 10^{-4}  $&$ 1.23\cdot 10^{-7}  $&$ 2.42\cdot 10^{-6}  $&$ 5.73\cdot 10^{-1}  $&$ 2.82\cdot 10^{-1}  $&$ 182 $\\
$690 $&$ 3.07\cdot 10^{-4}  $&$ 1.35\cdot 10^{-7}  $&$ 2.29\cdot 10^{-6}  $&$ 5.75\cdot 10^{-1}  $&$ 2.83\cdot 10^{-1}  $&$ 190 $\\
$700 $&$ 2.97\cdot 10^{-4}  $&$ 1.47\cdot 10^{-7}  $&$ 2.18\cdot 10^{-6}  $&$ 5.77\cdot 10^{-1}  $&$ 2.85\cdot 10^{-1}  $&$ 199 $\\
$710 $&$ 2.87\cdot 10^{-4}  $&$ 1.59\cdot 10^{-7}  $&$ 2.06\cdot 10^{-6}  $&$ 5.79\cdot 10^{-1}  $&$ 2.86\cdot 10^{-1}  $&$ 208 $\\
$720 $&$ 2.78\cdot 10^{-4}  $&$ 1.71\cdot 10^{-7}  $&$ 1.96\cdot 10^{-6}  $&$ 5.81\cdot 10^{-1}  $&$ 2.87\cdot 10^{-1}  $&$ 218 $\\
$730 $&$ 2.69\cdot 10^{-4}  $&$ 1.83\cdot 10^{-7}  $&$ 1.86\cdot 10^{-6}  $&$ 5.82\cdot 10^{-1}  $&$ 2.88\cdot 10^{-1}  $&$ 227 $\\
$740 $&$ 2.60\cdot 10^{-4}  $&$ 1.95\cdot 10^{-7}  $&$ 1.77\cdot 10^{-6}  $&$ 5.84\cdot 10^{-1}  $&$ 2.89\cdot 10^{-1}  $&$ 237 $\\
$750 $&$ 2.51\cdot 10^{-4}  $&$ 2.07\cdot 10^{-7}  $&$ 1.69\cdot 10^{-6}  $&$ 5.86\cdot 10^{-1}  $&$ 2.90\cdot 10^{-1}  $&$ 248 $\\
$760 $&$ 2.43\cdot 10^{-4}  $&$ 2.19\cdot 10^{-7}  $&$ 1.61\cdot 10^{-6}  $&$ 5.88\cdot 10^{-1}  $&$ 2.91\cdot 10^{-1}  $&$ 258 $\\
$770 $&$ 2.36\cdot 10^{-4}  $&$ 2.30\cdot 10^{-7}  $&$ 1.53\cdot 10^{-6}  $&$ 5.89\cdot 10^{-1}  $&$ 2.92\cdot 10^{-1}  $&$ 269 $\\
$780 $&$ 2.28\cdot 10^{-4}  $&$ 2.41\cdot 10^{-7}  $&$ 1.46\cdot 10^{-6}  $&$ 5.91\cdot 10^{-1}  $&$ 2.93\cdot 10^{-1}  $&$ 281 $\\
$790 $&$ 2.21\cdot 10^{-4}  $&$ 2.53\cdot 10^{-7}  $&$ 1.40\cdot 10^{-6}  $&$ 5.93\cdot 10^{-1}  $&$ 2.94\cdot 10^{-1}  $&$ 292 $\\
$800 $&$ 2.14\cdot 10^{-4}  $&$ 2.63\cdot 10^{-7}  $&$ 1.33\cdot 10^{-6}  $&$ 5.94\cdot 10^{-1}  $&$ 2.95\cdot 10^{-1}  $&$ 304 $\\
$810 $&$ 2.07\cdot 10^{-4}  $&$ 2.74\cdot 10^{-7}  $&$ 1.27\cdot 10^{-6}  $&$ 5.96\cdot 10^{-1}  $&$ 2.96\cdot 10^{-1}  $&$ 317 $\\
$820 $&$ 2.00\cdot 10^{-4}  $&$ 2.84\cdot 10^{-7}  $&$ 1.22\cdot 10^{-6}  $&$ 5.97\cdot 10^{-1}  $&$ 2.97\cdot 10^{-1}  $&$ 330 $\\
$830 $&$ 1.94\cdot 10^{-4}  $&$ 2.94\cdot 10^{-7}  $&$ 1.16\cdot 10^{-6}  $&$ 5.99\cdot 10^{-1}  $&$ 2.98\cdot 10^{-1}  $&$ 343 $\\
$840 $&$ 1.88\cdot 10^{-4}  $&$ 3.04\cdot 10^{-7}  $&$ 1.12\cdot 10^{-6}  $&$ 6.01\cdot 10^{-1}  $&$ 2.99\cdot 10^{-1}  $&$ 357 $\\
$850 $&$ 1.82\cdot 10^{-4}  $&$ 3.13\cdot 10^{-7}  $&$ 1.07\cdot 10^{-6}  $&$ 6.02\cdot 10^{-1}  $&$ 3.00\cdot 10^{-1}  $&$ 371 $\\
$860 $&$ 1.76\cdot 10^{-4}  $&$ 3.22\cdot 10^{-7}  $&$ 1.02\cdot 10^{-6}  $&$ 6.03\cdot 10^{-1}  $&$ 3.01\cdot 10^{-1}  $&$ 386 $\\
$870 $&$ 1.71\cdot 10^{-4}  $&$ 3.30\cdot 10^{-7}  $&$ 9.83\cdot 10^{-7}  $&$ 6.05\cdot 10^{-1}  $&$ 3.02\cdot 10^{-1}  $&$ 401 $\\
$880 $&$ 1.65\cdot 10^{-4}  $&$ 3.39\cdot 10^{-7}  $&$ 9.44\cdot 10^{-7}  $&$ 6.06\cdot 10^{-1}  $&$ 3.03\cdot 10^{-1}  $&$ 416 $\\
$890 $&$ 1.60\cdot 10^{-4}  $&$ 3.47\cdot 10^{-7}  $&$ 9.07\cdot 10^{-7}  $&$ 6.08\cdot 10^{-1}  $&$ 3.03\cdot 10^{-1}  $&$ 432 $\\
$900 $&$ 1.55\cdot 10^{-4}  $&$ 3.54\cdot 10^{-7}  $&$ 8.72\cdot 10^{-7}  $&$ 6.09\cdot 10^{-1}  $&$ 3.04\cdot 10^{-1}  $&$ 449 $\\
$910 $&$ 1.50\cdot 10^{-4}  $&$ 3.61\cdot 10^{-7}  $&$ 8.38\cdot 10^{-7}  $&$ 6.10\cdot 10^{-1}  $&$ 3.05\cdot 10^{-1}  $&$ 466 $\\
$920 $&$ 1.46\cdot 10^{-4}  $&$ 3.67\cdot 10^{-7}  $&$ 8.07\cdot 10^{-7}  $&$ 6.12\cdot 10^{-1}  $&$ 3.06\cdot 10^{-1}  $&$ 484 $\\
$930 $&$ 1.41\cdot 10^{-4}  $&$ 3.75\cdot 10^{-7}  $&$ 7.77\cdot 10^{-7}  $&$ 6.13\cdot 10^{-1}  $&$ 3.06\cdot 10^{-1}  $&$ 502 $\\
$940 $&$ 1.37\cdot 10^{-4}  $&$ 3.81\cdot 10^{-7}  $&$ 7.49\cdot 10^{-7}  $&$ 6.14\cdot 10^{-1}  $&$ 3.07\cdot 10^{-1}  $&$ 521 $\\
$950 $&$ 1.33\cdot 10^{-4}  $&$ 3.87\cdot 10^{-7}  $&$ 7.22\cdot 10^{-7}  $&$ 6.16\cdot 10^{-1}  $&$ 3.08\cdot 10^{-1}  $&$ 540 $\\
$960 $&$ 1.29\cdot 10^{-4}  $&$ 3.93\cdot 10^{-7}  $&$ 6.97\cdot 10^{-7}  $&$ 6.17\cdot 10^{-1}  $&$ 3.08\cdot 10^{-1}  $&$ 560 $\\
$970 $&$ 1.25\cdot 10^{-4}  $&$ 3.98\cdot 10^{-7}  $&$ 6.73\cdot 10^{-7}  $&$ 6.18\cdot 10^{-1}  $&$ 3.09\cdot 10^{-1}  $&$ 581 $\\
$980 $&$ 1.21\cdot 10^{-4}  $&$ 4.03\cdot 10^{-7}  $&$ 6.50\cdot 10^{-7}  $&$ 6.19\cdot 10^{-1}  $&$ 3.10\cdot 10^{-1}  $&$ 602 $\\
$990 $&$ 1.17\cdot 10^{-4}  $&$ 4.07\cdot 10^{-7}  $&$ 6.29\cdot 10^{-7}  $&$ 6.20\cdot 10^{-1}  $&$ 3.10\cdot 10^{-1}  $&$ 624 $\\
$1000 $&$ 1.14\cdot 10^{-4}  $&$ 4.12\cdot 10^{-7}  $&$ 6.08\cdot 10^{-7}  $&$ 6.21\cdot 10^{-1}  $&$ 3.11\cdot 10^{-1} $&$ 647 $\\
\hline
  \end{tabular}
\end{table}

\begin{landscape}
  \begin{table}
  \vspace{-\headsep}
  \caption{SM Higgs branching ratios for $4$-fermion final states in 
    the low-mass range. We list results for the specific final states
    $\Pep\Pem\Pep\Pem$ and $\Pep\Pem\PGmp\PGmm$, for final states with $4$ arbitrary charged
    leptons, $\Pep\PGne\Pem\PAGne$ and $\Pep\PGne\PGmm\PAGnGm$, and for
    final states $\Plp\Plm\PGnl\PAGnl$ with $2$ charged leptons
    plus $2$ neutrinos ($\PGnl$ represents any type of neutrinos).}
    \label{tab:PBR-lm}
  \centering
  \small
  \begin{tabular}{lcccccccc}
    \hline
    $\MH$ [GeV] &
    $\PH \rightarrow \Pep\Pem\Pep\Pem$ &
    $\PH \rightarrow \Pep\Pem\PGmp\PGmm$ &
    $\PH \rightarrow \Plp\Plm\Plp\Plm$ &
    $\PH \rightarrow \Plp\Plm\Plp\Plm$ &
    $\PH \rightarrow \Pep\PGne\Pem\PAGne$ &
    $\PH \rightarrow \Pep\PGne\PGmm\PAGnGm$ &
    $\PH \rightarrow \Plp\Plm\PGnl\PAGnl$ &
    $\PH \rightarrow \Plp\Plm\PGnl\PAGnl$ \\
    & & & $(\Pl=\Pe$ or $\PGm)$ & $(\Pl=\Pe, \PGm$ or $\PGt)$
    & & & $(\Pl=\Pe$ or $\PGm)$ & $(\Pl=\Pe, \PGm$ or $\PGt)$ \\
    \hline
$	90	$ & $	7.08 \cdot 10^{-7}	$ & $	9.39 \cdot 10^{-7}	$ & $	2.35 \cdot 10^{-6}	$ & $	4.94 \cdot 10^{-6}	$ & $	1.77 \cdot 10^{-5}	$ & $	2.45 \cdot 10^{-5}	$ & $	9.20 \cdot 10^{-5}	$ & $	2.12 \cdot 10^{-4}	 $ \\
$	95	$ & $	1.11 \cdot 10^{-6}	$ & $	1.49 \cdot 10^{-6}	$ & $	3.71 \cdot 10^{-6}	$ & $	7.80 \cdot 10^{-6}	$ & $	4.46 \cdot 10^{-5}	$ & $	5.54 \cdot 10^{-5}	$ & $	2.12 \cdot 10^{-4}	$ & $	4.85 \cdot 10^{-4}	 $ \\
$	100	$ & $	1.80 \cdot 10^{-6}	$ & $	2.51 \cdot 10^{-6}	$ & $	6.12 \cdot 10^{-6}	$ & $	1.30 \cdot 10^{-5}	$ & $	1.14 \cdot 10^{-4}	$ & $	1.30 \cdot 10^{-4}	$ & $	5.08 \cdot 10^{-4}	$ & $	1.15 \cdot 10^{-3}	 $ \\
$	105	$ & $	3.21 \cdot 10^{-6}	$ & $	4.78 \cdot 10^{-6}	$ & $	1.12 \cdot 10^{-5}	$ & $	2.40 \cdot 10^{-5}	$ & $	2.62 \cdot 10^{-4}	$ & $	2.85 \cdot 10^{-4}	$ & $	1.13 \cdot 10^{-3}	$ & $	2.56 \cdot 10^{-3}	 $ \\
$	110	$ & $	6.10 \cdot 10^{-6}	$ & $	9.78 \cdot 10^{-6}	$ & $	2.20 \cdot 10^{-5}	$ & $	4.77 \cdot 10^{-5}	$ & $	5.37 \cdot 10^{-4}	$ & $	5.66 \cdot 10^{-4}	$ & $	2.28 \cdot 10^{-3}	$ & $	5.13 \cdot 10^{-3}	 $ \\
$	115	$ & $	1.15 \cdot 10^{-5}	$ & $	1.95 \cdot 10^{-5}	$ & $	4.25 \cdot 10^{-5}	$ & $	9.31 \cdot 10^{-5}	$ & $	9.88 \cdot 10^{-4}	$ & $	1.02 \cdot 10^{-3}	$ & $	4.17 \cdot 10^{-3}	$ & $	9.31 \cdot 10^{-3}	 $ \\
$	120	$ & $	2.03 \cdot 10^{-5}	$ & $	3.60 \cdot 10^{-5}	$ & $	7.66 \cdot 10^{-5}	$ & $	1.69 \cdot 10^{-4}	$ & $	1.65 \cdot 10^{-3}	$ & $	1.68 \cdot 10^{-3}	$ & $	6.95 \cdot 10^{-3}	$ & $	1.55 \cdot 10^{-2}	 $ \\
$	125	$ & $	3.30 \cdot 10^{-5}	$ & $	5.98 \cdot 10^{-5}	$ & $	1.26 \cdot 10^{-4}	$ & $	2.79 \cdot 10^{-4}	$ & $	2.54 \cdot 10^{-3}	$ & $	2.55 \cdot 10^{-3}	$ & $	1.07 \cdot 10^{-2}	$ & $	2.36 \cdot 10^{-2}	 $ \\
$	130	$ & $	4.89 \cdot 10^{-5}	$ & $	9.03 \cdot 10^{-5}	$ & $	1.88 \cdot 10^{-4}	$ & $	4.18 \cdot 10^{-4}	$ & $	3.62 \cdot 10^{-3}	$ & $	3.59 \cdot 10^{-3}	$ & $	1.52 \cdot 10^{-2}	$ & $	3.35 \cdot 10^{-2}	 $ \\
$	135	$ & $	6.63 \cdot 10^{-5}	$ & $	1.24 \cdot 10^{-4}	$ & $	2.57 \cdot 10^{-4}	$ & $	5.71 \cdot 10^{-4}	$ & $	4.83 \cdot 10^{-3}	$ & $	4.75 \cdot 10^{-3}	$ & $	2.01 \cdot 10^{-2}	$ & $	4.44 \cdot 10^{-2}	 $ \\
$	140	$ & $	8.25 \cdot 10^{-5}	$ & $	1.56 \cdot 10^{-4}	$ & $	3.21 \cdot 10^{-4}	$ & $	7.15 \cdot 10^{-4}	$ & $	6.07 \cdot 10^{-3}	$ & $	5.94 \cdot 10^{-3}	$ & $	2.52 \cdot 10^{-2}	$ & $	5.57 \cdot 10^{-2}	 $ \\
$	145	$ & $	9.43 \cdot 10^{-5}	$ & $	1.79 \cdot 10^{-4}	$ & $	3.68 \cdot 10^{-4}	$ & $	8.21 \cdot 10^{-4}	$ & $	7.28 \cdot 10^{-3}	$ & $	7.11 \cdot 10^{-3}	$ & $	3.02 \cdot 10^{-2}	$ & $	6.66 \cdot 10^{-2}	 $ \\
$	150	$ & $	9.76 \cdot 10^{-5}	$ & $	1.87 \cdot 10^{-4}	$ & $	3.82 \cdot 10^{-4}	$ & $	8.53 \cdot 10^{-4}	$ & $	8.44 \cdot 10^{-3}	$ & $	8.23 \cdot 10^{-3}	$ & $	3.48 \cdot 10^{-2}	$ & $	7.70 \cdot 10^{-2}	 $ \\
$	155	$ & $	8.63 \cdot 10^{-5}	$ & $	1.66 \cdot 10^{-4}	$ & $	3.38 \cdot 10^{-4}	$ & $	7.57 \cdot 10^{-4}	$ & $	9.59 \cdot 10^{-3}	$ & $	9.39 \cdot 10^{-3}	$ & $	3.93 \cdot 10^{-2}	$ & $	8.71 \cdot 10^{-2}	 $ \\
$	160	$ & $	4.85 \cdot 10^{-5}	$ & $	9.36 \cdot 10^{-5}	$ & $	1.91 \cdot 10^{-4}	$ & $	4.26 \cdot 10^{-4}	$ & $	1.08 \cdot 10^{-2}	$ & $	1.07 \cdot 10^{-2}	$ & $	4.39 \cdot 10^{-2}	$ & $	9.80 \cdot 10^{-2}	 $ \\
$	165	$ & $	2.58 \cdot 10^{-5}	$ & $	5.00 \cdot 10^{-5}	$ & $	1.02 \cdot 10^{-4}	$ & $	2.27 \cdot 10^{-4}	$ & $	1.14 \cdot 10^{-2}	$ & $	1.13 \cdot 10^{-2}	$ & $	4.59 \cdot 10^{-2}	$ & $	1.03 \cdot 10^{-1}	 $ \\
$	170	$ & $	2.73 \cdot 10^{-5}	$ & $	5.32 \cdot 10^{-5}	$ & $	1.08 \cdot 10^{-4}	$ & $	2.42 \cdot 10^{-4}	$ & $	1.15 \cdot 10^{-2}	$ & $	1.14 \cdot 10^{-2}	$ & $	4.61 \cdot 10^{-2}	$ & $	1.03 \cdot 10^{-1}	 $ \\
$	175	$ & $	3.71 \cdot 10^{-5}	$ & $	7.28 \cdot 10^{-5}	$ & $	1.47 \cdot 10^{-4}	$ & $	3.30 \cdot 10^{-4}	$ & $	1.14 \cdot 10^{-2}	$ & $	1.13 \cdot 10^{-2}	$ & $	4.60 \cdot 10^{-2}	$ & $	1.03 \cdot 10^{-1}	 $ \\
$	180	$ & $	6.85 \cdot 10^{-5}	$ & $	1.36 \cdot 10^{-4}	$ & $	2.73 \cdot 10^{-4}	$ & $	6.12 \cdot 10^{-4}	$ & $	1.12 \cdot 10^{-2}	$ & $	1.10 \cdot 10^{-2}	$ & $	4.55 \cdot 10^{-2}	$ & $	1.01 \cdot 10^{-1}	 $ \\
$	185	$ & $	1.70 \cdot 10^{-4}	$ & $	3.38 \cdot 10^{-4}	$ & $	6.78 \cdot 10^{-4}	$ & $	1.52 \cdot 10^{-3}	$ & $	1.06 \cdot 10^{-2}	$ & $	9.95 \cdot 10^{-3}	$ & $	4.38 \cdot 10^{-2}	$ & $	9.56 \cdot 10^{-2}	 $ \\
$	190	$ & $	2.36 \cdot 10^{-4}	$ & $	4.72 \cdot 10^{-4}	$ & $	9.44 \cdot 10^{-4}	$ & $	2.12 \cdot 10^{-3}	$ & $	1.02 \cdot 10^{-2}	$ & $	9.26 \cdot 10^{-3}	$ & $	4.27 \cdot 10^{-2}	$ & $	9.18 \cdot 10^{-2}	 $ \\
$	195	$ & $	2.69 \cdot 10^{-4}	$ & $	5.37 \cdot 10^{-4}	$ & $	1.08 \cdot 10^{-3}	$ & $	2.42 \cdot 10^{-3}	$ & $	9.99 \cdot 10^{-3}	$ & $	8.92 \cdot 10^{-3}	$ & $	4.21 \cdot 10^{-2}	$ & $	8.99 \cdot 10^{-2}	 $ \\
$	200	$ & $	2.88 \cdot 10^{-4}	$ & $	5.75 \cdot 10^{-4}	$ & $	1.15 \cdot 10^{-3}	$ & $	2.59 \cdot 10^{-3}	$ & $	9.87 \cdot 10^{-3}	$ & $	8.73 \cdot 10^{-3}	$ & $	4.18 \cdot 10^{-2}	$ & $	8.89 \cdot 10^{-2}	 $ \\
$	210	$ & $	3.08 \cdot 10^{-4}	$ & $	6.17 \cdot 10^{-4}	$ & $	1.23 \cdot 10^{-3}	$ & $	2.78 \cdot 10^{-3}	$ & $	9.74 \cdot 10^{-3}	$ & $	8.52 \cdot 10^{-3}	$ & $	4.14 \cdot 10^{-2}	$ & $	8.77 \cdot 10^{-2}	 $ \\
$	220	$ & $	3.19 \cdot 10^{-4}	$ & $	6.38 \cdot 10^{-4}	$ & $	1.28 \cdot 10^{-3}	$ & $	2.87 \cdot 10^{-3}	$ & $	9.68 \cdot 10^{-3}	$ & $	8.41 \cdot 10^{-3}	$ & $	4.13 \cdot 10^{-2}	$ & $	8.71 \cdot 10^{-2}	 $ \\
$	230	$ & $	3.26 \cdot 10^{-4}	$ & $	6.52 \cdot 10^{-4}	$ & $	1.30 \cdot 10^{-3}	$ & $	2.93 \cdot 10^{-3}	$ & $	9.64 \cdot 10^{-3}	$ & $	8.34 \cdot 10^{-3}	$ & $	4.12 \cdot 10^{-2}	$ & $	8.68 \cdot 10^{-2}	 $ \\
$	240	$ & $	3.31 \cdot 10^{-4}	$ & $	6.61 \cdot 10^{-4}	$ & $	1.32 \cdot 10^{-3}	$ & $	2.97 \cdot 10^{-3}	$ & $	9.61 \cdot 10^{-3}	$ & $	8.29 \cdot 10^{-3}	$ & $	4.11 \cdot 10^{-2}	$ & $	8.65 \cdot 10^{-2}	 $ \\
$	250	$ & $	3.34 \cdot 10^{-4}	$ & $	6.68 \cdot 10^{-4}	$ & $	1.34 \cdot 10^{-3}	$ & $	3.01 \cdot 10^{-3}	$ & $	9.59 \cdot 10^{-3}	$ & $	8.26 \cdot 10^{-3}	$ & $	4.10 \cdot 10^{-2}	$ & $	8.63 \cdot 10^{-2}	 $ \\
$	260	$ & $	3.37 \cdot 10^{-4}	$ & $	6.74 \cdot 10^{-4}	$ & $	1.35 \cdot 10^{-3}	$ & $	3.03 \cdot 10^{-3}	$ & $	9.57 \cdot 10^{-3}	$ & $	8.23 \cdot 10^{-3}	$ & $	4.10 \cdot 10^{-2}	$ & $	8.62 \cdot 10^{-2}	 $ \\
$	270	$ & $	3.40 \cdot 10^{-4}	$ & $	6.79 \cdot 10^{-4}	$ & $	1.36 \cdot 10^{-3}	$ & $	3.06 \cdot 10^{-3}	$ & $	9.56 \cdot 10^{-3}	$ & $	8.21 \cdot 10^{-3}	$ & $	4.09 \cdot 10^{-2}	$ & $	8.60 \cdot 10^{-2}	 $ \\
$	280	$ & $	3.42 \cdot 10^{-4}	$ & $	6.83 \cdot 10^{-4}	$ & $	1.37 \cdot 10^{-3}	$ & $	3.07 \cdot 10^{-3}	$ & $	9.54 \cdot 10^{-3}	$ & $	8.18 \cdot 10^{-3}	$ & $	4.09 \cdot 10^{-2}	$ & $	8.59 \cdot 10^{-2}	 $ \\
$	290	$ & $	3.44 \cdot 10^{-4}	$ & $	6.87 \cdot 10^{-4}	$ & $	1.37 \cdot 10^{-3}	$ & $	3.09 \cdot 10^{-3}	$ & $	9.53 \cdot 10^{-3}	$ & $	8.16 \cdot 10^{-3}	$ & $	4.09 \cdot 10^{-2}	$ & $	8.58 \cdot 10^{-2}	 $ \\
$	300	$ & $	3.45 \cdot 10^{-4}	$ & $	6.90 \cdot 10^{-4}	$ & $	1.38 \cdot 10^{-3}	$ & $	3.11 \cdot 10^{-3}	$ & $	9.52 \cdot 10^{-3}	$ & $	8.15 \cdot 10^{-3}	$ & $	4.08 \cdot 10^{-2}	$ & $	8.57 \cdot 10^{-2}	 $ \\
    \hline
  \end{tabular}
  \end{table}
\end{landscape}

\begin{landscape}
  \begin{table}
  \vspace{-\headsep}
  \caption{SM Higgs branching ratios for $4$-fermion final states in 
    the intermediate-mass range. We list results for the specific final states
    $\Pep\Pem\Pep\Pem$ and $\Pep\Pem\PGmp\PGmm$, for final states with $4$ arbitrary charged
    leptons, $\Pep\PGne\Pem\PAGne$ and $\Pep\PGne\PGmm\PAGnGm$, and for
    final states $\Plp\Plm\PGnl\PAGnl$ with $2$ charged leptons
    plus $2$ neutrinos ($\PGnl$ represents any type of neutrinos).}
  \label{tab:PBR-im}
  \centering
  \small
  \begin{tabular}{lcccccccc}
\hline
    $\MH$ [GeV] &
    $\PH \rightarrow \Pep\Pem\Pep\Pem$ &
    $\PH \rightarrow \Pep\Pem\PGmp\PGmm$ &
    $\PH \rightarrow \Plp\Plm\Plp\Plm$ &
    $\PH \rightarrow \Plp\Plm\Plp\Plm$ &
    $\PH \rightarrow \Pep\PGne\Pem\PAGne$ &
    $\PH \rightarrow \Pep\PGne\PGmm\PAGnGm$ &
    $\PH \rightarrow \Plp\Plm\PGnl\PAGnl$ &
    $\PH \rightarrow \Plp\Plm\PGnl\PAGnl$ \\
    & & & $(\Pl=\Pe$ or $\PGm)$ & $(\Pl=\Pe, \PGm$ or $\PGt)$
    & & & $(\Pl=\Pe$ or $\PGm)$ & $(\Pl=\Pe, \PGm$ or $\PGt)$ \\
\hline
$	310	$ & $	3.47 \cdot 10^{-4}	$ & $	6.93 \cdot 10^{-4}	$ & $	1.39 \cdot 10^{-3}	$ & $	3.12 \cdot 10^{-3}	$ & $	9.51 \cdot 10^{-3}	$ & $	8.13 \cdot 10^{-3}	$ & $	4.08 \cdot 10^{-2}	$ & $	8.56 \cdot 10^{-2}	 $ \\
$	320	$ & $	3.48 \cdot 10^{-4}	$ & $	6.96 \cdot 10^{-4}	$ & $	1.39 \cdot 10^{-3}	$ & $	3.13 \cdot 10^{-3}	$ & $	9.50 \cdot 10^{-3}	$ & $	8.11 \cdot 10^{-3}	$ & $	4.08 \cdot 10^{-2}	$ & $	8.55 \cdot 10^{-2}	 $ \\
$	330	$ & $	3.49 \cdot 10^{-4}	$ & $	6.98 \cdot 10^{-4}	$ & $	1.40 \cdot 10^{-3}	$ & $	3.14 \cdot 10^{-3}	$ & $	9.49 \cdot 10^{-3}	$ & $	8.10 \cdot 10^{-3}	$ & $	4.07 \cdot 10^{-2}	$ & $	8.54 \cdot 10^{-2}	 $ \\
$	340	$ & $	3.50 \cdot 10^{-4}	$ & $	6.99 \cdot 10^{-4}	$ & $	1.40 \cdot 10^{-3}	$ & $	3.15 \cdot 10^{-3}	$ & $	9.47 \cdot 10^{-3}	$ & $	8.08 \cdot 10^{-3}	$ & $	4.07 \cdot 10^{-2}	$ & $	8.53 \cdot 10^{-2}	 $ \\
$	350	$ & $	3.45 \cdot 10^{-4}	$ & $	6.90 \cdot 10^{-4}	$ & $	1.38 \cdot 10^{-3}	$ & $	3.11 \cdot 10^{-3}	$ & $	9.33 \cdot 10^{-3}	$ & $	7.96 \cdot 10^{-3}	$ & $	4.01 \cdot 10^{-2}	$ & $	8.40 \cdot 10^{-2}	 $ \\
$	360	$ & $	3.34 \cdot 10^{-4}	$ & $	6.67 \cdot 10^{-4}	$ & $	1.33 \cdot 10^{-3}	$ & $	3.00 \cdot 10^{-3}	$ & $	8.99 \cdot 10^{-3}	$ & $	7.66 \cdot 10^{-3}	$ & $	3.86 \cdot 10^{-2}	$ & $	8.09 \cdot 10^{-2}	 $ \\
$	370	$ & $	3.23 \cdot 10^{-4}	$ & $	6.46 \cdot 10^{-4}	$ & $	1.29 \cdot 10^{-3}	$ & $	2.91 \cdot 10^{-3}	$ & $	8.68 \cdot 10^{-3}	$ & $	7.39 \cdot 10^{-3}	$ & $	3.73 \cdot 10^{-2}	$ & $	7.81 \cdot 10^{-2}	 $ \\
$	380	$ & $	3.15 \cdot 10^{-4}	$ & $	6.29 \cdot 10^{-4}	$ & $	1.26 \cdot 10^{-3}	$ & $	2.83 \cdot 10^{-3}	$ & $	8.43 \cdot 10^{-3}	$ & $	7.17 \cdot 10^{-3}	$ & $	3.62 \cdot 10^{-2}	$ & $	7.58 \cdot 10^{-2}	 $ \\
$	390	$ & $	3.08 \cdot 10^{-4}	$ & $	6.15 \cdot 10^{-4}	$ & $	1.23 \cdot 10^{-3}	$ & $	2.77 \cdot 10^{-3}	$ & $	8.22 \cdot 10^{-3}	$ & $	7.00 \cdot 10^{-3}	$ & $	3.53 \cdot 10^{-2}	$ & $	7.40 \cdot 10^{-2}	 $ \\
$	400	$ & $	3.03 \cdot 10^{-4}	$ & $	6.05 \cdot 10^{-4}	$ & $	1.21 \cdot 10^{-3}	$ & $	2.72 \cdot 10^{-3}	$ & $	8.06 \cdot 10^{-3}	$ & $	6.85 \cdot 10^{-3}	$ & $	3.46 \cdot 10^{-2}	$ & $	7.25 \cdot 10^{-2}	 $ \\
$	410	$ & $	2.99 \cdot 10^{-4}	$ & $	5.98 \cdot 10^{-4}	$ & $	1.20 \cdot 10^{-3}	$ & $	2.69 \cdot 10^{-3}	$ & $	7.93 \cdot 10^{-3}	$ & $	6.74 \cdot 10^{-3}	$ & $	3.41 \cdot 10^{-2}	$ & $	7.14 \cdot 10^{-2}	 $ \\
$	420	$ & $	2.96 \cdot 10^{-4}	$ & $	5.92 \cdot 10^{-4}	$ & $	1.18 \cdot 10^{-3}	$ & $	2.67 \cdot 10^{-3}	$ & $	7.83 \cdot 10^{-3}	$ & $	6.65 \cdot 10^{-3}	$ & $	3.37 \cdot 10^{-2}	$ & $	7.05 \cdot 10^{-2}	 $ \\
$	430	$ & $	2.94 \cdot 10^{-4}	$ & $	5.88 \cdot 10^{-4}	$ & $	1.18 \cdot 10^{-3}	$ & $	2.65 \cdot 10^{-3}	$ & $	7.76 \cdot 10^{-3}	$ & $	6.59 \cdot 10^{-3}	$ & $	3.34 \cdot 10^{-2}	$ & $	6.98 \cdot 10^{-2}	 $ \\
$	440	$ & $	2.93 \cdot 10^{-4}	$ & $	5.86 \cdot 10^{-4}	$ & $	1.17 \cdot 10^{-3}	$ & $	2.64 \cdot 10^{-3}	$ & $	7.70 \cdot 10^{-3}	$ & $	6.53 \cdot 10^{-3}	$ & $	3.31 \cdot 10^{-2}	$ & $	6.93 \cdot 10^{-2}	 $ \\
$	450	$ & $	2.92 \cdot 10^{-4}	$ & $	5.84 \cdot 10^{-4}	$ & $	1.17 \cdot 10^{-3}	$ & $	2.63 \cdot 10^{-3}	$ & $	7.66 \cdot 10^{-3}	$ & $	6.50 \cdot 10^{-3}	$ & $	3.30 \cdot 10^{-2}	$ & $	6.89 \cdot 10^{-2}	 $ \\
$	460	$ & $	2.92 \cdot 10^{-4}	$ & $	5.84 \cdot 10^{-4}	$ & $	1.17 \cdot 10^{-3}	$ & $	2.63 \cdot 10^{-3}	$ & $	7.63 \cdot 10^{-3}	$ & $	6.47 \cdot 10^{-3}	$ & $	3.29 \cdot 10^{-2}	$ & $	6.87 \cdot 10^{-2}	 $ \\
$	470	$ & $	2.92 \cdot 10^{-4}	$ & $	5.84 \cdot 10^{-4}	$ & $	1.17 \cdot 10^{-3}	$ & $	2.63 \cdot 10^{-3}	$ & $	7.62 \cdot 10^{-3}	$ & $	6.45 \cdot 10^{-3}	$ & $	3.28 \cdot 10^{-2}	$ & $	6.85 \cdot 10^{-2}	 $ \\
$	480	$ & $	2.92 \cdot 10^{-4}	$ & $	5.85 \cdot 10^{-4}	$ & $	1.17 \cdot 10^{-3}	$ & $	2.63 \cdot 10^{-3}	$ & $	7.61 \cdot 10^{-3}	$ & $	6.44 \cdot 10^{-3}	$ & $	3.28 \cdot 10^{-2}	$ & $	6.85 \cdot 10^{-2}	 $ \\
$	490	$ & $	2.93 \cdot 10^{-4}	$ & $	5.86 \cdot 10^{-4}	$ & $	1.17 \cdot 10^{-3}	$ & $	2.64 \cdot 10^{-3}	$ & $	7.60 \cdot 10^{-3}	$ & $	6.44 \cdot 10^{-3}	$ & $	3.28 \cdot 10^{-2}	$ & $	6.84 \cdot 10^{-2}	 $ \\
$	500	$ & $	2.94 \cdot 10^{-4}	$ & $	5.88 \cdot 10^{-4}	$ & $	1.18 \cdot 10^{-3}	$ & $	2.64 \cdot 10^{-3}	$ & $	7.61 \cdot 10^{-3}	$ & $	6.44 \cdot 10^{-3}	$ & $	3.28 \cdot 10^{-2}	$ & $	6.85 \cdot 10^{-2}	 $ \\
$	510	$ & $	2.95 \cdot 10^{-4}	$ & $	5.90 \cdot 10^{-4}	$ & $	1.18 \cdot 10^{-3}	$ & $	2.65 \cdot 10^{-3}	$ & $	7.62 \cdot 10^{-3}	$ & $	6.44 \cdot 10^{-3}	$ & $	3.28 \cdot 10^{-2}	$ & $	6.86 \cdot 10^{-2}	 $ \\
$	520	$ & $	2.96 \cdot 10^{-4}	$ & $	5.92 \cdot 10^{-4}	$ & $	1.18 \cdot 10^{-3}	$ & $	2.66 \cdot 10^{-3}	$ & $	7.63 \cdot 10^{-3}	$ & $	6.45 \cdot 10^{-3}	$ & $	3.29 \cdot 10^{-2}	$ & $	6.87 \cdot 10^{-2}	 $ \\
$	530	$ & $	2.97 \cdot 10^{-4}	$ & $	5.94 \cdot 10^{-4}	$ & $	1.19 \cdot 10^{-3}	$ & $	2.67 \cdot 10^{-3}	$ & $	7.65 \cdot 10^{-3}	$ & $	6.46 \cdot 10^{-3}	$ & $	3.30 \cdot 10^{-2}	$ & $	6.88 \cdot 10^{-2}	 $ \\
$	540	$ & $	2.98 \cdot 10^{-4}	$ & $	5.97 \cdot 10^{-4}	$ & $	1.19 \cdot 10^{-3}	$ & $	2.69 \cdot 10^{-3}	$ & $	7.67 \cdot 10^{-3}	$ & $	6.48 \cdot 10^{-3}	$ & $	3.30 \cdot 10^{-2}	$ & $	6.90 \cdot 10^{-2}	 $ \\
$	550	$ & $	3.00 \cdot 10^{-4}	$ & $	6.00 \cdot 10^{-4}	$ & $	1.20 \cdot 10^{-3}	$ & $	2.70 \cdot 10^{-3}	$ & $	7.69 \cdot 10^{-3}	$ & $	6.49 \cdot 10^{-3}	$ & $	3.31 \cdot 10^{-2}	$ & $	6.92 \cdot 10^{-2}	 $ \\
$	560	$ & $	3.01 \cdot 10^{-4}	$ & $	6.03 \cdot 10^{-4}	$ & $	1.20 \cdot 10^{-3}	$ & $	2.71 \cdot 10^{-3}	$ & $	7.71 \cdot 10^{-3}	$ & $	6.51 \cdot 10^{-3}	$ & $	3.32 \cdot 10^{-2}	$ & $	6.94 \cdot 10^{-2}	 $ \\
$	570	$ & $	3.03 \cdot 10^{-4}	$ & $	6.05 \cdot 10^{-4}	$ & $	1.21 \cdot 10^{-3}	$ & $	2.72 \cdot 10^{-3}	$ & $	7.73 \cdot 10^{-3}	$ & $	6.53 \cdot 10^{-3}	$ & $	3.33 \cdot 10^{-2}	$ & $	6.96 \cdot 10^{-2}	 $ \\
$	580	$ & $	3.04 \cdot 10^{-4}	$ & $	6.08 \cdot 10^{-4}	$ & $	1.22 \cdot 10^{-3}	$ & $	2.74 \cdot 10^{-3}	$ & $	7.76 \cdot 10^{-3}	$ & $	6.55 \cdot 10^{-3}	$ & $	3.35 \cdot 10^{-2}	$ & $	6.98 \cdot 10^{-2}	 $ \\
$	590	$ & $	3.06 \cdot 10^{-4}	$ & $	6.11 \cdot 10^{-4}	$ & $	1.22 \cdot 10^{-3}	$ & $	2.75 \cdot 10^{-3}	$ & $	7.79 \cdot 10^{-3}	$ & $	6.57 \cdot 10^{-3}	$ & $	3.36 \cdot 10^{-2}	$ & $	7.01 \cdot 10^{-2}	 $ \\
$	600	$ & $	3.07 \cdot 10^{-4}	$ & $	6.14 \cdot 10^{-4}	$ & $	1.23 \cdot 10^{-3}	$ & $	2.76 \cdot 10^{-3}	$ & $	7.81 \cdot 10^{-3}	$ & $	6.59 \cdot 10^{-3}	$ & $	3.37 \cdot 10^{-2}	$ & $	7.03 \cdot 10^{-2}	 $ \\
$	610	$ & $	3.09 \cdot 10^{-4}	$ & $	6.17 \cdot 10^{-4}	$ & $	1.23 \cdot 10^{-3}	$ & $	2.78 \cdot 10^{-3}	$ & $	7.84 \cdot 10^{-3}	$ & $	6.61 \cdot 10^{-3}	$ & $	3.38 \cdot 10^{-2}	$ & $	7.06 \cdot 10^{-2}	 $ \\
$	620	$ & $	3.10 \cdot 10^{-4}	$ & $	6.20 \cdot 10^{-4}	$ & $	1.24 \cdot 10^{-3}	$ & $	2.79 \cdot 10^{-3}	$ & $	7.87 \cdot 10^{-3}	$ & $	6.63 \cdot 10^{-3}	$ & $	3.39 \cdot 10^{-2}	$ & $	7.08 \cdot 10^{-2}	 $ \\
$	630	$ & $	3.12 \cdot 10^{-4}	$ & $	6.23 \cdot 10^{-4}	$ & $	1.25 \cdot 10^{-3}	$ & $	2.80 \cdot 10^{-3}	$ & $	7.90 \cdot 10^{-3}	$ & $	6.66 \cdot 10^{-3}	$ & $	3.41 \cdot 10^{-2}	$ & $	7.11 \cdot 10^{-2}	 $ \\
$	640	$ & $	3.13 \cdot 10^{-4}	$ & $	6.26 \cdot 10^{-4}	$ & $	1.25 \cdot 10^{-3}	$ & $	2.82 \cdot 10^{-3}	$ & $	7.93 \cdot 10^{-3}	$ & $	6.68 \cdot 10^{-3}	$ & $	3.42 \cdot 10^{-2}	$ & $	7.13 \cdot 10^{-2}	 $ \\
$	650	$ & $	3.15 \cdot 10^{-4}	$ & $	6.29 \cdot 10^{-4}	$ & $	1.26 \cdot 10^{-3}	$ & $	2.83 \cdot 10^{-3}	$ & $	7.96 \cdot 10^{-3}	$ & $	6.70 \cdot 10^{-3}	$ & $	3.43 \cdot 10^{-2}	$ & $	7.16 \cdot 10^{-2}	 $ \\
\hline
  \end{tabular}
  \end{table}
\end{landscape}

\begin{landscape}
  \begin{table}
  \vspace{-\headsep}
  \caption{SM Higgs branching ratios for $4$-fermion final states in 
    the high-mass range. We list results for the specific final states
    $\Pep\Pem\Pep\Pem$ and $\Pep\Pem\PGmp\PGmm$, for final states with $4$ arbitrary charged
    leptons, $\Pep\PGne\Pem\PAGne$ and $\Pep\PGne\PGmm\PAGnGm$, and for
    final states $\Plp\Plm\PGnl\PAGnl$ with $2$ charged leptons
    plus $2$ neutrinos ($\PGnl$ represents any type of neutrinos).}
  \label{tab:PBR-hm}
  \centering
  \small
  \begin{tabular}{lcccccccc}
\hline
    $\MH$ [GeV] &
    $\PH \rightarrow \Pep\Pem\Pep\Pem$ &
    $\PH \rightarrow \Pep\Pem\PGmp\PGmm$ &
    $\PH \rightarrow \Plp\Plm\Plp\Plm$ &
    $\PH \rightarrow \Plp\Plm\Plp\Plm$ &
    $\PH \rightarrow \Pep\PGne\Pem\PAGne$ &
    $\PH \rightarrow \Pep\PGne\PGmm\PAGnGm$ &
    $\PH \rightarrow \Plp\Plm\PGnl\PAGnl$ &
    $\PH \rightarrow \Plp\Plm\PGnl\PAGnl$ \\
    & & & $(\Pl=\Pe$ or $\PGm)$ & $(\Pl=\Pe, \PGm$ or $\PGt)$
    & & & $(\Pl=\Pe$ or $\PGm)$ & $(\Pl=\Pe, \PGm$ or $\PGt)$ \\
\hline
$	660	$ & $	3.16 \cdot 10^{-4}	$ & $	6.32 \cdot 10^{-4}	$ & $	1.26 \cdot 10^{-3}	$ & $	2.84 \cdot 10^{-3}	$ & $	7.98 \cdot 10^{-3}	$ & $	6.73 \cdot 10^{-3}	$ & $	3.44 \cdot 10^{-2}	$ & $	7.18 \cdot 10^{-2}	 $ \\
$	670	$ & $	3.17 \cdot 10^{-4}	$ & $	6.35 \cdot 10^{-4}	$ & $	1.27 \cdot 10^{-3}	$ & $	2.86 \cdot 10^{-3}	$ & $	8.01 \cdot 10^{-3}	$ & $	6.75 \cdot 10^{-3}	$ & $	3.46 \cdot 10^{-2}	$ & $	7.21 \cdot 10^{-2}	 $ \\
$	680	$ & $	3.19 \cdot 10^{-4}	$ & $	6.38 \cdot 10^{-4}	$ & $	1.28 \cdot 10^{-3}	$ & $	2.87 \cdot 10^{-3}	$ & $	8.04 \cdot 10^{-3}	$ & $	6.77 \cdot 10^{-3}	$ & $	3.47 \cdot 10^{-2}	$ & $	7.24 \cdot 10^{-2}	 $ \\
$	690	$ & $	3.20 \cdot 10^{-4}	$ & $	6.41 \cdot 10^{-4}	$ & $	1.28 \cdot 10^{-3}	$ & $	2.88 \cdot 10^{-3}	$ & $	8.07 \cdot 10^{-3}	$ & $	6.80 \cdot 10^{-3}	$ & $	3.48 \cdot 10^{-2}	$ & $	7.26 \cdot 10^{-2}	 $ \\
$	700	$ & $	3.22 \cdot 10^{-4}	$ & $	6.43 \cdot 10^{-4}	$ & $	1.29 \cdot 10^{-3}	$ & $	2.90 \cdot 10^{-3}	$ & $	8.10 \cdot 10^{-3}	$ & $	6.82 \cdot 10^{-3}	$ & $	3.50 \cdot 10^{-2}	$ & $	7.29 \cdot 10^{-2}	 $ \\
$	710	$ & $	3.23 \cdot 10^{-4}	$ & $	6.46 \cdot 10^{-4}	$ & $	1.29 \cdot 10^{-3}	$ & $	2.91 \cdot 10^{-3}	$ & $	8.13 \cdot 10^{-3}	$ & $	6.84 \cdot 10^{-3}	$ & $	3.51 \cdot 10^{-2}	$ & $	7.31 \cdot 10^{-2}	 $ \\
$	720	$ & $	3.24 \cdot 10^{-4}	$ & $	6.49 \cdot 10^{-4}	$ & $	1.30 \cdot 10^{-3}	$ & $	2.92 \cdot 10^{-3}	$ & $	8.15 \cdot 10^{-3}	$ & $	6.86 \cdot 10^{-3}	$ & $	3.52 \cdot 10^{-2}	$ & $	7.34 \cdot 10^{-2}	 $ \\
$	730	$ & $	3.26 \cdot 10^{-4}	$ & $	6.51 \cdot 10^{-4}	$ & $	1.30 \cdot 10^{-3}	$ & $	2.93 \cdot 10^{-3}	$ & $	8.18 \cdot 10^{-3}	$ & $	6.88 \cdot 10^{-3}	$ & $	3.53 \cdot 10^{-2}	$ & $	7.36 \cdot 10^{-2}	 $ \\
$	740	$ & $	3.27 \cdot 10^{-4}	$ & $	6.54 \cdot 10^{-4}	$ & $	1.31 \cdot 10^{-3}	$ & $	2.94 \cdot 10^{-3}	$ & $	8.21 \cdot 10^{-3}	$ & $	6.91 \cdot 10^{-3}	$ & $	3.54 \cdot 10^{-2}	$ & $	7.39 \cdot 10^{-2}	 $ \\
$	750	$ & $	3.28 \cdot 10^{-4}	$ & $	6.57 \cdot 10^{-4}	$ & $	1.31 \cdot 10^{-3}	$ & $	2.95 \cdot 10^{-3}	$ & $	8.24 \cdot 10^{-3}	$ & $	6.93 \cdot 10^{-3}	$ & $	3.56 \cdot 10^{-2}	$ & $	7.41 \cdot 10^{-2}	 $ \\
$	760	$ & $	3.29 \cdot 10^{-4}	$ & $	6.59 \cdot 10^{-4}	$ & $	1.32 \cdot 10^{-3}	$ & $	2.97 \cdot 10^{-3}	$ & $	8.26 \cdot 10^{-3}	$ & $	6.95 \cdot 10^{-3}	$ & $	3.57 \cdot 10^{-2}	$ & $	7.44 \cdot 10^{-2}	 $ \\
$	770	$ & $	3.31 \cdot 10^{-4}	$ & $	6.62 \cdot 10^{-4}	$ & $	1.32 \cdot 10^{-3}	$ & $	2.98 \cdot 10^{-3}	$ & $	8.29 \cdot 10^{-3}	$ & $	6.97 \cdot 10^{-3}	$ & $	3.58 \cdot 10^{-2}	$ & $	7.46 \cdot 10^{-2}	 $ \\
$	780	$ & $	3.32 \cdot 10^{-4}	$ & $	6.64 \cdot 10^{-4}	$ & $	1.33 \cdot 10^{-3}	$ & $	2.99 \cdot 10^{-3}	$ & $	8.32 \cdot 10^{-3}	$ & $	6.99 \cdot 10^{-3}	$ & $	3.59 \cdot 10^{-2}	$ & $	7.48 \cdot 10^{-2}	 $ \\
$	790	$ & $	3.33 \cdot 10^{-4}	$ & $	6.66 \cdot 10^{-4}	$ & $	1.33 \cdot 10^{-3}	$ & $	3.00 \cdot 10^{-3}	$ & $	8.34 \cdot 10^{-3}	$ & $	7.01 \cdot 10^{-3}	$ & $	3.60 \cdot 10^{-2}	$ & $	7.51 \cdot 10^{-2}	 $ \\
$	800	$ & $	3.34 \cdot 10^{-4}	$ & $	6.69 \cdot 10^{-4}	$ & $	1.34 \cdot 10^{-3}	$ & $	3.01 \cdot 10^{-3}	$ & $	8.36 \cdot 10^{-3}	$ & $	7.03 \cdot 10^{-3}	$ & $	3.61 \cdot 10^{-2}	$ & $	7.53 \cdot 10^{-2}	 $ \\
$	810	$ & $	3.35 \cdot 10^{-4}	$ & $	6.71 \cdot 10^{-4}	$ & $	1.34 \cdot 10^{-3}	$ & $	3.02 \cdot 10^{-3}	$ & $	8.39 \cdot 10^{-3}	$ & $	7.05 \cdot 10^{-3}	$ & $	3.62 \cdot 10^{-2}	$ & $	7.55 \cdot 10^{-2}	 $ \\
$	820	$ & $	3.36 \cdot 10^{-4}	$ & $	6.73 \cdot 10^{-4}	$ & $	1.35 \cdot 10^{-3}	$ & $	3.03 \cdot 10^{-3}	$ & $	8.41 \cdot 10^{-3}	$ & $	7.07 \cdot 10^{-3}	$ & $	3.63 \cdot 10^{-2}	$ & $	7.57 \cdot 10^{-2}	 $ \\
$	830	$ & $	3.38 \cdot 10^{-4}	$ & $	6.75 \cdot 10^{-4}	$ & $	1.35 \cdot 10^{-3}	$ & $	3.04 \cdot 10^{-3}	$ & $	8.44 \cdot 10^{-3}	$ & $	7.09 \cdot 10^{-3}	$ & $	3.64 \cdot 10^{-2}	$ & $	7.59 \cdot 10^{-2}	 $ \\
$	840	$ & $	3.39 \cdot 10^{-4}	$ & $	6.77 \cdot 10^{-4}	$ & $	1.35 \cdot 10^{-3}	$ & $	3.05 \cdot 10^{-3}	$ & $	8.46 \cdot 10^{-3}	$ & $	7.11 \cdot 10^{-3}	$ & $	3.65 \cdot 10^{-2}	$ & $	7.61 \cdot 10^{-2}	 $ \\
$	850	$ & $	3.40 \cdot 10^{-4}	$ & $	6.79 \cdot 10^{-4}	$ & $	1.36 \cdot 10^{-3}	$ & $	3.06 \cdot 10^{-3}	$ & $	8.49 \cdot 10^{-3}	$ & $	7.13 \cdot 10^{-3}	$ & $	3.66 \cdot 10^{-2}	$ & $	7.64 \cdot 10^{-2}	 $ \\
$	860	$ & $	3.41 \cdot 10^{-4}	$ & $	6.81 \cdot 10^{-4}	$ & $	1.36 \cdot 10^{-3}	$ & $	3.07 \cdot 10^{-3}	$ & $	8.51 \cdot 10^{-3}	$ & $	7.15 \cdot 10^{-3}	$ & $	3.67 \cdot 10^{-2}	$ & $	7.66 \cdot 10^{-2}	 $ \\
$	870	$ & $	3.42 \cdot 10^{-4}	$ & $	6.83 \cdot 10^{-4}	$ & $	1.37 \cdot 10^{-3}	$ & $	3.08 \cdot 10^{-3}	$ & $	8.53 \cdot 10^{-3}	$ & $	7.17 \cdot 10^{-3}	$ & $	3.68 \cdot 10^{-2}	$ & $	7.68 \cdot 10^{-2}	 $ \\
$	880	$ & $	3.43 \cdot 10^{-4}	$ & $	6.85 \cdot 10^{-4}	$ & $	1.37 \cdot 10^{-3}	$ & $	3.08 \cdot 10^{-3}	$ & $	8.55 \cdot 10^{-3}	$ & $	7.19 \cdot 10^{-3}	$ & $	3.69 \cdot 10^{-2}	$ & $	7.70 \cdot 10^{-2}	 $ \\
$	890	$ & $	3.44 \cdot 10^{-4}	$ & $	6.87 \cdot 10^{-4}	$ & $	1.37 \cdot 10^{-3}	$ & $	3.09 \cdot 10^{-3}	$ & $	8.57 \cdot 10^{-3}	$ & $	7.20 \cdot 10^{-3}	$ & $	3.70 \cdot 10^{-2}	$ & $	7.72 \cdot 10^{-2}	 $ \\
$	900	$ & $	3.45 \cdot 10^{-4}	$ & $	6.89 \cdot 10^{-4}	$ & $	1.38 \cdot 10^{-3}	$ & $	3.10 \cdot 10^{-3}	$ & $	8.60 \cdot 10^{-3}	$ & $	7.22 \cdot 10^{-3}	$ & $	3.71 \cdot 10^{-2}	$ & $	7.74 \cdot 10^{-2}	 $ \\
$	910	$ & $	3.46 \cdot 10^{-4}	$ & $	6.91 \cdot 10^{-4}	$ & $	1.38 \cdot 10^{-3}	$ & $	3.11 \cdot 10^{-3}	$ & $	8.62 \cdot 10^{-3}	$ & $	7.24 \cdot 10^{-3}	$ & $	3.72 \cdot 10^{-2}	$ & $	7.76 \cdot 10^{-2}	 $ \\
$	920	$ & $	3.47 \cdot 10^{-4}	$ & $	6.93 \cdot 10^{-4}	$ & $	1.39 \cdot 10^{-3}	$ & $	3.12 \cdot 10^{-3}	$ & $	8.64 \cdot 10^{-3}	$ & $	7.26 \cdot 10^{-3}	$ & $	3.73 \cdot 10^{-2}	$ & $	7.77 \cdot 10^{-2}	 $ \\
$	930	$ & $	3.47 \cdot 10^{-4}	$ & $	6.95 \cdot 10^{-4}	$ & $	1.39 \cdot 10^{-3}	$ & $	3.13 \cdot 10^{-3}	$ & $	8.66 \cdot 10^{-3}	$ & $	7.28 \cdot 10^{-3}	$ & $	3.74 \cdot 10^{-2}	$ & $	7.79 \cdot 10^{-2}	 $ \\
$	940	$ & $	3.48 \cdot 10^{-4}	$ & $	6.97 \cdot 10^{-4}	$ & $	1.39 \cdot 10^{-3}	$ & $	3.13 \cdot 10^{-3}	$ & $	8.68 \cdot 10^{-3}	$ & $	7.29 \cdot 10^{-3}	$ & $	3.75 \cdot 10^{-2}	$ & $	7.81 \cdot 10^{-2}	 $ \\
$	950	$ & $	3.49 \cdot 10^{-4}	$ & $	6.98 \cdot 10^{-4}	$ & $	1.40 \cdot 10^{-3}	$ & $	3.14 \cdot 10^{-3}	$ & $	8.70 \cdot 10^{-3}	$ & $	7.31 \cdot 10^{-3}	$ & $	3.76 \cdot 10^{-2}	$ & $	7.83 \cdot 10^{-2}	 $ \\
$	960	$ & $	3.50 \cdot 10^{-4}	$ & $	7.00 \cdot 10^{-4}	$ & $	1.40 \cdot 10^{-3}	$ & $	3.15 \cdot 10^{-3}	$ & $	8.72 \cdot 10^{-3}	$ & $	7.32 \cdot 10^{-3}	$ & $	3.77 \cdot 10^{-2}	$ & $	7.85 \cdot 10^{-2}	 $ \\
$	970	$ & $	3.51 \cdot 10^{-4}	$ & $	7.02 \cdot 10^{-4}	$ & $	1.40 \cdot 10^{-3}	$ & $	3.16 \cdot 10^{-3}	$ & $	8.74 \cdot 10^{-3}	$ & $	7.34 \cdot 10^{-3}	$ & $	3.77 \cdot 10^{-2}	$ & $	7.86 \cdot 10^{-2}	 $ \\
$	980	$ & $	3.52 \cdot 10^{-4}	$ & $	7.03 \cdot 10^{-4}	$ & $	1.41 \cdot 10^{-3}	$ & $	3.17 \cdot 10^{-3}	$ & $	8.76 \cdot 10^{-3}	$ & $	7.35 \cdot 10^{-3}	$ & $	3.78 \cdot 10^{-2}	$ & $	7.88 \cdot 10^{-2}	 $ \\
$	990	$ & $	3.53 \cdot 10^{-4}	$ & $	7.05 \cdot 10^{-4}	$ & $	1.41 \cdot 10^{-3}	$ & $	3.17 \cdot 10^{-3}	$ & $	8.78 \cdot 10^{-3}	$ & $	7.37 \cdot 10^{-3}	$ & $	3.79 \cdot 10^{-2}	$ & $	7.90 \cdot 10^{-2}	 $ \\
$	1000	$ & $	3.53 \cdot 10^{-4}	$ & $	7.07 \cdot 10^{-4}	$ & $	1.41 \cdot 10^{-3}	$ & $	3.18 \cdot 10^{-3}	$ & $	8.80 \cdot 10^{-3}	$ & $	7.39 \cdot 10^{-3}	$ & $	3.80 \cdot 10^{-2}	$ & $	7.91 \cdot 10^{-2}	 $ \\
\hline
  \end{tabular}
  \end{table}
\end{landscape}

  \begin{table}[h]
  \vspace{-\headsep}
  \caption{SM Higgs branching ratios for $4$-fermion final states in 
  the low- and intermediate-mass range. We list results for the specific final states for
  $2$ charged leptons plus $2$ quarks, $\Plp\PGnl\PQq\PAQq^{\prime}$
  (not including charge conjugate state),
  $2$ neutrinos plus $2$ quarks, $4$ quarks, as well as the result for
  arbitrary $4$ fermions, where $\PQq = \PQu\PQd\PQs\PQc\PQb$ and 
  $\PGnl$ represents any type of neutrinos.}
  \label{tab:PBR-lm2}
  \centering
  \small
  \begin{tabular}{lcccccc}
\hline
    $\MH$ [GeV] &
    $\PH \rightarrow \Plp\Plm\PQq\PAQq$ &
    $\PH \rightarrow \Plp\Plm\PQq\PAQq$ &
    $\PH \rightarrow \Plp\PGnl\PQq\PAQq^{\prime}$ &
    $\PH \rightarrow \PGnl\PAGnl\PQq\PAQq$ &
    $\PH \rightarrow \PQq\PQq\PQq\PQq$ &
    $\PH \rightarrow \Pf\Pf\Pf\Pf$ \\
    & $(\Pl=\Pe$ or $\PGm)$
    & $(\Pl=\Pe, \PGm$ or $\PGt)$
    & $(\Pl=\Pe$ or $\PGm)$ & & & \\
\hline
$	90	$ & $	3.95 \cdot 10^{-5}	$ & $	5.92 \cdot 10^{-5}	$ & $	3.06 \cdot 10^{-4}	$ & $	1.19 \cdot 10^{-4}	$ & $	1.06 \cdot 10^{-3}	$ & $	2.40 \cdot 10^{-3}	$ \\
$	95	$ & $	6.27 \cdot 10^{-5}	$ & $	9.41 \cdot 10^{-5}	$ & $	6.92 \cdot 10^{-4}	$ & $	1.89 \cdot 10^{-4}	$ & $	2.34 \cdot 10^{-3}	$ & $	5.22 \cdot 10^{-3}	$ \\
$	100	$ & $	1.05 \cdot 10^{-4}	$ & $	1.58 \cdot 10^{-4}	$ & $	1.62 \cdot 10^{-3}	$ & $	3.17 \cdot 10^{-4}	$ & $	5.39 \cdot 10^{-3}	$ & $	1.20 \cdot 10^{-2}	$ \\
$	105	$ & $	2.00 \cdot 10^{-4}	$ & $	3.00 \cdot 10^{-4}	$ & $	3.56 \cdot 10^{-3}	$ & $	6.01 \cdot 10^{-4}	$ & $	1.18 \cdot 10^{-2}	$ & $	2.61 \cdot 10^{-2}	$ \\
$	110	$ & $	4.08 \cdot 10^{-4}	$ & $	6.12 \cdot 10^{-4}	$ & $	7.06 \cdot 10^{-3}	$ & $	1.22 \cdot 10^{-3}	$ & $	2.36 \cdot 10^{-2}	$ & $	5.20 \cdot 10^{-2}	$ \\
$	115	$ & $	8.13 \cdot 10^{-4}	$ & $	1.22 \cdot 10^{-3}	$ & $	1.27 \cdot 10^{-2}	$ & $	2.44 \cdot 10^{-3}	$ & $	4.31 \cdot 10^{-2}	$ & $	9.46 \cdot 10^{-2}	$ \\
$	120	$ & $	1.50 \cdot 10^{-3}	$ & $	2.24 \cdot 10^{-3}	$ & $	2.09 \cdot 10^{-2}	$ & $	4.48 \cdot 10^{-3}	$ & $	7.19 \cdot 10^{-2}	$ & $	1.58 \cdot 10^{-1}	$ \\
$	125	$ & $	2.49 \cdot 10^{-3}	$ & $	3.73 \cdot 10^{-3}	$ & $	3.17 \cdot 10^{-2}	$ & $	7.45 \cdot 10^{-3}	$ & $	1.10 \cdot 10^{-1}	$ & $	2.42 \cdot 10^{-1}	$ \\
$	130	$ & $	3.75 \cdot 10^{-3}	$ & $	5.63 \cdot 10^{-3}	$ & $	4.47 \cdot 10^{-2}	$ & $	1.12 \cdot 10^{-2}	$ & $	1.57 \cdot 10^{-1}	$ & $	3.44 \cdot 10^{-1}	$ \\
$	135	$ & $	5.15 \cdot 10^{-3}	$ & $	7.73 \cdot 10^{-3}	$ & $	5.90 \cdot 10^{-2}	$ & $	1.54 \cdot 10^{-2}	$ & $	2.09 \cdot 10^{-1}	$ & $	4.57 \cdot 10^{-1}	$ \\
$	140	$ & $	6.48 \cdot 10^{-3}	$ & $	9.71 \cdot 10^{-3}	$ & $	7.38 \cdot 10^{-2}	$ & $	1.93 \cdot 10^{-2}	$ & $	2.62 \cdot 10^{-1}	$ & $	5.72 \cdot 10^{-1}	$ \\
$	145	$ & $	7.46 \cdot 10^{-3}	$ & $	1.12 \cdot 10^{-2}	$ & $	8.83 \cdot 10^{-2}	$ & $	2.22 \cdot 10^{-2}	$ & $	3.12 \cdot 10^{-1}	$ & $	6.81 \cdot 10^{-1}	$ \\
$	150	$ & $	7.76 \cdot 10^{-3}	$ & $	1.16 \cdot 10^{-2}	$ & $	1.02 \cdot 10^{-1}	$ & $	2.32 \cdot 10^{-2}	$ & $	3.57 \cdot 10^{-1}	$ & $	7.80 \cdot 10^{-1}	$ \\
$	155	$ & $	6.90 \cdot 10^{-3}	$ & $	1.03 \cdot 10^{-2}	$ & $	1.17 \cdot 10^{-1}	$ & $	2.06 \cdot 10^{-2}	$ & $	3.97 \cdot 10^{-1}	$ & $	8.69 \cdot 10^{-1}	$ \\
$	160	$ & $	3.89 \cdot 10^{-3}	$ & $	5.84 \cdot 10^{-3}	$ & $	1.33 \cdot 10^{-1}	$ & $	1.16 \cdot 10^{-2}	$ & $	4.33 \cdot 10^{-1}	$ & $	9.50 \cdot 10^{-1}	$ \\
$	165	$ & $	2.08 \cdot 10^{-3}	$ & $	3.12 \cdot 10^{-3}	$ & $	1.41 \cdot 10^{-1}	$ & $	6.21 \cdot 10^{-3}	$ & $	4.47 \cdot 10^{-1}	$ & $	9.82 \cdot 10^{-1}	$ \\
$	170	$ & $	2.21 \cdot 10^{-3}	$ & $	3.32 \cdot 10^{-3}	$ & $	1.41 \cdot 10^{-1}	$ & $	6.61 \cdot 10^{-3}	$ & $	4.50 \cdot 10^{-1}	$ & $	9.88 \cdot 10^{-1}	$ \\
$	175	$ & $	3.03 \cdot 10^{-3}	$ & $	4.54 \cdot 10^{-3}	$ & $	1.40 \cdot 10^{-1}	$ & $	9.05 \cdot 10^{-3}	$ & $	4.51 \cdot 10^{-1}	$ & $	9.91 \cdot 10^{-1}	$ \\
$	180	$ & $	5.65 \cdot 10^{-3}	$ & $	8.47 \cdot 10^{-3}	$ & $	1.37 \cdot 10^{-1}	$ & $	1.69 \cdot 10^{-2}	$ & $	4.53 \cdot 10^{-1}	$ & $	9.92 \cdot 10^{-1}	$ \\
$	185	$ & $	1.41 \cdot 10^{-2}	$ & $	2.11 \cdot 10^{-2}	$ & $	1.24 \cdot 10^{-1}	$ & $	4.21 \cdot 10^{-2}	$ & $	4.57 \cdot 10^{-1}	$ & $	9.94 \cdot 10^{-1}	$ \\
$	190	$ & $	1.97 \cdot 10^{-2}	$ & $	2.95 \cdot 10^{-2}	$ & $	1.15 \cdot 10^{-1}	$ & $	5.87 \cdot 10^{-2}	$ & $	4.59 \cdot 10^{-1}	$ & $	9.95 \cdot 10^{-1}	$ \\
$	195	$ & $	2.24 \cdot 10^{-2}	$ & $	3.36 \cdot 10^{-2}	$ & $	1.11 \cdot 10^{-1}	$ & $	6.69 \cdot 10^{-2}	$ & $	4.60 \cdot 10^{-1}	$ & $	9.96 \cdot 10^{-1}	$ \\
$	200	$ & $	2.40 \cdot 10^{-2}	$ & $	3.60 \cdot 10^{-2}	$ & $	1.08 \cdot 10^{-1}	$ & $	7.16 \cdot 10^{-2}	$ & $	4.61 \cdot 10^{-1}	$ & $	9.96 \cdot 10^{-1}	$ \\
$	210	$ & $	2.57 \cdot 10^{-2}	$ & $	3.86 \cdot 10^{-2}	$ & $	1.06 \cdot 10^{-1}	$ & $	7.68 \cdot 10^{-2}	$ & $	4.62 \cdot 10^{-1}	$ & $	9.97 \cdot 10^{-1}	$ \\
$	220	$ & $	2.66 \cdot 10^{-2}	$ & $	3.99 \cdot 10^{-2}	$ & $	1.05 \cdot 10^{-1}	$ & $	7.95 \cdot 10^{-2}	$ & $	4.63 \cdot 10^{-1}	$ & $	9.97 \cdot 10^{-1}	$ \\
$	230	$ & $	2.72 \cdot 10^{-2}	$ & $	4.07 \cdot 10^{-2}	$ & $	1.04 \cdot 10^{-1}	$ & $	8.12 \cdot 10^{-2}	$ & $	4.63 \cdot 10^{-1}	$ & $	9.98 \cdot 10^{-1}	$ \\
$	240	$ & $	2.76 \cdot 10^{-2}	$ & $	4.13 \cdot 10^{-2}	$ & $	1.03 \cdot 10^{-1}	$ & $	8.23 \cdot 10^{-2}	$ & $	4.63 \cdot 10^{-1}	$ & $	9.98 \cdot 10^{-1}	$ \\
$	250	$ & $	2.78 \cdot 10^{-2}	$ & $	4.18 \cdot 10^{-2}	$ & $	1.03 \cdot 10^{-1}	$ & $	8.32 \cdot 10^{-2}	$ & $	4.64 \cdot 10^{-1}	$ & $	9.98 \cdot 10^{-1}	$ \\
$	260	$ & $	2.81 \cdot 10^{-2}	$ & $	4.21 \cdot 10^{-2}	$ & $	1.02 \cdot 10^{-1}	$ & $	8.39 \cdot 10^{-2}	$ & $	4.64 \cdot 10^{-1}	$ & $	9.98 \cdot 10^{-1}	$ \\
$	270	$ & $	2.83 \cdot 10^{-2}	$ & $	4.24 \cdot 10^{-2}	$ & $	1.02 \cdot 10^{-1}	$ & $	8.45 \cdot 10^{-2}	$ & $	4.64 \cdot 10^{-1}	$ & $	9.98 \cdot 10^{-1}	$ \\
$	280	$ & $	2.85 \cdot 10^{-2}	$ & $	4.27 \cdot 10^{-2}	$ & $	1.02 \cdot 10^{-1}	$ & $	8.51 \cdot 10^{-2}	$ & $	4.64 \cdot 10^{-1}	$ & $	9.98 \cdot 10^{-1}	$ \\
$	290	$ & $	2.86 \cdot 10^{-2}	$ & $	4.30 \cdot 10^{-2}	$ & $	1.02 \cdot 10^{-1}	$ & $	8.56 \cdot 10^{-2}	$ & $	4.64 \cdot 10^{-1}	$ & $	9.99 \cdot 10^{-1}	$ \\
$	300	$ & $	2.88 \cdot 10^{-2}	$ & $	4.32 \cdot 10^{-2}	$ & $	1.01 \cdot 10^{-1}	$ & $	8.60 \cdot 10^{-2}	$ & $	4.64 \cdot 10^{-1}	$ & $	9.99 \cdot 10^{-1}	$ \\
$	310	$ & $	2.89 \cdot 10^{-2}	$ & $	4.34 \cdot 10^{-2}	$ & $	1.01 \cdot 10^{-1}	$ & $	8.64 \cdot 10^{-2}	$ & $	4.64 \cdot 10^{-1}	$ & $	9.99 \cdot 10^{-1}	$ \\
$	320	$ & $	2.90 \cdot 10^{-2}	$ & $	4.35 \cdot 10^{-2}	$ & $	1.01 \cdot 10^{-1}	$ & $	8.67 \cdot 10^{-2}	$ & $	4.64 \cdot 10^{-1}	$ & $	9.98 \cdot 10^{-1}	$ \\
$	330	$ & $	2.91 \cdot 10^{-2}	$ & $	4.37 \cdot 10^{-2}	$ & $	1.01 \cdot 10^{-1}	$ & $	8.70 \cdot 10^{-2}	$ & $	4.64 \cdot 10^{-1}	$ & $	9.98 \cdot 10^{-1}	$ \\
$	340	$ & $	2.92 \cdot 10^{-2}	$ & $	4.37 \cdot 10^{-2}	$ & $	1.01 \cdot 10^{-1}	$ & $	8.72 \cdot 10^{-2}	$ & $	4.64 \cdot 10^{-1}	$ & $	9.98 \cdot 10^{-1}	$ \\
$	350	$ & $	2.88 \cdot 10^{-2}	$ & $	4.32 \cdot 10^{-2}	$ & $	9.90 \cdot 10^{-2}	$ & $	8.61 \cdot 10^{-2}	$ & $	4.57 \cdot 10^{-1}	$ & $	9.83 \cdot 10^{-1}	$ \\
$	360	$ & $	2.78 \cdot 10^{-2}	$ & $	4.17 \cdot 10^{-2}	$ & $	9.53 \cdot 10^{-2}	$ & $	8.31 \cdot 10^{-2}	$ & $	4.41 \cdot 10^{-1}	$ & $	9.47 \cdot 10^{-1}	$ \\
$	370	$ & $	2.69 \cdot 10^{-2}	$ & $	4.04 \cdot 10^{-2}	$ & $	9.20 \cdot 10^{-2}	$ & $	8.05 \cdot 10^{-2}	$ & $	4.26 \cdot 10^{-1}	$ & $	9.15 \cdot 10^{-1}	$ \\
$	380	$ & $	2.62 \cdot 10^{-2}	$ & $	3.93 \cdot 10^{-2}	$ & $	8.92 \cdot 10^{-2}	$ & $	7.83 \cdot 10^{-2}	$ & $	4.13 \cdot 10^{-1}	$ & $	8.89 \cdot 10^{-1}	$ \\
$	390	$ & $	2.57 \cdot 10^{-2}	$ & $	3.85 \cdot 10^{-2}	$ & $	8.70 \cdot 10^{-2}	$ & $	7.66 \cdot 10^{-2}	$ & $	4.03 \cdot 10^{-1}	$ & $	8.67 \cdot 10^{-1}	$ \\
$	400	$ & $	2.52 \cdot 10^{-2}	$ & $	3.78 \cdot 10^{-2}	$ & $	8.52 \cdot 10^{-2}	$ & $	7.54 \cdot 10^{-2}	$ & $	3.96 \cdot 10^{-1}	$ & $	8.50 \cdot 10^{-1}	$ \\
$	410	$ & $	2.49 \cdot 10^{-2}	$ & $	3.74 \cdot 10^{-2}	$ & $	8.38 \cdot 10^{-2}	$ & $	7.44 \cdot 10^{-2}	$ & $	3.90 \cdot 10^{-1}	$ & $	8.37 \cdot 10^{-1}	$ \\
$	420	$ & $	2.47 \cdot 10^{-2}	$ & $	3.70 \cdot 10^{-2}	$ & $	8.27 \cdot 10^{-2}	$ & $	7.37 \cdot 10^{-2}	$ & $	3.85 \cdot 10^{-1}	$ & $	8.27 \cdot 10^{-1}	$ \\
$	430	$ & $	2.45 \cdot 10^{-2}	$ & $	3.68 \cdot 10^{-2}	$ & $	8.18 \cdot 10^{-2}	$ & $	7.32 \cdot 10^{-2}	$ & $	3.81 \cdot 10^{-1}	$ & $	8.20 \cdot 10^{-1}	$ \\
$	440	$ & $	2.44 \cdot 10^{-2}	$ & $	3.66 \cdot 10^{-2}	$ & $	8.12 \cdot 10^{-2}	$ & $	7.29 \cdot 10^{-2}	$ & $	3.79 \cdot 10^{-1}	$ & $	8.14 \cdot 10^{-1}	$ \\
$	450	$ & $	2.43 \cdot 10^{-2}	$ & $	3.65 \cdot 10^{-2}	$ & $	8.07 \cdot 10^{-2}	$ & $	7.27 \cdot 10^{-2}	$ & $	3.77 \cdot 10^{-1}	$ & $	8.10 \cdot 10^{-1}	$ \\
$	460	$ & $	2.43 \cdot 10^{-2}	$ & $	3.65 \cdot 10^{-2}	$ & $	8.04 \cdot 10^{-2}	$ & $	7.26 \cdot 10^{-2}	$ & $	3.76 \cdot 10^{-1}	$ & $	8.08 \cdot 10^{-1}	$ \\
$	470	$ & $	2.43 \cdot 10^{-2}	$ & $	3.65 \cdot 10^{-2}	$ & $	8.01 \cdot 10^{-2}	$ & $	7.26 \cdot 10^{-2}	$ & $	3.75 \cdot 10^{-1}	$ & $	8.06 \cdot 10^{-1}	$ \\
$	480	$ & $	2.43 \cdot 10^{-2}	$ & $	3.65 \cdot 10^{-2}	$ & $	8.00 \cdot 10^{-2}	$ & $	7.27 \cdot 10^{-2}	$ & $	3.75 \cdot 10^{-1}	$ & $	8.06 \cdot 10^{-1}	$ \\
$	490	$ & $	2.44 \cdot 10^{-2}	$ & $	3.66 \cdot 10^{-2}	$ & $	7.99 \cdot 10^{-2}	$ & $	7.29 \cdot 10^{-2}	$ & $	3.75 \cdot 10^{-1}	$ & $	8.06 \cdot 10^{-1}	$ \\
\hline
  \end{tabular}
  \end{table}

  \begin{table}[h]
  \vspace{-\headsep}
  \caption{SM Higgs branching ratios for $4$-fermion final states in 
  the high-mass range. We list results for the specific final states for
  $2$ charged leptons plus $2$ quarks, $\Plp\PGnl\PQq\PAQq^{\prime}$
  (not including charge conjugate state),
  $2$ neutrinos plus $2$ quarks, $4$ quarks, as well as the result for
  arbitrary $4$ fermions, where $\PQq = \PQu\PQd\PQs\PQc\PQb$ and 
  $\PGnl$ represents any type of neutrinos.}
  \label{tab:PBR-hm2}
  \centering
  \small
  \begin{tabular}{lcccccc}
\hline
    $\MH$ [GeV] &
    $\PH \rightarrow \Plp\Plm\PQq\PAQq$ &
    $\PH \rightarrow \Plp\Plm\PQq\PAQq$ &
    $\PH \rightarrow \Plp\PGnl\PQq\PAQq^{\prime}$ &
    $\PH \rightarrow \PGnl\PAGnl\PQq\PAQq$ &
    $\PH \rightarrow \PQq\PQq\PQq\PQq$ &
    $\PH \rightarrow \Pf\Pf\Pf\Pf$ \\
    & $(\Pl=\Pe$ or $\PGm)$
    & $(\Pl=\Pe, \PGm$ or $\PGt)$
    & $(\Pl=\Pe$ or $\PGm)$ & & & \\
\hline
$	500	$ & $	2.45 \cdot 10^{-2}	$ & $	3.67 \cdot 10^{-2}	$ & $	7.99 \cdot 10^{-2}	$ & $	7.31 \cdot 10^{-2}	$ & $	3.75 \cdot 10^{-1}	$ & $	8.06 \cdot 10^{-1}	$ \\
$	510	$ & $	2.45 \cdot 10^{-2}	$ & $	3.68 \cdot 10^{-2}	$ & $	8.00 \cdot 10^{-2}	$ & $	7.33 \cdot 10^{-2}	$ & $	3.76 \cdot 10^{-1}	$ & $	8.08 \cdot 10^{-1}	$ \\
$	520	$ & $	2.46 \cdot 10^{-2}	$ & $	3.70 \cdot 10^{-2}	$ & $	8.01 \cdot 10^{-2}	$ & $	7.36 \cdot 10^{-2}	$ & $	3.76 \cdot 10^{-1}	$ & $	8.09 \cdot 10^{-1}	$ \\
$	530	$ & $	2.47 \cdot 10^{-2}	$ & $	3.71 \cdot 10^{-2}	$ & $	8.02 \cdot 10^{-2}	$ & $	7.39 \cdot 10^{-2}	$ & $	3.77 \cdot 10^{-1}	$ & $	8.11 \cdot 10^{-1}	$ \\
$	540	$ & $	2.48 \cdot 10^{-2}	$ & $	3.73 \cdot 10^{-2}	$ & $	8.04 \cdot 10^{-2}	$ & $	7.42 \cdot 10^{-2}	$ & $	3.78 \cdot 10^{-1}	$ & $	8.13 \cdot 10^{-1}	$ \\
$	550	$ & $	2.50 \cdot 10^{-2}	$ & $	3.74 \cdot 10^{-2}	$ & $	8.06 \cdot 10^{-2}	$ & $	7.45 \cdot 10^{-2}	$ & $	3.80 \cdot 10^{-1}	$ & $	8.16 \cdot 10^{-1}	$ \\
$	560	$ & $	2.51 \cdot 10^{-2}	$ & $	3.76 \cdot 10^{-2}	$ & $	8.08 \cdot 10^{-2}	$ & $	7.48 \cdot 10^{-2}	$ & $	3.81 \cdot 10^{-1}	$ & $	8.18 \cdot 10^{-1}	$ \\
$	570	$ & $	2.52 \cdot 10^{-2}	$ & $	3.78 \cdot 10^{-2}	$ & $	8.10 \cdot 10^{-2}	$ & $	7.52 \cdot 10^{-2}	$ & $	3.82 \cdot 10^{-1}	$ & $	8.21 \cdot 10^{-1}	$ \\
$	580	$ & $	2.53 \cdot 10^{-2}	$ & $	3.80 \cdot 10^{-2}	$ & $	8.13 \cdot 10^{-2}	$ & $	7.55 \cdot 10^{-2}	$ & $	3.83 \cdot 10^{-1}	$ & $	8.24 \cdot 10^{-1}	$ \\
$	590	$ & $	2.54 \cdot 10^{-2}	$ & $	3.81 \cdot 10^{-2}	$ & $	8.15 \cdot 10^{-2}	$ & $	7.59 \cdot 10^{-2}	$ & $	3.85 \cdot 10^{-1}	$ & $	8.27 \cdot 10^{-1}	$ \\
$	600	$ & $	2.55 \cdot 10^{-2}	$ & $	3.83 \cdot 10^{-2}	$ & $	8.18 \cdot 10^{-2}	$ & $	7.63 \cdot 10^{-2}	$ & $	3.86 \cdot 10^{-1}	$ & $	8.30 \cdot 10^{-1}	$ \\
$	610	$ & $	2.57 \cdot 10^{-2}	$ & $	3.85 \cdot 10^{-2}	$ & $	8.21 \cdot 10^{-2}	$ & $	7.66 \cdot 10^{-2}	$ & $	3.88 \cdot 10^{-1}	$ & $	8.33 \cdot 10^{-1}	$ \\
$	620	$ & $	2.58 \cdot 10^{-2}	$ & $	3.87 \cdot 10^{-2}	$ & $	8.23 \cdot 10^{-2}	$ & $	7.70 \cdot 10^{-2}	$ & $	3.89 \cdot 10^{-1}	$ & $	8.37 \cdot 10^{-1}	$ \\
$	630	$ & $	2.59 \cdot 10^{-2}	$ & $	3.89 \cdot 10^{-2}	$ & $	8.26 \cdot 10^{-2}	$ & $	7.74 \cdot 10^{-2}	$ & $	3.91 \cdot 10^{-1}	$ & $	8.40 \cdot 10^{-1}	$ \\
$	640	$ & $	2.60 \cdot 10^{-2}	$ & $	3.91 \cdot 10^{-2}	$ & $	8.29 \cdot 10^{-2}	$ & $	7.77 \cdot 10^{-2}	$ & $	3.92 \cdot 10^{-1}	$ & $	8.43 \cdot 10^{-1}	$ \\
$	650	$ & $	2.62 \cdot 10^{-2}	$ & $	3.92 \cdot 10^{-2}	$ & $	8.32 \cdot 10^{-2}	$ & $	7.81 \cdot 10^{-2}	$ & $	3.94 \cdot 10^{-1}	$ & $	8.46 \cdot 10^{-1}	$ \\
$	660	$ & $	2.63 \cdot 10^{-2}	$ & $	3.94 \cdot 10^{-2}	$ & $	8.35 \cdot 10^{-2}	$ & $	7.84 \cdot 10^{-2}	$ & $	3.95 \cdot 10^{-1}	$ & $	8.49 \cdot 10^{-1}	$ \\
$	670	$ & $	2.64 \cdot 10^{-2}	$ & $	3.96 \cdot 10^{-2}	$ & $	8.37 \cdot 10^{-2}	$ & $	7.88 \cdot 10^{-2}	$ & $	3.96 \cdot 10^{-1}	$ & $	8.52 \cdot 10^{-1}	$ \\
$	680	$ & $	2.65 \cdot 10^{-2}	$ & $	3.98 \cdot 10^{-2}	$ & $	8.40 \cdot 10^{-2}	$ & $	7.91 \cdot 10^{-2}	$ & $	3.98 \cdot 10^{-1}	$ & $	8.55 \cdot 10^{-1}	$ \\
$	690	$ & $	2.66 \cdot 10^{-2}	$ & $	3.99 \cdot 10^{-2}	$ & $	8.43 \cdot 10^{-2}	$ & $	7.95 \cdot 10^{-2}	$ & $	3.99 \cdot 10^{-1}	$ & $	8.59 \cdot 10^{-1}	$ \\
$	700	$ & $	2.67 \cdot 10^{-2}	$ & $	4.01 \cdot 10^{-2}	$ & $	8.46 \cdot 10^{-2}	$ & $	7.98 \cdot 10^{-2}	$ & $	4.01 \cdot 10^{-1}	$ & $	8.62 \cdot 10^{-1}	$ \\
$	710	$ & $	2.69 \cdot 10^{-2}	$ & $	4.03 \cdot 10^{-2}	$ & $	8.48 \cdot 10^{-2}	$ & $	8.01 \cdot 10^{-2}	$ & $	4.02 \cdot 10^{-1}	$ & $	8.65 \cdot 10^{-1}	$ \\
$	720	$ & $	2.70 \cdot 10^{-2}	$ & $	4.04 \cdot 10^{-2}	$ & $	8.51 \cdot 10^{-2}	$ & $	8.05 \cdot 10^{-2}	$ & $	4.04 \cdot 10^{-1}	$ & $	8.68 \cdot 10^{-1}	$ \\
$	730	$ & $	2.71 \cdot 10^{-2}	$ & $	4.06 \cdot 10^{-2}	$ & $	8.54 \cdot 10^{-2}	$ & $	8.08 \cdot 10^{-2}	$ & $	4.05 \cdot 10^{-1}	$ & $	8.70 \cdot 10^{-1}	$ \\
$	740	$ & $	2.72 \cdot 10^{-2}	$ & $	4.08 \cdot 10^{-2}	$ & $	8.56 \cdot 10^{-2}	$ & $	8.11 \cdot 10^{-2}	$ & $	4.06 \cdot 10^{-1}	$ & $	8.73 \cdot 10^{-1}	$ \\
$	750	$ & $	2.73 \cdot 10^{-2}	$ & $	4.09 \cdot 10^{-2}	$ & $	8.59 \cdot 10^{-2}	$ & $	8.14 \cdot 10^{-2}	$ & $	4.08 \cdot 10^{-1}	$ & $	8.76 \cdot 10^{-1}	$ \\
$	760	$ & $	2.74 \cdot 10^{-2}	$ & $	4.11 \cdot 10^{-2}	$ & $	8.61 \cdot 10^{-2}	$ & $	8.17 \cdot 10^{-2}	$ & $	4.09 \cdot 10^{-1}	$ & $	8.79 \cdot 10^{-1}	$ \\
$	770	$ & $	2.75 \cdot 10^{-2}	$ & $	4.12 \cdot 10^{-2}	$ & $	8.64 \cdot 10^{-2}	$ & $	8.20 \cdot 10^{-2}	$ & $	4.10 \cdot 10^{-1}	$ & $	8.82 \cdot 10^{-1}	$ \\
$	780	$ & $	2.76 \cdot 10^{-2}	$ & $	4.14 \cdot 10^{-2}	$ & $	8.66 \cdot 10^{-2}	$ & $	8.23 \cdot 10^{-2}	$ & $	4.11 \cdot 10^{-1}	$ & $	8.85 \cdot 10^{-1}	$ \\
$	790	$ & $	2.77 \cdot 10^{-2}	$ & $	4.15 \cdot 10^{-2}	$ & $	8.69 \cdot 10^{-2}	$ & $	8.26 \cdot 10^{-2}	$ & $	4.12 \cdot 10^{-1}	$ & $	8.87 \cdot 10^{-1}	$ \\
$	800	$ & $	2.78 \cdot 10^{-2}	$ & $	4.17 \cdot 10^{-2}	$ & $	8.71 \cdot 10^{-2}	$ & $	8.28 \cdot 10^{-2}	$ & $	4.14 \cdot 10^{-1}	$ & $	8.90 \cdot 10^{-1}	$ \\
$	810	$ & $	2.79 \cdot 10^{-2}	$ & $	4.18 \cdot 10^{-2}	$ & $	8.74 \cdot 10^{-2}	$ & $	8.31 \cdot 10^{-2}	$ & $	4.15 \cdot 10^{-1}	$ & $	8.92 \cdot 10^{-1}	$ \\
$	820	$ & $	2.79 \cdot 10^{-2}	$ & $	4.19 \cdot 10^{-2}	$ & $	8.76 \cdot 10^{-2}	$ & $	8.34 \cdot 10^{-2}	$ & $	4.16 \cdot 10^{-1}	$ & $	8.95 \cdot 10^{-1}	$ \\
$	830	$ & $	2.80 \cdot 10^{-2}	$ & $	4.21 \cdot 10^{-2}	$ & $	8.78 \cdot 10^{-2}	$ & $	8.36 \cdot 10^{-2}	$ & $	4.17 \cdot 10^{-1}	$ & $	8.97 \cdot 10^{-1}	$ \\
$	840	$ & $	2.81 \cdot 10^{-2}	$ & $	4.22 \cdot 10^{-2}	$ & $	8.81 \cdot 10^{-2}	$ & $	8.39 \cdot 10^{-2}	$ & $	4.18 \cdot 10^{-1}	$ & $	9.00 \cdot 10^{-1}	$ \\
$	850	$ & $	2.82 \cdot 10^{-2}	$ & $	4.23 \cdot 10^{-2}	$ & $	8.83 \cdot 10^{-2}	$ & $	8.41 \cdot 10^{-2}	$ & $	4.19 \cdot 10^{-1}	$ & $	9.02 \cdot 10^{-1}	$ \\
$	860	$ & $	2.83 \cdot 10^{-2}	$ & $	4.24 \cdot 10^{-2}	$ & $	8.85 \cdot 10^{-2}	$ & $	8.44 \cdot 10^{-2}	$ & $	4.20 \cdot 10^{-1}	$ & $	9.04 \cdot 10^{-1}	$ \\
$	870	$ & $	2.84 \cdot 10^{-2}	$ & $	4.26 \cdot 10^{-2}	$ & $	8.87 \cdot 10^{-2}	$ & $	8.46 \cdot 10^{-2}	$ & $	4.21 \cdot 10^{-1}	$ & $	9.07 \cdot 10^{-1}	$ \\
$	880	$ & $	2.84 \cdot 10^{-2}	$ & $	4.27 \cdot 10^{-2}	$ & $	8.89 \cdot 10^{-2}	$ & $	8.48 \cdot 10^{-2}	$ & $	4.22 \cdot 10^{-1}	$ & $	9.09 \cdot 10^{-1}	$ \\
$	890	$ & $	2.85 \cdot 10^{-2}	$ & $	4.28 \cdot 10^{-2}	$ & $	8.91 \cdot 10^{-2}	$ & $	8.51 \cdot 10^{-2}	$ & $	4.23 \cdot 10^{-1}	$ & $	9.11 \cdot 10^{-1}	$ \\
$	900	$ & $	2.86 \cdot 10^{-2}	$ & $	4.29 \cdot 10^{-2}	$ & $	8.94 \cdot 10^{-2}	$ & $	8.53 \cdot 10^{-2}	$ & $	4.24 \cdot 10^{-1}	$ & $	9.13 \cdot 10^{-1}	$ \\
$	910	$ & $	2.87 \cdot 10^{-2}	$ & $	4.30 \cdot 10^{-2}	$ & $	8.96 \cdot 10^{-2}	$ & $	8.55 \cdot 10^{-2}	$ & $	4.25 \cdot 10^{-1}	$ & $	9.15 \cdot 10^{-1}	$ \\
$	920	$ & $	2.87 \cdot 10^{-2}	$ & $	4.31 \cdot 10^{-2}	$ & $	8.97 \cdot 10^{-2}	$ & $	8.57 \cdot 10^{-2}	$ & $	4.26 \cdot 10^{-1}	$ & $	9.17 \cdot 10^{-1}	$ \\
$	930	$ & $	2.88 \cdot 10^{-2}	$ & $	4.32 \cdot 10^{-2}	$ & $	8.99 \cdot 10^{-2}	$ & $	8.59 \cdot 10^{-2}	$ & $	4.27 \cdot 10^{-1}	$ & $	9.19 \cdot 10^{-1}	$ \\
$	940	$ & $	2.89 \cdot 10^{-2}	$ & $	4.33 \cdot 10^{-2}	$ & $	9.01 \cdot 10^{-2}	$ & $	8.61 \cdot 10^{-2}	$ & $	4.28 \cdot 10^{-1}	$ & $	9.21 \cdot 10^{-1}	$ \\
$	950	$ & $	2.90 \cdot 10^{-2}	$ & $	4.34 \cdot 10^{-2}	$ & $	9.03 \cdot 10^{-2}	$ & $	8.63 \cdot 10^{-2}	$ & $	4.29 \cdot 10^{-1}	$ & $	9.23 \cdot 10^{-1}	$ \\
$	960	$ & $	2.90 \cdot 10^{-2}	$ & $	4.35 \cdot 10^{-2}	$ & $	9.05 \cdot 10^{-2}	$ & $	8.65 \cdot 10^{-2}	$ & $	4.30 \cdot 10^{-1}	$ & $	9.25 \cdot 10^{-1}	$ \\
$	970	$ & $	2.91 \cdot 10^{-2}	$ & $	4.36 \cdot 10^{-2}	$ & $	9.07 \cdot 10^{-2}	$ & $	8.67 \cdot 10^{-2}	$ & $	4.30 \cdot 10^{-1}	$ & $	9.27 \cdot 10^{-1}	$ \\
$	980	$ & $	2.92 \cdot 10^{-2}	$ & $	4.37 \cdot 10^{-2}	$ & $	9.09 \cdot 10^{-2}	$ & $	8.69 \cdot 10^{-2}	$ & $	4.31 \cdot 10^{-1}	$ & $	9.29 \cdot 10^{-1}	$ \\
$	990	$ & $	2.92 \cdot 10^{-2}	$ & $	4.38 \cdot 10^{-2}	$ & $	9.10 \cdot 10^{-2}	$ & $	8.71 \cdot 10^{-2}	$ & $	4.32 \cdot 10^{-1}	$ & $	9.31 \cdot 10^{-1}	$ \\
$	1000	$ & $	2.93 \cdot 10^{-2}	$ & $	4.39 \cdot 10^{-2}	$ & $	9.12 \cdot 10^{-2}	$ & $	8.73 \cdot 10^{-2}	$ & $	4.33 \cdot 10^{-1}	$ & $	9.32 \cdot 10^{-1}	$ \\
\hline
  \end{tabular}
  \end{table}


\begin{table}
  \caption{MSSM Higgs branching ratio to $\PGtp\PGtm$ final states in the
$m_{\Ph}^{\rm max}$ scenario as a function of $\MA$ [GeV] and $\tb$.
The format in each cell is
$M_{\Ph}$ [GeV], BR($\Ph \to \PGtp\PGtm$) (upper line),
$M_{\PH}$ [GeV], BR($\PH \to \PGtp\PGtm$) (middle line),
$M_{\PA}$ [GeV], BR($\PA \to \PGtp\PGtm$) (lower line).
}
  \label{tab:BR-mssm}
  \centering
{
  \setlength{\tabcolsep}{3pt}
  \small
  \begin{tabular}{l|rc|rc|rc|rc|rc}\hline
{\small $\MA$} & \multicolumn{2}{c}{$\tb = 20$}
      & \multicolumn{2}{|c}{$\tb = 30$}
            & \multicolumn{2}{|c}{$\tb = 40$}
            & \multicolumn{2}{|c}{$\tb = 50$}
            & \multicolumn{2}{|c}{$\tb = 60$} \\
\hline\
$90 $&$   89.6 $&$ 1.09\cdot 10^{-1}  $&$  89.8 $&$ 1.17\cdot 10^{-1}  $&$  89.9 $&$ 1.25\cdot 10^{-1}  $&$
       89.9 $&$ 1.34\cdot 10^{-1}  $&$  89.9 $&$ 1.44\cdot 10^{-1}  $\\
   &$  130.6 $&$ 9.63\cdot 10^{-2}  $&$ 130.9 $&$ 1.05\cdot 10^{-1}  $&$ 130.6 $&$ 1.15\cdot 10^{-1}  $&$
      130.7 $&$ 1.20\cdot 10^{-1}  $&$ 130.9 $&$ 1.36\cdot 10^{-1}  $\\
   &$   90.0 $&$ 1.09\cdot 10^{-1}  $&$  90.0 $&$ 1.16\cdot 10^{-1}  $&$  90.0 $&$ 1.25\cdot 10^{-1}  $&$
       90.0 $&$ 1.33\cdot 10^{-1}  $&$  90.0 $&$ 1.43\cdot 10^{-1}  $\\
\hline
$100 $&$  99.4 $&$ 1.11\cdot 10^{-1}  $&$  99.7 $&$ 1.19\cdot 10^{-1}  $&$  99.9 $&$ 1.28\cdot 10^{-1}  $&$
       99.9 $&$ 1.37\cdot 10^{-1}  $&$  99.9 $&$ 1.47\cdot 10^{-1}  $\\
    &$ 130.8 $&$ 1.06\cdot 10^{-1}  $&$ 130.6 $&$ 1.16\cdot 10^{-1}  $&$ 130.6 $&$ 1.27\cdot 10^{-1}  $&$
      130.7 $&$ 1.38\cdot 10^{-1}  $&$ 130.9 $&$ 1.50\cdot 10^{-1}  $\\
    &$ 100.0 $&$ 1.11\cdot 10^{-1}  $&$ 100.0 $&$ 1.18\cdot 10^{-1}  $&$ 100.0 $&$ 1.27\cdot 10^{-1}  $&$
      100.0 $&$ 1.36\cdot 10^{-1}  $&$ 100.0 $&$ 1.45\cdot 10^{-1}  $\\
\hline
$110 $&$ 109.0 $&$ 1.13\cdot 10^{-1}  $&$ 109.6 $&$ 1.21\cdot 10^{-1}  $&$ 109.8 $&$ 1.30\cdot 10^{-1}  $&$
      109.8 $&$ 1.39\cdot 10^{-1}  $&$ 109.9 $&$ 1.50\cdot 10^{-1}  $\\
    &$ 131.1 $&$ 1.13\cdot 10^{-1}  $&$ 130.7 $&$ 1.23\cdot 10^{-1}  $&$ 130.7 $&$ 1.33\cdot 10^{-1}  $&$
      130.8 $&$ 1.45\cdot 10^{-1}  $&$ 130.9 $&$ 1.57\cdot 10^{-1}  $\\
    &$ 110.0 $&$ 1.12\cdot 10^{-1}  $&$ 110.0 $&$ 1.20\cdot 10^{-1}  $&$ 110.0 $&$ 1.29\cdot 10^{-1}  $&$
      110.0 $&$ 1.37\cdot 10^{-1}  $&$ 110.0 $&$ 1.47\cdot 10^{-1}  $\\
\hline
$120 $&$ 118.2 $&$ 1.14\cdot 10^{-1}  $&$ 119.1 $&$ 1.23\cdot 10^{-1}  $&$ 119.5 $&$ 1.32\cdot 10^{-1}  $&$
      119.7 $&$ 1.42\cdot 10^{-1}  $&$ 119.7 $&$ 1.52\cdot 10^{-1}  $\\
    &$ 132.0 $&$ 1.16\cdot 10^{-1}  $&$ 131.2 $&$ 1.25\cdot 10^{-1}  $&$ 131.0 $&$ 1.35\cdot 10^{-1}  $&$
      130.9 $&$ 1.45\cdot 10^{-1}  $&$ 131.0 $&$ 1.57\cdot 10^{-1}  $\\
    &$ 120.0 $&$ 1.14\cdot 10^{-1}  $&$ 120.0 $&$ 1.22\cdot 10^{-1}  $&$ 120.0 $&$ 1.30\cdot 10^{-1}  $&$
      120.0 $&$ 1.39\cdot 10^{-1}  $&$ 120.0 $&$ 1.49\cdot 10^{-1}  $\\
\hline
$130 $&$ 125.2 $&$ 1.15\cdot 10^{-1}  $&$ 126.9 $&$ 1.23\cdot 10^{-1}  $&$ 127.8 $&$ 1.33\cdot 10^{-1}  $&$
      128.4 $&$ 1.43\cdot 10^{-1}  $&$ 128.9 $&$ 1.54\cdot 10^{-1}  $\\
    &$ 134.9 $&$ 1.17\cdot 10^{-1}  $&$ 133.4 $&$ 1.25\cdot 10^{-1}  $&$ 132.6 $&$ 1.34\cdot 10^{-1}  $&$
      132.2 $&$ 1.43\cdot 10^{-1}  $&$ 131.8 $&$ 1.53\cdot 10^{-1}  $\\
    &$ 130.0 $&$ 1.16\cdot 10^{-1}  $&$ 130.0 $&$ 1.23\cdot 10^{-1}  $&$ 130.0 $&$ 1.32\cdot 10^{-1}  $&$
      130.0 $&$ 1.41\cdot 10^{-1}  $&$ 130.0 $&$ 1.51\cdot 10^{-1}  $\\
\hline
$140 $&$ 128.1 $&$ 1.13\cdot 10^{-1}  $&$ 129.3 $&$ 1.21\cdot 10^{-1}  $&$ 129.8 $&$ 1.29\cdot 10^{-1}  $&$
      130.2 $&$ 1.38\cdot 10^{-1}  $&$ 130.5 $&$ 1.48\cdot 10^{-1}  $\\
    &$ 142.1 $&$ 1.18\cdot 10^{-1}  $&$ 141.1 $&$ 1.26\cdot 10^{-1}  $&$ 140.6 $&$ 1.34\cdot 10^{-1}  $&$
      140.4 $&$ 1.43\cdot 10^{-1}  $&$ 140.3 $&$ 1.53\cdot 10^{-1}  $\\
    &$ 140.0 $&$ 1.17\cdot 10^{-1}  $&$ 140.0 $&$ 1.25\cdot 10^{-1}  $&$ 140.0 $&$ 1.33\cdot 10^{-1}  $&$
      140.0 $&$ 1.42\cdot 10^{-1}  $&$ 140.0 $&$ 1.52\cdot 10^{-1}  $\\
\hline
$150 $&$ 128.9 $&$ 1.10\cdot 10^{-1}  $&$ 129.7 $&$ 1.17\cdot 10^{-1}  $&$ 130.1 $&$ 1.24\cdot 10^{-1}  $&$
      130.4 $&$ 1.31\cdot 10^{-1}  $&$ 130.6 $&$ 1.39\cdot 10^{-1}  $\\
    &$ 151.2 $&$ 1.19\cdot 10^{-1}  $&$ 150.6 $&$ 1.27\cdot 10^{-1}  $&$ 150.4 $&$ 1.35\cdot 10^{-1}  $&$
      150.2 $&$ 1.44\cdot 10^{-1}  $&$ 150.1 $&$ 1.54\cdot 10^{-1}  $\\
    &$ 150.0 $&$ 1.18\cdot 10^{-1}  $&$ 150.0 $&$ 1.26\cdot 10^{-1}  $&$ 150.0 $&$ 1.35\cdot 10^{-1}  $&$
      150.0 $&$ 1.44\cdot 10^{-1}  $&$ 150.0 $&$ 1.54\cdot 10^{-1}  $\\
\hline
$160 $&$ 129.2 $&$ 1.07\cdot 10^{-1}  $&$ 129.9 $&$ 1.12\cdot 10^{-1}  $&$ 130.2 $&$ 1.18\cdot 10^{-1}  $&$
      130.4 $&$ 1.24\cdot 10^{-1}  $&$ 130.6 $&$ 1.30\cdot 10^{-1}  $\\
    &$ 160.9 $&$ 1.20\cdot 10^{-1}  $&$ 160.4 $&$ 1.28\cdot 10^{-1}  $&$ 160.3 $&$ 1.36\cdot 10^{-1}  $&$
      160.2 $&$ 1.46\cdot 10^{-1}  $&$ 160.1 $&$ 1.55\cdot 10^{-1}  $\\
    &$ 160.0 $&$ 1.19\cdot 10^{-1}  $&$ 160.0 $&$ 1.27\cdot 10^{-1}  $&$ 160.0 $&$ 1.36\cdot 10^{-1}  $&$
      160.0 $&$ 1.45\cdot 10^{-1}  $&$ 160.0 $&$ 1.55\cdot 10^{-1}  $\\
\hline
$170 $&$ 129.4 $&$ 1.03\cdot 10^{-1}  $&$ 123.0 $&$ 1.07\cdot 10^{-1}  $&$ 130.2 $&$ 1.12\cdot 10^{-1}  $&$
      130.4 $&$ 1.17\cdot 10^{-1}  $&$ 130.6 $&$ 1.23\cdot 10^{-1}  $\\
    &$ 170.7 $&$ 1.21\cdot 10^{-1}  $&$ 170.4 $&$ 1.29\cdot 10^{-1}  $&$ 170.2 $&$ 1.38\cdot 10^{-1}  $&$
      170.1 $&$ 1.47\cdot 10^{-1}  $&$ 170.1 $&$ 1.57\cdot 10^{-1}  $\\
    &$ 170.0 $&$ 1.21\cdot 10^{-1}  $&$ 170.0 $&$ 1.29\cdot 10^{-1}  $&$ 170.0 $&$ 1.37\cdot 10^{-1}  $&$
      170.0 $&$ 1.46\cdot 10^{-1}  $&$ 170.0 $&$ 1.56\cdot 10^{-1}  $\\
\hline
$180 $&$ 129.5 $&$ 9.90\cdot 10^{-2}  $&$ 130.0 $&$ 1.03\cdot 10^{-1}  $&$ 130.2 $&$ 1.07\cdot 10^{-1}  $&$
      130.4 $&$ 1.11\cdot 10^{-1}  $&$ 130.6 $&$ 1.15\cdot 10^{-1}  $\\
    &$ 180.6 $&$ 1.22\cdot 10^{-1}  $&$ 180.3 $&$ 1.30\cdot 10^{-1}  $&$ 180.2 $&$ 1.39\cdot 10^{-1}  $&$
      180.1 $&$ 1.48\cdot 10^{-1}  $&$ 180.1 $&$ 1.58\cdot 10^{-1}  $\\
    &$ 180.0 $&$ 1.20\cdot 10^{-1}  $&$ 180.0 $&$ 1.29\cdot 10^{-1}  $&$ 180.0 $&$ 1.38\cdot 10^{-1}  $&$
      180.0 $&$ 1.47\cdot 10^{-1}  $&$ 180.0 $&$ 1.57\cdot 10^{-1}  $\\
\hline
$190 $&$ 129.6 $&$ 9.55\cdot 10^{-2}  $&$ 130.0 $&$ 9.87\cdot 10^{-2}  $&$ 130.3 $&$ 1.02\cdot 10^{-1}  $&$
      130.4 $&$ 1.06\cdot 10^{-1}  $&$ 130.6 $&$ 1.09\cdot 10^{-1}  $\\
    &$ 190.5 $&$ 1.22\cdot 10^{-1}  $&$ 190.3 $&$ 1.31\cdot 10^{-1}  $&$ 190.2 $&$ 1.40\cdot 10^{-1}  $&$
      190.1 $&$ 1.49\cdot 10^{-1}  $&$ 190.0 $&$ 1.59\cdot 10^{-1}  $\\
    &$ 190.0 $&$ 1.20\cdot 10^{-1}  $&$ 190.0 $&$ 1.30\cdot 10^{-1}  $&$ 190.0 $&$ 1.39\cdot 10^{-1}  $&$
      190.0 $&$ 1.48\cdot 10^{-1}  $&$ 190.0 $&$ 1.58\cdot 10^{-1}  $\\
\hline
$200 $&$ 129.7 $&$ 9.23\cdot 10^{-2}  $&$ 130.1 $&$ 9.50\cdot 10^{-2}  $&$ 130.3 $&$ 9.78\cdot 10^{-2}  $&$
      130.4 $&$ 1.01\cdot 10^{-1}  $&$ 130.6 $&$ 1.04\cdot 10^{-1}  $\\
    &$ 200.5 $&$ 1.23\cdot 10^{-1}  $&$ 200.3 $&$ 1.32\cdot 10^{-1}  $&$ 200.2 $&$ 1.40\cdot 10^{-1}  $&$
      200.1 $&$ 1.50\cdot 10^{-1}  $&$ 200.0 $&$ 1.60\cdot 10^{-1}  $\\
    &$ 200.0 $&$ 1.21\cdot 10^{-1}  $&$ 200.0 $&$ 1.30\cdot 10^{-1}  $&$ 200.0 $&$ 1.40\cdot 10^{-1}  $&$
      200.0 $&$ 1.49\cdot 10^{-1}  $&$ 200.0 $&$ 1.59\cdot 10^{-1}  $\\
\hline
$220 $&$ 129.7 $&$ 8.68\cdot 10^{-2}  $&$ 130.1 $&$ 8.87\cdot 10^{-2}  $&$ 130.3 $&$ 9.08\cdot 10^{-2}  $&$
      130.4 $&$ 9.29\cdot 10^{-2}  $&$ 130.6 $&$ 9.51\cdot 10^{-2}  $\\
    &$ 220.4 $&$ 1.24\cdot 10^{-1}  $&$ 220.2 $&$ 1.33\cdot 10^{-1}  $&$ 220.1 $&$ 1.42\cdot 10^{-1}  $&$
      220.1 $&$ 1.52\cdot 10^{-1}  $&$ 220.0 $&$ 1.62\cdot 10^{-1}  $\\
    &$ 220.0 $&$ 1.22\cdot 10^{-1}  $&$ 220.0 $&$ 1.32\cdot 10^{-1}  $&$ 220.0 $&$ 1.41\cdot 10^{-1}  $&$
      220.0 $&$ 1.51\cdot 10^{-1}  $&$ 220.0 $&$ 1.61\cdot 10^{-1}  $\\
\hline
$240 $&$ 129.8 $&$ 8.24\cdot 10^{-2}  $&$ 130.1 $&$ 8.37\cdot 10^{-2}  $&$ 130.3 $&$ 8.53\cdot 10^{-2}  $&$
      130.4 $&$ 8.69\cdot 10^{-2}  $&$ 130.6 $&$ 8.85\cdot 10^{-2}  $\\
    &$ 240.4 $&$ 1.25\cdot 10^{-1}  $&$ 240.2 $&$ 1.35\cdot 10^{-1}  $&$ 240.2 $&$ 1.44\cdot 10^{-1}  $&$
      240.1 $&$ 1.53\cdot 10^{-1}  $&$ 240.0 $&$ 1.63\cdot 10^{-1}  $\\
    &$ 240.0 $&$ 1.21\cdot 10^{-1}  $&$ 240.0 $&$ 1.33\cdot 10^{-1}  $&$ 240.0 $&$ 1.43\cdot 10^{-1}  $&$
      240.0 $&$ 1.53\cdot 10^{-1}  $&$ 240.0 $&$ 1.63\cdot 10^{-1}  $\\
\hline
$260 $&$ 129.8 $&$ 7.88\cdot 10^{-2}  $&$ 130.1 $&$ 7.98\cdot 10^{-2}  $&$ 130.3 $&$ 8.10\cdot 10^{-2}  $&$
      130.4 $&$ 8.22\cdot 10^{-2}  $&$ 130.6 $&$ 8.35\cdot 10^{-2}  $\\
    &$ 260.4 $&$ 1.25\cdot 10^{-1}  $&$ 260.2 $&$ 1.35\cdot 10^{-1}  $&$ 260.2 $&$ 1.45\cdot 10^{-1}  $&$
      260.1 $&$ 1.55\cdot 10^{-1}  $&$ 260.0 $&$ 1.65\cdot 10^{-1}  $\\
    &$ 260.0 $&$ 1.17\cdot 10^{-1}  $&$ 260.0 $&$ 1.31\cdot 10^{-1}  $&$ 260.0 $&$ 1.42\cdot 10^{-1}  $&$
      260.0 $&$ 1.53\cdot 10^{-1}  $&$ 260.0 $&$ 1.63\cdot 10^{-1}  $\\
\hline
$280 $&$ 129.8 $&$ 7.60\cdot 10^{-2}  $&$ 130.1 $&$ 7.67\cdot 10^{-2}  $&$ 130.3 $&$ 7.76\cdot 10^{-2}  $&$
      130.4 $&$ 7.85\cdot 10^{-2}  $&$ 130.6 $&$ 7.95\cdot 10^{-2}  $\\
    &$ 280.4 $&$ 1.24\cdot 10^{-1}  $&$ 280.2 $&$ 1.36\cdot 10^{-1}  $&$ 280.2 $&$ 1.46\cdot 10^{-1}  $&$
      280.1 $&$ 1.56\cdot 10^{-1}  $&$ 280.0 $&$ 1.66\cdot 10^{-1}  $\\
    &$ 280.0 $&$ 1.16\cdot 10^{-1}  $&$ 280.0 $&$ 1.31\cdot 10^{-1}  $&$ 280.0 $&$ 1.43\cdot 10^{-1}  $&$
      280.0 $&$ 1.54\cdot 10^{-1}  $&$ 280.0 $&$ 1.64\cdot 10^{-1}  $\\
\hline
$300 $&$ 129.8 $&$ 7.36\cdot 10^{-2}  $&$ 130.1 $&$ 7.41\cdot 10^{-2}  $&$ 130.3 $&$ 7.48\cdot 10^{-2}  $&$
      130.4 $&$ 7.56\cdot 10^{-2}  $&$ 130.6 $&$ 7.64\cdot 10^{-2}  $\\
    &$ 300.4 $&$ 1.22\cdot 10^{-1}  $&$ 300.3 $&$ 1.36\cdot 10^{-1}  $&$ 300.3 $&$ 1.46\cdot 10^{-1}  $&$
      300.2 $&$ 1.56\cdot 10^{-1}  $&$ 300.1 $&$ 1.67\cdot 10^{-1}  $\\
    &$ 300.0 $&$ 1.06\cdot 10^{-1}  $&$ 300.0 $&$ 1.27\cdot 10^{-1}  $&$ 300.0 $&$ 1.41\cdot 10^{-1}  $&$
      300.0 $&$ 1.53\cdot 10^{-1}  $&$ 300.0 $&$ 1.65\cdot 10^{-1}   $\\
\hline
  \end{tabular}
}
\end{table}

\clearpage

\section{NLO Monte Carlo event generators\protect\footnote{M. Felcini, F. Krauss, 
F. Maltoni, P. Nason and J. Yu.}}
\newcommand\pperpmin{p_\perp^{\rm min}}

\def\fortran{Fortran\xspace}
\providecommand{\pythia}{{\sc Pythia}}
\def\phojet{Phojet\xspace}
\providecommand{\pythiasix}{{\sc Pythia~6}}
\providecommand{\pythiaeight}{{\sc Pythia~8}}
\providecommand{\herwig}{{\sc Herwig}}
\providecommand{\herwigsix}{{\sc Herwig~6}}
\providecommand{\herwigpp}{{\sc Herwig}\raisebox{0.1ex}{\small $++$}}
\providecommand{\sherpa}{{\sc Sherpa}}
\providecommand{\alpgen}{{\sc AlpGen}}
\def\helac{HELAC\xspace}
\def\charybdis{Charybdis\xspace}
\def\jimmy{Jimmy\xspace}
\def\rivet{Rivet\xspace}
\def\rivetgun{Rivetgun\xspace}
\def\professor{Professor\xspace}
\def\fastjet{FastJet\xspace}
\def\hepmc{HepMC\xspace}
\def\agile{AGILe\xspace}
\def\hztool{HZTool\xspace}
\def\numpy{NumPy\xspace}
\def\scipy{SciPy\xspace}
\def\minuit{Minuit\xspace}
\def\pyminuit{PyMinuit\xspace}
\def\python{Python\xspace}
\def\kT{\ensuremath{k_{\mathrm{T}}}} 
\newcommand\POWHEG{{\sc P\scalebox{0.9}{OWHEG}}\xspace}
\newcommand\POWHEGBOX{{\sc P\scalebox{0.9}{OWHEG} B\scalebox{0.9}{OX}}\xspace}
\newcommand\MCatNLO{{\sc M\scalebox{0.9}{C}@N\scalebox{0.9}{LO}}\xspace}
\newcommand\ARIADNE{{\sc A\scalebox{0.9}{RIADNE}}\xspace}
\newcommand\ADICIC{{\sc A\scalebox{0.9}{DICIC}++}\xspace}
\newcommand\AMEGIC{{\sc A\scalebox{0.9}{MEGIC}++}\xspace}
\newcommand\Comix{{\sc Comix}\xspace}
\newcommand\HNNLO{{\sc H\scalebox{0.9}{NNLO}}\xspace}
\newcommand\MGME{{\sc M\scalebox{0.9}{AD}G\scalebox{0.9}{RAPH/}M\scalebox{0.9}{AD}E\scalebox{0.9}{VENT}}\xspace}











\subsection{Introduction}

In recent years Monte Carlo event generators have been the subject of great
theoretical and practical developments, most significantly in the extension
of existing parton-shower simulations to consistently include exact
next-to-leading order (NLO) corrections~\cite{Frixione:2002ik,
Frixione:2003ei,Frixione:2005vw,Frixione:2006gn,Frixione:2007zp,Frixione:2008yi,
LatundeDada:2007jg,Nason:2004rx,Nason:2006hfa,Frixione:2007nu,Frixione:2007vw,
Frixione:2007nw,LatundeDada:2006gx,Hamilton:2008pd,Hamilton:2009za,
Alioli:2008gx,Alioli:2008tz,Alioli:2009je,LatundeDada:2008bv}  
and, separately, in the consistent combination of parton-shower simulations 
and high-multiplicity tree-level matrix-element generators~%
\cite{Mangano:2001xp,Catani:2001cc,Lonnblad:2001iq,Krauss:2002up,
Mrenna:2003if,Schalicke:2005nv,Alwall:2007fs,Hoeche:2009rj,Hamilton:2009ne}.  

In this note we aim at concisely reviewing the basic principles of the new 
generation of tools which are now available, underlying the most important 
improvements with respect to a more standard parton-shower approach.  We 
provide also guidelines for experimentalists on which tools to use for a 
given Higgs production channel, on the possible improvements/limitations and 
on how to perform a meaningful cross validation of the MC tools used in an 
experimental analysis vis-a-vis the best theoretical predictions available 
at a given moment (for example, at NNLO level).  As a result, we provide 
enough motivation for the new MC tools to be used as {\em default} analysis 
tools, both to better tune Higgs-boson searches and to perform precise 
measurements of its properties.  We also aim at providing guidelines for
how and when to use these tools.  We conclude by summarising the results 
and by commenting on the readiness of these theoretical tools for 
anticipated 
Higgs analyses, and by adding a wish-list for tools from the experimentalist 
point of view.   

\subsection{Embedding higher-order corrections into parton-shower
  Monte Carlo event generators}
\subsubsection{NLO cross sections}

Let us start by reminding the equation describing the calculation of 
next-to-leading order corrections in QCD for a $2\to n$ process; 
schematically it reads
\begin{equation}\label{eq:nlocorr}
{\rm d}\sigma^{\rm (NLO)}
=
{\rm d}\Phi_B\left[B(\Phi_B)+V(\Phi_B)\right]+
{\rm d}\Phi_R R(\Phi_R)\,,
\end{equation}
where $\Phi_B$ and $\Phi_R$ denote the phase-space elements related to the
$2\to n$ (Born level) and $2\to n+1$ (real-emission correction) kinematics;
$B(\Phi_B)$, $V(\Phi_B)$ denote the Born-level and the virtual contribution, 
while $R(\Phi_R)$ is the real-emission correction.

In this equation the virtual term contains soft and collinear divergences.
When integrated over the full real phase space, the real term generates 
soft and collinear divergences, too, and only when {\em infrared(IR)-safe} 
quantities are computed, these divergences cancel to yield a finite result.
IR-safe observables $O(\Phi)$ can be best understood by considering
the soft or collinear limit of $\Phi_R$, i.e.\ when the additional parton
has low energy or is parallel to another parton.  In this
limit, an IR-safe observable yields $\lim O(\Phi_R)=O(\Phi_B)$, where the 
Born-level configuration $\Phi_B$ is obtained from $\Phi_R$ by eliminating 
the soft particle (in case of soft singularities) or by merging the collinear 
particles (in case of collinear singularities).

Technically, singularities are often handled with the subtraction method, 
where the real phase space is parametrized in terms of an underlying Born 
phase space $\Phi_B$ and a radiation phase space $\Phi_{R|B}$. The only 
requirement upon this parametrization is that, in the singular limits, by 
merging collinear partons, or eliminating the soft parton, the real phase 
becomes equal to the underlying Born one.  Then the expectation value of
an IR-safe observable reads
\begin{eqnarray}
\int {\rm d}\sigma^{\rm (NLO)} O(\Phi)&=&
\int {\rm d}\Phi_B\left[B(\Phi_B)+V(\Phi_B)\right]O(\Phi_B)
+\int {\rm d}\Phi_R R(\Phi_R)\; O(\Phi_R)
\nonumber \\
&=&\int {\rm d}\Phi_B\left[B(\Phi_B)+V(\Phi_B)
+\int {\rm d}\Phi_{R|B}S(\Phi_R)\right]O(\Phi_B)
\nonumber \\
&+&\int {\rm d}\Phi_R \left[R(\Phi_R)\; O(\Phi_R)-S(\Phi_R)
O(\Phi_B)\right]\;. \label{eq:subtractionmethod}
\end{eqnarray}
The third member of the above equation is obtained by adding and subtracting
the same quantity from the two terms of the second member.  The terms 
$S(\Phi_{R|B})$ are the subtraction terms, which contain all soft and collinear 
singularities of the real-emission term.  Typically, using the universality 
of soft and collinear divergences, they are written in a factorised form as
\begin{equation}
    S(\Phi_R) = B(\Phi_B)\otimes \tilde S(\Phi_{R|B})\,,
\end{equation}  
where the $\tilde S(\Phi_{R|B})$ can be composed from universal, 
process-independent subtraction kernels with analytically known (divergent) 
integrals.  These integral, when summed and added to the virtual term, yield 
a finite result.  The second term of the last member of 
Eq.\,(\ref{eq:subtractionmethod}) is also finite if $O$ is an IR-safe 
observable, since by construction $S$ cancels all singularities in $R$ in 
the soft and collinear regions.

In the following we will always write the NLO corrections in the form
of Eq.\,(\ref{eq:nlocorr}), assuming that a subtraction procedure is
carried out in order to evaluate it explicitly.

\subsubsection{Parton shower (PS)}
Parton showers are able to dress a given Born process with all the dominant
(i.e.\ enhanced by collinear logarithms, and to some extent also soft ones)
QCD radiation processes at all orders in perturbation theory.
In particular, also the hardest radiation includes next-to-leading order
corrections, but only the dominant ones, i.e.\ those
given by the leading logarithms.
The cross section for the hardest emission in a shower -- often this is the 
first emission -- reads:
\begin{equation}
\label{Eq:1storder_in_PS}
{\rm d}\sigma^{\rm PS} = 
{\rm d}\Phi_B B(\Phi_B)
\left[\Delta(\pperpmin) + 
      {\rm d}\Phi_{R|B} \Delta(\pT(\Phi_{R|B}))
\frac{R^{\rm PS}(\Phi_R)}{B(\Phi_B)}   \right]\,,
\end{equation}
where $\Delta(\pT)$ denotes the Sudakov form factor
\begin{equation}
\Delta(\pT) = 
\exp\left[-\int\,{\rm d}\Phi_{R|B}\frac{R^{\rm PS}(\Phi_R)}{B(\Phi_B)} 
\Theta(\pT(\Phi_{R})-\pT)\right]\;.
\end{equation}
This Sudakov form factor can be understood as a no-emission probability
of secondary partons down to a resolution scale of $\pT$.
Here $R^{\rm PS}$ represents the PS approximation to the real cross section,
typically given schematically by a product of the underlying Born-level
term folded with a process-independent universal splitting function $P$:
\begin{equation}
R^{\rm PS}(\Phi)=P(\Phi_{R|B})\,B(\Phi_B).
\end{equation}
In this framework, $\Phi_{R|B}$ is often expressed in terms of three showering 
variables, like the virtuality $t$ in the splitting process\footnote{
           In more modern parton showers the transverse momentum
           of the splitting or the (scaled) opening angle serve
           as ordering variables instead of the virtuality, such
           a choice usually allows to catch more of the leading
           logarithmic terms.
}, the energy fraction of the splitting $z$ and the azimuth $\phi$.
The above definition of the Sudakov form factor, guarantees
that the square bracket in Eq.\ (\ref{Eq:1storder_in_PS}) integrates to unity,
a manifestation of the probabilistic nature of the parton shower.
Thus, integrating the shower cross section over the radiation variables
yields the total cross section, given at LO by the Born amplitude.
The corresponding radiation pattern consists of two parts: one 
given by the first term in the square bracket, where no further resolvable 
emission above the parton-shower cut-off $\pperpmin$ -- typically of the order
of $1$\UGeV\ -- emerges, and the other given by the second term in the square bracket describing the first emission, as determined by the splitting kernel.

It is important to stress that the real-emission cross section in a PS 
generator is only correct in the small angle and/or soft limit, where
$R^{\rm PS}$ is a reliable approximation of the complete matrix element.

While rather crude, the PS approximation is a very powerful one, due mainly
to the great flexibility and simplicity in the implementation of
 $2 \to 1$ and $2\to 2$ high-$Q^2$ processes. In addition, once 
augmented with a hadronisation model the simulation can easily provide a 
full description of a collision in terms of physical final states, i.e., 
hadrons, leptons and photons.  In the current terminology a generic Monte 
Carlo generator mainly refers to such tools, the most relevant examples of are 
\pythiasix\ and \pythiaeight~\cite{Sjostrand:2006za,Sjostrand:2007gs}, 
\herwig~\cite{Corcella:2000bw}, \herwigpp~\cite{Bahr:2008pv}, and 
\sherpa~\cite{Gleisberg:2003xi}.  

It should be noted here, however, that each of these tools employs a 
variety of the more advanced methods listed below, to enhance its accuracy 
in the description of the radiation pattern or the total cross sections,
thus going in some cases beyond the simple PS approach.

\subsubsection{Matrix-element correction (MEC)}
\label{mec}
A first improvement of the parton-shower approximation is to correct
the hardest emission with the
{\em exact first-order real-emission matrix element}.
This has traditionally been achieved by matrix-element corrections%
~\cite{Bengtsson:1986hr,Gustafson:1987rq,Seymour:1994we,
Seymour:1994df,Miu:1998ju,Lonnblad:1995ex}. 
These are provided by either replacing the approximate expression for the 
real-emission cross section $R^{\rm PS}$ with the exact one $R$, or by adding 
real-emission events with a cross section $R-R^{\rm PS}$, in order to 
compensate for the PS inaccuracies (including lack of complete coverage of 
phase space) in large-angle emissions.  Matrix-element corrections in PS 
have only been introduced for the most simple processes, such as Drell--Yan or
$\PW$ production, Higgs production in gluon fusion, or top decay, all
present in \pythiasix, \pythiaeight, \herwig, and \herwigpp.  

Decomposing the real-emission cross section into a singular and non-singular 
part, $R=R^{s}+R^{f}$ (both being non-negative), the first-emission 
cross section in this method can be written as
\begin{equation}\label{Eq:MEC}
{\rm d}\sigma^{\rm MEC} =
{\rm d}\Phi_B B(\Phi_B)
\left[\bar\Delta(\pperpmin)+\int\limits_{\pperpmin}{\rm d}\Phi_{R|B}
      \frac{R^{s}(\Phi_R)}{B(\Phi_B)}
      \bar\Delta(\pT(\Phi_{R|B}))\right]
+ {\rm d}\Phi_R R^{f}(\Phi_R)\,,
\end{equation}
where the modified Sudakov form factor is given by replacing the splitting 
function with the ratio of real-emission and Born-level matrix elements,
\begin{equation}
\bar \Delta(\pT) = 
\exp\left[-\int\limits_{\pperpmin}{\rm d}\Phi_{R|B}
  \frac{R^{s}(\Phi_R)}{B(\Phi_B)}\right]\,.
\end{equation}
Note that the term in the square bracket integrates to unity,
as in the usual PS case.  In the \pythia\ implementation $R^{f}=0$, while in \herwig, due to 
the lack of full phase-space coverage the term $R^{f}$ is introduced, with 
support only in the region of phase space that the parton shower cannot fill.
In the latter the hardest emission is not necessarily the first one,
since the ordering parameter is the splitting angle.
The correction is thus applied
to all branchings that are the hardest so far in the shower development.

While this method correctly reproduces the first-emission kinematics (formally
this is a next-to-leading order effect with respect to the Born-level 
cross section), 
it does not include the full next-to-leading order accuracy to the total 
cross section.

\subsubsection{NLO+PS}
\label{nlops}
Several proposals have been made for the full inclusion of complete NLO 
effects in PS generators. At this moment, only two of them have reached a
mature enough stage to be used in practice: \MCatNLO~\cite{Frixione:2002ik} and 
\POWHEG~\cite{Nason:2004rx}.  Both methods correct -- in different ways -- 
the real-emission matrix element to achieve an exact tree-level emission 
matrix element, even at large angle.  As we have seen in the previous 
subsection, this is what is also achieved with matrix-element corrections in 
parton showers, at least for the simplest processes listed earlier.  This, 
however, is not sufficient for the NLO accuracy, since the effect of 
virtual corrections also needs to be included.  In both methods, the real-emission 
cross section is split into a singular and non-singular part,
$R=R^{s}+R^{f}$.  One then computes the total NLO inclusive cross 
section, excluding the finite contribution, at fixed underlying Born 
kinematics, defined as
\begin{equation}\label{eq:nlocorr1}
\bar{B}^{s}=B(\Phi_B)+\left[V(\Phi_B)+
\int{\rm d}\Phi_{R|B} R^{s}(\Phi_{R|B})\right]\,,
\end{equation}
and uses the formula
\begin{equation}
\label{Eq:Powheg}
{\rm d}\sigma^{\rm NLO+PS} =
{\rm d}\Phi_B\bar B^{s}(\Phi_B)
\left[\Delta^s(\pperpmin) 
  +  {\rm d}\Phi_{R|B}
      \frac{R^{s}(\Phi_R)}{B(\Phi_B)}
      \Delta^s(\pT(\Phi))\right] 
+ {\rm d}\Phi_R R^{f}(\Phi_R)
\end{equation}
for the generation of the events.  In this formula, the term $\bar B$ can be 
understood as a local $K$-factor reweighting the soft matrix-element 
correction part of the simulation.  Clearly, employing the fact that 
the term in the first square bracket integrates to unity, as before, the 
cross section integrates to the full NLO cross section.  

\subsubsection{ MC@NLO}
\label{mcatnlo}
In \MCatNLO one chooses $R^s$ to be identically equal to the term $B\otimes P$
that the PS generator employs to generate emissions.  Within \MCatNLO,
$n\,$-body events are obtained using the $\bar{B}^s$ function, and then fed to 
the PS, which will generate the hardest emission according to 
Eq.\,(\ref{eq:nlocorr1}).  These are called ${\cal S}$ events in the \MCatNLO 
language. An appropriate number of events are also generated according to
the $R^f$  cross section, and are directly passed to the PS generator.
These are  called ${\cal H}$ events.

In \MCatNLO, $R^f=R-R^s$ is not positive definite, and it is thus necessary to
generate negative weighted events in this framework.

A library of \MCatNLO Higgs processes (gluon fusion, vector-boson associated
production, and charged Higgs associated with top) is available
at~\Bref{Frixione:2010wd}, which is interfaced to \herwig\ and \herwigpp.
A \MCatNLO interface to the virtuality-ordered \pythia\
shower for the $\PW$-production process has been recently
achieved~\cite{Torrielli:2010aw}.

\subsubsection{POWHEG}
\label{powheg}

In \POWHEG, one chooses $R^s\le R$, and in many cases even $R^s=R$, so that
the finite cross section $R^f$ vanishes. In this case, the hardest emission is
generated within \POWHEG itself, and the process is passed to the parton shower
only after the hardest radiation is generated.  Positive weighted events are 
obtained, since $R^f$ can always be chosen to be positive definite. In all cases
the chosen $R^s$ has exactly the same singularity structure as $R$,
so that $R^f$ always yield a finite contribution to the cross section.

In angular-ordered parton showers, the hardest emission is not necessarily the
first, so that, when interfacing \POWHEG{} to an angular order shower (\herwig\
and \herwigpp) soft coherence is spoiled unless truncated showers are
introduced. These are in fact generally needed when interfacing angular-ordered
parton showers to matrix-element generators. The effect of truncated showers has
been studied in the \herwigpp\ implementations of \POWHEG{} processes.
Implementations of Higgs production processes with the \POWHEG{} method are 
available in \herwigpp~\cite{Hamilton:2009za}, in the 
\POWHEGBOX~\cite{Alioli:2010xd} (interfaced to both \herwig\ and \pythia) and 
recently in \sherpa~\cite{Hoeche:2010pf}.

\subsubsection{Matrix-element merging (ME+PS)}
\label{meps}

Matrix-element merging~\cite{Catani:2001cc} aims at correcting as many large-angle 
emissions as possible with the corresponding tree-level accurate
prediction, rather than only {\em small-angle} accurate.
This is achieved by generating events up to a given 
(high) multiplicity using a matrix-element generator, with some internal
jet-resolution parameter $Q_{\rm cut}$ on the jet separation, such that 
practically all emissions above this scale are described by corresponding 
tree-level matrix elements.  Their contributions are corrected for running-coupling 
effects and by Sudakov form factors.  Radiation below $Q_{\rm cut}$ 
on the other hand is generated by a parton-shower program, which is 
required to veto radiation with separation larger than $Q_{\rm cut}$.  As 
far as the hardest emission is concerned, matrix-element merging is as 
accurate as matrix-element corrections (when these are available) or NLO+PS. 
Since they lack NLO virtual corrections, however, they do not reach NLO 
accuracy for inclusive quantities.
Nevertheless, they are capable to achieve leading-order accuracy for multiple
hard radiation, beyond the hardest only, while NLO+PS programs, relying on
the parton shower there are only accurate in the collinear and/or soft limit 
for these quantities.

Several merging schemes have been proposed, which include the CKKW 
scheme~\cite{Catani:2001cc,Krauss:2002up,Lonnblad:2001iq}
 and its improvements
~\cite{Hoeche:2009rj,Hamilton:2009ne},
the MLM matching~\cite{Mangano:2001xp},
and the $\kT$-MLM variation~\cite{Alwall:2008qv}. 
The MLM schemes have been implemented in several matrix element codes such
as \alpgen, \helac, and \MGME, through interfaces to \pythia/\herwig,
while \sherpa/\herwigpp\ have adopted the CKKW schemes and rely on their 
own parton showers. In \Bref{Alwall:2007fs} a detailed, although 
somewhat outdated description of each method has been given and a comparative 
study has been performed.

Basically all Higgs-boson production processes and their backgrounds are 
available in this method.

\subsubsection{MENLOPS}
\label{menlops}

The MENLOPS method \cite{Hamilton:2010wh,Hoeche:2010kg} aims at combining 
matrix-element merging and \POWHEG, in such a way that, besides having 
accurate LO matrix element for multi-jet generation, inclusive observables 
are also accurate at the NLO level. In essence, one introduces a MENLOPS 
separation scale, similar to the merging scale above. Events
with one extra jet (with respect to the basic process) above the hardest
scale are generated by \POWHEG, and events with more than one jet are
generated by the matrix-element merging method. The method works as
long as the fraction of events with more than one extra jet is of the
order of an NNLO correction. 

\subsection{Higgs production channels}

\label{HiggsChannels}

In this section we list and briefly discuss, channel by channel, the tools 
available for the simulation of Higgs production.

\subsubsection{Gluon fusion}

The gluon-fusion production process is implemented with all methods listed
in the previous section. Parton-shower codes based on the $2\to 1$ process
with the exact $\Mt$ dependence 
or in the large-$\Mt$ limit, and with matrix-element corrections, are 
available for this process.  Furthermore, ME+PS is available in the large-$\Mt$ limit in 
several matrix-element-based generators, such as \alpgen, \sherpa\ and \MGME.
The \MCatNLO\ and \POWHEG\ (\POWHEGBOX, \herwigpp\ and \sherpa) implementations 
use the large-$\Mt$ limit, but the Born term in the expression for the
$\bar{B}$ function is computed with the full $\Mt$ dependence. Several
variations of this approach are also available, the most realistic one
being the reweighting of the full $\bar{B}$ function with the ratio of
the Born term with exact top mass dependence, with respect to its
value in the large-$\Mt$ limit. Since the full NLO calculation with
finite $\Mt$ is available \cite{Djouadi:1991tka,Spira:1995rr},
comparison studies between Monte Carlos and this
NLO result can and should be carried out.

A comparison of several available Monte Carlo methods, together with the
bare NNLO calculation~\cite{Catani:2007vq}, was performed in the context of 
Les Houches 2009~\cite{Butterworth:2010ym}, page 58.  Details of the 
generation (parameters, inputs, cuts) of some key observables can be found 
there as well, allowing for a careful further validation and experimental 
work.  The analysis there was performed on the generated final
states and, with the exception of \HNNLO, after parton showering.  
Hadronisation effects were included for the \MCatNLO{} and \POWHEG{} results 
only.  \MGME and \sherpa\ results have been rescaled  to the \HNNLO result, 
while \herwigpp, \MCatNLO, and \POWHEG have not been rescaled.  In this study 
it was found that all methods gave a rather consistent behaviour, with only
few marginally problematic areas, displayed in \Fref{fig:ptHLesHouches}.
\begin{figure}[htb]
\centering
\includegraphics[width=.48\linewidth]{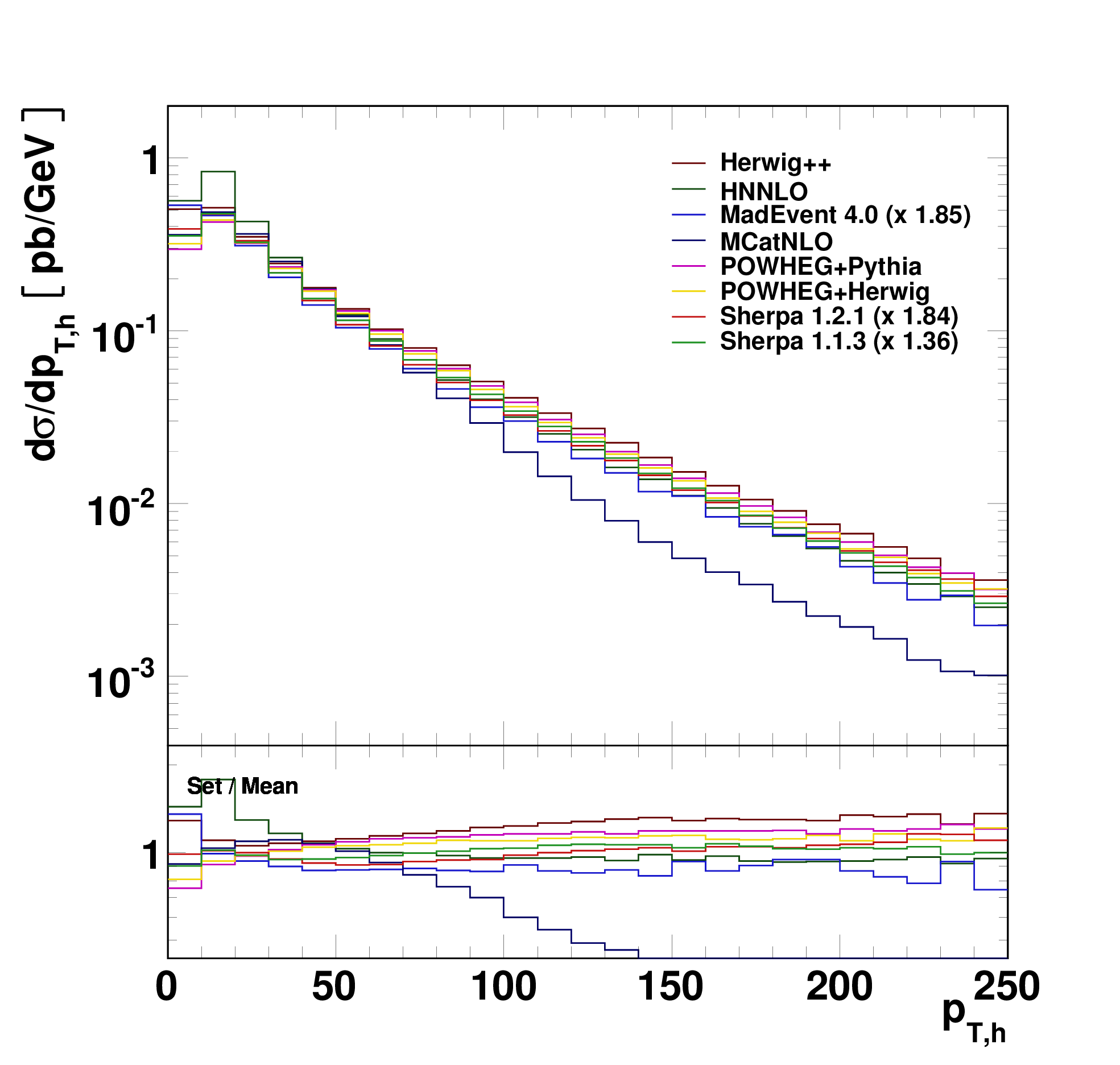}
\includegraphics[width=.48\linewidth]{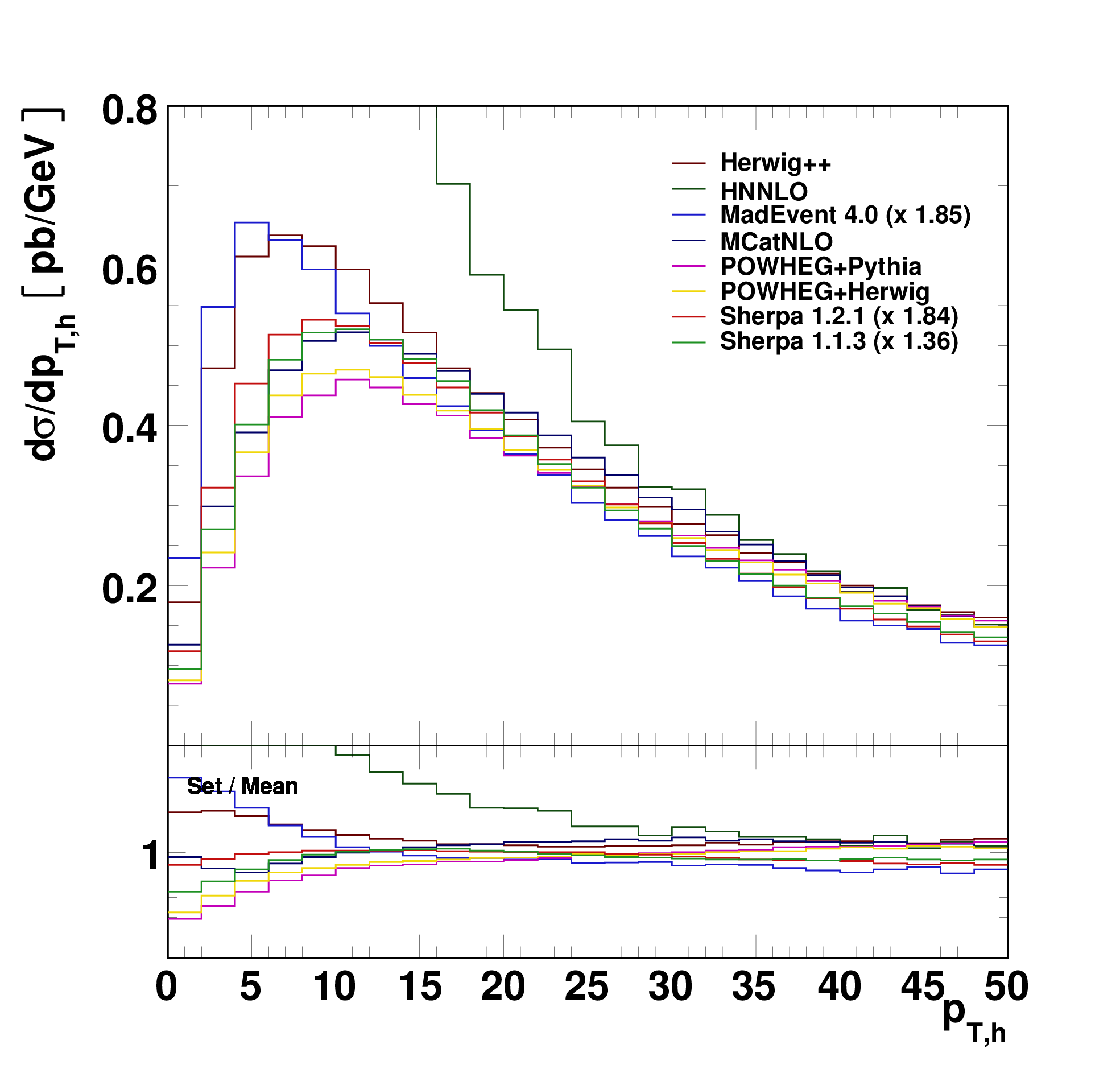}
\caption{Comparison of the $\pT$ spectrum of the
Higgs in various approaches, from \Bref{Butterworth:2010ym}.}
\label{fig:ptHLesHouches}
\end{figure}
In particular, the \MCatNLO{} result 
exhibits a much softer $\pT$ spectrum, for transverse
momenta above roughly one half of the Higgs-boson mass. The transverse momentum
region below $15$\UGeV\ (i.e.\ the Sudakov region) also displayed a somewhat 
different behaviour, being peaked around $5{-}6$\UGeV\ for \herwigpp{} and \MGME{}+\pythia{}
while for all the others the peak is slightly above $10$\UGeV. On the other hand,
this discrepancy cannot be considered fully conclusive, since hadronisation
effects -- although expected to be small -- were only included in the 
\MCatNLO and \POWHEG output.

In \Bref{Alioli:2008tz}, a further detailed study was carried out 
comparing results from \pythia{}, \POWHEG{}, and \MCatNLO{} with the fixed-order 
NNLO calculation, and with the NNLL resummed transverse-momentum
distribution of the Higgs boson.  
The findings of the study can be summarised in few points: 
\begin{itemize}
\item First of all, the \pythia{} (including MEC)
      distributions differ from the \POWHEG{} 
      ones by a $K$-factor that depends only mildly upon the Higgs rapidity. 
      This is explained by the fact that the first radiation in both
      is of the same accuracy, the only difference being that in \pythia{} 
      it is $B$ rather than $\bar{B}$ that appears in front of this formula.
\item This observation also clarifies why all methods but \MCatNLO{} yield 
      very similar transverse-momentum spectra.  We can understand the reason
      for this behaviour from Eq.~(\ref{eq:nlocorr1}).
      We see the reason for this: here the $K$-factor is applied to ${\cal S}$
      events only, but not to ${\cal H}$ events. These last ones populate 
      the region of $\pT$ above roughly $60$\UGeV, and 
      they are not amplified by the large $K$-factor present in Higgs 
      production.
\item The Higgs transverse-momentum distribution in \POWHEG{} shows
      very good agreement with the NNLO calculation. Again we should state 
      that the transverse-momentum spectrum of the Higgs is only computed
      at leading order in \POWHEG{} (being in fact part of the NLO correction
      to inclusive Higgs production). The presence of the full $K$-factor 
      in front of this distribution should thus be seen as an arbitrary 
      correction, trying to catch the true NNLO one. 
\item Similar observations also apply to ME+PS and \pythia{} results rescaled 
      to the full NLO cross section. It turns out, in this case, that the 
      NNLO calculation is in better agreement with these results.  It is 
      not clear whether this is a general pattern of NNLO corrections; however,
      while the same pattern is observed also in $\PW$ and $\PZ$ production
      it should be noted that these processes, being $s$-channel production
      as well are very similar to Higgs production. It would be interesting 
      to investigate whether this is also the case in processes for which 
      the NLO corrections to the production of an associated jet is also 
      known, like, for example, $\PQt\PAQt$ production.
\item In the Sudakov region the transverse-momentum distribution of \POWHEG{} 
      agrees well with the analytic NLL calculation of \Bref{Bozzi:2005wk}, 
      while the also available NNLL result is above \POWHEG{}, but with a 
      very similar shape (see \Fref{fig:HqTpwg}).
\begin{figure}[htb]
\centering
\includegraphics[width=.48\linewidth]{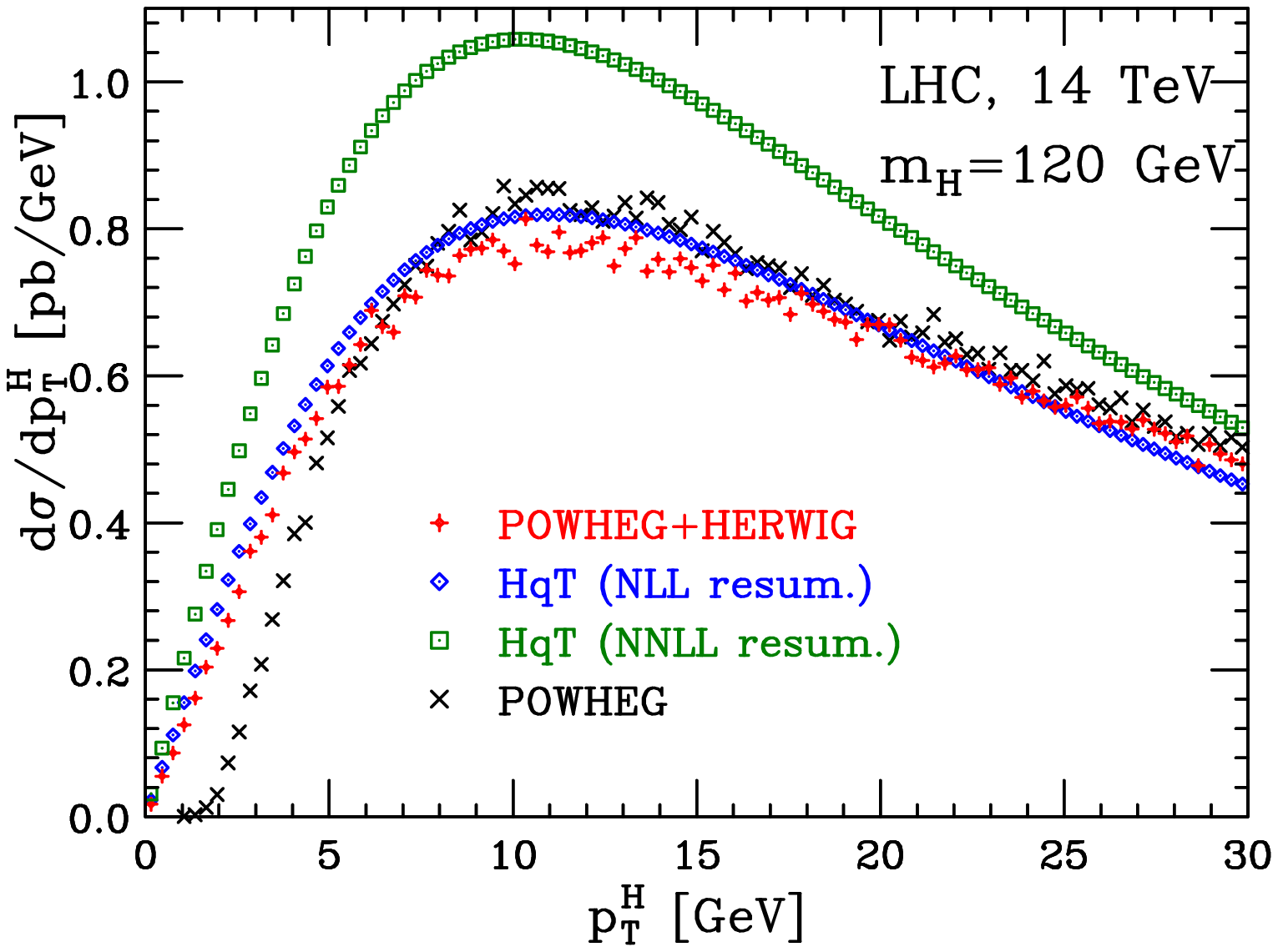}
\includegraphics[width=.48\linewidth]{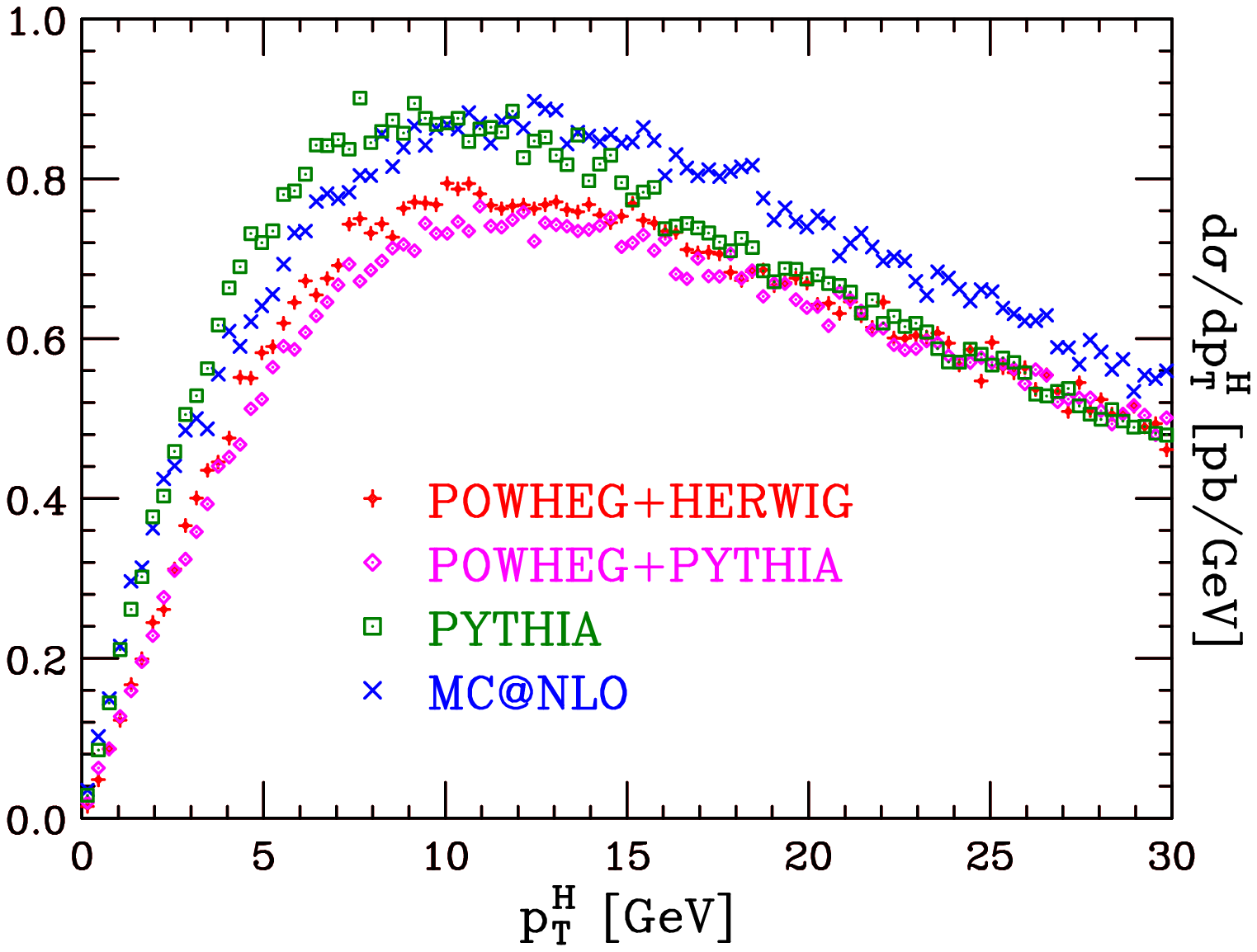}
\caption{Comparison of the Higgs $\pT$ spectrum
of {\protect \POWHEG}
compared with the analytical resummed results of \Bref{Bozzi:2005wk}
(on the left), and comparison of {\protect \POWHEG} interfaced with
{\protect \pythia},  {\protect \POWHEG} interfaced with
{\protect \herwig}, {\protect \MCatNLO} and {\protect \pythia}
standalone (on the right).}
\label{fig:HqTpwg}
\end{figure}
      This can be understood as being mostly an effect 
      due to the inclusion of NNLO terms in the analytic calculation,
      which globally increases the cross section.  In the same region, 
      \MCatNLO{} and \POWHEG{}, when both interfaced to \herwig{}, give very 
      similar results.  This region is the most likely to be important when a 
      jet veto is applied, and is affected by several physical effects of 
      perturbative and non-perturbative origin. These should be studied, 
      for example, using the \POWHEG{} and ME+PS methods, preferably 
      interfaced to different shower programs.
\end{itemize}

Detailed studies comparing \MCatNLO, the NNLO, and the NNLL results
for specific Higgs decay modes have been performed in
\Bref{Stockli:2005hz}
for the $\PGg\PGg$ channel, and in 
\Bref{Anastasiou:2008ik} for the $\PW^+\PW^-$ channel.
In both cases, a good agreement is found for the acceptance correction
found using the parton-level NNLO calculation and \MCatNLO.
This is quite remarkable, especially for the $\PW^+\PW^-$, where a jet veto
is an essential ingredient to suppress the large $\PQt\PAQt$ background.
Since the NNLL result only predict the inclusive transverse-momentum
distribution of the Higgs, it is used to validate the shower Monte Carlo
results. It is found that the NLL and also the NNLL results for the Higgs
cross section below a given transverse-momentum cut match
well with the shower results (consistently with what is displayed in
\Fref{fig:HqTpwg}), and also with the NNLO result. This is understandable,
since apparently large Sudakov logarithms are important for this distribution.
The shower Monte Carlo's (\MCatNLO and \herwig) both resum these logarithms
at the NLL level, and in the NNLO result one more logarithmic term is
included in the cross section with respect to the NLO one.

Hadronization and the underlying event are both likely to affect the
efficiency of the jet veto, by adding more activity to the event.
In \Bref{Anastasiou:2008ik} these effects are also
studied using \herwig\ and {\sc JIMMY}~\cite{Butterworth:1996zw}, for both a cone and 
a $\kT$ algorithm.
The results are reported here in \Fref{fig:UEeffect}.
\begin{figure}[htb]
\centering
\includegraphics[width=.48\linewidth]{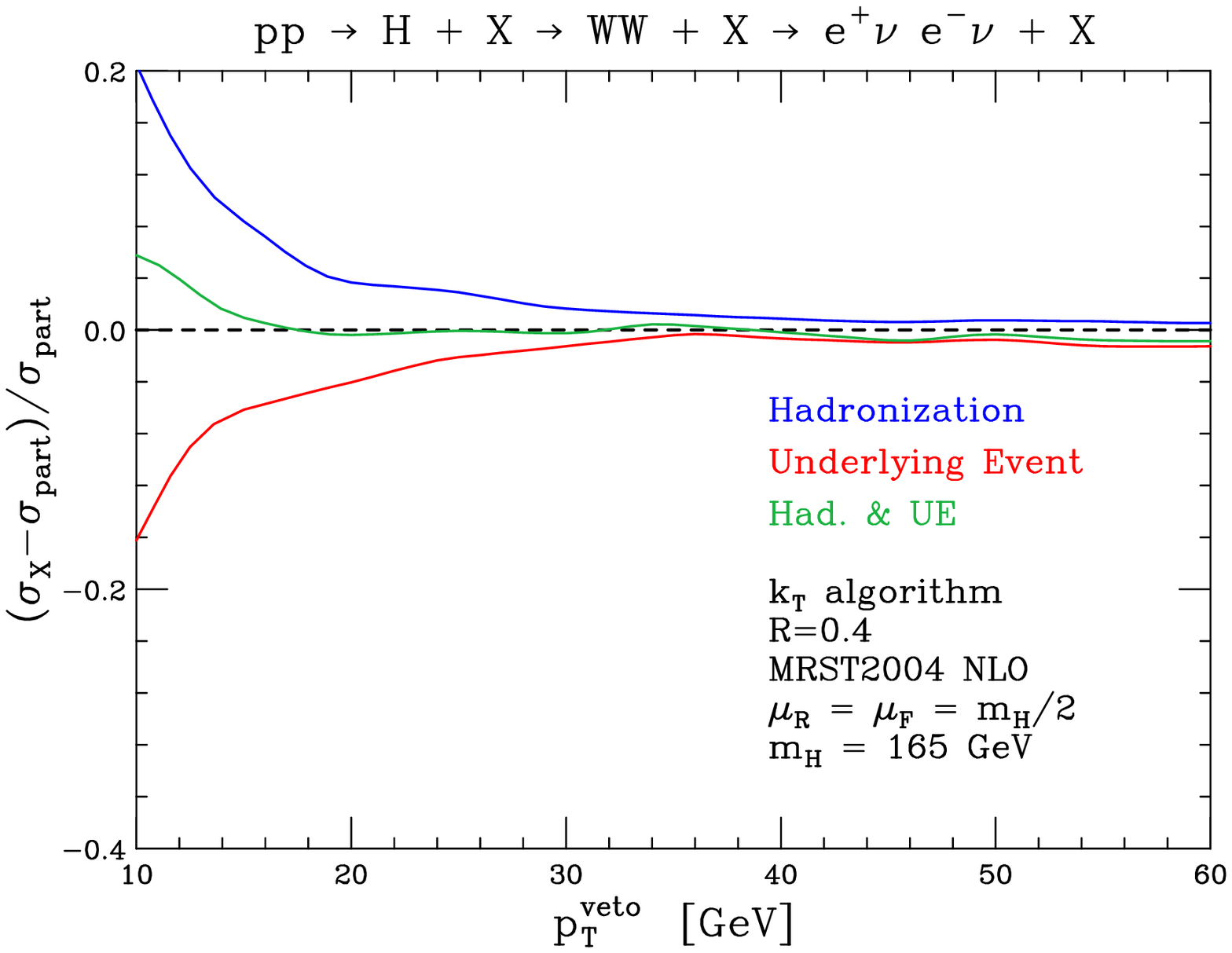}
\includegraphics[width=.48\linewidth]{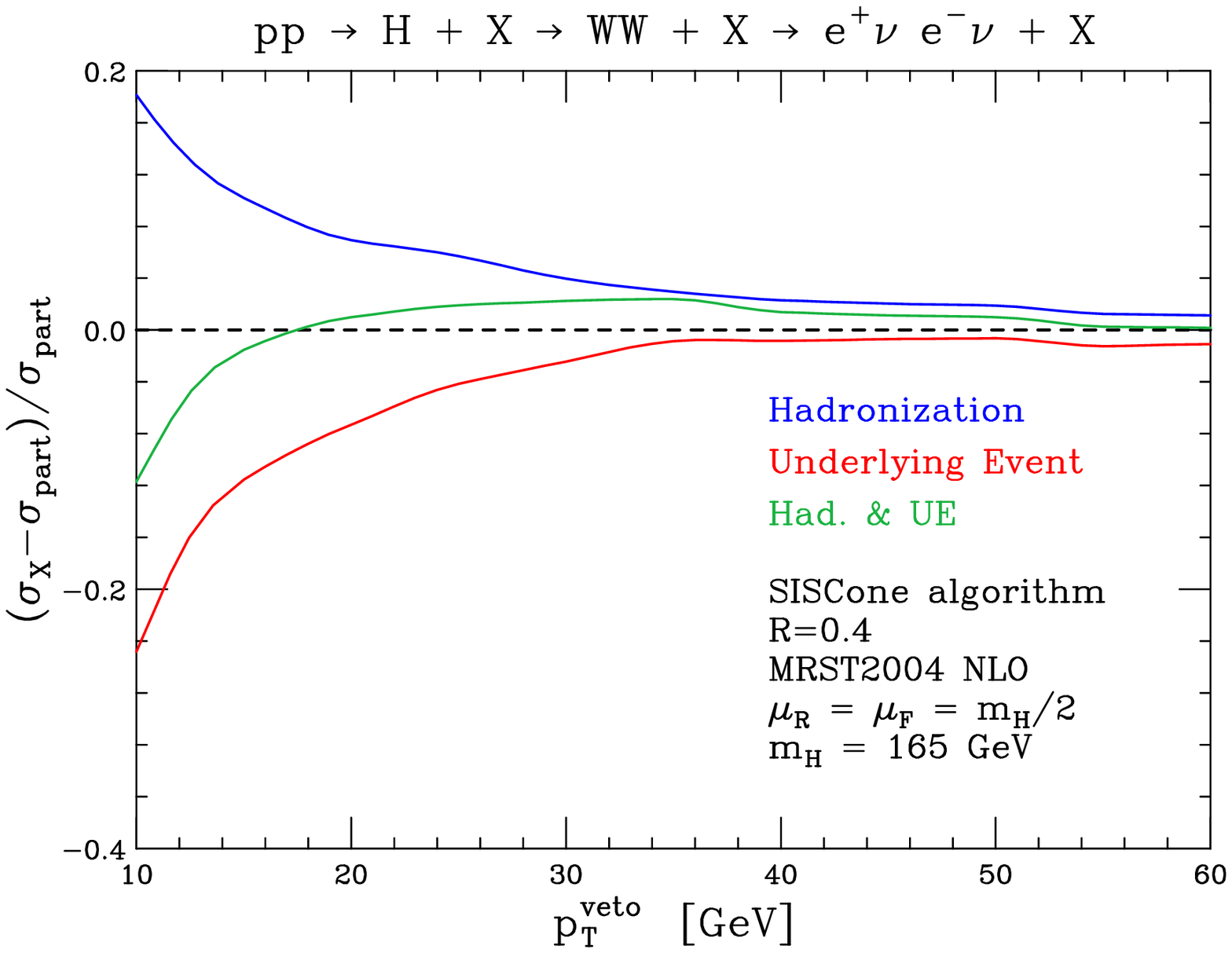}
\caption{ 
Difference of the cross section after signal cuts including the underlying
event and the hadronization model, with respect to the partonic cross section,
from \Bref{Anastasiou:2008ik}.
The cross section is shown as a function of the jet veto for both the
SISCone and the $\kT$ clustering algorithm.}
\label{fig:UEeffect}
\end{figure}
Effects of the order of $10\%$ are found for a jet veto of $25$\UGeV,
with the hadronization and the underlying event acting in opposite directions.
It would be desirable to extend these studies using other shower and
underlying event models.

\subsubsection{VBF}

Vector-boson fusion is available in the ME+PS approach in several matrix-element-based 
generators, such as \alpgen, \MGME\ and \sherpa\ as well as in 
the \POWHEGBOX{} implementation~\cite{Nason:2009ai}.   Another \POWHEG\
implementation will be available soon also in the \herwigpp\ event generator.

At variance with the gluon-fusion process, in VBF, NLO corrections are 
very small and extra QCD radiation in general is rather suppressed\footnote{
     It should be noted that for VBF the electroweak corrections are roughly 
     as large as the QCD ones, but there is currently no attempt to include
     them into a full simulation.
}. This feature is exploited experimentally to enhance the signal over the 
background by requiring a jet veto in the central region.  Due to angular 
ordering a simple PS approach is expected to work well for VBF; in 
order to estimate uncertainties related to details of the parton shower, a 
careful study invoking ME+PS and NLO+PS could be important.  A first study 
along these lines, using ME+PS, was presented in \Bref{DelDuca:2006hk}, 
where indeed substantial agreement between parton-level predictions and those 
merged with the PS on the key observables were found.

A more recent study compared the NLO fixed-order computation with the POWHEG 
implementation interfaced to both \pythia\ and \herwig~\cite{Nason:2009ai}.  
All the most relevant distributions are compared, from the most inclusive to 
the more exclusive. On the former, good agreement with fixed-order 
computations and only mild sensitivity to shower effects is found, confirming 
the small effects due to extra radiation.  However, for the more exclusive
observables, some discrepancies showed up.  As the probably most prominent 
example for such a discrepancy consider the efficiency of the central jet 
veto compared to fixed-order LO and NLO computations.  At low transverse 
momentum, where soft-resummation effects are important, fixed-order 
calculations cannot be reliable and resummation, realised by a parton-shower 
approach, becomes mandatory.  Thus, in this case sizable differences between 
fixed-order calculations and the parton shower are expected and indeed found.  
All in all, jet-veto effects show some appreciable dependence on resummation 
and therefore on the shower algorithm, at least for low $\pT$ veto, see 
Fig.~10 of \Bref{Nason:2009ai}.  

This may deserve a further, more careful study, also including the impact of 
e.g.\ PDFs and the underlying event.  On the same footing, it would be 
interesting to further compare the effect of jet vetoes or jet azimuthal 
correlations to fixed-order calculations that include one extra jet at 
NLO~\cite{Figy:2007kv} and to top-loop-induced 
$\PH + 2\,$jets~\cite{Campbell:2010cz}.

\subsubsection{VH}
Due to its comparably simple structure, Higgs associated production
with a vector boson is available at the NLO and NNLO level; while the former 
is fully worked out, including spin correlations in the decays, such 
differential distributions in general are lacking for the latter.  In 
addition, this process is also implemented in both the \MCatNLO and the 
\POWHEG approach.  A detailed discussion of these implementations, together 
with a tuned comparison with fixed-order results at NLO accuracy is available 
in \Bref{Hamilton:2009za}.  

In addition, standard ME+PS tools such as \alpgen, \MGME, and \sherpa\ are 
also capable to simulate this process.  It should be stressed, however, that 
for searches based on the boosted-Higgs idea of~\cite{Butterworth:2008iy} the 
impact of higher-order corrections to the Higgs decay into $\PQb$~quarks and
especially the impact of hard gluon radiation off the $\PQb$'s is a
crucial ingredient.  Therefore further studies on all possible levels
would be certainly most welcome.

\subsubsection{ttH}

As of today, no public code including either, at the parton-level NLO, or 
a full simulation in the \MCatNLO or \POWHEG frameworks, is available for 
Higgs production in association with top quarks.  However, ME+PS matching is 
available in several matrix-element-based generators, such as \alpgen, 
\sherpa, and \MGME.  

However, as in VH, an accurate simulation of the Higgs decay to $\PQb\PAQb$, 
which includes extra hard radiation at the matrix element level, is 
recommended.

\subsection{Modelling the Higgs boson in scenarios beyond the
Standard Model}

Accurate simulations of Higgs production in scenarios beyond the  
Standard Model will be needed both in the identification of the 
interesting 
signatures and also in the extraction of key information, such as masses,
couplings strengths, and CP structure. As a template one can consider a  
general two-Higgs doublet model, which encompasses the much more studied 
SUSY cases and displays five scalar particles, three neutrals and one charged
pair. Other extensions to higher representations, such as triplets,  
are also often considered.

The status of the MC tools in such models is quite different from that  
outlined above for the Standard Model (SM) Higgs.  As many new physics models are now 
easily implementable in matrix-element generators such as \MGME or in \sherpa, 
ME+PS predictions are de facto available for many interesting scenarios, 
including the minimal supersymmetric SM (MSSM) or extensions such as the 
next-to-MSSM. To our knowledge, however, no 
systematic comparisons with standard PS or fixed-order NLO calculations 
have been performed.

Regarding NLO+PS the availability for cenarios beyond the
Standard Model is much more limited. There 
are, however, notable exceptions. All processes in the new physics 
scenarios that can be obtained from those in the SM by simply rescaling the 
coupling strength and masses can be easily simulated. As an example, consider
VBF and VH for SUSY neutral Higgs production, where the cross sections for the
MSSM Higgs bosons $\Ph$ and 
$\PH$ can be obtained by such a simple rescaling. 

An example where simple rescaling will not work is the case of gluon fusion 
where a $\PQb$-quark loop could give a sizable contribution in the large-$\tan\beta$ 
scenarios. In this case, the usual approach of using the Higgs-effective
theory cannot be applied, and more work is needed (and would be welcome).  In 
addition, in SUSY scenario, effects from heavy coloured states would also 
play a role.

On a similar level, new calculations may be recycled for other channels. For
example, when a $\PQt\PAQt\PH$ calculation will be available in the 
NLO+PS framework,  with a few replacements this could be easily recycled for 
$\PQb\PAQb\PH$.  Extension to include the pseudo scalar 
$\PQt\PAQt\PA$ and $\PQb\PAQb\PA$ would also be straightforward.

Charged Higgs production is another example where a dedicated calculation was 
necessary.  Currently, the most promising mechanism is via the excitation of 
a top quark, either in association ($\PQt \PH^+$, heavy $\PH^+$) or via 
its decay  
($\PQt \to \PH + \PQb$, light $\PH^+$). The first processes is available 
in \MCatNLO ~\cite{Weydert:2009vr}. For the second one, $\PQt\PAQt$ 
production is available
both in \MCatNLO and in \POWHEG{}. However, the subsequent decay 
$\PQt \to \PH + \PQb$ can only be
simulated without spin correlations and NLO corrections at present.

\subsection{Currently used tools and wish list by the experimentalists}

The experimental collaborations use a collection of publicly available tools, which are 
properly versioned, maintained, well described, and referenced.  Typically
all multi-purpose or parton-level event generators fall into this category.
A list of currently used Monte Carlo generator tools for mass production of 
fully simulated and reconstructed events is presented in Table~\ref{tab:nlomctabone}.

\begin{table}[t]
\begin{center}
\caption{Combined set of Monte Carlo event generator tools currently used for mass production by the ATLAS and CMS collaborations.  Most of these tools are used by both the collaborations, few are used in one collaboration only.  The version numbers in the table represent the latest versions used but lower versions are also used in experiments due to validation and coordinated production schedules.}
\footnotesize{
\begin{tabular}{|c|c|c|c|c|} \hline
Type               &    Physics        &	Generator &	Version       &	Comments \\  
             &     processes       &	 &	      &	 \\  
\hline 
                       &                                       &                       &                      &                       \\  
Multi-purpose & 	EW, QCD,            &      PYTHIA6   &    6.423      & Standard tune D6T  \\
LO generators     & SM Higgs,                      &	                  &	                   &   with Q2 PS,            \\
                       &   MSSM Higgs                 &                       &                      &  PS hadronization  for                      \\  
                       & SUSY, exotica                &                        &                      & MADGRAPH, ALPGEN, \\   
                       &                                      &                        &                       &and TopRex   \\  
                       &    QCD di-jet                  &   PYTHIA8      &    8.145          &                                 \\  
                       &                                      &                        &                       &                       \\  
                       &  QCD studies                 &  HERWIG6    &     6.510        &  PS hadronization  for   \\  
                       &                                       & HERWIG++  &    2.4.2          & MADGRAPH, ALPGEN,  \\  
                       &                                       &                       &                      & MC@NLO and POWHEG          \\  
                       &                                       &                       &                      & interfaced to JIMMY  \\ 
                      &                                       &                       &                      &   (V4.31) for UE/MI                     \\  
                     &                          &                       &                      &                       \\     
\hline 
                     &                          &                       &                      &                       \\   
Dedicated  	     &  gg$\to VV$                     &  gg2ZZ,   gg2WW  &   1.0.0 &                  \\  
LO  generators       &          $\PQt\PAQt$                &   TopRex     &       4.11   &                       \\   
                     &                          &                       &                      &                       \\   
\hline 
                     &                          &                       &                      &                       \\    
 Multi-leg         & QCD,                    &  MADGRAPH  &  4.4.13        &        \\  
  matrix-element  & Q+jets, QQ+jets,       &                        &                 &                               \\  
 LO  generators    & $\gamma$+jets, $\gamma\gamma$+jets, &   &           &                              \\  
                       &  $V$+jets, $VV$+jets, &     SHERPA        &   1.2.2          &   \\        
                       & $\PQt\PAQt$, single top, $\PZ'$  &                       &                      &                       \\  
                       &                                       &                       &                      &                       \\  
                      &    $V$+jets, $V\PQb\PAQb$,            & ALPGEN &     2.13  &  \\  
                      &    QCD, $\PQt\PAQt$          &              &                 &                                      \\  
                     &                          &                       &                      &                       \\   
\hline 
                      &                          &                       &                      &                       \\   
NLO event      &DY, $\PW\PW$,                        & MC@NLO &   3.41/3.42      &                        \\  
generators     & $\PQt\PAQt$, single top,  &                       &                      &                      \\  
                       &    ggF Higgs                    &                      &                      &                       \\  
                       &                                       &                      &                      &                       \\  
                       & Drell--Yan, Higgs                       &  POWHEG      &                      &                       \\  
                      &                                       &                       &                      &                       \\   
\hline 
\end{tabular}
}
\label{tab:nlomctabone}
\end{center}
\end{table}

 
For some of the inclusive-cross-section calculators, however, are just privately communicated
and, thus, are not as systematically maintained as many of the event generators tools above.  
This issue needs to be addressed to achieve a more uniform prescription on how to incorporate 
and reference these calculations properly and make the results reproducible.
On a somewhat similar footing, it is important to reference also the standard
tools in a proper manner, by indicating both version number and tune identifier,
and by making clear which tools have been interfaced with each other\footnote{
    For example, for a simulation based on \pythiasix, the exact version number
    \pythia~6.xxx {\em plus} the properly documented underlying event tune, 
    like DW, must be indicated.  Similarly, when interfaced with \alpgen,
    the results should be called ``\alpgen~2.xxx+\pythiasix.yyy~(DW)'' or 
    similar.  
}.

At the same time, flexibility of cross-section calculator tools that 
allow detailed experimental cut and efficiency analyses and the
inclusion of fully exclusive final states into their results are critical.  
It would therefore be highly desirable to turn as many NLO calculations as 
possible into hadron-level event simulations.  A possible avenue, of course, 
would be to ask the authors of the respective NLO codes to facilitate turning 
their calculations into a full-fledged simulation, through either \POWHEG{} or
\MCatNLO{} techniques.  Automation of the whole process (NLO computation and 
interface to the parton showering) is also possible.


It should be stressed, however, that it is necessary for all theoretical 
tools to be maintained and remain accessible at all times.   
A commonly agreed central code repository, such as
the Hepforge database ({\tt http://www.hepforge.org/}) would be highly 
useful for improved accessibility and maintenance.

Based on the most immediate needs for the Higgs searches in the experiments, 
a wish list for theorists is also proposed as follows for the Higgs-signal 
MC and for background studies.
For Higgs signal MC generators, main processes that need urgent implementation are:
\begin{itemize}	
\item NLO corrections to $\PH\to \PQb\PAQb$  decay\footnote{
  An unreleased patch of HERWIG is available for NLO corrections to 
$\PH\to \PQb\PAQb$  decay.  For the experimental collaborations,  it       
 would be ideal to have this process implemented in the official 
\herwigpp\ release.
ME+PS corrections to $\PH\to \PQb\PAQb$  decay can be easily implemented 
also in \sherpa\ in which these corrections are already implemented for 
$\PQt$ decays.},
\item Finite $\Mt/\Mb$ in $\Pg\Pg\to \PH$ production  (especially relevant 
for SUSY Higgs),
\item $\Pp\Pp\to (\PQt\to \PH^+ \PQb) \PAQt$, including a treatment of  
$\PQt \to \PH^+ \PQb$ decay
with the
same level of accuracy achieved in $\PQt\to \PW+ \PQb$ in \MCatNLO{} and 
\POWHEG{},
\item $\Pp\Pp\to \PQt\PAQt \PH$ and $\Pp\Pp\to \PQb\PAQb \PH$,
\item $\Pp\Pp\to \PQb\PH$. 
\end{itemize} 

For background MC generators, main processes needed to be implemented at NLO are (listed in order of implementation complexity):
\begin{itemize}
\item $\PQq\qbar \to \PZ\PZ^*$  (now available without gamma interference),
\item $V\PQb\PAQb$,
\item $\PQt\PAQt \PQb\PAQb$, $\PQt\PAQt$+jets, and $V$+jets
In particular, $V$+jets processes are the most urgent since they have the largest expected cross section of the three.
It is however already possible to perform precise measurements of $V$+jets production with 
the LHC data and test the theoretical predictions given the expected high luminosity in 
the next year.  This will help understanding vector-boson production in association 
with jets for better understanding of the crucial background process to many Higgs search channels.
\end{itemize} 

Finally, for the optimization of the experimental Higgs searches and their interpretation, 
there are two main issues: signal and background predictions and the estimation of their 
uncertainties both inclusively and as a function of most important quantities used to 
separate signal from background in the experimental selections. 

For signal expectations, we have to rely on theoretical predictions.
It is recommended to use as much as possible all available higher-order MC generators, 
rather than using $K$-factors. 
In many cases NLO MC generators exist, as described in this section, thus there is 
no reason to use LO MC tools and apply $K$-factors NLO/LO to correct the 
LO predictions to NLO.
In some cases NNLO calculations exist for SM Higgs production ($\Pg\Pg\PH$ differential cross-section 
calculators) 
, but not a full event generator. 
In this case, in collaboration with the authors, the NLO MC team will provide values of 
the re-weighting factors NNLO/NLO for the Higgs momentum and pseudo-rapidity distributions.

The NLO MC team will also provide guidelines on how to estimate the theoretical uncertainties 
on signal production predictions as a function of most important measurable quantities 
used to discriminate signal from background in the experimental analyses (based on 
the compilation of the common set of variables and cuts by the experimental collaborations).

Future work of the NLO MC team will be devoted to the following open questions: 
Can we devise methods to test the signal predictions before the signal itself is measured, 
by using similar and already measurable SM processes? 
For example, how a precisely the measurement of the $\PQt\PAQt$ cross section 
may help to 
reduce the uncertainty on Higgs production via $\Pg\Pg$ fusion (such as 
$\Pg\Pg\PH$)? 
More generally what measurable processes may serve as the benchmark and 
the validation of the signal predictions?

For background predictions, a number of processes can and will be measured with the 
data collected during the current and coming years at the LHC.  
$V$+jets and $V\PQb\PAQb$ predictions are urgent because they can be studied already with the LHC data. 
For all background processes, experimentalists should review methods to measure 
these processes in some control regions where data statistics is abundant and 
to extrapolate the background predictions in some signal regions where data statistics 
is expected to be small. 
The theoretical predictions should give guidance and improve the precision of 
the extrapolated results.
For this, the NLO MC team will provide prescriptions on how to assign theoretical 
uncertainties to background predictions in the signal regions.

\subsection{Further issues and studies}
\subsubsection{Which tools to use?}
With the advent of better and more precise tools it becomes increasingly
important to understand which tool to pick for a given study, in order
to optimally use the best tools. Obviously it is very hard to find a 
solution that fits all eventualities, but we believe it is still worthwhile
to formulate a few guidelines:
\begin{itemize}
\item Never use one tool alone.\\
      Clearly, different tools have different accuracies and they may employ
      different approximations.  This in turn may lead to systematic  
      uncertainties, which can only be addressed by using different tools.
      Prime examples for this are uncertainties related to the underlying
      event or hadronisation, which involve a big amount of modelling.  While
      it is tempting to simply use only various tunes of the same generator
      it may be important to see if various models (e.g.\ \pythiaeight\ vs.\ 
      \herwigpp) lead to systematically different outcomes.  The same also 
      holds true for uncertainties related to parton showering etc.
\item Employ the accuracy dictated by your analysis:\\
      For very inclusive studies, like e.g.\ the rapidity distribution
      of the Higgs boson, NNLO accuracy is available and should be used.  
      For exclusive final states, the best accuracy available at the moment
      is given by the codes employing the \POWHEG and \MCatNLO techniques,
      which would yield NLO accuracy for relatively inclusive quantities
      such as the Higgs-boson rapidity, LO accuracy for more exclusive
      quantities such as the $\pT$ distribution of the first jet,
      and PS accuracy for all further jets.  In contrast ME+PS tools yield
      LO accuracy for all jets, as long as the corresponding MEs have been
      employed.  Thus, if an analysis relies on the correct description of 
      many jets, employing the ME+PS tools is preferred, while the NLO+PS 
      tools are the tools of choice for more inclusive quantities.  Conversely,
      this suggest that pure PS tools should typically not be used.      
\end{itemize} 

\subsubsection{Choice of parton distribution functions}
For hadron colliders, parton distribution functions (PDFs) play an important 
role in determining the outcome of theoretical predictions.  Common lore
suggests that the order of the PDFs must be consistent with the order of the 
predictions.  For leading-order predictions, LO PDFs should be used while 
for NLO predictions, NLO series of PDFs need to be used.  This simple picture,
however, is somewhat blurred by the parton showers, since they partially 
include higher-order effects, as discussed above.  


This issues is further obfuscated by the often large impact the PDF has on 
the underlying event simulation.  It is therefore important to ensure that
the PDFs are used in a conscientious way -- changing a PDF without changing 
other parameters may lead to huge and unphysical effects.  Therefore, in 
order to assess PDF uncertainties by comparing apples to apples, it would be 
paramount to have at hand various tunes for the underlying event etc.

\subsubsection{Interfacing codes}
Many of the tools discussed in this section are based on interfacing a
parton-level calculation at leading (\alpgen\ and \MGME) or next-to leading 
order (\POWHEGBOX and \MCatNLO) to a full event generator that includes the 
parton showers, the underlying-event simulation, and hadronisation.  While
most likely uncritical for very inclusive observables such as cross section,
a number of issues may arise for more exclusive observables such as jet 
production, jet spectra, and jet vetoes.  Since they have not been studied 
in details, it may be worth pointing out a few of these issues: 
\begin{itemize}
\item Typically, for NLO calculations, an NLO PDF is used, and the strong 
      coupling constant of this PDF is employed, to guarantee the theoretical
      consistency of the calculation.  In a similar fashion, already on the
      tree-level choices concerning PDFs and $\alphas$ are made.  When 
      interfacing such parton-level codes with their choice of PDFs etc.\ 
      with a parton-shower code such as \pythia\ or \herwig\ in a specific 
      tune, which also includes a definition of PDFs and $\alphas$.  
      This renders the evaluation of PDF and scale uncertainties a tricky
      task, for which no prescription has been defined yet.  
\item In this context it is worth noting that for ME+PS tools, which {\em sit} 
      on a full event generator, the merging algorithm often employs the
      Sudakov form factors of the underlying parton shower.  This may then
      lead to the counter-intuitive effect that a {\em harder} tune for the
      parton shower yields softer jets when interfaced with parton-level
      MEs.
\item As discussed above, quantities such as the central-jet veto in VBF
      depend on resummation properties -- typically the realm of the parton 
      showers in the simulations currently used by experimenters.  It would
      therefore be worthwhile to check for these effects in different
      interfaced codes, including the impact of the underlying event in 
      a more complete fashion.  
\end{itemize}

\subsection{Conclusions}

As we are entering the LHC running phase, we have available several very 
accurate {\em theoretical-style} predictions in the form of parton-level integrators that 
can output histograms for any IR-safe observable.  On the other hand, 
Monte Carlo event generators with NLO accuracy are now (or will be soon)  
available for all the relevant Higgs production processes. A systematic 
comparison between various implementations, PS programs and fixed or 
improved {\em theoretical-style} calculations is now possible. 

In this brief note we have listed the available tools and also given an 
example (taken from \Bref{Butterworth:2010ym}) on how such comparisons 
can be made.

The tunes for the various Monte Carlos need to be re-established as 
components relevant to assess, for instance, systematic uncertainties due to
PDFs.  We expect this to be a continuous process as the implemented order of 
calculations change and new codes and physics processes become available.  
Finally, it is important to establish 
\begin{itemize}
\item a consistent set of Standard Model parameters for MC tools  
      (the MC4LHC group is to provide a suggested set of these parameters),
\item a consistent and complete way to reference the tools used,
\item a common code repository,
\item procedures to cross compare and validate different codes and 
      implementations,
\item procedures to assess systematic uncertainties of the theoretical 
      predictions and simulations.  
\end{itemize}

\clearpage

\newcommand{\ssA}{{\scriptscriptstyle{A}}}
\newcommand{\ssS}{{\scriptscriptstyle{S}}}
\newcommand{\ssH}{{\scriptscriptstyle{H}}}
\newcommand{\ssh}{{\scriptscriptstyle{h}}}
\newcommand{\ssW}{{\scriptscriptstyle{W}}}
\newcommand{\ssT}{{\scriptscriptstyle{T}}}
\newcommand{\ssZ}{{\scriptscriptstyle{Z}}}
\newcommand{\bqas}{\begin{eqnarray*}}
\newcommand{\eqas}{\end{eqnarray*}}
\newcommand{\nl}{\nonumber\\}
\def\mnew{\mpar{\hfil NEW \hfil}\ignorespaces}
\newcommand{\lpar}{\left(}                            
\newcommand{\rpar}{\right)} 
\newcommand{\lrbr}{\left[}
\newcommand{\rrbr}{\right]}
\newcommand{\lcbr}{\left\{}
\newcommand{\rcbr}{\right\}} 
\newcommand{\rbrak}[1]{\lrbr#1\rrbr}
\newcommand{\bq}{\begin{equation}}                    
\newcommand{\eq}{\end{equation}}
\newcommand{\bqa}{\arraycolsep 0.14em\begin{eqnarray}}
\newcommand{\eqa}{\end{eqnarray}}
\newcommand{\ba}[1]{\begin{array}{#1}}
\newcommand{\ea}{\end{array}}
\newcommand{\ben}{\begin{enumerate}}
\newcommand{\een}{\end{enumerate}}
\newcommand{\bei}{\begin{itemize}}
\newcommand{\eei}{\end{itemize}}
\newcommand{\eqn}[1]{Eq.(\ref{#1})}
\newcommand{\eqns}[2]{Eqs.(\ref{#1})--(\ref{#2})}
\newcommand{\eqnss}[1]{Eqs.(\ref{#1})}
\newcommand{\eqnsc}[2]{Eqs.(\ref{#1}) and (\ref{#2})}
\newcommand{\eqnst}[3]{Eqs.(\ref{#1}), (\ref{#2}) and (\ref{#3})}
\newcommand{\eqnsf}[4]{Eqs.(\ref{#1}), (\ref{#2}), (\ref{#3}) and (\ref{#4})}
\newcommand{\eqnsv}[5]{Eqs.(\ref{#1}), (\ref{#2}), (\ref{#3}), (\ref{#4}) and (\ref{#5})}
\newcommand{\tbn}[1]{Tab.~\ref{#1}}
\newcommand{\tabn}[1]{Tab.~\ref{#1}}
\newcommand{\tbns}[2]{Tabs.~\ref{#1}--\ref{#2}}
\newcommand{\tabns}[2]{Tabs.~\ref{#1}--\ref{#2}}
\newcommand{\tbnsc}[2]{Tabs.~\ref{#1} and \ref{#2}}
\newcommand{\fig}[1]{Fig.~\ref{#1}}
\newcommand{\figs}[2]{Figs.~\ref{#1}--\ref{#2}}
\newcommand{\sect}[1]{Section~\ref{#1}}
\newcommand{\sects}[2]{Section~\ref{#1} and \ref{#2}}
\newcommand{\sectm}[2]{Section~\ref{#1} -- \ref{#2}}
\newcommand{\subsect}[1]{Subsection~\ref{#1}}
\newcommand{\subsectm}[2]{Subsection~\ref{#1} -- \ref{#2}}
\newcommand{\appendx}[1]{Appendix~\ref{#1}}
\newcommand{\hsp}{\hspace{.5mm}}
\def\negs{\hspace{-0.26in}}
\def\negsh{\hspace{-0.13in}}
\newcommand{\barq}{\overline q}
\newcommand{\barb}{\overline b}
\newcommand{\bmid}{\Bigr|}
\definecolor{Orange}{named}{Orange}
\definecolor{Purple}{named}{Purple}
\newcommand{\htr}[1]{{\color{red} #1}}
\newcommand{\htb}[1]{{\color{blue} #1}}
\newcommand{\htg}[1]{{\color{green} #1}}
\newcommand{\hto}[1]{{\color{Orange}  #1}}
\newcommand{\htp}[1]{{\color{purple}  #1}}
\definecolor{Lightblue}{cmyk}{0.9,0.1,0.1, 0.3}
\newcommand{\lbl}[1]{\color{Lightblue}#1\color{Lightblue}}
\providecommand{\lsim}
{\;\raisebox{-.3em}{$\stackrel{\displaystyle <}{\sim}$}\;}
\providecommand{\gsim}
{\;\raisebox{-.3em}{$\stackrel{\displaystyle >}{\sim}$}\;}
\section{Higgs pseudo-observables\footnote{S.~Heinemeyer and G.~Passarino.}}
\subsection{Introduction}
Recent years have witnessed dramatic advances in technologies for computing
production and decay of Higgs bosons, critical to the program of calculations
for collider physics.
The main goals of this Report have been to calculate inclusive cross sections 
for on-shell Higgs-boson production and Higgs-boson branching ratios (BRs). 
Therefore, Higgs-boson decays are considered, in the experimental analyses, 
as on-shell Higgs bosons decaying according to their BR's, including 
higher-order effects.
However, the quantities that can be directly measured in the (LHC)
experiments are cross sections, asymmetries, etc., called {\em Realistic
Observables} (RO, see below).
The obtained results depend on the specific set of experimental cuts that have
been applied and are influenced by detector effects and other details
of the experimental setup. In order to determine quantities like Higgs-boson
masses, partial widths, or couplings from the RO a deconvolution
procedure (unfolding some of the higher-order corrections, interference
contributions etc.) has to be applied. These secondary quantities are called
{\em Pseudo Observables} (PO, see below).
It should be kept in mind that the procedure of going from RO to the
PO results in a slight model dependence. 

\subsection{Formulation of the problem}
With respect to the measurement of Higgs-boson quantities at the LHC, 
some of the above mentioned aspects are mostly neglected so far.
Sophisticated issues, such as off-shell effects of Higgs and interference 
between background and signal, have not been included in experimental
analysis since 
it is assumed that they should not be so relevant for $\MH < 200\UGeV$,
within the SM. (The case of the most popular extension of the SM, the
Minimal Supersymmetric SM (MSSM) will be briefly discussed later.)
It should be noted that the statement on low importance of off-shellness for
the regime of low Higgs masses just comes from naive analysis of the
ratio $\GH/\MH$: for a SM Higgs boson below $200\,\UGeV$, the natural 
width is much below the experimental resolution; and on certain
assumptions about the vanishing of imaginary parts in the amplitudes.
However, in our opinion, one 
should analyze more carefully how much this ratio banishes the off-shellness, 
given an increase in $\Pg\Pg$ luminosity at small values of $x$.
In any case, the experimental strategy for searching a light Higgs boson 
has always been to produce an on-shell Higgs and model its decay in a
Monte Carlo (MC) generator.
No effort has been devoted to analyze how MCs such as {\sc Pythia}, 
{\sc Herwig},
or {\sc Sherpa} are treating the Higgs-boson width internally. Especially for 
heavier Higgs bosons, we expect studies that include the interference, but it is clear 
that most probably this will only be done at LO with MC generators for 
$\Pp \Pp \to n\,$fermions. 

There are few examples of theoretical studies as well: quite a while ago, 
\Bref{Dixon:2003yb} presented a study of the interference of a light 
Higgs boson with the continuum background in $\Pg\Pg \to \PH \to \PGg \PGg$. 
Although the effect turned out to be fairly small, there may be other cases 
where such interference effects might be sizable, maybe even in channels 
where there will be earlier sensitivity to SM Higgs.

While the implementation of higher-order corrections to Higgs production 
cross sections and Higgs decays does not represent a problem anymore, very
little effort has been devoted in analyzing the interference effect 
between Higgs-resonant and background diagrams. ATLAS and CMS studies have been 
done with full simulation, but without the interferences.

\subsection{Examples of pseudo-observables}
In the following we define the relevant quantities from a more general way of 
looking at this question.
Let us split {\em signal} ($S$) from {\em background} ($B$) at the 
diagrammatic level. In principle one could refer to some {\em idealized} 
experimental cross section, but here we advocate another road, the one to 
define universal quantities that have the same meaning in all schemes and 
models, see \Bref{Passarino:2010zz}.
Therefore, we go from {\em data} to {\em predictions} which are made of 
$\mid S\,\oplus\,B\mid^2$. Usually $S$ and $B$ come from different sources, and
$B$ is not always complete, e.g.\ the best prediction is reserved for $S$
(usually including as many loops as possible) while often $B$ is
only known at LO; furthermore, $S\,\otimes\,B$ is usually discarded. 

In order to pin down the theoretical uncertainty as much as possible all
calculations of $S$ are based on a consistent procedure; one does not use 
$\alphas$ at four loops in any LO calculation etc.  
In the end one is interested in the extraction of Higgs-boson masses,
widths etc.\ by comparing experimental measurements with theory predictions.
Therefore, the main question that we are going to address in this section is
about the meaning of any future comparison (theory versus data) where, 
for instance, $\Gamma(\PH \to \PGg\PGg)$ computed at $n$~loops is compared 
with something extracted from the data with much less precision and, sometimes, 
in a way that is not completely documented.
Without loss of generality and to continue our discussion it will be useful 
to introduce an elementary glossary of terms:

\begin{table}[h!]\centering
\setlength{\arraycolsep}{\tabcolsep}
\renewcommand\arraystretch{1.2}
\begin{tabular}{|l|l|}
\hline 
RD = & real data \\
RO = & going from {\em real data} to distributions with cuts 
             defines RO$_{\rm exp}$, \\
           & e.g.\ from diphoton pairs $(E, p)$ to 
             $M(\PGg\PGg)$; given a model, e.g.\ SM, \\
          &   RO$_{\rm th}$ can be computed\\
PO = & transform the {\em universal intuition} of a 
             {\em non-existing} quantity into an {\em archetype}, \\
{}         & e.g.\ $\sigma(\Pg\Pg \to \PH), \Gamma(\PH \to \PGg \PGg)$, 
             $\mbox{RO}_{\rm th}(\MH, 
             \Gamma(\PH \to \PGg \PGg), \dots)$ \\
{}         & fitted to $\mbox{RO}_{\rm exp}$ (e.g.\ RO${}_{\rm exp} 
             = M(\PGg\PGg)$) 
             defines and extracts $\MH$ etc. \\
\hline
\end{tabular}
\label{Gloss}
\end{table}
Examples of POs used at LEP can be found in \Bref{Bardin:1999gt}, for LHC
(e.g.\ $\Pg\Pg \to 4\,$fermions) see \Bref{Passarino:2010qk}.
In calculations performed to date the background 
$\Pp \Pp \to 4\,$f is generated at
LO, the production cross section (e.g.\ $\Pg\Pg \to \PH$) is known at NNLO, 
and the on-shell decay is known at NLO, including electroweak effects. Ideally 
one should extract the $\pT$ information from production and boost the decay 
rate computed in the Higgs rest frame in order to have a consistent matching in
production$\,\times\,$decay (P$\,\otimes\,$D). Next step is the 
replacement of P$\,\otimes\,$D with a Breit--Wigner, next-to-next the correct 
folding with a Dyson re-summed Higgs propagator. It is worth nothing that there 
is still a mismatch between background (LO) and signal (NNLO$\,\times\,$NLO) 
and that for this channel we have more than one unstable particle.
This leads us to consider the following, recommended, strategy: 
to go via idealized (model-independent?) RO distributions and from 
there then going to the POs, with the following steps:
\begin{itemize}

\item {Step 0)} Use a (new) MC tool -- the PO code -- to fit ROs;

\item {Step 1)} understand differences with a {\em standard} event generator 
plus detector simulation plus calibrating the method/event 
generator used (which differ from the PO code in its theoretical content);

\item {Step $\ge$ 2)} document the results of the analysis and understand 
implications.

\end{itemize}
\subsection{Experimental overview with theoretical eyes}
MC generators are usually selected for specific processes and used for all 
relevant final states. MC generators for Higgs production and decay, e.g.\ in 
CMS, are {\sc Pythia} and {\sc POWHEG}; for description and differences see 
\Bref{Alioli:2008tz}.
For Higgs production {\sc Pythia} is similar to {\sc POWHEG}, the main 
difference being normalization which is LO in {\sc Pythia} and NLO in 
{\sc POWHEG}.

The strategy of describing Higgs signal as production$\,\otimes\,$ decay
is based on the small value of width/mass (for a light Higgs) but also on the 
scalar nature of the Higgs resonance, i.e.\ there are no spin correlations, 
opposite to the case of $\PW/\PZ$ bosons. 
Therefore the typical experimental strategy for analyzing the Higgs signal is 
based on generating events with {\sc POWHEG}, storing them and using 
{\sc Pythia} for the remaining shower. 
The correct definition of production$\,\otimes\,$decay is better formulated
as follows: the MC {\em produces} a scalar resonance, the Higgs boson, with a
momentum distributed according to a Breit--Wigner where peak and width are
related to the on-shell mass and width of the Higgs boson. In other words
what has been done amounts to generating Higgs virtuality, ${\hat s}$, 
according to the replacement
\[ \delta\lpar {\hat s} - \MH^2\rpar\; \rightarrow\; \left\{
\begin{array}{ll}
\frac{1}{\pi}\,\frac{\MH\,\GH}
{\lpar {\hat s} - \MH^2\rpar^2 + \lpar \MH\,\GH\rpar^2} &
\mbox{MC@NLO} 
\\ \\
\frac{1}{\pi}\,\frac{{\hat s}\,\GH/\MH}
{\lpar {\hat s} - \MH^2\rpar^2 + 
\lpar {\hat s}\,\GH/\MH\rpar^2} &
\mbox{Pythia/POWHEG} 
\end{array}
\right.
\]
where $\MH, \GH$ are the on-shell mass and width.
Furthermore, Higgs-boson production (e.g.\ via $\Pg\Pg$ fusion) is also 
computed at ${\hat s}$, a procedure which does not guarantee gauge invariance 
at higher orders if the background is not included at the same order.
As a consequence of this approach, no Higgs-boson propagator appears and
the most important quantity at LHC -- the Higgs-boson {\em mass} -- appears only
through the peak position of the momentum distribution in Higgs production.

It is not the aim of this section to discuss how the shower is performed or
the NLO accuracy of the MC; we focus here on the treatment of the invariant-mass
distribution, e.g.\ the way a Breit--Wigner distribution is implemented. 
For instance, {\sc POWHEG} uses a running-width scheme for the resonance while 
{\sc MC@NLO} implements a fixed-width scheme, $\GH(\MH)$; therefore, the 
different treatments are sensitive to thresholds (e.g.\ $\PAQt \PQt$), a fact 
that becomes relevant for high Higgs-boson masses where, in any case, the whole 
procedure is ambiguous since the Higgs-boson width becomes larger and larger. 

The main point here is that both schemes are equally inadequate if the Higgs
boson is not light and propagator effects should be included. When talking
about NLO or NNLO effects most people visualize them as a lot of gluon lines
attached to the production triangle in $\Pg\Pg$ fusion; there is an often
forgot place where NLO  effects show up, the propagator function. The unusual
aspect of these corrections is that they manifest themselves in the denominator 
(the {\em propagator}), transforming a {\em bare} mass into a 
{\em complex pole}, a basic property of the $S$-matrix.
\subsection{Theoretical background}
Our review here will mention only the minimal material needed for the
description of the proposed solution.
To summarize, extraction of POs depends on many details, experimental cuts, 
detector effects etc., and requires deconvolution/unfolding.
There are also different priorities: from the theory side we need a {\em crystal
clear} definition, e.g.\ what is the correct definition of mass for an unstable
particle. 
The quest for a proper treatment of a relativistic description of unstable particles 
dates back to the sixties and to the work of Veltman~\cite{Veltman:1963th}; more 
recently the question has been readdressed by Sirlin and 
collaborators~\cite{Grassi:2000dz}.

The Higgs boson, as well as the $\PW$ or $\PZ$ bosons, are unstable 
particles; as such they should be removed from in/out bases in the
Hilbert space, without changing the unitarity of the theory. 
Concepts as the production of an unstable particle or its partial decay widths, 
not having a precise meaning, are only an approximation of a more complete 
description, see \Brefs{Actis:2006rc,Passarino:2010qk}.
From the experimental side priorities are on how to extract couplings
(can couplings be extracted?) etc. For a comprehensive analysis of the problem 
see \Bref{Duhrssen:2004cv}.

Concerning the definition of the Higgs-boson mass the object we have to deal
with is the complex pole of the Dyson re-summed propagator, 
whereas all MC implementations have been done with the on-shell mass definition.
In order to have these deviations under control it would be required
to (a)~investigate what is included in the MC tools actually used by the
experiments and (b)~to compare this to the results obtained from an MC tools
using the correct mass definition.
However, right now this cannot be done with realistic ATLAS/CMS distributions.
Hence, the strategy should be limited to: take latest ATLAS/CMS MC tools, use
(at most) a box detector (acceptance cuts, no resolutions) and try for a
closure test with state-of-the-art tools and document the findings.

There is no perfect solution to the problem but our suggestions are as follows.
As an example we take a process $i \to f$, 
e.g.\ $\Pg\Pg \to \PH \to \PGg\PGg$ that is already 
described by a two-loop set of diagrams, and parametrize the amplitude as
\bq
A\lpar i \to f\rpar = A_{\PH}\lpar i \to \PH \to f\rpar + 
A_{\rm back}\lpar i \to f\rpar,
\qquad
A_{\PH}\lpar i \to \PH \to f\rpar = 
\frac{S_i({\hat s})\,S_f({\hat s})}{{\hat s} - s_{\PH}},
\eq
where $s_{\PH}$ is a complex  quantity, the Higgs complex pole, usually 
parametrized as
\bq
s_{\PH} = \mu^2_{\PH} - i\,\mu_{\PH}\,\gamma_{\PH}.
\eq
It is the tough life of an unstable state whose energy (even in a non-relativistic
theory) is doomed to be complex. Kinematics, of course, is always real, and $s$ is
the corresponding invariant at the parton level. $S_{i,f}$ are the matrix
elements for the process $\Pg\Pg \to \PH^*$ and $\PH^* \to \PGg\PGg$. Theoretically 
speaking, these matrix elements alone are ill-defined quantities if $s$ is 
arbitrary and this reflects the intuition that only poles, their residues and 
non-resonant parts are well defined, e.g.\ they respect gauge invariance. 
Therefore, it is better to perform the following split in the amplitude:
\bqa
A_{\PH} &=& \frac{S_i(s_{\PH})\,S_f(s_{\PH})}{{\hat s} - s_{\PH}}
+ \frac{S_i({\hat s}) - S_i(s_{\PH})}{{\hat s} - s_{\PH}}\,S_f(s_{\PH})
+ \frac{S_f({\hat s}) - S_f(s_{\PH})}{{\hat s} - s_{\PH}}\,S_i(s_{\PH})
\nl
{}&+& \frac{\Bigl[ S_i({\hat s}) - S_i(s_{\PH})\Bigr]\,
            \Bigl[ S_f({\hat s}) - S_f(s_{\PH})\Bigr]}{{\hat s} - s_{\PH}}
= A_{\PH,{\rm signal}} + A_{\PH,{\rm non-res}},
\eqa
and to include $A_{\PH,{\rm non-res}}$ in $A_{\rm back}$, the latter given by all
diagrams contributing to $\Pp\Pp \to \PGg\PGg$ that are not $\PH\,$-resonant. They
can be classified as follows:
\begin{itemize}  
\item{LO{}} $\quad \qbar \PQq  \to \PGg\PGg$,
\item{beyond LO} $\quad \qbar \PQq \to \PGg\PGg$ and $\Pg\Pg \to \PGg\PGg$.
\end{itemize}
In case NLO is included one should worry about additional photons in the final
state and this influences, inevitably, the POs definition. After that, let us 
define 
\bq
\frac{1}{\hat s}\,\int\,dPS\,\bmid 
\frac{S_i(s_{\PH})\,S_f(s_{\PH})}{{\hat s} - s_{\PH}}\bmid^2 =
\frac{\mu^5_{\PH}}{{\hat s}\,\mid {\hat s} - s_{\PH}\mid^2}\,
\sigma_{\Pg\Pg \to \PH}(\mu_{\PH})\,\otimes\,\Gamma_{\PH \to \PGg\PGg}(\mu_{\PH}).
\eq
where the Higgs-boson mass is set (by convention) to $\mu_{\PH}$, but other
options are available as well. The phase space is always with real momenta
while the Mandelstam invariant is made complex through the substitution
${\hat s} \to s_{\PH}$, a procedure that can be genaralized to processes with
more final-state legs. At this point we have four parameters, all of them 
Pseudo-Observables,
\bq
\mu_{\PH}, \quad \gamma_{\PH}, \quad \sigma_{\Pg\Pg \to \PH}(\mu_{\PH}), \quad
\Gamma_{\PH \to \PGg\PGg}(\mu_{\PH}),
\eq
that we want to use in a fit to the (box-detector) experimental distribution
(of course, after folding with PDFs). These quantities are universal, 
uniquely defined, and in one-to-one correspondence with {\em corrected}
experimental data. After that one could start comparing the results of the fit
with a SM calculation. The way this calculation has to be performed is also
uniquely fixed.

The breakdown of a process into products of POs can be generalized to include 
unstable particles in the final state; an example is given by 
$\Pp\Pp \to 4\,$leptons; the amplitude can be written as
\bq
A\lpar \Pp\Pp \to 4\,{\rm l}\rpar =
A_{\rm back}\lpar \Pp\Pp \to 4\,{\rm l}\rpar +
A_{\PH}\lpar \Pp\Pp \to \PH \to \PZ\PZ \to 4\,\Pl\rpar +
A_{\PH}\lpar \Pp\Pp \to \PH \to 4\,\Pl\rpar.
\label{split}
\eq
The first and third amplitudes in \Eref{split} are subtracted by using SM 
(or MSSM) calculations while the second (triply resonant) can be parametrized 
in terms of POs and a fit to $M(\Pl\Pl\Pl\Pl)$ attempted. The (triply resonant) 
signal in $\Pg \Pg \to 4\,$l is split into a chain 
$\Pg\Pg \to \PH$ (production), 
$\PH \to \PZ\PZ$ (decay), and $\PZ \to {\bar l} l$ (decays) with a 
careful treatment of ($\PW/\PZ$) spin correlation. 
In this way we can also introduce the folowing PO: $\Gamma\lpar \PH \to \PZ\PZ\rpar$. 
It is worth noting that the introduction of complex poles allows us to split 
multi-leg processes into simple building blocks through the mechanism of
separating gauge-invariant parts, once again, the complex pole, its residue,
and the regular part.
How else can we stand against the temptation of introducing a quantity like 
$\Gamma(\PH \to \PZ\PZ)$ where three unstable particles occur in the in/out
states? 

For processes which are relevant for the LHC and, in particular, for 
$\PH \to \bbar \PQb$, $\PGg\PGg$, $\Pg\Pg$, and $\Pg\Pg \to \PH$ etc., it is 
possible to define three different schemes and compare their results. The 
schemes are:
\begin{itemize}

\item the RMRP scheme which is the usual on-shell scheme where all
  masses and all Mandelstam invariants are real;

\item the CMRP scheme~\cite{Actis:2008uh}, the complex-mass 
scheme~\cite{Denner:2005fg} with complex internal $\PW$ and $\PZ$ poles 
(extendable to top complex pole) but with real, external, on-shell Higgs, 
etc. legs and with the standard LSZ  wave-function renormalization;

\item the CMCP scheme, the (complete) complex-mass scheme with complex,
external, Higgs ($\PW, \PZ$, etc.) where the LSZ procedure is carried out at the 
Higgs complex pole (on the second Riemann sheet).

\end{itemize}
The introduction of three different schemes does not reflect a theoretical
uncertainty; only the CMCP scheme is fully consistent when one wants to
separate production and decay; therefore, comparisons only serve the purpose 
of quantifying deviations of more familiar schemes from the CMCP scheme.
Example of how to apply the ideas presented in this section can be found in
\Bref{Passarino:2010qk}.

The usual objection against moving Standard Model Higgs POs into 
the second Riemann sheet of the $S$-matrix is that a light Higgs boson, say 
below $140\,\UGeV$, has a very narrow width and the effects induced are tiny. 
Admittedly, it is a well taken point for all practical consequences but
one should remember that the Higgs-boson width rapidly increases after 
the opening of the $\PW\PW$ and $\PZ\PZ$ channels and, because of this, 
the on-shell treatment of an external Higgs particle becomes inadequate as 
a description of data if the Higgs boson is not (very) light. 
On top of all practical implications one should admit that it is hard to sustain 
a wrong theoretical description of experimental data.

It is also important to establish the proper connection between Higgs-boson  
propagator and Breit--Wigner distribution. Given the complex pole
$s_{\PH} = \mu^2_{\PH} - i\,\mu_{\PH}\,\gamma_{\PH}$, define new quantities
(up tho higher orders, HO) as follows:
\bq
{\overline M}^2_{\PH} = \mu^2_{\PH} + \gamma^2_{\PH} +
\hbox{HO},
\quad
\mu_{\PH}\,\gamma_{\PH} = {\overline M}_{\PH}\,{\overline\Gamma}_{\PH}\,
\lpar 1 - \frac{{\overline\Gamma}^2_{\PH}}{{\overline M}^2_{\PH}}\rpar +
\hbox{HO}.
\eq
At this order it can be shown that
\bq
\frac{1}{s - s_{\PH}} =
\Bigl( 1 + i\,\frac{{\overline\Gamma}_{\PH}}{{\overline M}_{\PH}} \Bigr)\,
\Bigr( s - {\overline M}^2_{\PH} + 
i\,\frac{{\overline\Gamma}_{\PH}}{{\overline M}_{\PH}}\,s \Bigr)^{-1},
\eq
which one should compare with the Breit--Wigner implementation in MC tools.
The practical recipe for introducing the Higgs complex pole in the
Higgs-resonant amplitude $\Pg \Pg \to \PH \to f$ is as follows:
\bq 
\sigma_{\Pg\Pg \to \PH}(\MH)\,
\frac{{\hat s}^2}{\lpar{\hat s} - \MH^2\rpar^2 + \lpar {\hat s}\,\GH/\MH\rpar^2}\,
\frac{\Gamma_{\PH \to f}(\MH)}{\MH}
\;\to\;
\sigma_{\Pg\Pg \to \PH}(s_{\PH})\,
\frac{{\hat s}^2}{\bmid {\hat s} - s_{\PH} \bmid^2}\,
\frac{\Gamma_{\PH \to f}(s_{\PH})}{s^{1/2}_{\PH}}.
\eq
It is worth noting that in any BSM scenario there will be interdependence among 
Higgs-boson masses and the simultaneous renormalization at the exact complex poles 
will also introduce consistency checks.

\subsection{Extensions of the SM}
Extensions of the SM allow for more complex Higgs sectors. Problems that can
be avoided in the SM can easily be encountered in new-physics models.
For instance, heavy SM-like Higgs bosons with a relatively large width
naturally occur in models with an additional $U(1)$~symmetry and a
corresponding $\PZ'$~boson. 

Here we briefly describe the situation in the MSSM, where the Higgs sector
consists of two Higgs doublets, leading to one light CP-even Higgs, $h$, one
heavy CP-even Higgs, $\PH$, one CP-odd Higgs, $\PA$ and two charged Higgs
bosons, $\PH^\pm$. At tree level the Higgs sector is described by $\MA$ and 
$\tan\be$ (the ratio of the two vacuum expectation values).
In general, concerning the determination of the MSSM parameters,
additional complications arise compared to the SM case. 
Firstly, the unfolding procedure often involves the assumption of the
SM. Using this data within the MSSM (or any other extension of the SM) is
obviously only justified if the new-physics contributions to the subtraction
terms and the implemented higher-order corrections are negligible. 
Secondly, the model dependence is
relatively small for masses (see below). For couplings (beyond the SM-like
gauge couplings), mixing angles, etc., on the other hand, the model dependence
is relatively large. In contrast to the SM, many of the MSSM parameters are
not closely related to one particular observable (for instance $\tan\be$),
resulting in a relatively large model dependence. Therefore the approach of
extracting PO with only a fairly small model dependence seems not to be
transferable ot the case of the MSSM. Eventually the MSSM parameters will have
to be determined in a global fit of the full MSSM to a large set of
observables, taking into account {\em consistently} higher-order corrections.

As mentioned above, the Higgs-boson masses in general constitute a
smaller problem, even compared to the SM case.
For large parts of the parameters space $\MA > 2 \MZ$, the light
CP-even Higgs boson is SM-like, while all other Higgs bosons are nearly mass
degenerate~\cite{Gunion:1989we}. Furthermore the upper limit of $M_{\ssh}$ is
about $135 \UGeV$~\cite{Degrassi:2002fi}.
Consequently, here the width of the $h$ is also SM-like and small. 
Exceptions can occur for low $M_{\ssA}$ and large $\tan\be$. 
Here the $\Ph \PQb\bar\PQb$ coupling can grow with $\tan\be$, so the width can
grow with $\tan^2\be$. 
On the other hand, in this part of the parameter space the $\Ph\PW\PW$ coupling
is reduced, so that the decay $\Ph \to \PW\PW^{(*)}$ contributes less than in
the SM. All in all for low $\MA$ and large $\tan\be$ one can find a strong 
enhancement with respect to the SM, but no large value of 
$\Gamma_{\Ph}/M_{\Ph}$. 

The situation is different for the $\PH$ and $\PA$. 
For heavy $\PH,\PA$, $\MA \gsim 150 \UGeV$, $\PH$ and $\PA$ have
no  substantial couplings to SM gauge bosons, so there is not the typical
growth with $\MH$. 
Again here the $\PH/\PA \PQb\bar\PQb$ coupling goes with $\tan\be$, leading to
an enhancement of the widths, but not to very large values of 
$\Gamma_{\PH/\PA}/M_{\PH/\PA}$ as in the SM for masses above $\sim 200 \UGeV$.
Only for intermediate masses $\MA \sim 150 \UGeV$ the enhancement in the
coupling to $\PQb$ quarks can overcompensate the reduced coupling to gauge
bosons, depending on $\tan\be$. 

\subsection{Conclusions}
In conclusion, the only purpose of this section has been to state the problem 
and the possible way to solutions, conventional but unique.
Therefore, the work in this section is quite plainly an interlude and an 
actuate all at the same time. In any case it is worth noting that one of 
the goals of LHC will be to discover or exclude a SM Higgs boson up to 
$600\UGeV$. 
Already at $500\UGeV$ the effect of using the complex pole instead of 
the on-shell mass on the $\Pg\Pg \to \PH$ cross section is large and comparable 
to higher-order QCD corrections.
Using on-shell Higgs-boson also for high values of $\MH$ can only be a very 
first step (i.e.\ a first guess, as taken elsewhere in this Report) and 
a truely quantitative analysis should do much better. 
But it is not the previous strategies that are important this time -- it is 
normal that in the start-up phase of a new machine, strategies will fall 
like autumn leaves -- what's significant here is that the LHC's performance 
significantly calls for further theoretical improvement. POs, they're the 
only things we can pay. 

\clearpage

\newcommand{\ssB}{{\scriptscriptstyle{B}}}
\newcommand{\ssR}{{\scriptscriptstyle{R}}}
\newcommand{\ssF}{{\scriptscriptstyle{F}}}
\newcommand{\ssL}{{\scriptscriptstyle{L}}}
\newcommand{\ssQ}{{\scriptscriptstyle{Q}}}
\newcommand{\ssE}{{\scriptscriptstyle{E}}}
\section{Parametric and theoretical uncertainties\footnote{A.~Denner,
S.~Dittmaier, S.~Forte and G.~Passarino.}}
\label{THUsection}
\subsection{Introduction}
In this note we address the following questions: definition of theoretical
uncertainties (THU) for LHC predictions, their statistical meaning, inclusion 
of parametric 
uncertainties (PU), their combination.
For the latter we want to stress that the solution (how to combine) relies on
some implicit assumptions; any variation in the assumptions leads to a 
somehow different solution. In this case intuition may still help 
to qualitatively guess how the value of the measurement is affected.

The first step that we need to do is establish the definition of PUs and THUs.
Following this we need to describe the issue (problem) of combination. 
\subsection{Parametric uncertainties}
In our attempt to encode an acceptable definition of theoretical uncertainty  
for observables at the LHC we differentiate parametric uncertainties -- those 
related to the value of input parameters -- from true theoretical uncertainties 
reflecting our lack of knowledge about higher orders in perturbation theory.
\begin{itemize}
\item{{\bf PU}}, Parametric uncertainties, will always be there, but 
eventually reduced when
more precise experiments produce improved results. They should not be mixed
with THU, but listed as

\bq
{\cal O} = x.xxx \pm 0.00y\;\hbox{(param)}\;{}^{+0.00a}_{-0.00b}\;\hbox{(th)}.
\eq
Ideally and assuming that the central value will not change significantly, the
better way of dealing with future improvements is as follows:
\begin{enumerate}
\item Produce for each observable ${\cal O}$, which is a
function of parameters $\{p_1\,\dots\,p_n\}$, the central value
\bq
{\cal O}\lpar p^c_1\,\dots\,p^c_n\rpar,
\eq
\item Provide derivatives
\bq
\frac{d {\cal O}}{d p_i}, \qquad \forall i.
\eq
\end{enumerate}
In this way users will not have to re-run codes as soon as an improved 
measurement of $p_i$ is available.
\end{itemize}
Here, the recommendation is that parametric errors cannot be neglected and 
calculations should include them in their final estimate. 

The main difference between PU and THU is that PU are distributed according to a 
known (usually Gaussian) distribution while the statistical interpretation of THU is
less clear, and they are arguably distributed according to a flat distribution. 
Sometimes the uncertainty on $\alphas$ (say) is added
in quadrature to the scale uncertainty (see Section \ref{scale}), which is  
questionable if the former is Gaussian and the latter is flat. 
It is worth mentioning that we are discussing essentially Standard Model PUs.
\subsection{THU, understanding the origin of the problem}
\label{orprob}
In this and the next section we are going to discuss two separate 
issues~\cite{Collins:1990bu} that are sometimes mixed:
\bei
\item What is the optimal choice for QCD scales?
\item Can one use scale variation to estimate higher-order corrections?
\eei

We begin by addressing the first question. Let us for a moment
concentrate on the uncertainty induced by 
variations of the renormalization scale, $\mu_{\ssR}$, and of the
factorization scale, $\mu_{\ssF}$.
The question is: Do we have a $\mu_{\ssR}$ problem in QED?
The answer is {\em yes}, but is it a real problem? This time the answer 
is {\em no}, because we have a {\em physical} subtraction point, $q^2 = 0$,
for photons with momentum transfer $q$, which defines the Thomson limit.
The next question is: Do we have a $\mu_{\ssR}$ problem in the electroweak 
(EW) theory?
Yes, but it is not a real problem since, once again, we have a 
{\em physical} subtraction point(s), since the electromagnetic coupling can still
be fixed in the Thomson limit and the weak mixing angle can be tied to
the ratio of the $\PW$- and $\PZ$-boson masses. Stated in other words, our 
calculations depend on $\mu_{\ssR}$, but once Lagrangian parameters (masses
and couplings) are replaced by data (according to the renormalization
program) this dependence disappears. Of course, in perturbation theory,
the numerical output depends on the set of data that we have chosen, 
therefore the next question will be: Do we have large logs in our radiative
corrections? The answer is {\em yes} for all cases where the coupling is related
to a EW gauge boson, i.e.\ $\PGg, \PW$, or $\PZ$, with momentum
transfer at the EW scale or higher.
In the EW part of our theory the first step of the 
solution will be: Use the $G_{\ssF}\,$-scheme, not $\alpha(0)$, which is 
equivalent to say {\em resum} large logarithms that are connected to the
running of $\alpha(q^2)$ from $q^2=0$ to the EW scale.
In the $G_{\ssF}$ scheme, $G_{\ssF}$ and the gauge-boson masses are used 
as input parameters and the electromagnetic coupling is derived according to
$\alpha_{G_{\ssF}} = \sqrt{2} G_{\ssF}\MW^2(1-\MW^2/\MZ^2)/\pi$.
Actually, the message would be: For an 
observable at a scale $s$ do not use an input parameter set at a scale 
$s_0 \muchless s$ unless you know how to resum large logarithms.
Of course, there may be more logarithms of large scale ratios
connected to effects other than the running of $\alpha$ in addition
(collinear photon radiation, EW Sudakov logarithms, etc.).
Furthermore, the $G_{\ssF}$  scheme should not be used for couplings that 
concern external photons where $\alpha(0)$ is appropriate.

What to do in QCD? Resummation is the keyword~\cite{Catani:2003zt} but, 
admittedly, apart from the running of the strong coupling it is not
always available. There, the most useful keyword will be 
{\em minimization}. To understand the problem consider a physical
observable which is affected by (large) QCD corrections. Since we have
no analogue of $G_{\ssF}$ in QCD, our LO calculation will always contain logarithms
$\ln(s/\mu_{\ssR})$ where $s$ is the scale where we want to study the process. 
Ideally, one should find a scale $s_0$ where some data is available and 
renormalization means the replacement
$\ln(s/\mu_{\ssR}) \to \ln(s/s_0)$
and $s_0$ should not be far away from $s$. This is not (yet) possible in 
QCD, so the question will translate into, {\em how do I choose} 
$\mu_{\ssR}$? The guideline will be {\em set} $\mu_{\ssR}$ {\em to} $s$,
i.e.\ to evolve the coupling to scale $s$ with renormalization
group equations, or,
in other words, make sure that you do not change much by going to the
next order. This is easy in a one-scale process but in any multi-scale
process one will have other additional large logarithms, say of
argument $s/s'$. What to do?
\begin{itemize}
\item Select $\mu_{\ssR}$ and $\mu_{\ssF}$, process by process, in such 
a way that when going from N${}^n$LO to $N^{n+1}$LO you minimize the
effect of the new corrections. 
In many cases a phase-space-dependent choice is needed in order to
achieve this in differential cross sections.
The recipe is the best simulation of a
subtraction at some physical point close to the relevant scale. In
jargon this is called {\em dynamical scale}.
\end{itemize}
\subsection{THU uncertainties}
In this section we will briefly discuss different sources of THU, starting with
QCD scale variation.
\subsubsection{QCD scale variation \label{scale}}
Once the dynamical scale has been selected (process by process) we can
address the second question mentioned in the beginning of
Section~\ref{orprob}. Namely,
how do we understand our approximation in terms of
scale variation, e.g.\ $s/n < \mu_{\ssR,\ssF} < n\,s$? The idea is as
follows: In the {\em full} theory there is no scale dependence
and order by order in perturbation theory we should be able to see this
asymptotic limit. Therefore, variation of the scale(s) is a pragmatic
way of understanding how far we are from controlling the theory. In
practice, this means {\em which value do I choose} for $n$?

The recommendation, in this case, should be as follows: at a given order
look for a plateau in the scale dependence and fix $n$ to be such that
the plateau is included. 
Therefore:
\begin{itemize}
\item Allow each calculation to set the range of scale variation (it is 
a matter of experience), but check that nobody is allowing for too small or
too large variations just to bring the {\em error} in the range 
foreseen by their religion.
\item Check that different calculations and different choices give
consistent results.
\item Drop extreme choices which are too far away from {\em common}
understanding of the problem. 
\end{itemize}
Renormalization scale and factorization scale have different origins and there
is no good argument according to which we should set 
$\mu_{\ssF} = \mu_{\ssR}$. Once again we invoke the {\em minimization
principle}, i.e.\ when going from N${}^n$LO to N$^{n+1}$LO the choice should 
minimize the effect of the new corrections.
Sometimes the estimate of the uncertainty is based on a diagonal scan, sometimes 
anti-diagonal directions are included. There is also another recipe, a 
two-dimensional scan with $1/n < \mu_{\ssR}/\mu_{\ssF} < n$. One should also 
mention that an independent variation of the two scales introduces large 
logarithmic corrections that are cancelled by the next order in perturbation theory.
Our recommendation here is for a one-dimensional scan, monitoring at the same time
large differences induced by the two-dimensional one.

A word of caution is needed at this point: there are examples where one can see
that it is easy to optimize the scale choice, but scale variation becomes a very 
poor way to estimate higher-order corrections (HO) (in fact at LO it misses even 
the order of magnitude).
Being pragmatic we should state that while there may be an optimal scale choice 
(i.e.\ one that minimizes higher-order corrections), one should be careful that 
this does not then bias the results of estimating higher orders by scale variation.
To be more specific, nobody would object to the suggestion that $\mu_{\ssR}$ and
$\mu_{\ssF}$ should be chosen in such a way that higher-order corrections are
minimized, but in practice the recipe is not always meaningful. It remains true that
if we do know the higher order, we will use it, and if we do not know it, we cannot 
estimate the scale which minimizes the difference. Therefore, what we are suggesting
here is to use the last two known orders for the search of stability and for
minimizing corrections (if reasonable), which is -- at best -- a 
{\em rule of thumb}.
Looking for a {\em plateau} simply means looking for a stationary point in the 
dependence of the observable on scale. If there is a stationary point it suggests 
greater reliability. How to trust a calculation if there is no stationary point 
remains an highly questionable point.
To summarize, one searches the region of the minimum of the
higher-order corrections, and for distributions
a dynamical scale that stays near the minimum in the whole
condsidered range, so that the $K$-factor does not drift away


\subsubsection{PDF}
For PDFs, theoretical uncertainties in the sense defined above are unknown and 
have never been estimated (see section~\ref{pdfsection} and  
Refs.~\cite{forterev,PDF4LHCwebpage,PDF4LHCwiki}). The known PDF uncertainties are
\bei
\item PDF uncertainties, which are propagated data uncertainties (PDF
  uncertainty, henceforth); 
\item parametric uncertainties, of which the one due to propagation of
  the uncertainty on the value of $\alphas$ ($\alphas$ uncertainty henceforth)
has been studied systematically
by several groups, while the uncertainties due to the value of the
heavy-quark masses are being gradually included.
\eei

While we refer to Section~\ref{pdfsection} for a more detailed
discussion, we note that the recommendation given there for the
determination of PDF uncertainties provides a result that already
includes the combination of the 
PDF uncertainty and the $\alphas$ uncertainty.
One option could have been to keep them separated but it was the PDF community 
recommendation to provide only the combination of these two. For
future studies it might be more advantageous to keep them separate in
that this would give more flexibility to the user.
It should
be understood clearly  that other parametric uncertainties are thus 
not included in this prescription.
It is interesting to note that the Gaussian behaviour of PDF
uncertainties has been checked explicitly within the HERAPDF
and NNPDF PDF determinations (see Section~3.2 in Ref.~\cite{Dittmar} and
Ref.~\cite{Ball:2010de}). 
\subsubsection{Other sources of THU}
Other potential sources of THU are:
\begin{itemize}
\item Pole masses vs.\ running masses? Whenever we know which mass (including its 
scale) is to be taken the uncertainty should not enter the game. This means that,
in general, we do not recommend inclusion of these effects in THU.

\item In the scheme production$\,\otimes\,$decay the Higgs shape is,
usually, represented by a Breit--Wigner distribution. Differences induced
by using fixed-width scheme vs.\ running width scheme should not enter the THU.
If the difference matters (e.g.\ at large values of the Higgs mass) one should
try to understand the difference and compare results with the complex-mass
scheme.

\item EW uncertainties; we have renormalization scheme dependence, but also an
uncertainty associated to inclusion of EW effect in a QCD NNLO 
calculation~\cite{Actis:2008ug}.
If the QCD $K$-factor is large it will make some difference to multiply
$\delta_{\mathrm{EW}}$ by the full $K$-factor (complete 
factorization~\cite{Anastasiou:2008tj}) or to include it 
only at LO (as the conservative recipe of partial factorization would suggest). 
The most conservative recipe for mixed EW--QCD effects is the vary between 
complete and partial factorization, but an estimate should be given in any case.

\item Full top mass dependence~\cite{Spira:1995rr} vs.\ large 
$\Mt\,$-approximation in the production $\sigma\lpar \PH \to \Pg\Pg\rpar$. 
The correct recipe is as follows: at NLO one should take the full top mass 
dependence and the estimate of the approximation at NNLO should go into the 
THU~\cite{Anastasiou:2009kn}.

\item Inclusion of the bottom-quark loop in gluon--gluon fusion, complete
factorization ($\mid \hbox{top} + \hbox{bottom}\mid^2$ with full NNLO $K$-factor) 
or partial one ($\mid \hbox{bottom}\mid^2$ and top--bottom interference with 
NLO $K$-factor)?

\item
Missing higher-order corrections not related to scale uncertainties.
Sometimes LO predictions lack some scale uncertainty 
that appears only in higher orders (e.g.\ no QCD renormalization scale in the
Drell--Yan process in LO),
sometimes new channels open in higher orders, which is also a systematic
effect that has nothing to do with scales 
(e.g.\ $\Pg\Pg$ channel for $\PW\PW$+jet production).
In particular, the THU resulting from missing higher-order EW corrections
cannot be estimated via scale uncertainties. NNLO EW corrections can be estimated 
to some extent based on the known structure and size of the NLO corrections 
combined with power counting of EW couplings and logarithms.
Here we are discussing mostly SM, at the moment no special recommendation is 
available for MSSM and one should include THU, whenever possible, by scaling the 
corresponding SM THU.
\end{itemize}
\subsection{How to combine THU}
The main question we want to discuss here is: {\em Are THU confidence intervals?} 
And also: {\em Do we have statistical meaning for THU?}
There are different opinions on the subject; some of us think that
THU {\em should} be confidence intervals, though of course being a distribution 
of true values they must be interpreted in a Bayesian sense. Obviously,
given that they refer to a distribution of the values there is no reason to think 
that they are Gaussian, and it might be more reasonable to take them as flat 
distributions. This said, they should be combined using the rules of Bayesian 
inference. The envelope method is then the correct rule to combine probability 
intervals from flat distributions.

No matter which opinion one has, it seems obvious that if THU come from flat 
distributions, then they should be added linearly, and if from Gaussian, in 
quadrature. It is more reasonable and more conservative to think that THU are 
flat, and thus to add them linearly. As already stated, PDF uncertainties are 
most certainly Gaussian uncertainties, they have been explicitly checked to be 
Gaussian, and should therefore be treated as such.
Therefore, there is enough evidence that the PDF + $\alphas$ uncertainty 
should be added in quadrature to all other PU. The way the total PU is then
combined with the THU comes down to the best way of combining a Gaussian and flat 
distributions (which is less obvious). Of course, whenever a
theoretical uncertainty dominates (typically the QCD
scale uncertainty, e.g.\ for gluon--gluon fusion) the problem becomes less
relevant.

If only one observable is needed each code should provide a set of options 
\bq
\{o_1,\,\dots\,,o_m\}, \qquad \hbox{with values}\quad
 o_i = \{o^1_i,\,\dots\,,o^k_i\}
\eq
where, for instance, $o_i$ is QCD $\mu_{\ssR}$ dependence and 
$\{o^1_i,o^2_i,o^3_i\}$ refer to 
$\mu_{\ssR} = \hbox{scale}/2\,,\,\hbox{scale}\,,\,2\,\hbox{scale}$.
After running the observable ${\cal O}$ over all options one determines 
${\cal O}_{\rm min}$, ${\cal O}_{\rm c}$, and ${\cal O}_{\rm max}$, where the 
central value is fixed by the author's taste, defining the {\em preferred} setup.

If several observables have to be combined one has to take into account that,
given $m$ options with multiple values, some of them are correlated, e.g.\ all 
options concerning production via $\Pg\Pg$ fusion should not be varied 
independently in all observables. This means do not compute ${\cal O}_i$ with 
$\hbox{scale}/2$ and ${\cal O}_j$ with $2\,\hbox{scale}$ if both come from 
$\Pg\Pg$ fusion. Even here we have two possibilities:
\begin{itemize}
\item Vary one option at the time and add the effects;
\item vary all options (taking into account correlations) and find the absolute 
minimum and maximum in the allowed range of variation.
\end{itemize}
The first choice has the virtue that experimentalists can decide, later on, on error
combination; the second one is more clean and reflects the true status of THU.

All this said, one has to face the problem of how to combine different
determinations of uncertainties, e.g., from different groups which
provide different uncertainty estimates for the same
observable. Assume for definiteness that
the two groups provide the probability
distribution  
for an observable ${\cal O}$ as $p_1({\cal O})$ and $p_2({\cal O})$,
for example by saying that the distributions are gaussian and
providing their means and standard deviations $m_i$ and $\sigma_i$.
In the case of statistical uncertainties, two attitudes are possible:
\begin{itemize}
\item The different determinations differ due to  statistical
  reasons.
In such case, the best value is found as the weighted
  average. In the above example, the combined determination $\bar p({\cal O})$
is a gaussian, with mean $m$ equal to the weighted average of the means
of the two starting distributions and standard deviation equal to the 
standard deviation of the mean,
$\sigma^2=\frac{\sigma_1^2+\sigma_2^2}{2}$. In this case, the error on
the combined determination is always smaller than the error on each of
the determinations that go into it.
\item The different determinations are exclusive, i.e.\ either one 
  or the other is correct, and they should be
  combined in a Bayesian way by assigning an a priori reliability to
  each of them. In this case  the combined
  probability is 
$\bar p({\cal O})=\frac{p_1({\cal O})+p_2({\cal O})}{2}$, and
  the error on
the combined determination
is not necessarily smaller, and in fact typically larger
  than the error of each of the determinations that go into it.
Indeed,  in practice, unless the
  probability distributions that are being combined are very
  inconsistent (e.g.\ if their respective means differ by many standard
  deviations) this Bayesian combination is very close to the envelope
  of the distributions which are being combined (compare the method
  for the combination of PDF
  uncertainties discussed in Section~\ref{pdfsection}).
\end{itemize}
The case of theoretical uncertainties is rather less obvious and it
will be discussed in the next section.
\subsubsection{Possibilities for option combination}
Consider a given observable ${\cal O}$ whose calculation is characterized by a set
of options $\{o_1,\,\dots\,,o_n\}$. A typical result, showing all components of
THU will be as follows:
\bq
{\cal O} = {\cal O}_{\rm c}\,{}^{+\,o^{\rm max}_1}_{-\,o^{\rm min}_1}\,[ o_1 ]\,
\cdots\,{}^{+\,o^{\rm max}_n}_{-\,o^{\rm min}_n}\,[ o_n ].
\eq
The question is on the combination of different sources. There are three options:
\bei
\item[L)] Linear combination:
\bq
{\cal O} = {\cal O}_{\rm c}\,{}^{+\,\Delta^{\ssL}_+ {\cal O}}_{-\,\Delta^{\ssL}_- {\cal O}},
\qquad
\Delta^{\ssL}_+ {\cal O} = \sum_{i=1}^n o^{\rm max}_i,
\quad
\Delta^{\ssL}_- {\cal O} = \sum_{i=1}^n o^{\rm min}_i,
\eq
\item[Q)] Quadratic combination:
\bq
{\cal O} = {\cal O}_{\rm c}\,{}^{+\,\Delta^{\ssQ}_+ {\cal O}}_{-\,\Delta^{\ssQ}_- {\cal O}},
\qquad
\Delta^{\ssQ}_+ {\cal O} = \Bigl[ \sum_{i=1}^n \bigl( o^{\rm max}_i\bigr)^2\Bigr]^{1/2},
\quad
\Delta^{\ssQ}_- {\cal O} = \Bigl[ \sum_{i=1}^n \bigl( o^{\rm min}_i\bigr)^2\Bigr]^{1/2},
\eq
\item[E)] Envelope combination:
\bq
{\cal O} = {\cal O}_{\rm c}\,{}^{+\,\Delta^{\ssE}_+ {\cal O}}_{-\,\Delta^{\ssE}_- {\cal O}},
\quad
\Delta^{\ssE}_+ {\cal O} = \max_{o_1,\,\dots\,,o_n}\,{\cal O}\lpar \{o\}\rpar -
{\cal O}_{\rm c},
\quad
\Delta^{\ssE}_- {\cal O} = {\cal O}_{\rm c} - 
\min_{o_1,\,\dots\,,o_n}\,{\cal O}\lpar \{o\}\rpar,
\eq
\end{itemize}
where ``c'' refers to the preferred setup for all options. A loop over all options
that are not correlated is indeed needed, all options that are independent should
be varied simultaneously. 
When a specific set of PDFs is used, it should be kept fixed at its
central value when computing the various THU. The PDF uncertainty
(which is not a THU as discussed above) is computed along with other statistical 
and parametric uncertainties.
Let us now assume that all options correspond to uncertainties which
are known to be THU. Clearly addition in quadrature is then not
appropriate. In an ideal case, all sources of THU should be recorded, with
correlated options not be varied independently but rather in the
correlated way discussed previously. The final ensuing uncorrelated
THUs can then be just combined linearly. Moreover, in the future, 
combination can be repeated when some uncertainty is reduced or some improved
strategy is found. To this purpose, all authors should provide 
information on each source of THU.

Unfortunately, this is not usually done, so, lacking 
detailed information, the problem of combining uncertainties arises.
To be more concrete, let us consider an observable ${\cal O}$ and two different 
predictions, ${\cal O}_{\ssA}$ and ${\cal O}_{\ssB}$, both with asymmetric 
error (with for definiteness ${\cal O}_{\ssA}>{\cal O}_{\ssB}$). The precise 
meaning of error here is not obvious; however, we can assume that
an error of $\pm\Delta_{\ssA}$ means that the observable has a constant
probability of being in the range ${\cal O}_{\ssA}-\Delta_{\ssA} < {\cal O} <
{\cal O}_{\ssA}+\Delta_{\ssA}$ (with the obvious generalization when the error is
asymmetric). 
Note that the standard deviation of such a probability
distribution is equal to $\sigma=\frac{\Delta}{\sqrt{3}}$.
The rationale for this choice is that if a calculation is performed by using the 
state-of-art of the present technology, the meaning of the error band
is then: {\em I don't know about higher-order 
effects, I haven't computed them, but I know that it is almost impossible 
that they will change my result more than what I have indicated}. Thus, the true 
result should be within the shown interval.

Given two predictions ${\cal O}_{\ssA,\ssB}$ with asymmetric errors
$\Delta A_{\pm}$ and $\Delta B_{\pm}$ the central value (this result rests on the 
assumption that the upper limit is due to $A$ and the lower due to $B$) can be defined 
as
\bq
\langle{\cal O}\rangle = \frac{1}{2}\,\Bigl( {\cal O}_{\ssA} + {\cal O}_{\ssB} 
+ \Delta B_+ - \Delta A_-\Bigr)
\label{bestcenter}
\eq
i.e.\ at the centre of the overlapping band (or at the centre of the gap in case
of no overlap). If the two uncertainties
are treated as completely independent and they are added linearly, the
width of the band then is 
\begin{equation}
W= \Delta A_+ + \Delta A_- + \Delta B_+ + \Delta B_-.
\label{bigband}
\end{equation}
This is a very conservative estimate, which contradicts
the above philosophy according to which the two intervals 
$\Delta A_+ + \Delta A_-$ and $\Delta B_+ + \Delta B_-$ should already 
be ``maximal'' ranges of variation. Furthermore, it neglects the fact that 
in practice the ranges of variation given by
different authors will include several common effects. 

One could then alternatively argue in the following way: the two determinations
provide each a maximal range of variation, however the two different
estimates of the range of variation include some common effects. The
total range of variation should then be smaller than $W$, Eq.~(\ref{bigband}), 
to account for these common effects. 
For example, suppose  $A$ does not include the $\beta\,$-effect and estimates 
the corresponding uncertainty, but includes the $\alpha\,$-effect. The 
opposite for $B$. 
If it is thought that a smaller $W$ reflects a genuine progress, then
an ideal solution would be that $A$ and $B$ include  both the $\alpha$- and 
$\beta$-effects. However, sometimes this is not possible, for example if 
$B$ considers the $\alpha\,$-effect to be wrong or questionable. 
Even so, a less conservative than just using Eq.~(\ref{bigband}) is possible. 
Namely, assuming that the possible inclusion of common effects is responsible 
for the fact that the two bands overlap, and the difference in central values 
is due to effects not included in both determinations, one should subtract 
from the uncertainty band the width
\bq
d = \max\{\, 0, \, {\cal O}_{\ssB} - {\cal O}_{\ssA} 
               + \Delta B_+ + \Delta A_- \}
\label{bigoverlap}
\eq
of the  region of overlap of the two bands, thereby getting an uncertainty band
with width
\bq
W^\prime= W - d.
\label{smallband}
\eq
If the bands have a nonzero overlap, so $d> 0$, $W^\prime$ Eq.~(\ref{smallband}) is 
just the envelope of the two bands, namely $W^\prime= W^{\prime\prime}$, with.
\bq
W^{\prime\prime}= \Delta A_+ + \Delta B_- + 
\Bigl( {\cal O}_{\ssA} - {\cal O}_{\ssB}\Bigr).
\label{envband}
\eq
If the bands do not overlap
the envelope Eq.~(\ref{envband}) is wider than the linear sum of the 
uncertainties Eq.~(\ref{bigband}): $W^{\prime\prime}>W$. In this case, the 
lack of overlap of the two bands suggests that either or both of
two determinations are missing some source of uncertainty, and the 
envelope prescription, which is now more conservative than the linear
sum, seems more advisable.

Hence we conclude that it is in general advisable to adopt
the envelope uncertainty  estimate Eq.~(\ref{envband}).
To formulate our recommendation in full generality we define
\bqa
O_- &=& \max\,\{{\cal O}_{\ssB} - \Delta B_-\,,
\,{\cal O}_{\ssA} - \Delta A_-\},
\nl
O_+ &=& \min\,\{{\cal O}_{\ssB} + \Delta B_+\,,
\,{\cal O}_{\ssA} + \Delta A_+\},
\nl
E_- &=& \min\,\{{\cal O}_{\ssB} - \Delta B_-\,,
\,{\cal O}_{\ssA} - \Delta A_-\},
\nl
E_+ &=& \max\,\{{\cal O}_{\ssB} + \Delta B_+\,,
\,{\cal O}_{\ssA} + \Delta A_+\}.
\label{genDEF}
\eqa
Our recommendation is thus to use as best prediction for the observable the
central value
\bq
\langle{\cal O}\rangle = \frac{1}{2}\,\Bigl( O_+ + O_-\Bigr),
\label{finalCV}
\eq
with an uncertainty band with envelope width
\bq
W^{\prime\prime} = E_+ - E_-
\eq
[generalization of Eq.~(\ref{envband})], namely:
\bq
{\cal O} = \;\langle{\cal O}\rangle^{+\{E_+ - \langle{\cal O}\rangle\}}_
                            {-\{\langle{\cal O}\rangle - E_-\}}.
\label{finalrec}
\eq
Finally, we note that a similar conclusion is reached if we assume
that the two determinations under discussion should be taken as
exclusive and with equal a priori probability. Indeed if $p_{\ssA}({\cal O})$ and
$p_{\ssB}({\cal O})$ are two flat probability distributions for the observable 
${\cal O}$,
then the combined distribution
$\bar p({\cal O})=(p_{\ssA}({\cal O})+p_{\ssB}({\cal O}))/2$ 
has an effect very similar 
to a flat distribution with width equal to the envelope of $p_{\ssA}({\cal O})$ and
$p_{\ssB}({\cal O})$, as long as the two bands at least touch each other. 

If the starting determinations distributions are inconsistent, i.e.\ 
${\cal O}_{\ssB}+\Delta  B_+ \muchless {\cal O}_{\ssA}-\Delta A_-$ none of 
these methods seems adequate, and in such case one should question the 
reliability of the results which are being combined. In all other cases, we 
conclude that the envelope method Eq.~(\ref{finalrec}) provides a conservative 
but not overly conservative way of combining THU, though it could overestimate 
a bit the combined THU.
\subsubsection{Conclusions}
The concept of THU and its use require few basic rules and an agreement within
the community:
\begin{itemize}

\item Sets of options in different calculations should be homogeneous; if one
calculation includes a {\em new} option its physical origin should be
motivated and its inclusion accepted, in which case all codes should include it.
If a calculation is based on options that inflate (or deflate) the THU without
a general consensus or a solid theoretical basis, it should not be included in 
the average. Controversial assumptions should be put on a waiting list and 
included only when the issue is fully clarified. Our recommendation for central 
value is the midpoint of the overlapping region, \eqn{finalCV}.

\item If different calculations include homogeneous sets of options the
difference between central values should be considered with particular care, 
unless the central value itself reflects a specific choice for the 
{\em preferred setup} with different choices in different calculations. If 
all options, including the preferred setup, are congruent then differences 
in the central values cannot be justified by THU.

\end{itemize}

To summarize our recommendations we suggest that:

\begin{itemize}

\item Parametric uncertainties that are distributed according to a 
known (usually Gaussian) distribution should be added in quadrature.

\item For the choice of central QCD scales we are suggesting to use the 
last two known orders for the search of stability and for
minimizing corrections (if reasonable), which is -- at best -- a 
{\em rule of thumb}. 
The corresponding theoretical uncertainty, which of course should be assessed 
by investigating the highest known order, 
is arguably distributed according to a flat distribution.
Problems related to incompatible data are more the rule than an exception
for THU and, in principle, THUs should be considered case by case; this is
this is particularly true whenever the two error bands are far apart,
and also the envelope (the standard method for 
incompatible statistical uncertainties) becomes questionable. 
In order to formulate a global recommendation we suggest that THU should be 
combined according to the envelope method: therefore, define the central value 
according to \eqn{finalCV} with uncertainty given by \eqn{finalrec}.

\item One should keep in mind that there are additional sources of THU, e.g.\ 
the THU resulting from missing higher-order EW corrections, that cannot be 
estimated via scale uncertainties. Therefore scale variation uncertainties
(SU) are a relevant portion of the global THU but do not exhaust the THU.
It is our recommendation that all sources of THU, not only SU, and their origin 
should always be documented.

\item The way the total PU is then combined with the THU comes down to the best 
way of combining a Gaussian and flat distributions.
As general rule that is sufficiently conservative only the linear combination
of those errors can be recommended.

\item To stress our point let us repeat that the PDF + $\alphas$ (plus
other parametric uncertainties, such as heavy-quark masses) are added in
quadrature to each other (i.e.\ if one wants to add heavy-quark mass effects, 
this has to be done in quadrature to PDF + $\alphas$, which is already the sum 
in quadrature of PDF + $\alphas$), but then they are added {\em only once} at 
the end to the THU.
Thus if one has two different estimates of the PDF + $\alphas$ uncertainty, 
$A$ and $B$, the recommendation is on averaging these two estimates (which is the
same as the uncertainty on two fully correlated measurements) before combining with the 
THU.
A remaining source of uncertainty, i.e. the scale dependence of PDF, cannot be
estimated at present.

\end{itemize}

\newpage
\section{Summary%
\footnote{S.~Dittmaier, C.~Mariotti, G.~Passarino and R.~Tanaka.}}
\label{se:summary}

The present document is the result of a workshop that started in
January 2010 as a new joint effort between ATLAS, CMS, LHCb, and the
theory community.
In this Report we have presented the state of the art for SM and MSSM
Higgs cross-section and branching-ratio calculations.

Here we summarize the Higgs production cross sections which are obtained
following the recommendation of \Sref{pdfsection} for the choice of
parton distribution functions (PDFs) and their combined uncertainty assessment
together with the one for the strong coupling constant $\alphas$ 
(PDF4LHC recipe).
Moreover, we combine this PDF + $\alphas$ uncertainty with the
theoretical uncertainty (THU) according to the prescription of
\Sref{THUsection}.
In detail, given two calculations ${\cal O}_{1,2}$ with THU uncertainties
$\Delta^{\rm THU}_{i,\pm}$ and PDF + $\alphas$ uncertainties (according
to the PDF4LHC recipe) $\Delta^{\rm PU}_{i,\pm}$, we
\begin{itemize}

\item define the corresponding central value as
\begin{equation}
\langle {\cal O}\rangle = \frac{1}{2}\,\Bigl(
O_+ + O_- \Bigr),
\end{equation}
where $O_{\pm}$ give the boundaries of the overlap,
\begin{equation}
O_+ = \min\{ {\cal O}_{1} + \Delta^{\rm THU}_{1,+},
              {\cal O}_{2} + \Delta^{\rm THU}_{2,+} \}, \qquad
O_- = \max\{ {\cal O}_{1} - \Delta^{\rm THU}_{1,-},
             {\cal O}_{2} - \Delta^{\rm THU}_{2,-} \},
\end{equation}
\item compute combined THU uncertainty
\begin{equation}
T^+ = E_{+} - \langle {\cal O}\rangle,
\quad
T^- = \langle {\cal O}\rangle - E_{-},
\end{equation}
where $E_{\pm}$ give the boundaries of the envelope,
\begin{equation}
E_+ = \max\{ {\cal O}_{1} + \Delta^{\rm THU}_{1,+},
             {\cal O}_{2} + \Delta^{\rm THU}_{2,+} \}, \qquad
E_- = \min\{ {\cal O}_{1} - \Delta^{\rm THU}_{1,-},
             {\cal O}_{2} - \Delta^{\rm THU}_{2,-} \},
\end{equation}
\item compute combined PDF + $\alphas$ uncertainty
\begin{equation}
P^{\pm} = \frac{1}{2}\,\Bigl( \Delta^{\rm PU}_{1,\pm} +
\Delta^{\rm PU}_{2,\pm}\Bigr)
\end{equation}
\item define total errors, $T^{\pm} + P^{\pm}$.

\end{itemize}

\begin{figure}[ht]
	\begin{center}
	\includegraphics[width=0.8\textwidth]{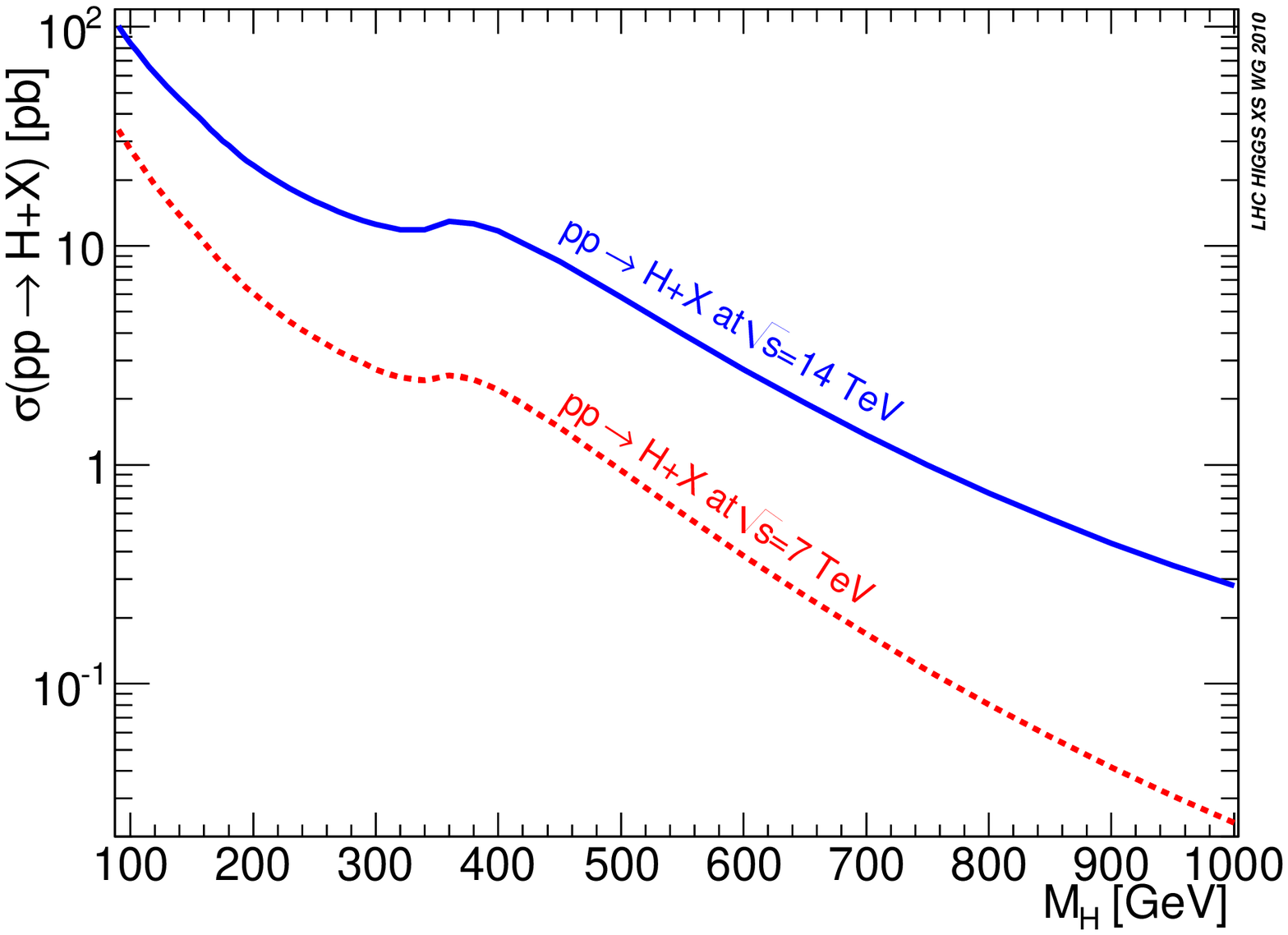}
	\caption{The total SM Higgs production cross section at
	$\sqrt{s} = 7$\UTeV and $14$\UTeV.}
	\label{fig:totalXS}
	\end{center}
\end{figure}

\begin{figure}[p]
	\begin{center}
	\includegraphics[width=0.8\textwidth]{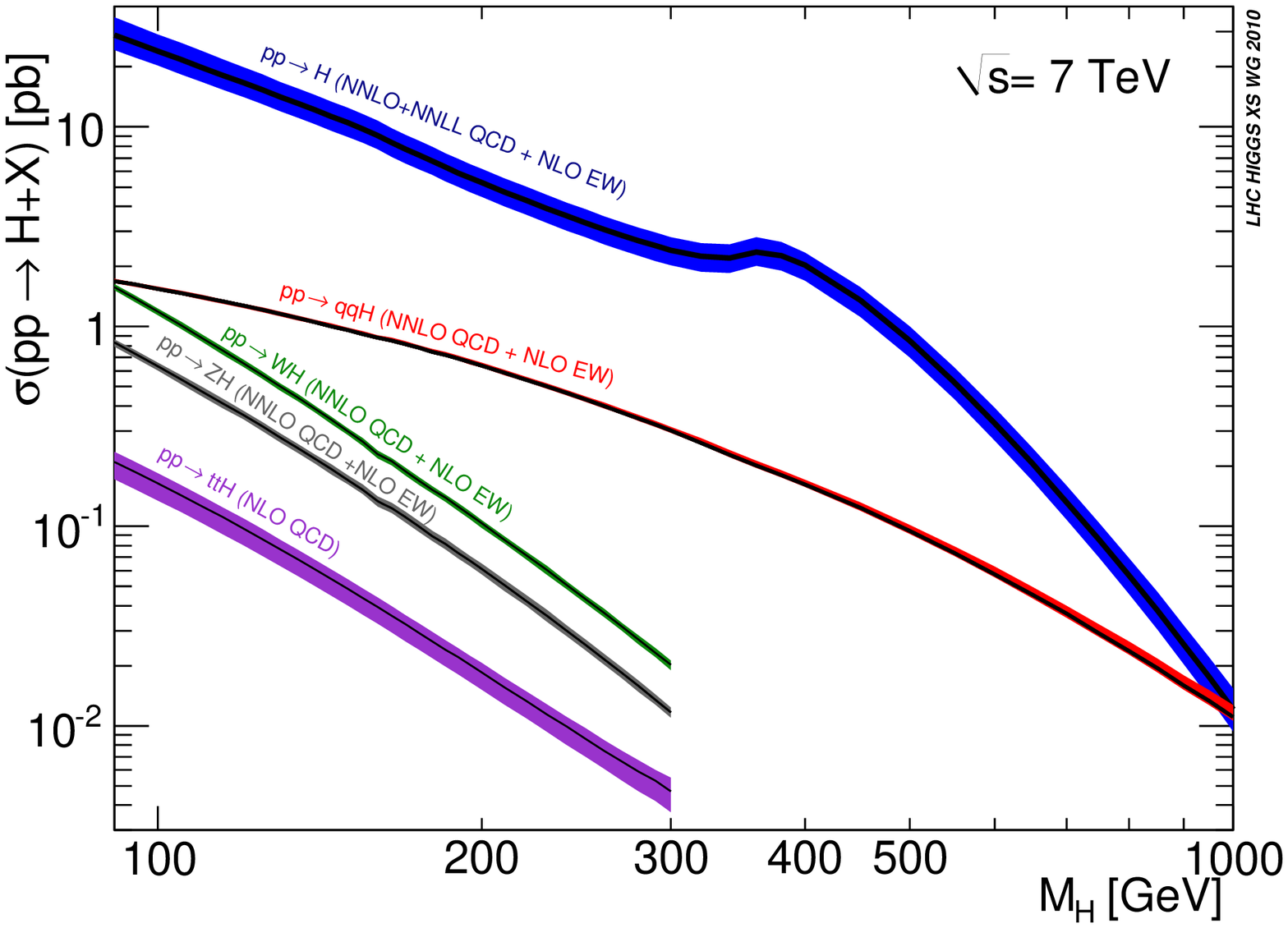}
	\caption{The SM Higgs production cross section at $\sqrt{s} = 7$\UTeV.}
	\label{fig:SMXS7TeV}
	\end{center}
\end{figure}

\begin{figure}[p]
	\begin{center}
	\includegraphics[width=0.8\textwidth]{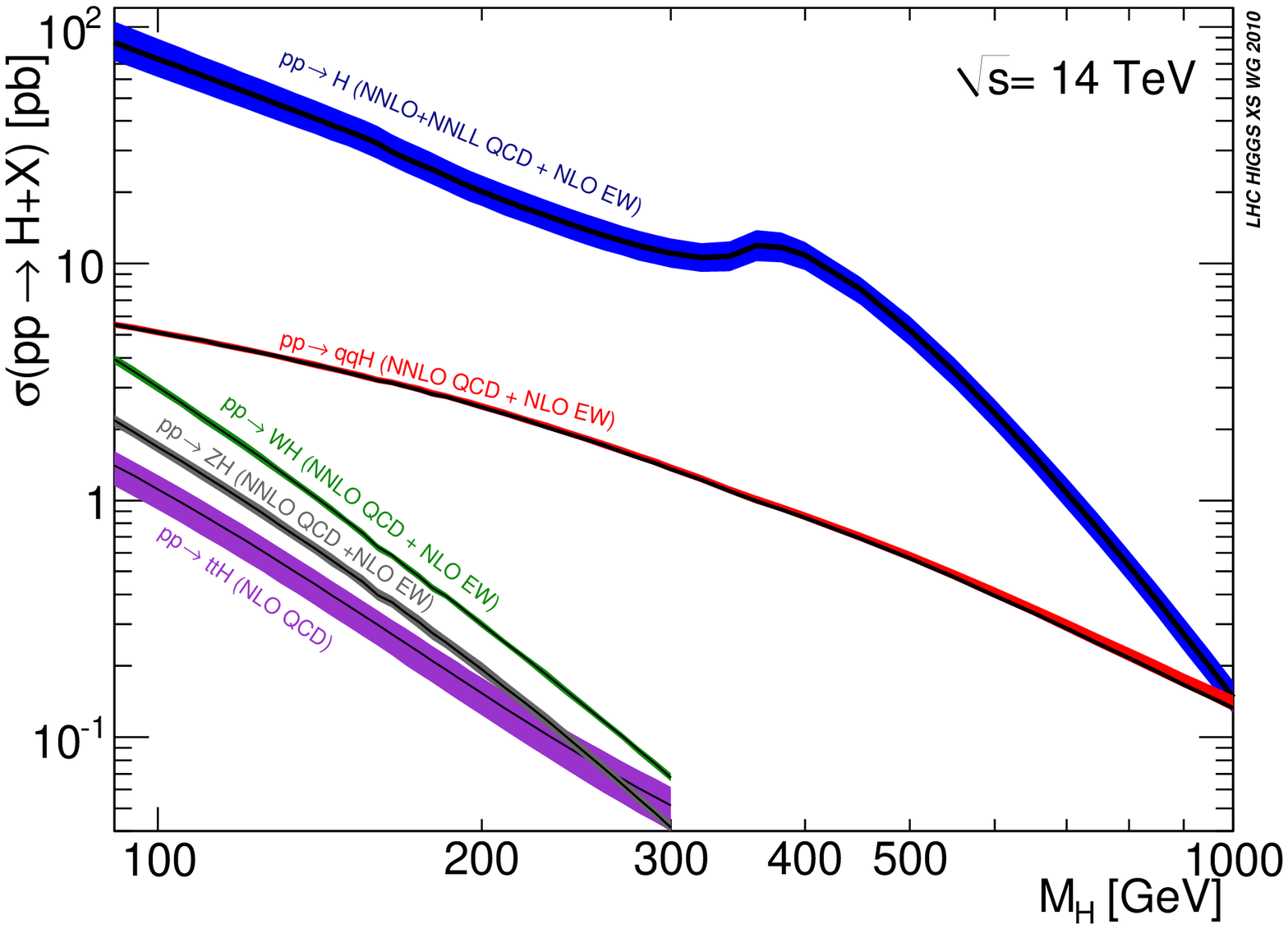}
	\caption{The SM Higgs production cross section at $\sqrt{s} = 14$\UTeV.}
	\label{fig:SMXS14TeV}
	\end{center}
\end{figure}

\begin{figure}[!h]
	\begin{center}
	\includegraphics[width=0.8\textwidth]{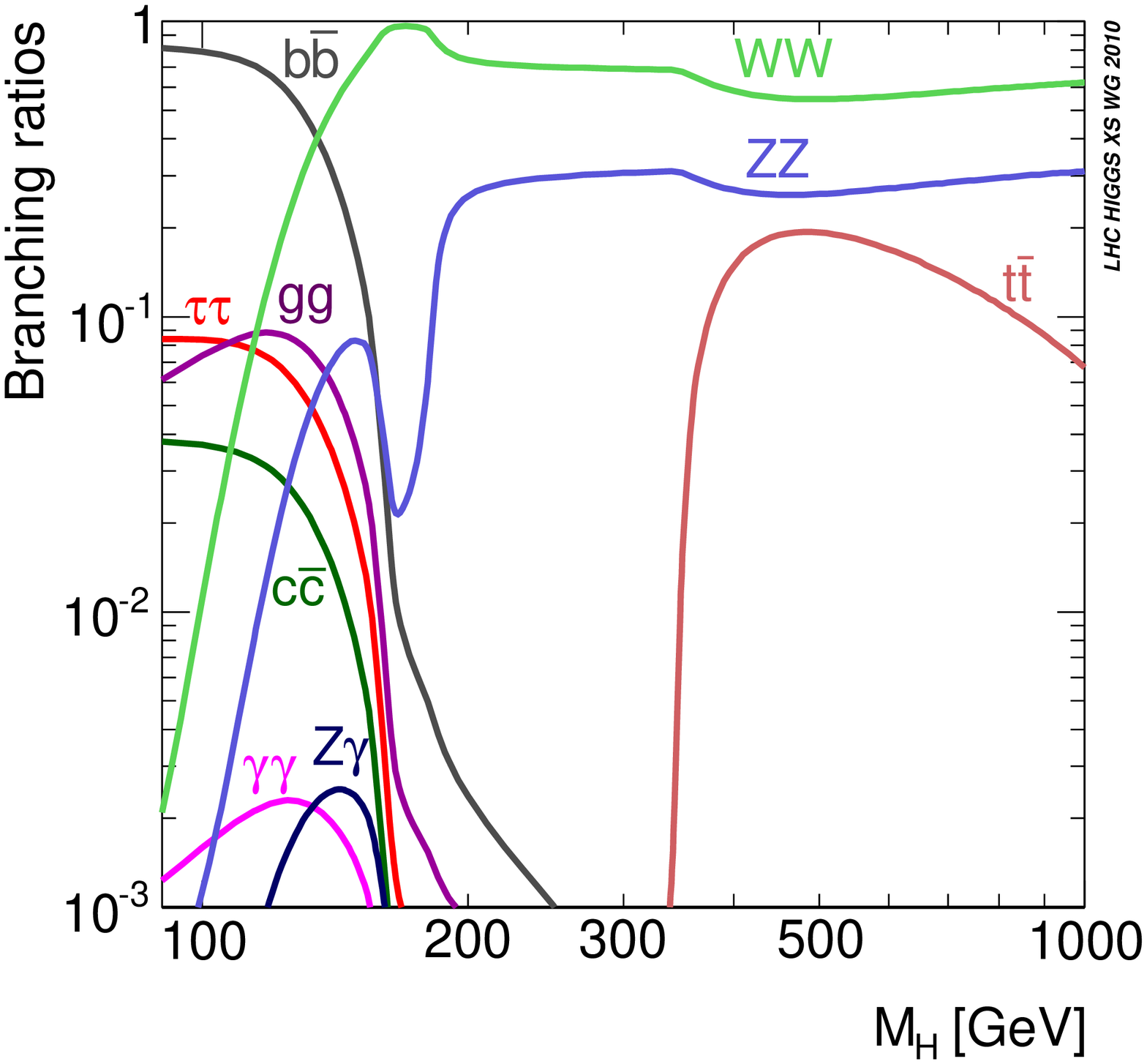}
	\caption{The SM Higgs branching ratios as a function of the Higgs-boson
              mass.}
	\label{fig:SMBRs}
	\end{center}
\end{figure}

The combined numbers in this Summary, for the $\Pg\Pg$-fusion process,
are based on the two predictions (ABPS and dFG) of Section~2; work is in
progress to understand the numerical impact of (possibly) remaining THUs with
the inclusion of other analyses (e.g.\ the BD calculation of Section~2.4).

The total cross section at $\sqrt{s} = 7\UTeV$ and $14\UTeV$ is shown in Fig.~\ref{fig:totalXS}.
The SM Higgs production cross sections for the individual channels are shown in
Fig.~\ref{fig:SMXS7TeV} 
for  $\sqrt{s} = 7$\UTeV\ and in
Fig.~\ref{fig:SMXS14TeV} 
for  $\sqrt{s} = 14$\UTeV, with the combined parametric and theoretical uncertainties,
PU + THU, illustrated by bands. 
The labels on the bands briefly indicated the type of radiative corrections that are
included in the predictions.
For details of the calculations and individual components of the error 
(THU, PDF + $\alphas$, etc.) we refer the reader to the main text, e.g.\
to \Sref{ggFsection} for the $\Pg\Pg$-fusion results.
In \Trefs{tab:XS7a}--\ref{tab:XS14b} the cross sections and associated
total errors for different production channels are summarized
together with the total inclusive Higgs production cross sections. 

The branching ratios for the SM Higgs boson are shown in Fig.~\ref{fig:SMBRs}.
Tables containing explicit numbers on partial widths, branching ratios, and on the 
total width can be found in \Sref{sec:BR} and Appendix~\ref{brappendix}.
As already pointed out in \Sref{sec:BR}, a full error analysis of the Higgs branching
ratio is still in progress (see \Bref{Baglio:2010ae} for a recent
independent analysis).

The results shown in this section will be regularly updated at our
webpage\footnote{\sl https://twiki.cern.ch/twiki/bin/view/LHCPhysics/CrossSections}.

Each experiment is recommended to use the common Standard Model input
parameters (\refA{sminput}),
the best known NNLO(NLO) cross sections and branching ratios
reported in this Report as common basis for Higgs physics at LHC.


Beyond the goals of this Report remains the agreement between NLO MC
predictions and NNLO calculations within the acceptance of the detectors.
The next step in the activities of this working group will be the
computation of cross sections that include acceptance cuts and differential
distributions for all final states that will be considered in the Higgs
search at the LHC.
Preferably this should be carried out with the same set of (benchmark) cuts
for ATLAS and CMS.
The goal is to understand how the $K$-factors from (N)LO to (N)NLO will
change after introduction of cuts and to compare the NNLO differential
distributions with the ones from Monte Carlo generators at NLO.

There is a final comment concerning the SM background: We plan to estimate
theoretical predictions for the most important backgrounds in the signal
regions.
This means that a {\em background control region} has to be defined, and
there the experiments will measure a given source of background, directly
from data.
The {\em control region} can be in the bulk of the background production
phase space, but can also be in the tail of the distributions.
Thus it is important to define the precision with which the SM background
will be measured and the theoretical precision available for that particular
region.
Then the background uncertainty should be extrapolated back to the
{\em signal region}, using available theoretical predictions and their
uncertainty.
It will be important to compute the interference between signal and background
and try to access this at NLO.
The (N)LO Monte Carlos will be used to simulate this background and
determine how the $K$-factor is changing with the chosen kinematic cuts.

\vfill



\begin{landscape}
\begin{table}
	\vspace{-\headsep}
	\begin{center}
	\caption{SM Higgs-boson production cross section at
	$\sqrt{s}=7$\UTeV: light Higgs boson.}
    \small
	\begin{tabular}{r|rc|rc|rc|rc|rc|r}
\hline
$\MH$ & \multicolumn{2}{c|}{ggF} & \multicolumn{2}{c|}{VBF} &
\multicolumn{2}{c|}{WH} & \multicolumn{2}{c|}{ZH} &
\multicolumn{2}{c|}{ttH} &  Total \\
$[\UGeVZ]$ & $~~~~\sigma [\UpbZ]$ & error [\%]
	       & $~~~~\sigma [\UpbZ]$ & error [\%]
	       & $~~~~\sigma [\UpbZ]$ & error [\%]
	       & $~~~~\sigma [\UpbZ]$ & error [\%]
	       & $~~~~\sigma [\UpbZ]$ & error [\%]
	       & $~~~~\sigma [\UpbZ]$ \\
\hline
 $  90$ & $  29.47$ & $ +22.9 \; -\!15.6$ & $  1.710$ & $ +2.7 \; -\!2.3$ & $  1.640$ & $ +3.3 \; -\!3.8$ & $ 0.8597$ & $ +3.9 \; -\!4.0$ & $  0.2162$ & $ +12.5 \; -\!18.1$ & $     33.90$ \\ 
 $  95$ & $  26.58$ & $ +21.9 \; -\!15.9$ & $  1.628$ & $ +2.5 \; -\!2.5$ & $  1.392$ & $ +3.3 \; -\!4.1$ & $ 0.7348$ & $ +4.6 \; -\!4.7$ & $  0.1880$ & $ +12.4 \; -\!18.0$ & $     30.52$ \\ 
 $ 100$ & $  24.02$ & $ +21.2 \; -\!15.6$ & $  1.546$ & $ +2.6 \; -\!2.4$ & $  1.186$ & $ +4.0 \; -\!3.9$ & $ 0.6313$ & $ +4.5 \; -\!4.6$ & $  0.1638$ & $ +12.3 \; -\!18.0$ & $     27.55$ \\ 
 $ 105$ & $  21.78$ & $ +20.8 \; -\!15.5$ & $  1.472$ & $ +2.5 \; -\!2.4$ & $  1.018$ & $ +3.8 \; -\!4.3$ & $ 0.5449$ & $ +5.0 \; -\!5.3$ & $  0.1433$ & $ +12.1 \; -\!17.9$ & $     24.96$ \\ 
 $ 110$ & $  19.84$ & $ +20.4 \; -\!15.3$ & $  1.398$ & $ +2.8 \; -\!2.3$ & $ 0.8754$ & $ +4.1 \; -\!4.5$ & $ 0.4721$ & $ +5.3 \; -\!5.3$ & $  0.1257$ & $ +12.1 \; -\!18.0$ & $     22.71$ \\ 
 $ 115$ & $  18.13$ & $ +20.0 \; -\!15.3$ & $  1.332$ & $ +2.5 \; -\!2.3$ & $ 0.7546$ & $ +4.3 \; -\!4.7$ & $ 0.4107$ & $ +5.5 \; -\!5.4$ & $  0.1106$ & $ +11.9 \; -\!17.8$ & $     20.74$ \\ 
 $ 120$ & $  16.63$ & $ +19.7 \; -\!15.1$ & $  1.269$ & $ +2.8 \; -\!2.5$ & $ 0.6561$ & $ +3.8 \; -\!4.1$ & $ 0.3598$ & $ +5.0 \; -\!4.7$ & $ 0.09756$ & $ +11.8 \; -\!17.8$ & $     19.01$ \\ 
 $ 125$ & $  15.31$ & $ +19.5 \; -\!15.1$ & $  1.211$ & $ +2.7 \; -\!2.4$ & $ 0.5729$ & $ +3.7 \; -\!4.3$ & $ 0.3158$ & $ +4.9 \; -\!5.1$ & $ 0.08634$ & $ +11.8 \; -\!17.8$ & $     17.50$ \\ 
 $ 130$ & $  14.12$ & $ +19.2 \; -\!15.1$ & $  1.154$ & $ +2.8 \; -\!2.3$ & $ 0.5008$ & $ +3.8 \; -\!4.3$ & $ 0.2778$ & $ +5.2 \; -\!5.1$ & $ 0.07658$ & $ +11.6 \; -\!17.7$ & $     16.13$ \\ 
 $ 135$ & $  13.08$ & $ +18.9 \; -\!15.0$ & $  1.100$ & $ +3.0 \; -\!2.2$ & $ 0.4390$ & $ +4.1 \; -\!3.8$ & $ 0.2453$ & $ +5.3 \; -\!5.0$ & $ 0.06810$ & $ +11.5 \; -\!17.6$ & $     14.93$ \\ 
 $ 140$ & $  12.13$ & $ +18.8 \; -\!14.9$ & $  1.052$ & $ +2.8 \; -\!2.2$ & $ 0.3857$ & $ +4.0 \; -\!4.0$ & $ 0.2172$ & $ +5.2 \; -\!5.3$ & $ 0.06072$ & $ +11.4 \; -\!17.6$ & $     13.85$ \\ 
 $ 145$ & $  11.27$ & $ +18.7 \; -\!14.9$ & $  1.004$ & $ +3.1 \; -\!2.1$ & $ 0.3406$ & $ +4.0 \; -\!4.6$ & $ 0.1930$ & $ +5.8 \; -\!5.8$ & $ 0.05435$ & $ +11.4 \; -\!17.6$ & $     12.86$ \\ 
 $ 150$ & $  10.50$ & $ +18.7 \; -\!14.9$ & $ 0.9617$ & $ +2.9 \; -\!2.2$ & $ 0.3001$ & $ +3.7 \; -\!4.1$ & $ 0.1713$ & $ +5.4 \; -\!5.2$ & $ 0.04869$ & $ +11.3 \; -\!17.5$ & $     11.98$ \\ 
 $ 155$ & $  9.795$ & $ +18.5 \; -\!15.0$ & $ 0.9180$ & $ +3.1 \; -\!2.1$ & $ 0.2646$ & $ +4.0 \; -\!4.3$ & $ 0.1525$ & $ +5.7 \; -\!5.2$ & $ 0.04374$ & $ +11.4 \; -\!17.7$ & $     11.17$ \\ 
 $ 160$ & $  9.080$ & $ +18.6 \; -\!15.0$ & $ 0.8787$ & $ +2.9 \; -\!2.3$ & $ 0.2291$ & $ +4.3 \; -\!4.5$ & $ 0.1334$ & $ +6.0 \; -\!5.7$ & $ 0.03942$ & $ +11.4 \; -\!17.7$ & $     10.36$ \\ 
 $ 165$ & $  8.319$ & $ +18.1 \; -\!14.7$ & $ 0.8517$ & $ +3.1 \; -\!2.1$ & $ 0.2107$ & $ +4.1 \; -\!4.3$ & $ 0.1233$ & $ +6.2 \; -\!5.8$ & $ 0.03559$ & $ +11.3 \; -\!17.7$ & $     9.540$ \\ 
 $ 170$ & $  7.729$ & $ +17.9 \; -\!14.9$ & $ 0.8173$ & $ +3.1 \; -\!2.2$ & $ 0.1883$ & $ +4.3 \; -\!4.5$ & $ 0.1106$ & $ +6.4 \; -\!6.1$ & $ 0.03219$ & $ +11.3 \; -\!17.6$ & $     8.877$ \\ 
 $ 175$ & $  7.211$ & $ +17.9 \; -\!14.8$ & $ 0.7814$ & $ +3.2 \; -\!2.1$ & $ 0.1689$ & $ +4.1 \; -\!4.9$ & $0.09950$ & $ +6.2 \; -\!6.0$ & $ 0.02918$ & $ +11.2 \; -\!17.6$ & $     8.290$ \\ 
 $ 180$ & $  6.739$ & $ +18.1 \; -\!14.7$ & $ 0.7480$ & $ +3.1 \; -\!2.4$ & $ 0.1521$ & $ +4.1 \; -\!4.1$ & $0.08917$ & $ +6.0 \; -\!5.7$ & $ 0.02652$ & $ +11.2 \; -\!17.6$ & $     7.755$ \\ 
 $ 185$ & $  6.295$ & $ +17.4 \; -\!15.0$ & $ 0.7193$ & $ +3.4 \; -\!2.2$ & $ 0.1387$ & $ +3.9 \; -\!4.4$ & $0.08139$ & $ +6.1 \; -\!5.8$ & $ 0.02414$ & $ +11.3 \; -\!17.7$ & $     7.259$ \\ 
 $ 190$ & $  5.896$ & $ +17.3 \; -\!15.0$ & $ 0.6925$ & $ +3.3 \; -\!2.2$ & $ 0.1253$ & $ +4.2 \; -\!4.4$ & $0.07366$ & $ +6.1 \; -\!6.0$ & $ 0.02206$ & $ +11.3 \; -\!17.7$ & $     6.810$ \\ 
 $ 195$ & $  5.551$ & $ +17.2 \; -\!15.1$ & $ 0.6643$ & $ +3.4 \; -\!2.5$ & $ 0.1138$ & $ +4.4 \; -\!4.3$ & $0.06699$ & $ +6.3 \; -\!5.9$ & $ 0.02016$ & $ +11.3 \; -\!17.7$ & $     6.416$ \\ 
 $ 200$ & $  5.249$ & $ +17.2 \; -\!15.2$ & $ 0.6371$ & $ +3.4 \; -\!2.3$ & $ 0.1032$ & $ +4.2 \; -\!4.8$ & $0.06096$ & $ +6.4 \; -\!6.0$ & $ 0.01849$ & $ +11.3 \; -\!17.8$ & $     6.069$ \\ 
 $ 210$ & $  4.723$ & $ +16.9 \; -\!15.3$ & $ 0.5869$ & $ +3.5 \; -\!2.4$ & $0.08557$ & $ +4.2 \; -\!4.4$ & $0.05068$ & $ +6.3 \; -\!6.2$ & $ 0.01562$ & $ +11.7 \; -\!18.1$ & $     5.462$ \\ 
 $ 220$ & $  4.288$ & $ +16.8 \; -\!15.3$ & $ 0.5420$ & $ +3.5 \; -\!2.5$ & $0.07142$ & $ +4.0 \; -\!4.6$ & $0.04235$ & $ +6.4 \; -\!6.1$ & $ 0.01330$ & $ +11.8 \; -\!18.2$ & $     4.957$ \\ 
 $ 230$ & $  3.908$ & $ +16.6 \; -\!15.5$ & $ 0.5011$ & $ +3.8 \; -\!2.4$ & $0.06006$ & $ +5.2 \; -\!5.2$ & $0.03560$ & $ +6.9 \; -\!6.7$ & $ 0.01143$ & $ +12.2 \; -\!18.4$ & $     4.516$ \\ 
 $ 240$ & $  3.581$ & $ +16.7 \; -\!15.4$ & $ 0.4641$ & $ +3.8 \; -\!2.5$ & $0.05075$ & $ +4.5 \; -\!4.7$ & $0.02999$ & $ +6.3 \; -\!6.2$ & $0.009873$ & $ +12.3 \; -\!18.6$ & $     4.136$ \\ 
 $ 250$ & $  3.312$ & $ +16.5 \; -\!15.6$ & $ 0.4304$ & $ +4.0 \; -\!2.6$ & $0.04308$ & $ +4.5 \; -\!4.7$ & $0.02540$ & $ +6.2 \; -\!5.8$ & $0.008593$ & $ +12.6 \; -\!18.8$ & $     3.819$ \\ 
 $ 260$ & $  3.072$ & $ +16.2 \; -\!15.9$ & $ 0.3988$ & $ +4.3 \; -\!2.4$ & $0.03674$ & $ +4.8 \; -\!4.7$ & $0.02158$ & $ +6.3 \; -\!6.2$ & $0.007524$ & $ +12.9 \; -\!18.9$ & $     3.537$ \\ 
 $ 270$ & $  2.864$ & $ +16.2 \; -\!15.8$ & $ 0.3715$ & $ +4.2 \; -\!2.6$ & $0.03146$ & $ +4.4 \; -\!4.7$ & $0.01839$ & $ +6.0 \; -\!6.0$ & $0.006636$ & $ +13.6 \; -\!19.4$ & $     3.292$ \\ 
 $ 280$ & $  2.696$ & $ +16.0 \; -\!16.2$ & $ 0.3461$ & $ +4.3 \; -\!2.7$ & $0.02700$ & $ +4.8 \; -\!5.4$ & $0.01575$ & $ +6.5 \; -\!6.2$ & $0.005889$ & $ +14.2 \; -\!19.9$ & $     3.091$ \\ 
 $ 290$ & $  2.546$ & $ +16.1 \; -\!16.1$ & $ 0.3226$ & $ +4.5 \; -\!2.6$ & $0.02333$ & $ +4.9 \; -\!5.0$ & $0.01355$ & $ +6.0 \; -\!5.8$ & $0.005256$ & $ +14.9 \; -\!20.3$ & $     2.911$ \\ 
 $ 300$ & $  2.418$ & $ +16.1 \; -\!16.1$ & $ 0.3010$ & $ +4.6 \; -\!2.7$ & $0.02018$ & $ +5.1 \; -\!5.4$ & $0.01169$ & $ +6.4 \; -\!6.2$ & $0.004719$ & $ +15.6 \; -\!20.9$ & $     2.756$ \\ 
\hline
	\end{tabular}
	\label{tab:XS7a}
	\end{center}
\end{table}
\end{landscape}


\begin{landscape}
\begin{table}
  \begin{center}
	\caption{SM Higgs-boson production cross section at 
	$\sqrt{s}=7$\UTeV: heavy Higgs boson.}    
    \small
	\begin{tabular}{r|rc|rc|r}
\hline
$\MH$ & \multicolumn{2}{c|}{ggF} & \multicolumn{2}{c|}{VBF} &  Total \\
$[\UGeVZ]$ & $~~~~\sigma [\UpbZ]$  & error [\%]
	       & $~~~~\sigma [\UpbZ]$  & error [\%]
	       & $~~~~\sigma [\UpbZ]$ \\
\hline
 $ 320$ & $  2.248$ & $ +16.3 \; -\!16.2$ & $ 0.2622$ & $ +4.9  \; -\!2.7$ &  $     2.510$ \\ 
 $ 340$ & $  2.199$ & $ +17.6 \; -\!15.7$ & $ 0.2286$ & $ +5.1  \; -\!2.9$ &  $     2.428$ \\ 
 $ 360$ & $  2.359$ & $ +19.1 \; -\!14.8$ & $ 0.2018$ & $ +5.3  \; -\!3.0$ &  $     2.561$ \\ 
 $ 380$ & $  2.263$ & $ +16.9 \; -\!15.8$ & $ 0.1807$ & $ +5.7  \; -\!3.0$ &  $     2.444$ \\ 
 $ 400$ & $  2.035$ & $ +15.3 \; -\!16.6$ & $ 0.1619$ & $ +5.9  \; -\!3.0$ &  $     2.197$ \\ 
 $ 450$ & $  1.356$ & $ +16.4 \; -\!17.4$ & $ 0.1235$ & $ +6.6  \; -\!3.2$ &  $     1.479$ \\ 
 $ 500$ & $ 0.8497$ & $ +17.6 \; -\!17.5$ & $0.09491$ & $ +7.2  \; -\!3.4$ &  $    0.9446$ \\ 
 $ 550$ & $ 0.5259$ & $ +18.4 \; -\!17.6$ & $0.07356$ & $ +7.9  \; -\!3.5$ &  $    0.5995$ \\ 
 $ 600$ & $ 0.3275$ & $ +19.3 \; -\!17.8$ & $0.05763$ & $ +8.6  \; -\!3.8$ &  $    0.3851$ \\ 
 $ 650$ & $ 0.2064$ & $ +19.8 \; -\!17.9$ & $0.04556$ & $ +9.3  \; -\!3.8$ &  $    0.2520$ \\ 
 $ 700$ & $ 0.1320$ & $ +20.5 \; -\!18.2$ & $0.03635$ & $ +9.9  \; -\!4.0$ &  $    0.1683$ \\ 
 $ 750$ & $0.08587$ & $ +21.4 \; -\!18.5$ & $0.02924$ & $ +10.7 \; -\!4.2$ &  $    0.1151$ \\ 
 $ 800$ & $0.05665$ & $ +22.1 \; -\!19.0$ & $0.02371$ & $ +11.3 \; -\!4.3$ &  $   0.08036$ \\ 
 $ 850$ & $0.03786$ & $ +23.1 \; -\!19.6$ & $0.01937$ & $ +11.9 \; -\!4.5$ &  $   0.05723$ \\ 
 $ 900$ & $0.02561$ & $ +24.3 \; -\!20.4$ & $0.01595$ & $ +12.6 \; -\!4.6$ &  $   0.04156$ \\ 
 $ 950$ & $0.01752$ & $ +25.5 \; -\!21.4$ & $0.01321$ & $ +13.4 \; -\!4.8$ &  $   0.03073$ \\ 
 $1000$ & $0.01210$ & $ +27.1 \; -\!22.6$ & $0.01103$ & $ +14.2 \; -\!4.9$ &  $   0.02313$ \\ 
\hline
	\end{tabular}
	\label{tab:XS7b}
	\end{center}
\end{table}
\end{landscape}


\begin{landscape}
\begin{table}
	\vspace{-\headsep}
	\begin{center}
	\caption{SM Higgs-boson production cross section at 
	$\sqrt{s}=14$\UTeV: light Higgs boson.}
    \small
	\begin{tabular}{r|rc|rc|rc|rc|rc|r}
\hline
$\MH$ & \multicolumn{2}{c|}{ggF} & \multicolumn{2}{c|}{VBF} &
\multicolumn{2}{c|}{WH} & \multicolumn{2}{c|}{ZH} &
\multicolumn{2}{c|}{ttH} &  Total \\
$[\UGeVZ]$ & $~~~~\sigma [\UpbZ]$ & error [\%]
	       & $~~~~\sigma [\UpbZ]$ & error [\%]
	       & $~~~~\sigma [\UpbZ]$ & error [\%]
	       & $~~~~\sigma [\UpbZ]$ & error [\%]
	       & $~~~~\sigma [\UpbZ]$ & error [\%]
	       & $~~~~\sigma [\UpbZ]$ \\
\hline
 $  90$ & $  87.55$ & $ +23.0 \; -\!16.4$ & $  5.569$ & $ +2.9 \; -\!3.0$ & $  4.090$ & $ +4.3 \; -\!4.6$ & $  2.245$ & $ +5.3 \; -\!5.7$ & $   1.449$ & $ +14.9 \; -\!18.0$ & $     100.9$ \\ 
 $  95$ & $  79.83$ & $ +22.3 \; -\!16.0$ & $  5.338$ & $ +3.0 \; -\!3.1$ & $  3.499$ & $ +4.4 \; -\!4.5$ & $  1.941$ & $ +5.2 \; -\!5.2$ & $   1.268$ & $ +14.8 \; -\!18.0$ & $     91.88$ \\ 
 $ 100$ & $  73.27$ & $ +21.5 \; -\!16.0$ & $  5.114$ & $ +2.8 \; -\!3.1$ & $  3.002$ & $ +4.5 \; -\!4.3$ & $  1.683$ & $ +5.7 \; -\!5.3$ & $   1.114$ & $ +14.8 \; -\!18.0$ & $     84.18$ \\ 
 $ 105$ & $  67.34$ & $ +21.1 \; -\!15.6$ & $  4.900$ & $ +3.2 \; -\!2.9$ & $  2.596$ & $ +4.1 \; -\!4.0$ & $  1.468$ & $ +5.4 \; -\!5.4$ & $  0.9816$ & $ +14.7 \; -\!18.0$ & $     77.29$ \\ 
 $ 110$ & $  62.16$ & $ +20.6 \; -\!15.3$ & $  4.750$ & $ +2.2 \; -\!3.9$ & $  2.246$ & $ +4.1 \; -\!4.6$ & $  1.283$ & $ +6.1 \; -\!5.6$ & $  0.8681$ & $ +14.8 \; -\!18.1$ & $     71.31$ \\ 
 $ 115$ & $  57.57$ & $ +20.2 \; -\!15.0$ & $  4.520$ & $ +2.9 \; -\!3.0$ & $  1.952$ & $ +4.5 \; -\!4.0$ & $  1.130$ & $ +6.2 \; -\!5.2$ & $  0.7699$ & $ +14.8 \; -\!18.1$ & $     65.94$ \\ 
 $ 120$ & $  53.49$ & $ +20.0 \; -\!14.8$ & $  4.361$ & $ +2.5 \; -\!3.5$ & $  1.710$ & $ +4.4 \; -\!4.1$ & $ 0.9967$ & $ +6.0 \; -\!5.4$ & $  0.6850$ & $ +14.7 \; -\!18.1$ & $     61.24$ \\ 
 $ 125$ & $  49.85$ & $ +19.6 \; -\!14.6$ & $  4.180$ & $ +2.8 \; -\!3.0$ & $  1.504$ & $ +4.1 \; -\!4.4$ & $ 0.8830$ & $ +6.4 \; -\!5.5$ & $  0.6113$ & $ +14.8 \; -\!18.2$ & $     57.03$ \\ 
 $ 130$ & $  46.55$ & $ +19.5 \; -\!14.4$ & $  4.029$ & $ +2.5 \; -\!3.1$ & $  1.324$ & $ +3.8 \; -\!3.7$ & $ 0.7846$ & $ +6.3 \; -\!5.2$ & $  0.5472$ & $ +14.8 \; -\!18.2$ & $     53.23$ \\ 
 $ 135$ & $  43.61$ & $ +19.1 \; -\!14.2$ & $  3.862$ & $ +3.1 \; -\!2.8$ & $  1.167$ & $ +3.5 \; -\!3.4$ & $ 0.6981$ & $ +5.9 \; -\!5.2$ & $  0.4910$ & $ +14.8 \; -\!18.2$ & $     49.83$ \\ 
 $ 140$ & $  40.93$ & $ +18.9 \; -\!13.9$ & $  3.732$ & $ +2.6 \; -\!3.3$ & $  1.034$ & $ +3.3 \; -\!3.8$ & $ 0.6256$ & $ +5.8 \; -\!5.2$ & $  0.4419$ & $ +14.8 \; -\!18.2$ & $     46.76$ \\ 
 $ 145$ & $  38.49$ & $ +18.8 \; -\!13.7$ & $  3.590$ & $ +3.0 \; -\!3.0$ & $ 0.9200$ & $ +3.8 \; -\!3.7$ & $ 0.5601$ & $ +6.7 \; -\!5.5$ & $  0.3989$ & $ +14.9 \; -\!18.3$ & $     43.96$ \\ 
 $ 150$ & $  36.27$ & $ +18.7 \; -\!13.5$ & $  3.460$ & $ +2.8 \; -\!3.0$ & $ 0.8156$ & $ +3.0 \; -\!3.3$ & $ 0.5016$ & $ +6.0 \; -\!4.7$ & $  0.3609$ & $ +14.9 \; -\!18.3$ & $     41.41$ \\ 
 $ 155$ & $  34.22$ & $ +18.6 \; -\!13.6$ & $  3.332$ & $ +2.9 \; -\!3.0$ & $ 0.7255$ & $ +3.5 \; -\!3.7$ & $ 0.4513$ & $ +6.5 \; -\!5.6$ & $  0.3275$ & $ +14.9 \; -\!18.4$ & $     39.06$ \\ 
 $ 160$ & $  32.10$ & $ +18.6 \; -\!13.7$ & $  3.198$ & $ +3.1 \; -\!2.8$ & $ 0.6341$ & $ +3.3 \; -\!3.6$ & $ 0.3986$ & $ +6.6 \; -\!5.5$ & $  0.2980$ & $ +15.0 \; -\!18.5$ & $     36.63$ \\ 
 $ 165$ & $  29.77$ & $ +17.8 \; -\!13.4$ & $  3.137$ & $ +3.0 \; -\!2.9$ & $ 0.5850$ & $ +2.6 \; -\!3.0$ & $ 0.3705$ & $ +6.4 \; -\!4.9$ & $  0.2718$ & $ +15.1 \; -\!18.5$ & $     34.13$ \\ 
 $ 170$ & $  27.93$ & $ +17.7 \; -\!13.3$ & $  3.033$ & $ +2.8 \; -\!3.0$ & $ 0.5260$ & $ +3.1 \; -\!3.5$ & $ 0.3355$ & $ +6.5 \; -\!5.4$ & $  0.2487$ & $ +15.7 \; -\!18.9$ & $     32.07$ \\ 
 $ 175$ & $  26.36$ & $ +17.7 \; -\!13.4$ & $  2.922$ & $ +3.5 \; -\!2.8$ & $ 0.4763$ & $ +3.4 \; -\!3.2$ & $ 0.3044$ & $ +6.6 \; -\!5.7$ & $  0.2279$ & $ +15.8 \; -\!18.9$ & $     30.29$ \\ 
 $ 180$ & $  24.92$ & $ +17.8 \; -\!13.4$ & $  2.805$ & $ +3.3 \; -\!2.8$ & $ 0.4274$ & $ +3.2 \; -\!3.4$ & $ 0.2744$ & $ +6.7 \; -\!5.8$ & $  0.2095$ & $ +15.8 \; -\!19.0$ & $     28.64$ \\ 
 $ 185$ & $  23.49$ & $ +17.2 \; -\!13.4$ & $  2.740$ & $ +2.8 \; -\!2.9$ & $ 0.3963$ & $ +2.9 \; -\!3.2$ & $ 0.2524$ & $ +6.1 \; -\!5.5$ & $  0.1930$ & $ +15.8 \; -\!19.0$ & $     27.07$ \\ 
 $ 190$ & $  22.21$ & $ +17.1 \; -\!13.2$ & $  2.652$ & $ +2.7 \; -\!2.9$ & $ 0.3600$ & $ +3.0 \; -\!3.4$ & $ 0.2301$ & $ +6.5 \; -\!5.9$ & $  0.1783$ & $ +16.0 \; -\!19.2$ & $     25.63$ \\ 
 $ 195$ & $  21.10$ & $ +17.0 \; -\!13.2$ & $  2.566$ & $ +2.9 \; -\!2.9$ & $ 0.3291$ & $ +3.0 \; -\!3.4$ & $ 0.2112$ & $ +6.4 \; -\!5.8$ & $  0.1650$ & $ +16.0 \; -\!19.2$ & $     24.37$ \\ 
 $ 200$ & $  20.16$ & $ +16.8 \; -\!13.2$ & $  2.472$ & $ +3.2 \; -\!2.7$ & $ 0.3004$ & $ +3.4 \; -\!3.5$ & $ 0.1936$ & $ +6.7 \; -\!6.1$ & $  0.1532$ & $ +16.2 \; -\!19.4$ & $     23.28$ \\ 
 $ 210$ & $  18.49$ & $ +16.6 \; -\!13.3$ & $  2.315$ & $ +3.2 \; -\!2.7$ & $ 0.2526$ & $ +2.8 \; -\!3.3$ & $ 0.1628$ & $ +6.5 \; -\!5.1$ & $  0.1329$ & $ +16.4 \; -\!19.5$ & $     21.35$ \\ 
 $ 220$ & $  17.08$ & $ +16.4 \; -\!13.3$ & $  2.171$ & $ +2.9 \; -\!3.1$ & $ 0.2138$ & $ +3.4 \; -\!3.3$ & $ 0.1380$ & $ +6.3 \; -\!5.6$ & $  0.1162$ & $ +16.7 \; -\!19.8$ & $     19.72$ \\ 
 $ 230$ & $  15.86$ & $ +16.3 \; -\!13.2$ & $  2.036$ & $ +3.2 \; -\!2.8$ & $ 0.1826$ & $ +3.9 \; -\!4.0$ & $ 0.1173$ & $ +7.0 \; -\!6.2$ & $  0.1025$ & $ +17.1 \; -\!20.0$ & $     18.30$ \\ 
 $ 240$ & $  14.82$ & $ +16.1 \; -\!13.2$ & $  1.918$ & $ +3.0 \; -\!2.7$ & $ 0.1561$ & $ +3.7 \; -\!3.8$ & $0.09996$ & $ +6.5 \; -\!5.9$ & $ 0.09109$ & $ +17.3 \; -\!20.3$ & $     17.09$ \\ 
 $ 250$ & $  13.92$ & $ +16.0 \; -\!13.2$ & $  1.807$ & $ +2.9 \; -\!3.0$ & $ 0.1343$ & $ +3.2 \; -\!3.7$ & $0.08540$ & $ +6.2 \; -\!5.5$ & $ 0.08156$ & $ +17.7 \; -\!20.5$ & $     16.03$ \\ 
 $ 260$ & $  13.14$ & $ +15.9 \; -\!13.3$ & $  1.711$ & $ +2.9 \; -\!3.7$ & $ 0.1161$ & $ +3.0 \; -\!3.5$ & $0.07341$ & $ +6.1 \; -\!5.2$ & $ 0.07351$ & $ +18.1 \; -\!20.8$ & $     15.11$ \\ 
 $ 270$ & $  12.47$ & $ +15.7 \; -\!13.1$ & $  1.606$ & $ +3.0 \; -\!2.9$ & $ 0.1009$ & $ +3.1 \; -\!3.2$ & $0.06325$ & $ +5.3 \; -\!4.7$ & $ 0.06667$ & $ +18.5 \; -\!21.1$ & $     14.31$ \\ 
 $ 280$ & $  11.90$ & $ +15.7 \; -\!13.1$ & $  1.514$ & $ +3.2 \; -\!2.7$ & $0.08781$ & $ +3.4 \; -\!3.6$ & $0.05474$ & $ +5.7 \; -\!5.0$ & $ 0.06081$ & $ +19.0 \; -\!21.4$ & $     13.62$ \\ 
 $ 290$ & $  11.43$ & $ +15.4 \; -\!13.2$ & $  1.436$ & $ +3.2 \; -\!2.8$ & $0.07714$ & $ +3.5 \; -\!3.8$ & $0.04769$ & $ +5.4 \; -\!4.7$ & $ 0.05575$ & $ +19.4 \; -\!21.7$ & $     13.05$ \\ 
 $ 300$ & $  11.05$ & $ +15.3 \; -\!13.0$ & $  1.358$ & $ +3.2 \; -\!2.9$ & $0.06755$ & $ +3.9 \; -\!3.8$ & $0.04156$ & $ +5.6 \; -\!5.2$ & $ 0.05133$ & $ +19.8 \; -\!21.9$ & $     12.57$ \\ 
\hline
	\end{tabular}
	\label{tab:XS14a}
	\end{center}
\end{table}
\end{landscape}


\begin{landscape}
\begin{table}
  \begin{center}
	\caption{SM Higgs-boson production cross section at 
	$\sqrt{s}=14$\UTeV: heavy Higgs boson.}    
    \small
	\begin{tabular}{r|rc|rc|r}
\hline
$\MH$ & \multicolumn{2}{c|}{ggF} & \multicolumn{2}{c|}{VBF} &  Total \\
$[\UGeVZ]$ & $~~~~\sigma [\UpbZ]$  & error [\%]
	       & $~~~~\sigma [\UpbZ]$  & error [\%]
	       & $~~~~\sigma [\UpbZ]$ \\
\hline
 $ 320$ & $  10.59$ & $ +15.4 \; -\!12.9$ & $  1.220$ & $ +3.2 \; -\!2.8$   & $      11.81$ \\ 
 $ 340$ & $  10.72$ & $ +15.9 \; -\!13.4$ & $  1.094$ & $ +3.3 \; -\!2.8$   & $      11.81$ \\ 
 $ 360$ & $  11.91$ & $ +16.5 \; -\!13.8$ & $ 0.9930$ & $ +3.3 \; -\!2.8$   & $      12.90$ \\ 
 $ 380$ & $  11.72$ & $ +15.3 \; -\!13.3$ & $ 0.9148$ & $ +3.4 \; -\!2.7$   & $      12.63$ \\ 
 $ 400$ & $  10.87$ & $ +13.2 \; -\!13.6$ & $ 0.8422$ & $ +3.6 \; -\!2.7$   & $      11.71$ \\ 
 $ 450$ & $  7.790$ & $ +12.6 \; -\!13.7$ & $ 0.6893$ & $ +3.8 \; -\!3.0$   & $      8.479$ \\ 
 $ 500$ & $  5.255$ & $ +13.7 \; -\!13.9$ & $ 0.5684$ & $ +4.0 \; -\!2.9$   & $      5.823$ \\ 
 $ 550$ & $  3.493$ & $ +14.2 \; -\!14.1$ & $ 0.4724$ & $ +4.4 \; -\!3.0$   & $      3.965$ \\ 
 $ 600$ & $  2.332$ & $ +14.5 \; -\!14.0$ & $ 0.3965$ & $ +4.7 \; -\!3.1$   & $      2.728$ \\ 
 $ 650$ & $  1.576$ & $ +14.5 \; -\!13.8$ & $ 0.3360$ & $ +4.9 \; -\!3.2$   & $      1.912$ \\ 
 $ 700$ & $  1.078$ & $ +15.2 \; -\!14.1$ & $ 0.2872$ & $ +5.2 \; -\!3.4$   & $      1.365$ \\ 
 $ 750$ & $ 0.7498$ & $ +15.5 \; -\!13.9$ & $ 0.2476$ & $ +5.6 \; -\!3.5$   & $     0.9974$ \\ 
 $ 800$ & $ 0.5280$ & $ +15.6 \; -\!14.0$ & $ 0.2155$ & $ +5.8 \; -\!3.7$   & $     0.7435$ \\ 
 $ 850$ & $ 0.3766$ & $ +15.9 \; -\!14.2$ & $ 0.1885$ & $ +6.3 \; -\!3.6$   & $     0.5651$ \\ 
 $ 900$ & $ 0.2723$ & $ +16.3 \; -\!14.5$ & $ 0.1666$ & $ +6.5 \; -\!3.8$   & $     0.4389$ \\ 
 $ 950$ & $ 0.1987$ & $ +16.8 \; -\!14.5$ & $ 0.1484$ & $ +6.6 \; -\!4.0$   & $     0.3471$ \\ 
 $1000$ & $ 0.1472$ & $ +16.8 \; -\!14.6$ & $ 0.1330$ & $ +7.0 \; -\!4.0$   & $     0.2802$ \\ 
\hline
	\end{tabular}
	\label{tab:XS14b}
	\end{center}
\end{table}
\end{landscape}

\clearpage

\begin{flushleft}
{\bf Acknowledgements}
\end{flushleft}

We are obliged to 
Ketevi Assamagan, 
Albert De Roeck,
Andrey Korytov,
Sandra Kortner, 
Bill Murray, 
Aleandro Nisati, 
Gigi Rolandi, and
Vivek Sharma
for their support and encouragement.

We would like to acknowledge the assistance of
Andrea Benaglia, 
Cristina Botta, 
Nicola De Filippis,  
David d'Enterria,
Pietro Govoni,
Claire Gwenlan,
Judith Katz,
and Andrea Massironi. 

We would like to thank Sergey Alekhin, Richard Ball, Johannes Bl\"umlein,
Jon Butterworth, Amanda Cooper-Sarkar, Sasha Glazov,
Alberto Guffanti, Pedro Jimenez-Delgado, Sven Moch, Pavel Nadolsky,
Ewald Reya, Juan Rojo, and Graeme Watt for discussions. 

We are obliged to CERN, in particular to the IT Department and to the
Theory Unit for the support with logistics, especially to Elena Gianolio for
technical assistance.

We also acknowledge partial support from the European Community's Marie-Curie
Research Training Network under contract MRTN-CT-2006-035505 `Tools
and Precision Calculations for Physics Discoveries at Colliders',
from the Science and Technology Facilities Council, 
from the US Department of Energy,
and
from the Forschungsschwerpunkt 101
by the Bundesministerium f\"ur Bildung und Forschung, Germany.

R.~S.~Thorne would also like to thank the IPPP Durham for the award of a
Research Associateship which helped make participation in the working
group possible. 

T.~Vickey acknowledges
the Oxford Oppenheimer Fund,
the Royal Society of the United Kingdom,
and the National Research Foundation of the Republic of South Africa.

O.~Brein and M.~Warsinsky acknowledge the support of the Initiative and 
Networking Fund of the Helmholtz Association, contract HA-101
(`Physics at the Terascale').

\bibliographystyle{atlasnote}
\bibliography{YRHXS}

\clearpage
\appendix
\section{The Standard Model input parameter set}
\label{sminput}

The production cross sections and decay branching ratios of the Higgs bosons depend on a large number of Standard Model parameters.
Unless otherwise specified, the following {\em default} parameter sets\footnote{These parameters can be found at {\sl https://twiki.cern.ch/twiki/bin/view/LHCPhysics/SMInputParameter }}
are used as listed in Table~\ref{tab:SMinput}.

The strong coupling constant $\alphas$ is in general taken to be the value from the PDF set used.
MSTW2008 determines the $\alphas$ value as part of its PDF fit: $\alphas(\MZ^2)=0.1394$ at LO,  $0.1202$ at NLO and $0.1171$ at NNLO.
The CTEQ collaboration uses the world  average values ($\alphas(\MZ^2)=0.130$ at LO and  $\alphas(\MZ^2)=0.118$ at NLO) for its PDF fits.
The NNPDF collaboration uses $\alphas(\MZ^2)=0.119$ at NLO.

\begin{table}[hb]
        \begin{center}
        \caption{The Standard Model input parameters for particle masses and widths for computing cross section and branching ratios as presented in this Report.}
        \begin{tabular}{|c|l|}
                \hline
                Parameter & Value$\pm$Error \\ \hline\hline
                electron mass & 0.510998910(13)\UMeV \\
                muon mass &  105.658367(4)\UMeV  \\
                tau mass &  1776.84(17)\UMeV  \\
                \PQu quark mass & 190\UMeV  \\
                \PQd quark mass & 190\UMeV  \\
                \PQs quark mass & 190\UMeV  \\
                \PQc quark mass & 1.40\UGeV  \\
                \PQb quark mass & 4.75\UGeV  \\
                \PQt quark mass & 172.5 $\pm$ 2.5\UGeV \\
                $\MSbar$ scheme \PQc mass & 1.28\UGeV \\
                $\MSbar$ scheme \PQb mass & 4.16\UGeV \\
                \PQc pole mass 1-loop & 1.41\UGeV\\
                \PQc pole mass 2-loop & 1.55\UGeV \\
                \PQb pole mass 1-loop & 4.49\UGeV \\
                \PQb pole mass 2-loop & 4.69\UGeV \\
                \PW mass & 80.398 $\pm$  0.025\UGeV \\
                $\GW$  & 2.141 $\pm$ 0.041\UGeV \\
                NLO $\GW$ & 2.08872\UGeV   \\
                \PZ mass &91.1876 $\pm$ 0.0021\UGeV  \\
                $\GZ$ & 2.4952 $\pm$ 0.0023\UGeV \\
                NLO $\GZ$ & 2.49595\UGeV   \\
                $\GF$ & 1.16637(1)  $\times$ 10$^{-5}$\UGeV$^{-2}$  \\ \hline \hline
        \end{tabular}
        \end{center}
        \label{tab:SMinput}
\end{table}

\clearpage

\section{SM Higgs-boson partial widths}
\label{brappendix}

SM Higgs boson partial widths\footnote{Full listings can be found at
  {\sl https://twiki.cern.ch/twiki/bin/view/LHCPhysics/CERNYellowReportPageBR}}
for all relevant 2-particle decay channels are listed in Tables~\ref{tab:Width-lm.part1}--\ref{tab:Width-hm.part2}.
In \Trefs{tab:PWidth-lm}--\ref{tab:PWidth-hm2} we list the
partial widths of the SM Higgs boson decaying into $4$-fermion final
states,
where leptons include
$\Pl=\Pe,\PGm,\PGt,\PGn_{\Pe},\PGn_{\PGm},\PGn_{\PGt}$,
and quarks $\PQq=\PQu,\PQd,\PQs,\PQc,\PQb$.
Since all fermion masses are neglected, the branching ratios are
identical for different flavours $\Pe$, $\PGm$ or $\PGt$.
We display results for various $4$-lepton final states ($\PH \to
\Pep\Pem\Pep\Pem$,
$\Pep\Pem\PGmp\PGmm$,
$\Plp\Plm\Plp\Plm$,
$\Pep\PGne\Pem\PAGne$,
$\Pep\PGne\PGmm\PAGnGm$,
$\Plp\Plm\PGnl\PAGnl$)
in \Trefs{tab:PWidth-lm}--\ref{tab:PWidth-hm}.
Results for final states with $2$ leptons
and $2$ quarks ($\PH \to \Plp\Plm\PQq\PAQq$,
$\Plp\PGnl\PQq\PAQq^{\prime}$, $\PGnl\PAGnl\PQq\PAQq$), 
$4$ arbitrary quarks ($\PH \to \PQq\PQq\PQq\PQq$),
and for all possible $4$-fermion final states ($\PH \to
\Pf\Pf\Pf\Pf$)
are provided in \Trefs{tab:PWidth-lm2}--\ref{tab:PWidth-hm2}.

\begin{table}
  \vspace{-\headsep}
  \caption{SM Higgs-boson partial widths [GeV] in fermionic final states, for the low- and intermediate-mass range.}
  \label{tab:Width-lm.part1}
  \centering
  \small
  \begin{tabular}{lccccccc}\hline
$\MH$ [GeV] & $\PH \rightarrow \PQb\PAQb$ &  $\PH \rightarrow \PGt \PGt$ 
& $\PH \rightarrow \PGm \PGm$ & $\PH \rightarrow \PQs \PAQs$ & 
$\PH \rightarrow \PQc \PAQc$
& $\PH \rightarrow \PQt \PAQt$ \\
\hline
$90 $&$ 1.79\cdot 10^{-3} $&$ 1.85\cdot 10^{-4} $&$ 6.43 \cdot 10^{-7} $&$ 1.36\cdot 10^{-6} $&$ 8.32\cdot 10^{-5} $&$ 0.00  $\\
$95 $&$ 1.87\cdot 10^{-3} $&$ 1.95\cdot 10^{-4} $&$ 6.79 \cdot 10^{-7} $&$ 1.42\cdot 10^{-6} $&$ 8.68\cdot 10^{-5} $&$ 0.00  $\\
$100 $&$ 1.95\cdot 10^{-3} $&$ 2.06\cdot 10^{-4} $&$ 7.14 \cdot 10^{-7} $&$ 1.48\cdot 10^{-6} $&$ 9.04\cdot 10^{-5} $&$ 0.00  $\\
$105 $&$ 2.02\cdot 10^{-3} $&$ 2.16\cdot 10^{-4} $&$ 7.50 \cdot 10^{-7} $&$ 1.54\cdot 10^{-6} $&$ 9.40\cdot 10^{-5} $&$ 0.00  $\\
$110 $&$ 2.10\cdot 10^{-3} $&$ 2.26\cdot 10^{-4} $&$ 7.85 \cdot 10^{-7} $&$ 1.60\cdot 10^{-6} $&$ 9.75\cdot 10^{-5} $&$ 0.00  $\\
$115 $&$ 2.18\cdot 10^{-3} $&$ 2.37\cdot 10^{-4} $&$ 8.21 \cdot 10^{-7} $&$ 1.66\cdot 10^{-6} $&$ 1.01\cdot 10^{-4} $&$ 0.00  $\\
$120 $&$ 2.25\cdot 10^{-3} $&$ 2.47\cdot 10^{-4} $&$ 8.56 \cdot 10^{-7} $&$ 1.72\cdot 10^{-6} $&$ 1.05\cdot 10^{-4} $&$ 0.00  $\\
$125 $&$ 2.33\cdot 10^{-3} $&$ 2.57\cdot 10^{-4} $&$ 8.92 \cdot 10^{-7} $&$ 1.77\cdot 10^{-6} $&$ 1.08\cdot 10^{-4} $&$ 0.00  $\\
$130 $&$ 2.40\cdot 10^{-3} $&$ 2.67\cdot 10^{-4} $&$ 9.28 \cdot 10^{-7} $&$ 1.83\cdot 10^{-6} $&$ 1.11\cdot 10^{-4} $&$ 0.00  $\\
$135 $&$ 2.48\cdot 10^{-3} $&$ 2.78\cdot 10^{-4} $&$ 9.63 \cdot 10^{-7} $&$ 1.89\cdot 10^{-6} $&$ 1.15\cdot 10^{-4} $&$ 0.00  $\\
$140 $&$ 2.55\cdot 10^{-3} $&$ 2.88\cdot 10^{-4} $&$ 9.99 \cdot 10^{-7} $&$ 1.94\cdot 10^{-6} $&$ 1.18\cdot 10^{-4} $&$ 0.00  $\\
$145 $&$ 2.63\cdot 10^{-3} $&$ 2.98\cdot 10^{-4} $&$ 1.03\cdot 10^{-6} $&$ 2.00\cdot 10^{-6} $&$ 1.22\cdot 10^{-4} $&$ 0.00  $\\
$150 $&$ 2.70\cdot 10^{-3} $&$ 3.08\cdot 10^{-4} $&$ 1.07\cdot 10^{-6} $&$ 2.05\cdot 10^{-6} $&$ 1.25\cdot 10^{-4} $&$ 0.00  $\\
$155 $&$ 2.77\cdot 10^{-3} $&$ 3.19\cdot 10^{-4} $&$ 1.11\cdot 10^{-6} $&$ 2.11\cdot 10^{-6} $&$ 1.28\cdot 10^{-4} $&$ 0.00  $\\
$160 $&$ 2.85\cdot 10^{-3} $&$ 3.29\cdot 10^{-4} $&$ 1.14\cdot 10^{-6} $&$ 2.17\cdot 10^{-6} $&$ 1.32\cdot 10^{-4} $&$ 0.00  $\\
$165 $&$ 2.92\cdot 10^{-3} $&$ 3.39\cdot 10^{-4} $&$ 1.18\cdot 10^{-6} $&$ 2.22\cdot 10^{-6} $&$ 1.35\cdot 10^{-4} $&$ 0.00  $\\
$170 $&$ 2.99\cdot 10^{-3} $&$ 3.50\cdot 10^{-4} $&$ 1.21\cdot 10^{-6} $&$ 2.27\cdot 10^{-6} $&$ 1.38\cdot 10^{-4} $&$ 0.00  $\\
$175 $&$ 3.06\cdot 10^{-3} $&$ 3.60\cdot 10^{-4} $&$ 1.25\cdot 10^{-6} $&$ 2.33\cdot 10^{-6} $&$ 1.42\cdot 10^{-4} $&$ 0.00  $\\
$180 $&$ 3.13\cdot 10^{-3} $&$ 3.70\cdot 10^{-4} $&$ 1.28\cdot 10^{-6} $&$ 2.38\cdot 10^{-6} $&$ 1.45\cdot 10^{-4} $&$ 0.00  $\\
$185 $&$ 3.20\cdot 10^{-3} $&$ 3.80\cdot 10^{-4} $&$ 1.32\cdot 10^{-6} $&$ 2.44\cdot 10^{-6} $&$ 1.48\cdot 10^{-4} $&$ 0.00  $\\
$190 $&$ 3.27\cdot 10^{-3} $&$ 3.91\cdot 10^{-4} $&$ 1.35\cdot 10^{-6} $&$ 2.49\cdot 10^{-6} $&$ 1.51\cdot 10^{-4} $&$ 0.00  $\\
$195 $&$ 3.34\cdot 10^{-3} $&$ 4.01\cdot 10^{-4} $&$ 1.39\cdot 10^{-6} $&$ 2.54\cdot 10^{-6} $&$ 1.55\cdot 10^{-4} $&$ 0.00  $\\
$200 $&$ 3.41\cdot 10^{-3} $&$ 4.11\cdot 10^{-4} $&$ 1.43\cdot 10^{-6} $&$ 2.59\cdot 10^{-6} $&$ 1.58\cdot 10^{-4} $&$ 0.00  $\\
$210 $&$ 3.55\cdot 10^{-3} $&$ 4.32\cdot 10^{-4} $&$ 1.50\cdot 10^{-6} $&$ 2.70\cdot 10^{-6} $&$ 1.64\cdot 10^{-4} $&$ 0.00  $\\
$220 $&$ 3.69\cdot 10^{-3} $&$ 4.53\cdot 10^{-4} $&$ 1.57\cdot 10^{-6} $&$ 2.80\cdot 10^{-6} $&$ 1.70\cdot 10^{-4} $&$ 0.00  $\\
$230 $&$ 3.83\cdot 10^{-3} $&$ 4.73\cdot 10^{-4} $&$ 1.64\cdot 10^{-6} $&$ 2.91\cdot 10^{-6} $&$ 1.77\cdot 10^{-4} $&$ 0.00  $\\
$240 $&$ 3.96\cdot 10^{-3} $&$ 4.94\cdot 10^{-4} $&$ 1.71\cdot 10^{-6} $&$ 3.01\cdot 10^{-6} $&$ 1.83\cdot 10^{-4} $&$ 0.00  $\\
$250 $&$ 4.10\cdot 10^{-3} $&$ 5.15\cdot 10^{-4} $&$ 1.78\cdot 10^{-6} $&$ 3.11\cdot 10^{-6} $&$ 1.89\cdot 10^{-4} $&$ 0.00  $\\
$260 $&$ 4.23\cdot 10^{-3} $&$ 5.35\cdot 10^{-4} $&$ 1.86\cdot 10^{-6} $&$ 3.21\cdot 10^{-6} $&$ 1.95\cdot 10^{-4} $&$ 2.45 \cdot 10^{-7}   $\\
$270 $&$ 4.36\cdot 10^{-3} $&$ 5.56\cdot 10^{-4} $&$ 1.93\cdot 10^{-6} $&$ 3.31\cdot 10^{-6} $&$ 2.02\cdot 10^{-4} $&$ 1.27\cdot 10^{-5}   $\\
$280 $&$ 4.50\cdot 10^{-3} $&$ 5.77\cdot 10^{-4} $&$ 2.00\cdot 10^{-6} $&$ 3.42\cdot 10^{-6} $&$ 2.08\cdot 10^{-4} $&$ 7.00\cdot 10^{-5}   $\\
$290 $&$ 4.63\cdot 10^{-3} $&$ 5.98\cdot 10^{-4} $&$ 2.07\cdot 10^{-6} $&$ 3.52\cdot 10^{-6} $&$ 2.14\cdot 10^{-4} $&$ 2.26\cdot 10^{-4}   $\\
$300 $&$ 4.76\cdot 10^{-3} $&$ 6.18\cdot 10^{-4} $&$ 2.14\cdot 10^{-6} $&$ 3.62\cdot 10^{-6} $&$ 2.20\cdot 10^{-4} $&$ 5.79\cdot 10^{-4}   $\\
$310 $&$ 4.90\cdot 10^{-3} $&$ 6.39\cdot 10^{-4} $&$ 2.22\cdot 10^{-6} $&$ 3.72\cdot 10^{-6} $&$ 2.26\cdot 10^{-4} $&$ 1.32\cdot 10^{-3}   $\\
$320 $&$ 5.02\cdot 10^{-3} $&$ 6.60\cdot 10^{-4} $&$ 2.29\cdot 10^{-6} $&$ 3.81\cdot 10^{-6} $&$ 2.32\cdot 10^{-4} $&$ 2.86\cdot 10^{-3}   $\\
$330 $&$ 5.16\cdot 10^{-3} $&$ 6.81\cdot 10^{-4} $&$ 2.36\cdot 10^{-6} $&$ 3.91\cdot 10^{-6} $&$ 2.38\cdot 10^{-4} $&$ 6.31\cdot 10^{-3}   $\\
$340 $&$ 5.29\cdot 10^{-3} $&$ 7.02\cdot 10^{-4} $&$ 2.43\cdot 10^{-6} $&$ 4.01\cdot 10^{-6} $&$ 2.44\cdot 10^{-4} $&$ 1.62\cdot 10^{-2}   $\\
$350 $&$ 5.42\cdot 10^{-3} $&$ 7.23\cdot 10^{-4} $&$ 2.50\cdot 10^{-6} $&$ 4.11\cdot 10^{-6} $&$ 2.50\cdot 10^{-4} $&$ 2.37\cdot 10^{-1}   $\\
$360 $&$ 5.55\cdot 10^{-3} $&$ 7.44\cdot 10^{-4} $&$ 2.58\cdot 10^{-6} $&$ 4.21\cdot 10^{-6} $&$ 2.56\cdot 10^{-4} $&$ 9.07\cdot 10^{-1}   $\\
$370 $&$ 5.68\cdot 10^{-3} $&$ 7.65\cdot 10^{-4} $&$ 2.65\cdot 10^{-6} $&$ 4.31\cdot 10^{-6} $&$ 2.62\cdot 10^{-4} $&$ 1.69  $\\
$380 $&$ 5.81\cdot 10^{-3} $&$ 7.86\cdot 10^{-4} $&$ 2.72\cdot 10^{-6} $&$ 4.41\cdot 10^{-6} $&$ 2.68\cdot 10^{-4} $&$ 2.54  $\\
$390 $&$ 5.94\cdot 10^{-3} $&$ 8.07\cdot 10^{-4} $&$ 2.80\cdot 10^{-6} $&$ 4.51\cdot 10^{-6} $&$ 2.74\cdot 10^{-4} $&$ 3.43  $\\
$400 $&$ 6.07\cdot 10^{-3} $&$ 8.28\cdot 10^{-4} $&$ 2.87\cdot 10^{-6} $&$ 4.61\cdot 10^{-6} $&$ 2.80\cdot 10^{-4} $&$ 4.34  $\\
$410 $&$ 6.20\cdot 10^{-3} $&$ 8.49\cdot 10^{-4} $&$ 2.94\cdot 10^{-6} $&$ 4.70\cdot 10^{-6} $&$ 2.86\cdot 10^{-4} $&$ 5.25  $\\
$420 $&$ 6.32\cdot 10^{-3} $&$ 8.70\cdot 10^{-4} $&$ 3.02\cdot 10^{-6} $&$ 4.80\cdot 10^{-6} $&$ 2.92\cdot 10^{-4} $&$ 6.16  $\\
$430 $&$ 6.45\cdot 10^{-3} $&$ 8.92\cdot 10^{-4} $&$ 3.09\cdot 10^{-6} $&$ 4.90\cdot 10^{-6} $&$ 2.98\cdot 10^{-4} $&$ 7.06  $\\
$440 $&$ 6.58\cdot 10^{-3} $&$ 9.13\cdot 10^{-4} $&$ 3.16\cdot 10^{-6} $&$ 4.99\cdot 10^{-6} $&$ 3.04\cdot 10^{-4} $&$ 7.96  $\\
$450 $&$ 6.71\cdot 10^{-3} $&$ 9.34\cdot 10^{-4} $&$ 3.24\cdot 10^{-6} $&$ 5.09\cdot 10^{-6} $&$ 3.09\cdot 10^{-4} $&$ 8.85  $\\
$460 $&$ 6.84\cdot 10^{-3} $&$ 9.55\cdot 10^{-4} $&$ 3.31\cdot 10^{-6} $&$ 5.19\cdot 10^{-6} $&$ 3.15\cdot 10^{-4} $&$ 9.73  $\\
$470 $&$ 6.96\cdot 10^{-3} $&$ 9.77\cdot 10^{-4} $&$ 3.38\cdot 10^{-6}
$&$ 5.29\cdot 10^{-6} $&$ 3.21\cdot 10^{-4} $&$ 10.6 $\\
$480 $&$ 7.09\cdot 10^{-3} $&$ 9.98\cdot 10^{-4} $&$ 3.46\cdot 10^{-6}
$&$ 5.38\cdot 10^{-6} $&$ 3.27\cdot 10^{-4} $&$ 11.4 $\\
$490 $&$ 7.22\cdot 10^{-3} $&$ 1.02\cdot 10^{-3} $&$ 3.53\cdot 10^{-6}
$&$ 5.48\cdot 10^{-6} $&$ 3.33\cdot 10^{-4} $&$ 12.3 $\\
\hline
  \end{tabular}
\end{table}

\begin{table}
  \vspace{-\headsep}
  \caption{SM Higgs-boson partial widths [GeV] in fermionic final states, for the high-mass range.}
  \label{tab:Width-hm.part1}
  \centering
  \small
  \begin{tabular}{lccccccc}\hline
$\MH$ [GeV] & $\PH \rightarrow \PQb\PAQb$ &  $\PH \rightarrow 
\PGt \PGt$ & $\PH \rightarrow \PGm \PGm$ & $\PH \rightarrow \PQs \PAQs$ & 
$\PH \rightarrow \PQc \PAQc$
& $\PH \rightarrow \PQt \PAQt$ \\
\hline
$500 $&$ 7.35\cdot 10^{-3} $&$ 1.04\cdot 10^{-3} $&$ 3.60\cdot 10^{-6} $&$ 5.57\cdot 10^{-6} $&$ 3.39\cdot 10^{-4} $&$ 13.1$\\
$510 $&$ 7.47\cdot 10^{-3} $&$ 1.06\cdot 10^{-3} $&$ 3.68\cdot 10^{-6} $&$ 5.67\cdot 10^{-6} $&$ 3.45\cdot 10^{-4} $&$ 13.9$\\
$520 $&$ 7.60\cdot 10^{-3} $&$ 1.08\cdot 10^{-3} $&$ 3.75\cdot 10^{-6} $&$ 5.77\cdot 10^{-6} $&$ 3.50\cdot 10^{-4} $&$ 14.8$\\
$530 $&$ 7.72\cdot 10^{-3} $&$ 1.10\cdot 10^{-3} $&$ 3.83\cdot 10^{-6} $&$ 5.86\cdot 10^{-6} $&$ 3.56\cdot 10^{-4} $&$ 15.5$\\
$540 $&$ 7.85\cdot 10^{-3} $&$ 1.13\cdot 10^{-3} $&$ 3.90\cdot 10^{-6} $&$ 5.96\cdot 10^{-6} $&$ 3.62\cdot 10^{-4} $&$ 16.3$\\
$550 $&$ 7.98\cdot 10^{-3} $&$ 1.15\cdot 10^{-3} $&$ 3.97\cdot 10^{-6} $&$ 6.05\cdot 10^{-6} $&$ 3.68\cdot 10^{-4} $&$ 17.1$\\
$560 $&$ 8.10\cdot 10^{-3} $&$ 1.17\cdot 10^{-3} $&$ 4.05\cdot 10^{-6} $&$ 6.15\cdot 10^{-6} $&$ 3.73\cdot 10^{-4} $&$ 17.8$\\
$570 $&$ 8.23\cdot 10^{-3} $&$ 1.19\cdot 10^{-3} $&$ 4.12\cdot 10^{-6} $&$ 6.24\cdot 10^{-6} $&$ 3.79\cdot 10^{-4} $&$ 18.6$\\
$580 $&$ 8.35\cdot 10^{-3} $&$ 1.21\cdot 10^{-3} $&$ 4.20\cdot 10^{-6} $&$ 6.34\cdot 10^{-6} $&$ 3.85\cdot 10^{-4} $&$ 19.3$\\
$590 $&$ 8.48\cdot 10^{-3} $&$ 1.23\cdot 10^{-3} $&$ 4.27\cdot 10^{-6} $&$ 6.43\cdot 10^{-6} $&$ 3.91\cdot 10^{-4} $&$ 20.1$\\
$600 $&$ 8.60\cdot 10^{-3} $&$ 1.25\cdot 10^{-3} $&$ 4.35\cdot 10^{-6} $&$ 6.53\cdot 10^{-6} $&$ 3.96\cdot 10^{-4} $&$ 20.8$\\
$610 $&$ 8.72\cdot 10^{-3} $&$ 1.28\cdot 10^{-3} $&$ 4.42\cdot 10^{-6} $&$ 6.62\cdot 10^{-6} $&$ 4.02\cdot 10^{-4} $&$ 21.5$\\
$620 $&$ 8.85\cdot 10^{-3} $&$ 1.30\cdot 10^{-3} $&$ 4.49\cdot 10^{-6} $&$ 6.71\cdot 10^{-6} $&$ 4.08\cdot 10^{-4} $&$ 22.2$\\
$630 $&$ 8.97\cdot 10^{-3} $&$ 1.32\cdot 10^{-3} $&$ 4.57\cdot 10^{-6} $&$ 6.81\cdot 10^{-6} $&$ 4.13\cdot 10^{-4} $&$ 22.9$\\
$640 $&$ 9.10\cdot 10^{-3} $&$ 1.34\cdot 10^{-3} $&$ 4.64\cdot 10^{-6} $&$ 6.90\cdot 10^{-6} $&$ 4.19\cdot 10^{-4} $&$ 23.6$\\
$650 $&$ 9.22\cdot 10^{-3} $&$ 1.36\cdot 10^{-3} $&$ 4.72\cdot 10^{-6} $&$ 6.99\cdot 10^{-6} $&$ 4.25\cdot 10^{-4} $&$ 24.2$\\
$660 $&$ 9.34\cdot 10^{-3} $&$ 1.38\cdot 10^{-3} $&$ 4.79\cdot 10^{-6} $&$ 7.08\cdot 10^{-6} $&$ 4.30\cdot 10^{-4} $&$ 24.9$\\
$670 $&$ 9.46\cdot 10^{-3} $&$ 1.40\cdot 10^{-3} $&$ 4.86\cdot 10^{-6} $&$ 7.18\cdot 10^{-6} $&$ 4.36\cdot 10^{-4} $&$ 25.6$\\
$680 $&$ 9.58\cdot 10^{-3} $&$ 1.43\cdot 10^{-3} $&$ 4.94\cdot 10^{-6} $&$ 7.27\cdot 10^{-6} $&$ 4.42\cdot 10^{-4} $&$ 26.2$\\
$690 $&$ 9.70\cdot 10^{-3} $&$ 1.45\cdot 10^{-3} $&$ 5.01\cdot 10^{-6} $&$ 7.36\cdot 10^{-6} $&$ 4.47\cdot 10^{-4} $&$ 26.8$\\
$700 $&$ 9.83\cdot 10^{-3} $&$ 1.47\cdot 10^{-3} $&$ 5.09\cdot 10^{-6} $&$ 7.45\cdot 10^{-6} $&$ 4.53\cdot 10^{-4} $&$ 27.5$\\
$710 $&$ 9.95\cdot 10^{-3} $&$ 1.49\cdot 10^{-3} $&$ 5.16\cdot 10^{-6} $&$ 7.54\cdot 10^{-6} $&$ 4.58\cdot 10^{-4} $&$ 28.1$\\
$720 $&$ 1.01\cdot 10^{-2} $&$ 1.51\cdot 10^{-3} $&$ 5.23\cdot 10^{-6} $&$ 7.64\cdot 10^{-6} $&$ 4.64\cdot 10^{-4} $&$ 28.7$\\
$730 $&$ 1.02\cdot 10^{-2} $&$ 1.53\cdot 10^{-3} $&$ 5.31\cdot 10^{-6} $&$ 7.73\cdot 10^{-6} $&$ 4.69\cdot 10^{-4} $&$ 29.4$\\
$740 $&$ 1.03\cdot 10^{-2} $&$ 1.55\cdot 10^{-3} $&$ 5.38\cdot 10^{-6} $&$ 7.82\cdot 10^{-6} $&$ 4.75\cdot 10^{-4} $&$ 29.9$\\
$750 $&$ 1.04\cdot 10^{-2} $&$ 1.57\cdot 10^{-3} $&$ 5.45\cdot 10^{-6} $&$ 7.91\cdot 10^{-6} $&$ 4.80\cdot 10^{-4} $&$ 30.6$\\
$760 $&$ 1.05\cdot 10^{-2} $&$ 1.60\cdot 10^{-3} $&$ 5.53\cdot 10^{-6} $&$ 8.00\cdot 10^{-6} $&$ 4.86\cdot 10^{-4} $&$ 31.2$\\
$770 $&$ 1.07\cdot 10^{-2} $&$ 1.62\cdot 10^{-3} $&$ 5.60\cdot 10^{-6} $&$ 8.08\cdot 10^{-6} $&$ 4.91\cdot 10^{-4} $&$ 31.7$\\
$780 $&$ 1.08\cdot 10^{-2} $&$ 1.64\cdot 10^{-3} $&$ 5.67\cdot 10^{-6} $&$ 8.17\cdot 10^{-6} $&$ 4.96\cdot 10^{-4} $&$ 32.3$\\
$790 $&$ 1.09\cdot 10^{-2} $&$ 1.66\cdot 10^{-3} $&$ 5.75\cdot 10^{-6} $&$ 8.26\cdot 10^{-6} $&$ 5.02\cdot 10^{-4} $&$ 32.9$\\
$800 $&$ 1.10\cdot 10^{-2} $&$ 1.68\cdot 10^{-3} $&$ 5.82\cdot 10^{-6} $&$ 8.35\cdot 10^{-6} $&$ 5.07\cdot 10^{-4} $&$ 33.5$\\
$810 $&$ 1.11\cdot 10^{-2} $&$ 1.70\cdot 10^{-3} $&$ 5.89\cdot 10^{-6} $&$ 8.44\cdot 10^{-6} $&$ 5.12\cdot 10^{-4} $&$ 34.1$\\
$820 $&$ 1.12\cdot 10^{-2} $&$ 1.72\cdot 10^{-3} $&$ 5.96\cdot 10^{-6} $&$ 8.52\cdot 10^{-6} $&$ 5.18\cdot 10^{-4} $&$ 34.6$\\
$830 $&$ 1.14\cdot 10^{-2} $&$ 1.74\cdot 10^{-3} $&$ 6.03\cdot 10^{-6} $&$ 8.61\cdot 10^{-6} $&$ 5.23\cdot 10^{-4} $&$ 35.2$\\
$840 $&$ 1.15\cdot 10^{-2} $&$ 1.76\cdot 10^{-3} $&$ 6.10\cdot 10^{-6} $&$ 8.69\cdot 10^{-6} $&$ 5.28\cdot 10^{-4} $&$ 35.7$\\
$850 $&$ 1.16\cdot 10^{-2} $&$ 1.78\cdot 10^{-3} $&$ 6.17\cdot 10^{-6} $&$ 8.78\cdot 10^{-6} $&$ 5.33\cdot 10^{-4} $&$ 36.2$\\
$860 $&$ 1.17\cdot 10^{-2} $&$ 1.80\cdot 10^{-3} $&$ 6.24\cdot 10^{-6} $&$ 8.86\cdot 10^{-6} $&$ 5.38\cdot 10^{-4} $&$ 36.8$\\
$870 $&$ 1.18\cdot 10^{-2} $&$ 1.82\cdot 10^{-3} $&$ 6.31\cdot 10^{-6} $&$ 8.95\cdot 10^{-6} $&$ 5.43\cdot 10^{-4} $&$ 37.3$\\
$880 $&$ 1.19\cdot 10^{-2} $&$ 1.84\cdot 10^{-3} $&$ 6.38\cdot 10^{-6} $&$ 9.03\cdot 10^{-6} $&$ 5.48\cdot 10^{-4} $&$ 37.8$\\
$890 $&$ 1.20\cdot 10^{-2} $&$ 1.86\cdot 10^{-3} $&$ 6.45\cdot 10^{-6} $&$ 9.11\cdot 10^{-6} $&$ 5.53\cdot 10^{-4} $&$ 38.3$\\
$900 $&$ 1.21\cdot 10^{-2} $&$ 1.88\cdot 10^{-3} $&$ 6.52\cdot 10^{-6} $&$ 9.19\cdot 10^{-6} $&$ 5.58\cdot 10^{-4} $&$ 38.9$\\
$910 $&$ 1.22\cdot 10^{-2} $&$ 1.90\cdot 10^{-3} $&$ 6.58\cdot 10^{-6} $&$ 9.27\cdot 10^{-6} $&$ 5.63\cdot 10^{-4} $&$ 39.4$\\
$920 $&$ 1.23\cdot 10^{-2} $&$ 1.92\cdot 10^{-3} $&$ 6.66\cdot 10^{-6} $&$ 9.35\cdot 10^{-6} $&$ 5.67\cdot 10^{-4} $&$ 39.9$\\
$930 $&$ 1.24\cdot 10^{-2} $&$ 1.94\cdot 10^{-3} $&$ 6.72\cdot 10^{-6} $&$ 9.43\cdot 10^{-6} $&$ 5.72\cdot 10^{-4} $&$ 40.4$\\
$940 $&$ 1.25\cdot 10^{-2} $&$ 1.96\cdot 10^{-3} $&$ 6.79\cdot 10^{-6} $&$ 9.51\cdot 10^{-6} $&$ 5.77\cdot 10^{-4} $&$ 40.8$\\
$950 $&$ 1.26\cdot 10^{-2} $&$ 1.98\cdot 10^{-3} $&$ 6.85\cdot 10^{-6} $&$ 9.58\cdot 10^{-6} $&$ 5.82\cdot 10^{-4} $&$ 41.3$\\
$960 $&$ 1.27\cdot 10^{-2} $&$ 2.00\cdot 10^{-3} $&$ 6.92\cdot 10^{-6} $&$ 9.66\cdot 10^{-6} $&$ 5.86\cdot 10^{-4} $&$ 41.8$\\
$970 $&$ 1.28\cdot 10^{-2} $&$ 2.01\cdot 10^{-3} $&$ 6.98\cdot 10^{-6} $&$ 9.74\cdot 10^{-6} $&$ 5.90\cdot 10^{-4} $&$ 42.2$\\
$980 $&$ 1.29\cdot 10^{-2} $&$ 2.03\cdot 10^{-3} $&$ 7.04\cdot 10^{-6} $&$ 9.81\cdot 10^{-6} $&$ 5.95\cdot 10^{-4} $&$ 42.7$\\
$990 $&$ 1.30\cdot 10^{-2} $&$ 2.05\cdot 10^{-3} $&$ 7.11\cdot 10^{-6} $&$ 9.88\cdot 10^{-6} $&$ 6.00\cdot 10^{-4} $&$ 43.2$\\
$1000 $&$ 1.31\cdot 10^{-2} $&$ 2.07\cdot 10^{-3} $&$ 7.17\cdot 10^{-6} $&$ 9.95\cdot 10^{-6} $&$ 6.04\cdot 10^{-4} $&$ 43.6$\\
\hline
  \end{tabular}
\end{table}

\begin{table}
  \vspace{-\headsep}
  \caption{SM Higgs-boson partial widths [GeV] in bosonic final states, for the low- and intermediate-mass range.}
  \label{tab:Width-lm.part2}
  \centering
  \small
  \begin{tabular}{lccccc}
\hline
$\MH$ [GeV] &
$\PH \rightarrow \Pg\Pg$  &  $\PH \rightarrow \PGg \PGg$ & $\PH 
\rightarrow \PZ\gamma$& $\PH \rightarrow \PW\PW$& $\PH \rightarrow \PZ\PZ$  \\
\hline
$90 $&$ 1.35\cdot 10^{-4} $&$ 2.71\cdot 10^{-6} $&$ 0.00 $&$ 4.60\cdot 10^{-6} $&$ 9.27 \cdot 10^{-7}   $\\
$95 $&$ 1.57\cdot 10^{-4} $&$ 3.26\cdot 10^{-6} $&$ 1.05 \cdot 10^{-8}  $&$ 1.10\cdot 10^{-5} $&$ 1.56\cdot 10^{-6}   $\\
$100 $&$ 1.81\cdot 10^{-4} $&$ 3.91\cdot 10^{-6} $&$ 1.22 \cdot 10^{-7} $&$ 2.72\cdot 10^{-5} $&$ 2.79\cdot 10^{-6}   $\\
$105 $&$ 2.08\cdot 10^{-4} $&$ 4.67\cdot 10^{-6} $&$ 4.54 \cdot 10^{-7} $&$ 6.36\cdot 10^{-5} $&$ 5.63\cdot 10^{-6}   $\\
$110 $&$ 2.38\cdot 10^{-4} $&$ 5.55\cdot 10^{-6} $&$ 1.11\cdot 10^{-6} $&$ 1.36\cdot 10^{-4} $&$ 1.24\cdot 10^{-5}   $\\
$115 $&$ 2.71\cdot 10^{-4} $&$ 6.59\cdot 10^{-6} $&$ 2.21\cdot 10^{-6} $&$ 2.68\cdot 10^{-4} $&$ 2.70\cdot 10^{-5}   $\\
$120 $&$ 3.06\cdot 10^{-4} $&$ 7.81\cdot 10^{-6} $&$ 3.88\cdot 10^{-6} $&$ 4.95\cdot 10^{-4} $&$ 5.57\cdot 10^{-5}   $\\
$125 $&$ 3.45\cdot 10^{-4} $&$ 9.27\cdot 10^{-6} $&$ 6.26\cdot 10^{-6} $&$ 8.73\cdot 10^{-4} $&$ 1.07\cdot 10^{-4}   $\\
$130 $&$ 3.87\cdot 10^{-4} $&$ 1.10\cdot 10^{-5} $&$ 9.55\cdot 10^{-6} $&$ 1.49\cdot 10^{-3} $&$ 1.95\cdot 10^{-4}   $\\
$135 $&$ 4.33\cdot 10^{-4} $&$ 1.31\cdot 10^{-5} $&$ 1.40\cdot 10^{-5} $&$ 2.47\cdot 10^{-3} $&$ 3.38\cdot 10^{-4}   $\\
$140 $&$ 4.82\cdot 10^{-4} $&$ 1.57\cdot 10^{-5} $&$ 2.01\cdot 10^{-5} $&$ 4.10\cdot 10^{-3} $&$ 5.62\cdot 10^{-4}   $\\
$145 $&$ 5.35\cdot 10^{-4} $&$ 1.91\cdot 10^{-5} $&$ 2.83\cdot 10^{-5} $&$ 6.86\cdot 10^{-3} $&$ 9.06\cdot 10^{-4}   $\\
$150 $&$ 5.93\cdot 10^{-4} $&$ 2.36\cdot 10^{-5} $&$ 4.00\cdot 10^{-5} $&$ 1.21\cdot 10^{-2} $&$ 1.43\cdot 10^{-3}   $\\
$155 $&$ 6.52\cdot 10^{-4} $&$ 3.02\cdot 10^{-5} $&$ 5.78\cdot 10^{-5} $&$ 2.41\cdot 10^{-2} $&$ 2.22\cdot 10^{-3}   $\\
$160 $&$ 7.10\cdot 10^{-4} $&$ 4.42\cdot 10^{-5} $&$ 9.56\cdot 10^{-5} $&$ 7.53\cdot 10^{-2} $&$ 3.44\cdot 10^{-3}   $\\
$165 $&$ 7.66\cdot 10^{-4} $&$ 5.66\cdot 10^{-5} $&$ 1.34\cdot 10^{-4} $&$ 2.36\cdot 10^{-1} $&$ 5.47\cdot 10^{-3}   $\\
$170 $&$ 8.29\cdot 10^{-4} $&$ 6.01\cdot 10^{-5} $&$ 1.52\cdot 10^{-4} $&$ 3.67\cdot 10^{-1} $&$ 8.98\cdot 10^{-3}   $\\
$175 $&$ 8.99\cdot 10^{-4} $&$ 6.13\cdot 10^{-5} $&$ 1.69\cdot 10^{-4} $&$ 4.80\cdot 10^{-1} $&$ 1.62\cdot 10^{-2}   $\\
$180 $&$ 9.72\cdot 10^{-4} $&$ 6.44\cdot 10^{-5} $&$ 1.87\cdot 10^{-4} $&$ 5.88\cdot 10^{-1} $&$ 3.80\cdot 10^{-2}   $\\
$185 $&$ 1.05\cdot 10^{-3} $&$ 6.73\cdot 10^{-5} $&$ 2.03\cdot 10^{-4} $&$ 7.02\cdot 10^{-1} $&$ 1.25\cdot 10^{-1}   $\\
$190 $&$ 1.13\cdot 10^{-3} $&$ 7.01\cdot 10^{-5} $&$ 2.20\cdot 10^{-4} $&$ 8.17\cdot 10^{-1} $&$ 2.18\cdot 10^{-1}   $\\
$195 $&$ 1.22\cdot 10^{-3} $&$ 7.28\cdot 10^{-5} $&$ 2.36\cdot 10^{-4} $&$ 9.36\cdot 10^{-1} $&$ 2.95\cdot 10^{-1}   $\\
$200 $&$ 1.31\cdot 10^{-3} $&$ 7.54\cdot 10^{-5} $&$ 2.51\cdot 10^{-4} $&$ 1.06 $&$ 3.66\cdot 10^{-1}   $\\
$210 $&$ 1.53\cdot 10^{-3} $&$ 8.01\cdot 10^{-5} $&$ 2.81\cdot 10^{-4} $&$ 1.33 $&$ 5.06\cdot 10^{-1}   $\\
$220 $&$ 1.77\cdot 10^{-3} $&$ 8.45\cdot 10^{-5} $&$ 3.10\cdot 10^{-4} $&$ 1.65 $&$ 6.54\cdot 10^{-1}   $\\
$230 $&$ 2.05\cdot 10^{-3} $&$ 8.86\cdot 10^{-5} $&$ 3.36\cdot 10^{-4} $&$ 2.00 $&$ 8.16\cdot 10^{-1}   $\\
$240 $&$ 2.37\cdot 10^{-3} $&$ 9.23\cdot 10^{-5} $&$ 3.62\cdot 10^{-4} $&$ 2.39 $&$ 9.97\cdot 10^{-1}   $\\
$250 $&$ 2.73\cdot 10^{-3} $&$ 9.58\cdot 10^{-5} $&$ 3.85\cdot 10^{-4} $&$ 2.83 $&$ 1.20  $\\
$260 $&$ 3.13\cdot 10^{-3} $&$ 9.90\cdot 10^{-5} $&$ 4.08\cdot 10^{-4} $&$ 3.32 $&$ 1.42  $\\
$270 $&$ 3.60\cdot 10^{-3} $&$ 1.02\cdot 10^{-4} $&$ 4.29\cdot 10^{-4} $&$ 3.87 $&$ 1.67  $\\
$280 $&$ 4.13\cdot 10^{-3} $&$ 1.05\cdot 10^{-4} $&$ 4.48\cdot 10^{-4} $&$ 4.47 $&$ 1.95  $\\
$290 $&$ 4.74\cdot 10^{-3} $&$ 1.07\cdot 10^{-4} $&$ 4.67\cdot 10^{-4} $&$ 5.12 $&$ 2.25  $\\
$300 $&$ 5.45\cdot 10^{-3} $&$ 1.09\cdot 10^{-4} $&$ 4.85\cdot 10^{-4} $&$ 5.83 $&$ 2.59  $\\
$310 $&$ 6.28\cdot 10^{-3} $&$ 1.12\cdot 10^{-4} $&$ 5.01\cdot 10^{-4} $&$ 6.60 $&$ 2.95  $\\
$320 $&$ 7.26\cdot 10^{-3} $&$ 1.14\cdot 10^{-4} $&$ 5.16\cdot 10^{-4} $&$ 7.44 $&$ 3.34  $\\
$330 $&$ 8.46\cdot 10^{-3} $&$ 1.16\cdot 10^{-4} $&$ 5.31\cdot 10^{-4} $&$ 8.32 $&$ 3.76  $\\
$340 $&$ 1.00\cdot 10^{-2} $&$ 1.18\cdot 10^{-4} $&$ 5.44\cdot 10^{-4} $&$ 9.26 $&$ 4.19  $\\
$350 $&$ 1.22\cdot 10^{-2} $&$ 1.16\cdot 10^{-4} $&$ 5.54\cdot 10^{-4} $&$ 10.3 $&$ 4.66  $\\
$360 $&$ 1.48\cdot 10^{-2} $&$ 1.07\cdot 10^{-4} $&$ 5.58\cdot 10^{-4} $&$ 11.4 $&$ 5.22  $\\
$370 $&$ 1.73\cdot 10^{-2} $&$ 9.82\cdot 10^{-5} $&$ 5.59\cdot 10^{-4} $&$ 12.7 $&$ 5.81  $\\
$380 $&$ 1.96\cdot 10^{-2} $&$ 8.92\cdot 10^{-5} $&$ 5.59\cdot 10^{-4} $&$ 14.1 $&$ 6.45  $\\
$390 $&$ 2.19\cdot 10^{-2} $&$ 8.05\cdot 10^{-5} $&$ 5.58\cdot 10^{-4} $&$ 15.5 $&$ 7.13  $\\
$400 $&$ 2.40\cdot 10^{-2} $&$ 7.22\cdot 10^{-5} $&$ 5.55\cdot 10^{-4} $&$ 17.0 $&$ 7.85  $\\
$410 $&$ 2.60\cdot 10^{-2} $&$ 6.45\cdot 10^{-5} $&$ 5.52\cdot 10^{-4} $&$ 18.6 $&$ 8.62  $\\
$420 $&$ 2.80\cdot 10^{-2} $&$ 5.73\cdot 10^{-5} $&$ 5.49\cdot 10^{-4} $&$ 20.2 $&$ 9.43  $\\
$430 $&$ 2.98\cdot 10^{-2} $&$ 5.06\cdot 10^{-5} $&$ 5.45\cdot 10^{-4}
$&$ 22.0 $&$ 10.3 $\\
$440 $&$ 3.16\cdot 10^{-2} $&$ 4.44\cdot 10^{-5} $&$ 5.42\cdot 10^{-4}
$&$ 23.9 $&$ 11.2 $\\
$450 $&$ 3.32\cdot 10^{-2} $&$ 3.88\cdot 10^{-5} $&$ 5.38\cdot 10^{-4}
$&$ 25.8 $&$ 12.2 $\\
$460 $&$ 3.48\cdot 10^{-2} $&$ 3.37\cdot 10^{-5} $&$ 5.33\cdot 10^{-4}
$&$ 27.9 $&$ 13.2 $\\
$470 $&$ 3.63\cdot 10^{-2} $&$ 2.90\cdot 10^{-5} $&$ 5.29\cdot 10^{-4}
$&$ 30.0 $&$ 14.2 $\\
$480 $&$ 3.78\cdot 10^{-2} $&$ 2.49\cdot 10^{-5} $&$ 5.25\cdot 10^{-4}
$&$ 32.3 $&$ 15.3 $\\
$490 $&$ 3.92\cdot 10^{-2} $&$ 2.12\cdot 10^{-5} $&$ 5.20\cdot 10^{-4}
$&$ 34.6 $&$ 16.5 $\\
\hline
  \end{tabular}
\end{table}

\begin{table}
  \vspace{-\headsep}
  \caption{SM Higgs-boson partial widths [GeV] in bosonic final states, for the low- and intermediate-mass range.}
  \label{tab:Width-hm.part2}
  \centering
  \small
  \begin{tabular}{lcccccc}
\hline
$\MH$ [GeV] &
$\PH \rightarrow \Pg\Pg$  &  $\PH \rightarrow \PGg \PGg$ & $\PH 
\rightarrow \PZ\gamma$& $\PH \rightarrow \PW\PW$& $\PH \rightarrow \PZ\PZ$ \\
\hline
$500 $&$ 4.05\cdot 10^{-2} $&$ 1.80\cdot 10^{-5} $&$ 5.16\cdot 10^{-4} $&$ 37.1 $&$ 17.7 $\\
$510 $&$ 4.18\cdot 10^{-2} $&$ 1.52\cdot 10^{-5} $&$ 5.11\cdot 10^{-4} $&$ 39.7 $&$ 19.0 $\\
$520 $&$ 4.31\cdot 10^{-2} $&$ 1.28\cdot 10^{-5} $&$ 5.07\cdot 10^{-4} $&$ 42.4 $&$ 20.4 $\\
$530 $&$ 4.42\cdot 10^{-2} $&$ 1.08\cdot 10^{-5} $&$ 5.02\cdot 10^{-4} $&$ 45.2 $&$ 21.8 $\\
$540 $&$ 4.54\cdot 10^{-2} $&$ 9.17\cdot 10^{-6} $&$ 4.98\cdot 10^{-4} $&$ 48.1 $&$ 23.2 $\\
$550 $&$ 4.65\cdot 10^{-2} $&$ 7.93\cdot 10^{-6} $&$ 4.93\cdot 10^{-4} $&$ 51.2 $&$ 24.7 $\\
$560 $&$ 4.76\cdot 10^{-2} $&$ 7.07\cdot 10^{-6} $&$ 4.89\cdot 10^{-4} $&$ 54.4 $&$ 26.3 $\\
$570 $&$ 4.85\cdot 10^{-2} $&$ 6.56\cdot 10^{-6} $&$ 4.84\cdot 10^{-4} $&$ 57.7 $&$ 28.0 $\\
$580 $&$ 4.95\cdot 10^{-2} $&$ 6.40\cdot 10^{-6} $&$ 4.80\cdot 10^{-4} $&$ 61.2 $&$ 29.7 $\\
$590 $&$ 5.05\cdot 10^{-2} $&$ 6.57\cdot 10^{-6} $&$ 4.75\cdot 10^{-4} $&$ 64.8 $&$ 31.5 $\\
$600 $&$ 5.14\cdot 10^{-2} $&$ 7.08\cdot 10^{-6} $&$ 4.71\cdot 10^{-4} $&$ 68.5 $&$ 33.4 $\\
$610 $&$ 5.23\cdot 10^{-2} $&$ 7.92\cdot 10^{-6} $&$ 4.67\cdot 10^{-4} $&$ 72.4 $&$ 35.4 $\\
$620 $&$ 5.31\cdot 10^{-2} $&$ 9.06\cdot 10^{-6} $&$ 4.63\cdot 10^{-4} $&$ 76.5 $&$ 37.4 $\\
$630 $&$ 5.40\cdot 10^{-2} $&$ 1.05\cdot 10^{-5} $&$ 4.59\cdot 10^{-4} $&$ 80.7 $&$ 39.5 $\\
$640 $&$ 5.48\cdot 10^{-2} $&$ 1.23\cdot 10^{-5} $&$ 4.55\cdot 10^{-4} $&$ 85.0 $&$ 41.7 $\\
$650 $&$ 5.56\cdot 10^{-2} $&$ 1.44\cdot 10^{-5} $&$ 4.51\cdot 10^{-4} $&$ 89.6 $&$ 44.0 $\\
$660 $&$ 5.63\cdot 10^{-2} $&$ 1.68\cdot 10^{-5} $&$ 4.47\cdot 10^{-4} $&$ 94.3 $&$ 46.3 $\\
$670 $&$ 5.71\cdot 10^{-2} $&$ 1.94\cdot 10^{-5} $&$ 4.44\cdot 10^{-4} $&$ 99.1 $&$ 48.8 $\\
$680 $&$ 5.78\cdot 10^{-2} $&$ 2.24\cdot 10^{-5} $&$ 4.40\cdot 10^{-4} $&$ 104 $&$ 51.3  $\\
$690 $&$ 5.84\cdot 10^{-2} $&$ 2.57\cdot 10^{-5} $&$ 4.36\cdot 10^{-4} $&$ 109 $&$ 53.9  $\\
$700 $&$ 5.91\cdot 10^{-2} $&$ 2.92\cdot 10^{-5} $&$ 4.33\cdot 10^{-4} $&$ 115 $&$ 56.7  $\\
$710 $&$ 5.98\cdot 10^{-2} $&$ 3.30\cdot 10^{-5} $&$ 4.30\cdot 10^{-4} $&$ 120 $&$ 59.5  $\\
$720 $&$ 6.04\cdot 10^{-2} $&$ 3.72\cdot 10^{-5} $&$ 4.27\cdot 10^{-4} $&$ 126 $&$ 62.4  $\\
$730 $&$ 6.11\cdot 10^{-2} $&$ 4.16\cdot 10^{-5} $&$ 4.24\cdot 10^{-4} $&$ 132 $&$ 65.5  $\\
$740 $&$ 6.16\cdot 10^{-2} $&$ 4.62\cdot 10^{-5} $&$ 4.20\cdot 10^{-4} $&$ 139 $&$ 68.6  $\\
$750 $&$ 6.22\cdot 10^{-2} $&$ 5.12\cdot 10^{-5} $&$ 4.18\cdot 10^{-4} $&$ 145 $&$ 71.9  $\\
$760 $&$ 6.28\cdot 10^{-2} $&$ 5.65\cdot 10^{-5} $&$ 4.15\cdot 10^{-4} $&$ 152 $&$ 75.2  $\\
$770 $&$ 6.34\cdot 10^{-2} $&$ 6.19\cdot 10^{-5} $&$ 4.12\cdot 10^{-4} $&$ 159 $&$ 78.7  $\\
$780 $&$ 6.39\cdot 10^{-2} $&$ 6.77\cdot 10^{-5} $&$ 4.10\cdot 10^{-4} $&$ 166 $&$ 82.3  $\\
$790 $&$ 6.45\cdot 10^{-2} $&$ 7.38\cdot 10^{-5} $&$ 4.08\cdot 10^{-4} $&$ 173 $&$ 86.1  $\\
$800 $&$ 6.50\cdot 10^{-2} $&$ 8.02\cdot 10^{-5} $&$ 4.06\cdot 10^{-4} $&$ 181 $&$ 89.9  $\\
$810 $&$ 6.55\cdot 10^{-2} $&$ 8.68\cdot 10^{-5} $&$ 4.04\cdot 10^{-4} $&$ 189 $&$ 93.9  $\\
$820 $&$ 6.60\cdot 10^{-2} $&$ 9.37\cdot 10^{-5} $&$ 4.01\cdot 10^{-4} $&$ 197 $&$ 98.0  $\\
$830 $&$ 6.65\cdot 10^{-2} $&$ 1.01\cdot 10^{-4} $&$ 4.00\cdot 10^{-4} $&$ 206 $&$ 102   $\\
$840 $&$ 6.70\cdot 10^{-2} $&$ 1.08\cdot 10^{-4} $&$ 3.98\cdot 10^{-4} $&$ 214 $&$ 107   $\\
$850 $&$ 6.75\cdot 10^{-2} $&$ 1.16\cdot 10^{-4} $&$ 3.96\cdot 10^{-4} $&$ 223 $&$ 111   $\\
$860 $&$ 6.80\cdot 10^{-2} $&$ 1.24\cdot 10^{-4} $&$ 3.95\cdot 10^{-4} $&$ 233 $&$ 116   $\\
$870 $&$ 6.84\cdot 10^{-2} $&$ 1.32\cdot 10^{-4} $&$ 3.94\cdot 10^{-4} $&$ 242 $&$ 121   $\\
$880 $&$ 6.88\cdot 10^{-2} $&$ 1.41\cdot 10^{-4} $&$ 3.93\cdot 10^{-4} $&$ 252 $&$ 126   $\\
$890 $&$ 6.93\cdot 10^{-2} $&$ 1.50\cdot 10^{-4} $&$ 3.92\cdot 10^{-4} $&$ 263 $&$ 131   $\\
$900 $&$ 6.97\cdot 10^{-2} $&$ 1.59\cdot 10^{-4} $&$ 3.91\cdot 10^{-4} $&$ 273 $&$ 137   $\\
$910 $&$ 7.01\cdot 10^{-2} $&$ 1.68\cdot 10^{-4} $&$ 3.91\cdot 10^{-4} $&$ 284 $&$ 142   $\\
$920 $&$ 7.05\cdot 10^{-2} $&$ 1.78\cdot 10^{-4} $&$ 3.90\cdot 10^{-4} $&$ 296 $&$ 148   $\\
$930 $&$ 7.09\cdot 10^{-2} $&$ 1.88\cdot 10^{-4} $&$ 3.90\cdot 10^{-4} $&$ 308 $&$ 154   $\\
$940 $&$ 7.13\cdot 10^{-2} $&$ 1.98\cdot 10^{-4} $&$ 3.90\cdot 10^{-4} $&$ 320 $&$ 160   $\\
$950 $&$ 7.17\cdot 10^{-2} $&$ 2.09\cdot 10^{-4} $&$ 3.90\cdot 10^{-4} $&$ 332 $&$ 166   $\\
$960 $&$ 7.21\cdot 10^{-2} $&$ 2.20\cdot 10^{-4} $&$ 3.90\cdot 10^{-4} $&$ 345 $&$ 173   $\\
$970 $&$ 7.25\cdot 10^{-2} $&$ 2.31\cdot 10^{-4} $&$ 3.91\cdot 10^{-4} $&$ 359 $&$ 180   $\\
$980 $&$ 7.29\cdot 10^{-2} $&$ 2.43\cdot 10^{-4} $&$ 3.92\cdot 10^{-4} $&$ 373 $&$ 187   $\\
$990 $&$ 7.32\cdot 10^{-2} $&$ 2.54\cdot 10^{-4} $&$ 3.92\cdot 10^{-4} $&$ 387 $&$ 194   $\\
$1000$&$ 7.36\cdot 10^{-2} $&$ 2.66\cdot 10^{-4} $&$ 3.94\cdot 10^{-4} $&$ 402 $&$ 201   $\\
\hline
  \end{tabular}
\end{table}


\begin{landscape}
  \begin{table}
    \vspace{-\headsep}
    \caption{SM Higgs-boson partial widths [GeV] for $4$-fermion final states for the low-mass range.
        We list results for the specific final states $\Pep\Pem\Pep\Pem$ and $\Pep\Pem\PGmp\PGmm$, for final
        states with $4$ arbitrary charged leptons, $\Pep\PGne\Pem\PAGne$ and $\Pep\PGne\PGmm\PAGnGm$, and for
            final states $\Plp\Plm\PGnl\PAGnl$ with $2$ charged
        leptons plus $2$ neutrinos ($\PGnl$ represents any type of neutrinos).}
    \label{tab:PWidth-lm}
      \centering
      \small
      \begin{tabular}{lcccccccc}
        \hline
        $\MH$ [GeV] &
        $\PH \rightarrow \Pep\Pem\Pep\Pem$ &
        $\PH \rightarrow \Pep\Pem\PGmp\PGmm$ &
        $\PH \rightarrow \Plp\Plm\Plp\Plm$ &
        $\PH \rightarrow \Plp\Plm\Plp\Plm$ &
        $\PH \rightarrow \Pep\PGne\Pem\PAGne$ &
        $\PH \rightarrow \Pep\PGne\PGmm\PAGnGm$ &
        $\PH \rightarrow \Plp\Plm\PGnl\PAGnl$ &
        $\PH \rightarrow \Plp\Plm\PGnl\PAGnl$ \\
        & & & $(\Pl=\Pe$ or $\PGm)$ & $(\Pl=\Pe, \PGm$ or $\PGt)$
        & & & $(\Pl=\Pe$ or $\PGm)$ & $(\Pl=\Pe, \PGm$ or $\PGt)$ \\
        \hline
$	90	$ & $	1.56 \cdot 10^{-9} 	$ & $	2.07 \cdot 10^{-9} 	$ & $	5.19 \cdot 10^{-9} 	$ & $	1.09 \cdot 10^{-8} 	$ & $	3.89 \cdot 10^{-8} 	$ & $	5.40 \cdot 10^{-8} 	$ & $	2.03 \cdot 10^{-7} 	$ & $	4.66 \cdot 10^{-7} 	 $ \\
$	95	$ & $	2.58 \cdot 10^{-9} 	$ & $	3.47 \cdot 10^{-9} 	$ & $	8.62 \cdot 10^{-9} 	$ & $	1.81 \cdot 10^{-8} 	$ & $	1.04 \cdot 10^{-7} 	$ & $	1.29 \cdot 10^{-7} 	$ & $	4.93 \cdot 10^{-7} 	$ & $	1.13 \cdot 10^{-6} 	 $ \\
$	100	$ & $	4.43 \cdot 10^{-9} 	$ & $	6.19 \cdot 10^{-9} 	$ & $	1.51 \cdot 10^{-8} 	$ & $	3.19 \cdot 10^{-8} 	$ & $	2.79 \cdot 10^{-7} 	$ & $	3.20 \cdot 10^{-7} 	$ & $	1.25 \cdot 10^{-6} 	$ & $	2.83 \cdot 10^{-6} 	 $ \\
$	105	$ & $	8.40 \cdot 10^{-9} 	$ & $	1.25 \cdot 10^{-8} 	$ & $	2.93 \cdot 10^{-8} 	$ & $	6.28 \cdot 10^{-8} 	$ & $	6.86 \cdot 10^{-7} 	$ & $	7.47 \cdot 10^{-7} 	$ & $	2.97 \cdot 10^{-6} 	$ & $	6.69 \cdot 10^{-6} 	 $ \\
$	110	$ & $	1.72 \cdot 10^{-8} 	$ & $	2.76 \cdot 10^{-8} 	$ & $	6.20 \cdot 10^{-8} 	$ & $	1.34 \cdot 10^{-7} 	$ & $	1.51 \cdot 10^{-6} 	$ & $	1.60 \cdot 10^{-6} 	$ & $	6.44 \cdot 10^{-6} 	$ & $	1.44 \cdot 10^{-5} 	 $ \\
$	115	$ & $	3.55 \cdot 10^{-8} 	$ & $	6.03 \cdot 10^{-8} 	$ & $	1.31 \cdot 10^{-7} 	$ & $	2.88 \cdot 10^{-7} 	$ & $	3.05 \cdot 10^{-6} 	$ & $	3.15 \cdot 10^{-6} 	$ & $	1.29 \cdot 10^{-5} 	$ & $	2.88 \cdot 10^{-5} 	 $ \\
$	120	$ & $	7.06 \cdot 10^{-8} 	$ & $	1.25 \cdot 10^{-7} 	$ & $	2.66 \cdot 10^{-7} 	$ & $	5.86 \cdot 10^{-7} 	$ & $	5.74 \cdot 10^{-6} 	$ & $	5.83 \cdot 10^{-6} 	$ & $	2.41 \cdot 10^{-5} 	$ & $	5.37 \cdot 10^{-5} 	 $ \\
$	125	$ & $	1.33 \cdot 10^{-7} 	$ & $	2.41 \cdot 10^{-7} 	$ & $	5.08 \cdot 10^{-7} 	$ & $	1.12 \cdot 10^{-6} 	$ & $	1.03 \cdot 10^{-5} 	$ & $	1.03 \cdot 10^{-5} 	$ & $	4.30 \cdot 10^{-5} 	$ & $	9.53 \cdot 10^{-5} 	 $ \\
$	130	$ & $	2.38 \cdot 10^{-7} 	$ & $	4.39 \cdot 10^{-7} 	$ & $	9.16 \cdot 10^{-7} 	$ & $	2.03 \cdot 10^{-6} 	$ & $	1.76 \cdot 10^{-5} 	$ & $	1.75 \cdot 10^{-5} 	$ & $	7.37 \cdot 10^{-5} 	$ & $	1.63 \cdot 10^{-4} 	 $ \\
$	135	$ & $	4.07 \cdot 10^{-7} 	$ & $	7.61 \cdot 10^{-7} 	$ & $	1.57 \cdot 10^{-6} 	$ & $	3.50 \cdot 10^{-6} 	$ & $	2.96 \cdot 10^{-5} 	$ & $	2.91 \cdot 10^{-5} 	$ & $	1.24 \cdot 10^{-4} 	$ & $	2.73 \cdot 10^{-4} 	 $ \\
$	140	$ & $	6.70 \cdot 10^{-7} 	$ & $	1.27 \cdot 10^{-6} 	$ & $	2.61 \cdot 10^{-6} 	$ & $	5.81 \cdot 10^{-6} 	$ & $	4.93 \cdot 10^{-5} 	$ & $	4.82 \cdot 10^{-5} 	$ & $	2.05 \cdot 10^{-4} 	$ & $	4.52 \cdot 10^{-4} 	 $ \\
$	145	$ & $	1.07 \cdot 10^{-6} 	$ & $	2.04 \cdot 10^{-6} 	$ & $	4.19 \cdot 10^{-6} 	$ & $	9.35 \cdot 10^{-6} 	$ & $	8.28 \cdot 10^{-5} 	$ & $	8.09 \cdot 10^{-5} 	$ & $	3.44 \cdot 10^{-4} 	$ & $	7.58 \cdot 10^{-4} 	 $ \\
$	150	$ & $	1.68 \cdot 10^{-6} 	$ & $	3.22 \cdot 10^{-6} 	$ & $	6.59 \cdot 10^{-6} 	$ & $	1.47 \cdot 10^{-5} 	$ & $	1.46 \cdot 10^{-4} 	$ & $	1.42 \cdot 10^{-4} 	$ & $	6.01 \cdot 10^{-4} 	$ & $	1.33 \cdot 10^{-3} 	 $ \\
$	155	$ & $	2.61 \cdot 10^{-6} 	$ & $	5.01 \cdot 10^{-6} 	$ & $	1.02 \cdot 10^{-5} 	$ & $	2.29 \cdot 10^{-5} 	$ & $	2.90 \cdot 10^{-4} 	$ & $	2.84 \cdot 10^{-4} 	$ & $	1.19 \cdot 10^{-3} 	$ & $	2.63 \cdot 10^{-3} 	 $ \\
$	160	$ & $	4.02 \cdot 10^{-6} 	$ & $	7.76 \cdot 10^{-6} 	$ & $	1.58 \cdot 10^{-5} 	$ & $	3.53 \cdot 10^{-5} 	$ & $	8.99 \cdot 10^{-4} 	$ & $	8.88 \cdot 10^{-4} 	$ & $	3.64 \cdot 10^{-3} 	$ & $	8.12 \cdot 10^{-3} 	 $ \\
$	165	$ & $	6.35 \cdot 10^{-6} 	$ & $	1.23 \cdot 10^{-5} 	$ & $	2.50 \cdot 10^{-5} 	$ & $	5.60 \cdot 10^{-5} 	$ & $	2.81 \cdot 10^{-3} 	$ & $	2.79 \cdot 10^{-3} 	$ & $	1.13 \cdot 10^{-2} 	$ & $	2.53 \cdot 10^{-2} 	 $ \\
$	170	$ & $	1.04 \cdot 10^{-5} 	$ & $	2.02 \cdot 10^{-5} 	$ & $	4.10 \cdot 10^{-5} 	$ & $	9.18 \cdot 10^{-5} 	$ & $	4.36 \cdot 10^{-3} 	$ & $	4.32 \cdot 10^{-3} 	$ & $	1.75 \cdot 10^{-2} 	$ & $	3.92 \cdot 10^{-2} 	 $ \\
$	175	$ & $	1.86 \cdot 10^{-5} 	$ & $	3.64 \cdot 10^{-5} 	$ & $	7.36 \cdot 10^{-5} 	$ & $	1.65 \cdot 10^{-4} 	$ & $	5.72 \cdot 10^{-3} 	$ & $	5.65 \cdot 10^{-3} 	$ & $	2.30 \cdot 10^{-2} 	$ & $	5.15 \cdot 10^{-2} 	 $ \\
$	180	$ & $	4.32 \cdot 10^{-5} 	$ & $	8.55 \cdot 10^{-5} 	$ & $	1.72 \cdot 10^{-4} 	$ & $	3.86 \cdot 10^{-4} 	$ & $	7.09 \cdot 10^{-3} 	$ & $	6.93 \cdot 10^{-3} 	$ & $	2.87 \cdot 10^{-2} 	$ & $	6.39 \cdot 10^{-2} 	 $ \\
$	185	$ & $	1.41 \cdot 10^{-4} 	$ & $	2.82 \cdot 10^{-4} 	$ & $	5.64 \cdot 10^{-4} 	$ & $	1.27 \cdot 10^{-3} 	$ & $	8.83 \cdot 10^{-3} 	$ & $	8.28 \cdot 10^{-3} 	$ & $	3.65 \cdot 10^{-2} 	$ & $	7.95 \cdot 10^{-2} 	 $ \\
$	190	$ & $	2.46 \cdot 10^{-4} 	$ & $	4.91 \cdot 10^{-4} 	$ & $	9.82 \cdot 10^{-4} 	$ & $	2.21 \cdot 10^{-3} 	$ & $	1.06 \cdot 10^{-2} 	$ & $	9.63 \cdot 10^{-3} 	$ & $	4.44 \cdot 10^{-2} 	$ & $	9.55 \cdot 10^{-2} 	 $ \\
$	195	$ & $	3.32 \cdot 10^{-4} 	$ & $	6.64 \cdot 10^{-4} 	$ & $	1.33 \cdot 10^{-3} 	$ & $	2.99 \cdot 10^{-3} 	$ & $	1.23 \cdot 10^{-2} 	$ & $	1.10 \cdot 10^{-2} 	$ & $	5.21 \cdot 10^{-2} 	$ & $	1.11 \cdot 10^{-1} 	 $ \\
$	200	$ & $	4.12 \cdot 10^{-4} 	$ & $	8.24 \cdot 10^{-4} 	$ & $	1.65 \cdot 10^{-3} 	$ & $	3.71 \cdot 10^{-3} 	$ & $	1.41 \cdot 10^{-2} 	$ & $	1.25 \cdot 10^{-2} 	$ & $	5.99 \cdot 10^{-2} 	$ & $	1.27 \cdot 10^{-1} 	 $ \\
$	210	$ & $	5.70 \cdot 10^{-4} 	$ & $	1.14 \cdot 10^{-3} 	$ & $	2.28 \cdot 10^{-3} 	$ & $	5.13 \cdot 10^{-3} 	$ & $	1.80 \cdot 10^{-2} 	$ & $	1.57 \cdot 10^{-2} 	$ & $	7.65 \cdot 10^{-2} 	$ & $	1.62 \cdot 10^{-1} 	 $ \\
$	220	$ & $	7.36 \cdot 10^{-4} 	$ & $	1.47 \cdot 10^{-3} 	$ & $	2.94 \cdot 10^{-3} 	$ & $	6.62 \cdot 10^{-3} 	$ & $	2.23 \cdot 10^{-2} 	$ & $	1.94 \cdot 10^{-2} 	$ & $	9.51 \cdot 10^{-2} 	$ & $	2.01 \cdot 10^{-1} 	 $ \\
$	230	$ & $	9.19 \cdot 10^{-4} 	$ & $	1.84 \cdot 10^{-3} 	$ & $	3.67 \cdot 10^{-3} 	$ & $	8.27 \cdot 10^{-3} 	$ & $	2.72 \cdot 10^{-2} 	$ & $	2.35 \cdot 10^{-2} 	$ & $	1.16 \cdot 10^{-1} 	$ & $	2.45 \cdot 10^{-1} 	 $ \\
$	240	$ & $	1.12 \cdot 10^{-3} 	$ & $	2.24 \cdot 10^{-3} 	$ & $	4.49 \cdot 10^{-3} 	$ & $	1.01 \cdot 10^{-2} 	$ & $	3.26 \cdot 10^{-2} 	$ & $	2.82 \cdot 10^{-2} 	$ & $	1.39 \cdot 10^{-1} 	$ & $	2.94 \cdot 10^{-1} 	 $ \\
$	250	$ & $	1.35 \cdot 10^{-3} 	$ & $	2.70 \cdot 10^{-3} 	$ & $	5.40 \cdot 10^{-3} 	$ & $	1.21 \cdot 10^{-2} 	$ & $	3.87 \cdot 10^{-2} 	$ & $	3.34 \cdot 10^{-2} 	$ & $	1.66 \cdot 10^{-1} 	$ & $	3.49 \cdot 10^{-1} 	 $ \\
$	260	$ & $	1.60 \cdot 10^{-3} 	$ & $	3.21 \cdot 10^{-3} 	$ & $	6.41 \cdot 10^{-3} 	$ & $	1.44 \cdot 10^{-2} 	$ & $	4.55 \cdot 10^{-2} 	$ & $	3.92 \cdot 10^{-2} 	$ & $	1.95 \cdot 10^{-1} 	$ & $	4.10 \cdot 10^{-1} 	 $ \\
$	270	$ & $	1.89 \cdot 10^{-3} 	$ & $	3.77 \cdot 10^{-3} 	$ & $	7.54 \cdot 10^{-3} 	$ & $	1.70 \cdot 10^{-2} 	$ & $	5.31 \cdot 10^{-2} 	$ & $	4.56 \cdot 10^{-2} 	$ & $	2.27 \cdot 10^{-1} 	$ & $	4.78 \cdot 10^{-1} 	 $ \\
$	280	$ & $	2.20 \cdot 10^{-3} 	$ & $	4.39 \cdot 10^{-3} 	$ & $	8.78 \cdot 10^{-3} 	$ & $	1.98 \cdot 10^{-2} 	$ & $	6.13 \cdot 10^{-2} 	$ & $	5.26 \cdot 10^{-2} 	$ & $	2.63 \cdot 10^{-1} 	$ & $	5.52 \cdot 10^{-1} 	 $ \\
$	290	$ & $	2.54 \cdot 10^{-3} 	$ & $	5.07 \cdot 10^{-3} 	$ & $	1.01 \cdot 10^{-2} 	$ & $	2.28 \cdot 10^{-2} 	$ & $	7.04 \cdot 10^{-2} 	$ & $	6.03 \cdot 10^{-2} 	$ & $	3.02 \cdot 10^{-1} 	$ & $	6.34 \cdot 10^{-1} 	 $ \\
$	300	$ & $	2.91 \cdot 10^{-3} 	$ & $	5.82 \cdot 10^{-3} 	$ & $	1.16 \cdot 10^{-2} 	$ & $	2.62 \cdot 10^{-2} 	$ & $	8.03 \cdot 10^{-2} 	$ & $	6.87 \cdot 10^{-2} 	$ & $	3.44 \cdot 10^{-1} 	$ & $	7.22 \cdot 10^{-1} 	 $ \\
        \hline
      \end{tabular}
  \end{table}
\end{landscape}


\begin{landscape}
  \begin{table}
    \vspace{-\headsep}
    \caption{SM Higgs-boson partial widths [GeV] for $4$-fermion final
        states for the intermedate-mass range.
        We list results for the specific final states $\Pep\Pem\Pep\Pem$ and $\Pep\Pem\PGmp\PGmm$, for final
        states with $4$ arbitrary charged leptons, $\Pep\PGne\Pem\PAGne$ and $\Pep\PGne\PGmm\PAGnGm$, and for
            final states $\Plp\Plm\PGnl\PAGnl$ with $2$ charged
        leptons plus $2$ neutrinos ($\PGnl$ represents any type of neutrinos).}
    \label{tab:PWidth-im}
      \centering
      \small
      \begin{tabular}{lcccccccc}
        \hline
        $\MH$ [GeV] &
        $\PH \rightarrow \Pep\Pem\Pep\Pem$ &
        $\PH \rightarrow \Pep\Pem\PGmp\PGmm$ &
        $\PH \rightarrow \Plp\Plm\Plp\Plm$ &
        $\PH \rightarrow \Plp\Plm\Plp\Plm$ &
        $\PH \rightarrow \Pep\PGne\Pem\PAGne$ &
        $\PH \rightarrow \Pep\PGne\PGmm\PAGnGm$ &
        $\PH \rightarrow \Plp\Plm\PGnl\PAGnl$ &
        $\PH \rightarrow \Plp\Plm\PGnl\PAGnl$ \\
        & & & $(\Pl=\Pe$ or $\PGm)$ & $(\Pl=\Pe, \PGm$ or $\PGt)$
        & & & $(\Pl=\Pe$ or $\PGm)$ & $(\Pl=\Pe, \PGm$ or $\PGt)$ \\
        \hline
$	310	$ & $	3.32 \cdot 10^{-3} 	$ & $	6.63 \cdot 10^{-3} 	$ & $	1.33 \cdot 10^{-2} 	$ & $	2.98 \cdot 10^{-2} 	$ & $	9.10 \cdot 10^{-2} 	$ & $	7.78 \cdot 10^{-2} 	$ & $	3.90 \cdot 10^{-1} 	$ & $	8.19 \cdot 10^{-1} 	 $ \\
$	320	$ & $	3.76 \cdot 10^{-3} 	$ & $	7.51 \cdot 10^{-3} 	$ & $	1.50 \cdot 10^{-2} 	$ & $	3.38 \cdot 10^{-2} 	$ & $	1.03 \cdot 10^{-1} 	$ & $	8.75 \cdot 10^{-2} 	$ & $	4.40 \cdot 10^{-1} 	$ & $	9.23 \cdot 10^{-1} 	 $ \\
$	330	$ & $	4.23 \cdot 10^{-3} 	$ & $	8.44 \cdot 10^{-3} 	$ & $	1.69 \cdot 10^{-2} 	$ & $	3.80 \cdot 10^{-2} 	$ & $	1.15 \cdot 10^{-1} 	$ & $	9.80 \cdot 10^{-2} 	$ & $	4.93 \cdot 10^{-1} 	$ & $	1.03 	 $ \\
$	340	$ & $	4.72 \cdot 10^{-3} 	$ & $	9.43 \cdot 10^{-3} 	$ & $	1.89 \cdot 10^{-2} 	$ & $	4.24 \cdot 10^{-2} 	$ & $	1.28 \cdot 10^{-1} 	$ & $	1.09 \cdot 10^{-1} 	$ & $	5.49 \cdot 10^{-1} 	$ & $	1.15 	 $ \\
$	350	$ & $	5.24 \cdot 10^{-3} 	$ & $	1.05 \cdot 10^{-2} 	$ & $	2.10 \cdot 10^{-2} 	$ & $	4.72 \cdot 10^{-2} 	$ & $	1.42 \cdot 10^{-1} 	$ & $	1.21 \cdot 10^{-1} 	$ & $	6.08 \cdot 10^{-1} 	$ & $	1.27 	 $ \\
$	360	$ & $	5.87 \cdot 10^{-3} 	$ & $	1.17 \cdot 10^{-2} 	$ & $	2.35 \cdot 10^{-2} 	$ & $	5.28 \cdot 10^{-2} 	$ & $	1.58 \cdot 10^{-1} 	$ & $	1.35 \cdot 10^{-1} 	$ & $	6.79 \cdot 10^{-1} 	$ & $	1.42 	 $ \\
$	370	$ & $	6.54 \cdot 10^{-3} 	$ & $	1.31 \cdot 10^{-2} 	$ & $	2.62 \cdot 10^{-2} 	$ & $	5.88 \cdot 10^{-2} 	$ & $	1.76 \cdot 10^{-1} 	$ & $	1.50 \cdot 10^{-1} 	$ & $	7.55 \cdot 10^{-1} 	$ & $	1.58 	 $ \\
$	380	$ & $	7.26 \cdot 10^{-3} 	$ & $	1.45 \cdot 10^{-2} 	$ & $	2.90 \cdot 10^{-2} 	$ & $	6.53 \cdot 10^{-2} 	$ & $	1.94 \cdot 10^{-1} 	$ & $	1.66 \cdot 10^{-1} 	$ & $	8.36 \cdot 10^{-1} 	$ & $	1.75 	 $ \\
$	390	$ & $	8.03 \cdot 10^{-3} 	$ & $	1.60 \cdot 10^{-2} 	$ & $	3.21 \cdot 10^{-2} 	$ & $	7.22 \cdot 10^{-2} 	$ & $	2.14 \cdot 10^{-1} 	$ & $	1.82 \cdot 10^{-1} 	$ & $	9.21 \cdot 10^{-1} 	$ & $	1.93 	 $ \\
$	400	$ & $	8.84 \cdot 10^{-3} 	$ & $	1.77 \cdot 10^{-2} 	$ & $	3.54 \cdot 10^{-2} 	$ & $	7.96 \cdot 10^{-2} 	$ & $	2.35 \cdot 10^{-1} 	$ & $	2.00 \cdot 10^{-1} 	$ & $	1.01 	$ & $	2.12 	 $ \\
$	410	$ & $	9.71 \cdot 10^{-3} 	$ & $	1.94 \cdot 10^{-2} 	$ & $	3.88 \cdot 10^{-2} 	$ & $	8.74 \cdot 10^{-2} 	$ & $	2.58 \cdot 10^{-1} 	$ & $	2.19 \cdot 10^{-1} 	$ & $	1.11 	$ & $	2.32 	 $ \\
$	420	$ & $	1.06 \cdot 10^{-2} 	$ & $	2.12 \cdot 10^{-2} 	$ & $	4.25 \cdot 10^{-2} 	$ & $	9.56 \cdot 10^{-2} 	$ & $	2.81 \cdot 10^{-1} 	$ & $	2.39 \cdot 10^{-1} 	$ & $	1.21 	$ & $	2.53 	 $ \\
$	430	$ & $	1.16 \cdot 10^{-2} 	$ & $	2.32 \cdot 10^{-2} 	$ & $	4.64 \cdot 10^{-2} 	$ & $	1.04 \cdot 10^{-1} 	$ & $	3.06 \cdot 10^{-1} 	$ & $	2.60 \cdot 10^{-1} 	$ & $	1.32 	$ & $	2.75 	 $ \\
$	440	$ & $	1.26 \cdot 10^{-2} 	$ & $	2.52 \cdot 10^{-2} 	$ & $	5.05 \cdot 10^{-2} 	$ & $	1.14 \cdot 10^{-1} 	$ & $	3.32 \cdot 10^{-1} 	$ & $	2.82 \cdot 10^{-1} 	$ & $	1.43 	$ & $	2.99 	 $ \\
$	450	$ & $	1.37 \cdot 10^{-2} 	$ & $	2.74 \cdot 10^{-2} 	$ & $	5.48 \cdot 10^{-2} 	$ & $	1.23 \cdot 10^{-1} 	$ & $	3.59 \cdot 10^{-1} 	$ & $	3.05 \cdot 10^{-1} 	$ & $	1.55 	$ & $	3.23 	 $ \\
$	460	$ & $	1.48 \cdot 10^{-2} 	$ & $	2.97 \cdot 10^{-2} 	$ & $	5.93 \cdot 10^{-2} 	$ & $	1.34 \cdot 10^{-1} 	$ & $	3.88 \cdot 10^{-1} 	$ & $	3.29 \cdot 10^{-1} 	$ & $	1.67 	$ & $	3.49 	 $ \\
$	470	$ & $	1.60 \cdot 10^{-2} 	$ & $	3.21 \cdot 10^{-2} 	$ & $	6.41 \cdot 10^{-2} 	$ & $	1.44 \cdot 10^{-1} 	$ & $	4.18 \cdot 10^{-1} 	$ & $	3.54 \cdot 10^{-1} 	$ & $	1.80 	$ & $	3.76 	 $ \\
$	480	$ & $	1.73 \cdot 10^{-2} 	$ & $	3.46 \cdot 10^{-2} 	$ & $	6.91 \cdot 10^{-2} 	$ & $	1.56 \cdot 10^{-1} 	$ & $	4.50 \cdot 10^{-1} 	$ & $	3.81 \cdot 10^{-1} 	$ & $	1.94 	$ & $	4.05 	 $ \\
$	490	$ & $	1.86 \cdot 10^{-2} 	$ & $	3.72 \cdot 10^{-2} 	$ & $	7.44 \cdot 10^{-2} 	$ & $	1.67 \cdot 10^{-1} 	$ & $	4.83 \cdot 10^{-1} 	$ & $	4.09 \cdot 10^{-1} 	$ & $	2.08 	$ & $	4.35 	 $ \\
$	500	$ & $	2.00 \cdot 10^{-2} 	$ & $	4.00 \cdot 10^{-2} 	$ & $	7.99 \cdot 10^{-2} 	$ & $	1.80 \cdot 10^{-1} 	$ & $	5.18 \cdot 10^{-1} 	$ & $	4.38 \cdot 10^{-1} 	$ & $	2.23 	$ & $	4.66 	 $ \\
$	510	$ & $	2.14 \cdot 10^{-2} 	$ & $	4.29 \cdot 10^{-2} 	$ & $	8.57 \cdot 10^{-2} 	$ & $	1.93 \cdot 10^{-1} 	$ & $	5.54 \cdot 10^{-1} 	$ & $	4.68 \cdot 10^{-1} 	$ & $	2.39 	$ & $	4.98 	 $ \\
$	520	$ & $	2.29 \cdot 10^{-2} 	$ & $	4.59 \cdot 10^{-2} 	$ & $	9.18 \cdot 10^{-2} 	$ & $	2.07 \cdot 10^{-1} 	$ & $	5.92 \cdot 10^{-1} 	$ & $	5.00 \cdot 10^{-1} 	$ & $	2.55 	$ & $	5.33 	 $ \\
$	530	$ & $	2.45 \cdot 10^{-2} 	$ & $	4.91 \cdot 10^{-2} 	$ & $	9.81 \cdot 10^{-2} 	$ & $	2.21 \cdot 10^{-1} 	$ & $	6.31 \cdot 10^{-1} 	$ & $	5.34 \cdot 10^{-1} 	$ & $	2.72 	$ & $	5.68 	 $ \\
$	540	$ & $	2.62 \cdot 10^{-2} 	$ & $	5.24 \cdot 10^{-2} 	$ & $	1.05 \cdot 10^{-1} 	$ & $	2.36 \cdot 10^{-1} 	$ & $	6.73 \cdot 10^{-1} 	$ & $	5.68 \cdot 10^{-1} 	$ & $	2.90 	$ & $	6.05 	 $ \\
$	550	$ & $	2.79 \cdot 10^{-2} 	$ & $	5.58 \cdot 10^{-2} 	$ & $	1.12 \cdot 10^{-1} 	$ & $	2.51 \cdot 10^{-1} 	$ & $	7.16 \cdot 10^{-1} 	$ & $	6.04 \cdot 10^{-1} 	$ & $	3.08 	$ & $	6.44 	 $ \\
$	560	$ & $	2.97 \cdot 10^{-2} 	$ & $	5.94 \cdot 10^{-2} 	$ & $	1.19 \cdot 10^{-1} 	$ & $	2.67 \cdot 10^{-1} 	$ & $	7.61 \cdot 10^{-1} 	$ & $	6.42 \cdot 10^{-1} 	$ & $	3.28 	$ & $	6.84 	 $ \\
$	570	$ & $	3.16 \cdot 10^{-2} 	$ & $	6.32 \cdot 10^{-2} 	$ & $	1.26 \cdot 10^{-1} 	$ & $	2.84 \cdot 10^{-1} 	$ & $	8.07 \cdot 10^{-1} 	$ & $	6.81 \cdot 10^{-1} 	$ & $	3.48 	$ & $	7.27 	 $ \\
$	580	$ & $	3.36 \cdot 10^{-2} 	$ & $	6.71 \cdot 10^{-2} 	$ & $	1.34 \cdot 10^{-1} 	$ & $	3.02 \cdot 10^{-1} 	$ & $	8.56 \cdot 10^{-1} 	$ & $	7.22 \cdot 10^{-1} 	$ & $	3.69 	$ & $	7.70 	 $ \\
$	590	$ & $	3.56 \cdot 10^{-2} 	$ & $	7.12 \cdot 10^{-2} 	$ & $	1.42 \cdot 10^{-1} 	$ & $	3.20 \cdot 10^{-1} 	$ & $	9.07 \cdot 10^{-1} 	$ & $	7.65 \cdot 10^{-1} 	$ & $	3.91 	$ & $	8.16 	 $ \\
$	600	$ & $	3.77 \cdot 10^{-2} 	$ & $	7.54 \cdot 10^{-2} 	$ & $	1.51 \cdot 10^{-1} 	$ & $	3.39 \cdot 10^{-1} 	$ & $	9.59 \cdot 10^{-1} 	$ & $	8.09 \cdot 10^{-1} 	$ & $	4.14 	$ & $	8.63 	 $ \\
$	610	$ & $	3.99 \cdot 10^{-2} 	$ & $	7.98 \cdot 10^{-2} 	$ & $	1.60 \cdot 10^{-1} 	$ & $	3.59 \cdot 10^{-1} 	$ & $	1.01 	$ & $	8.55 \cdot 10^{-1} 	$ & $	4.37 	$ & $	9.13 	 $ \\
$	620	$ & $	4.22 \cdot 10^{-2} 	$ & $	8.44 \cdot 10^{-2} 	$ & $	1.69 \cdot 10^{-1} 	$ & $	3.80 \cdot 10^{-1} 	$ & $	1.07 	$ & $	9.03 \cdot 10^{-1} 	$ & $	4.62 	$ & $	9.64 	 $ \\
$	630	$ & $	4.46 \cdot 10^{-2} 	$ & $	8.92 \cdot 10^{-2} 	$ & $	1.78 \cdot 10^{-1} 	$ & $	4.01 \cdot 10^{-1} 	$ & $	1.13 	$ & $	9.53 \cdot 10^{-1} 	$ & $	4.88 	$ & $	1.02 \cdot 10^{1} 	 $ \\
$	640	$ & $	4.71 \cdot 10^{-2} 	$ & $	9.41 \cdot 10^{-2} 	$ & $	1.88 \cdot 10^{-1} 	$ & $	4.24 \cdot 10^{-1} 	$ & $	1.19 	$ & $	1.00 	$ & $	5.14 	$ & $	1.07 \cdot 10^{1} 	 $ \\
$	650	$ & $	4.96 \cdot 10^{-2} 	$ & $	9.93 \cdot 10^{-2} 	$ & $	1.99 \cdot 10^{-1} 	$ & $	4.47 \cdot 10^{-1} 	$ & $	1.26 	$ & $	1.06 	$ & $	5.42 	$ & $	1.13 \cdot 10^{1} 	 $ \\
        \hline
      \end{tabular}
  \end{table}
\end{landscape}


\begin{landscape}
  \begin{table}
    \vspace{-\headsep}
    \caption{SM Higgs-boson partial widths [GeV] for $4$-fermion final states for the high-mass range.
        We list results for the specific final states $\Pep\Pem\Pep\Pem$ and $\Pep\Pem\PGmp\PGmm$, for final
        states with $4$ arbitrary charged leptons, $\Pep\PGne\Pem\PAGne$ and $\Pep\PGne\PGmm\PAGnGm$, and for
            final states $\Plp\Plm\PGnl\PAGnl$ with $2$ charged
        leptons plus $2$ neutrinos ($\PGnl$ represents any type of neutrinos).}
    \label{tab:PWidth-hm}
      \centering
      \small
      \begin{tabular}{lcccccccc}
        \hline
        $\MH$ [GeV] &
        $\PH \rightarrow \Pep\Pem\Pep\Pem$ &
        $\PH \rightarrow \Pep\Pem\PGmp\PGmm$ &
        $\PH \rightarrow \Plp\Plm\Plp\Plm$ &
        $\PH \rightarrow \Plp\Plm\Plp\Plm$ &
        $\PH \rightarrow \Pep\PGne\Pem\PAGne$ &
        $\PH \rightarrow \Pep\PGne\PGmm\PAGnGm$ &
        $\PH \rightarrow \Plp\Plm\PGnl\PAGnl$ &
        $\PH \rightarrow \Plp\Plm\PGnl\PAGnl$ \\
        & & & $(\Pl=\Pe$ or $\PGm)$ & $(\Pl=\Pe, \PGm$ or $\PGt)$
        & & & $(\Pl=\Pe$ or $\PGm)$ & $(\Pl=\Pe, \PGm$ or $\PGt)$ \\
        \hline
$	660	$ & $	5.23 \cdot 10^{-2} 	$ & $	1.05 \cdot 10^{-1} 	$ & $	2.09 \cdot 10^{-1} 	$ & $	4.71 \cdot 10^{-1} 	$ & $	1.32 	$ & $	1.11 	$ & $	5.70 	$ & $	1.19 \cdot 10^{1} 	 $ \\
$	670	$ & $	5.51 \cdot 10^{-2} 	$ & $	1.10 \cdot 10^{-1} 	$ & $	2.20 \cdot 10^{-1} 	$ & $	4.96 \cdot 10^{-1} 	$ & $	1.39 	$ & $	1.17 	$ & $	6.00 	$ & $	1.25 \cdot 10^{1} 	 $ \\
$	680	$ & $	5.80 \cdot 10^{-2} 	$ & $	1.16 \cdot 10^{-1} 	$ & $	2.32 \cdot 10^{-1} 	$ & $	5.22 \cdot 10^{-1} 	$ & $	1.46 	$ & $	1.23 	$ & $	6.31 	$ & $	1.32 \cdot 10^{1} 	 $ \\
$	690	$ & $	6.09 \cdot 10^{-2} 	$ & $	1.22 \cdot 10^{-1} 	$ & $	2.44 \cdot 10^{-1} 	$ & $	5.48 \cdot 10^{-1} 	$ & $	1.54 	$ & $	1.29 	$ & $	6.63 	$ & $	1.38 \cdot 10^{1} 	 $ \\
$	700	$ & $	6.40 \cdot 10^{-2} 	$ & $	1.28 \cdot 10^{-1} 	$ & $	2.56 \cdot 10^{-1} 	$ & $	5.76 \cdot 10^{-1} 	$ & $	1.61 	$ & $	1.36 	$ & $	6.96 	$ & $	1.45 \cdot 10^{1} 	 $ \\
$	710	$ & $	6.72 \cdot 10^{-2} 	$ & $	1.34 \cdot 10^{-1} 	$ & $	2.69 \cdot 10^{-1} 	$ & $	6.05 \cdot 10^{-1} 	$ & $	1.69 	$ & $	1.42 	$ & $	7.30 	$ & $	1.52 \cdot 10^{1} 	 $ \\
$	720	$ & $	7.05 \cdot 10^{-2} 	$ & $	1.41 \cdot 10^{-1} 	$ & $	2.82 \cdot 10^{-1} 	$ & $	6.35 \cdot 10^{-1} 	$ & $	1.77 	$ & $	1.49 	$ & $	7.66 	$ & $	1.60 \cdot 10^{1} 	 $ \\
$	730	$ & $	7.40 \cdot 10^{-2} 	$ & $	1.48 \cdot 10^{-1} 	$ & $	2.96 \cdot 10^{-1} 	$ & $	6.66 \cdot 10^{-1} 	$ & $	1.86 	$ & $	1.56 	$ & $	8.02 	$ & $	1.67 \cdot 10^{1} 	 $ \\
$	740	$ & $	7.75 \cdot 10^{-2} 	$ & $	1.55 \cdot 10^{-1} 	$ & $	3.10 \cdot 10^{-1} 	$ & $	6.98 \cdot 10^{-1} 	$ & $	1.95 	$ & $	1.64 	$ & $	8.41 	$ & $	1.75 \cdot 10^{1} 	 $ \\
$	750	$ & $	8.12 \cdot 10^{-2} 	$ & $	1.63 \cdot 10^{-1} 	$ & $	3.25 \cdot 10^{-1} 	$ & $	7.31 \cdot 10^{-1} 	$ & $	2.04 	$ & $	1.72 	$ & $	8.80 	$ & $	1.83 \cdot 10^{1} 	 $ \\
$	760	$ & $	8.51 \cdot 10^{-2} 	$ & $	1.70 \cdot 10^{-1} 	$ & $	3.40 \cdot 10^{-1} 	$ & $	7.66 \cdot 10^{-1} 	$ & $	2.13 	$ & $	1.79 	$ & $	9.21 	$ & $	1.92 \cdot 10^{1} 	 $ \\
$	770	$ & $	8.90 \cdot 10^{-2} 	$ & $	1.78 \cdot 10^{-1} 	$ & $	3.56 \cdot 10^{-1} 	$ & $	8.01 \cdot 10^{-1} 	$ & $	2.23 	$ & $	1.88 	$ & $	9.63 	$ & $	2.01 \cdot 10^{1} 	 $ \\
$	780	$ & $	9.31 \cdot 10^{-2} 	$ & $	1.86 \cdot 10^{-1} 	$ & $	3.73 \cdot 10^{-1} 	$ & $	8.38 \cdot 10^{-1} 	$ & $	2.33 	$ & $	1.96 	$ & $	1.01 \cdot 10^{1} 	$ & $	2.10 \cdot 10^{1} 	 $ \\
$	790	$ & $	9.74 \cdot 10^{-2} 	$ & $	1.95 \cdot 10^{-1} 	$ & $	3.89 \cdot 10^{-1} 	$ & $	8.76 \cdot 10^{-1} 	$ & $	2.44 	$ & $	2.05 	$ & $	1.05 \cdot 10^{1} 	$ & $	2.19 \cdot 10^{1} 	 $ \\
$	800	$ & $	1.02 \cdot 10^{-1} 	$ & $	2.04 \cdot 10^{-1} 	$ & $	4.07 \cdot 10^{-1} 	$ & $	9.16 \cdot 10^{-1} 	$ & $	2.55 	$ & $	2.14 	$ & $	1.10 \cdot 10^{1} 	$ & $	2.29 \cdot 10^{1} 	 $ \\
$	810	$ & $	1.06 \cdot 10^{-1} 	$ & $	2.13 \cdot 10^{-1} 	$ & $	4.25 \cdot 10^{-1} 	$ & $	9.57 \cdot 10^{-1} 	$ & $	2.66 	$ & $	2.24 	$ & $	1.15 \cdot 10^{1} 	$ & $	2.39 \cdot 10^{1} 	 $ \\
$	820	$ & $	1.11 \cdot 10^{-1} 	$ & $	2.22 \cdot 10^{-1} 	$ & $	4.44 \cdot 10^{-1} 	$ & $	9.99 \cdot 10^{-1} 	$ & $	2.77 	$ & $	2.33 	$ & $	1.20 \cdot 10^{1} 	$ & $	2.50 \cdot 10^{1} 	 $ \\
$	830	$ & $	1.16 \cdot 10^{-1} 	$ & $	2.32 \cdot 10^{-1} 	$ & $	4.63 \cdot 10^{-1} 	$ & $	1.04 	$ & $	2.90 	$ & $	2.43 	$ & $	1.25 \cdot 10^{1} 	$ & $	2.61 \cdot 10^{1} 	 $ \\
$	840	$ & $	1.21 \cdot 10^{-1} 	$ & $	2.42 \cdot 10^{-1} 	$ & $	4.83 \cdot 10^{-1} 	$ & $	1.09 	$ & $	3.02 	$ & $	2.54 	$ & $	1.30 \cdot 10^{1} 	$ & $	2.72 \cdot 10^{1} 	 $ \\
$	850	$ & $	1.26 \cdot 10^{-1} 	$ & $	2.52 \cdot 10^{-1} 	$ & $	5.04 \cdot 10^{-1} 	$ & $	1.13 	$ & $	3.15 	$ & $	2.65 	$ & $	1.36 \cdot 10^{1} 	$ & $	2.83 \cdot 10^{1} 	 $ \\
$	860	$ & $	1.31 \cdot 10^{-1} 	$ & $	2.63 \cdot 10^{-1} 	$ & $	5.26 \cdot 10^{-1} 	$ & $	1.18 	$ & $	3.28 	$ & $	2.76 	$ & $	1.42 \cdot 10^{1} 	$ & $	2.95 \cdot 10^{1} 	 $ \\
$	870	$ & $	1.37 \cdot 10^{-1} 	$ & $	2.74 \cdot 10^{-1} 	$ & $	5.48 \cdot 10^{-1} 	$ & $	1.23 	$ & $	3.42 	$ & $	2.87 	$ & $	1.48 \cdot 10^{1} 	$ & $	3.08 \cdot 10^{1} 	 $ \\
$	880	$ & $	1.43 \cdot 10^{-1} 	$ & $	2.85 \cdot 10^{-1} 	$ & $	5.71 \cdot 10^{-1} 	$ & $	1.28 	$ & $	3.56 	$ & $	2.99 	$ & $	1.54 \cdot 10^{1} 	$ & $	3.20 \cdot 10^{1} 	 $ \\
$	890	$ & $	1.49 \cdot 10^{-1} 	$ & $	2.97 \cdot 10^{-1} 	$ & $	5.94 \cdot 10^{-1} 	$ & $	1.34 	$ & $	3.71 	$ & $	3.12 	$ & $	1.60 \cdot 10^{1} 	$ & $	3.34 \cdot 10^{1} 	 $ \\
$	900	$ & $	1.55 \cdot 10^{-1} 	$ & $	3.09 \cdot 10^{-1} 	$ & $	6.19 \cdot 10^{-1} 	$ & $	1.39 	$ & $	3.86 	$ & $	3.24 	$ & $	1.67 \cdot 10^{1} 	$ & $	3.47 \cdot 10^{1} 	 $ \\
$	910	$ & $	1.61 \cdot 10^{-1} 	$ & $	3.22 \cdot 10^{-1} 	$ & $	6.44 \cdot 10^{-1} 	$ & $	1.45 	$ & $	4.02 	$ & $	3.37 	$ & $	1.73 \cdot 10^{1} 	$ & $	3.61 \cdot 10^{1} 	 $ \\
$	920	$ & $	1.68 \cdot 10^{-1} 	$ & $	3.35 \cdot 10^{-1} 	$ & $	6.71 \cdot 10^{-1} 	$ & $	1.51 	$ & $	4.18 	$ & $	3.51 	$ & $	1.80 \cdot 10^{1} 	$ & $	3.76 \cdot 10^{1} 	 $ \\
$	930	$ & $	1.74 \cdot 10^{-1} 	$ & $	3.49 \cdot 10^{-1} 	$ & $	6.98 \cdot 10^{-1} 	$ & $	1.57 	$ & $	4.35 	$ & $	3.65 	$ & $	1.88 \cdot 10^{1} 	$ & $	3.91 \cdot 10^{1} 	 $ \\
$	940	$ & $	1.81 \cdot 10^{-1} 	$ & $	3.63 \cdot 10^{-1} 	$ & $	7.26 \cdot 10^{-1} 	$ & $	1.63 	$ & $	4.52 	$ & $	3.80 	$ & $	1.95 \cdot 10^{1} 	$ & $	4.07 \cdot 10^{1} 	 $ \\
$	950	$ & $	1.89 \cdot 10^{-1} 	$ & $	3.77 \cdot 10^{-1} 	$ & $	7.54 \cdot 10^{-1} 	$ & $	1.70 	$ & $	4.70 	$ & $	3.95 	$ & $	2.03 \cdot 10^{1} 	$ & $	4.23 \cdot 10^{1} 	 $ \\
$	960	$ & $	1.96 \cdot 10^{-1} 	$ & $	3.92 \cdot 10^{-1} 	$ & $	7.84 \cdot 10^{-1} 	$ & $	1.76 	$ & $	4.88 	$ & $	4.10 	$ & $	2.11 \cdot 10^{1} 	$ & $	4.39 \cdot 10^{1} 	 $ \\
$	970	$ & $	2.04 \cdot 10^{-1} 	$ & $	4.08 \cdot 10^{-1} 	$ & $	8.15 \cdot 10^{-1} 	$ & $	1.83 	$ & $	5.08 	$ & $	4.26 	$ & $	2.19 \cdot 10^{1} 	$ & $	4.57 \cdot 10^{1} 	 $ \\
$	980	$ & $	2.12 \cdot 10^{-1} 	$ & $	4.24 \cdot 10^{-1} 	$ & $	8.47 \cdot 10^{-1} 	$ & $	1.91 	$ & $	5.27 	$ & $	4.43 	$ & $	2.28 \cdot 10^{1} 	$ & $	4.74 \cdot 10^{1} 	 $ \\
$	990	$ & $	2.20 \cdot 10^{-1} 	$ & $	4.40 \cdot 10^{-1} 	$ & $	8.80 \cdot 10^{-1} 	$ & $	1.98 	$ & $	5.48 	$ & $	4.60 	$ & $	2.37 \cdot 10^{1} 	$ & $	4.93 \cdot 10^{1} 	 $ \\
$	1000	$ & $	2.29 \cdot 10^{-1} 	$ & $	4.57 \cdot 10^{-1} 	$ & $	9.14 \cdot 10^{-1} 	$ & $	2.06 	$ & $	5.69 	$ & $	4.78 	$ & $	2.46 \cdot 10^{1} 	$ & $	5.12 \cdot 10^{1} 	 $ \\
        \hline
      \end{tabular}
  \end{table}
\end{landscape}

\begin{table}[h]
  \vspace{-\headsep}
  \caption{SM Higgs-boson partial widths [GeV] for $4$-fermion final states for the low- and intermediate-mass range.
    We list results for the specific final states for $2$ charged leptons plus $2$ quarks,
      $\Plp\PGnl\PQq\PAQq^{\prime}$ (not including charge conjugate state),
        $2$ neutrinos plus $2$ quarks, $4$ quarks, as well as the result for
        arbitrary $4$ fermions, where $\PQq = \PQu\PQd\PQs\PQc\PQb$ and
        $\PGnl$ represents any type of neutrinos.}
    \label{tab:PWidth-lm2}
  \centering
  \small
  \begin{tabular}{lcccccc}
    \hline
    $\MH$ [GeV] &
    $\PH \rightarrow \Plp\Plm\PQq\PAQq$ &
    $\PH \rightarrow \Plp\Plm\PQq\PAQq$ &
    $\PH \rightarrow \Plp\PGnl\PQq\PAQq^{\prime}$ &
    $\PH \rightarrow \PGnl\PAGnl\PQq\PAQq$ &
    $\PH \rightarrow \PQq\PQq\PQq\PQq$ &
    $\PH \rightarrow \Pf\Pf\Pf\Pf$ \\
    & $(\Pl=\Pe$ or $\PGm)$
    & $(\Pl=\Pe, \PGm$ or $\PGt)$
    & $(\Pl=\Pe$ or $\PGm)$ & & & \\
    \hline
$	90	$ & $	8.70 \cdot 10^{-8} 	$ & $	1.31 \cdot 10^{-7} 	$ & $	6.74 \cdot 10^{-7} 	$ & $	2.62 \cdot 10^{-7} 	$ & $	2.34 \cdot 10^{-6} 	$ & $	5.28 \cdot 10^{-6} 	 $ \\
$	95	$ & $	1.46 \cdot 10^{-7} 	$ & $	2.19 \cdot 10^{-7} 	$ & $	1.61 \cdot 10^{-6} 	$ & $	4.39 \cdot 10^{-7} 	$ & $	5.43 \cdot 10^{-6} 	$ & $	1.21 \cdot 10^{-5} 	 $ \\
$	100	$ & $	2.59 \cdot 10^{-7} 	$ & $	3.89 \cdot 10^{-7} 	$ & $	3.99 \cdot 10^{-6} 	$ & $	7.80 \cdot 10^{-7} 	$ & $	1.33 \cdot 10^{-5} 	$ & $	2.94 \cdot 10^{-5} 	 $ \\
$	105	$ & $	5.23 \cdot 10^{-7} 	$ & $	7.85 \cdot 10^{-7} 	$ & $	9.31 \cdot 10^{-6} 	$ & $	1.57 \cdot 10^{-6} 	$ & $	3.09 \cdot 10^{-5} 	$ & $	6.82 \cdot 10^{-5} 	 $ \\
$	110	$ & $	1.15 \cdot 10^{-6} 	$ & $	1.72 \cdot 10^{-6} 	$ & $	1.99 \cdot 10^{-5} 	$ & $	3.45 \cdot 10^{-6} 	$ & $	6.66 \cdot 10^{-5} 	$ & $	1.47 \cdot 10^{-4} 	 $ \\
$	115	$ & $	2.51 \cdot 10^{-6} 	$ & $	3.77 \cdot 10^{-6} 	$ & $	3.92 \cdot 10^{-5} 	$ & $	7.53 \cdot 10^{-6} 	$ & $	1.33 \cdot 10^{-4} 	$ & $	2.92 \cdot 10^{-4} 	 $ \\
$	120	$ & $	5.19 \cdot 10^{-6} 	$ & $	7.79 \cdot 10^{-6} 	$ & $	7.25 \cdot 10^{-5} 	$ & $	1.56 \cdot 10^{-5} 	$ & $	2.50 \cdot 10^{-4} 	$ & $	5.47 \cdot 10^{-4} 	 $ \\
$	125	$ & $	1.00 \cdot 10^{-5} 	$ & $	1.51 \cdot 10^{-5} 	$ & $	1.28 \cdot 10^{-4} 	$ & $	3.00 \cdot 10^{-5} 	$ & $	4.46 \cdot 10^{-4} 	$ & $	9.75 \cdot 10^{-4} 	 $ \\
$	130	$ & $	1.83 \cdot 10^{-5} 	$ & $	2.74 \cdot 10^{-5} 	$ & $	2.17 \cdot 10^{-4} 	$ & $	5.46 \cdot 10^{-5} 	$ & $	7.65 \cdot 10^{-4} 	$ & $	1.67 \cdot 10^{-3} 	 $ \\
$	135	$ & $	3.16 \cdot 10^{-5} 	$ & $	4.74 \cdot 10^{-5} 	$ & $	3.62 \cdot 10^{-4} 	$ & $	9.45 \cdot 10^{-5} 	$ & $	1.28 \cdot 10^{-3} 	$ & $	2.80 \cdot 10^{-3} 	 $ \\
$	140	$ & $	5.26 \cdot 10^{-5} 	$ & $	7.89 \cdot 10^{-5} 	$ & $	6.00 \cdot 10^{-4} 	$ & $	1.57 \cdot 10^{-4} 	$ & $	2.13 \cdot 10^{-3} 	$ & $	4.64 \cdot 10^{-3} 	 $ \\
$	145	$ & $	8.49 \cdot 10^{-5} 	$ & $	1.27 \cdot 10^{-4} 	$ & $	1.01 \cdot 10^{-3} 	$ & $	2.53 \cdot 10^{-4} 	$ & $	3.55 \cdot 10^{-3} 	$ & $	7.75 \cdot 10^{-3} 	 $ \\
$	150	$ & $	1.34 \cdot 10^{-4} 	$ & $	2.01 \cdot 10^{-4} 	$ & $	1.77 \cdot 10^{-3} 	$ & $	3.99 \cdot 10^{-4} 	$ & $	6.16 \cdot 10^{-3} 	$ & $	1.35 \cdot 10^{-2} 	 $ \\
$	155	$ & $	2.08 \cdot 10^{-4} 	$ & $	3.13 \cdot 10^{-4} 	$ & $	3.52 \cdot 10^{-3} 	$ & $	6.22 \cdot 10^{-4} 	$ & $	1.20 \cdot 10^{-2} 	$ & $	2.63 \cdot 10^{-2} 	 $ \\
$	160	$ & $	3.23 \cdot 10^{-4} 	$ & $	4.84 \cdot 10^{-4} 	$ & $	1.10 \cdot 10^{-2} 	$ & $	9.63 \cdot 10^{-4} 	$ & $	3.59 \cdot 10^{-2} 	$ & $	7.87 \cdot 10^{-2} 	 $ \\
$	165	$ & $	5.12 \cdot 10^{-4} 	$ & $	7.68 \cdot 10^{-4} 	$ & $	3.46 \cdot 10^{-2} 	$ & $	1.53 \cdot 10^{-3} 	$ & $	1.10 \cdot 10^{-1} 	$ & $	2.42 \cdot 10^{-1} 	 $ \\
$	170	$ & $	8.42 \cdot 10^{-4} 	$ & $	1.26 \cdot 10^{-3} 	$ & $	5.37 \cdot 10^{-2} 	$ & $	2.51 \cdot 10^{-3} 	$ & $	1.71 \cdot 10^{-1} 	$ & $	3.75 \cdot 10^{-1} 	 $ \\
$	175	$ & $	1.52 \cdot 10^{-3} 	$ & $	2.27 \cdot 10^{-3} 	$ & $	7.03 \cdot 10^{-2} 	$ & $	4.53 \cdot 10^{-3} 	$ & $	2.26 \cdot 10^{-1} 	$ & $	4.96 \cdot 10^{-1} 	 $ \\
$	180	$ & $	3.56 \cdot 10^{-3} 	$ & $	5.34 \cdot 10^{-3} 	$ & $	8.61 \cdot 10^{-2} 	$ & $	1.06 \cdot 10^{-2} 	$ & $	2.86 \cdot 10^{-1} 	$ & $	6.26 \cdot 10^{-1} 	 $ \\
$	185	$ & $	1.17 \cdot 10^{-2} 	$ & $	1.76 \cdot 10^{-2} 	$ & $	1.03 \cdot 10^{-1} 	$ & $	3.50 \cdot 10^{-2} 	$ & $	3.80 \cdot 10^{-1} 	$ & $	8.27 \cdot 10^{-1} 	 $ \\
$	190	$ & $	2.04 \cdot 10^{-2} 	$ & $	3.07 \cdot 10^{-2} 	$ & $	1.20 \cdot 10^{-1} 	$ & $	6.11 \cdot 10^{-2} 	$ & $	4.78 \cdot 10^{-1} 	$ & $	1.03 	 $ \\
$	195	$ & $	2.77 \cdot 10^{-2} 	$ & $	4.15 \cdot 10^{-2} 	$ & $	1.37 \cdot 10^{-1} 	$ & $	8.27 \cdot 10^{-2} 	$ & $	5.69 \cdot 10^{-1} 	$ & $	1.23 	 $ \\
$	200	$ & $	3.44 \cdot 10^{-2} 	$ & $	5.15 \cdot 10^{-2} 	$ & $	1.55 \cdot 10^{-1} 	$ & $	1.03 \cdot 10^{-1} 	$ & $	6.61 \cdot 10^{-1} 	$ & $	1.43 	 $ \\
$	210	$ & $	4.75 \cdot 10^{-2} 	$ & $	7.12 \cdot 10^{-2} 	$ & $	1.96 \cdot 10^{-1} 	$ & $	1.42 \cdot 10^{-1} 	$ & $	8.54 \cdot 10^{-1} 	$ & $	1.84 	 $ \\
$	220	$ & $	6.13 \cdot 10^{-2} 	$ & $	9.20 \cdot 10^{-2} 	$ & $	2.41 \cdot 10^{-1} 	$ & $	1.83 \cdot 10^{-1} 	$ & $	1.07 	$ & $	2.30 	 $ \\
$	230	$ & $	7.66 \cdot 10^{-2} 	$ & $	1.15 \cdot 10^{-1} 	$ & $	2.92 \cdot 10^{-1} 	$ & $	2.29 \cdot 10^{-1} 	$ & $	1.31 	$ & $	2.81 	 $ \\
$	240	$ & $	9.35 \cdot 10^{-2} 	$ & $	1.40 \cdot 10^{-1} 	$ & $	3.50 \cdot 10^{-1} 	$ & $	2.79 \cdot 10^{-1} 	$ & $	1.57 	$ & $	3.39 	 $ \\
$	250	$ & $	1.12 \cdot 10^{-1} 	$ & $	1.69 \cdot 10^{-1} 	$ & $	4.15 \cdot 10^{-1} 	$ & $	3.36 \cdot 10^{-1} 	$ & $	1.87 	$ & $	4.03 	 $ \\
$	260	$ & $	1.34 \cdot 10^{-1} 	$ & $	2.00 \cdot 10^{-1} 	$ & $	4.87 \cdot 10^{-1} 	$ & $	3.99 \cdot 10^{-1} 	$ & $	2.21 	$ & $	4.75 	 $ \\
$	270	$ & $	1.57 \cdot 10^{-1} 	$ & $	2.36 \cdot 10^{-1} 	$ & $	5.66 \cdot 10^{-1} 	$ & $	4.69 \cdot 10^{-1} 	$ & $	2.58 	$ & $	5.54 	 $ \\
$	280	$ & $	1.83 \cdot 10^{-1} 	$ & $	2.75 \cdot 10^{-1} 	$ & $	6.54 \cdot 10^{-1} 	$ & $	5.47 \cdot 10^{-1} 	$ & $	2.98 	$ & $	6.42 	 $ \\
$	290	$ & $	2.12 \cdot 10^{-1} 	$ & $	3.17 \cdot 10^{-1} 	$ & $	7.50 \cdot 10^{-1} 	$ & $	6.32 \cdot 10^{-1} 	$ & $	3.43 	$ & $	7.37 	 $ \\
$	300	$ & $	2.43 \cdot 10^{-1} 	$ & $	3.64 \cdot 10^{-1} 	$ & $	8.54 \cdot 10^{-1} 	$ & $	7.25 \cdot 10^{-1} 	$ & $	3.91 	$ & $	8.42 	 $ \\
$	310	$ & $	2.76 \cdot 10^{-1} 	$ & $	4.15 \cdot 10^{-1} 	$ & $	9.67 \cdot 10^{-1} 	$ & $	8.26 \cdot 10^{-1} 	$ & $	4.44 	$ & $	9.55 	 $ \\
$	320	$ & $	3.13 \cdot 10^{-1} 	$ & $	4.70 \cdot 10^{-1} 	$ & $	1.09 	$ & $	9.36 \cdot 10^{-1} 	$ & $	5.01 	$ & $	1.08 \cdot 10^{1} 	 $ \\
$	330	$ & $	3.52 \cdot 10^{-1} 	$ & $	5.28 \cdot 10^{-1} 	$ & $	1.22 	$ & $	1.05 	$ & $	5.62 	$ & $	1.21 \cdot 10^{1} 	 $ \\
$	340	$ & $	3.93 \cdot 10^{-1} 	$ & $	5.90 \cdot 10^{-1} 	$ & $	1.36 	$ & $	1.18 	$ & $	6.26 	$ & $	1.35 \cdot 10^{1} 	 $ \\
$	350	$ & $	4.37 \cdot 10^{-1} 	$ & $	6.56 \cdot 10^{-1} 	$ & $	1.50 	$ & $	1.31 	$ & $	6.94 	$ & $	1.49 \cdot 10^{1} 	 $ \\
$	360	$ & $	4.89 \cdot 10^{-1} 	$ & $	7.34 \cdot 10^{-1} 	$ & $	1.68 	$ & $	1.46 	$ & $	7.75 	$ & $	1.67 \cdot 10^{1} 	 $ \\
$	370	$ & $	5.45 \cdot 10^{-1} 	$ & $	8.18 \cdot 10^{-1} 	$ & $	1.86 	$ & $	1.63 	$ & $	8.62 	$ & $	1.85 \cdot 10^{1} 	 $ \\
$	380	$ & $	6.05 \cdot 10^{-1} 	$ & $	9.08 \cdot 10^{-1} 	$ & $	2.06 	$ & $	1.81 	$ & $	9.54 	$ & $	2.05 \cdot 10^{1} 	 $ \\
$	390	$ & $	6.69 \cdot 10^{-1} 	$ & $	1.00 	$ & $	2.27 	$ & $	2.00 	$ & $	1.05 \cdot 10^{1} 	$ & $	2.26 \cdot 10^{1} 	 $ \\
$	400	$ & $	7.37 \cdot 10^{-1} 	$ & $	1.11 	$ & $	2.49 	$ & $	2.20 	$ & $	1.16 \cdot 10^{1} 	$ & $	2.48 \cdot 10^{1} 	 $ \\
$	410	$ & $	8.09 \cdot 10^{-1} 	$ & $	1.21 	$ & $	2.72 	$ & $	2.42 	$ & $	1.27 \cdot 10^{1} 	$ & $	2.72 \cdot 10^{1} 	 $ \\
$	420	$ & $	8.85 \cdot 10^{-1} 	$ & $	1.33 	$ & $	2.97 	$ & $	2.64 	$ & $	1.38 \cdot 10^{1} 	$ & $	2.97 \cdot 10^{1} 	 $ \\
$	430	$ & $	9.66 \cdot 10^{-1} 	$ & $	1.45 	$ & $	3.22 	$ & $	2.88 	$ & $	1.50 \cdot 10^{1} 	$ & $	3.23 \cdot 10^{1} 	 $ \\
$	440	$ & $	1.05 	$ & $	1.58 	$ & $	3.50 	$ & $	3.14 	$ & $	1.63 \cdot 10^{1} 	$ & $	3.51 \cdot 10^{1} 	 $ \\
$	450	$ & $	1.14 	$ & $	1.71 	$ & $	3.78 	$ & $	3.41 	$ & $	1.77 \cdot 10^{1} 	$ & $	3.80 \cdot 10^{1} 	 $ \\
$	460	$ & $	1.24 	$ & $	1.85 	$ & $	4.08 	$ & $	3.69 	$ & $	1.91 \cdot 10^{1} 	$ & $	4.10 \cdot 10^{1} 	 $ \\
$	470	$ & $	1.33 	$ & $	2.00 	$ & $	4.40 	$ & $	3.99 	$ & $	2.06 \cdot 10^{1} 	$ & $	4.43 \cdot 10^{1} 	 $ \\
$	480	$ & $	1.44 	$ & $	2.16 	$ & $	4.73 	$ & $	4.30 	$ & $	2.22 \cdot 10^{1} 	$ & $	4.76 \cdot 10^{1} 	 $ \\
$	490	$ & $	1.55 	$ & $	2.32 	$ & $	5.08 	$ & $	4.63 	$ & $	2.38 \cdot 10^{1} 	$ & $	5.12 \cdot 10^{1} 	 $ \\
    \hline
  \end{tabular}
\end{table}

\begin{table}[h]
  \vspace{-\headsep}
  \caption{SM Higgs-boson partial widths [GeV] for $4$-fermion final states for the hig-mass range.
    We list results for the specific finalstates for $2$ charged leptons plus $2$ quarks,
      $\Plp\PGnl\PQq\PAQq^{\prime}$ (not including charge conjugate state),
        $2$ neutrinos plus $2$ quarks, $4$ quarks, as well as the result for
        arbitrary $4$ fermions, where $\PQq = \PQu\PQd\PQs\PQc\PQb$ and
        $\PGnl$ represents any type of neutrinos.}
    \label{tab:PWidth-hm2}
  \centering
  \small
  \begin{tabular}{lcccccc}
    \hline
    $\MH$ [GeV] &
    $\PH \rightarrow \Plp\Plm\PQq\PAQq$ &
    $\PH \rightarrow \Plp\Plm\PQq\PAQq$ &
    $\PH \rightarrow \Plp\PGnl\PQq\PAQq^{\prime}$ &
    $\PH \rightarrow \PGnl\PAGnl\PQq\PAQq$ &
    $\PH \rightarrow \PQq\PQq\PQq\PQq$ &
    $\PH \rightarrow \Pf\Pf\Pf\Pf$ \\
    & $(\Pl=\Pe$ or $\PGm)$
    & $(\Pl=\Pe, \PGm$ or $\PGt)$
    & $(\Pl=\Pe$ or $\PGm)$ & & & \\
    \hline
$	500	$ & $	1.66 	$ & $	2.50 	$ & $	5.44 	$ & $	4.97 	$ & $	2.55 \cdot 10^{1} 	$ & $	5.48 \cdot 10^{1} 	 $ \\
$	510	$ & $	1.78 	$ & $	2.68 	$ & $	5.82 	$ & $	5.33 	$ & $	2.73 \cdot 10^{1} 	$ & $	5.87 \cdot 10^{1} 	 $ \\
$	520	$ & $	1.91 	$ & $	2.87 	$ & $	6.21 	$ & $	5.70 	$ & $	2.92 \cdot 10^{1} 	$ & $	6.27 \cdot 10^{1} 	 $ \\
$	530	$ & $	2.04 	$ & $	3.06 	$ & $	6.62 	$ & $	6.10 	$ & $	3.12 \cdot 10^{1} 	$ & $	6.70 \cdot 10^{1} 	 $ \\
$	540	$ & $	2.18 	$ & $	3.27 	$ & $	7.05 	$ & $	6.51 	$ & $	3.32 \cdot 10^{1} 	$ & $	7.14 \cdot 10^{1} 	 $ \\
$	550	$ & $	2.32 	$ & $	3.49 	$ & $	7.50 	$ & $	6.94 	$ & $	3.53 \cdot 10^{1} 	$ & $	7.60 \cdot 10^{1} 	 $ \\
$	560	$ & $	2.47 	$ & $	3.71 	$ & $	7.97 	$ & $	7.38 	$ & $	3.76 \cdot 10^{1} 	$ & $	8.07 \cdot 10^{1} 	 $ \\
$	570	$ & $	2.63 	$ & $	3.94 	$ & $	8.46 	$ & $	7.85 	$ & $	3.99 \cdot 10^{1} 	$ & $	8.57 \cdot 10^{1} 	 $ \\
$	580	$ & $	2.79 	$ & $	4.19 	$ & $	8.97 	$ & $	8.33 	$ & $	4.23 \cdot 10^{1} 	$ & $	9.09 \cdot 10^{1} 	 $ \\
$	590	$ & $	2.96 	$ & $	4.44 	$ & $	9.49 	$ & $	8.84 	$ & $	4.48 \cdot 10^{1} 	$ & $	9.63 \cdot 10^{1} 	 $ \\
$	600	$ & $	3.14 	$ & $	4.71 	$ & $	1.00 \cdot 10^{1} 	$ & $	9.37 	$ & $	4.74 \cdot 10^{1} 	$ & $	1.02 \cdot 10^{2} 	 $ \\
$	610	$ & $	3.32 	$ & $	4.98 	$ & $	1.06 \cdot 10^{1} 	$ & $	9.91 	$ & $	5.02 \cdot 10^{1} 	$ & $	1.08 \cdot 10^{2} 	 $ \\
$	620	$ & $	3.51 	$ & $	5.27 	$ & $	1.12 \cdot 10^{1} 	$ & $	1.05 \cdot 10^{1} 	$ & $	5.30 \cdot 10^{1} 	$ & $	1.14 \cdot 10^{2} 	 $ \\
$	630	$ & $	3.71 	$ & $	5.56 	$ & $	1.18 \cdot 10^{1} 	$ & $	1.11 \cdot 10^{1} 	$ & $	5.59 \cdot 10^{1} 	$ & $	1.20 \cdot 10^{2} 	 $ \\
$	640	$ & $	3.91 	$ & $	5.87 	$ & $	1.25 \cdot 10^{1} 	$ & $	1.17 \cdot 10^{1} 	$ & $	5.89 \cdot 10^{1} 	$ & $	1.27 \cdot 10^{2} 	 $ \\
$	650	$ & $	4.13 	$ & $	6.19 	$ & $	1.31 \cdot 10^{1} 	$ & $	1.23 \cdot 10^{1} 	$ & $	6.21 \cdot 10^{1} 	$ & $	1.34 \cdot 10^{2} 	 $ \\
$	660	$ & $	4.35 	$ & $	6.53 	$ & $	1.38 \cdot 10^{1} 	$ & $	1.30 \cdot 10^{1} 	$ & $	6.54 \cdot 10^{1} 	$ & $	1.41 \cdot 10^{2} 	 $ \\
$	670	$ & $	4.58 	$ & $	6.87 	$ & $	1.45 \cdot 10^{1} 	$ & $	1.37 \cdot 10^{1} 	$ & $	6.88 \cdot 10^{1} 	$ & $	1.48 \cdot 10^{2} 	 $ \\
$	680	$ & $	4.82 	$ & $	7.23 	$ & $	1.53 \cdot 10^{1} 	$ & $	1.44 \cdot 10^{1} 	$ & $	7.23 \cdot 10^{1} 	$ & $	1.55 \cdot 10^{2} 	 $ \\
$	690	$ & $	5.06 	$ & $	7.60 	$ & $	1.60 \cdot 10^{1} 	$ & $	1.51 \cdot 10^{1} 	$ & $	7.60 \cdot 10^{1} 	$ & $	1.63 \cdot 10^{2} 	 $ \\
$	700	$ & $	5.32 	$ & $	7.99 	$ & $	1.68 \cdot 10^{1} 	$ & $	1.59 \cdot 10^{1} 	$ & $	7.98 \cdot 10^{1} 	$ & $	1.72 \cdot 10^{2} 	 $ \\
$	710	$ & $	5.59 	$ & $	8.38 	$ & $	1.77 \cdot 10^{1} 	$ & $	1.67 \cdot 10^{1} 	$ & $	8.37 \cdot 10^{1} 	$ & $	1.80 \cdot 10^{2} 	 $ \\
$	720	$ & $	5.87 	$ & $	8.80 	$ & $	1.85 \cdot 10^{1} 	$ & $	1.75 \cdot 10^{1} 	$ & $	8.78 \cdot 10^{1} 	$ & $	1.89 \cdot 10^{2} 	 $ \\
$	730	$ & $	6.15 	$ & $	9.23 	$ & $	1.94 \cdot 10^{1} 	$ & $	1.84 \cdot 10^{1} 	$ & $	9.20 \cdot 10^{1} 	$ & $	1.98 \cdot 10^{2} 	 $ \\
$	740	$ & $	6.45 	$ & $	9.67 	$ & $	2.03 \cdot 10^{1} 	$ & $	1.92 \cdot 10^{1} 	$ & $	9.64 \cdot 10^{1} 	$ & $	2.07 \cdot 10^{2} 	 $ \\
$	750	$ & $	6.75 	$ & $	1.01 \cdot 10^{1} 	$ & $	2.13 \cdot 10^{1} 	$ & $	2.01 \cdot 10^{1} 	$ & $	1.01 \cdot 10^{2} 	$ & $	2.17 \cdot 10^{2} 	 $ \\
$	760	$ & $	7.07 	$ & $	1.06 \cdot 10^{1} 	$ & $	2.22 \cdot 10^{1} 	$ & $	2.11 \cdot 10^{1} 	$ & $	1.06 \cdot 10^{2} 	$ & $	2.27 \cdot 10^{2} 	 $ \\
$	770	$ & $	7.40 	$ & $	1.11 \cdot 10^{1} 	$ & $	2.33 \cdot 10^{1} 	$ & $	2.21 \cdot 10^{1} 	$ & $	1.10 \cdot 10^{2} 	$ & $	2.37 \cdot 10^{2} 	 $ \\
$	780	$ & $	7.74 	$ & $	1.16 \cdot 10^{1} 	$ & $	2.43 \cdot 10^{1} 	$ & $	2.31 \cdot 10^{1} 	$ & $	1.15 \cdot 10^{2} 	$ & $	2.48 \cdot 10^{2} 	 $ \\
$	790	$ & $	8.09 	$ & $	1.21 \cdot 10^{1} 	$ & $	2.54 \cdot 10^{1} 	$ & $	2.41 \cdot 10^{1} 	$ & $	1.21 \cdot 10^{2} 	$ & $	2.59 \cdot 10^{2} 	 $ \\
$	800	$ & $	8.45 	$ & $	1.27 \cdot 10^{1} 	$ & $	2.65 \cdot 10^{1} 	$ & $	2.52 \cdot 10^{1} 	$ & $	1.26 \cdot 10^{2} 	$ & $	2.71 \cdot 10^{2} 	 $ \\
$	810	$ & $	8.83 	$ & $	1.32 \cdot 10^{1} 	$ & $	2.77 \cdot 10^{1} 	$ & $	2.63 \cdot 10^{1} 	$ & $	1.31 \cdot 10^{2} 	$ & $	2.83 \cdot 10^{2} 	 $ \\
$	820	$ & $	9.22 	$ & $	1.38 \cdot 10^{1} 	$ & $	2.89 \cdot 10^{1} 	$ & $	2.75 \cdot 10^{1} 	$ & $	1.37 \cdot 10^{2} 	$ & $	2.95 \cdot 10^{2} 	 $ \\
$	830	$ & $	9.62 	$ & $	1.44 \cdot 10^{1} 	$ & $	3.01 \cdot 10^{1} 	$ & $	2.87 \cdot 10^{1} 	$ & $	1.43 \cdot 10^{2} 	$ & $	3.08 \cdot 10^{2} 	 $ \\
$	840	$ & $	1.00 \cdot 10^{1} 	$ & $	1.51 \cdot 10^{1} 	$ & $	3.14 \cdot 10^{1} 	$ & $	2.99 \cdot 10^{1} 	$ & $	1.49 \cdot 10^{2} 	$ & $	3.21 \cdot 10^{2} 	 $ \\
$	850	$ & $	1.05 \cdot 10^{1} 	$ & $	1.57 \cdot 10^{1} 	$ & $	3.28 \cdot 10^{1} 	$ & $	3.12 \cdot 10^{1} 	$ & $	1.56 \cdot 10^{2} 	$ & $	3.35 \cdot 10^{2} 	 $ \\
$	860	$ & $	1.09 \cdot 10^{1} 	$ & $	1.64 \cdot 10^{1} 	$ & $	3.41 \cdot 10^{1} 	$ & $	3.25 \cdot 10^{1} 	$ & $	1.62 \cdot 10^{2} 	$ & $	3.49 \cdot 10^{2} 	 $ \\
$	870	$ & $	1.14 \cdot 10^{1} 	$ & $	1.71 \cdot 10^{1} 	$ & $	3.56 \cdot 10^{1} 	$ & $	3.39 \cdot 10^{1} 	$ & $	1.69 \cdot 10^{2} 	$ & $	3.63 \cdot 10^{2} 	 $ \\
$	880	$ & $	1.18 \cdot 10^{1} 	$ & $	1.78 \cdot 10^{1} 	$ & $	3.70 \cdot 10^{1} 	$ & $	3.53 \cdot 10^{1} 	$ & $	1.76 \cdot 10^{2} 	$ & $	3.78 \cdot 10^{2} 	 $ \\
$	890	$ & $	1.23 \cdot 10^{1} 	$ & $	1.85 \cdot 10^{1} 	$ & $	3.85 \cdot 10^{1} 	$ & $	3.68 \cdot 10^{1} 	$ & $	1.83 \cdot 10^{2} 	$ & $	3.94 \cdot 10^{2} 	 $ \\
$	900	$ & $	1.28 \cdot 10^{1} 	$ & $	1.93 \cdot 10^{1} 	$ & $	4.01 \cdot 10^{1} 	$ & $	3.83 \cdot 10^{1} 	$ & $	1.90 \cdot 10^{2} 	$ & $	4.10 \cdot 10^{2} 	 $ \\
$	910	$ & $	1.34 \cdot 10^{1} 	$ & $	2.00 \cdot 10^{1} 	$ & $	4.17 \cdot 10^{1} 	$ & $	3.99 \cdot 10^{1} 	$ & $	1.98 \cdot 10^{2} 	$ & $	4.27 \cdot 10^{2} 	 $ \\
$	920	$ & $	1.39 \cdot 10^{1} 	$ & $	2.09 \cdot 10^{1} 	$ & $	4.34 \cdot 10^{1} 	$ & $	4.15 \cdot 10^{1} 	$ & $	2.06 \cdot 10^{2} 	$ & $	4.44 \cdot 10^{2} 	 $ \\
$	930	$ & $	1.45 \cdot 10^{1} 	$ & $	2.17 \cdot 10^{1} 	$ & $	4.51 \cdot 10^{1} 	$ & $	4.31 \cdot 10^{1} 	$ & $	2.14 \cdot 10^{2} 	$ & $	4.62 \cdot 10^{2} 	 $ \\
$	940	$ & $	1.50 \cdot 10^{1} 	$ & $	2.26 \cdot 10^{1} 	$ & $	4.69 \cdot 10^{1} 	$ & $	4.49 \cdot 10^{1} 	$ & $	2.23 \cdot 10^{2} 	$ & $	4.80 \cdot 10^{2} 	 $ \\
$	950	$ & $	1.56 \cdot 10^{1} 	$ & $	2.35 \cdot 10^{1} 	$ & $	4.88 \cdot 10^{1} 	$ & $	4.66 \cdot 10^{1} 	$ & $	2.32 \cdot 10^{2} 	$ & $	4.99 \cdot 10^{2} 	 $ \\
$	960	$ & $	1.63 \cdot 10^{1} 	$ & $	2.44 \cdot 10^{1} 	$ & $	5.07 \cdot 10^{1} 	$ & $	4.85 \cdot 10^{1} 	$ & $	2.41 \cdot 10^{2} 	$ & $	5.18 \cdot 10^{2} 	 $ \\
$	970	$ & $	1.69 \cdot 10^{1} 	$ & $	2.53 \cdot 10^{1} 	$ & $	5.27 \cdot 10^{1} 	$ & $	5.04 \cdot 10^{1} 	$ & $	2.50 \cdot 10^{2} 	$ & $	5.39 \cdot 10^{2} 	 $ \\
$	980	$ & $	1.76 \cdot 10^{1} 	$ & $	2.63 \cdot 10^{1} 	$ & $	5.47 \cdot 10^{1} 	$ & $	5.23 \cdot 10^{1} 	$ & $	2.60 \cdot 10^{2} 	$ & $	5.59 \cdot 10^{2} 	 $ \\
$	990	$ & $	1.82 \cdot 10^{1} 	$ & $	2.74 \cdot 10^{1} 	$ & $	5.68 \cdot 10^{1} 	$ & $	5.44 \cdot 10^{1} 	$ & $	2.70 \cdot 10^{2} 	$ & $	5.81 \cdot 10^{2} 	 $ \\
$	1000	$ & $	1.89 \cdot 10^{1} 	$ & $	2.84 \cdot 10^{1} 	$ & $	5.90 \cdot 10^{1} 	$ & $	5.64 \cdot 10^{1} 	$ & $	2.80 \cdot 10^{2} 	$ & $	6.03 \cdot 10^{2} 	 $ \\
    \hline
  \end{tabular}
\end{table}



\end{document}